\documentclass[11pt]{article}
\usepackage{makeidx}
\usepackage{amsfonts}
\usepackage{amssymb}
\usepackage{graphicx}
\usepackage{amsmath}
\usepackage{geometry}
\usepackage{appendix}

\newtheorem{theorem}{Theorem}

\newtheorem{case}[theorem]{Case}

\newcommand{\func}[1]{\operatorname{#1}}

\DeclareMathSizes{10.95}{10}{7}{6}
\input{tcilatex}
\begin{document}

\title{A Statistical Field Perspective on Capital Allocation and Accumulation%
}
\author{Pierre Gosselin\thanks{%
Pierre Gosselin : Institut Fourier, UMR 5582 CNRS-UGA, Universit\'{e}
Grenoble Alpes, BP 74, 38402 St Martin d'H\`{e}res, France.\ E-Mail:
Pierre.Gosselin@univ-grenoble-alpes.fr} \and A\"{\i}leen Lotz\thanks{%
A\"{\i}leen Lotz: Cerca Trova, BP 114, 38001 Grenoble Cedex 1, France.\
E-mail: a.lotz@cercatrova.eu}}
\date{November 2023}
\maketitle

\begin{abstract}
This paper provides a general method to translate a standard economic model
with a large number of agents into a field-formalism model. This formalism
preserves the system's interactions and microeconomic features at the
individual level but reveals the emergence of collective states.

We apply this method to a simple microeconomic framework of investors and
firms. Both macro and micro aspects of the formalism are studied.

At the macro-scale, the field formalism shows that, in each sector, three
patterns of capital accumulation may emerge.\ A distribution of patterns
across sectors constitute a collective state. Any change in external
parameters or expectations in one sector will affect neighbouring sectors,
inducing transitions between collective states and generating permanent
fluctuations in patterns and flows of capital. Although changes in
expectations can cause abrupt changes in collective states, transitions may
be slow to occur. Due to its relative inertia, the real economy is bound to
be more affected by these constant variations than the financial markets.

At the micro-scale we compute the transition functions of individual agents
and study their probabilistic dynamics in a given collective state, as a
function of their initial state. We show that capital accumulation of an
individual agent depends on various factors. The probability associated with
each firm's trajectories is the result of several contradictory effects: the
firm tends to shift towards sectors with the greatest long-term return, but
must take into account the impact of its shift on its attractiveness for
investors throughout its trajectory. Since this trajectory depends largely
on the average capital of transition sectors, a firm's attractiveness during
its relocation depends on the relative level of capital in those sectors.
Moreover, the firm must also consider the effects of competition in the
intermediate sectors that tends to oust under-capitalized firm towards
sectors with lower average capital. For investors, capital allocation
depends on their short and long-term returns and investors will tend to
reallocate their capital to maximize both. The higher their level of
capital, the stronger the re-allocation will be.


\end{abstract}

\section{Introduction}

In large sets of agents, the dynamics of one agent never occurs in a vacuum,
but is impacted by the whole set of other agents' trajectories. In such
groups, the representative agent is a fiction: the set of trajectories gives
rise to collective states that will in turn condition individual dynamics.
These collective states can be studied analytically at a higher level by
using a statistical-field formalism (Gosselin, Lotz, Wambst 2020, 2021).

We have shown in previous papers how such a field-formalism allows to derive
the possible collective states of a given economic model while keeping track
of distributions and interactions among agents. It acts as a viability test
for a large range of classical model by revealing the inherent logic of each
system, which a standard economic model with representative agents cannot
do. Whereas a standard economic model determines the behavior of optimizing
agents, a field model will reveal how such a behavior, once generalized to a
large number of heterogeneous agents, would imply for the entire society.

The present paper develops and applies this method to a system composed of a
large number of firms and investors spread across numerous sectors and
studies the interactions between financial and physical capital as well as
the determinants of capital allocation among firms. Because it keeps track
of interactions at the macro level, the field formalism reveals what would
otherwise remain hidden: capital accumulation is a global mechanism, in
which each sector's average capital and number of firms depend on
neighbouring sectors. These global characteristics are encoded in the
potential collective states of the system, also referred to as background
states. Each of these states computes a possible equilibrium distribution of
capital and firms across sectors.

This emphasis on collective states does not hinder the extraction of
information about agents' individual dynamics. On the contrary, as the field
formalism captures agents' interactions at the micro level, it facilitates
the study of their probabilistic behavior within a given collective state
through the so-called transition functions. The field formalism serves a
dual purpose by providing an approach to collective backgrounds arising from
agents' interactions and capturing the diverse individual dynamics within 
\emph{\ }such background.

At the macro-level, this work shows that collective states depend on
external parameters and expectations, such as short-term and both absolute
and relative expected long-term returns.\ However, depending on external or
historical conditions, the interdependency between sectors induces multiple
collective states: in each sector, three patterns of accumulation emerge,
from low to high.\ Some are unstable: changes in exogeneous parameters or
expectations may induce complete portfolio reallocations,\ potentially
depleting some sectors. At a macro-timescale, any deviation from an
equilibrium average capital drives the sector towards the next stable
equilibrium, including zero, and if there is none, towards infinity. This
notion of instability is sector-relative and context-dependent: variations
of parameters may propagate from one sector to another.\ Sectors may change
pattern which induces transitions between collective states. The field
formalism thus allows to describe global transitions of the patterns
accumulation initiated by one local modification.

To study this systemic instability, we consider a dynamic system involving
average capital and endogenized long-term expected returns, so that average
capital per sector interacts both with neighbouring ones, and long-term
expected returns.\textbf{\ }This dynamic system differs from those in
standard economics: whereas in economics the dynamics are usually studied
around static equilibria, we consider the dynamic interactions between
potential equilibria and expected long-term returns.

Some solutions are oscillatory: changes in one or several sectors may
propagate over the whole space of sectors. We find, for each sector, the
conditions of stable or unstable oscillations for the system. Depending on
the sector's specific characteristics, oscillations in average capital and
expected long-term returns may dampen or increase.\ Some types of
expectations favour overall stability in equilibria, and others deter it.

Eventually, fluctuations in financial expectations impose their pace on the
real economy.\ The combination of expectations both highly sensitive to
exogenous conditions and highly reactive to variations in capital implies
large fluctuations of capital in the system at the possible expense of the
real economy.

At the micro-level, we derive the transition functions for individual agents
within a given collective state. These functions describe the probabilistic
dynamics of agents in the background field as a function of their initial
state.\textbf{\ }We demonstrate that several factors influence the
probability of each firm's relocation path.

First, firms tend to relocate in sectors with highest long-term returns.\
However, the path followed by the firm to realocate depends on the
characteristics of the transition sectors, that are themselves determined by
the collective state of the system. The attractiveness of the firm during
its relocation process depends on the average capital of the transition
sectors it stumbles into. Depending on the sector, investors may over or
underinvest in the firm.\ An under-capitalized firm may fail to attract
investors in and either end up being stuck in this sector or be repelled to
a less attractive one.

Second, competition along the transition sectors, depends on the background
state of the system and impact differently the firm's level or capital and
attractiveness. An overcapitalized firm facing many less-endowed
competitors, will oust them out of the sector. On the contrary, an
under-capitalized firm will be ousted out from its own sector and move
towards less-capitalized and denser sectors in average. A capital gain - or
loss - may follow. Under-capitalized firms tend to move towards lower than
average capitalized sectors, while over-capitalized firms tend to move
towards higher than average-capitalized sectors.

Third, investors' capital allocation depends on short and long-term
returns.\ Yet these returns are not independent: short-term returns,
dividends and stock prices variations are correlated to the long-term that
depend on growth expectations and stock prices expectations. Changes in
investors' capital allocation are therefore directly dependent on stock
prices' volatility and firms' dividends. Changes in growth expectations
impact stock prices and incite investors to reallocate capital to maximize
their returns. The higher their level of capital, the stronger the
reallocation will be.

The paper is organized as follows. The second section is a literature
review. In the first part, we sum up the field theoretic approach to
economic models. Section three presents the general method and translation
techniques to turn a microeconomic framework with a large number of agents
into a field model. In section four we expose the use of the field theoretic
formalism to compute both average quantities in the model and the agents'
transitions functions. These functions compute the probabilities of the
model to evolve from an initial to a final state. Section five details
technically the computation of these transition funcions in presence of a
given background field. Section six presents and translates a particular
microeconomic framework with two types of agents, firms and investors into a
field model.

The second part of this paper studies the collective aspect of this model.
In sections seven, we describe the resolution of the model and derive the
background field for the real economy, the number of firms par sector, the
background field for the financial markets and the number of investors per
sector.\ Ultimately, we derive the defining equation for average capital per
firm per sector and discuss the main properties of the solution. In section
eight, the model is extended to a dynamic system at the macro-time scale by
endogenizing the expected long-term revenue. This dynamic system presents
some oscillatory solutions whose stability depends on the various patterns
of accumulation. Section nine presents, analyses and synthesizes the results.

The third part of this work is devoted to study the individual agents
dynamics and interaction within a given background state. In section 10, we
derive the transition functions for one and two agents in this particular
model. The results are presented and discussed in section 11.

The fourth part of the paper consolidates and presents the main findings.
Section twelve discusses our results, their interpretation and their
consequences. Section thirteen concludes.

\section{Literature review}

Several branches of the economic literature seek to replace the
representative agent with a collection of heterogeneous ones. Among other
things, they differ in the way they model this collection of agents.

The first branch of the literature represents this collection of agents by
probability densities. This is the approach followed by mean field theory,
heterogeneous agents new Keynesian (HANK) models, and the
information-theoretic approach to economics.

Mean field theory studies the evolution of agents' density in the state
space of economic variables. It includes the interactions between agents and
the population as a whole but does not consider the direct interactions
between agents. This approach is thus at an intermediate scale between the
macro and micro scale: it does not aggregate agents but replaces them with
an overall probability distribution. Mean field theory has been applied to
game theory (Bensoussan et al. 2018, Lasry et al. 2010a, b) and economics
(Gomes et al. 2015). However, these mean fields are actually probability
distributions. In our formalism, the notion of fields refers to some
abstract complex functions defined on the state space and is similar to the
second-quantized-wave functions of quantum theory. Interactions between
agents are included at the individual level. Densities of agents are
recovered from these fields and depend directly on interactions.

Heterogeneous agents' new Keynesian (HANK) models use a probabilistic
treatment similar to mean fields theory.\ An equilibrium probability
distribution is derived from a set of optimizing heterogeneous agents\ in a
new Keynesian context (see Kaplan and Violante 2018 for an account). Our
approach, on the contrary, focuses on the direct interactions between agents
at the microeconomic level. We do not look for an equilibrium probability
distribution for each agent, but rather directly build a probability density
for the system of $N$ agents seen as a whole, that includes interactions,
and then translate this probability density in terms of fields. The states'
space we consider is thus much larger than those considered in the above
approaches. Because it is the space of all paths for a large number of
agents, it allows studying the agents' economic structural relations and the
emergence of the particular phases or collective states induced by these
specific micro-relations, that will in turn impact each agent's stochastic
dynamics at the microeconomic level. Other differences are worth mentioning.
While HANK models\ stress the role of an infinite number of
heterogeneously-behaved consumers, our formalism dwells on the relations
between physical and financial capital\footnote{%
Note that our formalism could also include heterogeneous consumers (see
Gosselin, Lotz, Wambst 2020).}. Besides, our formalism does not rely on
agents' rationality assumptions, since for a large number of agents,
behaviours, be they fully or partly rational, can be modeled as random.

The information theoretic approach to economics (see Yang 2018) considers
probabilistic states around the equilibrium.\ It is close to our
methodological stance: it replaces the Walrasian equilibrium with a
statistical equilibrium derived from an entropy maximisation program.\ Our
statistical weight is similar to the one they use, but is directly built
from microeconomic dynamic equations. The same difference stands for the
rational inattention theory (Sims 2006) in which non-gaussian density laws
are derived from limited information and constraints: our setting directly
includes constraints in the random description of an agent (Gosselin, Lotz,
Wambst 2020).

The differences highlighted above between these various approaches and our
work also manifest at the micro-scale in the description of agents'
dynamics. Actually, in the field framework, once the collective states have
been found, we can recover both the types of individual dynamics depending
on the initial conditions and the "effective" form of interactions between
two or more agents: At the individual level, agents are distributed along
some probability law. However, this probability law is directly conditioned
by the collective state of the system and the effective interactions.
Different collective states, given different parameters, yield different
individual dynamics. This approach allows for coming back and forth between
collective and individual aspects of the system. Different categories of
agents appear in the emerging collective state. Dynamics may present very
different patterns, given the collective state's form and the agents'
initial conditions.

\bigskip

A second branch of the literature is closest to our approach since it
considers the interacting system of agents in itself. It is the multi-agent
systems literature, notably agent-based models (see Gaffard Napoletano 2012,
Mandel et al. 2010 2012) and economic networks (Jackson 2010).

Agent-based models deal with the macroeconomic level, whereas network models
lower-scale phenomena such as contract theory, behaviour diffusion,
information sharing, or learning. In both settings, agents are typically
defined by and follow various sets of rules, leading to the emergence of
equilibria and dynamics otherwise inaccessible to the representative agent
setup. Both approaches are however highly numerical and model-dependent and
rely on microeconomic relations - such as ad-hoc reaction functions - that
may be too simplistic. Statistical fields theory on the contrary accounts
for transitions between scales. Macroeconomic patterns do not emerge from
the sole dynamics of a large set of agents: they are grounded in behaviours
and interaction structures. Describing these structures in terms of field
theory allows for the emergence of phases at the macro scale, and the study
of their impact at the individual level.

A third branch of the literature, Econophysics, is also related to ours
since it often considers the set of agents as a statistical system (for a
review, see Abergel et al. 2011a,b and references therein; or Lux 2008,
2016).\ But it tends to focus on empirical laws, rather than apply the full
potential of field theory to economic systems. In the same vein, Kleinert
(2009) uses path integrals to model stock prices' dynamics. Our approach, in
contrast, keeps track of usual microeconomic concepts, such as utility
functions, expectations, and forward-looking behaviours, and includes these
behaviours into the analytical treatment of multi-agent systems by
translating the main characteristics of optimizing agents in terms of
statistical systems. Closer to our approach, Bardoscia et al (2017) study a
general equilibrium model for a large economy in the context of statistical
mechanics, and show that phase transitions may occur in the system. Our
problematic is similar, but our use of field theory deals with a large class
of dynamic models.

The literature on interactions between finance and real economy or capital
accumulation takes place mainly in the context of DGSE models. (for a review
of the literature, see Cochrane 2006; for further developments see Grassetti
et al. 2022, Grosshans and Zeisberger 2018, B\"{o}hm et al. 2008, Caggese
and Orive, Bernanke e al. 1999, Campello et al. 2010, Holmstrom and Tirole
1997, Jermann, and Quadrini 2012, Khan Thomas 2013, Monacelli et al. 2011).
Theoretical models include several types of agents at the aggregated level.\
They describe the interactions between a few representative agents such as
producers for possibly several sectors, consumers, financial intermediaries,
etc. to determine interest rates, levels of production, and asset pricing,
in a context of ad-hoc anticipations.

Our formalism differs from this literature in three ways. First, we consider
several groups of a large number of agents to describe the emergence of
collective states and study the continuous space of sectors. Second, we
consider expected returns and the longer-term horizon as somewhat exogenous
or structural. Expected returns are a combination of elements, such as
technology, returns, productivity, sectoral capital stock, expectations, and
beliefs. These returns are also a function\ defined over the sectors' space:
the system's background fields are functionals of these expected returns.\
Taken together, the background fields of a field model describe an economic
configuration for a given environment of expected returns.\ As such,
expected returns are at first seen as exogenous functions. It is only in the
second step, when we consider the dynamics between capital accumulation and
expectations, that expectations may themselves be seen as endogenous.\textbf{%
\ }Even then, the form of relations between actual and expected variables
specified are general enough to derive some types of possible dynamics.

Last but not least, we do not seek individual or even aggregated dynamics,
but rather background fields that describe potential long-term equilibria
and may evolve with the structural parameters. For such a background,
agents' individual typical dynamics may nevertheless be retrieved through
Green functions (see GLW). These functions compute the transition
probabilities from one capital-sector point to another. But backgrounds
themselves may be considered as dynamical quantities. Structural or
long-term variations in the returns' landscape may modify the background and
in turn the individual dynamics. Expected returns themselves depend on and
interact with, capital accumulation.

\part*{Field formalism for economic system with large number of agents and
application to a firms-investors model}

In the first part of this work, we describe the field formalism for an
economic system, its application to derive the potential collective states
of the system and the individual dynamics within such collective states.
Ultimately we apply this formalism to translate a model with large number of
interacting investors and firms.

\section{General method of translation}

The formalism we propose transforms an economic model of dynamic agents into
a statistical field model.\ In classical models, each agent's dynamics is
described by an optimal path\ for\ some vector variable, say $A_{i}\left(
t\right) $, from an initial to a final point, up to some fluctuations.

But this system of agents could also be seen as probabilistic: each agent
could be described by a \emph{probability density }centered around the
classical optimal path, up to some idiosyncratic uncertainties\footnote{%
Because the number of possible paths is infinite, the probability of each
individual path is null.\ We, therefore, use the word "probability density"
rather than "probability".} \footnote{%
See Gosselin, Lotz and Wambst (2017, 2020, 2021).}. In this probabilistic
approach, each possible trajectory of the whole set of $N$ agents has a
specific probability. The classical model is therefore described by the set
of trajectories of the group of $N$ agents, each one being endowed with its
own probability, its statistical weight. The statistical weight is therefore
a function that associates a probability with each trajectory of the group.

This probabilistic approach can be translated into a more compact \emph{%
field formalism}\footnote{%
Ibid.} that preserves the essential information encoded in the model but
implements a change in perspective. A field model is a structure governed by
its own intrinsic rules that encapsulate the economic model chosen.\ This
field model contains all possible realizations that could arise from the
initial economic model, i.e. all the possible global outcomes, or collective
state, permitted by the economic model.\ So that, once constructed, the
field model provides a unique advantage over the standard economic model: it
allows to compute the probabilities of each of the possible outcomes for
each collective state of the economic model. These probabilities are
computed indirectly through the \emph{action functional} of the model, a
function that assigns a specific value to each realization of the field.
Technically, the random $N$\ agents' trajectories $\left\{ \mathbf{A}%
_{i}\left( t\right) \right\} $ are replaced by a field, a random variable
whose realizations are complex-valued functions $\Psi $ of the variables $%
\mathbf{A}$\textbf{,} and the statistical weight of the $N$\ agents'
trajectories $\left\{ \mathbf{A}_{i}\left( t\right) \right\} $ in the
probabilistic approach is translated into a statistical weight for each
realization $\Psi $. They encapsulate the collective states of the system.

Once the probabilities of each collective state computed, the most probable
collective state among all other collective states, can be found. In other
words, a field model allows to consider the true global outcome induced by
any standard economic model. This is what we will call the \emph{expression}
of the field model, more usually called the \emph{background field} of the
model.

This most probable realization of the field, the expression or background
field of the model, should not be seen as a final outcome resulting from a
trajectory, but rather as its most recurring realization. Actually, the
probability of the realizations of the model is peaked around the expression
of the field.\ This expression, which is characteristic of the system, will
determine the nature of individual trajectories within the structure, in the
same way as a biased dice would increase the probability of one event.\ The
field in itself is therefore static, insofar as each realization of the
system of agents only contributes to the emergence of the proper expression
of the field. However, studying variations in the parameters of the system
indirectly induce a time parameter at the field or macro level.

\subsection{Statistical weight and minimization functions for a classical
system}

In an economic framework with a large number of agents, each agent is
characterized by one or more stochastic dynamic equations. Some of these
equations result from the optimization of one or several objective
functions. Deriving the statistical weight from these equations is
straightforward: it associates, to each trajectory of the group of agents $%
\left\{ T_{i}\right\} $, a probability that is peaked around the set of
optimal trajectories of the system, of the form:%
\begin{equation}
W\left( s\left( \left\{ T_{i}\right\} \right) \right) =\exp \left( -s\left(
\left\{ T_{i}\right\} \right) \right)  \label{wdt}
\end{equation}%
where $s\left( \left\{ T_{i}\right\} \right) $ measures the distance between
the trajectories $\left\{ T_{i}\right\} $ and the optimal ones.

This paper considers two types of agents characterized by vector-variables $%
\left\{ \mathbf{A}_{i}\left( t\right) \right\} _{i=1,...N},$ and $\left\{ 
\mathbf{\hat{A}}_{l}\left( t\right) \right\} _{i=1,...\hat{N}}$
respectively, where $N$ and $\hat{N}$ are the number of agents of each type,
with vectors $\mathbf{A}_{i}\left( t\right) $\ and $\mathbf{\hat{A}}%
_{l}\left( t\right) $ of arbitrary dimension. For such a system, the
statistical weight writes:%
\begin{equation}
W\left( \left\{ \mathbf{A}_{i}\left( t\right) \right\} ,\left\{ \mathbf{\hat{%
A}}_{l}\left( t\right) \right\} \right) =\exp \left( -s\left( \left\{ 
\mathbf{A}_{i}\left( t\right) \right\} ,\left\{ \mathbf{\hat{A}}_{l}\left(
t\right) \right\} \right) \right)  \label{wdh}
\end{equation}

The optimal paths for the system are assumed to be described by the sets of
equations:%
\begin{equation}
\frac{d\mathbf{A}_{i}\left( t\right) }{dt}-\sum_{j,k,l...}f\left( \mathbf{A}%
_{i}\left( t\right) ,\mathbf{A}_{j}\left( t\right) ,\mathbf{A}_{k}\left(
t\right) ,\mathbf{\hat{A}}_{l}\left( t\right) ,\mathbf{\hat{A}}_{m}\left(
t\right) ...\right) =\epsilon _{i}\text{, }i=1...N  \label{gauche}
\end{equation}%
\begin{equation}
\frac{d\mathbf{\hat{A}}_{l}\left( t\right) }{dt}-\sum_{i,j,k...}\hat{f}%
\left( \mathbf{A}_{i}\left( t\right) ,\mathbf{A}_{j}\left( t\right) ,\mathbf{%
A}_{k}\left( t\right) ,\mathbf{\hat{A}}_{l}\left( t\right) ,\mathbf{\hat{A}}%
_{m}\left( t\right) ...\right) =\hat{\epsilon}_{l}\text{, }i=1...\hat{N}
\label{dnw}
\end{equation}%
where the $\epsilon _{i}$ and $\hat{\epsilon}_{i}$ are idiosynchratic random
shocks.\ These equations describe the general dynamics of the two types
agents, including their interactions with other agents. They may\ encompass
the dynamics of optimizing agents where interactions act as externalities so
that this set of equations is the full description of a system of
interacting agents\footnote{%
Expectations of agents could be included by replacing $\frac{d\mathbf{A}%
_{i}\left( t\right) }{dt}$ with $E\frac{d\mathbf{A}_{i}\left( t\right) }{dt}$%
, where $E$ is the expectation operator. This would amount to double some
variables by distinguishing "real variables" and expectations. However, for
our purpose, in the context of a large number of agents, at least in this
work, we discard as much as possible this possibility.}\footnote{%
A generalisation of equations (\ref{gauche}) and (\ref{dnw}), in which
agents interact at different times, and its translation in term of field is
presented in appendix 1.}\textbf{. }

For equations (\ref{gauche}) and (\ref{dnw}), the quadratic deviation at
time $t$ of any trajectory with respect to the optimal one for each type of
agent are:%
\begin{equation}
\left( \frac{d\mathbf{A}_{i}\left( t\right) }{dt}-\sum_{j,k,l...}f\left( 
\mathbf{A}_{i}\left( t\right) ,\mathbf{A}_{j}\left( t\right) ,\mathbf{A}%
_{k}\left( t\right) ,\mathbf{\hat{A}}_{l}\left( t\right) ,\mathbf{\hat{A}}%
_{m}\left( t\right) ...\right) \right) ^{2}  \label{pst}
\end{equation}%
and:%
\begin{equation}
\left( \frac{d\mathbf{\hat{A}}_{l}\left( t\right) }{dt}-\sum_{i,j,k...}\hat{f%
}\left( \mathbf{A}_{i}\left( t\right) ,\mathbf{A}_{j}\left( t\right) ,%
\mathbf{A}_{k}\left( t\right) ,\mathbf{\hat{A}}_{l}\left( t\right) ,\mathbf{%
\hat{A}}_{m}\left( t\right) ...\right) \right) ^{2}  \label{psh}
\end{equation}%
Since the function (\ref{wdh}) involves the deviations for all agents over
all trajectories, the function $s\left( \left\{ \mathbf{A}_{i}\left(
t\right) \right\} ,\left\{ \mathbf{\hat{A}}_{l}\left( t\right) \right\}
\right) $ is obtained by summing (\ref{pst}) and (\ref{psh}) over all
agents, and integrate over $t$. We thus find:%
\begin{eqnarray}
s\left( \left\{ \mathbf{A}_{i}\left( t\right) \right\} ,\left\{ \mathbf{\hat{%
A}}_{l}\left( t\right) \right\} \right) &=&\int dt\sum_{i}\left( \frac{d%
\mathbf{A}_{i}\left( t\right) }{dt}-\sum_{j,k,l...}f\left( \mathbf{A}%
_{i}\left( t\right) ,\mathbf{A}_{j}\left( t\right) ,\mathbf{A}_{k}\left(
t\right) ,\mathbf{\hat{A}}_{l}\left( t\right) ,\mathbf{\hat{A}}_{m}\left(
t\right) ...\right) \right) ^{2}  \label{prw} \\
&&+\int dt\sum_{l}\left( \frac{d\mathbf{\hat{A}}_{l}\left( t\right) }{dt}%
-\sum_{i,j,k...}\hat{f}\left( \mathbf{A}_{i}\left( t\right) ,\mathbf{A}%
_{j}\left( t\right) ,\mathbf{A}_{k}\left( t\right) ,\mathbf{\hat{A}}%
_{l}\left( t\right) ,\mathbf{\hat{A}}_{m}\left( t\right) ...\right) \right)
^{2}  \notag
\end{eqnarray}%
There is an alternate, more general, form to (\ref{prw}). We can assume that
the dynamical system is originally defined by some equations of type (\ref%
{gauche}) and (\ref{dnw}), plus some objective functions for agents $i$ and $%
l$, and that these agents aim at minimizing respectively:%
\begin{equation}
\sum_{j,k,l...}g\left( \mathbf{A}_{i}\left( t\right) ,\mathbf{A}_{j}\left(
t\right) ,\mathbf{A}_{k}\left( t\right) ,\mathbf{\hat{A}}_{l}\left( t\right)
,\mathbf{\hat{A}}_{m}\left( t\right) ...\right)  \label{glf}
\end{equation}%
and:%
\begin{equation}
\sum_{i,j,k..}\hat{g}\left( \mathbf{A}_{i}\left( t\right) ,\mathbf{A}%
_{j}\left( t\right) ,\mathbf{A}_{k}\left( t\right) ,\mathbf{\hat{A}}%
_{l}\left( t\right) ,\mathbf{\hat{A}}_{m}\left( t\right) ...\right)
\label{gln}
\end{equation}%
In the above equations, the objective functions depend on other agents'
actions seen as externalities\footnote{%
We may also assume intertemporal objectives, see (GLW).{}}. The functions (%
\ref{glf}) and (\ref{gln}) could themselves be considered as a measure of
the deviation of a trajectory from the optimum. Actually, the higher the
distance, the higher (\ref{glf}) and (\ref{gln}).

Thus, rather than describing the systm by a full system of dynamic
equations, we can consider some ad-hoc equations of type (\ref{gauche}) and (%
\ref{dnw}) and some objective functions (\ref{glf}) and (\ref{gln}) to write
the alternate form of $s\left( \left\{ \mathbf{A}_{i}\left( t\right)
\right\} ,\left\{ \mathbf{\hat{A}}_{l}\left( t\right) \right\} \right) $ as:%
\begin{eqnarray}
&&s\left( \left\{ \mathbf{A}_{i}\left( t\right) \right\} ,\left\{ \mathbf{%
\hat{A}}_{l}\left( t\right) \right\} \right)  \label{mNZ} \\
&=&\int dt\sum_{i}\left( \frac{d\mathbf{A}_{i}\left( t\right) }{dt}%
-\sum_{j,k,l...}f\left( \mathbf{A}_{i}\left( t\right) ,\mathbf{A}_{j}\left(
t\right) ,\mathbf{A}_{k}\left( t\right) ,\mathbf{\hat{A}}_{l}\left( t\right)
,\mathbf{\hat{A}}_{m}\left( t\right) ...\right) \right) ^{2}  \notag \\
&&+\int dt\sum_{l}\left( \frac{d\mathbf{\hat{A}}_{l}\left( t\right) }{dt}%
-\sum_{i,j,k...}\hat{f}\left( \mathbf{A}_{i}\left( t\right) ,\mathbf{A}%
_{j}\left( t\right) ,\mathbf{A}_{k}\left( t\right) ,\mathbf{\hat{A}}%
_{l}\left( t\right) ,\mathbf{\hat{A}}_{m}\left( t\right) ...\right) \right)
^{2}  \notag \\
&&+\int dt\sum_{i,j,k,l...}\left( g\left( \mathbf{A}_{i}\left( t\right) ,%
\mathbf{A}_{j}\left( t\right) ,\mathbf{A}_{k}\left( t\right) ,\mathbf{\hat{A}%
}_{l}\left( t\right) ,\mathbf{\hat{A}}_{m}\left( t\right) ...\right) +\hat{g}%
\left( \mathbf{A}_{i}\left( t\right) ,\mathbf{A}_{j}\left( t\right) ,\mathbf{%
A}_{k}\left( t\right) ,\mathbf{\hat{A}}_{l}\left( t\right) ,\mathbf{\hat{A}}%
_{m}\left( t\right) ...\right) \right)  \notag
\end{eqnarray}

In the sequel, we will refer to the various terms arising in equation (\ref%
{mNZ}) as the "minimization functions",\textbf{\ i.e}. the functions whose
minimization yield the dynamics equations of the system\footnote{%
A generalisation of equation (\ref{mNZ}), in which agents interact at
different times, and its translation in term of field is presented in
appendix 1.{}}.

\subsection{Translation techniques}

Once the statistical weight $W\left( s\left( \left\{ T_{i}\right\} \right)
\right) $ defined in (\ref{wdt}) iscomputed, it can be translated in terms
of field. To do so, and for each type $\alpha $ of agent, the sets of
trajectories $\left\{ \mathbf{A}_{\alpha i}\left( t\right) \right\} $ are
replaced by a field $\Psi _{\alpha }\left( \mathbf{A}_{\alpha }\right) $, a
random variable whose realizations are complex-valued functions $\Psi $ of
the variables $\mathbf{A}_{\alpha }$\footnote{%
In the following, we will use indifferently the term "field" and the
notation $\Psi $ for the random variable or any of its realization $\Psi $.}%
. The statistical weight for the whole set of fields $\left\{ \Psi _{\alpha
}\right\} $ has the form $\exp \left( -S\left( \left\{ \Psi _{\alpha
}\right\} \right) \right) $. The function $S\left( \left\{ \Psi _{\alpha
}\right\} \right) $\ is called the \emph{fields action functional}. It
represents the interactions among different types of agents. Ultimately, the
expression $\exp \left( -S\left( \left\{ \Psi _{\alpha }\right\} \right)
\right) $ is the statistical weight for the field\footnote{%
In general, one must consider the integral of $\exp \left( -S\left( \left\{
\Psi _{\alpha }\right\} \right) \right) $\ over the configurations $\left\{
\Psi _{\alpha }\right\} $. This integral is the partition function of the
system.} that computes the probability of any realization $\left\{ \Psi
_{\alpha }\right\} $\ of the field.

The form of $S\left( \left\{ \Psi _{\alpha }\right\} \right) $\ is obtained
directly from the classical description of our model. For two types of
agents, we start with expression (\ref{mNZ}). The various minimizations
functions involved in the definition of $s\left( \left\{ \mathbf{A}%
_{i}\left( t\right) \right\} ,\left\{ \mathbf{\hat{A}}_{l}\left( t\right)
\right\} \right) $ will be translated in terms of field and the sum of these
translations will produce finally the action functional $S\left( \left\{
\Psi _{\alpha }\right\} \right) $. The translation method can itself be
divided into two relatively simple processes, but varies slightly depending
on the type of terms that appear in the various minimization functions.

\subsubsection{Terms without temporal derivative}

In equation (\ref{mNZ}), the terms that involve indexed variables but no
temporal derivative terms are the easiest to translate.\ They are of the
form:%
\begin{equation*}
\sum_{i}\sum_{j,k,l,m...}g\left( \mathbf{A}_{i}\left( t\right) ,\mathbf{A}%
_{j}\left( t\right) ,\mathbf{A}_{k}\left( t\right) ,\mathbf{\hat{A}}%
_{l}\left( t\right) ,\mathbf{\hat{A}}_{m}\left( t\right) ...\right)
\end{equation*}%
These terms describe the whole set of interactions both among and between
two groups of agents. Here, agents are characterized by their variables $%
\mathbf{A}_{i}\left( t\right) ,\mathbf{A}_{j}\left( t\right) ,\mathbf{A}%
_{k}\left( t\right) $... and $\mathbf{\hat{A}}_{l}\left( t\right) ,\mathbf{%
\hat{A}}_{m}\left( t\right) $... respectively, for instance in our model
firms and investors.

In the field translation, agents of type $\mathbf{A}_{i}\left( t\right) $
and $\mathbf{\hat{A}}_{l}\left( t\right) $ are described by a field $\Psi
\left( \mathbf{A}\right) $ and $\hat{\Psi}\left( \mathbf{\hat{A}}\right) $,
respectively.

In a first step, the variables indexed $i$ such as $\mathbf{A}_{i}\left(
t\right) $ are replaced by variables $\mathbf{A}$ in the expression of $g$.
The variables indexed $j$, $k$, $l$, $m$..., such as $\mathbf{A}_{j}\left(
t\right) $, $\mathbf{A}_{k}\left( t\right) $, $\mathbf{\hat{A}}_{l}\left(
t\right) ,\mathbf{\hat{A}}_{m}\left( t\right) $... are replaced by $\mathbf{A%
}^{\prime },\mathbf{A}^{\prime \prime }$, $\mathbf{\hat{A}}$, $\mathbf{\hat{A%
}}^{\prime }$ , and so on for all the indices in the function. This yields
the expression:

\begin{equation*}
\sum_{i}\sum_{j,k,l,m...}g\left( \mathbf{A},\mathbf{A}^{\prime },\mathbf{A}%
^{\prime \prime },\mathbf{\hat{A},\hat{A}}^{\prime }...\right)
\end{equation*}%
In a second step, each sum is replaced by a weighted integration symbol: 
\begin{eqnarray*}
\sum_{i} &\rightarrow &\int \left\vert \Psi \left( \mathbf{A}\right)
\right\vert ^{2}d\mathbf{A}\text{, }\sum_{j}\rightarrow \int \left\vert \Psi
\left( \mathbf{A}^{\prime }\right) \right\vert ^{2}d\mathbf{A}^{\prime }%
\text{, }\sum_{k}\rightarrow \int \left\vert \Psi \left( \mathbf{A}^{\prime
\prime }\right) \right\vert ^{2}d\mathbf{A}^{\prime \prime } \\
\sum_{l} &\rightarrow &\int \left\vert \hat{\Psi}\left( \mathbf{\hat{A}}%
\right) \right\vert ^{2}d\mathbf{\hat{A}}\text{, }\sum_{m}\rightarrow \int
\left\vert \hat{\Psi}\left( \mathbf{\hat{A}}^{\prime }\right) \right\vert
^{2}d\mathbf{\hat{A}}^{\prime }
\end{eqnarray*}%
which leads to the translation:%
\begin{eqnarray}
&&\sum_{i}\sum_{j}\sum_{j,k...}g\left( \mathbf{A}_{i}\left( t\right) ,%
\mathbf{A}_{j}\left( t\right) ,\mathbf{A}_{k}\left( t\right) ,\mathbf{\hat{A}%
}_{l}\left( t\right) ,\mathbf{\hat{A}}_{m}\left( t\right) ...\right)  \notag
\\
&\rightarrow &\int g\left( \mathbf{A},\mathbf{A}^{\prime },\mathbf{A}%
^{\prime \prime },\mathbf{\hat{A},\hat{A}}^{\prime }...\right) \left\vert
\Psi \left( \mathbf{A}\right) \right\vert ^{2}\left\vert \Psi \left( \mathbf{%
A}^{\prime }\right) \right\vert ^{2}\left\vert \Psi \left( \mathbf{A}%
^{\prime \prime }\right) \right\vert ^{2}\times ...d\mathbf{A}d\mathbf{A}%
^{\prime }d\mathbf{A}^{\prime \prime }...  \label{tln} \\
&&\times \left\vert \hat{\Psi}\left( \mathbf{\hat{A}}\right) \right\vert
^{2}\left\vert \hat{\Psi}\left( \mathbf{\hat{A}}^{\prime }\right)
\right\vert ^{2}\times ...d\mathbf{\hat{A}}d\mathbf{\hat{A}}^{\prime }... 
\notag
\end{eqnarray}%
where the dots stand for the products of square fields and integration
symbols needed.

\subsubsection{Terms with temporal derivative}

In equation (\ref{mNZ}), the terms that involve a variable temporal
derivative are of the form:%
\begin{equation}
\sum_{i}\left( \frac{d\mathbf{A}_{i}^{\left( \alpha \right) }\left( t\right) 
}{dt}-\sum_{j,k,l,m...}f^{\left( \alpha \right) }\left( \mathbf{A}_{i}\left(
t\right) ,\mathbf{A}_{j}\left( t\right) ,\mathbf{A}_{k}\left( t\right) ,%
\mathbf{\hat{A}}_{l}\left( t\right) ,\mathbf{\hat{A}}_{m}\left( t\right)
...\right) \right) ^{2}  \label{edr}
\end{equation}%
This particular form represents the dynamics of the $\alpha $-th coordinate
of a variable $\mathbf{A}_{i}\left( t\right) $ as a function of the other
agents.

The method of translation is similar to the above, but the time derivative
adds an additional operation.

In a first step, we translate the terms without derivative inside the
parenthesis:%
\begin{equation}
\sum_{j,k,l,m...}f^{\left( \alpha \right) }\left( \mathbf{A}_{i}\left(
t\right) ,\mathbf{A}_{j}\left( t\right) ,\mathbf{A}_{k}\left( t\right) ,%
\mathbf{\hat{A}}_{l}\left( t\right) ,\mathbf{\hat{A}}_{m}\left( t\right)
...\right)  \label{ntr}
\end{equation}%
This type of term has already been translated in the previous paragraph, but
since there is no sum over $i$ in equation (\ref{ntr}), there should be no
integral over $\mathbf{A}$\textbf{,} nor factor $\left\vert \Psi \left( 
\mathbf{A}\right) \right\vert ^{2}$.

The translation of equation (\ref{ntr}) is therefore, as before:%
\begin{equation}
\int f^{\left( \alpha \right) }\left( \mathbf{A},\mathbf{A}^{\prime },%
\mathbf{A}^{\prime \prime },\mathbf{\hat{A},\hat{A}}^{\prime }...\right)
\left\vert \Psi \left( \mathbf{A}^{\prime }\right) \right\vert
^{2}\left\vert \Psi \left( \mathbf{A}^{\prime \prime }\right) \right\vert
^{2}d\mathbf{A}^{\prime }d\mathbf{A}^{\prime \prime }\left\vert \hat{\Psi}%
\left( \mathbf{\hat{A}}\right) \right\vert ^{2}\left\vert \hat{\Psi}\left( 
\mathbf{\hat{A}}^{\prime }\right) \right\vert ^{2}d\mathbf{\hat{A}}d\mathbf{%
\hat{A}}^{\prime }  \label{trn}
\end{equation}%
A free variable $\mathbf{A}$ remains, which will be integrated later, when
we account for the external sum $\sum_{i}$. We will call $\Lambda (\mathbf{A}%
)$ the expression obtained:%
\begin{equation}
\Lambda (\mathbf{A})=\int f^{\left( \alpha \right) }\left( \mathbf{A},%
\mathbf{A}^{\prime },\mathbf{A}^{\prime \prime },\mathbf{\hat{A},\hat{A}}%
^{\prime }...\right) \left\vert \Psi \left( \mathbf{A}^{\prime }\right)
\right\vert ^{2}\left\vert \Psi \left( \mathbf{A}^{\prime \prime }\right)
\right\vert ^{2}d\mathbf{A}^{\prime }d\mathbf{A}^{\prime \prime }\left\vert 
\hat{\Psi}\left( \mathbf{\hat{A}}\right) \right\vert ^{2}\left\vert \hat{\Psi%
}\left( \mathbf{\hat{A}}^{\prime }\right) \right\vert ^{2}d\mathbf{\hat{A}}d%
\mathbf{\hat{A}}^{\prime }  \label{bdt}
\end{equation}%
In a second step, we account for the derivative in time by using field
gradients. To do so, and as a rule, we replace :%
\begin{equation}
\sum_{i}\left( \frac{d\mathbf{A}_{i}^{\left( \alpha \right) }\left( t\right) 
}{dt}-\sum_{j}\sum_{j,k...}f^{\left( \alpha \right) }\left( \mathbf{A}%
_{i}\left( t\right) ,\mathbf{A}_{j}\left( t\right) ,\mathbf{A}_{k}\left(
t\right) ,\mathbf{\hat{A}}_{l}\left( t\right) ,\mathbf{\hat{A}}_{m}\left(
t\right) ...\right) \right) ^{2}  \label{inco}
\end{equation}%
by:%
\begin{equation}
\int \Psi ^{\dag }\left( \mathbf{A}\right) \left( -\nabla _{\mathbf{A}%
^{\left( \alpha \right) }}\left( \frac{\sigma _{\mathbf{A}^{\left( \alpha
\right) }}^{2}}{2}\nabla _{\mathbf{A}^{\left( \alpha \right) }}-\Lambda (%
\mathbf{A})\right) \right) \Psi \left( \mathbf{A}\right) d\mathbf{A}
\label{Trl}
\end{equation}%
The variance $\sigma _{\mathbf{A}^{\left( \alpha \right) }}^{2}$ reflects
the probabilistic nature of the model which is hidden behind the field
formalism. This variance represents the characteristic level of uncertainty
of the system's dynamics. It is a parameter of the model. Note also that in (%
\ref{Trl}), the integral over $\mathbf{A}$ reappears at the end, along with
the square of the field $\left\vert \Psi \left( \mathbf{A}\right)
\right\vert ^{2}$.\ This square is split into two terms, $\Psi ^{\dag
}\left( \mathbf{A}\right) $ and $\Psi \left( \mathbf{A}\right) $, with a
gradient operator inserted in between.

\subsection{Action functional}

The field description is ultimately obtained by summing all the terms
translated above and introducing a time dependency. This sum is called the
action functional. It is the sum of terms of the form (\ref{tln}) and (\ref%
{Trl}), and is denoted $S\left( \Psi ,\Psi ^{\dag }\right) $.

For example, in a system with two types of agents described by two fields $%
\Psi \left( \mathbf{A}\right) $and $\hat{\Psi}\left( \mathbf{\hat{A}}\right) 
$, the action functional has the form:%
\begin{eqnarray}
S\left( \Psi ,\Psi ^{\dag }\right) &=&\int \Psi ^{\dag }\left( \mathbf{A}%
\right) \left( -\nabla _{\mathbf{A}^{\left( \alpha \right) }}\left( \frac{%
\sigma _{\mathbf{A}^{\left( \alpha \right) }}^{2}}{2}\nabla _{\mathbf{A}%
^{\left( \alpha \right) }}-\Lambda _{1}(\mathbf{A})\right) \right) \Psi
\left( \mathbf{A}\right) d\mathbf{A}  \label{notime} \\
&&\mathbf{+}\int \hat{\Psi}^{\dag }\left( \mathbf{\hat{A}}\right) \left(
-\nabla _{\mathbf{\hat{A}}^{\left( \alpha \right) }}\left( \frac{\sigma _{%
\mathbf{\hat{A}}^{\left( \alpha \right) }}^{2}}{2}\nabla _{\mathbf{\hat{A}}%
^{\left( \alpha \right) }}-\Lambda _{2}(\mathbf{\hat{A}})\right) \right) 
\hat{\Psi}\left( \mathbf{\hat{A}}\right) d\mathbf{\hat{A}}  \notag \\
&&+\sum_{m}\int g_{m}\left( \mathbf{A},\mathbf{A}^{\prime },\mathbf{A}%
^{\prime \prime },\mathbf{\hat{A},\hat{A}}^{\prime }...\right) \left\vert
\Psi \left( \mathbf{A}\right) \right\vert ^{2}\left\vert \Psi \left( \mathbf{%
A}^{\prime }\right) \right\vert ^{2}\left\vert \Psi \left( \mathbf{A}%
^{\prime \prime }\right) \right\vert ^{2}\times ...d\mathbf{A}d\mathbf{A}%
^{\prime }d\mathbf{A}^{\prime \prime }...  \notag \\
&&\times \left\vert \hat{\Psi}\left( \mathbf{\hat{A}}\right) \right\vert
^{2}\left\vert \hat{\Psi}\left( \mathbf{\hat{A}}^{\prime }\right)
\right\vert ^{2}\times ...d\mathbf{\hat{A}}d\mathbf{\hat{A}}^{\prime }... 
\notag
\end{eqnarray}%
where the sequence of functions $g_{m}$\ describes the various types of
interactions in the system.

Note that the collective states described by the fields are structural
states of the system. The fields have their own dynamics at the macro-scale,
which will be discussed later in the paper. This is why the usual
microeconomic time variable used in standard models has disappeared in
formula (\ref{notime}). However, time dependency may at times be required in
fields, so that a time variable, written $\theta $ could be introduced by
replacing:%
\begin{eqnarray*}
\Psi \left( \mathbf{A}\right) &\rightarrow &\Psi \left( \mathbf{A},\theta
\right) \\
\hat{\Psi}\left( \mathbf{\hat{A}}\right) &\rightarrow &\hat{\Psi}\left( 
\mathbf{\hat{A}},\theta \right)
\end{eqnarray*}%
More about this point can be found in appendix 1.

\section{Use of the field model}

Once the field action functional $S$\ is found, we can use field theory to
study the system of agents.\ This can be done at two levels: the collective
and the individual level. At the collective level, the system is described
by the background fields of the system that condition average quantities of
economic variables of the system.

At the individual level, the field formalism allows to compute agents'
individual dynamics in the state defined by the background fields, through
the transition functions of the system.

\subsection{Collective level: background fields and averages}

At the collective level, the background fields of the system can be
computed. These background fields are the particular functions, $\Psi \left( 
\mathbf{A}\right) $\ and $\hat{\Psi}\left( \mathbf{\hat{A}}\right) $, and
their adjoints fields $\Psi ^{\dag }\left( \mathbf{A}\right) $\ and $\hat{%
\Psi}^{\dag }\left( \mathbf{\hat{A}}\right) $,\ that minimize the action
functional $S$. Once the background field(s) obtained, the associated
density of agents defined by a given $A$\ and a given $\hat{A}$\ are:%
\begin{equation}
\left\vert \Psi \left( \mathbf{A}\right) \right\vert ^{2}=\Psi ^{\dag
}\left( \mathbf{A}\right) \Psi \left( \mathbf{A}\right)  \label{DSNV}
\end{equation}%
and:%
\begin{equation}
\left\vert \hat{\Psi}\left( \mathbf{\hat{A}}\right) \right\vert ^{2}=\hat{%
\Psi}^{\dag }\left( \mathbf{\hat{A}}\right) \hat{\Psi}\left( \mathbf{\hat{A}}%
\right)  \label{DSTV}
\end{equation}%
respectively. With these density functions at hand, we can compute various
average quantities\ in the collective state. Actually, the averages for the
system in the state defined by $\Psi \left( \mathbf{A}\right) $ and $\hat{%
\Psi}\left( \mathbf{\hat{A}}\right) $\ of components $\left( \mathbf{A}%
\right) _{k}$ or $\left( \mathbf{\hat{A}}\right) _{l}$ are:%
\begin{equation*}
\left\langle \left( \mathbf{A}\right) _{k}\right\rangle =\frac{\int \left( 
\mathbf{A}\right) _{k}\left\vert \Psi \left( \mathbf{A}\right) \right\vert
^{2}d\mathbf{A}}{\int \left\vert \Psi \left( \mathbf{A}\right) \right\vert
^{2}d\mathbf{A}}
\end{equation*}%
\begin{equation*}
\left\langle \left( \mathbf{\hat{A}}\right) _{l}\right\rangle =\frac{\int
\left( \mathbf{\hat{A}}\right) \left\vert \hat{\Psi}\left( \mathbf{\hat{A}}%
\right) \right\vert ^{2}d\mathbf{\hat{A}}}{\int \left\vert \hat{\Psi}\left( 
\mathbf{\hat{A}}\right) \right\vert ^{2}d\mathbf{\hat{A}}}
\end{equation*}%
respectively. We can also define both partial densities and averages by
integrating some components and fixing the values of others, as will be
detailled in the particular model considered in the next sections.

\subsection{Individual level: agents transition functions and their field
expression}

\subsubsection{Transition functions in a classical framework}

In a classical perspective, the statistical weight (\ref{ST}) can be used to
compute the transition probabilities of the system, i.e. the probabilities
for any number of agents of both types to evolve from an initial state $%
\left\{ \mathbf{A}_{l}\right\} _{l=1,...},\left\{ \mathbf{\hat{A}}%
_{l}\right\} _{l=1,...}$\textbf{\ }to a final state in a given timespan.
These transition functions describe the dynamic of the agents of the system.

To do so, we first compute the integral of equation (\ref{ST}) over all
paths between the initial and the final points considered. Defining $\left\{ 
\mathbf{A}_{l}\left( s\right) \right\} _{l=1,...,N}$ and $\left\{ \mathbf{%
\hat{A}}_{l}\left( s\right) \right\} _{l=1,...,\hat{N}}$ the sets of paths
for agents of each type, where $N$\ and $\hat{N}$\ are the numbers of agents
of each type, we consider the set of $N+\hat{N}$\ independent paths written:

\begin{equation*}
\mathbf{Z}\left( s\right) =\left( \left\{ \mathbf{A}_{l}\left( s\right)
\right\} _{l=1,...,N},\left\{ \mathbf{\hat{A}}_{l}\left( s\right) \right\}
_{l=1,...,\hat{N}}\right)
\end{equation*}%
\ The weight (\ref{ST}) can now be written $\exp \left( -W\left( \mathbf{Z}%
\left( s\right) \right) \right) $.

The transition functions $T_{t}\left( \underline{\left( \mathbf{Z}\right) },%
\overline{\left( \mathbf{Z}\right) }\right) $\ compute the probability for
the $(N,\hat{N}$\ $)$\ agents to evolve from\ the initial points $Z\left(
0\right) \equiv \underline{\mathbf{Z}}$\ to the\ final points $Z\left(
t\right) \equiv \overline{\left( \mathbf{Z}\right) }$\ during a time span $t$%
. This probability is defined by:%
\begin{equation}
T_{t}\left( \underline{\mathbf{Z}},\overline{\left( \mathbf{Z}\right) }%
\right) =\frac{1}{\mathcal{N}}\int_{\substack{ \mathbf{Z}\left( 0\right)
\equiv \underline{\mathbf{Z}}  \\ \mathbf{Z}\left( t\right) \equiv \overline{%
\left( \mathbf{Z}\right) }}}\exp \left( -W\left( \mathbf{Z}\left( s\right)
\right) \right) \mathcal{D}\left( \mathbf{Z}\left( s\right) \right)
\label{tsn}
\end{equation}%
The integration symbol $D\mathbf{Z}\left( s\right) $\ covers all sets of $%
N\times \hat{N}$\ paths constrained by $\mathbf{Z}\left( 0\right) \equiv 
\underline{\mathbf{Z}}$\ and $\mathbf{Z}\left( t\right) \equiv \overline{%
\left( \mathbf{Z}\right) }$. The normalisation factor sets the total
probability defined by the weight (\ref{ST}) to $1$ and is equal to:%
\begin{equation*}
\mathcal{N=}\int \exp \left( -W\left( \mathbf{Z}\left( s\right) \right)
\right) \mathcal{D}\mathbf{Z}\left( s\right)
\end{equation*}%
The interpretation of (\ref{tsn}) is straightforward. Instead of studying
the full trajectory of one or several agents, we compute their probability
to evolve from one configuration to another, and in average, the usual
trajectory approach remains valid.

Equation (\ref{tsn}) can be generalized to define the transition functions
for $k\leqslant N$\ and $\hat{k}\leqslant \hat{N}$\ agents of each type. The
initial and final points respectively for this set of $k+\hat{k}$\ agents
are written:%
\begin{equation*}
\mathbf{Z}\left( 0\right) ^{\left[ k,\hat{k}\right] }\equiv \underline{%
\mathbf{Z}}^{\left[ k,\hat{k}\right] }
\end{equation*}%
and:%
\begin{equation*}
\mathbf{Z}\left( t\right) ^{\left[ k,\hat{k}\right] }\equiv \overline{\left( 
\mathbf{Z}\right) }^{\left[ k,\hat{k}\right] }
\end{equation*}%
The transition function for these agents is written:%
\begin{equation*}
T_{t}\left( \underline{\left( \mathbf{Z}\right) }^{\left[ k,\hat{k}\right] },%
\overline{\left( \mathbf{Z}\right) }^{\left[ k,\hat{k}\right] }\right)
\end{equation*}%
and the generalization of equation (\ref{tsn}) is: \ 
\begin{equation}
T_{t}\left( \underline{\left( \mathbf{Z}\right) }^{\left[ k,\hat{k}\right] },%
\overline{\left( \mathbf{Z}\right) }^{\left[ k,\hat{k}\right] }\right) =%
\frac{1}{\mathcal{N}}\int_{\substack{ \mathbf{Z}\left( 0\right) ^{\left[ k,%
\hat{k}\right] }=\underline{\left( \mathbf{Z}\right) }^{\left[ k,\hat{k}%
\right] }  \\ \mathbf{Z}\left( t\right) ^{\left[ k,\hat{k}\right] }=%
\overline{\left( \mathbf{Z}\right) }^{\left[ k,\hat{k}\right] }}}\exp \left(
-W\left( \left( \mathbf{Z}\left( s\right) \right) \right) \right) \mathcal{D}%
\left( \left( \mathbf{Z}\left( s\right) \right) \right)  \label{krtv}
\end{equation}%
The difference with (\ref{tsn}) is that only $k$\ paths are constrained by
their initial and final points.

Ultimately, the Laplace transform of $T_{t}\left( \underline{\left( Z\right) 
}^{\left[ k,\hat{k}\right] },\overline{\left( Z\right) }^{\left[ k,\hat{k}%
\right] }\right) $ computes the - time averaged - transition function for
agents with random lifespan of mean $\frac{1}{\alpha }$, up to a factor $%
\frac{1}{\alpha }$, and is given by:%
\begin{equation}
G_{\alpha }\left( \underline{\left( \mathbf{Z}\right) }^{\left[ k,\hat{k}%
\right] },\overline{\left( \mathbf{Z}\right) }^{\left[ k,\hat{k}\right]
}\right) =\int_{0}^{\infty }\exp \left( -\alpha t\right) T_{t}\left( 
\underline{\left( \mathbf{Z}\right) }^{\left[ k,\hat{k}\right] },\overline{%
\left( \mathbf{Z}\right) }^{\left[ k,\hat{k}\right] }\right) dt  \label{krvv}
\end{equation}%
This formulation of the transition functions is relatively intractable.
Therefore, we will now propose an alternative method based on the field
model.

\subsubsection{Field-theoretic expression}

The transition functions (\ref{krtv}) and (\ref{krvv}) can be retrieved
using the$\ $field theory transition functions - or\ Green functions, which
compute the probability for a variable number $\left( k,\hat{k}\right) $\ of
agents to transition from an initial state $\underline{\left( \mathbf{%
Z,\theta }\right) }^{\left[ k,\hat{k}\right] }$\ to a final state $\overline{%
\left( \mathbf{Z,\theta }\right) }^{\left[ k,\hat{k}\right] }$, where $%
\underline{\left( \mathbf{\theta }\right) }^{\left[ k,\hat{k}\right] }$ and
\ $\overline{\left( \mathbf{\theta }\right) }^{\left[ k,\hat{k}\right] }$ are%
\textbf{\ }vectors of initial and final times for $k+\hat{k}$\textbf{\ }%
agents respectively.

We will write: 
\begin{equation*}
T_{t}\left( \underline{\left( \mathbf{Z,\theta }\right) }^{\left[ k,\hat{k}%
\right] },\overline{\left( \mathbf{Z,\theta }\right) }^{\left[ k,\hat{k}%
\right] }\right)
\end{equation*}%
\ the transition function between $\underline{\left( \mathbf{Z},\mathbf{%
\theta }\right) }^{\left[ k,\hat{k}\right] }$\ and $\overline{\left( \mathbf{%
Z},\mathbf{\theta }\right) }^{\left[ k,\hat{k}\right] }$ with $\overline{%
\left( \mathbf{\theta }\right) }_{i}<t$, $\forall i$,\ \ and: 
\begin{equation*}
G_{\alpha }\left( \underline{\left( \mathbf{Z,\theta }\right) }^{\left[ k,%
\hat{k}\right] },\overline{\left( \mathbf{Z,\theta }\right) }^{\left[ k,\hat{%
k}\right] }\right)
\end{equation*}%
\ its Laplace transform. Setting $\underline{\left( \mathbf{\theta }\right) }%
_{i}=0$\ and $\overline{\left( \mathbf{\theta }\right) }_{i}=t$\ for $%
i=1,...,k+\hat{k}$, these functions\ reduce to (\ref{krtv}) or (\ref{krvv}):
the probabilistic formalism of the transition functions is thus a particular
case of the field formalism definition. In the sequel we therefore will use
the term transition function indiscriminately.

The computation of the transition functions relies on the fact that $\exp
\left( -S\left( \Psi \right) \right) $\ itself represents a statistical
weight for the system. Gosselin, Lotz, Wambst (2020) showed that $S\left(
\Psi \right) $\ can be modified in a straightforward manner to include
source terms:%
\begin{equation}
S\left( \Psi ,J\right) =S\left( \Psi \right) +\int \left( J\left( Z,\theta
\right) \Psi ^{\dag }\left( Z,\theta \right) +J^{\dag }\left( Z,\theta
\right) \Psi \left( Z,\theta \right) \right) d\left( Z,\theta \right)
\label{SwtS}
\end{equation}%
where $J\left( Z,\theta \right) $ is an arbitrary complex function, or
auxiliary field.

Introducing $J\left( Z,\theta \right) $ in $S\left( \Psi ,J\right) $ allows
to compute the transition functions by successive derivatives. Actually, we
can show that:%
\begin{equation}
G_{\alpha }\left( \underline{\left( \mathbf{Z},\theta \right) }^{\left[ k,%
\hat{k}\right] },\overline{\left( \mathbf{Z},\theta \right) }^{\left[ k,\hat{%
k}\right] }\right) =\left[ \prod\limits_{l=1}^{k}\left( \frac{\delta }{%
\delta J\left( \underline{\left( \mathbf{Z},\theta \right) }_{i_{l}}\right) }%
\frac{\delta }{\delta J^{\dag }\left( \overline{\left( \mathbf{Z},\theta
\right) }_{i_{l}}\right) }\right) \int \exp \left( -S\left( \Psi ,J\right)
\right) \mathcal{D}\Psi \mathcal{D}\Psi ^{\dag }\right] _{J=J^{\dag }=0}
\label{trnsgrtx}
\end{equation}%
where the notation $\mathcal{D}\Psi \mathcal{D}\Psi ^{\dag }$ denotes an
integration over the space of functions $\Psi \left( Z,\theta \right) $ and $%
\Psi ^{\dag }\left( Z,\theta \right) $, i.e. an integral in an infinite
dimensional space. Even though these integrals can only be computed in
simple cases, a series expansion of $G_{\alpha }\left( \underline{\left( 
\mathbf{Z},\theta \right) }^{\left[ k,\hat{k}\right] },\overline{\left( 
\mathbf{Z},\theta \right) }^{\left[ k,\hat{k}\right] }\right) $ can be found
using Feynman graphs techniques.

Once $G_{\alpha }\left( \underline{\left( \mathbf{Z},\theta \right) }^{\left[
k,\hat{k}\right] },\overline{\left( \mathbf{Z},\theta \right) }^{\left[ k,%
\hat{k}\right] }\right) $ is computed, the expression of $T_{t}\left( 
\underline{\left( \mathbf{Z},\theta \right) }^{\left[ k,\hat{k}\right] },%
\overline{\left( \mathbf{Z},\theta \right) }^{\left[ k,\hat{k}\right]
}\right) $ can be\ retrieved in principle by an inverse Laplace transform.
In field theory, formula (\ref{trnsgrtx}) shows that the transition functions%
\textbf{\ }(\ref{krvv}) are correlation functions of the field theory with
action $S\left( \Psi \right) $.

\section{Field-theoretic computations of transition functions}

The formula (\ref{trnsgrtx}) provides a precise and compact definition of
the transition functions for multiple agents in the system. However, in
practice, this formula is not directly applicable and does not shed much
light on the connection between the collective and microeconomic aspects of
the considered system. To calculate the dynamics of the agents, we will
proceed in three steps.\textbf{\ }

Firstly, we will minimize the system's action functional and determine the
background field, which represents the collective state of the system. Once
the background field is found, we will perform a series expansion of the
action functional around this background field, referred to as the effective
action of the system. It is with this effective action that we can compute
the transition functions for the state defined by the background field. We
will discover that each term in this expansion has an interpretation in
terms of a transition function.

Instead of directly computing the transition functions, we can consider a
series expansion of the action functional around a specific background field
of the system.

\subsection{Step 1: finding the background field}

For a particular type of agent, background fields are defined as the fields $%
\Psi _{0}\left( Z,\theta \right) $\ that maximize the statistical weight $%
\exp \left( -S\left( \Psi \right) \right) $ or, alternatively, minimize $%
S\left( \Psi \right) $: 
\begin{equation*}
\frac{\delta S\left( \Psi \right) }{\delta \Psi }\mid _{\Psi _{0}\left(
Z,\theta \right) }=0\text{, }\frac{\delta S\left( \Psi ^{\dag }\right) }{%
\delta \Psi ^{\dag }}\mid _{\Psi _{0}^{\dag }\left( Z,\theta \right) }=0
\end{equation*}%
The field $\Psi _{0}\left( Z,\theta \right) $\ represents the most probable
configuration, a specfic state of the entire system that influences the
dynamics of agents. It serves as the background state from which probability
transitions and average values can be computed. As we will see, the agents'
transitions explicitely depend on the chosen background field $\Psi
_{0}\left( Z,\theta \right) $, which represents the macroeconomic state in
which the agents evolve.

When considering two or more types of agents, the background field satisfies
the following condition:%
\begin{eqnarray*}
\frac{\delta S\left( \Psi ,\hat{\Psi}\right) }{\delta \Psi } &\mid &_{\Psi
_{0}\left( Z,\theta \right) }=0\text{, }\frac{\delta S\left( \Psi ,\hat{\Psi}%
\right) }{\delta \Psi ^{\dag }}\mid _{\Psi _{0}^{\dag }\left( Z,\theta
\right) }=0 \\
\frac{\delta S\left( \Psi ,\hat{\Psi}\right) }{\delta \hat{\Psi}} &\mid &_{%
\hat{\Psi}_{0}\left( Z,\theta \right) }=0\text{, }\frac{\delta S\left( \Psi ,%
\hat{\Psi}\right) }{\delta \hat{\Psi}^{\dag }}\mid _{\hat{\Psi}_{0}^{\dag
}\left( Z,\theta \right) }=0
\end{eqnarray*}

\subsection{Step 2: Series expansion around the background field}

In a given background state, the \emph{effective action}\footnote{%
Actually, this paper focuses on the \emph{classical} \emph{effective action}%
, which is an approximation sufficient for the computations at hand.} is the
series expansion of the field functional $S\left( \Psi \right) $\ around $%
\Psi _{0}\left( Z,\theta \right) $. We will present the expansion for one
type of agent, but generalizing it to two or several agents is
straightforward.

The series expansion around the background field simplifies the computations 
\textbf{of transition functions} and provides an interpretation of these
functions in terms of individual interactions within the collective state.\
To perform this series expansion, we decompose $\Psi $\ as: 
\begin{equation*}
\Psi =\Psi _{0}+\Delta \Psi
\end{equation*}%
and write the series expansion of the action functional: 
\begin{eqnarray}
S\left( \Psi \right) &=&S\left( \Psi _{0}\right) +\int \Delta \Psi ^{\dag
}\left( Z,\theta \right) O\left( \Psi _{0}\left( Z,\theta \right) \right)
\Delta \Psi \left( Z,\theta \right)  \label{SRP} \\
&&+\sum_{k>2}\int \prod\limits_{i=1}^{k}\Delta \Psi ^{\dag }\left(
Z_{i},\theta \right) O_{k}\left( \Psi _{0}\left( Z,\theta \right) ,\left(
Z_{i}\right) \right) \prod\limits_{i=1}^{k}\Delta \Psi \left( Z_{i},\theta
\right)  \notag
\end{eqnarray}%
The series expansion can be interpreted economically as follows. The first
term,\textbf{\ }$S\left( \Psi _{0}\right) $, describes the system of all
agents in a given macroeconomic state, $\Psi _{0}$. The other terms
potentially describe all the fluctuations or movements of the agents around
this macroeconomic state considered as given. Therefore, the expansion
around the background field represents the microeconomic reality of a
historical macroeconomic state. More precisely, it describes the range of
microeconomic possibilities allowed by a macroeconomic state.

The quadratic approximation is the first term of the expansion and can be
written as:%
\begin{equation}
S\left( \Psi \right) =S\left( \Psi _{0}\right) +\int \Delta \Psi ^{\dag
}\left( Z,\theta \right) O\left( \Psi _{0}\left( Z,\theta \right) \right)
\Delta \Psi \left( Z,\theta \right)  \label{kr}
\end{equation}%
\textbf{\ }It will allow us to find the transition functions of agents in
the historical macro state, where all interactions are averaged. The other
terms of the expansion allow us to detail the interactions within the
nebula, and are written as follows:%
\begin{equation*}
\sum_{k>2}\int \prod\limits_{i=1}^{k}\Delta \Psi ^{\dag }\left( Z_{i},\theta
\right) O_{k}\left( \Psi _{0}\left( Z,\theta \right) ,\left( Z_{i}\right)
\right) \prod\limits_{i=1}^{k}\Delta \Psi \left( Z_{i},\theta \right)
\end{equation*}%
\textbf{\ }They detail, given the historical macroeconomic state, how the
interactions of two or more agents can impact the dynamics of these agents.
Mathematically, this corresponds to correcting the transition probabilities
calculated in the quadratic approximation.

Here, we provide an interpretation of the third and fourth-order terms.

The third order introduces possibilities for an agent, during its
trajectory, to split into two, or conversely, for two agents to merge into
one. In other words, the third-order terms take into account or reveal, in
the historical macroeconomic environment, the possibilities for any agent to
undergo modifications along its trajectory. However, this assumption has
been excluded from our model.

The fourth order reveals that in the macroeconomic environment, due to the
presence of other agents and their tendency to occupy the same space, all
points in space will no longer have the same probabilities for an agent. In
fact, the fourth-order terms reveal the notion of geographical or sectoral
competition and potentially intertemporal competition. These terms describe
the interaction between two agents crossing paths, which leads to a
deviation of their trajectories due to the interaction.

We do not interpret higher-order terms, but the idea is similar. The
even-order terms (2n) describe interactions among n agents that modify their
trajectories. The odd-order terms modify the trajectories but also include
the possibility of agents reuniting or splitting into multiple agents. We
will see in more detail how these terms come into play in the transition
functions.

\subsection{Step 3: Computation of the transition functions}

\subsubsection{Quadratic approximation}

In the first approximation, transition functions in a given background field%
\textbf{\ }$\Psi _{0}\left( Z,\theta \right) $\textbf{\ }can be computed by
replacing $S\left( \Psi \right) $\ in (\ref{trnsgrtx}), with its quadratic
approximation (\ref{kr}). In formula (\ref{kr}), $O\left( \Psi _{0}\left(
Z,\theta \right) \right) $\ is a differential operator of second order. This
operator depends explicitly on $\Psi _{0}\left( Z,\theta \right) $. As a
consequence, transition functions and agent dynamics explicitly depend on
the collective state of the system.\ In this approximation, the formula for
the transition functions (\ref{trnsgrtx}) becomes:\ 
\begin{eqnarray}
G_{\alpha }\left( \underline{\left( Z,\theta \right) }^{\left[ k\right] },%
\overline{\left( Z,\theta \right) }^{\left[ k\right] }\right) &=&\left[
\prod\limits_{l=1}^{k}\left( \frac{\delta }{\delta J\left( \underline{\left(
Z,\theta \right) }_{i_{l}}\right) }\frac{\delta }{\delta J^{\dag }\left( 
\overline{\left( Z,\theta \right) }_{i_{l}}\right) }\right) \right. \\
&&\times \left. \int \exp \left( -\int \Delta \Psi ^{\dag }\left( Z,\theta
\right) O\left( \Psi _{0}\left( Z,\theta \right) \right) \Delta \Psi \left(
Z,\theta \right) \right) \mathcal{D}\Psi \mathcal{D}\Psi ^{\dag }\right]
_{J=J^{\dag }=0}  \notag
\end{eqnarray}%
Using this formula, we can show that the one-agent transition function is
given by:%
\begin{equation}
G_{\alpha }\left( \underline{\left( Z,\theta \right) }^{\left[ 1\right] },%
\overline{\left( Z,\theta \right) }^{\left[ 1\right] }\right) =O^{-1}\left(
\Psi _{0}\left( Z,\theta \right) \right) \left( \underline{\left( Z,\theta
\right) }^{\left[ 1\right] },\overline{\left( Z,\theta \right) }^{\left[ 1%
\right] }\right)  \label{rk}
\end{equation}%
where: 
\begin{equation*}
O^{-1}\left( \Psi _{0}\left( Z,\theta \right) \right) \left( \underline{%
\left( Z,\theta \right) }^{\left[ 1\right] },\overline{\left( Z,\theta
\right) }^{\left[ 1\right] }\right)
\end{equation*}%
\ is the kernel of the inverse operator $O^{-1}\left( \Psi _{0}\left(
Z,\theta \right) \right) $. It can be seen as the $\left( \underline{\left(
Z,\theta \right) }^{\left[ 1\right] },\overline{\left( Z,\theta \right) }^{%
\left[ 1\right] }\right) $\ matrix element of $O^{-1}\left( \Psi _{0}\left(
Z,\theta \right) \right) $\footnote{%
The differential operator $O\left( \Psi _{0}\left( Z,\theta \right) \right) $
can be seen as an infinite dimensional matrix indexed by the double
(infinite) entries $\left( \underline{\left( Z,\theta \right) }^{\left[ 1%
\right] },\overline{\left( Z,\theta \right) }^{\left[ 1\right] }\right) $.
With this description, the kernel $O^{-1}\left( \Psi _{0}\left( Z,\theta
\right) \right) \left( \underline{\left( Z,\theta \right) }^{\left[ 1\right]
},\overline{\left( Z,\theta \right) }^{\left[ 1\right] }\right) $ is the $%
\left( \underline{\left( Z,\theta \right) }^{\left[ 1\right] },\overline{%
\left( Z,\theta \right) }^{\left[ 1\right] }\right) $ element of the inverse
matrix.}.

More generally, the $k$-agents transition functions are the product of
individual transition functions:%
\begin{equation}
G_{\alpha }\left( \underline{\left( Z,\theta \right) }^{\left[ k\right] },%
\overline{\left( Z,\theta \right) }^{\left[ k\right] }\right)
=\prod\limits_{i=1}^{k}G_{\alpha }\left( \underline{\left( Z,\theta \right)
_{i}}^{\left[ 1\right] },\overline{\left( Z,\theta \right) _{i}}^{\left[ 1%
\right] }\right)  \label{gnr}
\end{equation}%
The above formula shows that in the quadratic approximation, the transition
probability from one state to another for a group is the product of
individual transition probabilities. In this approximation, the trajectories
of these agents are therefore independent. The agents do not interact with
each other and only interact with the environment described by the
background field.

The quadratic approximation must be corrected to account for individual
interactions within the group, by including higher-order terms in the
expansion of the action.

\subsubsection{Higher-order corrections}

To compute the effects of interactions between agents of a given group, we
consider terms of order greater than $2$ in the effective action. These
terms\ modify the transition functions. Writing the expansion:%
\begin{equation*}
\exp \left( -S\left( \Psi \right) \right) =\exp \left( -\left( S\left( \Psi
_{0}\right) +\int \Delta \Psi ^{\dag }\left( Z,\theta \right) O\left( \Psi
_{0}\left( Z,\theta \right) \right) \right) \right) \left(
1+\sum_{n\geqslant 1}\frac{A^{n}}{n!}\right)
\end{equation*}%
where:%
\begin{equation*}
A=\sum_{k>2}\int \prod\limits_{i=1}^{k}\Delta \Psi ^{\dag }\left(
Z_{i},\theta \right) O\left( \Psi _{0}\left( Z,\theta \right) ,\left(
Z_{i}\right) \right) \prod\limits_{i=1}^{k}\Delta \Psi \left( Z_{i},\theta
\right)
\end{equation*}%
is the sum of all possible interaction terms, leads to the series expansion
of (\ref{trnsgrtx}):%
\begin{eqnarray}
G_{\alpha }\left( \underline{\left( Z,\theta \right) }^{\left[ k\right] },%
\overline{\left( Z,\theta \right) }^{\left[ k\right] }\right) &=&\left[
\prod\limits_{l=1}^{k}\left( \frac{\delta }{\delta J\left( \underline{\left(
Z,\theta \right) }_{i_{l}}\right) }\frac{\delta }{\delta J^{\dag }\left( 
\overline{\left( Z,\theta \right) }_{i_{l}}\right) }\right) \right.
\label{trg} \\
&&\left. \int \exp \left( -\int \Delta \Psi ^{\dag }\left( Z,\theta \right)
O\left( \Psi _{0}\left( Z,\theta \right) \right) \Delta \Psi \left( Z,\theta
\right) \right) \left( 1+\sum_{n\geqslant 1}\frac{A^{n}}{n!}\right) \mathcal{%
D}\Psi \mathcal{D}\Psi ^{\dag }\right] _{J=J^{\dag }=0}  \notag
\end{eqnarray}%
These corrections can be computed using graphs' expansion. \ 

More precisely, the first term of the series:%
\begin{equation}
\left[ \prod\limits_{l=1}^{k}\left( \frac{\delta }{\delta J\left( \underline{%
\left( Z,\theta \right) }_{i_{l}}\right) }\frac{\delta }{\delta J^{\dag
}\left( \overline{\left( Z,\theta \right) }_{i_{l}}\right) }\right) \int
\exp \left( -\int \Delta \Psi ^{\dag }\left( Z,\theta \right) O\left( \Psi
_{0}\left( Z,\theta \right) \right) \Delta \Psi \left( Z,\theta \right)
\right) \mathcal{D}\Psi \mathcal{D}\Psi ^{\dag }\right] _{J=J^{\dag }=0}
\end{equation}%
is a transition function in the quadratic approximation. The other
contributions of the series expansion correct the approximated $n$ agents
transtns functions (\ref{gnr}).

Typically a contribution:%
\begin{eqnarray}
G_{\alpha }\left( \underline{\left( Z,\theta \right) }^{\left[ k\right] },%
\overline{\left( Z,\theta \right) }^{\left[ k\right] }\right) &=&\left[
\prod\limits_{l=1}^{k}\left( \frac{\delta }{\delta J\left( \underline{\left(
Z,\theta \right) }_{i_{l}}\right) }\frac{\delta }{\delta J^{\dag }\left( 
\overline{\left( Z,\theta \right) }_{i_{l}}\right) }\right) \right. \\
&&\left. \int \exp \left( -\int \Delta \Psi ^{\dag }\left( Z,\theta \right)
O\left( \Psi _{0}\left( Z,\theta \right) \right) \Delta \Psi \left( Z,\theta
\right) \right) \frac{A^{n}}{n!}\mathcal{D}\Psi \mathcal{D}\Psi ^{\dag }%
\right] _{J=J^{\dag }=0}  \notag
\end{eqnarray}%
can be depicted by a graph. The power $\frac{A^{n}}{n!}$ translates that
agents interact $n$ times along their path. The trajectories of each agent
of the group is broken $n$ times between its initial and final points. At
each time of interaction the trajectories of agents are deviated. To such a
graph is associated a probility that modifies the quadratic approximation
transition functions.

In the sequel we will only focus on the first order corrections to the
two-agents transition functions.

\section{Application to a microeconomic framework}

We will now present the microeconomic framework that will be turned into a
field model using our general method. We first describe the microeconomic
model, then derive the associated minimization function and the statistical
weight associated to the $N$ agents' set of trajectories. We will then
translate the minimization functions into an action functional and obtain
the statitical weight for the field model associated to the initial
microeconomic framework.

\subsection{Microeconomic setup}

To picture the interactions between the real and the financial economy, we
will consider two groups of agents, producers, and investors. In the
following, we will refer to producers or firms $i$ indistinctively, and use
the upper script $\symbol{94}$ for variables describing investors.

\subsubsection{Firms}

Producers are a set of firms operating each in a single sector, so that a
single firm with subsidiaries in different countries and/or offering
differentiated products can be modeled as a set of independent firms.
Similarly, a sector refers to a set of firms with similar productions, so
that sectors can be decomposed into sectors per country to account for local
specificities, or in several sectors.

Firms move across a vector space of sectors, which is of arbitrary
dimension. Firms are defined by their respective sector $X_{i}$\ and
physical capital $K_{i}$, two variables subject to dynamic changes. They may
change their capital stocks over time or altogether shift sectors.

Each firm produces a single differentiated good.\ However, in the following,
we will merely consider the return each producer may provide to its
investors.

The return of producer $i$ at time $t$, denoted $r_{i}$, depends on $K_{i}$, 
$X_{i}$ and on the level of competition in the sector.\ It is written: 
\begin{equation}
r_{i}=r\left( K_{i},X_{i}\right) -\gamma \sum_{j}\delta \left(
X_{i}-X_{j}\right) \frac{K_{j}}{K_{i}}  \label{dvd}
\end{equation}%
where $\delta \left( X_{i}-X_{j}\right) $ is the Dirac $\delta $ function
which is equal to $0$ for \ $X_{i}\neq X_{j}$.\ \ The first term in formula (%
\ref{dvd}) is an arbitrary function that depends on the sector and the level
of capital per firm in this sector.\ It represents the return of capital in
a specific sector $X_{i}$ under no competition. We deliberately keep the
form of $r\left( K_{i},X_{i}\right) $\ unspecified, since most of the
results of the model rely on the general properties of the functions
involved. When needed, we will give a standard Cobb-Douglas form to\ the
returns $r\left( K_{i},X_{i}\right) $.\ The second term in (\ref{dvd}) is
the decreasing return of capital. In any given sector, it is proportional to
both the number of competitors and the specific level of capital per firm
used.

We also assume that, for all $i$, firm $i$ has a market valuation defined by
both its price, $P_{i}$, and the variation of this price on financial
markets, $\dot{P}_{i}$.\ This variation is itself assumed to be a function
of an expected long-term return denoted $R_{i}$, or more precisely the
relative return $\bar{R}_{i}$\ of firm $i$\ against the whole set of firms: 
\begin{equation}
\frac{\dot{P}_{i}}{P_{i}}=F_{1}\left( \bar{R}_{i},\frac{\dot{K}_{i}}{K_{i}}%
\right)  \label{pr}
\end{equation}%
Formula (\ref{pr}) includes the main features of models of price dynamics.
In this equation, the time dependency of variables is implicit. Formula (\ref%
{pr}) reflects the impact of capital and location on the price of the firm
through its expected returns.\ It also reflects how variations in capital
impact its growth prospects, through competition and dividends (see (\ref%
{dvd})). Actually, the higher the capital of the firm, the lower impact of
competition and the higher the dividends.

We assume\textbf{\ }$R_{i}$ to have the general form:%
\begin{equation*}
R_{i}=R\left( K_{i},X_{i}\right)
\end{equation*}%
Expected long-term returns depend on the capital and sector in which the
firm operates, but also on external parameters, such as technology, ...
which are encoded in the shape of $R\left( K_{i},X_{i}\right) $.

The relative return $\bar{R}_{i}$ arising in (\ref{pr}) is defined by:%
\begin{equation}
\bar{R}_{i}=\bar{R}\left( K_{i},X_{i}\right) =\frac{R_{i}}{\sum_{l}R_{l}}
\label{RBR}
\end{equation}%
The function $F_{1}$ in (\ref{pr}) is arbitrary and reflects the preferences
of the market relatively to the firm's relative returns. We assume that
firms relocate in the sector space according to returns, in the direction of
the gradient of the expected long-term returns $R\left( K_{i},X_{i}\right) $%
, so that they chose the location $X_{i}$\ that minimizes at each continuous
time $t$\ the objective function:%
\begin{equation}
L_{i}\left( X_{i},\dot{X}_{i}\right) =\left( \dot{X}_{i}-\nabla
_{X}R_{i}H\left( K_{i}\right) \right) ^{2}+\tau \frac{K_{X_{i}}}{K_{i}}%
\sum_{j}\delta \left( X_{i}-X_{j}\right)  \label{dnp}
\end{equation}%
where $\dot{X}_{i}$\ stands for the continuous version of the discrete
variation, $X_{i}\left( t+1\right) -X_{i}\left( t\right) $, and $\delta
\left( X_{i}-X_{j}\right) $\ is again the Dirac-$\delta $-function. $%
K_{X_{i}}$ is the average capital per firm in sector $X_{i}$. The constant $%
\tau $\ measures the level of competition between firms, and describes the
cost incured to settling in a new sector\footnote{%
Formula (\ref{dnp}) represents the continuous version of the following
objective function, where the firm aims to maximize its expected long-term
revenue at each period:%
\begin{equation}
L=R\left( K_{i},X_{i}\right) -R\left( K_{i},X_{i}-\delta X_{i}\right) -\frac{%
1}{2}\left( \frac{1}{H\left( K_{i}\right) }\delta X_{i}\right) ^{2}+\tau 
\frac{K_{X_{i}}}{K_{i}}\sum_{j}\delta \left( X_{i}-X_{j}\right)  \label{shf}
\end{equation}%
Here, $X_{i}-\delta X_{i}$\ denotes the position of the agent in the
previous period. The term $\left( \frac{1}{H\left( K_{i}\right) }\delta
X_{i}\right) ^{2}$\ represents the cost of changing sector. The function $%
H\left( K_{i}\right) $\ is increasing with $K_{i}$,\ so that the higher the
firm's capital, the larger the shift. The term $\tau \sum_{j}\delta \left(
X_{i}-X_{j}\right) $ represents an externality increasing the cost of
shifting proportionally to the number of firms in sector $X_{i}$. The
accumulation of agents at any point in the space creates a repulsive force
that slows down the shift. In continuous time, formula (\ref{shf}) becomes
equivalent to formula (\ref{dnp}), up to a constant term.}. The inclusion of
the factor $\frac{K_{X_{i}}}{K_{i}}$ models that the\textbf{\ }lower a
firm's capital is compared to the sector average, the stronger the effect of
competition

Actually, when $\tau =0$, there are no repulsive forces and the move towards
the gradient of $R$ is given by the expression:

\begin{equation*}
\dot{X}_{i}=\nabla _{X_{i}}R_{i}H\left( K_{i}\right)
\end{equation*}%
When $\tau \neq 0$, repulsive forces deviate the trajectory.\ The dynamic
equation associated to the minimization of (\ref{dnp}) is given by the
general formula of the dynamic optimization:%
\begin{equation}
\frac{d}{dt}\frac{\partial }{\partial \dot{X}_{i}}L_{i}\left( X_{i},\dot{X}%
_{i}\right) =\frac{\partial }{\partial X_{i}}L_{i}\left( X_{i},\dot{X}%
_{i}\right)  \label{lgd}
\end{equation}%
This last equation does not need to be developed further, since formula (\ref%
{dnp}) is sufficient to switch to the field description of the system.

\subsubsection{Investors}

Each investor $j$ is defined by his level of capital $\hat{K}_{j}$ and his
position $\hat{X}_{j}$ in the sector space. Investors can invest in the
entire sector space, but tend to invest in sectors close to their position.

Besides, investors tend to diversify their capital: each investor $j$ chose
to allocate parts of his entire capital $\hat{K}_{j}$ between various firms $%
i$. The capital allocated by investor $j$ to firm $i$ is denoted $\hat{K}%
_{j}^{\left( i\right) }$, and given by:

\begin{equation}
\hat{K}_{j}^{\left( i\right) }\left( t\right) =\left( \hat{F}_{2}\left(
R_{i},\hat{X}_{j}\right) \hat{K}_{j}\right) \left( t\right)  \label{grandf2}
\end{equation}%
where:%
\begin{equation}
\hat{F}_{2}\left( R_{i},\hat{X}_{j}\right) =\frac{F_{2}\left( R_{i}\right)
G\left( X_{i}-\hat{X}_{j}\right) }{\sum_{l}F_{2}\left( R_{l}\right) G\left(
X_{l}-\hat{X}_{j}\right) }  \label{FRLV}
\end{equation}%
The function $F_{2}$ is arbitrary. It depends on the expected return of firm 
$i$ and on the distance between sectors $X_{i}$ and $\hat{X}_{j}$. The
function $\hat{F}_{2}\left( R\left( K_{i},X_{i}\right) ,\hat{X}_{j}\right) $%
\ is the relative version of $F_{2}$ and translates the dependency of
investments on firms' relative attractivity. Equation (\ref{grandf2}) is a
general form of risk averse portfolio allocation\footnote{%
Actually, an investor allocating capital exclusively in a sector $X_{i}$ and
optimizing the function:%
\begin{equation}
\frac{R_{i}}{\sum_{l}R_{l}}s_{j}-s_{j}^{2}Var\left( \frac{R_{i}}{%
\sum_{l}R_{l}}\right)  \label{tlr}
\end{equation}%
where the share $s_{j}$\ satisfies $\sum s_{j}=1$,\ will set $s_{j}=\frac{%
R_{i}}{\sum_{l}R_{l}}$. If we were to introduce the possibility of investing
in multiple sectors and consider more general preferences than this simple
quadratic function, we should introduce the functions $G\left( X_{i}-\hat{X}%
_{j}\right) $\ and $F_{2}\left( R_{i}\right) $\ in the solutions of (\ref%
{tlr}), leading to (\ref{grandf2}).}\textbf{.}

We now define $\varepsilon $ the time scale for capital accumulation.\ The
variation of capital\ of investor $j$ between $t$\ and $t+\varepsilon $\ is
the sum of two terms: the short-term returns $r_{i}$\ of the firms\ in which 
$j$\ invested, and the stock price variations of these same firms:

\begin{equation}
\hat{K}_{j}\left( t+\varepsilon \right) -\hat{K}_{j}\left( t\right)
=\sum_{i}\left( r_{i}+\frac{\dot{P}_{i}}{P_{i}}\right) \hat{K}_{j}^{\left(
i\right) }=\sum_{i}\left( r_{i}+F_{1}\left( \bar{R}_{i},\frac{\dot{K}%
_{i}\left( t\right) }{K_{i}\left( t\right) }\right) \right) \hat{K}%
_{j}^{\left( i\right) }  \label{fsn}
\end{equation}%
Incidentally, note that in equation (\ref{dnp}), the time scale of motions
within the sectors space was normalized to one. Here, on the contrary, we
define this motion time scale as $\varepsilon $, and assume $\varepsilon <<1$%
: the mobility in the sector space is lower than capital dynamics. To
rewrite (\ref{fsn}) on the same time-span as $\frac{dX_{i}}{dt}$, we write:

\begin{eqnarray*}
\hat{K}_{j}\left( t+1\right) -\hat{K}_{j}\left( t\right) &=&\sum_{k=1}^{%
\frac{1}{\varepsilon }}\hat{K}_{j}\left( t+k\varepsilon \right) -\hat{K}%
_{j}\left( t\right) \\
&=&\sum_{k=1}^{\frac{1}{\varepsilon }}\sum_{i}\left( r_{i}+\frac{\dot{P}_{i}%
}{P_{i}}\right) \hat{K}_{j}^{\left( i\right) }\left( t+k\varepsilon \right)
\\
&\simeq &\frac{1}{\varepsilon }\sum_{i}\left( r_{i}+F_{1}\left( \bar{R}_{i},%
\frac{\dot{K}_{i}\left( t\right) }{K_{i}\left( t\right) }\right) \right) 
\hat{K}_{j}^{\left( i\right) }
\end{eqnarray*}%
where the quantities in the sum have to be understood as averages over the
time span $\left[ t,t+1\right] $. Using equation (\ref{pr}), equation (\ref%
{fsn}) becomes in the continuous approximation: 
\begin{equation}
\frac{d}{dt}\hat{K}_{j}=\frac{1}{\varepsilon }\sum_{i}\left(
r_{i}+F_{1}\left( \frac{R_{i}}{\sum_{l}\delta \left( X_{l}-X_{i}\right) R_{l}%
},\frac{\dot{K}_{i}}{K_{i}}\right) \right) \hat{F}_{2}\left( R_{i},\hat{X}%
_{j}\right) \hat{K}_{j}  \label{nfs}
\end{equation}%
where $\frac{d}{dt}\hat{K}_{j}\left( t\right) =\hat{K}_{j}\left( t+1\right) -%
\hat{K}_{j}\left( t\right) $\ is now normalized to the time scale of $\frac{%
dX_{i}}{dt}$, i.e. $1$.

\subsubsection{Link between financial and physical capital}

The entire financial capital is, at any time, completely allocated by
investors between firms. For producers, there is no alternative source of
financing.\ Self-financing is discarded, since it would amount to
considering a producer and an investor as a single agent. The physical
capital of a any given firm is thus the sum of all capital allocated to this
specific firm by all its investors. Physical capital entirely depends on the
financial arbitrage of the financial sector. Firms do not own their capital:
they return it fully at the end of each period with a dividend, though
possibly negative. Investors then entirely reallocate their capital between
firms at the beginning of the next period. This generalisation of the
dividend irrelevance theory may not be fully accurate in the short-run but
holds in the long-run, since physical capital cannot last long without
investment.\ When investors choose not to finance a firm, this firm is bound
to disappear in the long run. Under these assumptions, the following
identity holds\textbf{:}%
\begin{equation}
K_{i}\left( t+\varepsilon \right) =\sum_{j}\hat{K}_{j}^{\left( i\right)
}=\sum_{j}\hat{F}_{2}\left( R_{i},\hat{X}_{j}\left( t\right) \right) \hat{K}%
_{j}\left( t\right)  \label{phs}
\end{equation}%
where $K_{i}$ stands for the physical capital of firm $i$ at time $t$, and $%
\sum_{j}\hat{K}_{j}^{\left( i\right) }$ for the sum of capital invested in
firm $i$ by investors $j$. Recall that the parameter $\varepsilon $ accounts
for the specific time scale of capital accumulation.\ It differs from that
of mobility within the sector space (\ref{dnp}), which is normalized to one.

The dynamic equation (\ref{phs}) rewrites:%
\begin{equation}
\frac{K_{i}\left( t+\varepsilon \right) -K_{i}\left( t\right) }{\varepsilon }%
=\frac{1}{\varepsilon }\left( \sum_{j}\hat{F}_{2}\left( R_{i},\hat{X}%
_{j}\left( t\right) \right) \hat{K}_{j}\left( t\right) -K_{i}\left( t\right)
\right)  \label{dnK}
\end{equation}%
Using the same token as in the derivation of (\ref{nfs}), its continuous
approximation becomes:%
\begin{equation}
\frac{d}{dt}K_{i}\left( t\right) +\frac{1}{\varepsilon }\left( K_{i}\left(
t\right) -\sum_{j}\hat{F}_{2}\left( R_{i},\hat{X}_{j}\left( t\right) \right) 
\hat{K}_{j}\left( t\right) \right) =0  \label{dnk}
\end{equation}%
where $\frac{d}{dt}K_{i}\left( t\right) $\ stands for $K_{i}\left(
t+1\right) -K_{i}\left( t\right) $.

\subsubsection{Capital allocation dynamics}

Investors allocate their capital among sectors, and may adjust their
portfolio to the returns of the sector or firms they are invested in. This
is modelled by a move along the sectors' space in the direction of the
gradient of $R\left( K_{i},X_{i}\right) $.

Investor $j$ capital reallocation is described by a dynamic equation for $%
\hat{X}_{j}$ :

\begin{equation}
\frac{d}{dt}\hat{X}_{j}-\frac{1}{\sum_{i}\delta \left( X_{i}-\hat{X}%
_{j}\right) }\sum_{i}\left( \nabla _{\hat{X}}F_{0}\left( R\left( K_{i},\hat{X%
}_{j}\right) \right) +\nu \nabla _{\hat{X}}F_{1}\left( \bar{R}\left( K_{i},%
\hat{X}_{j}\right) ,\frac{\dot{K}_{i}}{K_{i}}\right) \right) =0  \label{prd}
\end{equation}%
The factor $\sum_{i}\delta \left( X_{i}-\hat{X}_{j}\right) $ is the firms'
density in sector $\hat{X}_{j}$.\ The more competitors in a sector, the
slower the reallocation. The term $\nabla _{\hat{X}}F_{0}\left( R\left(
K_{i},\hat{X}_{j}\right) \right) $ models the investors' propensity to
reallocate in sectors with highest long-term returns. The term $\nu \nabla _{%
\hat{X}}F_{1}\left( \bar{R}\left( K_{i},\hat{X}_{j}\right) \right) $
describes the investors' preference for stocks displaying the highest
price's increase\footnote{%
As for (\ref{dnp}), equation (\ref{prd}) can be justified by an objective
function that would depend on returns and that includes a cost for any
sector shift, to translate the loss of information and connections induced
by the shift.}.

\subparagraph{Remark on the model}

Note that consumers do not appear in this framework. This choice of a
partial model is deliberate and aims to focus solely on the allocation of
capital among a large number of investors and firms.

In the framework of the representative agent, this choice of a partial,
consumer-free model could be problematic. A model with a single firm and one
investor would be useless: capital would be invested in the single firm and
would grow through dividends. In this context, introducing a consumer that
can arbitrage between consumption and long-term returns becomes relevant.

Here we rather study how investment varies and how capital is allocated
among multiple firms under varying exogeneous conditions, for which our
partial model will provide ample results\footnote{%
Adding a consumer would modify the model by introducing in (\ref{grandf2}) a
factor lower than one and depending on the sector long-term returns. This
factor directly arises from the arbitrage between investment and
consumption, but appears irrelevant in the present context. Additionally,
the firm's dividend would be influenced by the consumption of the good
produced by the firm. Once again, a general model would deviate from our
intended purpose.}.

\subsection{Minimization functions}

To find the statistical weight associated to the trajectories of the system,
we must write the minimization functions of the system's equations. Recall
that they are functions whose minimization yields the dynamic equations of
the system.

We have seen that the dynamics of the variable $X_{i}$ comes from the
minimization of the function:%
\begin{equation*}
\left( \frac{dX_{i}}{dt}-\nabla _{X}R\left( K_{i},X_{i}\right) H\left(
K_{i}\right) \right) ^{2}+\tau \frac{K_{X_{i}}}{K_{i}}\sum_{j}\delta \left(
X_{i}-X_{j}\right)
\end{equation*}%
We simply re-use this function and sum over the whole set of agents and
yield the minimization function for the capital allocation dynamics:

\begin{equation}
s_{1}=\sum_{i}\left( \frac{dX_{i}}{dt}-\nabla _{X}R\left( K_{i},X_{i}\right)
H\left( K_{i}\right) \right) ^{2}+\sum_{i}\tau \frac{K_{X_{i}}}{K_{i}}%
\sum_{j}\delta \left( X_{i}-X_{j}\right)  \label{minX}
\end{equation}%
The dynamics of $K_{i},$ $\hat{K}_{i}$ et $\hat{X}_{i}$ are not the result
of a minimization, but their associated quadratic functions (\ref{mNZ}) can
easily be found and yield the following minimization functions:

for the physical capital $K_{i}$:

\begin{equation}
s_{2}=\sum_{i}\left( \frac{d}{dt}K_{i}+\frac{1}{\varepsilon }\left(
K_{i}-\sum_{j}\hat{F}_{2}\left( R\left( K_{i}\left( t\right) ,X_{i}\left(
t\right) \right) ,\hat{X}_{j}\left( t\right) \right) \hat{K}_{j}\left(
t\right) \right) \right) ^{2}  \label{minK}
\end{equation}%
for the financial capital $\hat{K}_{i}$:%
\begin{equation}
s_{3}=\sum_{j}\left( \frac{d}{dt}\hat{K}_{j}-\frac{1}{\varepsilon }\left(
\sum_{i}\left( r_{i}+F_{1}\left( \frac{R\left( K_{i},X_{i}\right) }{%
\sum_{l}\delta \left( X_{l}-X_{i}\right) R\left( K_{l},X_{l}\right) },\frac{%
\dot{K}_{i}\left( t\right) }{K_{i}\left( t\right) }\right) \right) \hat{F}%
_{2}\left( R\left( K_{i}\left( t\right) ,X_{i}\left( t\right) \right) ,\hat{X%
}_{j}\left( t\right) \right) \hat{K}_{j}\right) \right) ^{2}
\label{minKchap}
\end{equation}%
and ultimately, for financial capital allocation $\hat{X}_{i}$:%
\begin{equation}
s_{4}=\sum_{i}\left( \frac{d}{dt}\hat{X}_{j}-\frac{1}{\sum_{i}\delta \left(
X_{i}-\hat{X}_{j}\right) }\sum_{i}\left( \nabla _{\hat{X}}F_{0}\left(
R\left( K_{i},\hat{X}_{j}\right) \right) +\nu \nabla _{\hat{X}}F_{1}\left( 
\bar{R}\left( K_{i},\hat{X}_{j}\right) \right) \right) \right) ^{2}
\label{minXchap}
\end{equation}%
As a consequence, the statistical weight associated to the trajectories is
simply:%
\begin{equation}
W\left( \left\{ K_{i}\left( t\right) ,X_{i}\left( t\right) \right\} ,\left\{ 
\hat{K}_{i}\left( t\right) ,\hat{X}_{j}\left( t\right) \right\} \right)
=\exp \left( -\int dt\left( s_{1}+s_{2}+s_{3}+s_{4}\right) \right)
\label{WGH}
\end{equation}

\subsection{Translation in terms of fields}

To translate the previous microeconomic framework into a field model, we
must translate the minimization functions (\ref{minX}), (\ref{minK}) for the
firms, and (\ref{minKchap}) and (\ref{minXchap}) for investors in terms of
four functionals of the fields \footnote{%
The term functional refers to a function of a function, i.e. a function
whose argument is itself a function.}. The sum of these four functionals is
the "field action functional"\footnote{%
Details about the probabilistic step will be given as a reminder along the
text and in appendix 1.} that determine the statistical weight for the field.

We will apply the general method developed in section 3 and start with
firms, by translating first (\ref{minX}) and (\ref{minK}).

\subsubsection{Real Economy}

In both\textbf{\ }capital allocation dynamics (\ref{minX}) and capital
accumulation dynamics (\ref{minK}), time derivatives appear.\ However, one
of them, equation (\ref{minX}), includes time-independent terms and is thus
of the form (\ref{mNZ}), the other, equation (\ref{minX}) is of the type (%
\ref{mnZ}). Based on the translation rules, appendix 1.3 computes the
translation of the various minimization functions.

Using the translation (\ref{bdt}) of (\ref{ntr})-type term, the minimization
function of physical capital allocation (\ref{minX}) translates into:%
\begin{eqnarray}
S_{1} &=&-\int \Psi ^{\dag }\left( K,X\right) \nabla _{X}\left( \frac{\sigma
_{X}^{2}}{2}\nabla _{X}-\nabla _{X}R\left( K,X\right) H\left( K\right)
\right) \Psi \left( K,X\right) dKdX  \label{SN} \\
&&+\tau \frac{K_{X}}{K}\int \left\vert \Psi \left( K^{\prime },X\right)
\right\vert ^{2}\left\vert \Psi \left( K,X\right) \right\vert ^{2}dK^{\prime
}dKdX  \notag
\end{eqnarray}%
where $K_{X}$ is the average capital per firm in sector $X$. We show below
that $K_{X}$ has the field expression:%
\begin{equation*}
K_{X}=\frac{\int K\left\vert \Psi \left( K,X\right) \right\vert ^{2}dK}{%
\left\Vert \Psi \left( X\right) \right\Vert ^{2}}
\end{equation*}

Similarly, the minimization function of the physical capital (\ref{minK}),
translates into:

\begin{equation}
S_{2}=-\int \Psi ^{\dag }\left( K,X\right) \nabla _{K}\left( \frac{\sigma
_{K}^{2}}{2}\nabla _{K}+\frac{1}{\varepsilon }\left( K-\int \hat{F}%
_{2}\left( R\left( K,X\right) ,\hat{X}\right) \hat{K}\left\vert \hat{\Psi}%
\left( \hat{K},\hat{X}\right) \right\vert ^{2}d\hat{K}d\hat{X}\right)
\right) \Psi \left( K,X\right)  \label{SD}
\end{equation}%
with:%
\begin{equation}
\hat{F}_{2}\left( R\left( K,X\right) ,\hat{X}\right) =\frac{F_{2}\left(
R\left( K,X\right) \right) G\left( X-\hat{X}\right) }{\int F_{2}\left(
R\left( K^{\prime },X^{\prime }\right) \right) G\left( X^{\prime }-\hat{X}%
\right) \left\vert \Psi \left( K^{\prime },X^{\prime }\right) \right\vert
^{2}d\left( K^{\prime },X^{\prime }\right) }  \label{KDR}
\end{equation}

\subsubsection{Financial markets}

The financial capital dynamics (\ref{minKchap}) and the financial capital
allocation (\ref{minXchap}) both include a time derivative and are thus of
type (\ref{edr}). The application of the translation rules is
straightforward.

Using the general translation formula of expression (\ref{inco}) in (\ref%
{Trl}), the minimization function (\ref{minKchap}) for the financial capital
dynamics translates into: 
\begin{eqnarray}
S_{3} &=&-\int \hat{\Psi}^{\dag }\left( \hat{K},\hat{X}\right) \nabla _{\hat{%
K}}\left( \frac{\sigma _{\hat{K}}^{2}}{2}\nabla _{\hat{K}}-\frac{\hat{K}}{%
\varepsilon }\int \left( r\left( K,X\right) -\gamma \frac{\int K^{\prime
}\left\Vert \Psi \left( K^{\prime },X\right) \right\Vert ^{2}}{K}\right.
\right.  \label{ST} \\
&&\left. \left. +F_{1}\left( \bar{R}\left( K,X\right) ,\Gamma \left(
K,X\right) \right) \right) \hat{F}_{2}\left( R\left( K,X\right) ,\hat{X}%
\right) \left\Vert \Psi \left( K,X\right) \right\Vert ^{2}d\left( K,X\right)
\right) \hat{\Psi}\left( \hat{K},\hat{X}\right)  \notag
\end{eqnarray}%
where:%
\begin{eqnarray}
\bar{R}\left( K,X\right) &=&\frac{R\left( K,X\right) }{\int R\left(
K^{\prime },X^{\prime }\right) \left\Vert \Psi \left( K^{\prime },X^{\prime
}\right) \right\Vert ^{2}d\left( K^{\prime },X^{\prime }\right) }
\label{RPN} \\
\Gamma \left( K,X\right) &=&\frac{\int \hat{F}_{2}\left( R\left( K,X\right) ,%
\hat{X}\right) \hat{K}\left\vert \hat{\Psi}\left( \hat{K},\hat{X}\right)
\right\vert ^{2}d\hat{K}d\hat{X}}{K}-1  \label{Gmm}
\end{eqnarray}

and the function for financial capital allocation (\ref{minXchap})
translates into:%
\begin{eqnarray}
S_{4} &=&-\int \hat{\Psi}^{\dag }\left( \hat{K},\hat{X}\right)  \label{SQ} \\
&&\times \left( \nabla _{\hat{X}}\sigma _{\hat{X}}^{2}\nabla _{\hat{X}}-\int
\left( \frac{\nabla _{\hat{X}}F_{0}\left( R\left( K,\hat{X}\right) \right)
+\nu \nabla _{\hat{X}}F_{1}\left( \bar{R}\left( K,X\right) ,\Gamma \left(
K,X\right) \right) }{\int \left\Vert \Psi \left( K^{\prime },\hat{X}\right)
\right\Vert ^{2}dK^{\prime }}\right) \left\Vert \Psi \left( K,\hat{X}\right)
\right\Vert ^{2}dK\right) \hat{\Psi}\left( \hat{K},\hat{X}\right)  \notag
\end{eqnarray}

\subsubsection{Fields action functional and statistical weight}

We can now find the statistical weight of any realization of the fields.
Actually, the action functional of the system is the sum of the
contributions (\ref{SN}),(\ref{SD}),(\ref{ST}),(\ref{SQ}):%
\begin{equation*}
S\left( \Psi ,\hat{\Psi}\right) =S_{1}+S_{2}+S_{3}+S_{4}
\end{equation*}%
and the field statistical weight for the realization $\left( \Psi ,\hat{\Psi}%
\right) $ is:%
\begin{equation*}
\exp \left( -S\left( \Psi ,\hat{\Psi}\right) \right)
\end{equation*}%
With no loss of generality, we can find a more compact form for the action $%
S $\ by assuming that investors invest in only one sector, so that:%
\begin{equation}
G\left( X-\hat{X}\right) =\delta \left( X-\hat{X}\right)  \label{smp}
\end{equation}%
We can thus write the action functional $S$: 
\begin{eqnarray}
S &=&-\int \Psi ^{\dag }\left( K,X\right) \left( \nabla _{X}\left( \frac{%
\sigma _{X}^{2}}{2}\nabla _{X}-\nabla _{X}R\left( K,X\right) H\left(
K\right) \right) -\tau \frac{K_{X}}{K}\left( \int \left\vert \Psi \left(
K^{\prime },X\right) \right\vert ^{2}dK^{\prime }\right) \right.  \label{fcn}
\\
&&+\left. \nabla _{K}\left( \frac{\sigma _{K}^{2}}{2}\nabla _{K}+u\left(
K,X,\Psi ,\hat{\Psi}\right) \right) \right) \Psi \left( K,X\right) dKdX 
\notag \\
&&-\int \hat{\Psi}^{\dag }\left( \hat{K},\hat{X}\right) \left( \nabla _{\hat{%
K}}\left( \frac{\sigma _{\hat{K}}^{2}}{2}\nabla _{\hat{K}}-\hat{K}f\left( 
\hat{X},\Psi ,\hat{\Psi}\right) \right) +\nabla _{\hat{X}}\left( \frac{%
\sigma _{\hat{X}}^{2}}{2}\nabla _{\hat{X}}-g\left( K,X,\Psi ,\hat{\Psi}%
\right) \right) \right) \hat{\Psi}\left( \hat{K},\hat{X}\right)  \notag
\end{eqnarray}%
where each line corresponds to one $S_{i}$ and where, to simplify, we have
defined:%
\begin{eqnarray}
u\left( K,X,\Psi ,\hat{\Psi}\right) &=&\frac{1}{\varepsilon }\left( K-\int 
\hat{F}_{2}\left( R\left( K,X\right) \right) \hat{K}\left\vert \hat{\Psi}%
\left( \hat{K},X\right) \right\vert ^{2}d\hat{K}\right)  \label{fcs} \\
f\left( \hat{X},\Psi ,\hat{\Psi}\right) &=&\frac{1}{\varepsilon }\int \left(
r\left( K,X\right) -\frac{\gamma \int K^{\prime }\left\vert \Psi \left(
K,X\right) \right\vert ^{2}}{K}+F_{1}\left( \bar{R}\left( K,X\right) ,\Gamma
\left( K,X\right) \right) \right) \\
&&\times \hat{F}_{2}\left( R\left( K,X\right) \right) \left\vert \Psi \left(
K,\hat{X}\right) \right\vert ^{2}dK  \label{fcS} \\
g\left( K,\hat{X},\Psi ,\hat{\Psi}\right) &=&\int \frac{\nabla _{\hat{X}%
}F_{0}\left( R\left( K,\hat{X}\right) \right) +\nu \nabla _{\hat{X}%
}F_{1}\left( \bar{R}\left( K,\hat{X}\right) ,\Gamma \left( K,X\right)
\right) }{\int \left\vert \Psi \left( K^{\prime },\hat{X}\right) \right\vert
^{2}dK^{\prime }}\left\vert \Psi \left( K,\hat{X}\right) \right\vert ^{2}dK
\label{fCS}
\end{eqnarray}%
The expression for $\bar{R}\left( K,X\right) $ is still given by (\ref{RPN}%
). Under our assumption, the functions $\hat{F}_{2}$ and\textbf{\ }$\Gamma $
become:%
\begin{eqnarray}
\hat{F}_{2}\left( R\left( K,X\right) \right) &=&\frac{F_{2}\left( R\left(
K,X\right) \right) }{\int F_{2}\left( R\left( K^{\prime },X\right) \right)
\left\vert \Psi \left( K^{\prime },X\right) \right\vert ^{2}dK^{\prime }}
\label{FH} \\
\Gamma \left( K,X\right) &=&\frac{\int \hat{F}_{2}\left( R\left( K,X\right)
\right) \hat{K}\left\vert \hat{\Psi}\left( \hat{K},X\right) \right\vert ^{2}d%
\hat{K}}{K}-1  \label{GM}
\end{eqnarray}

Despite its compact and abstract form, equation (\ref{fcn}) encompasses the
main elements of our microeconomic framework. Recall that function $H\left(
K_{X}\right) $\ encompasses the determinants of the firms' mobility across
the sector space. We will specify this function below as a function of
expected long term-returns and capital.

Function $u$\ describes the evolution of capital of a firm, located at $X$.
This dynamics depends on the relative value of a function $F_{2}$ that is
itself a function of the firms' expected returns $R\left( K,X\right) $.
Investors allocate their capital based on their expectations of the firms'
long-term returns.

Function $f$\ describes the returns of investors located at $\hat{X}$,\ and
investing in sector $X$\ a capital $K$. These returns depend on short-term
dividends $r\left( K,X\right) $, the field-equivalent cost of capital\textbf{%
\ }$\frac{\gamma \int K^{\prime }\left\Vert \Psi \left( K,X\right)
\right\Vert ^{2}}{K}$\textbf{,\ }and a function $F_{1}$ that depends on
firms' expected long-term stock valuations.\ These valuations themselves
depend on the relative attractivity of a firm expected long-term returns
vis-a-vis its competitors.

Function $g$\ describes investors' shifts across the sectors' space. They
are driven by the gradient of expected long-term returns and stocks
valuations, who themselves depend on the firms' relative expected long-term
returns.

Note that we do not introduce a time variable at this stage. Our purpose is
to find the collective states of the system, which can be considered as
static in a first step. It is only when we will study the impact of
exogeneous parameters on the collective states that we will introduce a
macro time scale.

\subsection{Field model and averages}

As detailed above, once the field action functional $S$ is found, we can use
field theory to study the system of agents, both at the collective and
individual levels. At the collective level we can compute the averages of
the system in a given background field. The individual level, described by
the transition functions will be studied in the third part.

Recall that the background fields emerging at the collective level are
particular functions, $\Psi \left( K,X\right) $\ and $\hat{\Psi}\left( \hat{K%
},\hat{X}\right) $, and their adjoints fields $\Psi ^{\dag }\left(
K,X\right) $\ and $\hat{\Psi}^{\dag }\left( \hat{K},\hat{X}\right) $,\ that
minimize the functional $S$.

Once the background fields are obtained, the associated number of firms and
investors per sector for a given average capital $K$ can be computed. They
are given by:%
\begin{equation}
\left\vert \Psi \left( K,X\right) \right\vert ^{2}=\Psi ^{\dag }\left(
K,X\right) \Psi \left( K,X\right)  \label{DSN}
\end{equation}%
and:%
\begin{equation}
\left\vert \hat{\Psi}\left( \hat{K},\hat{X}\right) \right\vert ^{2}=\hat{\Psi%
}^{\dag }\left( \hat{K},\hat{X}\right) \hat{\Psi}\left( \hat{K},\hat{X}%
\right)  \label{DST}
\end{equation}%
With these two density functions at hand, various average quantities\ in the
collective state can be computed.

The number of producers $\left\Vert \Psi \left( X\right) \right\Vert ^{2}$\
and investors $\left\Vert \hat{\Psi}\left( \hat{X}\right) \right\Vert ^{2}$
in sectors are computed using the formula:%
\begin{eqnarray}
\left\Vert \Psi \left( X\right) \right\Vert ^{2} &\equiv &\int \left\vert
\Psi \left( K,X\right) \right\vert ^{2}dK  \label{Nx} \\
\left\Vert \hat{\Psi}\left( \hat{X}\right) \right\Vert ^{2} &\equiv &\int
\left\vert \hat{\Psi}\left( \hat{K},\hat{X}\right) \right\vert ^{2}d\hat{K}
\label{Nxh}
\end{eqnarray}%
The total invested capital $\hat{K}_{X}$\ in sector $X$ is defined by a
partial average:%
\begin{equation}
\hat{K}_{\hat{X}}=\int \hat{K}\left\vert \hat{\Psi}\left( \hat{K},X\right)
\right\vert ^{2}d\hat{K}=\int \hat{K}\left\vert \hat{\Psi}\left( \hat{X}%
\right) \right\vert ^{2}d\hat{K}  \label{Khx}
\end{equation}%
and the average invested capital per firm in sector $X$ defined by:\textbf{\ 
}%
\begin{equation}
K_{X}=\frac{\int \hat{K}\left\vert \hat{\Psi}\left( \hat{K},X\right)
\right\vert ^{2}d\hat{K}}{\left\Vert \Psi \left( X\right) \right\Vert ^{2}}
\label{kx}
\end{equation}%
Note that, given our assumptions, the total physical capital is equal to the
total capital invested:%
\begin{equation*}
\int K\left\vert \Psi \left( K,X\right) \right\vert ^{2}dK=\int \hat{K}%
\left\vert \hat{\Psi}\left( \hat{K},\hat{X}\right) \right\vert ^{2}d\hat{K}
\end{equation*}%
so that $K_{X}$ is also equal to the average physical capital per firm for
sector $X$, i.e. :%
\begin{equation}
K_{X}=\frac{\int K\left\vert \Psi \left( K,X\right) \right\vert ^{2}dK}{%
\left\Vert \Psi \left( X\right) \right\Vert ^{2}}  \label{KX}
\end{equation}%
\textbf{\ }In the following, we will use both expressions (\ref{kx}) or (\ref%
{KX}) alternately for $K_{X}$.

\bigskip

Ultimately, the distributions of invested capital per investor and of
capital per firm, given a collective state and a sector $X$,\textbf{\ } are $%
\frac{\left\vert \hat{\Psi}\left( \hat{K},X\right) \right\vert ^{2}}{%
\left\Vert \hat{\Psi}\left( \hat{X}\right) \right\Vert ^{2}}$\ and $\frac{%
\left\vert \Psi \left( K,X\right) \right\vert ^{2}}{\left\Vert \Psi \left(
X\right) \right\Vert ^{2}}$,\ respectively.

Gathering equations (\ref{Nx}), (\ref{Nxh}) and (\ref{kx}), each collective
state is singularly determined by the collection of data that characterizes
each sector: the number of firms, investors, the average capital, and the
distribution of capital. All these quantities allow the study of capital
allocation among sectors and its dependency in the parameters of the system,
such as expected long-term and short-term returns, and any other parameter.
This "static" point of view, will be extended by introducing some
fluctuations in the expectations, leading to a dynamic of the average
capital at the macro-level. In the following, we solve the system for the
background fields and compute the average associated quantities.

\part*{System at the macro level: background fields and equilibria}

In this part, we consider the study of our economic framework at the
macro-scale. Starting with the action functional (\ref{fcn}), we derive the
background fields $\Psi \left( K,X\right) $\ and $\hat{\Psi}\left( \hat{K},%
\hat{X}\right) $, and their adjoints fields $\Psi ^{\dag }\left( K,X\right) $%
\ and $\hat{\Psi}^{\dag }\left( \hat{K},\hat{X}\right) $,\ that minimize the
functional $S$. This allow to derive the potential equilibria of the system.
We show that at each point of the sector space, several patterns of
accumulation may appear, leading to an infinite number of potential
background states. Thes pattern are not independent, and describe the system
as a whole that may experience global transitions in patterns of accumulation

\section{Resolution}

Now that the initial framework has been translated into a proper field
formalism, we can solve the field model.\ Average capital per sector
(defined in (\ref{kx}) and (\ref{Khx})) depends on the background fields $%
\Psi \left( K,X\right) $ and $\hat{\Psi}\left( \hat{K},X\right) $ and their
conjugate $\Psi ^{\dag }\left( K,X\right) $ and $\hat{\Psi}^{\dag }\left( 
\hat{K},\hat{X}\right) $\ that minimize the field action $S$.

To study the influence of investment and financial allocation on the
dynamics of the real economy, we must express the quantities relevant to
firms as functions of financial quantities.

The order of resolution will thus be the following: we will first minimize\
the $\left( K,X\right) $\ part of the fields action (\ref{fcn}), i.e. $%
S_{1}+S_{2}$, to find the real economy background fields $\Psi \left(
K,X\right) $ and $\Psi ^{\dag }\left( K,X\right) $ and the number of firms $%
\left\vert \Psi \left( K,X\right) \right\vert ^{2}$\ as functions of the
financial sectors' background fields $\Psi ^{\dag }\left( K,X\right) $%
\textbf{\ }and\textbf{\ }$\hat{\Psi}^{\dag }\left( \hat{K},\hat{X}\right) $
and investors' variables. We will then minimize $S_{3}+S_{4}$, and find the
minimal configuration of the investors' field $\hat{\Psi}\left( \hat{K},\hat{%
X}\right) $ and $\hat{\Psi}^{\dag }\left( \hat{K},\hat{X}\right) $.

\subsection{\textbf{Background field for the real economy }}

To compute\footnote{%
For detailed computations of this subsection, see appendix 2.} the field of
the real economy $\Psi \left( K,X\right) $ as a function of the field of the
financial sector $\hat{\Psi}\left( \hat{K},\hat{X}\right) $. We first
minimize $S_{1}+S_{2}$, i.e. the real economy part of equation (\ref{fcn}): 
\begin{eqnarray}
S_{1}+S_{2} &=&-\int \Psi ^{\dag }\left( K,X\right) \left( \nabla _{X}\left( 
\frac{\sigma _{X}^{2}}{2}\nabla _{X}-\nabla _{X}R\left( K,X\right) H\left(
K\right) \right) -\tau \frac{K_{X}}{K}\left( \int \left\vert \Psi \left(
K^{\prime },X\right) \right\vert ^{2}dK^{\prime }\right) \right. \\
&&+\left. \nabla _{K}\left( \frac{\sigma _{K}^{2}}{2}\nabla _{K}+u\left(
K,X,\Psi ,\hat{\Psi}\right) \right) \right) \Psi \left( K,X\right) dKdX 
\notag
\end{eqnarray}

Since, in this second part of the paper, we are interested in studying the
collective states, we will consider that in first approximation $\frac{K_{X}%
}{K}\simeq 1$. This corresponds to consider that in average the fluctuations
of capital in one sector $X$ is relatively low with respect to the average $%
K_{X}$. This assumptions will be removed in the third part, when we consider
individual dynamics. As a consequence, we consider the replacement:%
\begin{equation*}
\tau \frac{K_{X}}{K}\rightarrow \tau
\end{equation*}

For relatively slow fluctuations in $X$, and up to an exponential change of
variable in the fields, we show in appendix 2.1 that the background fields $%
\Psi \left( K,X\right) $ and $\Psi ^{\dag }\left( K,X\right) $\ decompose as
a product:

\begin{equation}
\Psi \left( K,X\right) =\Psi ^{\dag }\left( K,X\right) =\Psi \left( X\right)
\Psi _{1}\left( K-K_{X}\right)  \label{psf}
\end{equation}%
where $K_{X}$, the average invested capital per firm in sector $X$, is given
by (\ref{KX}) and the functions $\Psi \left( X\right) $ and $\Psi _{1}\left(
K-K_{X}\right) $ satisfy \ the following differential equations:%
\begin{eqnarray}
&&0=\left( -\frac{\sigma _{X}^{2}}{2}\nabla _{X}^{2}+\frac{\left( \nabla
_{X}R\left( X\right) H\left( K_{X}\right) \right) ^{2}}{2\sigma _{X}^{2}}+%
\frac{\nabla _{X}^{2}R\left( K,X\right) }{2}H\left( K\right) +2\tau
\left\vert \Psi \left( X\right) \right\vert ^{2}\right) \Psi \left( X\right)
\label{mnt} \\
&&+D\left( \left\Vert \Psi \right\Vert ^{2}\right) \left( \int \left\Vert
\Psi \left( X\right) \right\Vert ^{2}-N\right) +\int \mu \left( X\right)
\left\Vert \Psi \left( X\right) \right\Vert ^{2}  \notag
\end{eqnarray}%
for $\Psi \left( X\right) $, and:%
\begin{equation}
0=-\nabla _{K}^{2}\Psi _{1}\left( K-K_{X}\right) +\left( K-\frac{F_{2}\left(
R\left( K,X\right) \right) K_{X}}{F_{2}\left( R\left( K_{X},X\right) \right) 
}\right) ^{2}\Psi _{1}\left( K-K_{X}\right) +\gamma \left( X\right) \Psi
_{1}\left( K-K_{X}\right)  \label{cpf}
\end{equation}%
for $\Psi _{1}\left( K-K_{X}\right) $.

The constants $D\left( \left\Vert \Psi \right\Vert ^{2}\right) $, $\mu
\left( X\right) $ and $\gamma \left( X\right) $ arising in (\ref{mnt}) and (%
\ref{cpf})\ are Lagrange multipliers\footnote{%
Incidentally, note that, to keep track of the dependency of the Lagrange
multiplier in $\left\Vert \Psi \right\Vert ^{2}$ in the above, we have
chosen the notation $D\left( \left\Vert \Psi \right\Vert ^{2}\right) $.}
that implement the constraints: 
\begin{equation*}
\int \left\Vert \Psi \left( X\right) \right\Vert ^{2}=N\text{, }\left\Vert
\Psi \left( X\right) \right\Vert ^{2}\geqslant 0\text{, }\left\Vert \Psi
_{1}\left( K-K_{X}\right) \right\Vert ^{2}=1
\end{equation*}%
where $N$ is the total number of firms of the system.\smallskip

The intuition that the background field can be decomposed as a product,
presented in equation (\ref{psf}), is straightforward: in the space of
sectors, firms relocate more slowly than capital accumulates, so that firms
are first described the position $X$ of their sector, and second, their
capital in this sector, distributed around the average capital of the
sector, $K_{X}$.\ This is translated in the decomposition (\ref{psf}) of $%
\Psi \left( K,X\right) $ by the two factors $\Psi \left( X\right) $ and $%
\Psi _{1}\left( K-K_{X}\right) $.

\smallskip

The function $\Psi _{1}\left( K-K_{X}\right) $, involved in the definitions (%
\ref{psf}) of the background fields $\Psi \left( K,X\right) $ describes the
fluctuations\textbf{\ }of capital in a given sector $X$ around an average
value $K_{X}$\footnote{%
It is computed in appendix 2.1.2.}:%
\begin{equation}
\Psi _{1}\left( K-K_{X}\right) =\mathcal{N}\exp \left( -\left( K-\frac{%
F_{2}\left( R\left( K,X\right) \right) K_{X}}{F_{2}\left( R\left(
K_{X},X\right) \right) }\right) ^{2}\right)  \label{fpc}
\end{equation}%
where $\mathcal{N}$ is a normalization factor. The capital accumulated by a
firm in a sector $X$ is centered around the average capital $K_{X}$ in this
sector, weighted by a factor $\frac{F_{2}\left( R\left( K,X\right) \right) }{%
F_{2}\left( R\left( K_{X},X\right) \right) }$. This factor depends on the
firm's expected long-term return.\ It is relative to the average expected
long-term return of the whole sector $X$ described by the function $%
F_{2}\left( R\left( K_{X},X\right) \right) $\footnote{%
See discussion below equation (\ref{grandf2}).}.

\smallskip

Equation (\ref{mnt}) can be solved for\ the $X$-dependent part of the
background field $\Psi \left( X\right) $\footnote{%
A method of resolution of\ (\ref{mnt}) and two examples for particular forms
of the function $H\left( K\right) $ are presented in appendix 2.2.}. From
this solution, we can deduce the number of firms $\left\Vert \Psi \left(
X\right) \right\Vert ^{2}$\ in sector $X$. However, when fluctuations in
capital allocation $\sigma _{X}^{2}$ are small, we can express directly $%
\left\Vert \Psi \left( X\right) \right\Vert ^{2}$\ as a function\textbf{\ }%
of the financial variables.

This number of firms is given by:%
\begin{equation}
\left\Vert \Psi \left( X\right) \right\Vert ^{2}=\frac{D\left( \left\Vert
\Psi \right\Vert ^{2}\right) }{2\tau }-\frac{1}{4\tau }\left( \left( \nabla
_{X}R\left( X\right) \right) ^{2}+\frac{\sigma _{X}^{2}\nabla
_{X}^{2}R\left( K_{X},X\right) }{H\left( K_{X}\right) }\right) \left( 1-%
\frac{H^{\prime }\left( \hat{K}_{X}\right) K_{X}}{H\left( \hat{K}_{X}\right) 
}\right) H^{2}\left( K_{X}\right)  \label{psl}
\end{equation}%
provided that the rhs of equation (\ref{psl}) is positive.\ It is $0$
otherwise.

The Lagrange multiplier $D\left( \left\Vert \Psi \right\Vert ^{2}\right) $
is obtained by integration of (\ref{mnt}) and yields: 
\begin{equation}
ND\left( \left\Vert \Psi \right\Vert ^{2}\right) =2\tau \int \left\vert \Psi
\left( X\right) \right\vert ^{4}+\frac{1}{2}\int \left( \nabla _{X}R\left(
X\right) H\left( K_{X}\right) \right) ^{2}\left\Vert \Psi \left( X\right)
\right\Vert ^{2}  \label{pnq}
\end{equation}%
Formula (\ref{psl}) will be used extensively in the sequel to compute $K_{X}$%
, the average physical capital per firm in sector $X$.

\subsection{Background field for the financial markets}

We have computed the background fields for firms, $\Psi \left( K,X\right) $
and $\Psi ^{\dag }\left( X,K\right) $, and the number of firms by minimizing 
$S_{1}+S_{2}$. We can now compute\textbf{\ }the background fields $\hat{\Psi}%
\left( \hat{K},\hat{X}\right) $\ and $\hat{\Psi}^{\dag }\left( \hat{K},\hat{X%
}\right) $ for investors along with the number of investors \ $\left\vert 
\hat{\Psi}\left( \hat{X},\hat{K}\right) \right\vert ^{2}$\ by minimizing $%
S_{3}+S_{4}$.

\smallskip

We first rewrite the field action for investors, $S_{3}+S_{4}$.\ Inserting
the number of firms $\left\Vert \Psi \left( X\right) \right\Vert ^{2}$,
formula (\ref{psl}), reduces $\ S_{3}+S_{4}$ to:%
\begin{equation}
S_{3}+S_{4}=-\int \hat{\Psi}^{\dag }\left( \hat{K},\hat{X}\right) \left(
\nabla _{\hat{K}}\left( \frac{\sigma _{\hat{K}}^{2}}{2}\nabla _{\hat{K}}-%
\hat{K}f\left( \hat{X}\right) \right) +\nabla _{\hat{X}}\left( \frac{\sigma
_{\hat{X}}^{2}}{2}\nabla _{\hat{X}}-g\left( \hat{X}\right) \right) \right) 
\hat{\Psi}\left( \hat{K},\hat{X}\right)  \label{tsm}
\end{equation}%
where $f\left( K_{\hat{X}},\hat{X}\right) $ is the short-term return%
\footnote{%
See appendix 3.1.1.}:

\begin{equation}
f\left( K_{\hat{X}},\hat{X}\right) =\frac{1}{\varepsilon }\left( r\left( K_{%
\hat{X}},\hat{X}\right) -\gamma \left\Vert \Psi \left( \hat{X}\right)
\right\Vert ^{2}+F_{1}\left( \bar{R}\left( K_{\hat{X}},\hat{X}\right)
\right) \right)  \label{fcf}
\end{equation}%
and $g\left( K_{\hat{X}},\hat{X}\right) $\ describes investors capital
reallocation:%
\begin{equation}
g\left( K_{\hat{X}},\hat{X}\right) =\left( \nabla _{\hat{X}}F_{0}\left(
R\left( K_{\hat{X}},\hat{X}\right) \right) +\nu \nabla _{\hat{X}}F_{1}\left( 
\bar{R}\left( K_{\hat{X}},\hat{X}\right) \right) \right)  \label{fcg}
\end{equation}%
which depends on long-term returns $R\left( K_{\hat{X}},\hat{X}\right) $ and
their relative value $\bar{R}\left( K_{\hat{X}},\hat{X}\right) $ given by:%
\begin{equation}
\bar{R}\left( K_{\hat{X}},\hat{X}\right) =\frac{R\left( K_{\hat{X}},\hat{X}%
\right) }{\int R\left( K_{X^{\prime }}^{\prime },X^{\prime }\right)
\left\Vert \Psi \left( X^{\prime }\right) \right\Vert ^{2}dX^{\prime }}
\label{VRN}
\end{equation}%
Recall that $F_{1}$ measures the share of capital reallocation that depends
on stock prices variation:%
\begin{equation}
F_{1}\left( \bar{R}\left( K_{\hat{X}},\hat{X}\right) \right) =F_{1}\left( 
\bar{R}\left( K_{\hat{X}},\hat{X}\right) ,\Gamma =0\right)  \label{VRT}
\end{equation}%
In the sequel, any function $h\left( K_{\hat{X}},\hat{X}\right) $ and its
partial derivatives $h\left( K_{\hat{X}},\hat{X}\right) $ will be written $%
h\left( \hat{X}\right) $, $\nabla _{K_{\hat{X}}}h\left( \hat{X}\right) $ and 
$\nabla _{\hat{X}}h\left( \hat{X}\right) $, respectively.

\smallskip

Using a change of variable:%
\begin{equation*}
\hat{\Psi}\rightarrow \exp \left( \frac{1}{\sigma _{\hat{X}}^{2}}\int
g\left( \hat{X}\right) d\hat{X}+\frac{\hat{K}^{2}}{\sigma _{\hat{K}}^{2}}%
f\left( \hat{X}\right) \right) \hat{\Psi}
\end{equation*}%
the minimization of $S_{3}+S_{4}$, equation (\ref{tsm}) yields\footnote{%
See appendix 3.1.2.} the equation for $\hat{\Psi}$:%
\begin{equation}
0=\left( \frac{\sigma _{\hat{X}}^{2}\nabla _{\hat{X}}^{2}}{2}-\frac{\left(
g\left( \hat{X}\right) \right) ^{2}}{2\sigma _{\hat{X}}^{2}}-\frac{\nabla _{%
\hat{X}}g\left( \hat{X}\right) }{2}\right) \hat{\Psi}+\left( \nabla _{\hat{K}%
}\left( \frac{\sigma _{\hat{K}}^{2}\nabla _{\hat{K}}}{2}-\hat{K}f\left( \hat{%
X}\right) \right) -F\left( \hat{X}\right) \hat{K}-\hat{\lambda}\right) \hat{%
\Psi}  \label{hqn}
\end{equation}%
and the equation for its conjugate $\hat{\Psi}^{\dag }$: 
\begin{equation}
0=\left( \frac{\sigma _{\hat{X}}^{2}\nabla _{\hat{X}}^{2}}{2}-\frac{\left(
g\left( \hat{X}\right) \right) ^{2}}{2\sigma _{\hat{X}}^{2}}-\frac{\nabla _{%
\hat{X}}g\left( \hat{X}\right) }{2}\right) \hat{\Psi}^{\dag }+\left( \left( 
\frac{\sigma _{\hat{K}}^{2}\nabla _{\hat{K}}}{2}+\hat{K}f\left( \hat{X}%
\right) \right) \nabla _{\hat{K}}-F\left( \hat{X}\right) \hat{K}-\hat{\lambda%
}\right) \hat{\Psi}^{\dag }  \label{nqh}
\end{equation}%
with:%
\begin{equation}
F\left( \hat{X}\right) =\nabla _{K_{\hat{X}}}\left( \frac{\left( g\left( 
\hat{X}\right) \right) ^{2}}{2\sigma _{\hat{X}}^{2}}+\frac{1}{2}\nabla _{%
\hat{X}}g\left( \hat{X}\right) +f\left( \hat{X}\right) \right) \frac{%
\left\Vert \hat{\Psi}\left( \hat{X}\right) \right\Vert ^{2}}{\left\Vert \Psi
\left( \hat{X}\right) \right\Vert ^{2}}+\frac{\left\langle \hat{K}%
^{2}\right\rangle _{\hat{X}}\nabla _{K_{\hat{X}}}f^{2}\left( \hat{X}\right) 
}{\sigma _{\hat{K}}^{2}\left\Vert \Psi \left( \hat{X}\right) \right\Vert ^{2}%
}  \label{Fct}
\end{equation}%
where $\left\langle \hat{K}^{2}\right\rangle _{\hat{X}}$\ denotes the
average of $\hat{K}^{2}$\ in sector $\hat{X}$\ (see appendix 3.1.2) and\ $%
\left\Vert \hat{\Psi}\left( \hat{X}\right) \right\Vert ^{2}=\int \left\vert 
\hat{\Psi}\left( \hat{X},\hat{K}\right) \right\vert ^{2}d\hat{K}$.

A Lagrange multiplier $\hat{\lambda}$ has been included in the system of
equations (\ref{hqn}) and (\ref{nqh}) to implement the constraint for $\hat{%
\Psi}$\ and $\hat{\Psi}^{\dag }$:%
\begin{equation}
\int \left\vert \hat{\Psi}\left( \hat{X},\hat{K}\right) \right\vert ^{2}d%
\hat{X}d\hat{K}=\hat{N}  \label{nbg}
\end{equation}

Incidentally, note that the function $F\left( \hat{X},K_{\hat{X}}\right) $\
arising in the minimization equations (\ref{hqn}) and (\ref{nqh}) describes
the impact of individual variations on the collective state (the field $\hat{%
\Psi}$).\ It can be neglected in first approximation.

\smallskip

Appendix 3.1.3 \ computes the solutions for the investors' background fields
(equations (\ref{hqn}) and (\ref{nqh})). We find an infinite number of
solutions for $\hat{\Psi}_{\hat{\lambda}}$ and $\hat{\Psi}_{\hat{\lambda}%
}^{\dag }$ parametrized by $\hat{\lambda}\in 
\mathbb{R}
$, which translates the fact that $S_{3}+S_{4}$ has an infinite number of
local minima.

However, the eigenvalue $\left\vert \hat{\lambda}\right\vert $\ computed in
appendix 3.1.4.2 has a lower bound $M$\footnote{%
This lower bound is reminiscent of the fact that the Lagrange multiplier $%
\lambda $ is the eigenvalue of the second order operator arising in equation
(\ref{nqh}), and that this operator is bounded from below.} defined by:%
\begin{equation}
M=\max_{\hat{X}}\left( A\left( \hat{X}\right) \right)  \label{MDN}
\end{equation}%
where:%
\begin{equation}
A\left( \hat{X}\right) =\frac{\left( g\left( \hat{X}\right) \right) ^{2}}{%
\sigma _{\hat{X}}^{2}}+f\left( \hat{X}\right) +\frac{1}{2}\sqrt{f^{2}\left( 
\hat{X}\right) }+\nabla _{\hat{X}}g\left( \hat{X}\right) -\frac{\sigma _{%
\hat{K}}^{2}F^{2}\left( \hat{X}\right) }{2f^{2}\left( \hat{X}\right) }
\label{DFT}
\end{equation}%
and that $\hat{\Psi}_{-M}$ is the global minimum of $S_{3}+S_{4}$. The
background fields are thus $\hat{\Psi}_{-M}$ and its adjoint $\hat{\Psi}%
_{-M}^{\dagger }$.

For these background fields, the number of agents with capital $\hat{K}$
invested in sector $\hat{X}$ is:%
\begin{equation*}
\left\vert \hat{\Psi}_{-M}\left( \hat{K},\hat{X}\right) \right\vert ^{2}=%
\hat{\Psi}_{-M}^{\dag }\left( \hat{X},\hat{K}\right) \hat{\Psi}_{-M}\left( 
\hat{K},\hat{X}\right)
\end{equation*}%
We find:%
\begin{equation}
\left\vert \hat{\Psi}_{-M}\left( \hat{K},\hat{X}\right) \right\vert
^{2}\simeq C\left( \bar{p}\right) \exp \left( -\frac{\sigma _{X}^{2}\hat{K}%
^{4}\left( f^{\prime }\left( X\right) \right) ^{2}}{96\sigma _{\hat{K}%
}^{2}\left\vert f\left( \hat{X}\right) \right\vert }\right) D_{p\left( \hat{X%
}\right) }^{2}\left( \left( \frac{\left\vert f\left( \hat{X}\right)
\right\vert }{\sigma _{\hat{K}}^{2}}\right) ^{\frac{1}{2}}\left( \hat{K}+%
\frac{\sigma _{\hat{K}}^{2}F\left( \hat{X}\right) }{f^{2}\left( \hat{X}%
\right) }\right) \right)  \label{dns}
\end{equation}%
where $D_{p}$ is the parabolic cylinder function with parameter $p\left( 
\hat{X}\right) $ and:%
\begin{equation}
p\left( \hat{X}\right) =\frac{M-A\left( \hat{X}\right) }{\sqrt{f^{2}\left( 
\hat{X}\right) }}  \label{PDF}
\end{equation}%
The constant $C\left( \bar{p}\right) $ ensures that the constraint (\ref{nbg}%
) is satisfied\footnote{%
Its expression is given in appendix 3.1.3.}.\textbf{\ }

Section 9.2 will show that $p\left( \hat{X}\right) $\ encompasses the
relative expected returns of sector $X$\ vis-\`{a}-vis its neighbouring
sectors.

\subsection{Average capital per firm per sector}

Now that the number of firms and investors per sector are computed, we can
determine the average capital invested per firm in sector $\hat{X}$, i.e. $%
K_{\hat{X}}$.

We first rewrite the defining equation of $K_{\hat{X}}$ (\ref{kx}) as:%
\begin{equation}
K_{\hat{X}}\left\Vert \Psi \left( \hat{X}\right) \right\Vert ^{2}=\int \hat{K%
}\left\vert \hat{\Psi}\left( \hat{K},\hat{X}\right) \right\vert ^{2}d\hat{K}
\label{ctc}
\end{equation}%
and evaluate this equation for the background field (\ref{dns}):%
\begin{equation}
K_{X}\left\Vert \Psi \left( X\right) \right\Vert ^{2}=\int \hat{K}\left\vert 
\hat{\Psi}_{-M}\left( \hat{K},\hat{X}\right) \right\vert ^{2}d\hat{K}
\label{prK}
\end{equation}

Equation (\ref{prK}) allows to find the average capital $K_{\hat{X}}$.
Actually,\ both the densities of agents $\left\Vert \Psi \left( \hat{X}%
\right) \right\Vert ^{2}$\ and $\left\vert \hat{\Psi}_{-M}\left( \hat{K},%
\hat{X}\right) \right\vert ^{2}$, equations (\ref{psl}) and (\ref{dns}), are
functions of $K_{\hat{X}}$, so that equation (\ref{prK}) is itself an
equation for $K_{\hat{X}}$.

\smallskip

From this general equation, we can find the average capital at point $\hat{X}
$. Appendix 3.1.4.2 computes the integral (\ref{prK}) using the financial
background field (\ref{dns}).

In the sequel, we will write $p\left( \hat{X}\right) $ defined in (\ref{PDF}%
) as:%
\begin{equation}
p\equiv p\left( \hat{X}\right)  \label{fRM}
\end{equation}%
and equation (\ref{prK}) becomes ultimately:%
\begin{equation}
K_{\hat{X}}\left\Vert \Psi \left( \hat{X}\right) \right\Vert ^{2}\left\vert
f\left( \hat{X}\right) \right\vert =C\left( \bar{p}\right) \sigma _{\hat{K}%
}^{2}\hat{\Gamma}\left( p+\frac{1}{2}\right)  \label{qtk}
\end{equation}%
with:%
\begin{eqnarray}
\hat{\Gamma}\left( p+\frac{1}{2}\right) &=&\exp \left( -\frac{\sigma
_{X}^{2}\sigma _{\hat{K}}^{2}\left( p+\frac{1}{2}\right) ^{2}\left(
f^{\prime }\left( X\right) \right) ^{2}}{96\left\vert f\left( \hat{X}\right)
\right\vert ^{3}}\right)  \label{Gmh} \\
&&\times \left( \frac{\Gamma \left( -\frac{p+1}{2}\right) \Gamma \left( 
\frac{1-p}{2}\right) -\left( \Gamma \left( -\frac{p}{2}\right) \right) ^{2}}{%
2^{p+2}\Gamma \left( -p-1\right) \Gamma \left( -p\right) }+p\frac{\Gamma
\left( -\frac{p}{2}\right) \Gamma \left( \frac{2-p}{2}\right) -\left( \Gamma
\left( -\frac{p-1}{2}\right) \right) ^{2}}{2^{p+1}\Gamma \left( -p\right)
\Gamma \left( -p+1\right) }\right)  \notag
\end{eqnarray}%
and where $\Gamma $ is the Gamma function.

\smallskip This final form of the capital equation, (\ref{qtk}),\ will be
central to our following computations. However, it involves some functions,
such as $f$, that have a general form, and functions of the unknown variable 
$K_{\hat{X}}$\ (see for instance equation (\ref{fcf})). Thus, it cannot, in
general, be solved analytically.

\subsection{Solving for average capital}

Except for some particular cases, the final form of the capital equation (%
\ref{qtk}) cannot be solved analytically. Several approaches can nonetheless
be used to approximate its solutions and study their behaviours. The
detailed results and their derivation are given in appendix 4.

A first and most general approach studies, for each sector,\ the variation
of average capital per firm $K_{\hat{X}}$\ with respect to any parameter of
the system. This is done by studying the differential form of the capital
equation (\ref{qtk}) while keeping very general forms for the
parameter-functions $f$ and $g$. This approach allows to study, on a sector,
the influence of its neighbours. It is depicted by the variation of $K_{\hat{%
X}}$ with respect to the sector's relative expected returns. It reveals
stable and unstable equilibria in the system but does not yield the sectors'
precise levels of capital.

A second approach expands the capital equation (\ref{qtk}) around particular
solutions. These particular solutions are the average capital in sectors
where accumulation is the strongest. This approach confirms the existence of
both stable and unstable equilibria, which correspond to several possible
average capital in a given sector: depending on initial configurations, an
infinite number of collective states may arise\footnote{%
This point will be developed in section 8.}.

A third approach provides approximate solutions to the capital equation (\ref%
{qtk}) for standard forms of the parameter-functions. The existence of
multiple solutions is confirmed, along with the associated stability
analysis. Combined, these three approaches confirm and complete each other.

Equation (\ref{qtk}) has in general several solutions per sector that can be
ranked by their level of average capital; low average, high or very high.
For each of these levels of capital, the solutions of equation (\ref{qtk})
may be stable or unstable. A stable average capital is one that, when
slightly modified, comes back to its initial values, wheras an unstable
solution does not. An unstable average capital may be interpreted as a
threshold of capital accumulation for firms in the sector. The solutions of (%
\ref{qtk}) depend on short-term returns, long-term returns, absolute and
relative, and the dependency of the solutions of (\ref{qtk}) depends on the
stability of this solution.This results will be interpreted in section 9.

\section{Dynamic \textbf{average capital}}

So far, we have determined and studied how average capital per firm and per
sector react to changes in parameters. However these same parameters may
vary over time, and so should average capital values. We thus introduce a
macro timescale and design a dynamic model in which average capital and
expectations in long-term returns interact and vary over time.

\subsection{A\textbf{verage capital} and long-run expected returns}

We consider how modifications in parameters generate the dynamics for $K_{%
\hat{X}}$. Assuming that some time-dependent parameters modify expected
long-term returns $R\left( X\right) $, average capital $K_{\hat{X}}$ becomes
a function of the time variable $\theta $. To find how the average physical
capital per firm in sector $\hat{X}$, $K_{\hat{X}}$, evolves over time, we
must define the equation for $K_{\hat{X}}$, (\ref{qtk}), and compute its
variation with respect to $\theta $, using the fact that the functions $%
\left\Vert \Psi \left( \hat{X}\right) \right\Vert ^{2}$\ and $\hat{\Gamma}%
\left( p+\frac{1}{2}\right) $\ both depend on time $\theta $\ through $K_{%
\hat{X}}$\ and $R\left( X\right) $. The variations of these two functions
with respect to the dynamic variables $K_{\hat{X}}$ and $R\left( X\right) $
are computed in appendix 5.1.\ We show that, when $C\left( \bar{p}\right) $
constant, the variation of (\ref{qtk}) writes:%
\begin{equation}
k\frac{\nabla _{\theta }K_{\hat{X}}}{K_{\hat{X}}}+l\frac{\nabla _{\theta
}R\left( \hat{X}\right) }{R\left( \hat{X}\right) }-2m\frac{\nabla _{\hat{X}%
}\nabla _{\theta }R\left( \hat{X}\right) }{\nabla _{\hat{X}}R\left( \hat{X}%
\right) }+n\frac{\nabla _{\hat{X}}^{2}\nabla _{\theta }R\left( \hat{X}%
\right) }{\nabla _{\hat{X}}^{2}R\left( \hat{X}\right) }=-C_{3}\left( p,\hat{X%
}\right) \frac{\nabla _{\theta }r\left( \hat{X}\right) }{f\left( \hat{X}%
\right) }  \label{krv}
\end{equation}%
where coefficients $k$, $l$, $m$ and $n$ are computed in appendix 5.1.

To make the system self-consistent, and since $K_{\hat{X}}$ already depends
on $R$, we merely need to introduce an endogenous dynamics for $R$.

To do so, we assume that $R$ depends on $K_{\hat{X}},\hat{X}$ and $\nabla
_{\theta }K_{\hat{X}}$, and that this dependency has the form of a diffusion
process\footnote{%
See appendix 5.2.}. This leads to write $R$ as a function $R\left( K_{\hat{X}%
},\hat{X},\nabla _{\theta }K_{\hat{X}}\right) $. The variation of $R$ is of
the form:%
\begin{eqnarray}
\nabla _{\theta }R\left( \theta ,\hat{X}\right) &=&a_{0}\left( \hat{X}%
\right) \nabla _{\theta }K_{\hat{X}}+b\left( \hat{X}\right) \nabla _{\hat{X}%
}^{2}\nabla _{\theta }K_{\hat{X}}+c\left( \hat{X}\right) \nabla _{\theta
}\left( \nabla _{\theta }K_{\hat{X}}\right) +d\left( \hat{X}\right) \nabla
_{\theta }^{2}\left( \nabla _{\theta }K_{\hat{X}}\right)  \label{rvn} \\
&&+f\left( \hat{X}\right) \nabla _{\hat{X}}^{2}\left( \nabla _{\theta
}R\left( \theta ,\hat{X}\right) \right) +h\left( \hat{X}\right) \nabla
_{\theta }^{2}\left( \nabla _{\theta }R\left( \theta ,\hat{X}\right) \right)
\notag \\
&&+u\left( \hat{X}\right) \nabla _{\hat{X}}\nabla _{\theta }\left( \nabla
_{\theta }K_{\hat{X}}\right) +v\left( \hat{X}\right) \nabla _{\hat{X}}\nabla
_{\theta }\left( \nabla _{\theta }R\left( \theta ,\hat{X}\right) \right) 
\notag
\end{eqnarray}%
We can also assume that the coefficients in the expansion are slowly
varying, since they are obtained by computing averages.

The dynamics (\ref{rvn}) corresponds to a diffusion process: expected
returns in one sector depend on the variations of capital and returns in
neighbouring sectors.

To find the intrinsic dynamics for $K_{\hat{X}}$, we assume that the
exogenous variation $\frac{\nabla _{\theta }r\left( \hat{X}\right) }{r\left(
K_{\hat{X}},\hat{X}\right) }$ is null, and that the system of equations (\ref%
{krvv}) and (\ref{rvn}) yields the dynamics for $\nabla _{\theta }K_{\hat{X}%
} $ and $\nabla _{\theta }R\left( \theta ,\hat{X}\right) $. Approximating
these dynamics to the first order in derivatives, we find in appendix 5.2
the following matricial equation:%
\begin{equation}
0=M_{1}\left( 
\begin{array}{c}
\nabla _{\theta }K_{\hat{X}} \\ 
\nabla _{\theta }R%
\end{array}%
\right) -M_{2}\left( 
\begin{array}{c}
\nabla _{\theta }K_{\hat{X}} \\ 
\nabla _{\theta }R%
\end{array}%
\right) -M_{3}\left( 
\begin{array}{c}
\nabla _{\theta }K_{\hat{X}} \\ 
\nabla _{\theta }R%
\end{array}%
\right)  \label{dqk}
\end{equation}

\subsection{Oscillatory solutions}

We look for oscillating solutions of (\ref{dqk}) of the type:%
\begin{equation}
\left( 
\begin{array}{c}
\nabla _{\theta }K_{\hat{X}} \\ 
\nabla _{\theta }R\left( \hat{X}\right)%
\end{array}%
\right) =\exp \left( i\Omega \left( \hat{X}\right) \theta +iG\left( \hat{X}%
\right) \hat{X}\right) \left( 
\begin{array}{c}
\nabla _{\theta }K_{0} \\ 
\nabla _{\theta }R_{0}%
\end{array}%
\right)  \label{dqK}
\end{equation}%
with slowly varying $G\left( \hat{X}\right) $ and $\Omega \left( \hat{X}%
\right) $.\ We are then led to the relation between $\Omega \left( \hat{X}%
\right) $ and $G\left( \hat{X}\right) $: 
\begin{eqnarray}
0 &=&\frac{k}{K_{\hat{X}}}\left( 1-ieG-ig\Omega \right) +\left( \frac{l}{%
R\left( \hat{X}\right) }-i\frac{2m}{\nabla _{\hat{X}}R\left( \hat{X}\right) }%
G\right) \left( a_{0}+iaG+ic\Omega \right)  \label{FRQ} \\
&&-\frac{l}{R\left( \hat{X}\right) }\left( d\Omega ^{2}+bG^{2}+u\Omega
G\right) +\frac{k}{K_{\hat{X}}}\left( e\Omega ^{2}+fG^{2}+v\Omega G\right) 
\notag
\end{eqnarray}%
We will limit our study to the first order terms which yields the expression
for $\Omega $ as a function of the parameters involved in (\ref{krv}) and (%
\ref{rvn}). Appendix 5.3 computes the expression of $\Omega $.

It also derives the condition of stability for the oscillations. When: 
\begin{equation}
\frac{lc}{R\left( \hat{X}\right) }\left( \frac{k}{K_{\hat{X}}}+\frac{a_{0}l}{%
R\left( \hat{X}\right) }\right) +\frac{4m^{2}ca_{0}}{\left( \nabla _{\hat{X}%
}R\left( \hat{X}\right) \right) ^{2}}G^{2}>0  \label{sln}
\end{equation}%
oscillations are dampened and return to the steady state. Otherwise,
oscillations are diverging: the system settles on another steady state, i.e.
another background state. Appendix 5.4 studies the condition (\ref{sln}) as
a function of the parameter functions $f\left( \hat{X}\right) $\ and $%
R\left( \hat{X}\right) $, the level of average capital $K_{\hat{X}}$, and
the coefficients arising in the expectations formations. The results are
presented in the next section.

\section{Results and nterpretations}

For each sector, the equation defining its average capital, equation (\ref%
{qtk}), accepts several solutions, so that each sector could present several
average capital. We will discuss the stability of these solutions, before
detailing the determinants of average capital, and number of firms and
investors per sector. We will then describe the three patterns of capital
accumulation that emerge and their possible transitions. Ultimately we will
study how average capital per sector interacts with endogeneized long-term
returns expectations. We will end up this section by providing a synthesis
of the main results.

\subsection{Stability of average capital}

Each average value of capital solving equation (\ref{qtk}) can be either
stable or unstable. A stable average capital, when modified, will naturally
return to its initial value.\ An unstable one will not: it will merely act
as a potentially varying capital accumulation threshold. Once modified, an
unstable average capital will settle at another equilibrium level.

The average capital per firm per sector defined in equation (\ref{qtk}) acts
as a fixed point of a dynamic equation\footnote{%
The definitions of the parameters are given in appendix 4.1.1.1.
\par
.}\footnote{%
See section 7.2.1.2 and appendix 4.1.1.1} whose (in)stability\ depends the
sector's parameters. An unstable average capital will act as a potential
threshold for capital accumulation in a sector: any deviation in this
threshold will set firms above or below the threshold, initiate a path
towards a new equilibrium and ultimately shift average capital. There is
therefore an intrinsic transition dynamics of average capital per sector
that is driven by instability and exogenous variations in the system's
parameters.

Average capital is potentially unstable in a sector when the following
condition is met:%
\begin{equation}
\left\vert B\left( \hat{X}\right) \right\vert \equiv \left\vert k\left(
p\right) \frac{\partial p}{\partial K_{\hat{X}}}-\left( \frac{\partial \ln
f\left( \hat{X},K_{\hat{X}}\right) }{\partial K_{\hat{X}}}+\frac{\partial
\ln \left\vert \Psi \left( \hat{X},K_{\hat{X}}\right) \right\vert ^{2}}{%
\partial K_{\hat{X}}}+l\left( \hat{X},K_{\hat{X}}\right) \right) \right\vert
>1  \label{nNT}
\end{equation}%
This instability depends on the four parameters of $\left\vert B\left( \hat{X%
}\right) \right\vert $: directly through short-term returns, dividends and
price fluctuations, $\frac{\partial \ln f\left( \hat{X},K_{\hat{X}}\right) }{%
\partial K_{\hat{X}}}$, and the net flow of firms entering the sector, $%
\frac{\partial \ln \left\vert \Psi \left( \hat{X},K_{\hat{X}}\right)
\right\vert ^{2}}{\partial K_{\hat{X}}}$; indirectly through a change in the
background field induced either by a variation in short-term returns $%
f\left( \hat{X}\right) $ via $l\left( \hat{X},K_{\hat{X}}\right) $, or by
the modification of the relative return of sector $\hat{X}$ which depends on
the shape of the returns around $\hat{X}$ via $k\left( p\right) \frac{%
\partial p}{\partial K_{\hat{X}}}$.\ Any variation in these parameters will
affect the system as a whole and may reshape the collective state through a
change in the background field. Altogether, these modifications may magnify
or dampen changes in a sector's average capital and impact the stability of
the system.

\subsection{Determinants of capital accumulation}

We will describe the determinants of capital accumulation, before studying
the number of firms and investors per sector.

\subsubsection{Average capital per sector}

The average capital in a sector $\hat{X}$ is determined by short-term
returns, $f\left( \hat{X}\right) $ - dividends and price fluctuations - and
by the growth prospects of the firm, its expected long-term returns, $%
R\left( \hat{X}\right) $. These returns are not fully independent since the
price fluctuations in short-term returns are driven by expected long-term
returns.\textbf{\ }

Besides, average capital in a sector depends on expected long-term returns
of neighbouring sectors. This dependency is measured by $p\left( \hat{X}%
\right) $\ defined in (\ref{PDF})\footnote{%
See explanation and derivation\textbf{\ }in appendix 4.1.1.}. For positive
short-term returns, which is the case here\footnote{%
See explanation\textbf{\ }in appendix 4.1.3.2.}, it writes: 
\begin{equation}
p(\hat{X})=\frac{M-\left( \frac{\left( g\left( \hat{X}\right) \right) ^{2}}{%
\sigma _{\hat{X}}^{2}}+\nabla _{\hat{X}}g\left( \hat{X}\right) -\frac{\sigma
_{\hat{K}}^{2}F^{2}\left( \hat{X}\right) }{2f^{2}\left( \hat{X}\right) }%
\right) }{f\left( \hat{X}\right) }-\frac{3}{2}  \label{nPRV}
\end{equation}%
Except for the normalization by the short-term return $f\left( \hat{X}%
\right) $ of sector $\hat{X}$, the function $p(\hat{X})$\ is composed of
three terms.

The two first terms, $\frac{\left( g\left( \hat{X}\right) \right) ^{2}}{%
\sigma _{\hat{X}}^{2}}$ \footnote{%
This term is directly proportional to the gradient of expected long-term
returns $\nabla R\left( \hat{X}\right) $. See the definition of the
parameter function $g$, equation (\ref{fcg}), and (\ref{VRN}).} and $\nabla
_{\hat{X}}g\left( \hat{X}\right) $\footnote{%
This term is proportional to the second derivative $\nabla ^{2}R\left( \hat{X%
}\right) $\ of $R\left( \hat{X}\right) $.}, measure the variations of
expected returns across sectors, i.e. the value of expected returns in
sector $\hat{X}$\ relative to its neighbours. The last term, $\frac{\sigma _{%
\hat{K}}^{2}F^{2}\left( \hat{X},K_{\hat{X}}\right) }{2f^{2}\left( \hat{X}%
\right) }$, is a smoothing factor between neighbours' sectors. It can be
neglected in the first approximation\footnote{%
See the discussion following equation (\ref{Fct}). This term will also be
discussed in section 7.2.2.}. The parameter $p(\hat{X})$ is a local maximum
when $R\left( \hat{X}\right) $\ is itself a local maximum\footnote{%
Actually, $p(\hat{X})$ is maximal for sectors such that $\nabla R\left( \hat{%
X}\right) =0$ and $\nabla ^{2}R\left( \hat{X}\right) <0$. It is thus}, so
that it describes the expected long-term returns of a sector relative to its
neighbours.\ The higher $p\left( \hat{X}\right) $, the more attractive is
sector $\hat{X}$\ relative to its neighbours\footnote{%
Note that the parameter $p\left( \hat{X}\right) $ is normalized by
short-term returns. It computes the ratio of relative attractivity to
short-term returns. This allows to consider these two variables separately.}.

\smallskip Taken altogether, the three parameters $R\left( \hat{X}\right) $, 
$f\left( \hat{X}\right) $, $p\left( \hat{X}\right) $ are the main
determinants of average capital in a sector. However, their influence on
capital will depend on the stability of the sector. In stable sectors,
average capital values can be understood as equilibria. In unstable ones,
they are potential thresholds for the capital accumulation of individual
firms. In stable sectors, average capital is increasing in short-term
returns\ $f\left( \hat{X}\right) $, expected long-term returns $R\left( \hat{%
X}\right) $, and in the sector's relative attractivity $p\left( \hat{X}%
\right) $, respectively. The higher the short and long-term returns, the
higher the capital accumulation.\ Besides, any increase in relative returns
will attract capital from neighbouring sectors and increases\ the sector
average capital. In unstable sectors, average capital is decreasing in these
same variables, and any increase in short- or expected long-term, be they
absolute or relative, returns reduces the amount of capital required to
initiate the capital accumulation process for the individual firms.

\subsubsection{Firms per sector}

Various parameters determine how firms and investors shift across the
sectors' space.

The number of firms per sector defined in equation (\ref{psl}) depends on
expected long-term returns: 
\begin{equation*}
V(X)=\left( \nabla _{X}R\left( X\right) \right) ^{2}+\frac{\sigma
_{X}^{2}\nabla _{X}^{2}R\left( X\right) }{H\left( K_{X}\right) }
\end{equation*}%
where $\nabla _{X}R\left( X\right) $ is the gradient of expected long-term
returns along the sectors space, and $\nabla _{X}^{2}R\left( K_{X},X\right) $
is the Laplacian, i.e. the generalisation of the second derivative of $%
R\left( K_{X},X\right) $ with respect to the sectors' space. The number of
firms is a decreasing function of $V(X)$\footnote{%
See equation (\ref{psl}).}.

When expected returns are minimal, $\nabla _{X}R\left( X\right) =0$ and $%
\nabla _{X}^{2}R\left( K_{X},X\right) >0$, average capital\ is low, and a
large number of small firms provide short-term returns through dividends.

When returns in sector $X$, $R\left( X\right) $, are at a local maximum, $%
\nabla _{X}R\left( X\right) =0$ and $\nabla _{X}^{2}R\left( K_{X},X\right)
<0 $, the sector exhibit both a large number of firms and a high level of
capital $K_{X}$ per firm, but this equilibrium is unstable.

Incidentally, competition ensures that sectors with low or minimal expected
returns are not completely depleted.

When $\nabla _{X}R\left( X\right) \neq 0$, the sector is "transitory". It is
surrounded by neighbouring sectors, with both lower and higher expected
returns. Firms head towards sectors with higher returns. The greater the
discrepancy between neighbouring returns $\nabla _{X}R\left( X\right) $, the
faster firms leave the sector.

\subsubsection{Investors per sector}

The average number of investors in a sector\footnote{%
See formula (\ref{dns}).} is an increasing function of the sector short-term
returns and relative long-term attractivity $p$\footnote{%
See equation (\ref{fRM}).}. All else equal, an increase in short-term
returns or an improvement of the sector's relative long-term attractivity
increases the number of investors and, in turn, firms' disposable capital.

The number of investors in a given sector increases with its relative
attractivity $p\left( \hat{X}\right) $ defined in equation (\ref{PDF}). The
first term in (\ref{PDF}) is the sector's relative attractivity towards its
neighbours, normalized by its short-term returns $f\left( \hat{X}\right) $.\
The second term is a factor that smoothes differences between sectors. It is
negatively correlated to the variations of the sectors' relative
attractivity. Investors and capital will increase in sectors\ surrounded by
significantly more attractive sectors, i.e. sectors with higher average
capital and investors\footnote{%
A close inspection of equation (\ref{Fct}) reveals that this term contains
the -squared- contributions of short-term returns, $f\left( \hat{X}\right) $%
,\ and the sector's relative attractivity:\ $\frac{\left( g\left( \hat{X}%
\right) \right) ^{2}}{2\sigma _{\hat{X}}^{2}}+\frac{1}{2}\nabla _{\hat{X}%
}g\left( \hat{X}\right) $. Both contributions are proportional to the
gradient of $R$ with respect to $K_{\hat{X}}$\ . When this gradient is
non-zero, indicating that an increase in capital may enhance either the
sector's relative attractiveness or short-term returns, the correction $%
\frac{\sigma _{\hat{K}}^{2}F^{2}\left( \hat{X}\right) }{2\sigma _{\hat{X}%
}^{2}\left( \sqrt{f^{2}\left( \hat{X}\right) }\right) ^{3}}$\ amplifies $%
\frac{A\left( \hat{X}\right) }{f\left( \hat{X}\right) },$\ and consequently $%
K_{\hat{X}}$,\ in most cases.}: the whole system tends to reach stable
configurations, and capital discrepancies are reduced between close
neighbours\footnote{%
Derivation of the minimization equation in appendix 3.1.2 shows that the
term $F\left( \hat{X}\right) $\ arises as a backreaction of the whole system
with respect to modifications at one point of the thread.}.

\subsection{Capital accumulation}

In each sector, several average capital may exist, and three patterns of
capital accumulation arise, defined by their average capital, number of
firms, and long- and short-term returns. These parameters will determine the
stability of the pattern\textbf{. }Shocks will shift unstable patterns to
another one. Any deviation of average capital above or below an unstable
equilibrium value will drive firms away from this equilibrium and ultimately
shift average capital towards another equilibrium. These transitions provide
bridges between patterns of capital. Due to a change in external conditions,
sectors may move from one pattern to another.

\subsubsection{First pattern: low capital, high short-term returns driven by
dividends only}

These are sectors where growth prospects are subdued, with a relatively
large number of low-capitalized firms. Because firms are small, marginal
productivity is high and firms attract capital with short-term returns
through dividends, but lack the capital to move towards growth sectors.

These sectors are stable to small fluctuations in growth prospects: any
increase in expected long-term returns will only shift moderately investment
and average capital. They are unstable to short-term returns: any increase
in $f\left( \hat{X}\right) $ will drive dividends higher and attract
investors. Average capital will accumulate and reach a stable pattern-2
equilibrium, with more firms and a higher average capital.

However, an adverse shock lowering short-term returns will increase the
threshold of capital accumulation and drive the equilibrium towards $0$.
Producers remain in the sector, but their very lack of capital will prevent
them to shifting towards more attractive sectors in the long run (see
appendix 3.3).

\subsubsection{Second pattern: intermediate-to-high level of capital,
short-term returns, long-term expectations}

These sectors have moderate growth prospects, so that any increase in
short-term, i.e. dividends and stock prices or long-term returns, increases
their relative attractivity $p\left( \hat{X}\right) $ and attracts investors
and capital.\ Locally, the higher the relative attractivity of the sector,
the higher the capital accumulation. The relatively high number of firms in
the sector is a decreasing function of average capital: competition favours
higher average capital, and concentration of firms. This is the most
standard pattern of capital allocation. It is stable to variations in
average capital, except when average capital is high and the firms' density
is low.

In this case, any deviation of average capital above its equilibrium
increases the threshold and drives the sector backward to a stable pattern 2
equilibrium, i.e. a sector with a large number of average capitalized firms.
The lower capital per firm reduces competition and attracts new firms into
the sector.

On the contrary, any deviation of average capital below its equilibrium
reduces the threshold and favours capital accumulation. The sector is driven
towards a stable pattern 3 equilibrium, with a small number of very
capitalized firms (see description of this pattern below).

\subsubsection{Third pattern: high capital, long-term returns, and relative
attractivity}

These are sectors where growth prospects are extremely high. Capital
accumulation is driven by expectations of long-term returns sustained by
ever-higher levels of investment. These are the most attractive sectors. Two
cases arise.

When expected long-term returns are not maximal, the sector stabilizes with
very few firms with very high capital arises. This extension of pattern 2
corresponds to a few large oligopolistic groups.

When expected long-term returns are maximal, the sector's attractivity
allows a large number of firms with high capital to coexist. All else equal,
these firms could grow indefinitely, so that such equilibria are bound to be
unstable\footnote{%
See appendix 4.1.}.\smallskip This describes bubble-like, unstable sectors.

An adverse shock drives these unstable sectors towards a stable pattern 3:
average capital is approximatively maintained, but an increase in
competition evicts the less capitalized firms and the total number of firms
is reduced to a small set.

On the contrary, a positive shock reduces the threshold of capital
accumulation. Most firms can accumulate without bound, which attracts even
higher capital. Capital accumulation is modified in all sectors, which may
transform the whole economic landscape. Total available capital is reduced,
which modifies the stability conditions for all sectors. Low-capitalized
sectors may become unstable and disappear, whereas others may accumulate
capital. All in all, the system may end with a reduced sectors space%
\footnote{%
See appendix 4.3 for technical details.}.

\subsubsection*{Global instability}

Another source of instability stems from the constraint imposed by the model
on the total number of investors.

In our model, we have assumed a fixed number of agents, that are spread
across sectors. This hypothesis binds the dynamics of the whole set of
sectors. If this constraint were to be lifted, the sectors would be
independent and could each reach a stable average capital, given their own
short and expected long-term returns.

However, the number of agents in a sector is dependent on the whole system's
characteristics. Thus there can only be global equilibria for the system.
Any change in the parameters induces a perturbation $\delta \Psi \left( \hat{%
X},K_{\hat{X}}\right) $ that destabilizes the whole system as a whole: the
equilibrium is globally unstable\footnote{%
The mechanism of this instability is detailed in appendix 3.3.1.}. Relaxing
the condition on the number of agents amounts to replacing the average
capital equation (\ref{qtk}) by\footnote{%
The derivation is given in appendix 4.1.2.}:%
\begin{equation}
K_{\hat{X}}\left\Vert \Psi \left( \hat{X}\right) \right\Vert ^{2}\left\vert
f\left( \hat{X}\right) \right\vert =C\left( \bar{p}\right) \sigma _{\hat{K}%
}^{2}\hat{\Gamma}\left( \frac{1}{2}\right) =C\left( \bar{p}\right) \sigma _{%
\hat{K}}^{2}\exp \left( -\frac{\sigma _{X}^{2}\sigma _{\hat{K}}^{2}\left(
f^{\prime }\left( X\right) \right) ^{2}}{384\left\vert f\left( \hat{X}%
\right) \right\vert ^{3}}\right)  \label{gbd}
\end{equation}%
This equation has at least one locally stable solution. The solutions of the
modified average capital equation (\ref{gbd}) do no longer directly depend
on a sector's relative characteristics, but rather on the returns $f\left( 
\hat{X}\right) $ and on the number of firms in the sector, $\left\Vert \Psi
\left( \hat{X}\right) \right\Vert ^{2}$\footnote{%
An intermediate situation between (\ref{qtk}) and (\ref{gbd}) could also be
considered: it would be to assume a constant number of agents in some
regions of the sector space.}\footnote{%
Alternatively, limiting the number of investors per sector can be achieved
through some public regulation to maintain a constant flow of investment in
the sector.}.

\subsection{Dynamic capital accumulation}

The dynamic system (\ref{dqk}) propagates shocks in capital and expectations
across the system\footnote{%
See appendix 5.3.}: assuming a shock on average capital or long-term returns
in a given sector, the interactions between average capital and expected
long-term returns induce some volatility around the equilibrium values $K_{%
\hat{X}}$ and $R\left( \hat{X}\right) $. The fluctuations of long-term
returns directly impact average capital and expected return in neighbouring
sectors through the induced variation of relative expected returns, which
initiates the propagation of the initial perturbation to the whole system.

This propagation is described by the oscillating solutions (\ref{dqK}). For
a given sector $\hat{X}$, the velocity of oscillations in average capital
and expected returns are measured by the frequency $\Omega \left( \hat{X}%
\right) $, that depends on the sector's characteristics, These oscillations
may be dampening (stable oscillations) or widening (unstable oscillations).
Three main parameters determine which type of oscillations a sector may
experience\footnote{%
We have already given the condition for dampening oscillations in (\ref{sln}%
). See appendix 5.4.}.

\begin{enumerate}
\item[1.] The elasticity of expected long-term returns with respect to
variations of capital, $c$, that arises in equation (\ref{rvn}), determines
two relevant forms of expectations. When expectations are highly reactive to
variations of capital, $c>0$, and when expected long-term returns increase
with any acceleration in capital accumulation, expected long-term returns
depend positively on the variations of average capital $K_{\hat{X}}$. When
expectations are moderately reactive to variations in the capital, $c<0$,
expected long-term returns depend negatively on the variations of average
capital $K_{\hat{X}}$.

\item[2.] The neighbouring sectors' discrepancy in capital fluctuations at a
given time, $G$. It\ arises in the oscillatory solutions (\ref{dqK}) and
measures the inhomogeneity\textbf{\ }between sectors.

\item[3.] Last but not least, the sector average level of capital $K_{\hat{X}%
}$ impacts the type of fluctuations experienced by the sector.
\end{enumerate}

Our results are the following.

\subsubsection{Low average capital sectors}

When average capital is very low in a sector, the sole relevant parameter to
the fluctuations is the reactivity $c$ of the expected return $R\left( \hat{X%
}\right) $ to an increase in capital\footnote{%
See appendix 5.4.}.

Two cases arise.

When long-term returns strongly react to capital fluctuations, $c>0$,
oscillations are unstable. When they react mildly, $c<0$, oscillations are
stable.

In the first case, expected long-term returns and average capital variations
are positively correlated, and any increase in capital will amplify expected
returns that will in turn increase capital. In the second case, expected
long-term returns and average capital variations are negatively correlated,
which will induce dampening oscillations and stabilize the system.

These results show that for expectations mildly reactive to variations of
capital, some equilibria with relatively low capital are possible and
resilient to oscillations in expectations, a niche effect may exist for some
sectors.

\subsubsection{High average capital sectors}

In high average capital sectors, be they stable or unstable, here again only
expectations reactivity to capital increase, i.e. $c$, matters. Oscillations
are dampening for $c>0$ and explosive for $c<0$. Highly reactive
expectations, $c>0$, will amplify fluctuations of capital and expected
returns:

In the stable case, fluctuations that would otherwise be destabilizing for
sectors with low capital may stabilize or maintain sectors with both stable
and high levels of capital. A large reactivity between expectations and
capital will allow for an intrinsic high level of capital to consolidate.
Fluctuations will moderately impact these high-capitalized sectors: for
instance, considering an initial increase in returns only, i.e. $\delta
R\left( \hat{X}\right) >0$, will induce a net outflow of capital towards
less capitalized sectors with an higher increase in relative returns, while
a decreasing return, i.e. $\delta R\left( \hat{X}\right) <0$, will induce a
net inflow of capital dampening the sector's fluctuations.

In the unstable case, an initial increase in capital increases expected
long-term returns, while at the same time, the negative correlation between
variations in investment and expected return\ lowers the average capital. An
increase in capital will improve the sector profitability, lowering the
capital threshold in capital. To put it differently, an initial increase in
the average capital amplifies the expected return, which reduces $K_{\hat{X}%
} $ and offsets the initial increase in capital.

When expectations are mildly reactive, i.e. $c<0$, the mechanism of
dampening oscillations that arises for $c>0$ is impaired. In the unstable
case, for instance, an initial increase in the threshold $K_{\hat{X}}$,
impacts only moderately the sector expected returns, and does not offset the
initial increase in capital.

\subsubsection{Intermediate average capital sectors}

In intermediate capital sectors, oscillations depend both on the reactivity
of expectations to an increase in capital, $c$, and discrepancy between
sectors, $G$\footnote{%
See appendix 5.4.2.}.

Midly reactive expectations, $c<0$, and a moderate discrepancy between
neighbouring sectors, $G<<1$, oscillations are dampening for sectors with
relatively low average capital. The analysis of the first case applies to
the extent that indeed some homogeneity in capital between the neighbouring
sectors exists.

Strongly reactive expectations, $c>0$, and a large discrepancy between
neighbouring sectors, $G>>1$, oscillations are dampening for relatively high
average capital sectors. The analysis of the second case applies to a
locally dominating sector.

\subsubsection{The role of expectations in average capital fluctuations}

For each sector, the threshold between dampening and explosive oscillations
depends on the parameters of the system. Mildly reactive expectations only
favour low- to high-capital sectors (patterns 1 and 2). Very reactive
expectations will only favour very high-capital sectors (pattern 3) where
oscillations will be felt as relatively weak, in absolute value, attracting
further capital.

Recall that in extreme cases of pattern 3, both maximal capital and returns
act as thresholds that repel low-capital firms and propel high-capital firms
to ever higher accumulation. Oscillations in these thresholds generate a
high global instability: a constantly oscillating threshold crowds firms out
of the sector.

To conclude, the dynamics for average capital and expected returns merely
reflect fluctuations in the background fields i.e. the collective states.
These fluctuations may destabilize the patterns in some sectors and
ultimately switch the collective state and modify the patterns' landscape.

\subsection{Synthesis of the results}

Let us briefly synthesize our results before discussing them. They can be
regrouped along three main axes.

\subsubsection{Capital allocation}

We have shown that capital allocation by producers and investors differ and
interact: these interactions impact the form of the collective state and
average capital per sector. The main determinants of capital allocation are
short-term returns, expected long-term returns and the sector's relative
attractivity. Short-term returns are composed of dividends, and are driven
by marginal productivity and variations in stock prices, themselves driven
by expectations of long-term returns. Expected long-term returns describe
growth prospects, and.the sector's relative attractivity measures the growth
prospects of a sector relative to its neighbouring sectors.

Firms tend to relocate in sectors with relatively higher long-term returns
at a speed that depends on their capital endowment. However, they can be
crowded out by competitors. The higher the firm's capital, the higher the
power to overcome competitors. Eventually, firms with the highest capital
concentrate in sectors that have the highest expected long-term returns,
while the rest relocates in neighbouring and likely lowest expected-return
sectors.

Capital allocation depends on short-term returns, dividends and price
fluctuations, and expected long-term returns. But since price fluctuations
are driven by expected long-term returns, short and long-term returns are
not independent. The financial capital allocation also depends on the
sector's relative attractivity, which measures the expected returns of a
sector relative to its neighbours. However financial capital is volatile.
High short-term returns are an incentive, but the relative attractivity of
sectors lures investors. Financial capital allocation thus depends on the
ratio of sectors' relative attractivity to short-term returns. Since this
ratio depends on expectations, it is subject to fluctuations, which in turn
impact the collective state.

\subsubsection{Capital accumulation}

Three stationary\footnote{%
The values of average capital are stationary results: agents accumulate and
shift from sectors to other ones, but, in average, the density of firms and
average capital per firm per sector are constant.} patterns of capital
accumulation may emerge in each sector. A pattern is characterized by the
combination of the firms' average capitalization, the number of firms in the
sector, and the type of returns these firms may provide to their investors.
The emergence of a given pattern depends on the parameters of this sector:

The first pattern associates a large number of low-capitalised firms.
Dividends are determinant in this pattern; the lack of capital, combined
with the prospects of competition with better-capitalized firms prevent
firms to shift to neighbouring and more profitable sectors.

The second pattern associates a relatively high number of average-to-high
capitalised firms and a combination of short and long-term returns. This
combination lures intermediate-to-high capital investors in the sector.

In the third pattern, high expected long-term returns generate massive
inflows of capital toward highly-capitalized firms. In this pattern, firms
with the highest expected returns could theoretically accumulate endlessly.
Actually, this accumulation is limited by the amount of available capital.

In each pattern, some sectors are stable, others are unstable. Transitions
between patterns occur through exogenous shocks. In pattern 1, some sectors
may disappear, whereas in pattern 3, some may grow endlessly and the large
amounts of capital they drive may modify the whole system's landscape.

\subsubsection{Collective states}

We have shown how statistical field theory can describe a microeconomic
framework in terms of collective states\textbf{\ }of sectors composed of a
large number of firms.

Each collective state encodes the data characterizing each sector: number of
firms, number of investors, average capital, and distribution of capital.
These data are theoretical averages over long-term periods, not
instantaneous empirical averages.

The collective states are not arbitrary: they directly result from the
agents' interactions, and are the most likely stable states of the system.
Other states do exist, but they are unstable. A particular collective state
can be described by its distribution into patterns of capital accumulation -
type 1, 2, or 3 - across sectors. Besides, sectors are connected and benefit
from the relative attractiveness of their neighbours: this smoothing effect
between sectors materialises in mergers and acquisitions.

The multiple combinations of accumulation patterns in each sector may yield
an infinite number of possible collective states.\ It does not follow that
all combinations are possible: sector patterns depend on the relative
attractivity of both the sector and its neighbours'. There are also
constraints: for instance, massive inflows of capital are needed for the
emergence of the pattern 3, which is only driven by high expected long-term
returns, while niche effects merely occur for relatively highly productive
firms. However, a potentially infinite range of collective states may exist.

The selection of a particular collective state and its sectoral patterns is
ultimately determined by exogenous conditions. Structural changes, such as
an extra-loose monetary policy or the choice of a pension system are
external conditions that modify collective states.

The existence of multiple collective states has a dynamic implication. When
parameters vary, a given collective state may switch to another: a change in
expectations may, for instance, induce variations in average capital and in
turn, induce changes in sectors' patterns of capital accumulation. To study
these possible switches, we introduced a dynamic interaction between average
capital and expected long-term\ returns, now endogenized. This dynamic
interaction depends both on the patterns of accumulation and the way
expectations are formed.

In this dynamics system, average capital and expectations present some
oscillatory patterns that may dampen equilibria or drive them towards other
equilibria. Expectations highly reactive to capital variations stabilize
high-capital configurations.\ They drive low-to-moderate capital sectors
towards zero or higher capital, depending on their initial conditions.
Inversely, expectations moderately reactive to capital variations stabilize
low-to-moderate capital configurations, and drive high-capital sectors
towards lower capital equilibria. Amplifying oscillations may modify some
sectors' pattern: the ensuing reallocation of capital across the whole
sectors' space may initiate a transition in collective states. The mechanism
of transition and its implications are discussed below.

\part*{System at the individual level: effective action and transition
functions}

This third part focuses on the micro-scale of the system. In a given
background state, we can derive the individual dynamics for agents, firms
and investors. This approach stresses the dependence of individual dynamics
in the collective states described by the background fields.

\section{Computation of transition functions}

We use the results of section 5 to compute the agents' transition functions.
To do so, we compute the effective action of the system which is given by
the series expansion of the action around the background fields. These
background fields where computed in the second part of this work (see
sections 7.1 and 7.2).

\subsection{Effective action expansion}

\subsubsection{Second-order expansion of effective action}

Consider the field action:%
\begin{equation*}
S=S_{1}+S_{2}+S_{3}+S_{4}
\end{equation*}%
where the $S_{i}$ are defined by equations (\ref{SN}),(\ref{SD}),(\ref{ST})
and (\ref{SQ})\footnote{%
Recall that at the individual level, we use again the full interaction term $%
\tau \frac{K_{X}}{K}$.}. Expanding the action $S$ to the second-order around
the background field will allow us to compute the transition functions of
individual agents in the background, without taking into account individual
interactions. We can rewrite the fields as follows:%
\begin{eqnarray*}
\Psi \left( K,X\right) &=&\Psi _{0}\left( K,X\right) +\Delta \Psi \left(
Z,\theta \right) \\
\hat{\Psi}\left( \hat{K},\hat{X}\right) &=&\hat{\Psi}_{0}\left( \hat{K},\hat{%
X}\right) +\Delta \hat{\Psi}\left( Z,\theta \right)
\end{eqnarray*}%
where $\Psi _{0}\left( K,X\right) ,\hat{\Psi}_{0}\left( \hat{K},\hat{X}%
\right) $ are the background fields. This yields the quadratic approximation:%
\begin{equation}
S\left( \Psi ,\hat{\Psi}\right) =S\left( \Psi _{0},\hat{\Psi}_{0}\right)
+\int \left( \Delta \Psi ^{\dag }\left( Z,\theta \right) ,\Delta \hat{\Psi}%
^{\dag }\left( Z,\theta \right) \right) \left( Z,\theta \right) O\left( \Psi
_{0}\left( Z,\theta \right) \right) \left( 
\begin{array}{c}
\Delta \Psi \left( Z,\theta \right) \\ 
\Delta \hat{\Psi}\left( Z,\theta \right)%
\end{array}%
\right)
\end{equation}%
with:%
\begin{equation}
O\left( \Psi _{0}\left( Z,\theta \right) \right) =\left( 
\begin{array}{cc}
\frac{\delta ^{2}S\left( \Psi \right) }{\delta \Psi ^{\dag }\delta \Psi } & 
\frac{\delta ^{2}S\left( \Psi \right) }{\delta \Psi ^{\dag }\left( Z,\theta
\right) \delta \hat{\Psi}} \\ 
\frac{\delta ^{2}S\left( \Psi \right) }{\delta \hat{\Psi}^{\dag }\delta \Psi 
} & \frac{\delta ^{2}S\left( \Psi \right) }{\delta \hat{\Psi}^{\dag }\delta 
\hat{\Psi}}%
\end{array}%
\right) _{\substack{ \Psi \left( Z,\theta \right) =\Psi _{0}\left( Z,\theta
\right)  \\ \hat{\Psi}\left( Z,\theta \right) =\hat{\Psi}_{0}\left( Z,\theta
\right) }}  \label{prtv}
\end{equation}%
The anti-diagonal terms in equation (\ref{prtv}) involve crossed derivatives
with respect to both the fields of the real economy and the financial
economy. These terms represent the interactions between the two economies.
However, as explained in (Gosselin Lotz Wambst 2022), the cross-dependency
between $\Psi \left( Z,\theta \right) $\ and $\hat{\Psi}\left( Z,\theta
\right) $\ is relatively weak, since these interactions are taken into
account by the background fields. In first approximation, the minimization
of $S\left( \Psi \right) $ can be separated between $S_{1}+S_{2}$ and $%
S_{3}+S_{4}$. Therefore, we can write:%
\begin{equation}
O\left( \Psi _{0}\left( Z,\theta \right) \right) \simeq \left( 
\begin{array}{cc}
\frac{\delta ^{2}\left( S_{1}+S_{2}\right) }{\delta \Psi ^{\dag }\delta \Psi 
} & 0 \\ 
0 & \frac{\delta ^{2}\left( S_{3}\left( \Psi \right) +S_{4}\left( \Psi
\right) \right) }{\delta \hat{\Psi}^{\dag }\delta \hat{\Psi}}%
\end{array}%
\right) _{\substack{ \Psi \left( Z,\theta \right) =\Psi _{0}\left( Z,\theta
\right)  \\ \hat{\Psi}\left( Z,\theta \right) =\hat{\Psi}_{0}\left( Z,\theta
\right) }}  \label{Dfr}
\end{equation}%
The second-order expansion then becomes:%
\begin{eqnarray}
S\left( \Psi ,\hat{\Psi}\right) &=&S\left( \Psi _{0},\hat{\Psi}_{0}\right)
+\Delta S_{2}\left( \Psi ,\hat{\Psi}\right)  \label{ftv} \\
&=&S\left( \Psi _{0},\hat{\Psi}_{0}\right) +\int \Delta \Psi ^{\dag }\left(
K,X\right) \frac{\delta ^{2}\left( S_{1}+S_{2}\right) }{\delta \Psi ^{\dag
}\left( Z,\theta \right) \delta \Psi \left( Z,\theta \right) }\Delta \Psi
\left( K,\theta \right)  \notag \\
&&+\int \Delta \hat{\Psi}^{\dag }\left( Z,\theta \right) \frac{\delta
^{2}\left( S_{3}\left( \Psi \right) +S_{4}\left( \Psi \right) \right) }{%
\delta \hat{\Psi}^{\dag }\left( Z,\theta \right) \delta \hat{\Psi}\left(
Z,\theta \right) }\Delta \hat{\Psi}\left( Z,\theta \right)  \notag
\end{eqnarray}%
Computing the second order derivatives involved in (\ref{ftv}), and using
the definition of the background fields (see appendix 6) leads to the
formulas:%
\begin{eqnarray*}
\frac{\delta ^{2}\left( S_{1}+S_{2}\right) }{\delta \Psi ^{\dag }\left(
Z,\theta \right) \delta \Psi \left( Z,\theta \right) } &=&-\frac{\sigma
_{X}^{2}}{2}\nabla _{X}^{2}-\frac{\sigma _{K}^{2}}{2}\nabla _{K}^{2}+\left(
D\left( \left\Vert \Psi \right\Vert ^{2}\right) +2\tau \frac{\left\vert \Psi
\left( X\right) \right\vert ^{2}\left( K_{X}-K\right) }{K}\right) \\
&&+\frac{1}{2\sigma _{K}^{2}}\left( K-\hat{F}_{2}\left( R\left( K,X\right)
\right) K_{X}\right) ^{2}+\frac{1-\nabla _{K}\hat{F}_{2}\left( R\left(
K,X\right) \right) K_{X}}{2}
\end{eqnarray*}%
\begin{eqnarray*}
&&\frac{\delta ^{2}\left( S_{3}\left( \Psi \right) +S_{4}\left( \Psi \right)
\right) }{\delta \hat{\Psi}^{\dag }\left( Z,\theta \right) \delta \hat{\Psi}%
\left( Z,\theta \right) } \\
&=&\left( -\frac{\sigma _{\hat{X}}^{2}}{2}\nabla _{\hat{X}}^{2}+\frac{\left(
g\left( \hat{X}\right) \right) ^{2}+\sigma _{\hat{X}}^{2}\left( f\left( \hat{%
X}\right) +\nabla _{\hat{X}}g\left( \hat{X},K_{\hat{X}}\right) -\frac{\sigma
_{\hat{K}}^{2}F^{2}\left( \hat{X},K_{\hat{X}}\right) }{2f^{2}\left( \hat{X}%
\right) }\right) }{\sigma _{\hat{X}}^{2}\sqrt{f^{2}\left( \hat{X}\right) }}%
\right. \\
&&\left. -\frac{\sigma _{\hat{K}}^{2}}{2\sqrt{f^{2}\left( \hat{X}\right) }}%
\nabla _{\hat{K}}^{2}+\left( \frac{\sqrt{f^{2}\left( \hat{X}\right) }\left( 
\hat{K}+\frac{\sigma _{\hat{K}}^{2}F\left( \hat{X},K_{\hat{X}}\right) }{%
f^{2}\left( \hat{X}\right) }\right) ^{2}}{4\sigma _{\hat{K}}^{2}}\right)
\right)
\end{eqnarray*}

\subsubsection{Fourth-order corrections}

Calculating the fourth-order corrections to the effective action is
sufficient for deriving the main aspects of the interactions in a given
background field. We show in appendix 7 that the third-order terms can be
neglected, and that the series expansion of the action to the fourth-order
writes:%
\begin{eqnarray}
S\left( \Psi ,\hat{\Psi}\right) &=&S\left( \Psi _{0},\hat{\Psi}_{0}\right) \\
&&+\int \Delta \Psi ^{\dag }\left( K,X\right) \frac{\delta ^{2}\left(
S_{1}+S_{2}\right) }{\delta \Psi ^{\dag }\left( Z,\theta \right) \delta \Psi
\left( Z,\theta \right) }\Delta \Psi \left( K,\theta \right)  \notag \\
&&+\int \Delta \hat{\Psi}^{\dag }\left( Z,\theta \right) \frac{\delta
^{2}\left( S_{3}\left( \Psi \right) +S_{4}\left( \Psi \right) \right) }{%
\delta \hat{\Psi}^{\dag }\left( Z,\theta \right) \delta \hat{\Psi}\left(
Z,\theta \right) }\Delta \hat{\Psi}\left( Z,\theta \right) +\Delta
S_{4}\left( \Psi ,\hat{\Psi}\right)  \notag
\end{eqnarray}%
with: 
\begin{eqnarray}
&&\Delta S_{4}\left( \Psi ,\hat{\Psi}\right)  \label{dtc} \\
&\simeq &2\tau \int \left\vert \Delta \Psi \left( K^{\prime },X\right)
\right\vert ^{2}dK^{\prime }\left\vert \Delta \Psi \left( K,X\right)
\right\vert ^{2}dKdX  \notag \\
&&-\Delta \Psi ^{\dag }\left( K,X\right) \Delta \Psi ^{\dag }\left(
K^{\prime },X^{\prime }\right) \nabla _{K}\frac{\delta ^{2}u\left( K,X,\Psi ,%
\hat{\Psi}\right) }{\delta \Psi \left( K^{\prime },X\right) \delta \Psi
^{\dagger }\left( K^{\prime },X\right) }\Delta \Psi \left( K^{\prime
},X^{\prime }\right) \Delta \Psi \left( K,X\right)  \notag \\
&&-\Delta \Psi ^{\dagger }\left( K,\theta \right) \Delta \hat{\Psi}^{\dagger
}\left( \hat{K},\theta \right) \nabla _{K}\frac{\delta ^{2}u\left( K,X,\Psi ,%
\hat{\Psi}\right) }{\delta \hat{\Psi}\left( \hat{K},\hat{X}\right) \delta 
\hat{\Psi}^{\dagger }\left( \hat{K},\hat{X}\right) }\Delta \hat{\Psi}\left( 
\hat{K},\theta \right) \Delta \Psi \left( K,\theta \right)  \notag \\
&&-\Delta \hat{\Psi}^{\dag }\left( \hat{K},\hat{X}\right) \Delta \Psi ^{\dag
}\left( K^{\prime },\theta \right) \left\{ \nabla _{\hat{K}}\frac{\hat{K}%
\delta ^{2}f\left( \hat{X},\Psi ,\hat{\Psi}\right) }{\delta \Psi \left(
K^{\prime },X\right) \delta \Psi ^{\dagger }\left( K^{\prime },X\right) }%
+\nabla _{\hat{X}}\frac{\delta ^{2}g\left( \hat{X},\Psi ,\hat{\Psi}\right) }{%
\delta \Psi \left( K^{\prime },X\right) \delta \Psi ^{\dagger }\left(
K^{\prime },X\right) }\right\} \Delta \Psi \left( K^{\prime },X^{\prime
}\right) \Delta \hat{\Psi}\left( \hat{K},\hat{X}\right)  \notag
\end{eqnarray}%
Computing the terms involved in (\ref{dtc}) (see appendix 7) allows us to
interpret the various terms arising in the correction to the action.

The first term in the right-hand side of (\ref{dtc}) describes the direct
repulsive interaction between firms due to competition in a given sector.

The second term describes the indirect competition between firms through
capital allocation by investors, since:%
\begin{equation}
\frac{\delta ^{2}u\left( K,X,\Psi ,\hat{\Psi}\right) }{\delta \Psi \left(
K^{\prime },X\right) \delta \Psi ^{\dagger }\left( K^{\prime },X\right) }=-%
\frac{1}{\varepsilon }\hat{F}_{2}\left( s,R\left( K,X\right) \right) \hat{F}%
_{2}\left( s,R\left( K^{\prime },X^{\prime }\right) \right) \hat{K}_{X}
\label{duf}
\end{equation}%
and this term involves the relative attractiveness of two firms with capital 
$K$ and $K^{\prime }$ respectively in sector $X$. \ 

The third term represents the firms-investors direct interactions through
investment, since: 
\begin{equation}
\frac{\delta ^{2}u\left( K,X,\Psi ,\hat{\Psi}\right) }{\delta \hat{\Psi}%
\left( \hat{K},\hat{X}\right) \delta \hat{\Psi}^{\dagger }\left( \hat{K},%
\hat{X}\right) }=\frac{1}{\varepsilon }\hat{F}_{2}\left( s,R\left(
K,X\right) \right) \hat{K}  \label{duh}
\end{equation}%
is the relative attractiveness of a firm with capital $K^{\prime }$ at
sector $X$.

The last term describes the variation of investement due to the relative
short-term and long-term return of a given firm. Specifically, we have: 
\begin{equation}
\frac{\delta ^{2}f\left( \hat{X},\Psi ,\hat{\Psi}\right) }{\delta \Psi
\left( K^{\prime },X\right) \delta \Psi ^{\dagger }\left( K^{\prime
},X\right) }\simeq \frac{1}{\varepsilon }\left( \Delta f\left( K^{\prime },%
\hat{X},\Psi ,\hat{\Psi}\right) -\gamma \frac{K^{\prime }}{K_{X}}\right)
\label{dtf}
\end{equation}%
and:%
\begin{equation}
\frac{\delta ^{2}g\left( \hat{X},\Psi ,\hat{\Psi}\right) }{\delta \Psi
\left( K^{\prime },X\right) \delta \Psi ^{\dagger }\left( K^{\prime
},X\right) }=\frac{1}{\int \left\Vert \Psi \left( K^{\prime },\hat{X}\right)
\right\Vert ^{2}dK^{\prime }}\Delta \left( g\left( K^{\prime },\hat{X},\Psi ,%
\hat{\Psi}\right) \right)  \label{dtg}
\end{equation}%
with:%
\begin{equation*}
\Delta f\left( K^{\prime },\hat{X},\Psi ,\hat{\Psi}\right) =f\left(
K^{\prime },\hat{X},\Psi ,\hat{\Psi}\right) -f\left( \hat{X},\Psi ,\hat{\Psi}%
\right)
\end{equation*}%
and:%
\begin{equation*}
\Delta g\left( K^{\prime },\hat{X},\Psi ,\hat{\Psi}\right) =g\left(
K^{\prime },\hat{X},\Psi ,\hat{\Psi}\right) -g\left( \hat{X},\Psi ,\hat{\Psi}%
\right)
\end{equation*}%
are the relative short-term return and long-term return for firm with
capital $K^{\prime }$ at sector $\hat{X}$ respectively.

\subsection{One agent transition functions}

Following section 5.3.1, we consider first the "free" transition functions
that are given by the inverse operator of:%
\begin{equation}
\left( O\left( \Psi _{0}\left( Z,\theta \right) \right) +\alpha \right) ^{-1}
\label{lpc}
\end{equation}%
Given (\ref{Dfr}), the inverse (\ref{lpc}) reduces to:%
\begin{equation*}
\left( 
\begin{array}{cc}
\left( \frac{\delta ^{2}\left( S_{1}+S_{2}\right) }{\delta \Psi ^{\dag
}\delta \Psi }+\alpha \right) ^{-1} & 0 \\ 
0 & \left( \frac{\delta ^{2}\left( S_{3}\left( \Psi \right) +S_{4}\left(
\Psi \right) \right) }{\delta \hat{\Psi}^{\dag }\delta \hat{\Psi}}+\alpha
\right) ^{-1}%
\end{array}%
\right) _{\substack{ \Psi \left( Z,\theta \right) =\Psi _{0}\left( Z,\theta
\right)  \\ \hat{\Psi}\left( Z,\theta \right) =\hat{\Psi}_{0}\left( Z,\theta
\right) }}
\end{equation*}%
This implies that the transition functions can be computed independently for
the individual firms and investors. We will write:%
\begin{equation*}
G_{1}\left( \left( K_{f},X_{f}\right) ,\left( X_{i},K_{i}\right) ,\alpha
\right)
\end{equation*}%
the transition probability for a firm between an initial state $\left(
X_{i},K_{i}\right) $ and a final state $\left( K_{f},X_{f}\right) $ during
an average timespan $\alpha ^{-1}$ and:%
\begin{equation*}
G_{2}\left( \left( \hat{K}_{f},\hat{X}_{f}\right) ,\left( \hat{X}_{i},\hat{K}%
_{i}\right) ,\alpha \right)
\end{equation*}%
the transition probability for a firm between an initial state $\left( \hat{X%
}_{i},\hat{K}_{i}\right) $ and a final state $\left( \hat{K}_{f},\hat{X}%
_{f}\right) $ average timespan $\alpha ^{-1}$. Appendix 8 computes these
transition functions. We find the following results.

\paragraph{One firm transition function}

\begin{eqnarray}
&&G_{1}\left( \left( K_{f},X_{f}\right) ,\left( X_{i},K_{i}\right) \right)
\label{Gn} \\
&=&\exp \left( D\left( \left( K_{f},X_{f}\right) ,\left( X_{i},K_{i}\right)
\right) -\alpha _{eff}\left( \Psi ,\left( K_{f},X_{f}\right) ,\left(
X_{i},K_{i}\right) \right) \sqrt{\frac{\left( X_{f}-X_{i}\right) ^{2}}{%
2\sigma _{X}^{2}}+\frac{\left( K_{f}^{\prime }-K_{i}^{\prime }\right) ^{2}}{%
2\sigma _{K}^{2}}}\right)  \notag
\end{eqnarray}%
where:%
\begin{equation}
D\left( \left( K_{f},X_{f}\right) ,\left( X_{i},K_{i}\right) \right)
=D_{1}+D_{2}+D_{3}  \label{frdv}
\end{equation}%
with:%
\begin{equation}
D_{1}=\int_{X_{i}}^{X_{f}}\frac{\nabla _{X}R\left( K_{X},X\right) }{\sigma
_{X}^{2}}H\left( K_{X}\right)  \label{DTN}
\end{equation}%
\begin{equation}
D_{2}=-\int_{K_{i}}^{K_{f}}\left( K-\hat{F}_{2}\left( s,R\left( K,\bar{X}%
\right) \right) K_{\bar{X}}\right) dK  \label{DTT}
\end{equation}%
\begin{equation}
D_{3}=\int_{K_{i}}^{K_{f}}\left( \left( \frac{X_{f}-X_{i}}{2}\right) \nabla
_{X}\hat{F}_{2}\left( s,R\left( K,\bar{X}\right) \right) K_{\bar{X}}\right)
dK  \label{DTR}
\end{equation}%
\begin{eqnarray}
&&\alpha _{eff}\left( \Psi ,\left( K_{f},X_{f}\right) ,\left(
X_{i},K_{i}\right) \right)  \label{frtv} \\
&=&\alpha +D\left( \left\Vert \Psi \right\Vert ^{2}\right) +\tau \left( 
\frac{\left\vert \Psi \left( X_{f}\right) \right\vert ^{2}\left(
K_{X_{f}}-K_{f}\right) }{K_{f}}-\frac{\left\vert \Psi \left( X_{i}\right)
\right\vert ^{2}\left( K_{X_{i}}-K_{i}\right) }{K_{i}}\right) +\frac{\sigma
_{K}^{2}}{2}K_{f}^{\prime }K_{i}^{\prime }  \notag
\end{eqnarray}%
and:%
\begin{eqnarray*}
K_{i}^{\prime } &=&K_{i}-\hat{F}_{2}\left( s,R\left( K_{X_{i}},X_{i}\right)
\right) K_{X_{i}} \\
K_{f}^{\prime } &=&K_{f}-\hat{F}_{2}\left( s,R\left( K_{X_{f}},X_{f}\right)
\right) K_{X_{f}}
\end{eqnarray*}

\paragraph{One investor transition function}

\begin{eqnarray}
&&G_{2}\left( \left( \hat{K}_{f},\hat{X}_{f}\right) ,\left( \hat{X}_{i},\hat{%
K}_{i}\right) \right)  \label{Gt} \\
&=&\exp \left( D^{\prime }\left( \left( \hat{K}_{f},\hat{X}_{f}\right)
,\left( \hat{X}_{i},\hat{K}_{i}\right) \right) \right)  \notag \\
&&\times \exp \left( -\alpha _{eff}^{\prime }\left( \left( \hat{K}_{f},\hat{X%
}_{f}\right) ,\left( \hat{X}_{i},\hat{K}_{i}\right) \right) \left\vert
\left( \hat{K}_{f}+\frac{\sigma _{\hat{K}}^{2}F\left( \hat{X}_{f},K_{\hat{X}%
_{f}}\right) }{f^{2}\left( \hat{X}_{f}\right) }\right) -\left( \hat{K}_{i}+%
\frac{\sigma _{\hat{K}}^{2}F\left( \hat{X}_{i},K_{\hat{X}_{i}}\right) }{%
f^{2}\left( \hat{X}_{i}\right) }\right) \right\vert \right)  \notag
\end{eqnarray}%
with:%
\begin{equation}
D^{\prime }\left( \left( \hat{K}_{f},\hat{X}_{f}\right) ,\left( \hat{X}_{i},%
\hat{K}_{i}\right) \right) =\frac{1}{\sigma _{\hat{X}}^{2}}\int_{\hat{X}%
_{i}}^{\hat{X}_{f}}g\left( \hat{X}\right) d\hat{X}+\frac{\hat{K}_{f}^{2}}{%
\sigma _{\hat{K}}^{2}}f\left( \hat{X}_{f}\right) -\frac{\hat{K}_{i}^{2}}{%
\sigma _{\hat{K}}^{2}}f\left( \hat{X}_{i}\right)  \label{Frd}
\end{equation}%
and:%
\begin{eqnarray}
&&\alpha _{eff}^{\prime }\left( \left( \hat{K}_{f},\hat{X}_{f}\right)
,\left( \hat{X}_{i},\hat{K}_{i}\right) \right)  \label{Frt} \\
&=&\left( \alpha +\frac{\sigma _{\hat{X}}^{2}}{2}\left( \hat{K}_{f}+\frac{%
\sigma _{\hat{K}}^{2}F\left( \hat{X}_{f},K_{\hat{X}_{f}}\right) }{%
f^{2}\left( \hat{X}_{f}\right) }\right) \left( \hat{K}_{i}+\frac{\sigma _{%
\hat{K}}^{2}F\left( \hat{X}_{i},K_{\hat{X}_{i}}\right) }{f^{2}\left( \hat{X}%
_{i}\right) }\right) \right) \sqrt{\frac{\left\vert f\left( \frac{\hat{X}%
_{f}+\hat{X}_{i}}{2}\right) \right\vert }{2\sigma _{\hat{X}}^{2}}}+g^{\left(
R\right) }\left( \hat{X}\right)  \notag
\end{eqnarray}%
with:%
\begin{equation*}
g^{\left( R\right) }\left( \hat{X}\right) =\int_{\hat{X}_{i}}^{\hat{X}_{f}}%
\frac{\left( g\left( \hat{X}\right) \right) ^{2}+\sigma _{\hat{X}}^{2}\left(
f\left( \hat{X}\right) +\nabla _{\hat{X}}g\left( \hat{X},K_{\hat{X}}\right) -%
\frac{\sigma _{\hat{K}}^{2}F^{2}\left( \hat{X},K_{\hat{X}}\right) }{%
2f^{2}\left( \hat{X}\right) }\right) }{\left\Vert \hat{X}_{f}-\hat{X}%
_{i}\right\Vert \sigma _{\hat{X}}^{2}\sqrt{f^{2}\left( \hat{X}\right) }}
\end{equation*}

\subsection{Two agents transition functions and Interactions between agents}

To study the agents interactions within the background field we consider the
two-agent transition functions. There are three of them. One for two firms:%
\begin{equation*}
G_{11}\left( \left[ \left( K_{f},X_{f}\right) ,\left( K_{f},X_{f}\right)
^{\prime }\right] ,\left[ \left( X_{i},K_{i}\right) ,\left(
X_{i},K_{i}\right) ^{\prime }\right] \right)
\end{equation*}%
one for one firm and one investor:%
\begin{equation*}
G_{12}\left( \left[ \left( K_{f},X_{f}\right) ,\left( \hat{K}_{f},\hat{X}%
_{f}\right) \right] ,\left[ \left( X_{i},K_{i}\right) ,\left( \hat{X}_{i},%
\hat{K}_{i}\right) \right] \right)
\end{equation*}%
and one for for two investors:%
\begin{equation*}
G_{22}\left( \left[ \left( \hat{K}_{f},\hat{X}_{f}\right) ,\left( \hat{K}%
_{f},\hat{X}_{f}\right) ^{\prime }\right] ,\left[ \left( \hat{X}_{i},\hat{K}%
_{i}\right) ,\left( \hat{X}_{i},\hat{K}_{i}\right) ^{\prime }\right] \right)
\end{equation*}%
If we neglect the terms of order greater than $2$ in the effective action,
the transition functions reduce to simple products:%
\begin{eqnarray*}
&&G_{11}\left( \left[ \left( K_{f},X_{f}\right) ,\left( K_{f},X_{f}\right)
^{\prime }\right] ,\left[ \left( X_{i},K_{i}\right) ,\left(
X_{i},K_{i}\right) ^{\prime }\right] \right) \\
&=&G_{1}\left( \left( K_{f},X_{f}\right) ,\left( X_{i},K_{i}\right) \right)
G_{1}\left( \left( K_{f},X_{f}\right) ^{\prime },\left( X_{i},K_{i}\right)
^{\prime }\right)
\end{eqnarray*}%
\begin{eqnarray*}
&&G_{12}\left( \left[ \left( K_{f},X_{f}\right) ,\left( \hat{K}_{f},\hat{X}%
_{f}\right) \right] ,\left[ \left( X_{i},K_{i}\right) ,\left( \hat{X}_{i},%
\hat{K}_{i}\right) \right] \right) \\
&=&G_{1}\left( \left( K_{f},X_{f}\right) ,\left( X_{i},K_{i}\right) \right)
G_{2}\left( \left( \hat{K}_{f},\hat{X}_{f}\right) ,\left( \hat{X}_{i},\hat{K}%
_{i}\right) \right)
\end{eqnarray*}%
\begin{eqnarray*}
&&G_{22}\left( \left[ \left( \hat{K}_{f},\hat{X}_{f}\right) ,\left( \hat{K}%
_{f},\hat{X}_{f}\right) ^{\prime }\right] ,\left[ \left( \hat{X}_{i},\hat{K}%
_{i}\right) ,\left( \hat{X}_{i},\hat{K}_{i}\right) ^{\prime }\right] \right)
\\
&=&G_{2}\left( \left( \hat{K}_{f},\hat{X}_{f}\right) ,\left( \hat{X}_{i},%
\hat{K}_{i}\right) \right) G_{2}\left( \left( \hat{K}_{f},\hat{X}_{f}\right)
^{\prime },\left( \hat{X}_{i},\hat{K}_{i}\right) ^{\prime }\right)
\end{eqnarray*}%
In first approximation, agents behave independently, solely influenced by
the given background state.

To take into account agents interactions we write the expansion:%
\begin{equation*}
\exp \left( -S\left( \Psi \right) \right) =\exp \left( -\left( S\left( \Psi
_{0},\hat{\Psi}_{0}\right) +\Delta S_{2}\left( \Psi ,\hat{\Psi}\right)
\right) \right) \left( 1+\sum_{n\geqslant 1}\frac{\left( -\Delta S_{4}\left(
\Psi ,\hat{\Psi}\right) \right) ^{n}}{n!}\right)
\end{equation*}%
as explained in section 5.3.2, the series produces corrective terms to the
transition functions. Appendix 9 presents the computations and compute the
transitions in the approximations of paths that cross each other one time at
some $X$. In this approximation, we find the following formula:

\subsubsection{Firm-firm transition function:}

\begin{eqnarray}
&&G_{11}\left( \left[ \left( K_{f},X_{f}\right) ,\left( K_{f},X_{f}\right)
^{\prime }\right] ,\left[ \left( X_{i},K_{i}\right) ,\left(
X_{i},K_{i}\right) ^{\prime }\right] \right)  \label{nng} \\
&\simeq &G_{1}\left( \left( K_{f},X_{f}\right) ,\left( X_{i},K_{i}\right)
\right) G_{1}\left( \left( K_{f},X_{f}\right) ^{\prime },\left(
X_{i},K_{i}\right) ^{\prime }\right)  \notag \\
&&-\left( 2\tau -\nabla _{K}\frac{\delta ^{2}u\left( \bar{K},\bar{X},\Psi ,%
\hat{\Psi}\right) }{\delta \Psi \left( \bar{K}^{\prime },\bar{X}\right)
\delta \Psi ^{\dagger }\left( K^{\prime },\bar{X}\right) }\right) \hat{G}%
_{1}\left( \left( K_{f},X_{f}\right) ,\left( X_{i},K_{i}\right) \right) \hat{%
G}_{1}\left( \left( K_{f},X_{f}\right) ^{\prime },\left( X_{i},K_{i}\right)
^{\prime }\right)  \notag
\end{eqnarray}

\subsubsection{Firm-investor transition function:}

\begin{eqnarray}
&&G_{12}\left( \left[ \left( K_{f},X_{f}\right) ,\left( \hat{K}_{f},\hat{X}%
_{f}\right) ^{\prime }\right] ,\left[ \left( X_{i},K_{i}\right) ,\left( \hat{%
X},\hat{K}_{i}\right) ^{\prime }\right] \right)  \label{ntg} \\
&\simeq &G_{1}\left( \left( K_{f},X_{f}\right) ,\left( X_{i},K_{i}\right)
\right) G_{2}\left( \left( \hat{K}_{f},\hat{X}_{f}\right) ^{\prime },\left( 
\hat{X},\hat{K}_{i}\right) ^{\prime }\right)  \notag \\
&&+\left( \nabla _{K}\frac{\delta ^{2}u\left( \bar{K},\bar{X},\Psi ,\hat{\Psi%
}\right) }{\delta \hat{\Psi}\left( \overline{\hat{K},\hat{X}}\right) \delta 
\hat{\Psi}^{\dagger }\left( \overline{\hat{K},\hat{X}}\right) }+\nabla _{%
\hat{K}}\frac{\hat{K}\delta ^{2}f\left( \overline{\hat{X}},\Psi ,\hat{\Psi}%
\right) }{\delta \Psi \left( \overline{K^{\prime },X}\right) \delta \Psi
^{\dagger }\left( \overline{K^{\prime },X}\right) }+\nabla _{\hat{X}}\frac{%
\delta ^{2}g\left( \overline{\hat{X}},\Psi ,\hat{\Psi}\right) }{\delta \Psi 
\overline{\left( K^{\prime },X\right) }\delta \Psi ^{\dagger }\left( 
\overline{K^{\prime },X}\right) }\right)  \notag \\
&&\times \hat{G}_{1}\left( \left( K_{f},X_{f}\right) ,\left(
X_{i},K_{i}\right) \right) \hat{G}_{2}\left( \left( \hat{K}_{f},\hat{X}%
_{f}\right) ^{\prime },\left( \hat{X},\hat{K}_{i}\right) ^{\prime }\right) 
\notag
\end{eqnarray}

\subsubsection{Investor-investor transition function:}

\begin{eqnarray}
&&G_{22}\left( \left[ \left( \hat{K}_{f},\hat{X}_{f}\right) ,\left( \hat{K}%
_{f},\hat{X}_{f}\right) ^{\prime }\right] ,\left[ \left( \hat{X},\hat{K}%
_{i}\right) ,\left( \hat{X},\hat{K}_{i}\right) ^{\prime }\right] \right)
\label{ttg} \\
&\simeq &G_{2}\left( \left( \hat{K}_{f},\hat{X}_{f}\right) ,\left( \hat{X},%
\hat{K}_{i}\right) \right) G_{2}\left( \left( \hat{K}_{f},\hat{X}_{f}\right)
^{\prime },\left( \hat{X},\hat{K}_{i}\right) ^{\prime }\right)  \notag
\end{eqnarray}%
with:%
\begin{eqnarray*}
\left( \bar{X},\bar{K}\right) &=&\frac{\left( K_{f},X_{f}\right) +\left(
X_{i},K_{i}\right) }{2} \\
\left( \bar{X},\bar{K}\right) ^{\prime } &=&\frac{\left( K_{f},X_{f}\right)
^{\prime }+\left( X_{i},K_{i}\right) ^{\prime }}{2}
\end{eqnarray*}%
The derivatives are given in (\ref{duf}), (\ref{duh}), (\ref{dtf}), (\ref%
{dtg}) and:%
\begin{equation*}
\hat{G}_{i}\left( \left( K_{f},X_{f}\right) ,\left( X,K\right) \right) \hat{G%
}_{j}\left( \left( K_{f},X_{f}\right) ^{\prime },\left( X,K\right) ^{\prime
}\right)
\end{equation*}%
is the transition function computed on paths that cross once.

\section{Results and interpretations}

\subsection{One-agent transition functions}

We present a synthesis of the results for firms and investors transition
functions. Some technical details are given in appendix 10.

\subsubsection{Firms transition function}

For a given background state, the probability of transition for a firm
between two states $K_{i},X_{i}$ and $K_{f},X_{f}$, over an average time of $%
1/\alpha $, is given by $G_{1}$ (see \ref{Gn}). This formula computes the
probability that a firm initialy endowed with a capital $K_{i}$ in sector $%
X_{i}$ will relocate to sector $X_{f}$ with capital $K_{f}$. The transition
probability is the result of competing effects, as it is composed of several
interdependent terms of similar magnitude. Firm transitions occur over the
medium to long term but at a slower time scale than transitions for
investors. Firms remain in each transitory sector long enough to resettle,
and for investors to adjust the capital allocated between firms. Thus, in
each transitory sector, firm capital evolves depending on the
characteristics of the firm, the sector, and investors expectations.

\paragraph{Attractiveness and sectors shifts}

The drift term $D$ in formula (\ref{frdv}) is the average transition of a
firm between its initial and final points $\left( X_{i},K_{i}\right) $ and $%
\left( K_{f},X_{f}\right) $, respectively. This term is usually different
from zero because firms tend to shift sectors, and their capital evolves.\
This tendency for a firm to evolve depends both on the transitory sectors
and the background field, i.e., the entire landscape in which the transition
occurs. In addition, fluctuations around the drift term can alter a firm's
trajectory, contributing to the probabilistic nature of the transition.

The drift term of equation (\ref{frdv}) is composed of three interacting
contributions, $D_{1}$,\textbf{\ }$D_{2}$\textbf{\ }$\ $and $D_{3}$.\textbf{%
\ }

The first component $D_{1}$\textbf{\ }shows that firms tend to relocate to
sectors with higher long-term returns, shifts which in turn modify their
present and future attractiveness to investors.

The second component $D_{2}$\ shows that the shift alters the capital of the
firm. Specifically, the amount of investment that investors are willing to
make in the firm, $\hat{F}_{2}\left( R\left( K,\bar{X}\right) \right) K_{%
\bar{X}}$ depends on three key parameters: the average capital of the new
sector,\ $K_{\bar{X}}$, the absolute average return on capital in the
sector,\ $R\left( K,\bar{X}\right) $, and the propensity of investors, $\hat{%
F}_{2}$ to invest in the firm based on its given capital compared to the
average capital of firms in the sector.

When a firm begins the process of relocating to a nearby sector, its
capitalization may differ from that of firms already present in that sector,
which in turn affects its attractiveness to investors, represented by\ $\hat{%
F}_{2}$. The shape of $\hat{F}_{2}$ reflects the propensity of investors to
invest in the firm.\ When $\hat{F}_{2}$ is concave, this propensity
marginally decreases, while a convex shape results in a marginal increase.%
\textbf{\ }

The equilibrium capital of the firm in the new sector is $\hat{F}_{2}\left(
s,R\left( K,\bar{X}\right) \right) K_{\bar{X}}$.\ When a firm relocates, its
capital may turn out to be below or above this equilibrium level. For each
of these cases, two possibilities arise depending on the shape $s$ of $\hat{F%
}_{2}$.

If $\hat{F}_{2}$ is concave, the marginal propensity of investors to invest
is decreasing: once the firm has entered the sector, its capital will
converge towards the sector's average capital.\ It will either increase or
decrease, depending on whether its initial level of capital is above or
below the equilibrium capital, respectively. If $\hat{F}_{2}\left( s,R\left(
K,\bar{X}\right) \right) $ is convex, the marginal propensity of investors
to invest is increasing: the dynamics of its capital accumulation is
unstable.\ Investors will tend to over or underinvest in the firm.

The third contribution $D_{3}$ reflects the firm's relative attractiveness
in different transition sectors. If the the firm's relative attractiveness
is reduced during the shift, such that it attracts less capital than the
average capital of the transition sector, it may become stuck in an
intermediate sector.

\paragraph{Impact of competition}

The coefficient \ $\alpha _{eff}\ $defined in equation (\ref{frtv})
represents the inverse of the average mobility of a firm. This mobility
depends on the competition in transitional sectors which is captured by the
two first terms on the right-hand side of (\ref{frtv}).

The first term, $D\left( \left\Vert \Psi \right\Vert ^{2}\right) $ is a
constant that characterizes the background state of the firms and is
correlated with the total number of firms in the space of sectors.\ As
competition increases, $\alpha _{eff}$ rises and firms' mobility decreases.

The second term measures the local competition that firms face as they move
through the sector space. It is determined by the density of agents in the
sector, multiplied by the variation, along the path, of the firm's excess
capital with respect to the average capital of the sector. A
well-capitalized firm facing many less-capitalized competitors will repel
them and create its own market share.\textbf{\ }Relocation will occur
towards sectors that are denser and less capitalized. Under-capitalized
firms will be forced out of their sectors and into denser, less capitalized
sectors. The relocation process may result in a capital gain or loss.
However, holding capital constant, initially under-capitalized firms will
tend to move towards sectors with lower average capital, whereas
over-capitalized firms tend to move towards sectors with higher average
capital.

\paragraph{Stabilization terms:}

The square-rooted term multiplying $\alpha _{eff}$ is written: 
\begin{equation}
\sqrt{\frac{\left( X_{f}-X_{i}\right) ^{2}}{2\sigma _{X}^{2}}+\frac{\left( 
\tilde{K}_{f}-\tilde{K}_{i}\right) ^{2}}{2\sigma _{K}^{2}}}  \label{abc}
\end{equation}%
and the last term in the right-hand side of equation (\ref{frdv}) both
describe random oscillations around a path of zero marginal capital demand.
Changes in equity, investments, for instance, may modify (\ref{abc}). and
the oscillations are of magnitude $\frac{\sigma _{K}^{2}}{2}$. However,
these oscillations do not necessarily imply a return to the initial point.
The larger the deviation from the average, the more likely firms are to
deviate from the average, and possibly shift to a new trajectory. Therefore,
a capital increase above the average may induce a shift in sector, which in
turn may modify the firm's accumulation and prospects. Thus, oscillations do
not prevent trends and may even initiate them.\ However, such "random
shifts" may prove disadvantageous as they could harm the firm's position and
reduce its capital compared to the sector.

\paragraph{Possible paths}

Overall, what are the possible dynamics for a firm in terms of capital and
sector? If a firm experiences capital growth in a sector where the investor
propensity, $F_{2}$, is concave, the accumulation of its capital could cause
the firm to shift to a higher-return sector, but this may result in the firm
being perceived as less attractive by investors in this new sector.

This shift can lead to a change in the firm's attractiveness to investors, $%
F_{2}$. The growth or decline of the firm in the new sector will depend on
both its capital level and the shape of\textbf{\ }$F_{2}$.\textbf{\ }These
factors will also determine the speed\textbf{\ }of this change. If the
firm's capital level gradually declines, it may have time to react and
reposition itself. However, if the decline in capital is sudden, the firm
may not have enough resources to reposition itself.\ The new sector may turn
out to be a capital trap.

The patterns of possible trajectories are various and may be irregular, with
some transitions occurring at a constant rate, while others may involve
discontinuities and sudden increases or reductions in capital, depending on
the characteristics of the landscape such as expected returns in sectors,
density of firms, and other background factors.

\subsubsection{Investors transition functions}

\paragraph{Drift term}

Short-term and long-term returns are the two parameters that determine
investors' capital allocation. Short-term returns include the firm's
dividends and increase with the value of its shares, while long-term returns
reflect the market's expectations for the firm's future growth potential,
which in turn affect expectations for higher dividends and share price
appreciation. Both types of returns are captured in the drift term $%
D^{\prime }$, which is defined in equation (\ref{Frd}). The most likely
paths are those that maximize both short-term and long-term returns.

However, these returns are not independent, since faltering share prices in
the short-term impact long-term returns expectations, and vice versa.

Ideally, to maximize their capital, investors seek both higher short-term
and long-term returns. Therefore, capital allocation within and across
sectors will depend on firms share prices volatility and dividends.

A sector in which share prices increase tends to attract capital, since
investors can maximize both short-term and long-term returns: an increase in
share prices sustains the firm's growth expectations. Investors tend to move
towards the next local maximum of long-term returns while also maximizing
their short-term return. In this case, there is no trade-off between the two
objectives.

In a sector where stock prices fall or remain stagnant, investors are faced
with a trade-off between short- and long-term returns. When stock prices no
longer support long-term earnings expectations, capital allocation is
determined by short-term dividends. Capital reallocation will depend on the
level of capital held by investors. While investors may consider long-term
expectationsthey must also generate short-term returns to maintain their
capital. An investor who ignores dividends in a context of falling share
prices would eventually see his capital depleted, which could hinder or
impair his ability to reallocate capital in the long term.

\paragraph{Stabilization terms:}

Similarly to firms, investors have an effective inverse mobility $\alpha
_{eff}^{\prime }$, defined in equation (\ref{Frt}). This formula shows that
mobility $\frac{1}{\alpha _{eff}^{\prime }}$ decreases with the average
short-term return along the path\ : the higher the returns, the lower the
incentive to switch from one sector to another. Similarly, mobility
increases with $g^{\left( R\right) }\left( \hat{X}\right) $, which measures
the relative long-term return of the sectors along the path. The higher this
relative return, the lower the incentive to switch to another sector.

Moreover, $\frac{1}{\alpha _{eff}^{\prime }}$ decreases with the final level
of capital $\hat{K}_{f}$ increases, impairing the firm's capacity to reach
high levels of capital. Conversely, $\frac{1}{\alpha _{eff}^{\prime }}$
decreases with the initial capital $\hat{K}_{i}$ decreases, indicating that
investors with high capitalization are less likely to experience significant
capital losses. This is supported by the factor multiplying $\alpha
_{eff}^{\prime }$: 
\begin{equation*}
\left\vert \left( \hat{K}_{f}+\frac{\sigma _{\hat{K}}^{2}F\left( \hat{X}%
_{f},K_{\hat{X}_{f}}\right) }{f^{2}\left( \hat{X}_{f}\right) }\right)
-\left( \hat{K}_{i}+\frac{\sigma _{\hat{K}}^{2}F\left( \hat{X}_{i},K_{\hat{X}%
_{i}}\right) }{f^{2}\left( \hat{X}_{i}\right) }\right) \right\vert
\end{equation*}%
As a result, the probability for an investor to deviate significantly\textbf{%
\ }from its initial capital value, apart from the smoothing term which can
be neglected, is relatively low.

\subsection{Two-agent transition functions}

First, it should be noted that the transition function $G_{22}$, as defined
in equation (\ref{ttg}), does not include any interaction corrections.
Specifically, the transition probability for two investors is simply the
product of their individual transition probabilities. In our model,
investors do not directly interact with each other, but only through their
investments in various firms. Only two transition functions are affected by
these indirect interactions.

\subsubsection{Firm-firm interactions}

First the transition $G_{11}$ is modified by the term:%
\begin{equation*}
I=2\tau -\nabla _{K}\frac{\delta ^{2}u\left( \bar{K},\bar{X},\Psi ,\hat{\Psi}%
\right) }{\delta \Psi \left( \bar{K}^{\prime },\bar{X}\right) \delta \Psi
^{\dagger }\left( K^{\prime },\bar{X}\right) }
\end{equation*}%
The interaction $I$ measures the interactions between two firms in the same
sector. The first contribution to $I$ describes a direct competition between
firms in a given sector, whereas the second term describes the competition
of the firms to attract investors that share their investments between the
two firms. Given that the 2-agents transition functions are modified by (see
(\ref{nng})):%
\begin{equation*}
\left( 2\tau -\nabla _{K}\frac{\delta ^{2}u\left( \bar{K},\bar{X},\Psi ,\hat{%
\Psi}\right) }{\delta \Psi \left( \bar{K}^{\prime },\bar{X}\right) \delta
\Psi ^{\dagger }\left( K^{\prime },\bar{X}\right) }\right) \hat{G}_{1}\left(
\left( K_{f},X_{f}\right) ,\left( X_{i},K_{i}\right) \right) \hat{G}%
_{1}\left( \left( K_{f},X_{f}\right) ^{\prime },\left( X_{i},K_{i}\right)
^{\prime }\right)
\end{equation*}%
and since $I>0$, the contribution to the green function of paths crossing at
some point are underweighted. The competition between the two firms repell
them from the sector where they interact. If we consider that the competion
factor $\tau $ is capital-dependent (see (\ref{tfc})), the less capitalized
firm is relatively more repelled than the more capitalized one.

\subsubsection{Firm-investor interactions}

Second, the firm-investor transition function $G_{12}$ is modified by the
term: 
\begin{equation*}
\left( \nabla _{K}\frac{\delta ^{2}u\left( \bar{K},\bar{X},\Psi ,\hat{\Psi}%
\right) }{\delta \hat{\Psi}\left( \overline{\hat{K},\hat{X}}\right) \delta 
\hat{\Psi}^{\dagger }\left( \overline{\hat{K},\hat{X}}\right) }+\left\{
\nabla _{\hat{K}}\frac{\hat{K}\delta ^{2}f\left( \overline{\hat{X}},\Psi ,%
\hat{\Psi}\right) }{\delta \Psi \left( \overline{K^{\prime },X}\right)
\delta \Psi ^{\dagger }\left( \overline{K^{\prime },X}\right) }+\nabla _{%
\hat{X}}\frac{\delta ^{2}g\left( \overline{\hat{X}},\Psi ,\hat{\Psi}\right) 
}{\delta \Psi \overline{\left( K^{\prime },X\right) }\delta \Psi ^{\dagger
}\left( \overline{K^{\prime },X}\right) }\right\} \right)
\end{equation*}%
Given (\ref{duh}), (\ref{dth}), (\ref{dtg}), this term depends mainly on
three contributions: 
\begin{equation*}
\nabla _{K}\frac{F_{2}\left( s,R\left( K^{\prime },X\right) \right) \hat{K}}{%
\int F_{2}\left( s,R\left( K^{\prime },X\right) \right) \left\Vert \Psi
\left( K^{\prime },X\right) \right\Vert ^{2}dK^{\prime }}
\end{equation*}%
\begin{equation*}
\nabla _{\hat{K}}\Delta f\left( K^{\prime },\hat{X},\Psi ,\hat{\Psi}\right)
\end{equation*}%
\begin{equation*}
\nabla _{\hat{X}}\Delta \left( g\left( K^{\prime },\hat{X},\Psi ,\hat{\Psi}%
\right) \right)
\end{equation*}%
each of this contribution describes the relative perspectives of the firm in
his path through the sectors.

The first one represents the gradient of firm"s attractiveness with respect
to capital. The investor willl decide to invest or not depending on the
marginal gain of long term returns of the firm.

The second term represents the marginal short-term return of an investment
of the firm, and the third one measures the reltive attractiveness of the
firm with respect to his neighbours (see Gosselin Lotz Wambst 2022).

The interaction between the firm and the investor is a combination of these
three quantities.

When the combination of these term is positive, the firm has positive
perspective either in terms of short or long term returns, or relatively to
his neighbors.

In this case the associated corrections to the path crossing at some points
is positive and this paths will be overweighted: in probability, this
translates by the fact that paths in which a firm presents above average
perspectives in his capital accumaltion and shift in sectors, will be
favoured by an increase in investment. The firm will take advantage from its
interaction with the investor, except if this one experiences, for any
reason, an decrease of capital. On the contrary, a firm perceived as moving
toward lower perspective will experience in average a decrease in
investment. This decrease will be dampened if its investor has itself low
capital to invest. Some mixed situation may arise: good short term
perspective, but uncertain long term expectations may cancel or compensate
each other.

\part*{Discussion of the main outcomes}

Field formalism presents an alternate point of view about the economic
reality that surround us. In such a formalism, representative agents do not
exist, only collective states do. They emerge from the interactions of a
large number of agents, and condition the behaviours and the economic
activity. In this context, agents only randomly carry out the possible
trajectories authorized by the system.

Collective states can be multiple and present transitions.\ The economic
dynamic is not limited to fluctuations around an average trajectory which
would be a dynamic equilibrium, but rather by dynamic transitions between
collective states, which completely condition the fluctuations. apparent
individual dynamics. The collective states dynamics depend on the form of
short-term and long-term return functions, that are exogenous, and more
broadly on a whole landscape of technological and economic conditions. But
as a system, they have their own internal dynamics: the system is not inert.
We have considered these two types of variations in the paper.

First, collective states are sensitive to structural changes. Any such
change in expectations, economic and/or monetary conditions may alter
expected returns and in turn impact the collective state. Unstable type-3
sectors are particularly sensitive to these changes in long-term growth,
inflation, and interest rates. Higher expectations in these sectors attract
investment which in turn increase expectations. This seemingly endless
expected growth spirals until the outlook flattens or deteriorates. This
would be the growth model of a company whose ever-broadening range of
products fuels higher expected long-term returns and stock price increases.
Type-1 and -2 sectors attract capital through dividends and, although only
partially and for high capital type-2 sectors, expected returns. Under
higher expectations, these sectors are relatively less attractive than the
nearby type-3 sector.\ They may nonetheless survive in the long-term
provided their short-term returns and dividends are high enough. This may be
done by cutting costs or investment, at the expense of future growth.
Moreover, advert signaling may emerge: an increase in dividends can be
interpreted as faltering growth prospects. Conversely, any increase in
long-term uncertainty impact expected returns and drive sector-3 capital
towards other\ patterns. External shocks, inflation, and monetary policy
impact expectations, reduce long-term investment and either drive capital
out of sectors 3 to sector 1 or 2 or favour other pattern-3 sectors.

Second, any deviation of capital from its equilibrium value may initiate
oscillations in the collective state of the entire system. A temporary
deviation will induce an unstable redistribution of capital, growth
expectations and returns, and generate intersectoral capital reallocation
and global oscillations that can either dampen or drive the system toward a
new collective state. There are thus potential transitions between
collective states, which occur at a slower, larger timescale than that of
market fluctuations.\ In the long run, once the transition has occured, both
sectors' averages and patterns may have changed: pattern 2 may morph in,
say, pattern 3 stable or unstable, or sectors may simply disappear.
Concretely, any significant modification in average capital in a sector
could induce oscillations and initiate a transition.

Moreover, once endogenous expectations are introduced, they react to
variations in the capital: collective states of mixed 1-2-3 patterns are
difficult to maintain. Highly reactive expectations favour pattern 3:
expected returns magnify capital accumulation at the expense of other
patterns. Mildly reactive expectations favour patterns 1 and 2: their
oscillations, which are actually induced by uncertainties, dampen.\textbf{\ }%
Type-3 sectors on the contrary experience strong fluctuations in capital :
attracting capital is less effective with fading expectations. The threshold
in capital accumulation shifts upwards and the least-profitable firms are
ousted from the sector.\ The recent evolution in performances between value
and growth investment strategies exemplifies these shifts in investors'
sentiment between expected growth and real returns. In periods of
uncertainties, fluctuations affect capital accumulation in growth sectors
and today's tech companies, and strengthen more dividend-driven investments.
Note however that the most profitable and best-capitalized firms, that
remain above the threshold, maintain relatively high levels of capital.\
Here our versatile notion of firms\footnote{%
We modeled a single company as a set of independent firms. Similarly, the
notion of sector merely refers to a group of entities with similar
activities.} proves convenient: any firm that accumulates enough capital to
be able to buy back, in periods of volatility, its own stocks is actually
acting as an autonomous investor. When volatility is high, the most likely
investors for the best-capitalized firms are, first and foremost, the
best-capitalized companies themselves. They react, so to speak, as pools of
closely held investors.\ In other words, provided firms have high enough
capital, they can always cushion the impact of price fluctuations and
adverse shocks through buybacks. Similarly, they also could choose to
acquire companies in their sector or in neighbouring sectors.

Fluctuations in financial expectations impose their pace on the real
economy.\ Expected returns are both exogenous and endogenous. Being
exogenous, they may change quickly. Expected returns theoretically reflect
long-term perspectives, but actually rely on short-term sentiments: any
incoming information, change in the global economic outlook or adverse shock
will modify long-term expectations and shift capital from sector to sector.
But expected returns are also endogenous.\ Being expectations, they react to
changes within the system. When high levels of capital seek to maximize
returns, expectations react strongly to capital changes. Expectations both
highly sensitive to exogenous conditions and highly reactive to variations
in capital induce large fluctuations of capital in the system. Creating or
inflating expectations may attract capital, at times unduly.\ When
expectations can no longer be worked on, the sole remaining tool to reduce
capital outflows is a high dividend policy, which may be done at the expense
of the labour force, capital expenditures and future growth.

At the level of individual dynamics, macro fluctuations condition agents
transitions. In fact, the field formalism encompasses both macro and
microeconomic elements: the macro scale keeps track of the entire set of
agents and, in turn, influences the microeconomic scale, allowing for
two-level interpretations. We showed that individual dynamics heavily rely
on underlying macroeconomic parameters such as average capital and the
number of firms per sector. Consequently, our model also describes the
microeconomic impact of the present macroeconomic states.

In the face of macroeconomic fluctuations, investors may experience capital
losses. However they can always shield their capital by reallocating it to
more profitable or stable sectors. In doing so, they may amplify global
capital fluctuations for firms, which are unable to react at the same pace.
Financial risk is therefore limited in our model. Investors can always
reposition themselves and, as a result, do not bear the same risk as firms
that move to attract investors. The primary burden of risk falls on the
firms themselves, not the financial sector. Our model demonstrates that
investors do not experience the eviction phenomenon that firms do. However,
investors may face eviction from certain investment sectors if their capital
no longer allows them to invest in sectors perceived as the most promising,
based on returns and share prices.

We posit that firms have a natural inclination to switch sectors. Indeed,
firms tend to change due to the continuous evolution and transformation of
sectors and the changing economic environment. In our model, the historical
development of a sector is not depicted by a specific variable, but rather
by firms shifts between closely related sectors. In the shift, the initial
sector is the past state of the sector, and the final sector its present
state. Thus, firms transitions captures both firm reorientations and their
adaptation to an evolving environment.

Attracting investors is crucial to firms and can be achieved through
continuous expansion. However, firms face higher uncertainty and risk than
investors. Specifically, firms face two distinct risks:

First, the individual risk associated with seeking higher returns. Switching
to more attractive sectors may expose firms to higher competition and
faltering investors sentiment. For example, a firm shifting to a
high-capitalized sector will experience a stronger competition and weaker
prospects, potentially deterring any present or additional investment. When
these two phenomena combine, they may induce a substantial loss of capital,
and trap the firm in the sector, evict it towards less-capitalized and
less-attractive ones, and impact its ability to position itself for future
sectoral changes and transformations.

Second, the global risk, caused by exogenous and macro fluctuations.\ This
risk can alter sectoral growth prospects and, consequently, affect
individual dynamics. Our model captures these potential instabilities at the
individual level.\ Within a sector, sub-sectors may emerge, some presenting
more promising opportunities than others. However, the entire sector can be
impacted. Even though, on average, the collective state may exhibit some
stability, fluctuations among a set of similar firms can be substantial at
the individual dynamics level. Consequently, fluctuations in this context
magnify the uncertainty at the individual level, making it difficult to
identify and capitalize on profitable shifts while also increasing the risk
of making detrimental moves. To sum up, both collective and individual
results suggest that firms with high initial capitalization are generally
less exposed to market fluctuations. Note incidentally that these risks may
be amplified by swift financial reallocation in the face of global
uncertainties.

Therefore, firms can undergo sharp changes in dynamics due to variations in
the landscape of expected returns, reactivity of expectations, relative
attractiveness compared to neighboring sectors, or the number of competing
firms.

The present paper also advocates that field formalism, in addition to mixing
macro and micro analysis, provides some precise insights about the
structures of interactions inside the macroeconomic state. The technique of
series expansion of the effective action induces emerging interactions that
are not detected in the classical formalism, such as indirect emerging
competition among agents. More precisely, interactions between firms within
a sector reveal phenomena of specialization and eviction. Competition is at
first determined by the firms relative levels of capital. This is the direct
form of competition. The firm with the highest capital is more likely to
evict its competitors. However, field formalism reveals that competition
also revolves around attracting investor capital. This is the indirect form
of competition. A firm that successfully differentiates itself within a
sector, through specialization, has the potential to attract capital and
mitigate or reverse a possible eviction. However, specialization makes the
firm dependent on its investors. If investors suffer capital losses, the
firm is directly impacted.

Interactions between firms and their investors detail the impact of
investment at the individual level. A firm that attracts more investors will
be better positioned in the sector, as it enjoys a stronger position,
whereas its competitors will be compelled to reorient themselves. To attract
investors, a firm needs to demonstrate a high growth potential, which may
favor new entrants in a sector, provided they have the necessary capital to
position themselves, or better growth prospects. Note that from this
perspective the concept of comparative advantage is not relevant in our
model. Indeed, given that changes are inevitable within sectors, any
comparative advantage is bound to be swept away, potentially even by
relatively distant and unexpected causes. Actually, exogenous fluctuations,
such as the perception of the sector and the firm within it (by investors),
as well as competition among firms to retain their position and attract
investors, create inherent instability within a specific sector.
Specializing in a single sector exposes firms to the risk of eventual
eviction, forming a trap.

\section{Conclusion}

This paper has shown how a statistical field model could be constructed from
a simple microeconomic model. Using a simple economic framework involving
two types of agents, firms and investors, we have studied the impact
financial capital could have on physical capital allocation. and shown the
complexity of the collective states reached in this very simple case. We
have examined how variations in external parameters could induce transitions
in these collective states.

At the individual level, we derived the probabilistic dynamics of agents in
this environment. We identified several types of dynamics for producers,
depending on the firms' landscape, returns, and the firms' and sectors'
relative attractiveness. A firm's dynamics also depends on its initial
sector and level of capital, and may exhibit turning points Modifications of
the macroeconomic state may lead to significant fluctuations in a firm's
growth trajectory.

However, in this work, to examine the impact of financial allocation, we
concentrated on interactions between firms and between firms and investors.
Investors only interact indirectly, through firms. For a more comprehensive
study, we will include interactions between investors in a subsequent work,
where they invest in each other.

\section*{References}


\begin{description}
\item Abergel F, Chakraborti A, Muni Toke I and Patriarca M (2011a)
Econophysics review: I. Empirical facts, Quantitative Finance, Vol. 11, No.
7, 991-1012

\item Abergel F, Chakraborti A, Muni Toke I and Patriarca M (2011b)
Econophysics review: II. Agent-based models, Quantitative Finance, Vol. 11,
No. 7, 1013-1041

\item Bardoscia M., Livan G., Marsili M. (2017), Statistical mechanics of
complex economies, Journal of Statistical Mechanics: Theory and Experiment,
Volume 2017

\item Bernanke B., Gertler, M. and S. Gilchrist (1999), The financial
accelerator in a quantitative business cycle framework, Chapter 21 in
Handbook of Macroeconomics, 1999, vol. 1, Part C, pp 1341-1393

\item Bensoussan A, Frehse J, Yam P (2018) Mean Field Games and Mean Field
Type Control Theory. Springer, New York

\item B\"{o}hm, V., Kikuchi, T., Vachadze, G.: Asset pricing and
productivity growth: the role of consumption scenarios. Comput. Econ. 32,
163--181 (2008)

\item Caggese A, Orive A P, The Interaction between Household and Firm
Dynamics and the Amplification of Financial Shocks. Barcelona GSE Working
Paper Series, Working Paper n%
${{}^o}$
866, 2015

\item Campello, M., Graham, J. and Harvey, C.R. (2010). The Real Effects of
Financial Constraints: Evidence from a Financial Crisis, Journal of
Financial Economics, vol. 97(3), 470-487.

\item Gaffard JL and Napoletano M Editors (2012): Agent-based models and
economic policy. OFCE, Paris

\item Gomes DA, Nurbekyan L, Pimentel EA (2015) Economic Models and
Mean-Field Games Theory, Publica\c{c}\~{o}es Matem\'{a}ticas do IMPA, 30o Col%
\'{o}quio Brasileiro de Matem\'{a}tica, Rio de Janeiro

\item Gosselin P, Lotz A and Wambst M (2017) A Path Integral Approach to
Interacting Economic Systems with Multiple Heterogeneous Agents. IF\_PREPUB.
2017. hal-01549586v2

\item Gosselin P, Lotz A and Wambst M (2020) A Path Integral Approach to
Business Cycle Models with Large Number of Agents. Journal of Economic
Interaction and Coordination volume 15, pages 899--942

\item Gosselin P, Lotz A and Wambst M (2021) A statistical field approach to
capital accumulation. Journal of Economic Interaction and Coordination 16,
pages 817--908 (2021)

\item Grassetti, F., Mammana, C. \& Michetti, E. A dynamical model for real
economy and finance. Math Finan Econ (2022).
https://doi.org/10.1007/s11579-021-00311-3

\item Grosshans, D., Zeisberger, S.: All's well that ends well? on the
importance of how returns are achieved. J. Bank. Finance 87, 397--410 (2018)

\item Holmstrom, B., and Tirole, J. (1997). Financial intermediation,
loanable funds, and the
\end{description}

real sector. Quarterly Journal of Economics, 663-691.

\begin{description}
\item Jackson M (2010) Social and Economic Networks. Princeton University
Press, Princeton

\item Jermann, U.J. and Quadrini, V., (2012). "Macroeconomic Effects of
Financial Shocks," American Economic Review, Vol. 102, No. 1.

\item Khan, A., and Thomas, J. K. (2013). "Credit Shocks and Aggregate
Fluctuations in an Economy with Production Heterogeneity," Journal of
Political Economy, 121(6), 1055-1107.

\item Kaplan G, Violante L (2018) Microeconomic Heterogeneity and
Macroeconomic Shocks, Journal of Economic Perspectives, Vol. 32, No. 3,
167-194

\item Kleinert H (1989) Gauge fields in condensed matter Vol. I , Superflow
and vortex lines, Disorder Fields, Phase Transitions, Vol. II, Stresses and
defects, Differential Geometry, Crystal Melting. World Scientific, Singapore

\item Kleinert H (2009) Path Integrals in Quantum Mechanics, Statistics,
Polymer Physics, and Financial Markets 5th edition. World Scientific,
Singapore

\item Krugman P (1991) Increasing Returns and Economic Geography. Journal of
Political Economy, 99(3), 483-499

\item Lasry JM, Lions PL, Gu\'{e}ant O (2010a) Application of Mean Field
Games to Growth Theory \newline
https://hal.archives-ouvertes.fr/hal-00348376/document

\item Lasry JM, Lions PL, Gu\'{e}ant O (2010b) Mean Field Games and
Applications. Paris-Princeton lectures on Mathematical Finance, Springer.%
\textbf{\ }https://hal.archives-ouvertes.fr/hal-01393103

\item Lux T (2008) Applications of Statistical Physics in Finance and
Economics. Kiel Institute for the World Economy (IfW), Kiel

\item Lux T (2016) Applications of Statistical Physics Methods in Economics:
Current state and perspectives. Eur. Phys. J. Spec. Top. (2016) 225: 3255.
https://doi.org/10.1140/epjst/e2016-60101-x

\item Mandel A, Jaeger C, F\"{u}rst S, Lass W, Lincke D, Meissner F,
Pablo-Marti F, Wolf S (2010). Agent-based dynamics in disaggregated growth
models. Documents de travail du Centre d'Economie de la Sorbonne. Centre
d'Economie de la Sorbonne, Paris

\item Mandel A (2012) Agent-based dynamics in the general equilibrium model.
Complexity Economics 1, 105--121

\item Monacelli, T., Quadrini, V. and A. Trigari (2011). "Financial Markets
and Unemployment," NBER Working Papers 17389, National Bureau of Economic
Research.

\item Sims C A (2006) Rational inattention: Beyond the Linear Quadratic
Case, American Economic Review, vol. 96, no. 2, 158-163

\item Yang J (2018) Information theoretic approaches to economics, Journal
of Economic Survey, Vol. 32, No. 3, 940-960

\item Cochrane, J.H. (ed.): Financial Markets and the Real Economy,
International Library of Critical Writings in Financial Economics, vol. 18.
Edward Elgar (2006)
\end{description}

\pagebreak

\section*{Appendix 1 From large number of agents to field formalism}

This appendix summarizes the most useful steps of the method developed in
Gosselin, Lotz and Wambst (2017, 2020, 2021),\ to switch from the
probabilistic description of the model to the field theoretic formalism and
summarizes the translation of a generalization of (\ref{mNZ}) involving
different time variables. By convention and unless otherwise mentioned, the
symbol of integration $\int $\ refers to all the variables involved.

\subsection*{A1.1 Probabilistic formalism}

The probabilistic formalism for a system with $N$ identical economic agents
in interaction is based on the minimization functions described in the text.
Classically, the dynamics derives through the optimization problem of these
functions. The probabilistic formalism relies on the contrary on the fact,
that, due to uncertainties, shocks... agents do not optimize fully these
functions. Moreover, given the large number of agents, there may be some
discrepancy between agents minimization functions, and this fact may be
translated in an uncertainty of behavior around some average minimization,
or objective function.

We thus assume that each agent chooses for his action a path randomly
distributed around the optimal path. The agent's behavior can be described
as a weight that is an exponential of the intertemporal utility, that
concentrates the probability around the optimal path. This feature models
some internal uncertainty as well as non-measurable shocks. Gathering all
agents, it yields a probabilistic description of the system in terms of a
probabilistic weight.

In general, this weight includes utility functions and internalizes
forward-looking behaviors, such as intertemporal budget constraints and
interactions among agents. These interactions may for instance arise through
constraints, since income flows depend on other agents demand. The
probabilistic description then allows to compute the transition functions of
the system, and in turn compute the probability for a system to evolve from
an initial state to a final state within a given time span. They have the
form of Euclidean path integrals.

In the context of the present paper, we have seen that the minimization
functions for the system considered in this work have the form:%
\begin{eqnarray}
&&\int dt\left( \sum_{i}\left( \frac{d\mathbf{A}_{i}\left( t\right) }{dt}%
-\sum_{j,k,l...}f\left( \mathbf{A}_{i}\left( t\right) ,\mathbf{A}_{j}\left(
t\right) ,\mathbf{A}_{k}\left( t\right) ,\mathbf{A}_{l}\left( t\right)
...\right) \right) ^{2}\right.  \label{mnz} \\
&&\left. +\sum_{i}\left( \sum_{j,k,l...}g\left( \mathbf{A}_{i}\left(
t\right) ,\mathbf{A}_{j}\left( t\right) ,\mathbf{A}_{k}\left( t\right) ,%
\mathbf{A}_{l}\left( t\right) ...\right) \right) \right)  \notag
\end{eqnarray}%
The minimization of this function will yield a dynamic equation for $N$
agents in interaction described by a set of dynamic variables $\mathbf{A}%
_{i}\left( t\right) $ during a given timespan $T$.

The probabilistic description is straightforwardly obtained from (\ref{mnz}%
). The probability associated to a configuration $\left( \mathbf{A}%
_{i}\left( t\right) \right) _{\substack{ i=1,...,N  \\ 0\leqslant t\leqslant
T }}$ \ is directly given by:%
\begin{eqnarray}
&&\mathcal{N}\exp \left( -\frac{1}{\sigma ^{2}}\int dt\left( \sum_{i}\left( 
\frac{d\mathbf{A}_{i}\left( t\right) }{dt}-\sum_{j,k,l...}f\left( \mathbf{A}%
_{i}\left( t\right) ,\mathbf{A}_{j}\left( t\right) ,\mathbf{A}_{k}\left(
t\right) ,\mathbf{A}_{l}\left( t\right) ...\right) \right) ^{2}\right.
\right.  \label{prz} \\
&&\left. \left. +\sum_{i}\left( \sum_{j,k,l...}g\left( \mathbf{A}_{i}\left(
t\right) ,\mathbf{A}_{j}\left( t\right) ,\mathbf{A}_{k}\left( t\right) ,%
\mathbf{A}_{l}\left( t\right) ...\right) \right) \right) \right)  \notag
\end{eqnarray}%
where $\mathcal{N}$ is a normalization factor and $\sigma ^{2}$ is a
variance whose magnitude describes the amplitude of deviations around the
optimal path.

As in the paper, the system is in general modelled by several equations, and
thus, several minimization function. The overall system is thus described by
several functions, and the minimization function of the whole system is
described by the set of functions:%
\begin{eqnarray}
&&\int dt\left( \sum_{i}\left( \frac{d\mathbf{A}_{i}\left( t\right) }{dt}%
-\sum_{j,k,l...}f^{\left( \alpha \right) }\left( \mathbf{A}_{i}\left(
t\right) ,\mathbf{A}_{j}\left( t\right) ,\mathbf{A}_{k}\left( t\right) ,%
\mathbf{A}_{l}\left( t\right) ...\right) \right) ^{2}\right.  \label{znm} \\
&&\left. +\sum_{i}\left( \sum_{j,k,l...}g^{\left( \alpha \right) }\left( 
\mathbf{A}_{i}\left( t\right) ,\mathbf{A}_{j}\left( t\right) ,\mathbf{A}%
_{k}\left( t\right) ,\mathbf{A}_{l}\left( t\right) ...\right) \right) \right)
\notag
\end{eqnarray}%
where $\alpha $ runs over the set equations describing the system's
dynamics. The associated weight is then:%
\begin{eqnarray}
&&\mathcal{N}\exp \left( -\left( \sum_{i,\alpha }\frac{1}{\sigma _{\alpha
}^{2}}\int dt\left( \frac{d\mathbf{A}_{i}\left( t\right) }{dt}%
-\sum_{j,k,l...}f^{\left( \alpha \right) }\left( \mathbf{A}_{i}\left(
t\right) ,\mathbf{A}_{j}\left( t\right) ,\mathbf{A}_{k}\left( t\right) ,%
\mathbf{A}_{l}\left( t\right) ...\right) \right) ^{2}\right. \right.
\label{pnz} \\
&&\left. \left. +\sum_{i,\alpha }\left( \sum_{j,k,l...}g^{\left( \alpha
\right) }\left( \mathbf{A}_{i}\left( t\right) ,\mathbf{A}_{j}\left( t\right)
,\mathbf{A}_{k}\left( t\right) ,\mathbf{A}_{l}\left( t\right) ...\right)
\right) \right) \right)  \notag
\end{eqnarray}

The appearance of the sum of minimization functions in the probabilistic
weight (\ref{pnz}) translates the hypothesis that the deviations with
respect to the optimization of the functions (\ref{znm}) are assumed to be
independent.

For a large number of agents, the system described by (\ref{pnz}) involves a
large number of variables $K_{i}\left( t\right) $, $P_{i}\left( t\right) $
and $X_{i}\left( t\right) $\ that are difficult to handle. To overcome this
difficulty, we consider the space $H$\ of complex functions defined on the
space of a single agent's actions. The space $H$ describes the collective
behavior of the system. Each function $\Psi $ of $H$ encodes a particular
state of the system. We then associate to\ each function $\Psi $ of $H$ a
statistical weight, i.e. a probability describing the state encoded in $\Psi 
$. This probability is written $\exp \left( -S\left( \Psi \right) \right) $,
where $S\left( \Psi \right) $ is a functional, i.e. the function of the
function $\Psi $. The form of $S\left( \Psi \right) $ is derived directly
from the form of (\ref{pnz}) as detailed in the text. As seen from (\ref{pnz}%
), this translation can in fact be directly obtained from the sum of
"classical" minimization functions weighted by the factors $\frac{1}{\sigma
_{\alpha }^{2}}$:%
\begin{eqnarray*}
&&\sum_{i,\alpha }\frac{1}{\sigma _{\alpha }^{2}}\int dt\left( \frac{d%
\mathbf{A}_{i}\left( t\right) }{dt}-\sum_{j,k,l...}f^{\left( \alpha \right)
}\left( \mathbf{A}_{i}\left( t\right) ,\mathbf{A}_{j}\left( t\right) ,%
\mathbf{A}_{k}\left( t\right) ,\mathbf{A}_{l}\left( t\right) ...\right)
\right) ^{2} \\
&&+\sum_{i,\alpha }\left( \sum_{j,k,l...}g^{\left( \alpha \right) }\left( 
\mathbf{A}_{i}\left( t\right) ,\mathbf{A}_{j}\left( t\right) ,\mathbf{A}%
_{k}\left( t\right) ,\mathbf{A}_{l}\left( t\right) ...\right) \right)
\end{eqnarray*}%
This is this shortcut we used in the text.

\subsection*{A1.2 Interactions between agents at different times}

A straightforward generalization of (\ref{mNZ}) involve agents interactions
at different times. The terms considered have the form:%
\begin{eqnarray}
&&\sum_{i}\left( \frac{d\mathbf{A}_{i}\left( t\right) }{dt}%
-\sum_{j,k,l...}\int f\left( \mathbf{A}_{i}\left( t_{i}\right) ,\mathbf{A}%
_{j}\left( t_{j}\right) ,\mathbf{A}_{k}\left( t_{k}\right) ,\mathbf{A}%
_{l}\left( t_{l}\right) ...,\mathbf{t}\right) \mathbf{dt}\right) ^{2}
\label{gR} \\
&&+\sum_{i}\sum_{j,k,l...}\int g\left( \mathbf{A}_{i}\left( t_{i}\right) ,%
\mathbf{A}_{j}\left( t_{j}\right) ,\mathbf{A}_{k}\left( t_{k}\right) ,%
\mathbf{A}_{l}\left( t_{l}\right) ...,\mathbf{t}\right) \mathbf{dt}  \notag
\end{eqnarray}%
where $\mathbf{t}$\ stands for $\left( t_{i},t_{j},t_{k},t_{l}...\right) $
and $\mathbf{dt}$ stands for $dt_{i}dt_{j}dt_{k}dt_{l}...$

The translation is straightforward. We introduce a time variable $\theta $
on the field side and the fields write $\left\vert \Psi \left( \mathbf{A}%
,\theta \right) \right\vert ^{2}$ and $\left\vert \hat{\Psi}\left( \mathbf{%
\hat{A}},\hat{\theta}\right) \right\vert ^{2}$. The second term in (\ref{gR}%
) becomes: 
\begin{eqnarray}
&&\sum_{i}\sum_{j}\sum_{j,k...}\int g\left( \mathbf{A}_{i}\left(
t_{i}\right) ,\mathbf{A}_{j}\left( t_{j}\right) ,\mathbf{A}_{k}\left(
t_{k}\right) ,\mathbf{A}_{l}\left( t_{l}\right) ...,\mathbf{t}\right) 
\mathbf{dt}  \notag \\
&\rightarrow &\int g\left( \mathbf{A},\mathbf{A}^{\prime },\mathbf{A}%
^{\prime \prime },\mathbf{\hat{A},\hat{A}}^{\prime }...,\mathbf{\theta ,\hat{%
\theta}}\right) \left\vert \Psi \left( \mathbf{A},\theta \right) \right\vert
^{2}\left\vert \Psi \left( \mathbf{A}^{\prime },\theta ^{\prime }\right)
\right\vert ^{2}\left\vert \Psi \left( \mathbf{A}^{\prime \prime },\theta
^{\prime \prime }\right) \right\vert ^{2}d\mathbf{A}d\mathbf{A}^{\prime }d%
\mathbf{A}^{\prime \prime } \\
&&\times \left\vert \hat{\Psi}\left( \mathbf{\hat{A}},\hat{\theta}\right)
\right\vert ^{2}\left\vert \hat{\Psi}\left( \mathbf{\hat{A}}^{\prime },\hat{%
\theta}^{\prime }\right) \right\vert ^{2}d\mathbf{\hat{A}}d\mathbf{\hat{A}}%
^{\prime }\mathbf{d\theta d\hat{\theta}}  \notag
\end{eqnarray}%
where $\mathbf{\theta }$ and $\mathbf{\hat{\theta}}$ are the multivariables $%
\left( \theta ,\theta ^{\prime },\theta ^{\prime \prime }...\right) $ and $%
\left( \hat{\theta},\hat{\theta}^{\prime }...\right) $ respectively and $%
\mathbf{d\theta d\hat{\theta}}$ stands for $d\theta d\theta ^{\prime
}d\theta ^{\prime \prime }...$ and $d\hat{\theta}d\hat{\theta}^{\prime }...$

Similarly, the first term in (\ref{gR}) translates as:%
\begin{eqnarray}
&&\sum_{i}\left( \frac{d\mathbf{A}_{i}\left( t\right) }{dt}%
-\sum_{j,k,l...}\int f\left( \mathbf{A}_{i}\left( t_{i}\right) ,\mathbf{A}%
_{j}\left( t_{j}\right) ,\mathbf{A}_{k}\left( t_{k}\right) ,\mathbf{A}%
_{l}\left( t_{l}\right) ...,\mathbf{t}\right) \mathbf{dt}\right) ^{2} \\
&\rightarrow &\int \Psi ^{\dag }\left( \mathbf{A},\theta \right) \left(
-\nabla _{\mathbf{A}^{\left( \alpha \right) }}\left( \frac{\sigma _{\mathbf{A%
}^{\left( \alpha \right) }}^{2}}{2}\nabla _{\mathbf{A}^{\left( \alpha
\right) }}-\Lambda (\mathbf{A},\theta )\right) \right) \Psi \left( \mathbf{A}%
,\theta \right) d\mathbf{A}d\theta
\end{eqnarray}%
by:%
\begin{eqnarray}
\Lambda (\mathbf{A},\theta ) &=&\int f^{\left( \alpha \right) }\left( 
\mathbf{A},\mathbf{A}^{\prime },\mathbf{A}^{\prime \prime },\mathbf{\hat{A},%
\hat{A}}^{\prime }...,\mathbf{\theta ,\hat{\theta}}\right) \left\vert \Psi
\left( \mathbf{A}^{\prime },\theta ^{\prime }\right) \right\vert
^{2}\left\vert \Psi \left( \mathbf{A}^{\prime \prime },\theta ^{\prime
\prime }\right) \right\vert ^{2}d\mathbf{A}^{\prime }d\mathbf{A}^{\prime
\prime } \\
&&\times \left\vert \hat{\Psi}\left( \mathbf{\hat{A}},\theta \right)
\right\vert ^{2}\left\vert \hat{\Psi}\left( \mathbf{\hat{A}}^{\prime
},\theta ^{\prime \prime }\right) \right\vert ^{2}d\mathbf{\hat{A}}d\mathbf{%
\hat{A}}^{\prime }\mathbf{d\bar{\theta}d\hat{\theta}}  \notag
\end{eqnarray}%
with $\mathbf{d\bar{\theta}}=d\theta ^{\prime }d\theta ^{\prime \prime }$.

Ultimately, as in the text, additional terms (\ref{fct}):%
\begin{eqnarray}
&&\Psi ^{\dag }\left( \mathbf{A},\theta \right) \left( -\nabla _{\theta
}\left( \frac{\sigma _{\theta }^{2}}{2}\nabla _{\theta }-1\right) \right)
\Psi \left( \mathbf{A},\theta \right) \\
&&+\hat{\Psi}^{\dag }\left( \mathbf{\hat{A}},\theta \right) \left( -\nabla
_{\theta }\left( \frac{\sigma _{\theta }^{2}}{2}\nabla _{\theta }-1\right)
\right) \hat{\Psi}\left( \mathbf{\hat{A}},\theta \right) +\alpha \left\vert
\Psi \left( \mathbf{A}\right) \right\vert ^{2}+\alpha \left\vert \hat{\Psi}%
\left( \mathbf{\hat{A}}\right) \right\vert ^{2}  \notag
\end{eqnarray}%
are included to the action functional to take into account for the time
variable.

\subsection*{A1.3 Translation of the minimization functions}

\subsubsection*{Real economy}

\paragraph*{Translation of the minimization function: Physical capital
allocation}

Let us start by translating in terms of fields the expression (\ref{minX}):

\begin{equation}
\sum_{i}\left( \left( \frac{dX_{i}}{dt}-\nabla _{X}R\left(
K_{i},X_{i}\right) H\left( K_{i}\right) \right) ^{2}+\tau \frac{K_{X_{i}}}{%
K_{i}}\sum_{j}\delta \left( X_{i}-X_{j}\right) \right)  \label{mnd}
\end{equation}%
To do so, we first consider the last term $\tau \frac{K_{X_{i}}}{K_{i}}%
\sum_{i}\sum_{j}\delta \left( X_{i}-X_{j}\right) $. This term contains no
derivative. The form of the translation is given by formula (\ref{tln}).
Since the expression contains two indices, both of them are summed.\ 

The first step of the translation is to replace $X_{i}$ and $X_{j}$ by two
variables $X$ et $X^{\prime }$, and substitute:

\begin{equation*}
\tau \frac{K_{X_{i}}}{K_{i}}\delta \left( X_{i}-X_{j}\right) \rightarrow
\tau \frac{K_{X}}{K}\delta \left( X-X^{\prime }\right)
\end{equation*}%
where $K_{X}$ is the average capital per firm in sector $X$. The sum over $i$
and the sum over $j$ are then replaced directly by the integrals $\int
\left\vert \Psi \left( K,X\right) \right\vert ^{2}d\left( K,X\right) $, $%
\int \left\vert \Psi \left( K^{\prime },X^{\prime }\right) \right\vert
^{2}d\left( K^{\prime },X^{\prime }\right) $, which leads to the following
translation:

\begin{eqnarray}
\tau \frac{K_{X_{i}}}{K_{i}}\sum_{i}\sum_{j}\delta \left( X_{i}-X_{j}\right)
&\rightarrow &\int \left\vert \Psi \left( K,X\right) \right\vert ^{2}d\left(
K,X\right) \int \left\vert \Psi \left( K^{\prime },X^{\prime }\right)
\right\vert ^{2}d\left( K^{\prime },X^{\prime }\right) \tau \frac{K_{X}}{K}%
\delta \left( X-X^{\prime }\right)  \notag \\
&=&\int \tau \frac{K_{X}}{K}\left\vert \Psi \left( K,X\right) \right\vert
^{2}\left\vert \Psi \left( K^{\prime },X\right) \right\vert ^{2}d\left(
K,X\right) dK^{\prime }  \label{tnl}
\end{eqnarray}%
To translate the first term in formula (\ref{mnd}): 
\begin{equation}
\sum_{i}\left( \frac{dX_{i}}{dt}-\nabla _{X}R\left( K_{i},X_{i}\right)
H\left( K_{i}\right) \right) ^{2}  \label{dnm}
\end{equation}%
We use the translation (\ref{Trl}) of a type-(\ref{inco}) expression. The
gradient term appearing in equation (\ref{Trl}) is $\nabla _{X}$. We thus
obtain the translation: 
\begin{eqnarray}
&&\sum_{i}\left( \frac{dX_{i}}{dt}-\nabla _{X}R\left( K_{i},X_{i}\right)
H\left( K_{i}\right) \right) ^{2}  \label{grl} \\
&\rightarrow &\int \Psi ^{\dag }\left( K,X\right) \left( -\nabla _{X}\left( 
\frac{\sigma _{X}^{2}}{2}\nabla _{X}+\Lambda (X,K)\right) \right) \Psi
\left( K,X\right) dKdX  \notag
\end{eqnarray}%
Note that the variance $\sigma _{X}^{2}$ reflects the probabilistic nature
of the model hidden behind the field formalism. This $\sigma _{X}^{2}$
represents the characteristic level of uncertainty of the sectors space
dynamics. It is a parameter of the model. The term $\Lambda (X,K)$ is the
translation of the term $-\nabla _{X}R\left( K_{i},X_{i}\right) H\left(
K_{i}\right) $ in the parenthesis of (\ref{dnm}). This term is a function of
one sole index "$i$". In that case, the term $\Lambda $ is simply obtained
by replacing $\left( K_{i},X_{i}\right) $ by $\left( K,X\right) $.\ We use
the translation (\ref{bdt}) of (\ref{ntr})-type term, so that $\Lambda $
writes:%
\begin{equation*}
\Lambda (X,K)=-\nabla _{X}R\left( K,X\right) H\left( K\right)
\end{equation*}%
and the translation of expression (\ref{dnm}) is: 
\begin{eqnarray}
&&\sum_{i}\left( \frac{dX_{i}}{dt}-\nabla _{X}R\left( K_{i},X_{i}\right)
H\left( K_{i}\right) \right) ^{2}  \label{ttn} \\
&\rightarrow &\int \Psi ^{\dag }\left( K,X\right) \left( -\nabla _{X}\left( 
\frac{\sigma _{X}^{2}}{2}\nabla _{X}-\nabla _{X}R\left( K,X\right) H\left(
K\right) \right) \right) \Psi \left( K,X\right) dKdX  \notag
\end{eqnarray}%
Using equations (\ref{tnl}) and (\ref{ttn}), the translation of (\ref{mnd})
is thus:%
\begin{eqnarray}
S_{1} &=&-\int \Psi ^{\dag }\left( K,X\right) \nabla _{X}\left( \frac{\sigma
_{X}^{2}}{2}\nabla _{X}-\nabla _{X}R\left( K,X\right) H\left( K\right)
\right) \Psi \left( K,X\right) dKdX  \label{sn} \\
&&+\tau \frac{K_{X}}{K}\int \left\vert \Psi \left( K^{\prime },X\right)
\right\vert ^{2}\left\vert \Psi \left( K,X\right) \right\vert ^{2}dK^{\prime
}dKdX  \notag
\end{eqnarray}

\paragraph*{Translation of the minimization function: Physical capital}

We can now turn to the translation of the second equation (\ref{minK}):

\begin{equation}
\sum_{i}\left( \frac{d}{dt}K_{i}+\frac{1}{\varepsilon }\left( K_{i}\left(
t\right) -\frac{F_{2}\left( R\left( K_{i}\left( t\right) ,X_{i}\left(
t\right) \right) \right) G\left( X_{i}\left( t\right) -\hat{X}_{j}\left(
t\right) \right) }{\sum_{l}F_{2}\left( R\left( K_{l}\left( t\right)
,X_{l}\left( t\right) \right) \right) G\left( X_{l}\left( t\right) -\hat{X}%
_{j}\left( t\right) \right) }\hat{K}_{j}\left( t\right) \right) \right) ^{2}
\label{mnn}
\end{equation}%
To detail the computations, we have kept the expanded formula (\ref{FRLV})
for $F_{2}\left( R\left( K_{i}\left( t\right) ,X_{i}\left( t\right) \right) ,%
\hat{X}_{j}\left( t\right) \right) $ Once again, we use the translation (\ref%
{bdt}) of (\ref{ntr})-type term, and start by building the field functional
associated to the term inside the square:%
\begin{equation*}
K_{i}\left( t\right) -\sum_{j}\frac{F_{2}\left( R\left( K_{i}\left( t\right)
,X_{i}\left( t\right) \right) \right) G\left( X_{i}\left( t\right) -\hat{X}%
_{j}\left( t\right) \right) }{\sum_{l}F_{2}\left( R\left( K_{l}\left(
t\right) ,X_{l}\left( t\right) \right) \right) G\left( X_{l}\left( t\right) -%
\hat{X}_{j}\left( t\right) \right) }\hat{K}_{j}\left( t\right)
\end{equation*}%
We replace:%
\begin{eqnarray*}
\left( K_{i}\left( t\right) ,X_{i}\left( t\right) \right) &\rightarrow
&\left( K,X\right) \\
\left( K_{l}\left( t\right) ,X_{l}\left( t\right) \right) &\rightarrow
&\left( K^{\prime },X^{\prime }\right) \\
\left( \hat{K}_{j}\left( t\right) ,\hat{X}_{j}\left( t\right) \right)
&\rightarrow &\left( \hat{K},\hat{X}\right)
\end{eqnarray*}%
and:%
\begin{equation}
K_{i}\left( t\right) -\sum_{j}\frac{F_{2}\left( R\left( K_{i}\left( t\right)
,X_{i}\left( t\right) \right) \right) G\left( X_{i}\left( t\right) -\hat{X}%
_{j}\right) }{\sum_{l}F_{2}\left( R\left( K_{l}\left( t\right) ,X_{l}\left(
t\right) \right) \right) G\left( X_{l}\left( t\right) -\hat{X}_{j}\right) }%
\hat{K}_{j}\left( t\right) \rightarrow K-\sum_{j}\frac{F_{2}\left( R\left(
K,X\right) \right) G\left( X-\hat{X}\right) }{\sum_{l}F_{2}\left( R\left(
K^{\prime },X^{\prime }\right) \right) G\left( X^{\prime }-\hat{X}\right) }%
\hat{K}  \label{tr}
\end{equation}%
The sum over $l$ is then replaced by an integral $\int \left\vert \Psi
\left( K^{\prime },X^{\prime }\right) \right\vert ^{2}d\left( K^{\prime
},X^{\prime }\right) $:%
\begin{eqnarray}
&&K_{i}\left( t\right) -\sum_{j}\frac{F_{2}\left( R\left( K_{i}\left(
t\right) ,X_{i}\left( t\right) \right) \right) G\left( X_{i}\left( t\right) -%
\hat{X}_{j}\right) }{\sum_{l}F_{2}\left( R\left( K_{l}\left( t\right)
,X_{l}\left( t\right) \right) \right) G\left( X_{l}\left( t\right) -\hat{X}%
_{j}\right) }\hat{K}_{j}\left( t\right)  \label{lrt} \\
&\rightarrow &K-\sum_{j}\frac{F_{2}\left( R\left( K,X\right) \right) G\left(
X-\hat{X}\right) }{\int \left\vert \Psi \left( K^{\prime },X^{\prime
}\right) \right\vert ^{2}d\left( K^{\prime },X^{\prime }\right) F_{2}\left(
R\left( K^{\prime },X^{\prime }\right) \right) G\left( X^{\prime }-\hat{X}%
_{j}\right) }\hat{K}  \notag
\end{eqnarray}%
Recall that investors' variables are denoted with an upper script $\symbol{94%
}$.

Finally, the sum over $j$ and the second field are replaced by $\int
\left\vert \hat{\Psi}\left( \hat{K},\hat{X}\right) \right\vert ^{2}d\left( 
\hat{K},\hat{X}\right) $. After introducing the characteristic factor $\frac{%
1}{\varepsilon }$ of the capital accumulation time scale (see (\ref{dnK})),
the translation becomes: 
\begin{eqnarray}
&&\frac{1}{\varepsilon }\left( K_{i}\left( t\right) -\sum_{j}\frac{%
F_{2}\left( R\left( K_{i}\left( t\right) ,X_{i}\left( t\right) \right)
\right) G\left( X_{i}\left( t\right) -\hat{X}_{j}\right) }{%
\sum_{l}F_{2}\left( R\left( K_{l}\left( t\right) ,X_{l}\left( t\right)
\right) \right) G\left( X_{l}\left( t\right) -\hat{X}_{j}\right) }\hat{K}%
_{j}\left( t\right) \right)  \notag \\
&\rightarrow &\frac{1}{\varepsilon }\left( K-\int \left\vert \hat{\Psi}%
\left( \hat{K},\hat{X}\right) \right\vert ^{2}d\left( \hat{K},\hat{X}\right) 
\frac{F_{2}\left( R\left( K,X\right) \right) G\left( X-\hat{X}\right) \hat{K}%
}{\int \left\vert \Psi \left( K^{\prime },X^{\prime }\right) \right\vert
^{2}d\left( K^{\prime },X^{\prime }\right) F_{2}\left( R\left( K^{\prime
},X^{\prime }\right) \right) G\left( X^{\prime }-\hat{X}\right) }\right) 
\notag \\
&\equiv &\Lambda \left( K,X\right)  \label{dml}
\end{eqnarray}%
Using the translation (\ref{Trl}) of (\ref{inco})-type term, we are led to
the translation of (\ref{mnn}). Since the square (\ref{mnn}) includes a
derivative $\frac{d}{dt}K_{i}$, the expression starts with a gradient with
respect to $K$, and we have:%
\begin{eqnarray}
&&\sum_{i}\left( \frac{d}{dt}K_{i}+\frac{1}{\varepsilon }\left(
K_{i}-\sum_{j}\frac{F_{2}\left( R\left( K_{i}\left( t\right) ,X_{i}\left(
t\right) \right) \right) G\left( X_{i}\left( t\right) -\hat{X}_{j}\right) }{%
\sum_{l}F_{2}\left( R\left( K_{l}\left( t\right) ,X_{l}\left( t\right)
\right) \right) G\left( X_{l}\left( t\right) -\hat{X}_{j}\right) }\hat{K}%
_{j}\left( t\right) \right) \right) ^{2}  \label{sq} \\
&\rightarrow &\int \Psi ^{\dag }\left( K,X\right) \left( -\nabla _{K}\left( 
\frac{\sigma _{K}^{2}}{2}\nabla _{K}+\Lambda \left( K,X\right) \right)
\right) \Psi \left( K,X\right) dKdX  \notag
\end{eqnarray}%
where, here again, the variance $\sigma _{K}^{2}$ reflects the probabilistic
nature of the model that is hidden behind the field formalism. Recall that
it represents the characteristic level of uncertainty in the dynamics of
capital.

Inserting result (\ref{dml}) in equation (\ref{sq}), the translation of (\ref%
{mnn)}) becomes:

\begin{equation}
S_{2}=-\int \Psi ^{\dag }\left( K,X\right) \nabla _{K}\left( \frac{\sigma
_{K}^{2}}{2}\nabla _{K}+\frac{1}{\varepsilon }\left( K-\int \hat{F}%
_{2}\left( R\left( K,X\right) ,\hat{X}\right) \hat{K}\left\vert \hat{\Psi}%
\left( \hat{K},\hat{X}\right) \right\vert ^{2}d\hat{K}d\hat{X}\right)
\right) \Psi \left( K,X\right)  \label{sd}
\end{equation}%
with:%
\begin{equation*}
\hat{F}_{2}\left( R\left( K,X\right) ,\hat{X}\right) =\frac{F_{2}\left(
R\left( K,X\right) \right) G\left( X-\hat{X}\right) }{\int F_{2}\left(
R\left( K,X\right) \right) G\left( X-\hat{X}\right) \left\vert \Psi \left(
K,X\right) \right\vert ^{2}}
\end{equation*}%
as quoted in the text.

\subsubsection*{Financial markets}

The functions to be translated are those of the financial capital dynamics (%
\ref{minKchap}) and of the financial capital allocation (\ref{minXchap}).\
Both expressions include a time derivative and are thus of type (\ref{edr}).
As for the real economy, the application of the translation rules is
straightforward.

\paragraph*{Translation of the minimization function: Financial capital
dynamics}

We consider the function (\ref{minKchap}): 
\begin{equation}
\sum_{j}\left( \frac{d}{dt}\hat{K}_{j}-\frac{1}{\varepsilon }\left(
\sum_{i}\left( r_{i}+F_{1}\left( \frac{R\left( K_{i},X_{i}\right) }{%
\sum_{l}\delta \left( X_{l}-X_{i}\right) R\left( K_{l},X_{l}\right) },\frac{%
\dot{K}_{i}\left( t\right) }{K_{i}\left( t\right) }\right) \right) \frac{%
F_{2}\left( R\left( K_{i},X_{i}\right) \right) G\left( X_{i}-\hat{X}%
_{j}\right) }{\sum_{l}F_{2}\left( R\left( K_{l},X_{l}\right) \right) G\left(
X_{l}-\hat{X}_{j}\right) }\hat{K}_{j}\right) \right) ^{2}  \label{minKd}
\end{equation}%
which translates, using the general translation formula of expression (\ref%
{inco}) in (\ref{Trl}), into:%
\begin{equation*}
\int \hat{\Psi}^{\dag }\left( \hat{K},\hat{X}\right) \left( -\nabla _{\hat{K}%
}\left( \frac{\sigma _{\hat{K}}^{2}}{2}\nabla _{\hat{K}}+\Lambda \left( \hat{%
K},\hat{X}\right) \right) \right) \hat{\Psi}\left( \hat{K},\hat{X}\right) d%
\hat{K}d\hat{X}
\end{equation*}%
The function $\Lambda \left( \hat{K},\hat{X}\right) $ is obtained, as
before, by translating the term following the derivative in the function (%
\ref{minKd}):%
\begin{equation}
\frac{1}{\varepsilon }\sum_{i}\left( r_{i}+F_{1}\left( \frac{R\left(
K_{i},X_{i}\right) }{\sum_{l}\delta \left( X_{l}-X_{i}\right) R\left(
K_{l},X_{l}\right) },\frac{\dot{K}_{i}\left( t\right) }{K_{i}\left( t\right) 
}\right) \right) \frac{F_{2}\left( R\left( K_{i},X_{i}\right) \right)
G\left( X_{i}-\hat{X}_{j}\right) }{\sum_{l}F_{2}\left( R\left(
K_{l},X_{l}\right) \right) G\left( X_{l}-\hat{X}_{j}\right) }\hat{K}%
_{j}\rightarrow \Lambda \left( \hat{K},\hat{X}\right)  \label{rkt}
\end{equation}%
First, we use the price dynamics equation (\ref{pr}) at the zero-th order in
fluctuations to translate the capital dynamics $\frac{\dot{K}_{i}\left(
t\right) }{K_{i}\left( t\right) }$:%
\begin{eqnarray*}
\frac{\dot{K}_{i}\left( t\right) }{K_{i}\left( t\right) } &=&\sum_{j}\frac{%
F_{2}\left( R\left( K_{i}\left( t\right) ,X_{i}\left( t\right) \right)
\right) G\left( X_{i}\left( t\right) -\hat{X}_{j}\right) }{%
K_{i}\sum_{l}F_{2}\left( R\left( K_{l}\left( t\right) ,X_{l}\left( t\right)
\right) \right) G\left( X_{l}\left( t\right) -\hat{X}_{j}\right) }\hat{K}%
_{j}\left( t\right) -K_{i}\left( t\right) \\
&\rightarrow &\Gamma \left( K,X\right)
\end{eqnarray*}%
where:%
\begin{eqnarray}
\Gamma \left( K,X\right) &=&\frac{\int \frac{F_{2}\left( R\left( K,X\right)
\right) G\left( X-\hat{X}\right) }{\int F_{2}\left( R\left( K,X\right)
\right) G\left( X-\hat{X}\right) \left\Vert \Psi \left( K,X\right)
\right\Vert ^{2}}\hat{K}\left\Vert \hat{\Psi}\left( \hat{K},\hat{X}\right)
\right\Vert ^{2}d\left( \hat{K},\hat{X}\right) -K}{K}  \label{mg} \\
&=&\int \frac{F_{2}\left( R\left( K,X\right) \right) G\left( X-\hat{X}%
\right) }{K\int F_{2}\left( R\left( K,X\right) \right) G\left( X-\hat{X}%
\right) \left\Vert \Psi \left( K,X\right) \right\Vert ^{2}}\hat{K}\left\Vert 
\hat{\Psi}\left( \hat{K},\hat{X}\right) \right\Vert ^{2}d\left( \hat{K},\hat{%
X}\right) -1  \notag
\end{eqnarray}%
Then, using the translation (\ref{bdt}) of (\ref{ntr}), we translate
expression (\ref{rkt}) by replacing: 
\begin{eqnarray*}
\left( K_{i},X_{i}\right) &\rightarrow &\left( K,X\right) \\
\left( K_{l},X_{l}\right) &\rightarrow &\left( K^{\prime },X^{\prime }\right)
\\
\left( \hat{K}_{j},\hat{X}_{j}\right) &\rightarrow &\left( \hat{K},\hat{X}%
\right)
\end{eqnarray*}%
We also replace the sums by integrals times the appropriate square of field,
which yields:%
\begin{eqnarray*}
\Lambda \left( \hat{K},\hat{X}\right) &=&-\frac{\hat{K}}{\varepsilon }\int
\left( r\left( K,X\right) -\gamma \frac{\int K^{\prime }\left\Vert \Psi
\left( K^{\prime },X\right) \right\Vert ^{2}}{K}+F_{1}\left( \frac{R\left(
K,X\right) }{\int R\left( K^{\prime },X^{\prime }\right) \left\Vert \Psi
\left( K^{\prime },X^{\prime }\right) \right\Vert ^{2}d\left( K^{\prime
},X^{\prime }\right) },\Gamma \left( K,X\right) \right) \right) \\
&&\times \frac{F_{2}\left( R\left( K,X\right) \right) G\left( X-\hat{X}%
\right) }{\int F_{2}\left( R\left( K^{\prime },X^{\prime }\right) \right)
G\left( X^{\prime }-\hat{X}\right) \left\Vert \Psi \left( K^{\prime
},X^{\prime }\right) \right\Vert ^{2}d\left( K^{\prime },X^{\prime }\right) }%
\left\Vert \Psi \left( K,X\right) \right\Vert ^{2}d\left( K,X\right)
\end{eqnarray*}%
Ultimately, the translation of (\ref{minKchap}) is: 
\begin{eqnarray*}
S_{3} &=&-\int \hat{\Psi}^{\dag }\left( \hat{K},\hat{X}\right) \nabla _{\hat{%
K}}\left( \frac{\sigma _{\hat{K}}^{2}}{2}\nabla _{\hat{K}}-\frac{\hat{K}}{%
\varepsilon }\int \left( r\left( K,X\right) -\gamma \frac{\int K^{\prime
}\left\Vert \Psi \left( K^{\prime },X\right) \right\Vert ^{2}}{K}\right.
\right. \\
&&\left. +F_{1}\left( \frac{R\left( K,X\right) }{\int R\left( K^{\prime
},X^{\prime }\right) \left\Vert \Psi \left( K^{\prime },X^{\prime }\right)
\right\Vert ^{2}d\left( K^{\prime },X^{\prime }\right) },\Gamma \left(
K,X\right) \right) \right) \\
&&\times \left. \frac{F_{2}\left( R\left( K,X\right) \right) G\left( X-\hat{X%
}\right) }{\int F_{2}\left( R\left( K^{\prime },X^{\prime }\right) \right)
G\left( X^{\prime }-\hat{X}\right) \left\Vert \Psi \left( K^{\prime
},X^{\prime }\right) \right\Vert ^{2}d\left( K^{\prime },X^{\prime }\right) }%
\left\Vert \Psi \left( K,X\right) \right\Vert ^{2}d\left( K,X\right) \right) 
\hat{\Psi}\left( \hat{K},\hat{X}\right)
\end{eqnarray*}%
Using expressions (\ref{KDR}) and (\ref{RPN}) yields the expression of the
text.

\paragraph{Translation of the minimization function: Financial capital
allocation}

The translation of the function for financial capital allocation (\ref%
{minXchap}) follows the previous pattern. We obtain:%
\begin{eqnarray*}
S_{4} &=&-\int \hat{\Psi}^{\dag }\left( \hat{K},\hat{X}\right) \nabla _{\hat{%
X}}\left( \sigma _{\hat{X}}^{2}\nabla _{\hat{X}}\right. -\int \left( \nabla
_{\hat{X}}F_{0}\left( R\left( K,\hat{X}\right) \right) +\nu \nabla _{\hat{X}%
}F_{1}\left( \frac{R\left( K,\hat{X}\right) }{\int R\left( K^{\prime
},X^{\prime }\right) \left\Vert \Psi \left( K^{\prime },X^{\prime }\right)
\right\Vert ^{2}d\left( K^{\prime },X^{\prime }\right) }\right) \right) \\
&&\times \left. \frac{\left\Vert \Psi \left( K,\hat{X}\right) \right\Vert
^{2}dK}{\int \left\Vert \Psi \left( K^{\prime },\hat{X}\right) \right\Vert
^{2}dK^{\prime }}\right) \hat{\Psi}\left( \hat{K},\hat{X}\right)
\end{eqnarray*}%
and (\ref{RPN}) yields the formula quoted in the text.

\section*{Appendix 2 expression of $\Psi \left( K,X\right) $ as function of
financial variables}

\subsection*{A2.1 \textbf{Finding }$\Psi \left( K,X\right) $: principle}

In this paragraph, we give the principle of resolution for $\Psi \left(
K,X\right) $ for an arbitrary function $H$. The full resolution for some
particular cases is given below. Given a particular state $\hat{\Psi}$, we
aim at minimizing the action functional $S_{1}+S_{2}+S_{3}+S_{4}$. However,
given our assumptions, the action functional $S_{3}+S_{4}$ depends on $\Psi
\left( K,X\right) $, through average quantities, and moreover, we have
assumed that physical capital dynamics depends on financial accumulation.
Consequently, we can neglect, in first approximation, the impact of $\Psi
\left( K,X\right) $ on $S_{3}+S_{4}$ and consider rather the minimization of 
$S_{1}+S_{2}$ which is given by:

\begin{eqnarray}
S_{1}+S_{2} &=&-\int \Psi ^{\dag }\left( K,X\right) \left( \nabla _{X}\left( 
\frac{\sigma _{X}^{2}}{2}\nabla _{X}-\nabla _{X}R\left( K,X\right) H\left(
K\right) \right) -\tau \frac{K_{X}}{K}\left( \int \left\vert \Psi \left(
K^{\prime },X\right) \right\vert ^{2}dK^{\prime }\right) \right.  \label{prt}
\\
&&+\left. \nabla _{K}\left( \frac{\sigma _{K}^{2}}{2}\nabla _{K}+u\left(
K,X,\Psi ,\hat{\Psi}\right) \right) \right) \Psi \left( K,X\right) dKdX 
\notag
\end{eqnarray}%
with:%
\begin{equation}
u\left( K,X,\Psi ,\hat{\Psi}\right) =\frac{1}{\varepsilon }\left( K-\int 
\frac{F_{2}\left( R\left( K,X\right) \right) G\left( X-\hat{X}\right) }{\int
F_{2}\left( R\left( K,X\right) \right) G\left( X-\hat{X}\right) \left\Vert
\Psi \left( K,X\right) \right\Vert ^{2}}\hat{K}\left\Vert \hat{\Psi}\left( 
\hat{K},\hat{X}\right) \right\Vert ^{2}d\hat{K}d\hat{X}\right)  \label{lpr}
\end{equation}%
and:%
\begin{equation*}
\Gamma \left( K,X\right) =\int \frac{F_{2}\left( R\left( K,X\right) \right)
G\left( X-\hat{X}\right) }{K\int F_{2}\left( R\left( K,X\right) \right)
G\left( X-\hat{X}\right) \left\Vert \Psi \left( K,X\right) \right\Vert ^{2}}%
\hat{K}\left\Vert \hat{\Psi}\left( \hat{K},\hat{X}\right) \right\Vert
^{2}d\left( \hat{K},\hat{X}\right) -1
\end{equation*}%
This is done in two steps. First, we find $\Psi \left( X\right) $, the
background field for $X$ when $K$ determined by $X$. We then find the
corrections to the particular cases considered and compute $\Psi \left(
K,X\right) $.

\subsubsection*{A2.1.1 \textbf{Particular case}: $K$ determined by $X$}

A simplification arises, assuming $K$ adapting to $X$. We assume that in
first approximation $K$ is a function of $X$, written $K_{X}$: 
\begin{equation}
K=K_{X}=\int \frac{F_{2}\left( R\left( K_{X},X\right) \right) G\left( X-\hat{%
X}\right) }{\int F_{2}\left( R\left( K_{X^{\prime }}^{\prime },X^{\prime
}\right) \right) G\left( X^{\prime }-\hat{X}\right) \left\Vert \Psi \left(
X^{\prime }\right) \right\Vert ^{2}dX^{\prime }}\hat{K}\left\Vert \hat{\Psi}%
\left( \hat{K},\hat{X}\right) \right\Vert ^{2}d\left( \hat{K},\hat{X}\right)
\label{kvr}
\end{equation}%
This means that for any sector $X$, the capital of all agents in this sector
are equal. At the individual level, this corresponds to set $\frac{d}{dt}%
K_{i}\left( t\right) =0$. The level of capital adapts faster than the motion
in sector space and reaches quickly its equilibrium value. Incindently, (\ref%
{kvr}) implies that $\Gamma \left( K,X\right) =0$. Actually, using (\ref{kvr}%
):%
\begin{eqnarray*}
\Gamma \left( K,X\right) &=&\int \frac{F_{2}\left( R\left( K,X\right)
\right) G\left( X-\hat{X}\right) }{K\int F_{2}\left( R\left( K,X\right)
\right) G\left( X-\hat{X}\right) \left\Vert \Psi \left( K,X\right)
\right\Vert ^{2}}\hat{K}\left\Vert \hat{\Psi}\left( \hat{K},\hat{X}\right)
\right\Vert ^{2}d\left( \hat{K},\hat{X}\right) -1 \\
&=&\int \frac{F_{2}\left( R\left( K,X\right) \right) G\left( X-\hat{X}%
\right) \hat{K}\left\Vert \hat{\Psi}\left( \hat{K},X\right) \right\Vert ^{2}d%
\hat{K}}{\int F_{2}\left( R\left( K_{X},X\right) \right) G\left( X-\hat{X}%
\right) \hat{K}\left\Vert \hat{\Psi}\left( \hat{K},\hat{X}\right)
\right\Vert ^{2}d\left( \hat{K},\hat{X}\right) }-1 \\
&=&0
\end{eqnarray*}

\paragraph*{A2.1.1.1 Justification of approximation (\protect\ref{kvr})}

Approximation (\ref{kvr}) justifies in the following way. When $F_{2}$ is
slowly varying with $K$, we perform the following change of variable in (\ref%
{prt}):%
\begin{eqnarray*}
\Psi &\rightarrow &\Psi \exp \left( -\frac{\int u\left( K,X,\Psi ,\hat{\Psi}%
\right) dK}{\sigma _{K}^{2}}\right) \simeq \Psi \exp \left( -\frac{1}{%
2\sigma _{K}^{2}}\varepsilon u^{2}\left( K,X,\Psi ,\hat{\Psi}\right) \right)
\\
\Psi ^{\dag } &\rightarrow &\Psi ^{\dag }\exp \left( \frac{1}{\sigma _{K}^{2}%
}\int u\left( K,X,\Psi ,\hat{\Psi}\right) dK\right) \simeq \Psi ^{\dag }\exp
\left( -\frac{1}{2\sigma _{K}^{2}}\varepsilon u^{2}\left( K,X,\Psi ,\hat{\Psi%
}\right) \right)
\end{eqnarray*}%
and this replaces $S_{2}$ in (\ref{prt}) by:%
\begin{equation}
-\int \Psi ^{\dag }\left( K,X\right) \left( \frac{\sigma _{K}^{2}}{2}\nabla
_{K}^{2}-\frac{u^{2}}{2\sigma _{K}^{2}}\left( K,X,\Psi ,\hat{\Psi}\right) +%
\frac{1}{2}\nabla _{K}u\left( K,X,\Psi ,\hat{\Psi}\right) \right) \Psi
\left( K,X\right) dKdX  \label{trd}
\end{equation}%
The change of variable modifies $S_{1}$ in (\ref{prt}). Actually, the
derivative $\nabla _{X}$ acts on $\exp \left( -\frac{1}{2\sigma _{K}^{2}}%
u^{2}\left( K,X,\Psi ,\hat{\Psi}\right) \right) $ and the term:%
\begin{equation*}
-\int \Psi ^{\dag }\left( K,X\right) \nabla _{X}\left( \frac{\sigma _{X}^{2}%
}{2}\nabla _{X}-\nabla _{X}R\left( K,X\right) H\left( K\right) \right) \Psi
\left( K,X\right) dKdX
\end{equation*}%
becomes:%
\begin{eqnarray}
&&-\int \Psi ^{\dag }\left( X\right) \nabla _{X}\left( \frac{\sigma _{X}^{2}%
}{2}\nabla _{X}-\nabla _{X}R\left( K,X\right) H\left( K\right) \right) \Psi
\left( X\right) dKdX  \label{trf} \\
&&+\varepsilon \int \Psi ^{\dag }\left( K,X\right) \left( \frac{\sigma
_{X}^{2}}{2\sigma _{K}^{2}}u\nabla _{X}u\right) \nabla _{X}\Psi \left(
K,X\right) dKdX+\varepsilon \int \Psi ^{\dag }\left( K,X\right) \left( \frac{%
\sigma _{X}^{2}}{2\sigma _{K}^{2}}\left( \left( \nabla _{X}u\right)
^{2}+u\nabla _{X}^{2}u\right) \right) \Psi \left( K,X\right) dKdX  \notag \\
&&-\int \Psi ^{\dag }\left( K,X\right) \left( \varepsilon \frac{u\nabla _{X}u%
}{\sigma _{K}^{2}}\nabla _{X}R\left( K,X\right) H\left( K\right)
+\varepsilon ^{2}\frac{\sigma _{X}^{2}}{2\sigma _{K}^{4}}\left( u\nabla
_{X}u\right) ^{2}\right) \Psi \left( K,X\right) dKdX  \notag
\end{eqnarray}%
Using that $u$ is of order $\frac{1}{\varepsilon }$ (see(\ref{lpr})), the
minimum of $S_{1}+S_{2}$ is obtained when the potential:%
\begin{eqnarray}
&&\int \Psi ^{\dag }\left( K,X\right) \left( \frac{u^{2}}{2\sigma _{K}^{2}}-%
\frac{1}{2}\nabla _{K}u\right) \Psi \left( K,X\right) dKdX  \label{ptl} \\
&&+\varepsilon \int \Psi ^{\dag }\left( K,X\right) \left( \frac{\sigma
_{X}^{2}}{2\sigma _{K}^{2}}\left( \left( \nabla _{X}u\right) ^{2}+u\nabla
_{X}^{2}u\right) \right) \Psi \left( K,X\right) dKdX  \notag \\
&&-\int \Psi ^{\dag }\left( K,X\right) \left( \varepsilon \frac{u\nabla _{X}u%
}{\sigma _{K}^{2}}\nabla _{X}R\left( K,X\right) H\left( K\right)
+\varepsilon ^{2}\frac{\sigma _{X}^{2}}{2\sigma _{K}^{4}}\left( u\nabla
_{X}u\right) ^{2}\right) \Psi \left( K,X\right) dKdX  \notag
\end{eqnarray}%
is nul. The dominant term in (\ref{ptl}) for $\varepsilon <<1$ is: 
\begin{equation}
\int \Psi ^{\dag }\left( K,X\right) \left( \frac{u^{2}}{2\sigma _{K}^{2}}%
-\varepsilon ^{2}\frac{\sigma _{X}^{2}}{2\sigma _{K}^{4}}\left( u\nabla
_{X}u\right) ^{2}\right) \Psi \left( K,X\right) dKdX  \label{trm}
\end{equation}%
For $\sigma _{X}^{2}<<\sigma _{K}^{2}$ it implies that the minimum for $%
S_{1}+S_{2}$ is obtained for: 
\begin{equation*}
u\left( K,X,\Psi ,\hat{\Psi}\right) \simeq 0
\end{equation*}%
with solution (\ref{kvr}).

\paragraph*{A2.1.1.2 Rewriting the action $S_{1}+S_{2}$}

With our choice $G\left( X-\hat{X}\right) =\delta \left( X-\hat{X}\right) $
we find:%
\begin{equation}
K_{X}=\frac{\int \hat{K}\left\Vert \hat{\Psi}\left( \hat{K},X\right)
\right\Vert ^{2}d\hat{K}}{\left\Vert \Psi \left( X\right) \right\Vert ^{2}}
\label{xK}
\end{equation}%
and $\Psi \left( K,X\right) $ becomes a function $\Psi \left( X\right) $:%
\begin{equation*}
\Psi \left( K,X\right) \rightarrow \Psi \left( X\right)
\end{equation*}%
To find the action for $\Psi \left( X\right) $ we evaluate (\ref{ptl}) using 
$u\left( K_{X},X,\Psi ,\hat{\Psi}\right) =0$, and compute the first term in (%
\ref{trm}) for $\Psi \left( X\right) =\Psi \left( K_{X},X\right) \delta
\left( u\right) $ by replacing:%
\begin{equation*}
\delta \left( u\right) \rightarrow \frac{\exp \left( -\varepsilon
u^{2}\right) }{\sqrt{2\pi }\varepsilon }
\end{equation*}%
We obtain: 
\begin{eqnarray*}
-\int \Psi ^{\dag }\left( K,X\right) \left( \frac{\sigma _{K}^{2}}{2}\nabla
_{K}^{2}\right) \Psi \left( K,X\right) dKdX &=&\frac{\sigma _{K}^{2}}{2}\int
\left\vert \Psi \left( X\right) \right\vert ^{2}dX\int \frac{\exp \left(
-\varepsilon u^{2}\right) }{\sqrt{2\pi \varepsilon }}\nabla _{K}^{2}\frac{%
\exp \left( -\varepsilon u^{2}\right) }{\sqrt{2\pi \varepsilon }}dK \\
&\simeq &\frac{\sigma _{K}^{2}}{2\varepsilon }\int \left\vert \Psi \left(
X\right) \right\vert ^{2}dX
\end{eqnarray*}%
and the action $S_{1}$ restricted to the variable $X$\ is given by:%
\begin{eqnarray*}
S_{1} &=&\int \Psi ^{\dag }\left( X\right) \left( -\nabla _{X}\left( \frac{%
\sigma _{X}^{2}}{2}\nabla _{X}-\left( \nabla _{X}R\left( X\right) H\left(
K_{X}\right) \right) \right) +\tau \frac{K_{X}}{K}\left\vert \Psi \left(
X\right) \right\vert ^{2}\right) \Psi \left( X\right) \\
&&+\int \Psi ^{\dag }\left( K,X\right) \left( \frac{\sigma _{X}^{2}}{4\sigma
_{K}^{2}}\left( \nabla _{X}u\left( K_{X},X,\Psi ,\hat{\Psi}\right) \right)
^{2}\right) \Psi \left( K,X\right) dKdX \\
&&+\int \left( \frac{\sigma _{K}^{2}}{2\varepsilon }-\frac{1}{2}\nabla
_{K}u\left( K_{X},X,\Psi ,\hat{\Psi}\right) \right) \left\vert \Psi \left(
X\right) \right\vert ^{2}dX
\end{eqnarray*}%
In our order of approximation $\nabla _{K}u\left( K_{X},X,\Psi ,\hat{\Psi}%
\right) \simeq \varepsilon $. Ultimately, for $\sigma _{X}^{2}<<\sigma
_{K}^{2}$, action $S_{1}$ reduces to:

\begin{equation}
S_{1}=\int \Psi ^{\dag }\left( X\right) \left( -\nabla _{X}\left( \frac{%
\sigma _{X}^{2}}{2}\nabla _{X}-\left( \nabla _{X}R\left( X\right) H\left(
K_{X}\right) \right) \right) +\tau \frac{K_{X}}{K}\left\vert \Psi \left(
X\right) \right\vert ^{2}+\frac{\sigma _{K}^{2}-1}{2\varepsilon }\right)
\Psi \left( X\right)  \label{cnt}
\end{equation}%
and we look for $\Psi \left( X\right) $ minimizing (\ref{cnt}). As explained
in the text, we will consider at the collective level that we can replace:%
\begin{equation*}
\tau \frac{K_{X}}{K}\rightarrow \tau
\end{equation*}

\paragraph*{A2.1.1.3 Minimization of (\protect\ref{cnt})}

To minimize (\ref{cnt}), we assume for the sake of simplicity, that for $%
i\neq j$:%
\begin{equation*}
\left\vert \nabla _{X_{i}}\nabla _{X_{j}}R\left( X\right) \right\vert
<<\left\vert \nabla _{X_{i}}^{2}R\left( X\right) \right\vert
\end{equation*}%
which is the case for example if $R\left( X\right) $ is a function with
separated variables : $R\left( X\right) =\sum R_{i}\left( X_{i}\right) $.
This can be also realized if locally, one chooses the variables $X_{i}$ to
diagonalize $\nabla _{X_{i}}\nabla _{X_{j}}R\left( X\right) $ at some points
in the sector space.

We then perform the change of variables:%
\begin{equation*}
\exp \left( \int^{X}\frac{\nabla _{X}R\left( X\right) }{\sigma
_{X}^{2}\left\Vert \nabla _{X}R\left( X\right) \right\Vert }H\left(
K_{X}\right) \right) \Psi \left( X\right) \rightarrow \Psi \left( X\right)
\end{equation*}%
and:%
\begin{equation*}
\exp \left( -\int^{X}\nabla _{X}R\left( X\right) H\left( K_{X}\right)
\right) \Psi ^{\dag }\left( X\right) \rightarrow \Psi ^{\dag }\left( X\right)
\end{equation*}%
so that (\ref{cnt}) becomes:%
\begin{equation}
\int \Psi ^{\dag }\left( X\right) \left( -\frac{\sigma _{X}^{2}}{2}\nabla
_{X}^{2}+\frac{1}{2\sigma _{X}^{2}}\left( \nabla _{X}R\left( X\right)
H\left( K_{X}\right) \right) ^{2}+\frac{\nabla _{X}^{2}R\left(
K_{X},X\right) }{2}H\left( K_{X}\right) +\tau \left\vert \Psi \left(
X\right) \right\vert ^{2}+\frac{\sigma _{K}^{2}-1}{2\varepsilon }\right)
\Psi \left( X\right)  \label{ntc}
\end{equation}%
which is of second order in derivatives with a potential:%
\begin{equation*}
\tau \left\Vert \Psi \left( X\right) \right\Vert ^{4}+\frac{1}{2\sigma
_{X}^{2}}\int \left( \nabla _{X}R\left( X\right) H\left( K_{X}\right)
\right) ^{2}\left\Vert \Psi \left( X\right) \right\Vert ^{2}
\end{equation*}%
We assume the number of agents fixed equal to $N$. We must minimize (\ref%
{ntc}) with the constraint $\left\Vert \Psi \left( X\right) \right\Vert
^{2}\geqslant 0$ and $\int \left\Vert \Psi \left( X\right) \right\Vert
^{2}=N $. We thus replace (\ref{ntc}) by:%
\begin{eqnarray}
&&\int \Psi ^{\dag }\left( X\right) \left( -\frac{\sigma _{X}^{2}\nabla
_{X}^{2}}{2}+\frac{\left( \nabla _{X}R\left( X\right) H\left( K_{X}\right)
\right) ^{2}}{2\sigma _{X}^{2}}+\frac{\nabla _{X}^{2}R\left( K_{X},X\right) 
}{2}H\left( K_{X}\right) +\tau \left\vert \Psi \left( X\right) \right\vert
^{2}+\frac{\sigma _{K}^{2}-1}{2\varepsilon }\right) \Psi \left( X\right) 
\notag \\
&&+D\left( \left\Vert \Psi \right\Vert ^{2}\right) \left( \int \left\Vert
\Psi \left( X\right) \right\Vert ^{2}-N\right) +\int \mu \left( X\right)
\left\Vert \Psi \left( X\right) \right\Vert ^{2}  \label{ntC}
\end{eqnarray}%
we have written $D\left( \left\Vert \Psi \right\Vert ^{2}\right) $ the
Lagrange multiplier for $\int \left\Vert \Psi \left( X\right) \right\Vert
^{2}$,$\ $to keep track of its dependency multiplier in $\left\Vert \Psi
\right\Vert ^{2}$. By a redefinition $D\left( \left\Vert \Psi \right\Vert
^{2}\right) -\frac{\sigma _{K}^{2}-1}{2\varepsilon }\rightarrow D\left(
\left\Vert \Psi \right\Vert ^{2}\right) $, $\frac{D\left( \left\Vert \Psi
\right\Vert ^{2}\right) }{D\left( \left\Vert \Psi \right\Vert ^{2}\right) -%
\frac{\sigma _{K}^{2}}{2\varepsilon }}N\rightarrow N$ we can write (\ref{ntC}%
) as: 
\begin{eqnarray}
&&\int \Psi ^{\dag }\left( X\right) \left( -\frac{\sigma _{X}^{2}}{2}\nabla
_{X}^{2}+\frac{1}{2\sigma _{X}^{2}}\left( \nabla _{X}R\left( X\right)
H\left( K_{X}\right) \right) ^{2}+\frac{H\left( K_{X}\right) \nabla
_{X}^{2}R\left( K_{X},X\right) }{2}+\tau \left\vert \Psi \left( X\right)
\right\vert ^{2}\right) \Psi \left( X\right)  \label{rdc} \\
&&+D\left( \left\Vert \Psi \right\Vert ^{2}\right) \left( \int \left\Vert
\Psi \left( X\right) \right\Vert ^{2}-N\right) +\int \mu \left( X\right)
\left\Vert \Psi \left( X\right) \right\Vert ^{2}  \notag
\end{eqnarray}%
Introducing the change of variable for $\nabla _{X}R\left( X\right) $ for
the sake of simplicity:%
\begin{equation}
\left( \nabla _{X}R\left( X\right) \right) ^{2}+\sigma _{X}^{2}\frac{\nabla
_{X}^{2}R\left( K_{X},X\right) }{H\left( K_{X}\right) }\rightarrow \left(
\nabla _{X}R\left( X\right) \right) ^{2}  \label{dfg}
\end{equation}%
the minimization of the potential yields, for $\sigma _{X}^{2}<<1$:%
\begin{eqnarray}
&&iD\left( \left\Vert \Psi \right\Vert ^{2}\right) +\mu \left( X\right)
\label{bnq} \\
&=&2\tau \left\Vert \Psi \left( X\right) \right\Vert ^{2}-\frac{H^{\prime
}\left( \frac{\int \hat{K}\left\Vert \hat{\Psi}\left( \hat{K},X\right)
\right\Vert ^{2}d\hat{K}}{\left\Vert \Psi \left( X\right) \right\Vert ^{2}}%
\right) }{2\sigma _{X}^{2}H\left( \frac{\int \hat{K}\left\Vert \hat{\Psi}%
\left( \hat{K},X\right) \right\Vert ^{2}d\hat{K}}{\left\Vert \Psi \left(
X\right) \right\Vert ^{2}}\right) }  \notag \\
&&\times \left( \nabla _{X}R\left( X\right) H\left( \frac{\int \hat{K}%
\left\Vert \hat{\Psi}\left( \hat{K},X\right) \right\Vert ^{2}d\hat{K}}{%
\left\Vert \Psi \left( X\right) \right\Vert ^{2}}\right) \right) ^{2}\frac{%
\int \hat{K}\left\Vert \hat{\Psi}\left( \hat{K},X\right) \right\Vert ^{2}d%
\hat{K}}{\left\Vert \Psi \left( X\right) \right\Vert ^{4}}\left\Vert \Psi
\left( X\right) \right\Vert ^{2}  \notag \\
&&+\frac{1}{2\sigma _{X}^{2}}\left( \nabla _{X}R\left( X\right) H\left( 
\frac{\int \hat{K}\left\Vert \hat{\Psi}\left( \hat{K},X\right) \right\Vert
^{2}d\hat{K}}{\left\Vert \Psi \left( X\right) \right\Vert ^{2}}\right)
\right) ^{2}  \notag
\end{eqnarray}%
Moreover, multiplying (\ref{bnq}) by $\left\Vert \Psi \left( X\right)
\right\Vert ^{2}$ and integrating yields:

\begin{eqnarray}
D\left( \left\Vert \Psi \right\Vert ^{2}\right) N &=&2\tau \int \left\vert
\Psi \left( X\right) \right\vert ^{4}  \label{tgn} \\
&&-\int \frac{H^{\prime }\left( \frac{\int \hat{K}\left\Vert \hat{\Psi}%
\left( \hat{K},X\right) \right\Vert ^{2}d\hat{K}}{\left\Vert \Psi \left(
X\right) \right\Vert ^{2}}\right) }{2\sigma _{X}^{2}H\left( \frac{\int \hat{K%
}\left\Vert \hat{\Psi}\left( \hat{K},X\right) \right\Vert ^{2}d\hat{K}}{%
\left\Vert \Psi \left( X\right) \right\Vert ^{2}}\right) }\left( \nabla
_{X}R\left( X\right) H\left( \frac{\int \hat{K}\left\Vert \hat{\Psi}\left( 
\hat{K},X\right) \right\Vert ^{2}d\hat{K}}{\left\Vert \Psi \left( X\right)
\right\Vert ^{2}}\right) \right) ^{2}\int \hat{K}\left\Vert \hat{\Psi}\left( 
\hat{K},X\right) \right\Vert ^{2}d\hat{K}  \notag \\
&&+\frac{1}{2\sigma _{X}^{2}}\int \left( \nabla _{X}R\left( X\right) H\left( 
\frac{\int \hat{K}\left\Vert \hat{\Psi}\left( \hat{K},X\right) \right\Vert
^{2}d\hat{K}}{\left\Vert \Psi \left( X\right) \right\Vert ^{2}}\right)
\right) ^{2}\left\Vert \Psi \left( X\right) \right\Vert ^{2}  \notag \\
&\simeq &2\tau \int \left\vert \Psi \left( X\right) \right\vert ^{4}+\frac{1%
}{2\sigma _{X}^{2}}\int \left( \nabla _{X}R\left( X\right) H\left( \frac{%
\int \hat{K}\left\Vert \hat{\Psi}\left( \hat{K},X\right) \right\Vert ^{2}d%
\hat{K}}{\left\Vert \Psi \left( X\right) \right\Vert ^{2}}\right) \right)
^{2}\left\Vert \Psi \left( X\right) \right\Vert ^{2}  \notag
\end{eqnarray}%
Note that in first approximation, for $H^{\prime }<<1$, (\ref{bnq}) and (\ref%
{tgn}) become:%
\begin{equation}
D\left( \left\Vert \Psi \right\Vert ^{2}\right) +\mu \left( X\right) =2\tau
\left\Vert \Psi \left( X\right) \right\Vert ^{2}+\frac{1}{2\sigma _{X}^{2}}%
\left( \nabla _{X}R\left( X\right) \right) ^{2}H^{2}\left( \frac{\int \hat{K}%
\left\Vert \hat{\Psi}\left( \hat{K},X\right) \right\Vert ^{2}d\hat{K}}{%
\left\Vert \Psi \left( X\right) \right\Vert ^{2}}\right)  \label{psn}
\end{equation}%
and:%
\begin{equation}
ND\left( \left\Vert \Psi \right\Vert ^{2}\right) =2\tau \int \left\vert \Psi
\left( X\right) \right\vert ^{4}+\frac{1}{2\sigma _{X}^{2}}\int \left(
\nabla _{X}R\left( X\right) H\left( \frac{\int \hat{K}\left\Vert \hat{\Psi}%
\left( \hat{K},X\right) \right\Vert ^{2}d\hat{K}}{\left\Vert \Psi \left(
X\right) \right\Vert ^{2}}\right) \right) ^{2}\left\Vert \Psi \left(
X\right) \right\Vert ^{2}  \label{snp}
\end{equation}

\paragraph*{A2.1.1.4 Resolution of (\protect\ref{psn}) and (\protect\ref{snp}%
)}

Two cases arise in the resolution:

\subparagraph{Case 1: $\left\Vert \Psi \left( X\right) \right\Vert ^{2}>0$}

For $\left\Vert \Psi \left( X\right) \right\Vert ^{2}>0$, (\ref{bnq})
writes: 
\begin{subequations}
\begin{equation}
D\left( \left\Vert \Psi \right\Vert ^{2}\right) =2\tau \left\Vert \Psi
\left( X\right) \right\Vert ^{2}+\frac{1}{2\sigma _{X}^{2}}\left( \nabla
_{X}R\left( X\right) \right) ^{2}H^{2}\left( \frac{\hat{K}_{X}}{\left\Vert
\Psi \left( X\right) \right\Vert ^{2}}\right) \left( 1-\frac{H^{\prime
}\left( \hat{K}_{X}\right) }{H\left( \hat{K}_{X}\right) }\frac{\hat{K}_{X}}{%
\left\Vert \Psi \left( X\right) \right\Vert ^{2}}\right)  \label{nbq}
\end{equation}%
with: 
\end{subequations}
\begin{equation}
\hat{K}_{X}=\int \hat{K}\left\Vert \hat{\Psi}\left( \hat{K},X\right)
\right\Vert ^{2}d\hat{K}=K_{X}\left\Vert \Psi \left( X\right) \right\Vert
^{2}  \label{Kx}
\end{equation}%
Note that restoring the initial variable:%
\begin{equation}
\left( \nabla _{X}R\left( X\right) \right) ^{2}\rightarrow \left( \nabla
_{X}R\left( X\right) \right) ^{2}+\sigma _{X}^{2}\frac{\nabla
_{X}^{2}R\left( K_{X},X\right) }{H\left( K_{X}\right) }  \label{stg}
\end{equation}%
yields (\ref{psl}) in the text.

Given the setup, we can assume that%
\begin{equation*}
H^{2}\left( \frac{\hat{K}_{X}}{\left\Vert \Psi \left( X\right) \right\Vert
^{2}}\right) \left( 1-\frac{H^{\prime }\left( \hat{K}_{X}\right) }{H\left( 
\hat{K}_{X}\right) }\frac{\hat{K}_{X}}{\left\Vert \Psi \left( X\right)
\right\Vert ^{2}}\right)
\end{equation*}%
is a decreasing function of $\left\Vert \Psi \left( X\right) \right\Vert
^{2} $. Assume a minimum $\Psi _{0}\left( X\right) $ for the right hand side
of (\ref{nbq}). It leads to a condition for $D\left( \left\Vert \Psi
\right\Vert ^{2}\right) $:%
\begin{equation}
D\left( \left\Vert \Psi \right\Vert ^{2}\right) >2\tau \left\Vert \Psi
_{0}\left( X\right) \right\Vert ^{2}+\frac{1}{2\sigma _{X}^{2}}\left( \nabla
_{X}R\left( X\right) \right) ^{2}H^{2}\left( \frac{\hat{K}_{X}}{\left\Vert
\Psi _{0}\left( X\right) \right\Vert ^{2}}\right) \left( 1-\frac{H^{\prime
}\left( \hat{K}_{X}\right) }{H\left( \hat{K}_{X}\right) }\frac{\hat{K}_{X}}{%
\left\Vert \Psi _{0}\left( X\right) \right\Vert ^{2}}\right)  \label{cd}
\end{equation}%
and the solution of (\ref{nbq}) writes:%
\begin{equation}
\left\Vert \Psi \left( X,\left( \nabla _{X}R\left( X\right) \right) ^{2},%
\frac{\hat{K}_{X}}{\hat{K}_{X,0}}\right) \right\Vert ^{2}  \label{sp}
\end{equation}%
where $\hat{K}_{X,0}$ is a constant representing some average to normalize $%
\frac{\hat{K}_{X}}{\hat{K}_{X,0}}$ as a dimensionless number.

\subparagraph{Case 2 $\left\Vert \Psi \left( X\right) \right\Vert ^{2}=0$}

On the other hand, if:%
\begin{equation}
D\left( \left\Vert \Psi \right\Vert ^{2}\right) <2\tau \left\Vert \Psi
_{0}\left( X\right) \right\Vert ^{2}+\frac{1}{2\sigma _{X}^{2}}\left( \nabla
_{X}R\left( X\right) \right) ^{2}H^{2}\left( \frac{\hat{K}_{X}}{\left\Vert
\Psi _{0}\left( X\right) \right\Vert ^{2}}\right) \left( 1-\frac{H^{\prime
}\left( \hat{K}_{X}\right) }{H\left( \hat{K}_{X}\right) }\frac{\hat{K}_{X}}{%
\left\Vert \Psi _{0}\left( X\right) \right\Vert ^{2}}\right)  \label{dc}
\end{equation}%
the solution of (\ref{nbq}) is $\left\Vert \Psi \left( X\right) \right\Vert
^{2}=0$

\subparagraph{Gathering both cases}

The value of $\left\Vert \Psi \right\Vert ^{2}$ thus depends on the
conditions (\ref{cd}) and (\ref{dc}). To compute the value of $D\left(
\left\Vert \Psi \right\Vert ^{2}\right) $ we integrate (\ref{nbq}) over $%
V/V_{0}$ with $V_{0}$ locus where $\left\Vert \Psi \left( X\right)
\right\Vert ^{2}=0$. $V_{0}$ will be then defined by (\ref{dc}) once $%
D\left( \left\Vert \Psi \right\Vert ^{2}\right) $ found. For $H$ slowly
varying, we can replace $\frac{\hat{K}_{X}}{\left\Vert \Psi \left( X\right)
\right\Vert ^{2}}$ by:%
\begin{equation*}
\frac{\int \hat{K}\left\Vert \hat{\Psi}\left( \hat{K},X\right) \right\Vert
^{2}d\hat{K}dX}{\int \left\Vert \Psi \left( X\right) \right\Vert ^{2}dX}=%
\frac{\int \hat{K}\left\Vert \hat{\Psi}\left( \hat{K},X\right) \right\Vert
^{2}d\hat{K}dX}{N}
\end{equation*}%
so that the integration of (\ref{dc}) over $X$ yields:

\begin{eqnarray*}
D\left( \left\Vert \Psi \right\Vert ^{2}\right) \left( V-V_{0}\right)
&\simeq &2\tau N+\frac{1}{2\sigma _{X}^{2}}\int \left( \nabla _{X}R\left(
X\right) \right) ^{2}H^{2}\left( \frac{\int \hat{K}\left\Vert \hat{\Psi}%
\left( \hat{K},X\right) \right\Vert ^{2}d\hat{K}dX}{N}\right) \\
&&\times \left( 1-\frac{H^{\prime }\left( \frac{\int \hat{K}\left\Vert \hat{%
\Psi}\left( \hat{K},X\right) \right\Vert ^{2}d\hat{K}dX}{N}\right) }{H\left( 
\frac{\int \hat{K}\left\Vert \hat{\Psi}\left( \hat{K},X\right) \right\Vert
^{2}d\hat{K}dX}{N}\right) }\frac{\int \hat{K}\left\Vert \hat{\Psi}\left( 
\hat{K},X\right) \right\Vert ^{2}d\hat{K}dX}{N}\right) \\
&=&2\tau N+\frac{1}{2\sigma _{X}^{2}}\left( \nabla _{X}R\left( X\right)
\right) ^{2}H^{2}\left( \frac{\left\langle \hat{K}\right\rangle }{N}\right)
\left( 1-\frac{H^{\prime }\left( \frac{\left\langle \hat{K}\right\rangle }{N}%
\right) }{H\left( \frac{\left\langle \hat{K}\right\rangle }{N}\right) }\frac{%
\left\langle \hat{K}\right\rangle }{N}\right)
\end{eqnarray*}%
Consequently:%
\begin{equation*}
D\left( \left\Vert \Psi \right\Vert ^{2}\right) \simeq 2\tau \frac{N}{V-V_{0}%
}+\frac{1}{2\sigma _{X}^{2}}\left\langle \left( \nabla _{X}R\left( X\right)
\right) ^{2}\right\rangle _{V/V_{0}}H^{2}\left( \frac{\left\langle \hat{K}%
\right\rangle }{N}\right) \left( 1-\frac{H^{\prime }\left( \frac{%
\left\langle \hat{K}\right\rangle }{N}\right) }{H\left( \frac{\left\langle 
\hat{K}\right\rangle }{N}\right) }\frac{\left\langle \hat{K}\right\rangle }{N%
}\right)
\end{equation*}%
and $V_{0}$ is defined by (\ref{dc}):%
\begin{eqnarray}
&&2\tau \frac{N}{V-V_{0}}+\frac{1}{2\sigma _{X}^{2}}\left\langle \left(
\nabla _{X}R\left( X\right) \right) ^{2}\right\rangle _{V/V_{0}}H^{2}\left( 
\frac{\left\langle \hat{K}\right\rangle }{N}\right) \left( 1-\frac{H^{\prime
}\left( \frac{\left\langle \hat{K}\right\rangle }{N}\right) }{H\left( \frac{%
\left\langle \hat{K}\right\rangle }{N}\right) }\frac{\left\langle \hat{K}%
\right\rangle }{N}\right)  \label{qvn} \\
&<&2\tau \left\Vert \Psi _{0}\left( X\right) \right\Vert ^{2}+\frac{1}{%
2\sigma _{X}^{2}}\left( \nabla _{X}R\left( X\right) \right) ^{2}H^{2}\left( 
\frac{\hat{K}_{X}}{\left\Vert \Psi _{0}\left( X\right) \right\Vert ^{2}}%
\right) \left( 1-\frac{H^{\prime }\left( \hat{K}_{X}\right) }{H\left( \hat{K}%
_{X}\right) }\frac{\hat{K}_{X}}{\left\Vert \Psi _{0}\left( X\right)
\right\Vert ^{2}}\right)  \notag
\end{eqnarray}%
On $V/V_{0}$, $\left\Vert \Psi \right\Vert ^{2}$ is given by (\ref{sp}) and
on $V_{0}$, $\left\Vert \Psi \right\Vert ^{2}=0$.

Below, we give explicitaly the form of $\Psi \left( X\right) $ form two
different form of the function $H$.

\subsubsection*{A2.1.2 \textbf{Introducing the }$K$ dependency}

\paragraph*{A2.1.2.1 First order condition}

To go beyond approximation (\ref{kvr}) and solve for the field $\Psi \left(
K,X\right) $ that minimizes (\ref{prt}), we come back to the full system for 
$K$ and $X$: 
\begin{eqnarray}
&&\int \Psi ^{\dag }\left( K,X\right) \left( \left( -\nabla _{X}\left( \frac{%
\sigma _{X}^{2}}{2}\nabla _{X}-\left( \frac{\nabla _{X}R\left( K,X\right) }{%
\left\Vert \nabla _{X}R\left( K,X\right) \right\Vert }\right) H\left(
K\right) +\tau \left\vert \Psi \left( K,X\right) \right\vert ^{2}\right)
\right) \right.  \label{cmk} \\
&&\left. -\nabla _{K}\left( \frac{\sigma _{K}^{2}}{2}\nabla _{K}+u\left(
K,X,\Psi ,\hat{\Psi}\right) \right) -\frac{1}{2}\nabla _{K}u\left( K,X,\Psi ,%
\hat{\Psi}\right) \right) \Psi \left( K,X\right)  \notag
\end{eqnarray}%
with $u\left( K,X,\Psi ,\hat{\Psi}\right) $ given by (\ref{lpr}). We then
look for a minimum of (\ref{cmk}) of the form:%
\begin{equation}
\Psi \left( K,X\right) =\Psi \left( X\right) \Psi _{1}\left( K-K_{X}\right)
\label{dcp}
\end{equation}%
with $K_{X}$ given in (\ref{xK}):%
\begin{equation}
K_{X}=\frac{\int \hat{K}\left\Vert \hat{\Psi}\left( \hat{K},X\right)
\right\Vert ^{2}d\hat{K}}{\left\Vert \Psi \left( X\right) \right\Vert ^{2}}
\label{xkdd}
\end{equation}%
and $\Psi _{1}$ peaked around $0$ and of norm $1$. When $H\left( K\right) $
is slowly varying around $K_{X}$, the minimization of (\ref{cmk}) for $\Psi
_{1}\left( K-K_{X}\right) $ writes:%
\begin{equation}
\nabla _{K}\left( \frac{\sigma _{K}^{2}}{2}\nabla _{K}+u\left( K,X,\Psi ,%
\hat{\Psi}\right) +\frac{1}{2}\nabla _{K}u\left( K,X,\Psi ,\hat{\Psi}\right)
\right) \Psi _{1}\left( K-K_{X}\right) =0  \label{qnx}
\end{equation}%
Then, using that, in first approximation:%
\begin{equation*}
\int F_{2}\left( R\left( K^{\prime },X\right) \right) \left\Vert \Psi \left(
K^{\prime },X\right) \right\Vert ^{2}dK^{\prime }\simeq F_{2}\left( R\left(
K_{X},X\right) \right) \left\Vert \Psi \left( X\right) \right\Vert ^{2}
\end{equation*}%
Equation (\ref{qnx}) becomes:%
\begin{equation}
\nabla _{K}\left( \frac{\sigma _{K}^{2}}{2}\nabla _{K}+K-\frac{F_{2}\left(
R\left( K,X\right) \right) K_{X}}{F_{2}\left( R\left( K_{X},X\right) \right) 
}\right) \Psi _{1}\left( K-K_{X}\right) =0  \label{tsl}
\end{equation}

\paragraph*{A2.1.2.2 Solving (\protect\ref{tsl})}

To solve the first order condition (\ref{tsl}) we perform the change of
variable:%
\begin{equation*}
\Psi _{1}\left( K-K_{X}\right) \rightarrow \exp \left( \frac{1}{\sigma
_{K}^{2}}\int \left[ K-\frac{F_{2}\left( R\left( K,X\right) \right) K_{X}}{%
F_{2}\left( R\left( K_{X},X\right) \right) }\right] dK\right) \Psi
_{1}\left( K-K_{X}\right)
\end{equation*}%
and (\ref{tsl}) is transformed into%
\begin{equation}
-\frac{\sigma _{K}^{2}}{2}\nabla _{K}^{2}\Psi _{1}\left( K-K_{X}\right) +%
\frac{1}{2\sigma _{K}^{2}}\left( K-\frac{F_{2}\left( R\left( K,X\right)
\right) K_{X}}{F_{2}\left( R\left( K_{X},X\right) \right) }\right) ^{2}\Psi
_{1}\left( K-K_{X}\right) =0  \label{scnv}
\end{equation}%
This equation can be solved by implementing the constraint:%
\begin{equation*}
\int \left\Vert \Psi _{1}\left( K-K_{X}\right) \right\Vert ^{2}=1
\end{equation*}%
and we find:%
\begin{eqnarray*}
&&\Psi _{1}\left( K-K_{X}\right) \simeq \mathcal{N}\exp \left( -\frac{1}{%
\sigma _{K}^{2}}\left( K-\frac{F_{2}\left( R\left( K,X\right) \right) K_{X}}{%
F_{2}\left( R\left( K_{X},X\right) \right) }\right) ^{2}\right) \\
&\simeq &\mathcal{N}\exp \left( -\frac{1}{\sigma _{K}^{2}}\left(
K-K_{X}-\left( K-K_{X}\right) \frac{\partial _{K}R\left( K_{X},X\right)
F_{2}^{\prime }\left( R\left( K_{X},X\right) \right) }{F_{2}\left( R\left(
K_{X},X\right) \right) }K_{X}\right) ^{2}\right) \\
&=&\mathcal{N}\exp \left( -\frac{1}{\sigma _{K}^{2}}\left( 1-\frac{\partial
_{K}R\left( K_{X},X\right) F_{2}^{\prime }\left( R\left( K,X\right) \right) 
}{F_{2}\left( R\left( K_{X},X\right) \right) }K_{X}\right) ^{2}\left(
K-K_{X}\right) ^{2}\right)
\end{eqnarray*}%
with the normalization factor $\mathcal{N}$\ given by:%
\begin{equation*}
\mathcal{N}=\sqrt{\frac{c}{\sigma _{K}^{2}\left( 1-\frac{\partial
_{K}R\left( K_{X},X\right) F_{2}^{\prime }\left( R\left( K,X\right) \right) 
}{F_{2}\left( R\left( K_{X},X\right) \right) }K_{X}\right) ^{2}}}
\end{equation*}

\paragraph*{A2.1.2.3 Expression for the density of firms $\left\Vert \Psi
\left( K,X\right) \right\Vert ^{2}$}

Having found $\Psi _{1}$, and using (\ref{sp}) and (\ref{dcp}) we obtain the
expression for $\left\Vert \Psi \left( K,X\right) \right\Vert ^{2}$:%
\begin{eqnarray}
\left\Vert \Psi \left( K,X\right) \right\Vert ^{2} &=&\mathcal{N}\left\Vert
\Psi \right\Vert ^{2}\left( X,\left( \nabla _{X}R\left( X\right) \right)
^{2},\frac{\hat{K}_{X}}{\hat{K}_{X,0}}\right)  \label{PSc} \\
&&\times \exp \left( -\frac{1}{\sigma _{K}^{2}}\left( K-\frac{F_{2}\left(
R\left( K,X\right) \right) }{F_{2}\left( R\left( K_{X},X\right) \right)
\left\Vert \Psi \left( X\right) \right\Vert ^{2}}\int \hat{K}\left\Vert \hat{%
\Psi}\left( \hat{K},X\right) \right\Vert ^{2}d\hat{K}\right) ^{2}\right) 
\notag \\
&=&\left\Vert \Psi \right\Vert ^{2}\left( X,\left( \nabla _{X}R\left(
X\right) \right) ^{2},\frac{\hat{K}_{X}}{\hat{K}_{X,0}}\right) \frac{c\exp
\left( -\frac{1}{\sigma _{K}^{2}}\left( 1-\frac{\partial _{K}R\left(
K_{X},X\right) F_{2}^{\prime }\left( R\left( K,X\right) \right) }{%
F_{2}\left( R\left( K_{X},X\right) \right) }K_{X}\right) ^{2}\left(
K-K_{X}\right) ^{2}\right) }{\frac{1}{\sigma _{K}^{2}}\left( 1-\frac{%
\partial _{K}R\left( K_{X},X\right) F_{2}^{\prime }\left( R\left( K,X\right)
\right) }{F_{2}\left( R\left( K_{X},X\right) \right) }K_{X}\right) ^{2}} 
\notag
\end{eqnarray}%
for $X\in V/V_{0}$ and $\left\Vert \Psi \left( K,X\right) \right\Vert ^{2}=0$
otherwise.

As stated in the text, note that the form of the exponential in (\ref{PSc})
implies that:%
\begin{equation*}
\int K\left\Vert \Psi \left( K,X\right) \right\Vert ^{2}d\hat{K}=\int \hat{K}%
\left\Vert \hat{\Psi}\left( \hat{K},X\right) \right\Vert ^{2}d\hat{K}
\end{equation*}

\subsection*{A2.2 \textbf{Examples}}

We solve the minimization for $\Psi \left( K,X\right) $ for two particular
forms of the function $H\left( K\right) $.

\subsubsection*{A2.2.1 Example 1}

We compute $\Psi \left( K,X\right) $ for the specific function: 
\begin{equation*}
H\left( y\right) =\left( \frac{y}{1+y}\right) ^{\varsigma }\text{, }%
H^{\prime }\left( y\right) =\varsigma \frac{\left( \frac{y}{y+1}\right)
^{\varsigma }}{y\left( y+1\right) }
\end{equation*}%
We use the simplified equations (\ref{psn}) and (\ref{snp}) that yield:%
\begin{equation*}
D\left( \left\Vert \Psi \right\Vert ^{2}\right) +\mu \left( X\right) =\tau
\left\Vert \Psi \left( X\right) \right\Vert ^{2}+\frac{\frac{1}{\sigma
_{X}^{2}}\left( \nabla _{X}R\left( X\right) \right) ^{2}\left( \left( \frac{%
\int \hat{K}\left\Vert \hat{\Psi}\left( \hat{K},X\right) \right\Vert ^{2}d%
\hat{K}}{\left\Vert \Psi \left( X\right) \right\Vert ^{2}}\right)
^{\varsigma }\right) ^{2}\left( 1-\varsigma \frac{1}{\left( \frac{\int \hat{K%
}\left\Vert \hat{\Psi}\left( \hat{K},X\right) \right\Vert ^{2}d\hat{K}}{%
\left\Vert \Psi \left( X\right) \right\Vert ^{2}}+\left\langle \hat{K}%
\right\rangle \right) }\right) }{\left( \left\langle \hat{K}\right\rangle +%
\frac{\int \hat{K}\left\Vert \hat{\Psi}\left( \hat{K},X\right) \right\Vert
^{2}d\hat{K}}{\left\Vert \Psi \left( X\right) \right\Vert ^{2}}\right)
^{2\varsigma }}
\end{equation*}%
or equivalently:%
\begin{eqnarray*}
D\left( \left\Vert \Psi \right\Vert ^{2}\right) +\mu \left( X\right) &=&\tau
\left\Vert \Psi \left( X\right) \right\Vert ^{2} \\
&&+\frac{\frac{1}{\sigma _{X}^{2}}\left( \nabla _{X}R\left( X\right) \right)
^{2}\left( \int \hat{K}\left\Vert \hat{\Psi}\left( \hat{K},X\right)
\right\Vert ^{2}d\hat{K}\right) ^{2\varsigma }}{\left( \left\langle \hat{K}%
\right\rangle \left\Vert \Psi \left( X\right) \right\Vert ^{2}+\int \hat{K}%
\left\Vert \hat{\Psi}\left( \hat{K},X\right) \right\Vert ^{2}d\hat{K}\right)
^{2\varsigma +1}} \\
&&\times \left( \int \hat{K}\left\Vert \hat{\Psi}\left( \hat{K},X\right)
\right\Vert ^{2}d\hat{K}+\left( 1-\varsigma \right) \left\langle \hat{K}%
\right\rangle \left\Vert \Psi \left( X\right) \right\Vert ^{2}\right)
\end{eqnarray*}%
For $\varsigma \simeq \frac{1}{2}$, this reduces to:%
\begin{equation*}
D\left( \left\Vert \Psi \right\Vert ^{2}\right) +\mu \left( X\right) =\tau
\left\Vert \Psi \left( X\right) \right\Vert ^{2}+\frac{\frac{1}{\sigma
_{X}^{2}}\left( \nabla _{X}R\left( X\right) \right) ^{2}\hat{K}_{X}\left( 
\hat{K}_{X}+\frac{1}{2}\left\langle \hat{K}\right\rangle \left\Vert \Psi
\left( X\right) \right\Vert ^{2}\right) }{\left( \left\langle \hat{K}%
\right\rangle \left\Vert \Psi \left( X\right) \right\Vert ^{2}+\hat{K}%
_{X}\right) ^{2}}
\end{equation*}%
and for $\left\langle \hat{K}\right\rangle \left\Vert \Psi \left( X\right)
\right\Vert ^{2}<<\hat{K}_{X}$ this becomes:%
\begin{equation}
D\left( \left\Vert \Psi \right\Vert ^{2}\right) +\mu \left( X\right) \simeq
\tau \left\Vert \Psi \left( X\right) \right\Vert ^{2}+\frac{\frac{1}{\sigma
_{X}^{2}}\left( \nabla _{X}R\left( X\right) \right) ^{2}\hat{K}_{X}}{\left(
\left\langle \hat{K}\right\rangle \left\Vert \Psi \left( X\right)
\right\Vert ^{2}+\hat{K}_{X}\right) }  \label{prn}
\end{equation}%
Two cases arise.

When $\frac{1}{\sigma _{X}^{2}}\left( \nabla _{X}R\left( X\right) \right)
^{2}<<\tau $: 
\begin{eqnarray}
\left\Vert \Psi \left( X\right) \right\Vert ^{2} &=&\frac{\left( D\left(
\left\Vert \Psi \right\Vert ^{2}\right) -\tau \frac{\hat{K}_{X}}{%
\left\langle \hat{K}\right\rangle }\right) +\sqrt{\left( D\left( \left\Vert
\Psi \right\Vert ^{2}\right) -\tau \frac{\hat{K}_{X}}{\left\langle \hat{K}%
\right\rangle }\right) ^{2}-4\tau \frac{\hat{K}_{X}}{\left\langle \hat{K}%
\right\rangle }\left( \frac{\left( \nabla _{X}R\left( X\right) \right) ^{2}}{%
\sigma _{X}^{2}}-D\left( \left\Vert \Psi \right\Vert ^{2}\right) \right) }}{%
2\tau }  \label{psp} \\
&=&\frac{4\tau \frac{\hat{K}_{X}}{\left\langle \hat{K}\right\rangle }\left( 
\frac{1}{\sigma _{X}^{2}}\left( \nabla _{X}R\left( X\right) \right)
^{2}-D\left( \left\Vert \Psi \right\Vert ^{2}\right) \right) }{2\tau \left(
\left( D\left( \left\Vert \Psi \right\Vert ^{2}\right) -\tau \frac{\hat{K}%
_{X}}{\left\langle \hat{K}\right\rangle }\right) -\sqrt{\left( D\left(
\left\Vert \Psi \right\Vert ^{2}\right) -\tau \frac{\hat{K}_{X}}{%
\left\langle \hat{K}\right\rangle }\right) ^{2}-4\tau \frac{\hat{K}_{X}}{%
\left\langle \hat{K}\right\rangle }\left( \frac{\left( \nabla _{X}R\left(
X\right) \right) ^{2}}{\sigma _{X}^{2}}-D\left( \left\Vert \Psi \right\Vert
^{2}\right) \right) }\right) }  \notag
\end{eqnarray}%
This is positive on the set: 
\begin{equation}
\left\{ \left( D\left( \left\Vert \Psi \right\Vert ^{2}\right) -\tau \frac{%
\hat{K}_{X}}{\left\langle \hat{K}\right\rangle }\right) >0\right\} \cup
\left\{ \frac{1}{\sigma _{X}^{2}}\left( \nabla _{X}R\left( X\right) \right)
^{2}-D\left( \left\Vert \Psi \right\Vert ^{2}\right) <0\right\}  \label{dbc}
\end{equation}%
To detail these two conditions, we write (\ref{prn}) for $\left\Vert \Psi
\left( X\right) \right\Vert ^{2}>0$:%
\begin{equation*}
D\left( \left\Vert \Psi \right\Vert ^{2}\right) \simeq \tau \left\Vert \Psi
\left( X\right) \right\Vert ^{2}+\frac{\frac{1}{\sigma _{X}^{2}}\left(
\nabla _{X}R\left( X\right) \right) ^{2}\frac{\hat{K}_{X}}{\left\langle \hat{%
K}\right\rangle }}{\left( \left\Vert \Psi \left( X\right) \right\Vert ^{2}+%
\frac{\hat{K}_{X}}{\left\langle \hat{K}\right\rangle }\right) }
\end{equation*}%
which is equivalent to:%
\begin{equation*}
\frac{\frac{1}{\sigma _{X}^{2}}\left( \nabla _{X}R\left( X\right) \right)
^{2}-D\left( \left\Vert \Psi \right\Vert ^{2}\right) }{\left( \left\Vert
\Psi \left( X\right) \right\Vert ^{2}+\frac{\hat{K}_{X}}{\left\langle \hat{K}%
\right\rangle }\right) }\frac{\hat{K}_{X}}{\left\langle \hat{K}\right\rangle 
}=\frac{-\tau \left\Vert \Psi \left( X\right) \right\Vert ^{2}+D\left(
\left\Vert \Psi \right\Vert ^{2}\right) -\tau \frac{\hat{K}_{X}}{%
\left\langle \hat{K}\right\rangle }}{\left( \left\Vert \Psi \left( X\right)
\right\Vert ^{2}+\frac{\hat{K}_{X}}{\left\langle \hat{K}\right\rangle }%
\right) }\left\Vert \Psi \left( X\right) \right\Vert ^{2}
\end{equation*}%
Then, we have the implication:%
\begin{equation}
\frac{1}{\sigma _{X}^{2}}\left( \nabla _{X}R\left( X\right) \right)
^{2}-D\left( \left\Vert \Psi \right\Vert ^{2}\right) >0\Rightarrow D\left(
\left\Vert \Psi \right\Vert ^{2}\right) -\tau \frac{\hat{K}_{X}}{%
\left\langle \hat{K}\right\rangle }>0  \label{mpn}
\end{equation}%
This implies that (\ref{dbc}) is always satisfied, and formula (\ref{psp})
is valid for all $X$.

The second case arises when $\frac{1}{\sigma _{X}^{2}}\left( \nabla
_{X}R\left( X\right) \right) ^{2}<<\tau $. In this case, the solution is:%
\begin{equation*}
\left\Vert \Psi \left( X\right) \right\Vert ^{2}=\frac{\left( D\left(
\left\Vert \Psi \right\Vert ^{2}\right) -\tau \frac{\hat{K}_{X}}{%
\left\langle \hat{K}\right\rangle }\right) -\sqrt{\left( D\left( \left\Vert
\Psi \right\Vert ^{2}\right) -\tau \frac{\hat{K}_{X}}{\left\langle \hat{K}%
\right\rangle }\right) ^{2}-4\tau \frac{\hat{K}_{X}}{\left\langle \hat{K}%
\right\rangle }\left( \frac{\left( \nabla _{X}R\left( X\right) \right) ^{2}}{%
\sigma _{X}^{2}}-D\left( \left\Vert \Psi \right\Vert ^{2}\right) \right) }}{%
2\tau }
\end{equation*}%
This solution is valid, i.e. $\left\Vert \Psi \left( X\right) \right\Vert
^{2}>0$, under the conditions: 
\begin{equation}
\left\{ D\left( \left\Vert \Psi \right\Vert ^{2}\right) -\tau \frac{\hat{K}%
_{X}}{\left\langle \hat{K}\right\rangle }>0\right\} \cap \left\{ \frac{1}{%
\sigma _{X}^{2}}\left( \nabla _{X}R\left( X\right) \right) ^{2}-D\left(
\left\Vert \Psi \right\Vert ^{2}\right) >0\right\}  \label{cdt}
\end{equation}%
and $\left\Vert \Psi \right\Vert ^{2}=0$ for:%
\begin{equation*}
\left\{ D\left( \left\Vert \Psi \right\Vert ^{2}\right) -\tau \frac{\hat{K}%
_{X}}{\left\langle \hat{K}\right\rangle }<0\right\} \cup \left\{ \frac{1}{%
\sigma _{X}^{2}}\left( \nabla _{X}R\left( X\right) \right) ^{2}-D\left(
\left\Vert \Psi \right\Vert ^{2}\right) <0\right\}
\end{equation*}%
To detail these two conditions, we use the implication (\ref{mpn}) that is
equivalent to:%
\begin{equation*}
D\left( \left\Vert \Psi \right\Vert ^{2}\right) -\tau \frac{\hat{K}_{X}}{%
\left\langle \hat{K}\right\rangle }<0\Rightarrow \frac{1}{\sigma _{X}^{2}}%
\left( \nabla _{X}R\left( X\right) \right) ^{2}-D\left( \left\Vert \Psi
\right\Vert ^{2}\right) <0
\end{equation*}%
Consequently, $\left\Vert \Psi \left( X\right) \right\Vert ^{2}=0$ only if:%
\begin{equation}
\frac{1}{\sigma _{X}^{2}}\left( \nabla _{X}R\left( X\right) \right)
^{2}-D\left( \left\Vert \Psi \right\Vert ^{2}\right) <0  \label{ncl}
\end{equation}%
We find $D\left( \left\Vert \Psi \right\Vert ^{2}\right) $ by integration
of: 
\begin{equation}
D\left( \left\Vert \Psi \right\Vert ^{2}\right) +\mu \left( X\right) \simeq
\tau \left\Vert \Psi \left( X\right) \right\Vert ^{2}+\frac{\frac{1}{\sigma
_{X}^{2}}\left( \nabla _{X}R\left( X\right) \right) ^{2}\hat{K}_{X}}{\left(
\left\langle \hat{K}\right\rangle \left\Vert \Psi \left( X\right)
\right\Vert ^{2}+\hat{K}_{X}\right) }  \label{psd}
\end{equation}%
and this leads to:%
\begin{eqnarray*}
\int_{V/V_{0}}D\left( \left\Vert \Psi \right\Vert ^{2}\right) &\simeq &\tau
N+\int_{V/V_{0}}\frac{\frac{1}{\sigma _{X}^{2}}\left( \nabla _{X}R\left(
X\right) \right) ^{2}\frac{\hat{K}_{X}}{\left\langle \hat{K}\right\rangle }}{%
\left( \left\Vert \Psi \left( X\right) \right\Vert ^{2}+\frac{\hat{K}_{X}}{%
\left\langle \hat{K}\right\rangle }\right) } \\
&\simeq &\tau N+\frac{1}{2}\int_{V/V_{0}}\frac{1}{\sigma _{X}^{2}}\left(
\nabla _{X}R\left( X\right) \right) ^{2}=\tau N+\frac{1}{2}\left(
V-V_{0}\right) \left\langle \frac{1}{\sigma _{X}^{2}}\left( \nabla
_{X}R\left( X\right) \right) ^{2}\right\rangle _{V/V_{0}}
\end{eqnarray*}%
we thus have:%
\begin{equation}
D\left( \left\Vert \Psi \right\Vert ^{2}\right) \simeq \frac{\tau N}{V-V_{0}}%
+\frac{1}{2}\left\langle \frac{1}{\sigma _{X}^{2}}\left( \nabla _{X}R\left(
X\right) \right) ^{2}\right\rangle _{V/V_{0}}  \label{cdn}
\end{equation}%
and $V_{0}$ is defined using (\ref{ncl}). It is the set of points $X$\ such
that:

\begin{equation}
\frac{\tau N}{V-V_{0}}+\frac{1}{2}\left\langle \frac{1}{\sigma _{X}^{2}}%
\left( \nabla _{X}R\left( X\right) \right) ^{2}\right\rangle _{V/V_{0}}-%
\frac{1}{\sigma _{X}^{2}}\left( \nabla _{X}R\left( X\right) \right) ^{2}>0
\label{bnc}
\end{equation}%
Similarly, the set $V/V_{0}$ is defined by: 
\begin{equation}
\frac{\tau N}{V-V_{0}}+\frac{1}{2}\left\langle \frac{1}{\sigma _{X}^{2}}%
\left( \nabla _{X}R\left( X\right) \right) ^{2}\right\rangle _{V/V_{0}}-%
\frac{1}{\sigma _{X}^{2}}\left( \nabla _{X}R\left( X\right) \right) ^{2}<0
\label{cnb}
\end{equation}%
To each function $R\left( X\right) $ and any $d>0$, we associate two
functions that depend on the form of $\frac{1}{\sigma _{X}^{2}}\left( \nabla
_{X}R\left( X\right) \right) ^{2}$ over the whole space. First, $v\left(
V-V_{0}\right) $ is a decreasing function of $V-V_{0}$, defined by:%
\begin{equation}
V\left( \frac{1}{\sigma _{X}^{2}}\left( \nabla _{X}R\left( X\right) \right)
^{2}>v\left( V-V_{0}\right) \right) =V-V_{0}  \label{vft}
\end{equation}%
Second, for every $d\geqslant 0$, the function $h\left( d\right) $ is given
by: 
\begin{equation}
h\left( d\right) =\frac{1}{\int_{\nabla _{X}R\left( X\right) >d}dX}%
\int_{\nabla _{X}R\left( X\right) >d}\frac{1}{\sigma _{X}^{2}}\left( \nabla
_{X}R\left( X\right) \right) ^{2}dX  \label{hcf}
\end{equation}%
This is an increasing function of $d$.

Thus, we can rewrite (\ref{cnb}) as:%
\begin{equation}
\frac{\tau N}{V-V_{0}}+\frac{1}{2}\left\langle \frac{1}{\sigma _{X}^{2}}%
\left( \nabla _{X}R\left( X\right) \right) ^{2}\right\rangle
_{V/V_{0}}=v\left( V-V_{0}\right)  \label{rmf}
\end{equation}%
and moreover, by integration of (\ref{cnb}) over $V/V_{0}$:%
\begin{equation}
\left\langle \frac{1}{\sigma _{X}^{2}}\left( \nabla _{X}R\left( X\right)
\right) ^{2}\right\rangle _{V/V_{0}}=h\left( \frac{\tau N}{V-V_{0}}+\frac{1}{%
2}\left\langle \frac{1}{\sigma _{X}^{2}}\left( \nabla _{X}R\left( X\right)
\right) ^{2}\right\rangle _{V/V_{0}}\right)  \label{frm}
\end{equation}%
Equations (\ref{rmf}) and (\ref{frm}) combine as:%
\begin{equation}
2\left( v\left( V-V_{0}\right) -\frac{\tau N}{V-V_{0}}\right) =h\left(
v\left( V-V_{0}\right) \right)  \label{lvt}
\end{equation}%
which is an equation depending on the form of $R\left( X\right) $. If it has
a solution, the set on which $\left\Vert \Psi \left( X\right) \right\Vert
^{2}=0$ is defined by:%
\begin{equation*}
\frac{1}{\sigma _{X}^{2}}\left( \nabla _{X}R\left( X\right) \right)
^{2}<v\left( V-V_{0}\right)
\end{equation*}%
and $D\left( \left\Vert \Psi \right\Vert ^{2}\right) $ is given by 
\begin{equation*}
D\left( \left\Vert \Psi \right\Vert ^{2}\right) \simeq v\left( V-V_{0}\right)
\end{equation*}%
Once the solution of (\ref{lvt}) is known, the constant $D\left( \left\Vert
\Psi \right\Vert ^{2}\right) $ is given by (\ref{cdn}) and:

\begin{equation}
\left\Vert \Psi \left( X\right) \right\Vert ^{2}=\frac{2\frac{\hat{K}_{X}}{%
\left\langle \hat{K}\right\rangle }\left( \frac{1}{\sigma _{X}^{2}}\left(
\nabla _{X}R\left( X\right) \right) ^{2}-D\left( \left\Vert \Psi \right\Vert
^{2}\right) \right) }{D\left( \left\Vert \Psi \right\Vert ^{2}\right) -\tau 
\frac{\hat{K}_{X}}{\left\langle \hat{K}\right\rangle }+\sqrt{\left( D\left(
\left\Vert \Psi \right\Vert ^{2}\right) -\tau \frac{\hat{K}_{X}}{%
\left\langle \hat{K}\right\rangle }\right) ^{2}-4\tau \frac{\hat{K}_{X}}{%
\left\langle \hat{K}\right\rangle }\left( \frac{1}{\sigma _{X}^{2}}\left(
\nabla _{X}R\left( X\right) \right) ^{2}-D\left( \left\Vert \Psi \right\Vert
^{2}\right) \right) }}  \label{spp}
\end{equation}%
for $X\in V/V_{0}$.

\subsubsection*{A2.2.2 Example 2}

We choose $H\left( y\right) =y$ and equations (\ref{psn}) and (\ref{snp})
yield:%
\begin{equation*}
D\left( \left\Vert \Psi \right\Vert ^{2}\right) \simeq \tau \left\Vert \Psi
\left( X\right) \right\Vert ^{2}+\frac{1}{\sigma _{X}^{2}}\left( \nabla
_{X}R\left( X\right) \right) ^{2}\frac{\hat{K}_{X}}{\left\Vert \Psi \left(
X\right) \right\Vert ^{2}}
\end{equation*}%
If:%
\begin{equation}
D\left( \left\Vert \Psi \right\Vert ^{2}\right) >2\sqrt{\tau \frac{1}{\sigma
_{X}^{2}}\left( \nabla _{X}R\left( X\right) \right) ^{2}\hat{K}_{X}}
\label{dcn}
\end{equation}

then:%
\begin{equation*}
\left\Vert \Psi \left( X\right) \right\Vert ^{2}=\frac{1}{2\tau }\left(
D\left( \left\Vert \Psi \right\Vert ^{2}\right) -\sqrt{\left( D\left(
\left\Vert \Psi \right\Vert ^{2}\right) \right) ^{2}-4\hat{K}_{X}\frac{1}{%
\sigma _{X}^{2}}\left( \nabla _{X}R\left( X\right) \right) ^{2}\tau }\right)
>0
\end{equation*}%
To solve (\ref{cdn}) and to find $V_{0}$, we compute $D\left( \left\Vert
\Psi \right\Vert ^{2}\right) $ by integrating (\ref{psd}) and (\ref{cdn}) is
still valid:%
\begin{equation}
D\left( \left\Vert \Psi \right\Vert ^{2}\right) \simeq \frac{\tau N}{V-V_{0}}%
+\frac{1}{2}\left\langle \frac{1}{\sigma _{X}^{2}}\left( \nabla _{X}R\left(
X\right) \right) ^{2}\right\rangle _{V/V_{0}}  \label{dnc}
\end{equation}%
We proceed as in the previous paragraph to find $D\left( \left\Vert \Psi
\right\Vert ^{2}\right) $ and $V_{0}$. Using (\ref{dnc}), (\ref{dcn})
becomes:%
\begin{equation}
\frac{1}{4\tau \hat{K}_{X}}\left( \frac{\tau N}{V-V_{0}}+\frac{1}{2}%
\left\langle \frac{1}{\sigma _{X}^{2}}\left( \nabla _{X}R\left( X\right)
\right) ^{2}\right\rangle _{V/V_{0}}\right) ^{2}>\frac{1}{\sigma _{X}^{2}}%
\left( \nabla _{X}R\left( X\right) \right) ^{2}  \label{qln}
\end{equation}%
Definitions (\ref{vft}) and (\ref{hcf}) allow to rewrite (\ref{dnc}) and (%
\ref{qln}):%
\begin{equation*}
\frac{1}{4\tau \hat{K}_{X}}\left( \frac{\tau N}{V-V_{0}}+\frac{1}{2}%
\left\langle \frac{1}{\sigma _{X}^{2}}\left( \nabla _{X}R\left( X\right)
\right) ^{2}\right\rangle _{V/V_{0}}\right) ^{2}=v\left( V-V_{0}\right)
\end{equation*}%
\begin{equation*}
\left\langle \frac{1}{\sigma _{X}^{2}}\left( \nabla _{X}R\left( X\right)
\right) ^{2}\right\rangle _{V/V_{0}}=h\left( v\left( V-V_{0}\right) \right)
\end{equation*}%
that reduce to an equation for $V-V_{0}$:%
\begin{equation*}
2\left( 2\sqrt{\tau v\left( V-V_{0}\right) \hat{K}_{X}}-\frac{\tau N}{V-V_{0}%
}\right) =h\left( v\left( V-V_{0}\right) \right)
\end{equation*}%
If it has a solution, the set on which $\left\Vert \Psi \left( X\right)
\right\Vert ^{2}=0$ is defined by:%
\begin{equation*}
\frac{1}{\sigma _{X}^{2}}\left( \nabla _{X}R\left( X\right) \right)
^{2}<v\left( V-V_{0}\right)
\end{equation*}%
and $D\left( \left\Vert \Psi \right\Vert ^{2}\right) $ is given by 
\begin{equation*}
D\left( \left\Vert \Psi \right\Vert ^{2}\right) \simeq 2\sqrt{\tau v\left(
V-V_{0}\right) \hat{K}_{X}}
\end{equation*}

\section*{Appendix 3. Computation of the background field $\hat{\Psi}\left( 
\hat{K},\hat{X}\right) $ and average capital $\hat{K}_{X}$}

\subsection*{A3.1 System for $\hat{\Psi}\left( \hat{K},\hat{X}\right) $}

\subsubsection*{A3.1.1 \textbf{Replacing quantities depending on }$\left(
K,X\right) $}

\textbf{\ }Having found $\Psi \left( K,X\right) $, we can rewrite an action
functional for $\hat{\Psi}\left( \hat{K},\hat{X}\right) $. To do so, we
first replace the quantities depending on $\Psi \left( K,X\right) $ in the
action (\ref{fcn}). Given the form of this function we can use the
approximation $K\simeq K_{X}$: at the collective level, the relevant
quantity, from the point of view of investors are the sectors.

Using that:%
\begin{equation*}
\frac{R\left( K,X\right) }{\int R\left( K^{\prime },X^{\prime }\right)
\left\Vert \Psi \left( K^{\prime },X^{\prime }\right) \right\Vert
^{2}d\left( K^{\prime },X^{\prime }\right) }\simeq \frac{R\left( K,X\right) 
}{\int R\left( K_{X^{\prime }}^{\prime },X^{\prime }\right) \left\Vert \Psi
\left( X^{\prime }\right) \right\Vert ^{2}dX^{\prime }}
\end{equation*}%
we first start by rewriting $F_{1}$ and we have: \textbf{\ }%
\begin{equation*}
F_{1}\left( \frac{R\left( K,X\right) }{\int R\left( K^{\prime },X^{\prime
}\right) \left\Vert \Psi \left( K^{\prime },X^{\prime }\right) \right\Vert
^{2}d\left( K^{\prime },X^{\prime }\right) },\Gamma \left( K,X\right)
\right) \simeq F_{1}\left( \frac{R\left( K_{X},X\right) }{\int R\left(
K_{X^{\prime }}^{\prime },X^{\prime }\right) \left\Vert \Psi \left(
X^{\prime }\right) \right\Vert ^{2}dX^{\prime }},\Gamma \left( K,X\right)
\right)
\end{equation*}%
As explained in appendix 1, when $K\simeq K_{X}$, we also have:%
\begin{equation*}
\Gamma \left( K,X\right) =\int \frac{F_{2}\left( R\left( K,X\right) \right) 
}{K_{X}F_{2}\left( R\left( K_{X},X\right) \right) \left\Vert \Psi \left(
K_{X},X\right) \right\Vert }\hat{K}\left\Vert \hat{\Psi}\left( \hat{K}%
,X\right) \right\Vert ^{2}d\hat{K}-1=0
\end{equation*}%
Then, we rewrite the expression involving $F_{2}$ in (\ref{fcn}): 
\begin{eqnarray*}
\frac{F_{2}\left( R\left( K,\hat{X}\right) \right) }{\int F_{2}\left(
R\left( K^{\prime },\hat{X}\right) \right) \left\Vert \Psi \left( K^{\prime
},\hat{X}\right) \right\Vert ^{2}dK^{\prime }}\left\Vert \Psi \left( K,\hat{X%
}\right) \right\Vert ^{2} &\simeq &\frac{F_{2}\left( R\left( K,\hat{X}%
\right) \right) }{F_{2}\left( R\left( K_{\hat{X}},\hat{X}\right) \right)
\left\Vert \Psi \left( \hat{X}\right) \right\Vert ^{2}}\left\Vert \Psi
\left( K,\hat{X}\right) \right\Vert ^{2} \\
&=&\frac{F_{2}\left( R\left( K,\hat{X}\right) \right) \left\Vert \Psi
_{0}\left( K-K_{\hat{X}}\right) \right\Vert ^{2}}{F_{2}\left( R\left( K_{%
\hat{X}},\hat{X}\right) \right) }
\end{eqnarray*}%
and the $\hat{\Psi}\left( \hat{K},\hat{X}\right) $ part of the action
functional (\ref{fcn}) writes:%
\begin{eqnarray}
&&S_{3}+S_{4}=-\int \hat{\Psi}^{\dag }\left( \hat{K},\hat{X}\right) \left(
\nabla _{\hat{K}}\left( \frac{\sigma _{\hat{K}}^{2}}{2}\nabla _{\hat{K}}-%
\hat{K}f\left( K,X,\Psi ,\hat{\Psi}\right) \right) \right.  \label{mts} \\
&&\left. +\nabla _{\hat{X}}\left( \frac{\sigma _{\hat{X}}^{2}}{2}\nabla _{%
\hat{X}}-g\left( K,X,\Psi ,\hat{\Psi}\right) \right) \right) \hat{\Psi}%
\left( \hat{K},\hat{X}\right)  \notag
\end{eqnarray}%
where:%
\begin{eqnarray}
f\left( \hat{X},\Psi ,\hat{\Psi}\right) &=&\frac{1}{\varepsilon }\int \left(
\nabla _{K}R\left( K,X\right) -\gamma \frac{\int K^{\prime }\left\Vert \Psi
\left( K^{\prime },X\right) \right\Vert ^{2}}{K}+F_{1}\left( \frac{R\left(
K,X\right) }{\int R\left( K^{\prime },X^{\prime }\right) \left\Vert \Psi
\left( K^{\prime },X^{\prime }\right) \right\Vert ^{2}d\left( K^{\prime
},X^{\prime }\right) }\right) \right)  \notag \\
&&\times \frac{F_{2}\left( R\left( K,\hat{X}\right) \right) \left\Vert \Psi
_{0}\left( K-K_{\hat{X}}\right) \right\Vert ^{2}}{F_{2}\left( R\left( K_{%
\hat{X}},\hat{X}\right) \right) }dK  \label{hfr} \\
g\left( \hat{X},\Psi ,\hat{\Psi}\right) &=&\int \left( \frac{\nabla _{\hat{X}%
}F_{0}\left( R\left( K,\hat{X}\right) \right) }{\left\Vert \nabla _{\hat{X}%
}R\left( K,\hat{X}\right) \right\Vert }+\nu \nabla _{\hat{X}}F_{1}\left( 
\frac{R\left( K,\hat{X}\right) }{\int R\left( K^{\prime },X^{\prime }\right)
\left\Vert \Psi \left( K^{\prime },X^{\prime }\right) \right\Vert
^{2}d\left( K^{\prime },X^{\prime }\right) },\Gamma \left( K,X\right)
\right) \right)  \notag \\
&&\times \frac{\left\Vert \Psi \left( K,\hat{X}\right) \right\Vert ^{2}dK}{%
\int \left\Vert \Psi \left( K^{\prime },\hat{X}\right) \right\Vert
^{2}dK^{\prime }}  \label{hgr}
\end{eqnarray}%
Another simplification arises for the function $F_{2}\left( R\left( K,\hat{X}%
\right) \right) $. Actually:

\begin{eqnarray*}
&&\frac{F_{2}\left( R\left( K,\hat{X}\right) \right) }{\int F_{2}\left(
R\left( K^{\prime },\hat{X}\right) \right) \left\Vert \Psi \left( K^{\prime
},\hat{X}\right) \right\Vert ^{2}dK^{\prime }}\left\Vert \Psi \left( K,\hat{X%
}\right) \right\Vert ^{2} \\
&\simeq &\frac{F_{2}\left( R\left( K,\hat{X}\right) \right) }{\int
F_{2}\left( R\left( K_{\hat{X}},\hat{X}\right) \right) \left\Vert \Psi
\left( \hat{X}\right) \right\Vert ^{2}}\left\Vert \Psi \left( K,\hat{X}%
\right) \right\Vert ^{2} \\
&\simeq &\frac{F_{2}\left( R\left( K,\hat{X}\right) \right) }{F_{2}\left(
R\left( K_{\hat{X}},\hat{X}\right) \right) }\left\Vert \Psi \left( K-K_{\hat{%
X}}\right) \right\Vert ^{2}
\end{eqnarray*}%
and by integration in (\ref{hfr}) and (\ref{hgr}), we have:%
\begin{eqnarray}
f\left( \hat{X},\Psi ,\hat{\Psi}\right) &=&\frac{1}{\varepsilon }\left(
r\left( K_{\hat{X}},\hat{X}\right) -\gamma \left\Vert \Psi \left( \hat{X}%
\right) \right\Vert ^{2}+F_{1}\left( \frac{R\left( K_{\hat{X}},\hat{X}%
\right) }{\int R\left( K_{X^{\prime }}^{\prime },X^{\prime }\right)
\left\Vert \Psi \left( X^{\prime }\right) \right\Vert ^{2}dX^{\prime }}%
\right) \right)  \label{ncf} \\
g\left( \hat{X},\Psi ,\hat{\Psi}\right) &=&\frac{\nabla _{\hat{X}%
}F_{0}\left( R\left( K_{\hat{X}},\hat{X}\right) \right) }{\left\Vert \nabla
_{\hat{X}}R\left( K_{\hat{X}},\hat{X}\right) \right\Vert }+\nu \nabla _{\hat{%
X}}F_{1}\left( \frac{R\left( K_{\hat{X}},\hat{X}\right) }{\int R\left(
K_{X^{\prime }}^{\prime },X^{\prime }\right) \left\Vert \Psi \left(
X^{\prime }\right) \right\Vert ^{2}dX^{\prime }}\right)  \notag
\end{eqnarray}%
In the sequel, for the sake of simplicity, we will write $f\left( \hat{X}%
\right) $ and $g\left( \hat{X}\right) $ for $f\left( \hat{X},K_{\hat{X}%
}\right) $ and $g\left( \hat{X},K_{\hat{X}}\right) $ respectively. We then
perform the following change of variable in (\ref{mts}): 
\begin{eqnarray*}
\hat{\Psi} &\rightarrow &\exp \left( \frac{1}{\sigma _{\hat{X}}^{2}}\int
g\left( \hat{X}\right) d\hat{X}\right) \hat{\Psi} \\
\hat{\Psi}^{\dag } &\rightarrow &\exp \left( \frac{1}{\sigma _{\hat{X}}^{2}}%
\int g\left( \hat{X}\right) d\hat{X}\right) \hat{\Psi}^{\dag }
\end{eqnarray*}%
so that (\ref{mts}) becomes:%
\begin{eqnarray}
&&S_{3}+S_{4}=-\int \hat{\Psi}^{\dag }\left( \frac{\sigma _{\hat{X}}^{2}}{2}%
\nabla _{\hat{X}}^{2}-\frac{1}{2\sigma _{\hat{X}}^{2}}\left( g\left( \hat{X}%
,K_{\hat{X}}\right) \right) ^{2}-\frac{1}{2}\nabla _{\hat{X}}g\left( \hat{X}%
,K_{\hat{X}}\right) \right) \hat{\Psi}  \label{stmv} \\
&&-\int \hat{\Psi}^{\dag }\left( \nabla _{\hat{K}}\left( \frac{\sigma _{\hat{%
K}}^{2}}{2}\nabla _{\hat{K}}-\hat{K}f\left( \hat{X},K_{\hat{X}}\right)
\right) \right) \hat{\Psi}  \notag
\end{eqnarray}%
This action functional for $\hat{\Psi}$ will be minimized in the next
paragraph. Note that we should also include to (\ref{stmv}), the action
functional $S_{1}+S_{2}$ evaluated at the background field $\Psi $, since
this one depends on $\hat{\Psi}$. However, we have seen that at the
background field $\Psi $, for $K\simeq K_{X}$, $u\left( K,X,\Psi ,\hat{\Psi}%
\right) \simeq 0$ and the action functional $S_{1}+S_{2}$ defined in (\ref%
{prt}) reduces to:

\begin{equation}
S_{1}+S_{2}\simeq \int \Psi ^{\dag }\left( X\right) \left( -\nabla
_{X}\left( \frac{\sigma _{X}^{2}}{2}\nabla _{X}-\left( \nabla _{X}R\left(
X\right) H\left( K_{X}\right) \right) \right) +\tau \left\vert \Psi \left(
X\right) \right\vert ^{2}+\frac{\sigma _{K}^{2}-1}{2\varepsilon }\right)
\Psi \left( X\right)
\end{equation}%
and this depends on through $K_{X}$. Then, due to the first order condition
for $\Psi \left( X\right) $, one has:%
\begin{equation*}
\frac{\delta }{\delta \hat{\Psi}}\left( S_{1}+S_{2}\right) =\frac{\delta
K_{X}}{\delta \hat{\Psi}}\frac{\partial }{\partial K_{X}}\left(
S_{1}+S_{2}\right)
\end{equation*}%
We have assumed previously that $H\left( K_{X}\right) $ is slowly varying.
Moreover, due to is definition:%
\begin{equation*}
\frac{\delta K_{X}}{\delta \hat{\Psi}\left( \hat{K},X\right) }=\frac{\hat{K}%
}{\left\Vert \Psi \left( X\right) \right\Vert ^{2}}
\end{equation*}%
In most of the cases, this reduces to:%
\begin{equation*}
\frac{\delta K_{X}}{\delta \hat{\Psi}\left( \hat{K},X\right) }\simeq \frac{%
\hat{K}}{D\left( \left\Vert \Psi \right\Vert ^{2}\right) }<<\hat{K}
\end{equation*}%
Consequently, we can assume that $\frac{\delta }{\delta \hat{\Psi}}\left(
S_{1}+S_{2}\right) $ will be negligible with respect to the other quantities
in the minimization with respect to $\hat{\Psi}\left( \hat{K},X\right) $.
The rationale for this approximation is the following. The field action $%
S_{1}+S_{2}$ for $\Psi \left( X\right) $ depends on the global quantity $%
\int \hat{K}\left\Vert \hat{\Psi}\left( \hat{K},X\right) \right\Vert ^{2}d%
\hat{K}$ that represents the total investment in sector $X$. While
minimizing the field action $S_{1}+S_{2}$ with respect to $\hat{\Psi}\left( 
\hat{K},X\right) $, we compute the change in this action with respect to an
individual variation $\hat{\Psi}\left( \hat{K},X\right) $, and the impact of
this variation is, consequently, negligible.

\subsubsection*{A3.1.2 \textbf{Minimization for} $\hat{\Psi}\left( \hat{K},%
\hat{X}\right) $}

Adding the Lagrange multiplier $\hat{\lambda}$ implementing the constraint $%
\int \left\Vert \hat{\Psi}\left( \hat{K},\hat{X}\right) \right\Vert ^{2}=%
\hat{N}$ , the minimization of (\ref{stm}) with the functions given by (\ref%
{ncf}) leads to the first order conditions:

\begin{eqnarray}
0 &=&\left( \frac{\sigma _{\hat{X}}^{2}\nabla _{\hat{X}}^{2}}{2}-\frac{%
\left( g\left( \hat{X},K_{\hat{X}}\right) \right) ^{2}}{2\sigma _{\hat{X}%
}^{2}}-\frac{\nabla _{\hat{X}}g\left( \hat{X},K_{\hat{X}}\right) }{2}\right) 
\hat{\Psi}+\nabla _{\hat{K}}\left( \frac{\sigma _{\hat{K}}^{2}\nabla _{\hat{K%
}}}{2}-\hat{K}f\left( \hat{X},K_{\hat{X}}\right) -\hat{\lambda}\right) \hat{%
\Psi}  \label{drf} \\
&&-\left( \int \hat{\Psi}^{\dag }\frac{\delta }{\delta \hat{\Psi}^{\dag }}%
\left( \frac{1}{2\sigma _{\hat{X}}^{2}}\left( g\left( \hat{X},K_{\hat{X}%
}\right) \right) ^{2}+\frac{1}{2}\nabla _{\hat{X}}g\left( \hat{X},K_{\hat{X}%
}\right) \right) \hat{\Psi}\right) -\left( \int \hat{\Psi}^{\dag }\nabla _{%
\hat{K}}\frac{\delta }{\delta \hat{\Psi}^{\dag }}\left( \hat{K}f\left( \hat{X%
},K_{\hat{X}}\right) \right) \hat{\Psi}\right)  \notag
\end{eqnarray}%
Using that:%
\begin{equation*}
\frac{\delta }{\delta \hat{\Psi}^{\dag }}K_{\hat{X}}=\frac{\hat{K}}{%
\left\Vert \Psi \left( \hat{X}\right) \right\Vert ^{2}}\hat{\Psi}
\end{equation*}%
equation (\ref{drf}) becomes: 
\begin{eqnarray}
0 &=&\left( \frac{\sigma _{\hat{X}}^{2}}{2}\nabla _{\hat{X}}^{2}-\frac{1}{%
2\sigma _{\hat{X}}^{2}}\left( g\left( \hat{X},K_{\hat{X}}\right) \right)
^{2}-\frac{1}{2}\nabla _{\hat{X}}g\left( \hat{X},K_{\hat{X}}\right) \right) 
\hat{\Psi}  \label{fdr} \\
&&+\left( \nabla _{\hat{K}}\left( \frac{\sigma _{\hat{K}}^{2}}{2}\nabla _{%
\hat{K}}-\hat{K}f\left( \hat{X},K_{\hat{X}}\right) \right) -\hat{\lambda}%
\right) \hat{\Psi}-F\left( \hat{X},K_{\hat{X}}\right) \hat{K}\hat{\Psi} 
\notag
\end{eqnarray}%
with:%
\begin{equation}
F\left( \hat{X},K_{\hat{X}}\right) =\frac{\left\langle \nabla _{K_{\hat{X}%
}}\left( \frac{\left( g\left( \hat{X},K_{\hat{X}}\right) \right) ^{2}}{%
2\sigma _{\hat{X}}^{2}}+\frac{1}{2}\nabla _{\hat{X}}g\left( \hat{X},K_{\hat{X%
}}\right) \right) \right\rangle }{\left\Vert \Psi \left( \hat{X}\right)
\right\Vert ^{2}}+\frac{\left\langle \nabla _{\hat{K}}\left( \hat{K}\nabla
_{K_{\hat{X}}}f\left( \hat{X},K_{\hat{X}}\right) \right) \right\rangle }{%
\left\Vert \Psi \left( \hat{X}\right) \right\Vert ^{2}}  \label{dnf}
\end{equation}%
The brackets in (\ref{dnf}) are given by:%
\begin{eqnarray}
&&\left\langle \nabla _{K_{\hat{X}}}\left( \frac{\left( g\left( \hat{X},K_{%
\hat{X}}\right) \right) ^{2}}{2\sigma _{\hat{X}}^{2}}+\frac{1}{2}\nabla _{%
\hat{X}}g\left( \hat{X},K_{\hat{X}}\right) \right) \right\rangle  \notag \\
&=&\int \hat{\Psi}^{\dag }\left( \hat{X},\hat{K}\right) \nabla _{K_{\hat{X}%
}}\left( \frac{\left( g\left( \hat{X},K_{\hat{X}}\right) \right) ^{2}}{%
2\sigma _{\hat{X}}^{2}}+\frac{1}{2}\nabla _{\hat{X}}g\left( \hat{X},K_{\hat{X%
}}\right) \right) \hat{\Psi}\left( \hat{X},\hat{K}\right) d\hat{K}  \notag \\
&\equiv &\nabla _{K_{\hat{X}}}\left( \frac{\left( g\left( \hat{X},K_{\hat{X}%
}\right) \right) ^{2}}{2\sigma _{\hat{X}}^{2}}+\frac{1}{2}\nabla _{\hat{X}%
}g\left( \hat{X},K_{\hat{X}}\right) \right) \left\Vert \hat{\Psi}\left( \hat{%
X}\right) \right\Vert ^{2}  \notag \\
&&\left\langle \nabla _{\hat{K}}\left( \hat{K}\nabla _{K_{\hat{X}}}f\left( 
\hat{X},K_{\hat{X}}\right) \right) \right\rangle  \notag \\
&=&\int \hat{\Psi}^{\dag }\left( \hat{X},K_{\hat{X}}\right) \nabla _{\hat{K}%
}\left( \hat{K}\nabla _{K_{\hat{X}}}f\left( \hat{X},K_{\hat{X}}\right)
\right) \hat{\Psi}\left( \hat{X},K_{\hat{X}}\right) d\hat{K}  \notag \\
&=&-\nabla _{K_{\hat{X}}}f\left( \hat{X},K_{\hat{X}}\right) \int \left( \hat{%
K}\nabla _{\hat{K}}\left\Vert \hat{\Psi}\left( \hat{X},K_{\hat{X}}\right)
\right\Vert ^{2}-\frac{2\hat{K}^{2}}{\sigma _{\hat{K}}^{2}}f\left( \hat{X}%
\right) \left\Vert \hat{\Psi}\left( \hat{X},K_{\hat{X}}\right) \right\Vert
^{2}\right) d\hat{K}  \notag \\
&=&\nabla _{K_{\hat{X}}}f\left( \hat{X},K_{\hat{X}}\right) \left\Vert \hat{%
\Psi}\left( \hat{X}\right) \right\Vert ^{2}+\frac{\nabla _{K_{\hat{X}%
}}f^{2}\left( \hat{X},K_{\hat{X}}\right) }{\sigma _{\hat{K}}^{2}}%
\left\langle \hat{K}^{2}\right\rangle _{\hat{X}}  \label{smt}
\end{eqnarray}%
Where the average $\left\langle \hat{K}^{2}\right\rangle _{\hat{X}}$ is
defined by:%
\begin{equation*}
\left\langle \hat{K}^{2}\right\rangle _{\hat{X}}=\int \left\Vert \hat{\Psi}%
\left( \hat{X},\hat{K}\right) \right\Vert ^{2}d\hat{K}
\end{equation*}%
The previous expression (\ref{smt}) for $F\left( \hat{X},K_{\hat{X}}\right) $
can also be rewritten as: 
\begin{eqnarray}
F\left( \hat{X},K_{\hat{X}}\right) &=&\frac{\left\langle \nabla _{K_{\hat{X}%
}}\left( \frac{\left( g\left( \hat{X},K_{\hat{X}}\right) \right) ^{2}}{%
2\sigma _{\hat{X}}^{2}}+\frac{1}{2}\nabla _{\hat{X}}g\left( \hat{X},K_{\hat{X%
}}\right) \right) \right\rangle }{\left\Vert \Psi \left( \hat{X}\right)
\right\Vert ^{2}}+\frac{\left\langle \nabla _{\hat{K}}\left( \hat{K}\nabla
_{K_{\hat{X}}}f\left( \hat{X},K_{\hat{X}}\right) \right) \right\rangle }{%
\left\Vert \Psi \left( \hat{X}\right) \right\Vert ^{2}}  \label{frl} \\
&=&\nabla _{K_{\hat{X}}}\left( \frac{\left( g\left( \hat{X},K_{\hat{X}%
}\right) \right) ^{2}}{2\sigma _{\hat{X}}^{2}}+\frac{1}{2}\nabla _{\hat{X}%
}g\left( \hat{X},K_{\hat{X}}\right) +f\left( \hat{X},K_{\hat{X}}\right)
\right) \frac{\left\Vert \hat{\Psi}\left( \hat{X}\right) \right\Vert ^{2}}{%
\left\Vert \Psi \left( \hat{X}\right) \right\Vert ^{2}}  \notag \\
&&+\frac{\nabla _{K_{\hat{X}}}f^{2}\left( \hat{X},K_{\hat{X}}\right) }{%
\sigma _{\hat{K}}^{2}\left\Vert \Psi \left( \hat{X}\right) \right\Vert ^{2}}%
\left\langle \hat{K}^{2}\right\rangle _{\hat{X}}  \notag
\end{eqnarray}%
It will be useful to rewrite the last term as:%
\begin{equation}
\frac{\nabla _{K_{\hat{X}}}f^{2}\left( \hat{X},K_{\hat{X}}\right) }{\sigma _{%
\hat{K}}^{2}\left\Vert \Psi \left( \hat{X}\right) \right\Vert ^{2}}%
\left\langle \hat{K}^{2}\right\rangle _{\hat{X}}\simeq \frac{\nabla _{K_{%
\hat{X}}}f^{2}\left( \hat{X},K_{\hat{X}}\right) }{\sigma _{\hat{K}}^{2}}%
\left\langle \hat{K}\right\rangle _{\hat{X}}^{2}=\frac{\nabla _{K_{\hat{X}%
}}f^{2}\left( \hat{X},K_{\hat{X}}\right) }{\sigma _{\hat{K}}^{2}}\frac{%
\left\Vert \Psi \left( \hat{X}\right) \right\Vert ^{2}}{\left\Vert \hat{\Psi}%
\left( \hat{X}\right) \right\Vert ^{2}}  \label{lfr}
\end{equation}%
Consequently:%
\begin{eqnarray}
F\left( \hat{X},K_{\hat{X}}\right) &=&\nabla _{K_{\hat{X}}}\left( \frac{%
\left( g\left( \hat{X},K_{\hat{X}}\right) \right) ^{2}}{2\sigma _{\hat{X}%
}^{2}}+\frac{1}{2}\nabla _{\hat{X}}g\left( \hat{X},K_{\hat{X}}\right)
+f\left( \hat{X},K_{\hat{X}}\right) \right) \frac{\left\Vert \hat{\Psi}%
\left( \hat{X}\right) \right\Vert ^{2}}{\left\Vert \Psi \left( \hat{X}%
\right) \right\Vert ^{2}}  \label{fnd} \\
&&+\frac{\nabla _{K_{\hat{X}}}f^{2}\left( \hat{X},K_{\hat{X}}\right) }{%
\sigma _{\hat{K}}^{2}}\frac{\left\Vert \Psi \left( \hat{X}\right)
\right\Vert ^{2}}{\left\Vert \hat{\Psi}\left( \hat{X}\right) \right\Vert ^{2}%
}  \notag
\end{eqnarray}%
We also have an equation for $\hat{\Psi}^{\dag }$ similar to (\ref{fdr}):%
\begin{eqnarray}
0 &=&\left( \frac{\sigma _{\hat{X}}^{2}}{2}\nabla _{\hat{X}}^{2}-\frac{1}{%
2\sigma _{\hat{X}}^{2}}\left( g\left( \hat{X},K_{\hat{X}}\right) \right)
^{2}-\frac{1}{2}\nabla _{\hat{X}}g\left( \hat{X},K_{\hat{X}}\right) \right) 
\hat{\Psi}^{\dag }  \label{fdp} \\
&&+\left( \left( \frac{\sigma _{\hat{K}}^{2}}{2}\nabla _{\hat{K}}+\hat{K}%
f\left( \hat{X},K_{\hat{X}}\right) \right) \nabla _{\hat{K}}-\hat{\lambda}%
\right) \hat{\Psi}-F\left( \hat{X},K_{\hat{X}}\right) \hat{K}\hat{\Psi}%
^{\dag }  \notag
\end{eqnarray}

\subsubsection*{A3.1.3 Resolution of (\protect\ref{fdr})}

\paragraph{A3.1.3.1 \textbf{zeroth order in} $\protect\sigma _{X}^{2}$}

We consider $\sigma _{X}^{2}<<1$ (which means that fluctuation in $X<<$
fluctuation in $K$). Thus (\ref{fdr}) writes at the lowest order:

\begin{equation}
\left( \nabla _{\hat{K}}\left( \frac{\sigma _{\hat{K}}^{2}}{2}\nabla _{\hat{K%
}}-\hat{K}f\left( \hat{X},K_{\hat{X}}\right) \right) -\frac{\left( g\left( 
\hat{X}\right) \right) ^{2}}{2\sigma _{\hat{X}}^{2}}-\frac{\nabla _{\hat{X}%
}g\left( \hat{X},K_{\hat{X}}\right) }{2}-F\left( \hat{X},K_{\hat{X}}\right) 
\hat{K}-\hat{\lambda}\right) \hat{\Psi}=0  \label{knq}
\end{equation}%
Performing the change of variable:%
\begin{equation*}
\hat{\Psi}\rightarrow \exp \left( \frac{\hat{K}^{2}}{\sigma _{\hat{K}}^{2}}%
f\left( \hat{X}\right) \right) \hat{\Psi}
\end{equation*}%
leads to the equation for $\hat{K}$: 
\begin{equation}
\frac{\sigma _{\hat{K}}^{2}}{2}\nabla _{\hat{K}}^{2}\hat{\Psi}-\left( \frac{%
\hat{K}^{2}}{2\sigma _{\hat{K}}^{2}}f^{2}\left( \hat{X}\right) +F\left( \hat{%
X},K_{\hat{X}}\right) \hat{K}+\frac{1}{2}f\left( \hat{X},K_{\hat{X}}\right) +%
\frac{\left( g\left( \hat{X}\right) \right) ^{2}}{2\sigma _{\hat{X}}^{2}}+%
\frac{1}{2}\nabla _{\hat{X}}g\left( \hat{X},K_{\hat{X}}\right) +\hat{\lambda}%
\right) \hat{\Psi}\simeq 0  \label{nqk}
\end{equation}%
This equation can be normalized by dividing by $f^{2}\left( \hat{X}\right) $%
: 
\begin{equation*}
\frac{\sigma _{\hat{K}}^{2}\nabla _{\hat{K}}^{2}\hat{\Psi}}{2f^{2}\left( 
\hat{X}\right) }-\left( \frac{\hat{K}^{2}}{2\sigma _{\hat{K}}^{2}}+\frac{%
F\left( \hat{X},K_{\hat{X}}\right) \hat{K}}{f^{2}\left( \hat{X}\right) }+%
\frac{\frac{f\left( \hat{X},K_{\hat{X}}\right) }{2}+\frac{\left( g\left( 
\hat{X}\right) \right) ^{2}}{2\sigma _{\hat{X}}^{2}}+\frac{1}{2}\nabla _{%
\hat{X}}g\left( \hat{X},K_{\hat{X}}\right) +\hat{\lambda}}{f^{2}\left( \hat{X%
}\right) }\right) \hat{\Psi}\simeq 0
\end{equation*}%
We then define:%
\begin{equation*}
y=\frac{\hat{K}+\frac{\sigma _{\hat{K}}^{2}F\left( \hat{X},K_{\hat{X}%
}\right) }{f^{2}\left( \hat{X}\right) }}{\sqrt{\sigma _{\hat{K}}^{2}}}\left(
f^{2}\left( \hat{X}\right) \right) ^{\frac{1}{4}}
\end{equation*}%
and (\ref{fdr}) is transformed into:%
\begin{equation}
\nabla _{y}^{2}\hat{\Psi}-\left( \frac{y^{2}}{4}+\frac{\left( g\left( \hat{X}%
\right) \right) ^{2}+\sigma _{\hat{X}}^{2}\left( f\left( \hat{X}\right)
+\nabla _{\hat{X}}g\left( \hat{X},K_{\hat{X}}\right) -\frac{\sigma _{\hat{K}%
}^{2}F^{2}\left( \hat{X},K_{\hat{X}}\right) }{2f^{2}\left( \hat{X}\right) }+%
\hat{\lambda}\right) }{\sigma _{\hat{X}}^{2}\sqrt{f^{2}\left( \hat{X}\right) 
}}\right) \Psi \simeq 0  \label{lgm}
\end{equation}%
Solutions of (\ref{lgm}) are obtained by rewriting (\ref{lgm}):%
\begin{equation*}
\hat{\Psi}^{\prime \prime }+\left( p\left( \hat{X},\hat{\lambda}\right) +%
\frac{1}{2}-\frac{1}{4}y^{2}\right) \hat{\Psi}
\end{equation*}%
$\allowbreak \allowbreak $where:%
\begin{equation}
p\left( \hat{X},\hat{\lambda}\right) =-\frac{\left( g\left( \hat{X}\right)
\right) ^{2}+\sigma _{\hat{X}}^{2}\left( f\left( \hat{X}\right) +\nabla _{%
\hat{X}}g\left( \hat{X},K_{\hat{X}}\right) -\frac{\sigma _{\hat{K}%
}^{2}F^{2}\left( \hat{X},K_{\hat{X}}\right) }{2f^{2}\left( \hat{X}\right) }+%
\hat{\lambda}\right) }{\sigma _{\hat{X}}^{2}\sqrt{f^{2}\left( \hat{X}\right) 
}}-\frac{1}{2}  \label{pxd}
\end{equation}%
The solution of (\ref{lgm}) is thus:%
\begin{equation}
\hat{\Psi}_{\hat{\lambda},C}^{\left( 0\right) }\left( \hat{X},\hat{K}\right)
=\sqrt{C}D_{p\left( \hat{X},\hat{\lambda}\right) }\left( \left( \left\vert
f\left( \hat{X}\right) \right\vert \right) ^{\frac{1}{2}}\frac{\left( \hat{K}%
+\frac{\sigma _{\hat{K}}^{2}F\left( \hat{X},K_{\hat{X}}\right) }{f^{2}\left( 
\hat{X}\right) }\right) }{\sigma _{\hat{K}}}\right)  \label{slt}
\end{equation}%
where $D_{p}$ denotes the parabolic cylinder function with parameter $p$ and 
$C$ is a normalization constant that will be computed as a function of $%
\lambda $ using the constraint $\int \left\Vert \hat{\Psi}\left( \hat{K},%
\hat{X}\right) \right\Vert ^{2}=\hat{N}$.

A similar equation to (\ref{knq}) can be obtained for $\hat{\Psi}^{\dag }$.
The equivalent of (\ref{drf}) is (\ref{fdp}):%
\begin{eqnarray}
0 &=&\left( \frac{\sigma _{\hat{X}}^{2}}{2}\nabla _{\hat{X}}^{2}-\frac{1}{%
2\sigma _{\hat{X}}^{2}}\left( g\left( \hat{X},K_{\hat{X}}\right) \right)
^{2}-\frac{1}{2}\nabla _{\hat{X}}g\left( \hat{X},K_{\hat{X}}\right) \right) 
\hat{\Psi}  \label{cjt} \\
&&+\left( \left( \frac{\sigma _{\hat{K}}^{2}}{2}\nabla _{\hat{K}}+\hat{K}%
f\left( \hat{X},K_{\hat{X}}\right) \right) \nabla _{\hat{K}}-\hat{\lambda}%
\right) \hat{\Psi}-F\left( \hat{X},K_{\hat{X}}\right) \hat{K}\hat{\Psi} 
\notag
\end{eqnarray}%
The change of variable:%
\begin{equation*}
\hat{\Psi}^{\dag }\rightarrow \exp \left( -\frac{\hat{K}^{2}}{\sigma _{\hat{K%
}}^{2}}f\left( \hat{X}\right) \right) \hat{\Psi}^{\dag }
\end{equation*}%
and the approximation $\sigma _{\hat{X}}^{2}<<1$ lead ultimately to:%
\begin{equation}
\frac{\sigma _{\hat{K}}^{2}}{2}\nabla _{\hat{K}}^{2}\hat{\Psi}^{\dag
}-\left( \frac{\hat{K}^{2}}{2\sigma _{\hat{K}}^{2}}f^{2}\left( \hat{X}%
\right) +\frac{1}{2}\nabla _{\hat{X}}f\left( \hat{X},K_{\hat{X}}\right) +%
\frac{1}{2\sigma _{\hat{X}}^{2}}\left( g\left( \hat{X}\right) \right) ^{2}+%
\frac{1}{2}\nabla _{\hat{X}}g\left( \hat{X},K_{\hat{X}}\right) +F\left( \hat{%
X},K_{\hat{X}}\right) +\hat{\lambda}\right) \hat{\Psi}^{\dag }\simeq 0
\label{fbm}
\end{equation}%
which is the same equation as (\ref{nqk}). Consequently, the solutions of (%
\ref{fbm}) write:%
\begin{equation}
\hat{\Psi}_{\lambda ,C}^{\left( 0\right) \dag }\left( \hat{X},\hat{K}\right)
=\hat{\Psi}_{\lambda ,C}^{\left( 0\right) }\left( \hat{X},\hat{K}\right) =%
\sqrt{C}D_{p\left( \hat{X},\hat{\lambda}\right) }\left( \left( \left\vert
f\left( \hat{X}\right) \right\vert \right) ^{\frac{1}{2}}\frac{\left( \hat{K}%
+\frac{\sigma _{\hat{K}}^{2}F\left( \hat{X},K_{\hat{X}}\right) }{f^{2}\left( 
\hat{X}\right) }\right) }{\sigma _{\hat{K}}}\right)  \label{tls}
\end{equation}%
To conclude this section, we detail the expressions for $\frac{\sigma _{\hat{%
K}}^{2}F\left( \hat{X},K_{\hat{X}}\right) }{f^{2}\left( \hat{X}\right) }$
and $\frac{\sigma _{\hat{K}}^{2}F^{2}\left( \hat{X},K_{\hat{X}}\right) }{%
2f^{2}\left( \hat{X}\right) }$. \ Given the expression for $F\left( \hat{X}%
,K_{\hat{X}}\right) $ in (\ref{fnd}), the term $\frac{\sigma _{\hat{K}%
}^{2}F\left( \hat{X},K_{\hat{X}}\right) }{f^{2}\left( \hat{X}\right) }$
arising in (\ref{slt}) and (\ref{tls}) 
\begin{eqnarray}
\frac{\sigma _{\hat{K}}^{2}F\left( \hat{X},K_{\hat{X}}\right) }{f^{2}\left( 
\hat{X},K_{\hat{X}}\right) } &=&\frac{\sigma _{\hat{K}}^{2}}{f^{2}\left( 
\hat{X}\right) }\nabla _{K_{\hat{X}}}\left( \frac{\left( g\left( \hat{X},K_{%
\hat{X}}\right) \right) ^{2}}{2\sigma _{\hat{X}}^{2}}+\frac{1}{2}\nabla _{%
\hat{X}}g\left( \hat{X},K_{\hat{X}}\right) +f\left( \hat{X},K_{\hat{X}%
}\right) \right) \frac{\left\Vert \hat{\Psi}\left( \hat{X}\right)
\right\Vert ^{2}}{\left\Vert \Psi \left( \hat{X}\right) \right\Vert ^{2}} 
\notag \\
&&+\frac{\nabla _{K_{\hat{X}}}f^{2}\left( \hat{X},K_{\hat{X}}\right) }{%
f^{2}\left( \hat{X},K_{\hat{X}}\right) }\frac{\left\Vert \Psi \left( \hat{X}%
\right) \right\Vert ^{2}}{\left\Vert \hat{\Psi}\left( \hat{X}\right)
\right\Vert ^{2}}  \notag \\
&\simeq &\frac{\nabla _{K_{\hat{X}}}f\left( \hat{X},K_{\hat{X}}\right) }{%
f\left( \hat{X},K_{\hat{X}}\right) }\frac{\left\Vert \Psi \left( \hat{X}%
\right) \right\Vert ^{2}}{\left\Vert \hat{\Psi}\left( \hat{X}\right)
\right\Vert ^{2}}  \label{tmd}
\end{eqnarray}%
$\frac{\sigma _{\hat{K}}^{2}F^{2}\left( \hat{X},K_{\hat{X}}\right) }{%
2f^{2}\left( \hat{X}\right) }$ arising in the definition (\ref{pxd}) of $%
p\left( \hat{X},\hat{\lambda}\right) $ is equal to: 
\begin{eqnarray}
\frac{\sigma _{\hat{K}}^{2}F^{2}\left( \hat{X},K_{\hat{X}}\right) }{%
2f^{2}\left( \hat{X}\right) } &=&\frac{\sigma _{\hat{K}}^{2}}{2}\left(
\left( \frac{\nabla _{K_{\hat{X}}}\left( g\left( \hat{X},K_{\hat{X}}\right)
\right) ^{2}+\sigma _{\hat{X}}^{2}\left( \nabla _{\hat{X}}^{2}g\left( \hat{X}%
,K_{\hat{X}}\right) +\nabla _{K_{\hat{X}}}f\left( \hat{X},K_{\hat{X}}\right)
\right) }{2\sigma _{\hat{X}}^{2}f\left( \hat{X},K_{\hat{X}}\right) }\right) 
\frac{\left\Vert \hat{\Psi}\left( \hat{X}\right) \right\Vert ^{2}}{%
\left\Vert \Psi \left( \hat{X}\right) \right\Vert ^{2}}\right.  \notag \\
&&\left. +2\nabla _{K_{\hat{X}}}f\left( \hat{X},K_{\hat{X}}\right) \frac{%
\left\Vert \Psi \left( \hat{X}\right) \right\Vert ^{2}}{\left\Vert \hat{\Psi}%
\left( \hat{X}\right) \right\Vert ^{2}}\right) ^{2}  \label{kmd}
\end{eqnarray}%
and this simplifies as:%
\begin{equation}
\frac{\sigma _{\hat{K}}^{2}F^{2}\left( \hat{X},K_{\hat{X}}\right) }{%
2f^{2}\left( \hat{X}\right) }\simeq 2\sigma _{\hat{K}}^{2}\left( \nabla _{K_{%
\hat{X}}}f\left( \hat{X},K_{\hat{X}}\right) \frac{\left\Vert \Psi \left( 
\hat{X}\right) \right\Vert ^{2}}{\left\Vert \hat{\Psi}\left( \hat{X}\right)
\right\Vert ^{2}}\right) ^{2}  \label{kmdp}
\end{equation}%
since:%
\begin{eqnarray*}
&&\frac{\nabla _{K_{\hat{X}}}\left( g\left( \hat{X},K_{\hat{X}}\right)
\right) ^{2}+\sigma _{\hat{X}}^{2}\left( \nabla _{\hat{X}}^{2}g\left( \hat{X}%
,K_{\hat{X}}\right) +\nabla _{K_{\hat{X}}}f\left( \hat{X},K_{\hat{X}}\right)
\right) }{2\sigma _{\hat{X}}^{2}f\left( \hat{X},K_{\hat{X}}\right) }\frac{%
\left\Vert \hat{\Psi}\left( \hat{X}\right) \right\Vert ^{2}}{\left\Vert \Psi
\left( \hat{X}\right) \right\Vert ^{2}} \\
&\sim &\frac{\left( g\left( \hat{X},K_{\hat{X}}\right) \right) ^{2}+\sigma _{%
\hat{X}}^{2}\left( \nabla _{\hat{X}}^{2}g\left( \hat{X},K_{\hat{X}}\right)
+\nabla _{K_{\hat{X}}}f\left( \hat{X},K_{\hat{X}}\right) \right) }{2\sigma _{%
\hat{X}}^{2}f\left( \hat{X},K_{\hat{X}}\right) }\left( \frac{\left\Vert \hat{%
\Psi}\left( \hat{X}\right) \right\Vert ^{2}}{K_{\hat{X}}\left\Vert \Psi
\left( \hat{X}\right) \right\Vert ^{2}}\right) <<1
\end{eqnarray*}

\paragraph*{A3.1.3.2 \textbf{Corrections in} $\protect\sigma _{X}^{2}$:}

To introduce the corrections in $\sigma _{X}^{2}$ in (\ref{fdr}) we factor
the solution as: 
\begin{eqnarray*}
\hat{\Psi}_{\lambda ,C}\left( \hat{K},\hat{X}\right) &=&\sqrt{C}\exp \left( 
\frac{\hat{K}^{2}}{\sigma _{\hat{K}}^{2}}f\left( \hat{X}\right) \right)
D_{p\left( \hat{X},\hat{\lambda}\right) }\left( \left( \frac{\left\vert
f\left( \hat{X}\right) \right\vert }{\sigma _{\hat{K}}^{2}}\right) ^{\frac{1%
}{2}}\left( \hat{K}+\frac{\sigma _{\hat{K}}^{2}F\left( \hat{X},K_{\hat{X}%
}\right) }{f^{2}\left( \hat{X}\right) }\right) \right) \hat{\Psi}^{\left(
1\right) }\left( \hat{K},\hat{X}\right) \\
&\equiv &\hat{\Psi}_{\lambda ,C}^{\left( 0\right) }\left( \hat{K},\hat{X}%
\right) \hat{\Psi}^{\left( 1\right) }\left( \hat{K},\hat{X}\right)
\end{eqnarray*}%
and we look for $\hat{\Psi}^{\left( 1\right) }$ of the form:%
\begin{equation}
\hat{\Psi}^{\left( 1\right) }=\exp \left( \sigma _{X}^{2}h\left( K,X\right)
\right)  \label{cps}
\end{equation}%
Introducing the postulated form in (\ref{fdr}) we are led to:

\begin{equation*}
\frac{\sigma _{X}^{2}}{2}\nabla _{\hat{X}}^{2}\left( \hat{\Psi}^{\left(
1\right) }\hat{\Psi}_{\lambda ,C}^{\left( 0\right) }\right) +\left( \frac{%
\sigma _{\hat{K}}^{2}}{2}\nabla _{\hat{K}}^{2}\hat{\Psi}^{\left( 1\right)
}\right) \hat{\Psi}_{\lambda ,C}^{\left( 0\right) }+\left( \nabla _{\hat{K}}%
\hat{\Psi}^{\left( 1\right) }\right) \left( \sigma _{\hat{K}}^{2}\nabla _{%
\hat{K}}\hat{\Psi}_{\lambda ,C}^{\left( 0\right) }-\hat{K}f\left( \hat{X}%
\right) \hat{\Psi}_{\lambda ,C}^{\left( 0\right) }\right) =0
\end{equation*}%
Written in terms of $h\left( \hat{K},\hat{X}\right) $, this equation becomes
at the first order in $\sigma _{X}^{2}$: 
\begin{equation}
\frac{\nabla _{\hat{X}}^{2}\hat{\Psi}_{\lambda ,C}^{\left( 0\right) }}{\hat{%
\Psi}_{\lambda ,C}^{\left( 0\right) }}+\sigma _{\hat{K}}^{2}\nabla _{\hat{K}%
}^{2}h\left( \hat{K},\hat{X}\right) +2\left( \nabla _{\hat{K}}h\left( \hat{K}%
,\hat{X}\right) \right) \left( \sigma _{\hat{K}}^{2}\frac{\nabla _{\hat{K}}%
\hat{\Psi}_{\lambda ,C}^{\left( 0\right) }}{\hat{\Psi}_{\lambda ,C}^{\left(
0\right) }}-\hat{K}f\left( \hat{X}\right) \right) =0  \label{qnh}
\end{equation}%
The solution of\ (\ref{qnh}) is of the type:

\begin{equation*}
\nabla _{\hat{K}}\left( h\left( K,X\right) \right) =C\left( \hat{K},X\right)
\exp \left( -2\int \left( \frac{\nabla _{\hat{K}}\hat{\Psi}_{\lambda
,C}^{\left( 0\right) }}{\hat{\Psi}_{\lambda ,C}^{\left( 0\right) }}-\frac{%
\hat{K}f\left( \hat{X}\right) }{\sigma _{\hat{K}}^{2}}\right) d\hat{K}%
\right) =C\left( \hat{K},X\right) \exp \left( -\left( 2\ln \hat{\Psi}%
_{\lambda ,C}^{\left( 0\right) }-\frac{\hat{K}^{2}}{\sigma _{\hat{K}}^{2}}%
f\left( \hat{X}\right) \right) \right)
\end{equation*}%
where $C\left( X\right) $\ satisfies: 
\begin{equation*}
C^{\prime }\left( \hat{K},X\right) =-\frac{\nabla _{\hat{X}}^{2}\hat{\Psi}%
_{\lambda ,C}^{\left( 0\right) }}{\hat{\Psi}_{\lambda ,C}^{\left( 0\right)
}\sigma _{\hat{K}}^{2}}\exp \left( 2\ln \hat{\Psi}_{\lambda ,C}^{\left(
0\right) }-\frac{\hat{K}^{2}f\left( \hat{X}\right) }{\sigma _{\hat{K}}^{2}}%
\right) =-\frac{\nabla _{\hat{X}}^{2}\hat{\Psi}_{\lambda ,C}^{\left(
0\right) }}{\hat{\Psi}_{\lambda ,C}^{\left( 0\right) }}\left( \hat{\Psi}%
_{\lambda ,C}^{\left( 0\right) }\right) ^{2}\exp \left( -\frac{\hat{K}%
^{2}f\left( \hat{X}\right) }{\sigma _{\hat{K}}^{2}}\right)
\end{equation*}%
and the solution of (\ref{qnh}) is:%
\begin{equation*}
\nabla _{\hat{K}}\left( h\left( K,X\right) \right) =\exp \left( -\left( 2\ln 
\hat{\Psi}_{\lambda ,C}^{\left( 0\right) }-\frac{\hat{K}^{2}}{\sigma _{\hat{K%
}}^{2}}f\left( \hat{X}\right) \right) \right) \left( C-\int \frac{\nabla _{%
\hat{X}}^{2}\hat{\Psi}_{\lambda ,C}^{\left( 0\right) }}{\hat{\Psi}_{\lambda
,C}^{\left( 0\right) }\sigma _{\hat{K}}^{2}}\left( \hat{\Psi}_{\lambda
,C}^{\left( 0\right) }\right) ^{2}\exp \left( -\frac{\hat{K}^{2}}{\sigma _{%
\hat{K}}^{2}}f\left( \hat{X}\right) \right) d\hat{K}\right)
\end{equation*}%
letting $C=0$, we obtain:%
\begin{equation}
\nabla _{\hat{K}}\left( h\left( K,X\right) \right) =-\frac{1}{\sigma _{\hat{K%
}}^{2}\left( \hat{\Psi}_{\lambda ,C}^{\left( 0\right) }\right) ^{2}}\exp
\left( \frac{\hat{K}^{2}}{\sigma _{\hat{K}}^{2}}f\left( \hat{X}\right)
\right) \left( \int \frac{\nabla _{\hat{X}}^{2}\hat{\Psi}_{\lambda
,C}^{\left( 0\right) }}{\hat{\Psi}_{\lambda ,C}^{\left( 0\right) }}\left( 
\hat{\Psi}_{\lambda ,C}^{\left( 0\right) }\right) ^{2}\exp \left( -\frac{%
\hat{K}^{2}}{\sigma _{\hat{K}}^{2}}f\left( \hat{X}\right) \right) d\hat{K}%
\right)  \label{drg}
\end{equation}%
To compute $h\left( K,X\right) $, we must estimate $\frac{\nabla _{\hat{X}%
}^{2}\hat{\Psi}_{\lambda ,C}^{\left( 0\right) }}{\hat{\Psi}_{\lambda
,C}^{\left( 0\right) }}$ in (\ref{drg}). To do so, we write, for $%
\varepsilon <<1$, i.e. $\left\vert f\left( \hat{X}\right) \right\vert >>1$:

\begin{eqnarray*}
&&\exp \left( \frac{\hat{K}^{2}}{\sigma _{\hat{K}}^{2}}f\left( \hat{X}%
\right) \right) D_{p\left( \hat{X},\hat{\lambda}\right) }\left( \left( \frac{%
\left\vert f\left( \hat{X}\right) \right\vert }{\sigma _{\hat{K}}^{2}}%
\right) ^{\frac{1}{2}}\left( \hat{K}+\frac{\sigma _{\hat{K}}^{2}F\left( \hat{%
X},K_{\hat{X}}\right) }{f^{2}\left( \hat{X}\right) }\right) \right) \\
&\simeq &\exp \left( \frac{\hat{K}^{2}}{\sigma _{\hat{K}}^{2}}f\left( \hat{X}%
\right) -\frac{\left( \hat{K}+\frac{\sigma _{\hat{K}}^{2}F\left( \hat{X},K_{%
\hat{X}}\right) }{f^{2}\left( \hat{X}\right) }\right) ^{2}\left\vert f\left( 
\hat{X}\right) \right\vert }{4\sigma _{\hat{K}}^{2}}\right) \left( \left( 
\frac{\left\vert f\left( \hat{X}\right) \right\vert }{\sigma _{\hat{K}}^{2}}%
\right) ^{\frac{1}{2}}\left( \hat{K}+\frac{\sigma _{\hat{K}}^{2}F\left( \hat{%
X},K_{\hat{X}}\right) }{f^{2}\left( \hat{X}\right) }\right) \right)
^{p\left( \hat{X},\hat{\lambda}\right) } \\
&=&\exp \left( \frac{\hat{K}^{2}}{\sigma _{\hat{K}}^{2}}f\left( \hat{X}%
\right) -\frac{\left( \hat{K}+\frac{\sigma _{\hat{K}}^{2}F\left( \hat{X},K_{%
\hat{X}}\right) }{f^{2}\left( \hat{X}\right) }\right) ^{2}\left\vert f\left( 
\hat{X}\right) \right\vert }{4\sigma _{\hat{K}}^{2}}\right) \\
&&\times \exp \left( \left( p\left( \hat{X},\hat{\lambda}\right) \right) \ln
\left( \left( \frac{\left\vert f\left( \hat{X}\right) \right\vert }{\sigma _{%
\hat{K}}^{2}}\right) ^{\frac{1}{2}}\left( \hat{K}+\frac{\sigma _{\hat{K}%
}^{2}F\left( \hat{X},K_{\hat{X}}\right) }{f^{2}\left( \hat{X}\right) }%
\right) \right) \right)
\end{eqnarray*}%
which allows to compute the successives derivatives of $\hat{\Psi}$. We
find, for $f>0$:%
\begin{eqnarray}
\frac{\nabla _{\hat{X}}^{2}\hat{\Psi}_{\lambda ,C}^{\left( 0\right) }}{\hat{%
\Psi}_{\lambda ,C}^{\left( 0\right) }} &\simeq &\left( \frac{-f^{\prime
}\sigma _{\hat{X}}^{2}\hat{\lambda}-g^{2}f^{\prime }+2fgg^{\prime }}{\sigma
_{\hat{X}}^{2}f^{2}}\ln \left( \left( \hat{K}+\frac{\sigma _{\hat{K}%
}^{2}F\left( \hat{X},K_{\hat{X}}\right) }{f^{2}\left( \hat{X}\right) }%
\right) \left( \frac{f\left( \hat{X}\right) }{\sigma _{\hat{K}}^{2}}\right)
^{\frac{1}{2}}\right) \right.  \label{dvt} \\
&&+\frac{1}{2}\left( \frac{\left( g\left( \hat{X}\right) \right) ^{2}+\sigma
_{\hat{X}}^{2}\left( f\left( \hat{X}\right) +\nabla _{\hat{X}}g\left( \hat{X}%
,K_{\hat{X}}\right) -\frac{\sigma _{\hat{K}}^{2}F^{2}\left( \hat{X},K_{\hat{X%
}}\right) }{2f^{2}\left( \hat{X}\right) }+\hat{\lambda}\right) }{\sigma _{%
\hat{X}}^{2}\sqrt{f^{2}\left( \hat{X}\right) }}+\frac{1}{2}\right) \frac{%
f^{\prime }}{f}  \notag \\
&&\left. +\frac{\hat{K}^{2}-\left( \frac{\hat{K}+\frac{\sigma _{\hat{K}%
}^{2}F\left( \hat{X},K_{\hat{X}}\right) }{f^{2}\left( \hat{X}\right) }}{2}%
\right) ^{2}}{\sigma _{\hat{K}}^{2}}f^{\prime }\right) ^{2}  \notag \\
&\simeq &\left( \frac{\left( 4\hat{K}^{2}-\left( \hat{K}+\frac{\sigma _{\hat{%
K}}^{2}F\left( \hat{X},K_{\hat{X}}\right) }{f^{2}\left( \hat{X}\right) }%
\right) ^{2}\right) f^{\prime }\left( X\right) }{4\sigma _{\hat{K}}^{2}}%
\right) ^{2}  \notag
\end{eqnarray}%
The same approximation is valid for $f<0$ and we find for this case:%
\begin{equation*}
\frac{\nabla _{\hat{X}}^{2}\hat{\Psi}_{\lambda ,C}^{\left( 0\right) }}{\hat{%
\Psi}_{\lambda ,C}^{\left( 0\right) }}\simeq \left( \frac{\left( 4\hat{K}%
^{2}+\left( \hat{K}+\frac{\sigma _{\hat{K}}^{2}F\left( \hat{X},K_{\hat{X}%
}\right) }{f^{2}\left( \hat{X}\right) }\right) ^{2}\right) f^{\prime }\left(
X\right) }{4\sigma _{\hat{K}}^{2}}\right) ^{2}
\end{equation*}%
Then, introducing $\mp $ to account for the sign of $-f$, (\ref{drg})
becomes:%
\begin{eqnarray}
&\nabla _{\hat{K}}\left( h\left( K,X\right) \right) =&-\frac{1}{\sigma _{%
\hat{K}}^{2}\left( \hat{\Psi}_{\lambda ,C}^{\left( 0\right) }\right) ^{2}}%
\exp \left( \frac{\hat{K}^{2}}{\sigma _{\hat{K}}^{2}}f\left( \hat{X}\right)
\right) \int \left( \frac{\left( 4\hat{K}^{2}\mp \left( \hat{K}+\frac{\sigma
_{\hat{K}}^{2}F\left( \hat{X},K_{\hat{X}}\right) }{f^{2}\left( \hat{X}%
\right) }\right) ^{2}\right) f^{\prime }\left( X\right) }{4\sigma _{\hat{K}%
}^{2}}\right) ^{2} \\
&&\times \left( \hat{\Psi}_{\lambda ,C}^{\left( 0\right) }\right) ^{2}\exp
\left( -\frac{\hat{K}^{2}}{\sigma _{\hat{K}}^{2}}f\left( \hat{X}\right)
\right) d\hat{K}  \notag \\
&\simeq &-\frac{1}{\sigma _{\hat{K}}^{2}\left( \hat{\Psi}_{\lambda
,C}^{\left( 0\right) }\right) ^{2}}\exp \left( \frac{\hat{K}^{2}}{\sigma _{%
\hat{K}}^{2}}f\left( \hat{X}\right) \right) \int \left( \frac{\left( 4\hat{K}%
^{2}\mp \left( \hat{K}+\frac{\sigma _{\hat{K}}^{2}F\left( \hat{X},K_{\hat{X}%
}\right) }{f^{2}\left( \hat{X}\right) }\right) ^{2}\right) f^{\prime }\left(
X\right) }{4\sigma _{\hat{K}}^{2}}\right) ^{2}  \notag \\
&&\times \exp \left( \frac{\hat{K}^{2}f\left( \hat{X}\right) -\frac{1}{2}%
\left( \hat{K}+\frac{\sigma _{\hat{K}}^{2}F\left( \hat{X},K_{\hat{X}}\right) 
}{f^{2}\left( \hat{X}\right) }\right) ^{2}\left\vert f\left( \hat{X}\right)
\right\vert }{\sigma _{\hat{K}}^{2}}\right) d\hat{K} \\
&\simeq &-\frac{1}{\sigma _{\hat{K}}^{2}\left( \hat{\Psi}_{\lambda
,C}^{\left( 0\right) }\right) ^{2}}\exp \left( \frac{\hat{K}^{2}}{\sigma _{%
\hat{K}}^{2}}f\left( \hat{X}\right) \right) \int \left( \sigma _{\hat{K}}^{2}%
\frac{\left( \hat{K}^{2}\mp \frac{1}{4}\left( \hat{K}+\frac{\sigma _{\hat{K}%
}^{2}F\left( \hat{X},K_{\hat{X}}\right) }{f^{2}\left( \hat{X}\right) }%
\right) ^{2}\right) f^{\prime }\left( X\right) }{\left( 2\hat{K}f\left( \hat{%
X}\right) -\left( \hat{K}+\frac{\sigma _{\hat{K}}^{2}F\left( \hat{X},K_{\hat{%
X}}\right) }{f^{2}\left( \hat{X}\right) }\right) \left\vert f\left( \hat{X}%
\right) \right\vert \right) ^{2}}\right) ^{2}  \notag \\
&&\times \partial _{\hat{K}}^{4}\exp \left( \frac{\hat{K}^{2}f\left( \hat{X}%
\right) -\frac{1}{2}\left( \hat{K}+\frac{\sigma _{\hat{K}}^{2}F\left( \hat{X}%
,K_{\hat{X}}\right) }{f^{2}\left( \hat{X}\right) }\right) ^{2}\left\vert
f\left( \hat{X}\right) \right\vert }{\sigma _{\hat{K}}^{2}}\right) d\hat{K}
\end{eqnarray}%
Assuming $\frac{\sigma _{\hat{K}}^{2}F\left( \hat{X},K_{\hat{X}}\right) }{%
f^{2}\left( \hat{X}\right) }<<1$, we have ultimately:%
\begin{eqnarray*}
\nabla _{\hat{K}}\left( h\left( K,X\right) \right) &\simeq &-\frac{1}{\sigma
_{\hat{K}}^{2}\left( \hat{\Psi}_{\lambda ,C}^{\left( 0\right) }\right) ^{2}}%
\exp \left( \frac{\hat{K}^{2}}{\sigma _{\hat{K}}^{2}}f\left( \hat{X}\right)
\right) \left( \sigma _{\hat{K}}^{2}\frac{\left( \hat{K}^{2}\mp \frac{1}{4}%
\left( \hat{K}+\frac{\sigma _{\hat{K}}^{2}F\left( \hat{X},K_{\hat{X}}\right) 
}{f^{2}\left( \hat{X}\right) }\right) ^{2}\right) f^{\prime }\left( X\right) 
}{2\hat{K}f\left( \hat{X}\right) -\left( \hat{K}+\frac{\sigma _{\hat{K}%
}^{2}F\left( \hat{X},K_{\hat{X}}\right) }{f^{2}\left( \hat{X}\right) }%
\right) \left\vert f\left( \hat{X}\right) \right\vert }\right) ^{2} \\
&&\times \partial _{\hat{K}}^{3}\exp \left( \frac{\hat{K}^{2}f\left( \hat{X}%
\right) -\frac{1}{2}\left( \hat{K}+\frac{\sigma _{\hat{K}}^{2}F\left( \hat{X}%
,K_{\hat{X}}\right) }{f^{2}\left( \hat{X}\right) }\right) ^{2}\left\vert
f\left( \hat{X}\right) \right\vert }{\sigma _{\hat{K}}^{2}}\right) \\
&=&-\frac{\left( \frac{\left( \hat{K}^{2}-\frac{1}{4}\left( \hat{K}+\frac{%
\sigma _{\hat{K}}^{2}F\left( \hat{X},K_{\hat{X}}\right) }{f^{2}\left( \hat{X}%
\right) }\right) ^{2}\right) f^{\prime }\left( X\right) }{\sigma _{\hat{K}%
}^{2}}\right) ^{2}}{2\hat{K}f\left( \hat{X}\right) -\left( \hat{K}+\frac{%
\sigma _{\hat{K}}^{2}F\left( \hat{X},K_{\hat{X}}\right) }{f^{2}\left( \hat{X}%
\right) }\right) \left\vert f\left( \hat{X}\right) \right\vert } \\
&=&-\frac{\left( \left( \hat{K}^{2}\mp \frac{1}{4}\left( \hat{K}+\frac{%
\sigma _{\hat{K}}^{2}F\left( \hat{X},K_{\hat{X}}\right) }{f^{2}\left( \hat{X}%
\right) }\right) ^{2}\right) f^{\prime }\left( X\right) \right) ^{2}}{\left(
\sigma _{\hat{K}}^{2}\right) ^{2}\left( 2\hat{K}f\left( \hat{X}\right)
-\left( \hat{K}+\frac{\sigma _{\hat{K}}^{2}F\left( \hat{X},K_{\hat{X}%
}\right) }{f^{2}\left( \hat{X}\right) }\right) \left\vert f\left( \hat{X}%
\right) \right\vert \right) }
\end{eqnarray*}

Replacing in first approximation $\hat{K}$ by $\frac{\left\Vert \Psi \left( 
\hat{X}\right) \right\Vert ^{2}\hat{K}_{\hat{X}}}{\left\Vert \hat{\Psi}%
\left( \hat{X}\right) \right\Vert ^{2}}$ in (\ref{dvt}), and using (\ref{drg}%
) and (\ref{cps}) leads to: 
\begin{equation*}
\hat{\Psi}^{\left( 1\right) }\left( \hat{X}\right) =\sqrt{C}\exp \left(
-\int \frac{\left( \left( \hat{K}^{2}\mp \frac{1}{4}\left( \hat{K}+\frac{%
\sigma _{\hat{K}}^{2}F\left( \hat{X},K_{\hat{X}}\right) }{f^{2}\left( \hat{X}%
\right) }\right) ^{2}\right) f^{\prime }\left( X\right) \right) ^{2}}{\left(
\sigma _{\hat{K}}^{2}\right) ^{2}\left( 2\hat{K}f\left( \hat{X}\right)
-\left( \hat{K}+\frac{\sigma _{\hat{K}}^{2}F\left( \hat{X},K_{\hat{X}%
}\right) }{f^{2}\left( \hat{X}\right) }\right) \left\vert f\left( \hat{X}%
\right) \right\vert \right) }d\hat{K}\right)
\end{equation*}%
with $C$ a constant to be computed using the normalization condition.

To find $\Psi ^{\dag }$, we need also $\hat{\Psi}^{\left( 1\right) \dag }$.
Writing:%
\begin{equation*}
\hat{\Psi}^{\left( 1\right) \dag }=\exp \left( \sigma _{X}^{2}g\left(
K,X\right) \right)
\end{equation*}%
with a function $g\left( K,X\right) $ that satisfies:%
\begin{equation*}
\frac{\nabla _{\hat{X}}^{2}\hat{\Psi}_{\lambda ,C}^{\left( 0\right) \dag }}{%
\hat{\Psi}_{\lambda ,C}^{\left( 0\right) }}+\sigma _{\hat{K}}^{2}\nabla _{%
\hat{K}}^{2}g\left( \hat{K},\hat{X}\right) +2\left( \nabla _{\hat{K}}g\left( 
\hat{K},\hat{X}\right) \right) \left( \sigma _{\hat{K}}^{2}\frac{\nabla _{%
\hat{K}}\hat{\Psi}_{\lambda ,C}^{\left( 0\right) \dag }}{\hat{\Psi}_{\lambda
,C}^{\left( 0\right) }}+\hat{K}f\left( \hat{X}\right) \right) =0
\end{equation*}%
with:%
\begin{equation*}
\hat{\Psi}_{\lambda ,C}^{\left( 0\right) \dag }=\exp \left( -\frac{\hat{K}%
^{2}}{\sigma _{\hat{K}}^{2}}f\left( \hat{X}\right) \right) D_{p\left( \hat{X}%
,\hat{\lambda}\right) }\left( \left( \frac{\left\vert f\left( \hat{X}\right)
\right\vert }{\sigma _{\hat{K}}^{2}}\right) ^{\frac{1}{2}}\left( \hat{K}+%
\frac{\sigma _{\hat{K}}^{2}F\left( \hat{X},K_{\hat{X}}\right) }{f^{2}\left( 
\hat{X}\right) }\right) \right)
\end{equation*}%
we find:%
\begin{equation*}
\nabla _{\hat{K}}\left( g\left( K,X\right) \right) =-\frac{\nabla _{\hat{X}%
}^{2}\hat{\Psi}_{\lambda ,C}^{\left( 0\right) \dag }}{\hat{\Psi}_{\lambda
,C}^{\left( 0\right) }}\exp \left( -\frac{\hat{K}^{2}}{\sigma _{\hat{K}}^{2}}%
f\left( \hat{X}\right) \right) \left( \int \frac{\nabla _{\hat{X}}^{2}\hat{%
\Psi}_{\lambda ,C}^{\left( 0\right) \dag }}{\hat{\Psi}_{\lambda ,C}^{\left(
0\right) \dag }}\left( \hat{\Psi}_{\lambda ,C}^{\left( 0\right) \dag
}\right) ^{2}\exp \left( \frac{\hat{K}^{2}}{\sigma _{\hat{K}}^{2}}f\left( 
\hat{X}\right) \right) d\hat{K}\right)
\end{equation*}%
and:

\begin{equation*}
\hat{\Psi}^{\left( 1\right) \dag }\left( \hat{X}\right) =\sqrt{C}\exp \left(
\int \frac{\left( \left( \hat{K}^{2}\pm \frac{1}{4}\left( \hat{K}+\frac{%
\sigma _{\hat{K}}^{2}F\left( \hat{X},K_{\hat{X}}\right) }{f^{2}\left( \hat{X}%
\right) }\right) ^{2}\right) f^{\prime }\left( X\right) \right) ^{2}}{\sigma
_{\hat{K}}^{2}\left( 2\hat{K}f\left( \hat{X}\right) +\left( \hat{K}+\frac{%
\sigma _{\hat{K}}^{2}F\left( \hat{X},K_{\hat{X}}\right) }{f^{2}\left( \hat{X}%
\right) }\right) \left\vert f\left( \hat{X}\right) \right\vert \right) }d%
\hat{K}\right)
\end{equation*}%
where $\pm $ accounts for the sign of $f$.\ 

Ultimately, coming back to the initial definition of the fields we obtain
for $\hat{\Psi}_{\lambda ,C}\left( \hat{K},\hat{X}\right) $ and $\hat{\Psi}%
_{\lambda ,C}^{\dag }\left( \hat{K},\hat{X}\right) $:%
\begin{eqnarray*}
\hat{\Psi}_{\lambda ,C}\left( \hat{K},\hat{X}\right) &=&\sqrt{C}\exp \left(
-\sigma _{X}^{2}\int \frac{\left( \left( \hat{K}^{2}\mp \frac{1}{4}\left( 
\hat{K}+\frac{\sigma _{\hat{K}}^{2}F\left( \hat{X},K_{\hat{X}}\right) }{%
f^{2}\left( \hat{X}\right) }\right) ^{2}\right) f^{\prime }\left( X\right)
\right) ^{2}}{\left( \sigma _{\hat{K}}^{2}\right) ^{2}\left( 2\hat{K}f\left( 
\hat{X}\right) -\left( \hat{K}+\frac{\sigma _{\hat{K}}^{2}F\left( \hat{X},K_{%
\hat{X}}\right) }{f^{2}\left( \hat{X}\right) }\right) \left\vert f\left( 
\hat{X}\right) \right\vert \right) }d\hat{K}\right) \\
&&\times \exp \left( \frac{1}{\sigma _{\hat{X}}^{2}}\int g\left( \hat{X}%
\right) d\hat{X}+\frac{\hat{K}^{2}}{\sigma _{\hat{K}}^{2}}f\left( \hat{X}%
\right) \right) D_{p\left( \hat{X},\hat{\lambda}\right) }\left( \hat{K}%
\left( \frac{\left\vert f\left( \hat{X}\right) \right\vert }{\sigma _{\hat{K}%
}^{2}}\right) ^{\frac{1}{2}}\right) \\
\hat{\Psi}_{\lambda ,C}^{\dag }\left( \hat{K},\hat{X}\right) &=&\sqrt{C}\exp
\left( \sigma _{X}^{2}\int \frac{\left( \left( \hat{K}^{2}\pm \frac{1}{4}%
\left( \hat{K}+\frac{\sigma _{\hat{K}}^{2}F\left( \hat{X},K_{\hat{X}}\right) 
}{f^{2}\left( \hat{X}\right) }\right) ^{2}\right) f^{\prime }\left( X\right)
\right) ^{2}}{\left( \sigma _{\hat{K}}^{2}\right) ^{2}\left( 2\hat{K}f\left( 
\hat{X}\right) +\left( \hat{K}+\frac{\sigma _{\hat{K}}^{2}F\left( \hat{X},K_{%
\hat{X}}\right) }{f^{2}\left( \hat{X}\right) }\right) \left\vert f\left( 
\hat{X}\right) \right\vert \right) }d\hat{K}\right) \\
&&\times \exp \left( -\left( \frac{1}{\sigma _{\hat{X}}^{2}}\int g\left( 
\hat{X}\right) d\hat{X}+\frac{\hat{K}^{2}}{\sigma _{\hat{K}}^{2}}f\left( 
\hat{X}\right) \right) \right) D_{p\left( \hat{X},\hat{\lambda}\right)
}\left( \hat{K}\left( \frac{\left\vert f\left( \hat{X}\right) \right\vert }{%
\sigma _{\hat{K}}^{2}}\right) ^{\frac{1}{2}}\right)
\end{eqnarray*}

\paragraph{A3.1.3.3 Computation of $\left\Vert \hat{\Psi}\left( \hat{K},\hat{%
X}\right) \right\Vert ^{2}$}

As a consequence of the previsous result, we can compute $\left\Vert \hat{%
\Psi}_{\lambda ,C}\left( \hat{K},\hat{X}\right) \right\Vert ^{2}$. We start
with $\hat{\Psi}^{\left( 1\right) \dag }\hat{\Psi}^{\left( 1\right) }$. We
have: 
\begin{eqnarray*}
\hat{\Psi}^{\left( 1\right) \dag }\hat{\Psi}^{\left( 1\right) } &=&C\exp
\left( -\sigma _{X}^{2}\int \left( \frac{\left( \left( \hat{K}^{2}\mp \frac{1%
}{4}\left( \hat{K}+\frac{\sigma _{\hat{K}}^{2}F\left( \hat{X},K_{\hat{X}%
}\right) }{f^{2}\left( \hat{X}\right) }\right) ^{2}\right) f^{\prime }\left(
X\right) \right) ^{2}}{\left( \sigma _{\hat{K}}^{2}\right) ^{2}\left( 2\hat{K%
}f\left( \hat{X}\right) -\left( \hat{K}+\frac{\sigma _{\hat{K}}^{2}F\left( 
\hat{X},K_{\hat{X}}\right) }{f^{2}\left( \hat{X}\right) }\right) \left\vert
f\left( \hat{X}\right) \right\vert \right) }\right. \right. \\
&&\left. \left. -\frac{\left( \left( \hat{K}^{2}\pm \frac{1}{4}\left( \hat{K}%
+\frac{\sigma _{\hat{K}}^{2}F\left( \hat{X},K_{\hat{X}}\right) }{f^{2}\left( 
\hat{X}\right) }\right) ^{2}\right) f^{\prime }\left( X\right) \right) ^{2}}{%
\left( \sigma _{\hat{K}}^{2}\right) ^{2}\left( 2\hat{K}f\left( \hat{X}%
\right) +\left( \hat{K}+\frac{\sigma _{\hat{K}}^{2}F\left( \hat{X},K_{\hat{X}%
}\right) }{f^{2}\left( \hat{X}\right) }\right) \left\vert f\left( \hat{X}%
\right) \right\vert \right) }d\hat{K}\right) \right) \\
&=&C\exp \left( -\sigma _{X}^{2}\int \left( \frac{\left( \left( \hat{K}^{2}-%
\frac{1}{4}\left( \hat{K}+\frac{\sigma _{\hat{K}}^{2}F\left( \hat{X},K_{\hat{%
X}}\right) }{f^{2}\left( \hat{X}\right) }\right) ^{2}\right) f^{\prime
}\left( X\right) \right) ^{2}}{\sigma _{\hat{K}}^{2}\left( 2\hat{K}%
\left\vert f\left( \hat{X}\right) \right\vert -\left( \hat{K}+\frac{\sigma _{%
\hat{K}}^{2}F\left( \hat{X},K_{\hat{X}}\right) }{f^{2}\left( \hat{X}\right) }%
\right) \left\vert f\left( \hat{X}\right) \right\vert \right) }\right.
\right. \\
&&-\left. \left. -\frac{\left( \left( \hat{K}^{2}+\frac{1}{4}\left( \hat{K}+%
\frac{\sigma _{\hat{K}}^{2}F\left( \hat{X},K_{\hat{X}}\right) }{f^{2}\left( 
\hat{X}\right) }\right) ^{2}\right) f^{\prime }\left( X\right) \right) ^{2}}{%
\left( \sigma _{\hat{K}}^{2}\right) ^{2}\left( 2\hat{K}\left\vert f\left( 
\hat{X}\right) \right\vert +\left( \hat{K}+\frac{\sigma _{\hat{K}%
}^{2}F\left( \hat{X},K_{\hat{X}}\right) }{f^{2}\left( \hat{X}\right) }%
\right) \left\vert f\left( \hat{X}\right) \right\vert \right) }d\hat{K}%
\right) \right)
\end{eqnarray*}%
And for $\frac{\sigma _{\hat{K}}^{2}F\left( \hat{X},K_{\hat{X}}\right) }{%
f^{2}\left( \hat{X}\right) }<<1$:%
\begin{eqnarray*}
\hat{\Psi}^{\left( 1\right) \dag }\hat{\Psi}^{\left( 1\right) } &\simeq
&C\exp \left( -\sigma _{X}^{2}\int \left( \frac{\left( \frac{3}{4}\hat{K}%
^{2}f^{\prime }\left( X\right) \right) ^{2}}{\left( \sigma _{\hat{K}%
}^{2}\right) ^{2}\hat{K}\left\vert f\left( \hat{X}\right) \right\vert }-%
\frac{\left( \frac{5}{4}\hat{K}^{2}f^{\prime }\left( X\right) \right) ^{2}}{%
3\left( \sigma _{\hat{K}}^{2}\right) ^{2}\hat{K}f\left( \hat{X}\right) }%
\right) d\hat{K}\right) \\
&=&C\exp \left( -\frac{\sigma _{X}^{2}\hat{K}^{4}\left( f^{\prime }\left(
X\right) \right) ^{2}}{96\left( \sigma _{\hat{K}}^{2}\right) ^{2}\left\vert
f\left( \hat{X}\right) \right\vert }\right)
\end{eqnarray*}%
Gathering the previous results, we obtain the norm of $\left\Vert \hat{\Psi}%
_{\lambda ,C}\left( \hat{K},\hat{X}\right) \right\Vert ^{2}$:%
\begin{equation}
\left\Vert \hat{\Psi}_{\lambda ,C}\left( \hat{K},\hat{X}\right) \right\Vert
^{2}\simeq C\exp \left( -\frac{\sigma _{X}^{2}\hat{K}^{4}\left( f^{\prime
}\left( X\right) \right) ^{2}}{96\left( \sigma _{\hat{K}}^{2}\right)
^{2}\left\vert f\left( \hat{X}\right) \right\vert }\right) D_{p\left( \hat{X}%
,\hat{\lambda}\right) }^{2}\left( \left( \frac{\left\vert f\left( \hat{X}%
\right) \right\vert }{\sigma _{\hat{K}}^{2}}\right) ^{\frac{1}{2}}\left( 
\hat{K}+\frac{\sigma _{\hat{K}}^{2}F\left( \hat{X},K_{\hat{X}}\right) }{%
f^{2}\left( \hat{X}\right) }\right) \right)  \label{spc}
\end{equation}

with:%
\begin{eqnarray}
f\left( \hat{X},K_{\hat{X}}\right) &=&\left( r\left( K_{\hat{X}},\hat{X}%
\right) -\gamma \left\Vert \Psi \left( \hat{X}\right) \right\Vert
^{2}+F_{1}\left( \frac{R\left( K_{\hat{X}},\hat{X}\right) }{\int R\left(
K_{X^{\prime }}^{\prime },X^{\prime }\right) \left\Vert \Psi \left(
X^{\prime }\right) \right\Vert ^{2}dX^{\prime }}\right) \right)  \label{ftf}
\\
g\left( \hat{X},K_{\hat{X}}\right) &=&\left( \frac{\nabla _{\hat{X}%
}F_{0}\left( R\left( K_{\hat{X}},\hat{X}\right) \right) }{\left\Vert \nabla
_{\hat{X}}R\left( K_{\hat{X}},\hat{X}\right) \right\Vert }+\nu \nabla _{\hat{%
X}}F_{1}\left( \frac{R\left( K_{\hat{X}},\hat{X}\right) }{\int R\left(
K_{X^{\prime }}^{\prime },X^{\prime }\right) \left\Vert \Psi \left(
X^{\prime }\right) \right\Vert ^{2}dX^{\prime }}\right) \right)  \label{gft}
\end{eqnarray}%
The solutions are parametrized by $C$ and $\hat{\lambda}$ and $\hat{K}_{\hat{%
X}}$. Using the constraint $\left\Vert \hat{\Psi}\left( \hat{K},\hat{X}%
\right) \right\Vert ^{2}=\hat{N}$ will reduce the solutions to a
one-parameter set of solutions. The computation of the average capital over
this set will lead to the defining equation for $\hat{K}_{\hat{X}}$.

Replacing in first approximation $\hat{K}$ by its average $\frac{\left\Vert
\Psi \left( \hat{X}\right) \right\Vert ^{2}\hat{K}_{\hat{X}}}{\left\Vert 
\hat{\Psi}\left( \hat{X}\right) \right\Vert ^{2}}$ in the first term yields:%
\begin{equation}
\left\Vert \hat{\Psi}_{\lambda ,C}\left( \hat{K},\hat{X}\right) \right\Vert
^{2}\simeq C\exp \left( -\frac{\sigma _{X}^{2}\left( \frac{\left\Vert \Psi
\left( \hat{X}\right) \right\Vert ^{2}\hat{K}_{\hat{X}}}{\left\Vert \hat{\Psi%
}\left( \hat{X}\right) \right\Vert ^{2}}\right) ^{4}\left( f^{\prime }\left(
X\right) \right) ^{2}}{96\left( \sigma _{\hat{K}}^{2}\right) ^{2}\left\vert
f\left( \hat{X}\right) \right\vert }\right) D_{p\left( \hat{X},\hat{\lambda}%
\right) }^{2}\left( \left( \frac{\left\vert f\left( \hat{X}\right)
\right\vert }{\sigma _{\hat{K}}^{2}}\right) ^{\frac{1}{2}}\left( \hat{K}+%
\frac{\sigma _{\hat{K}}^{2}F\left( \hat{X},K_{\hat{X}}\right) }{f^{2}\left( 
\hat{X}\right) }\right) \right)  \label{bgH}
\end{equation}

\subsubsection*{A3.1.4 Estimation of $\ S_{3}\left( \hat{\Psi}_{\hat{\protect%
\lambda}}\left( \hat{K},\hat{X}\right) \right) +S_{4}\left( \hat{\Psi}_{\hat{%
\protect\lambda}}\left( \hat{K},\hat{X}\right) \right) $}

For later purposes, we compute an estimation of $S_{3}\left( \hat{\Psi}_{%
\hat{\lambda}}\left( \hat{K},\hat{X}\right) \right) +S_{4}\left( \hat{\Psi}_{%
\hat{\lambda}}\left( \hat{K},\hat{X}\right) \right) $ for any background
field $\hat{\Psi}_{\hat{\lambda}}\left( \hat{K},\hat{X}\right) $. We
multiply (\ref{hqn})by $\hat{\Psi}_{\hat{\lambda}}^{\dagger }\left( \hat{K},%
\hat{X}\right) $ on the left and integrate the equation over $\hat{K}$ and $%
\hat{X}$. It yields:%
\begin{equation*}
0=S_{3}\left( \hat{\Psi}_{\hat{\lambda}}\left( \hat{K},\hat{X}\right)
\right) +S_{4}\left( \hat{\Psi}_{\hat{\lambda}}\left( \hat{K},\hat{X}\right)
\right) -\hat{\lambda}\int \left\Vert \hat{\Psi}_{\hat{\lambda}}\left( \hat{K%
},\hat{X}\right) \right\Vert ^{2}d\hat{K}d\hat{X}-\int F\left( \hat{X},K_{%
\hat{X}}\right) \hat{K}\left\Vert \hat{\Psi}_{\hat{\lambda}}\left( \hat{K},%
\hat{X}\right) \right\Vert ^{2}d\hat{K}d\hat{X}
\end{equation*}

Using the constraint about the number of investors:%
\begin{equation*}
\int \left\Vert \hat{\Psi}_{\hat{\lambda}}\left( \hat{K},\hat{X}\right)
\right\Vert ^{2}d\hat{K}=\hat{N}
\end{equation*}%
we find:%
\begin{equation*}
S_{3}\left( \hat{\Psi}_{\hat{\lambda}}\left( \hat{K},\hat{X}\right) \right)
+S_{4}\left( \hat{\Psi}_{\hat{\lambda}}\left( \hat{K},\hat{X}\right) \right)
=\hat{\lambda}\hat{N}+\int F\left( \hat{X},K_{\hat{X}}\right) \hat{K}%
\left\Vert \hat{\Psi}_{\hat{\lambda}}\left( \hat{K},\hat{X}\right)
\right\Vert ^{2}d\hat{K}d\hat{X}
\end{equation*}%
Moreover, equation (\ref{Fct}) implies\footnote{%
All averages in the next formula are computed in state $\hat{\Psi}_{\hat{%
\lambda}}\left( \hat{K},\hat{X}\right) $.
\par
{}}:%
\begin{eqnarray}
&&\int F\left( \hat{X},K_{\hat{X}}\right) \hat{K}\left\Vert \hat{\Psi}_{\hat{%
\lambda}}\left( \hat{K},\hat{X}\right) \right\Vert ^{2}d\hat{K}d\hat{X} \\
&=&\int K_{\hat{X}}\nabla _{K_{\hat{X}}}\left( \frac{\left( g\left( \hat{X}%
,K_{\hat{X}}\right) \right) ^{2}}{2\sigma _{\hat{X}}^{2}}+\frac{1}{2}\nabla
_{\hat{X}}g\left( \hat{X},K_{\hat{X}}\right) +f\left( \hat{X},K_{\hat{X}%
}\right) \right) \left\Vert \hat{\Psi}\left( \hat{X}\right) \right\Vert ^{2}d%
\hat{X}  \notag \\
&&+\int K_{\hat{X}}\frac{\nabla _{K_{\hat{X}}}f^{2}\left( \hat{X},K_{\hat{X}%
}\right) }{\sigma _{\hat{K}}^{2}}\left\langle \hat{K}^{2}\right\rangle _{%
\hat{X}}d\hat{X}  \notag
\end{eqnarray}%
In our applications the involved functions are roughly power functions in $%
K_{\hat{X}}$, and consequently, the integral $\int F\left( \hat{X},K_{\hat{X}%
}\right) \hat{K}\left\Vert \hat{\Psi}_{\hat{\lambda}}\left( \hat{K},\hat{X}%
\right) \right\Vert ^{2}d\hat{K}d\hat{X}$ is of order:%
\begin{equation}
\int \left( \frac{\left( g\left( \hat{X},K_{\hat{X}}\right) \right) ^{2}}{%
2\sigma _{\hat{X}}^{2}}+\frac{1}{2}\nabla _{\hat{X}}g\left( \hat{X},K_{\hat{X%
}}\right) +f\left( \hat{X},K_{\hat{X}}\right) \right) \left\Vert \hat{\Psi}%
\left( \hat{X}\right) \right\Vert ^{2}d\hat{X}+\int \frac{f^{2}\left( \hat{X}%
,K_{\hat{X}}\right) }{\sigma _{\hat{K}}^{2}}\left\langle \hat{K}%
^{2}\right\rangle _{\hat{X}}d\hat{X}
\end{equation}%
Since $\left\langle \hat{K}^{2}\right\rangle _{\hat{X}}\simeq K_{\hat{X}}^{2}%
\frac{\left\Vert \Psi \left( \hat{X}\right) \right\Vert ^{2}}{\left\Vert 
\hat{\Psi}\left( \hat{X}\right) \right\Vert ^{2}}$, the second term in (\ref%
{ntrM}) is negligible if we assume $\frac{\left\Vert \Psi \left( \hat{X}%
\right) \right\Vert ^{2}}{\left\Vert \hat{\Psi}\left( \hat{X}\right)
\right\Vert ^{2}}<<1$, i.e. the number of firms is smaller than the number
of investors. Consequently, (\ref{ntrM}) reduces to:%
\begin{eqnarray*}
\int \left( \frac{\left( g\left( \hat{X},K_{\hat{X}}\right) \right) ^{2}}{%
2\sigma _{\hat{X}}^{2}}+\frac{1}{2}\nabla _{\hat{X}}g\left( \hat{X},K_{\hat{X%
}}\right) +f\left( \hat{X},K_{\hat{X}}\right) \right) \left\Vert \hat{\Psi}%
\left( \hat{X}\right) \right\Vert ^{2}d\hat{X} &\lesssim &\int M\left\Vert 
\hat{\Psi}\left( \hat{X}\right) \right\Vert ^{2}d\hat{X} \\
&=&M\hat{N}
\end{eqnarray*}%
where $M$ is the lowest bound for $\left\vert \hat{\lambda}\right\vert $,
computed below in (\ref{mdf}) and (\ref{mdF}). Our previous estimation
relies on $\frac{\sigma _{\hat{K}}^{2}F^{2}\left( \hat{X},K_{\hat{X}}\right) 
}{2f^{2}\left( \hat{X}\right) }<<1$,which is true for $f^{2}\left( \hat{X}%
\right) >>1$. As a consequence:%
\begin{equation}
S_{3}\left( \hat{\Psi}_{\hat{\lambda}}\left( \hat{K},\hat{X}\right) \right)
+S_{4}\left( \hat{\Psi}_{\hat{\lambda}}\left( \hat{K},\hat{X}\right) \right)
=\left( \hat{\lambda}+M\right) \hat{N}=-\left( \left\vert \hat{\lambda}%
\right\vert -M\right) \hat{N}  \label{stT}
\end{equation}

\subsubsection*{A3.1.4 \textbf{Identification of} $K_{\hat{X}}$ and $%
\left\Vert \Psi \left( \hat{X}\right) \right\Vert ^{2}$:}

\paragraph*{A3.1.4.1 Formula depending on $\hat{\protect\lambda}$ and $C$}

In this paragraph, we compute the average capital $K_{\hat{X}}$ and the
density of investors $\left\Vert \hat{\Psi}\left( \hat{X}\right) \right\Vert
^{2}$ at $\hat{X}$ that are defined by using (\ref{Kx}): 
\begin{eqnarray}
K_{\hat{X}}\left\Vert \Psi \left( \hat{X}\right) \right\Vert ^{2}
&=&\int_{0}^{\infty }\hat{K}C\exp \left( -\frac{\sigma _{X}^{2}u\left( \hat{X%
},\hat{K}_{\hat{X}}\right) }{\left( \sigma _{\hat{K}}^{2}\right) ^{2}}\right)
\label{nmk} \\
&&\times D_{p\left( \hat{X},\hat{\lambda}\right) }^{2}\left( \left( \frac{%
\left\vert f\left( \hat{X}\right) \right\vert }{\sigma _{\hat{K}}^{2}}%
\right) ^{\frac{1}{2}}\left( \hat{K}+\frac{\sigma _{\hat{K}}^{2}F\left( \hat{%
X},K_{\hat{X}}\right) }{f^{2}\left( \hat{X}\right) }\right) \right) d\hat{K}
\notag
\end{eqnarray}%
and:%
\begin{eqnarray*}
\left\Vert \hat{\Psi}\left( \hat{X}\right) \right\Vert ^{2}
&=&C\int_{0}^{\infty }\exp \left( -\frac{\sigma _{X}^{2}u\left( \hat{X},\hat{%
K}_{\hat{X}}\right) }{\left( \sigma _{\hat{K}}^{2}\right) ^{2}}\right) \\
&&\times D_{p\left( \hat{X},\hat{\lambda}\right) }^{2}\left( \left( \frac{%
\left\vert f\left( \hat{X}\right) \right\vert }{\sigma _{\hat{K}}^{2}}%
\right) ^{\frac{1}{2}}\left( \hat{K}+\frac{\sigma _{\hat{K}}^{2}F\left( \hat{%
X},K_{\hat{X}}\right) }{f^{2}\left( \hat{X}\right) }\right) \right) d\hat{K}
\end{eqnarray*}%
with:%
\begin{equation}
u\left( \hat{X},\hat{K}_{\hat{X}}\right) =\frac{\left( \frac{\left\Vert \Psi
\left( \hat{X}\right) \right\Vert ^{2}\hat{K}_{\hat{X}}}{\left\Vert \hat{\Psi%
}\left( \hat{X}\right) \right\Vert ^{2}}\right) ^{4}\left( f^{\prime }\left(
X\right) \right) ^{2}}{96\left\vert f\left( \hat{X}\right) \right\vert }
\label{dnl}
\end{equation}%
Note that in these formulas, $K_{\hat{X}}$ and $\left\Vert \hat{\Psi}\left( 
\hat{X}\right) \right\Vert ^{2}$ depend implicitely of $\hat{\lambda}$ since
they have been computed in the state defined by the background field $\hat{%
\Psi}_{\lambda ,C}\left( \hat{K},\hat{X}\right) $. In the sequel, for the
sake of simplicity, $\hat{\Psi}_{\lambda ,C}\left( \hat{K},\hat{X}\right) $,
the indices $\lambda $ and $C$ may be omitted.

We will also need $\frac{K_{\hat{X}}\left\Vert \Psi \left( \hat{X}\right)
\right\Vert ^{2}}{\left\Vert \hat{\Psi}\left( \hat{X}\right) \right\Vert ^{2}%
}$ that arises in (\ref{dnl}): 
\begin{equation*}
\frac{K_{\hat{X}}\left\Vert \Psi \left( \hat{X}\right) \right\Vert ^{2}}{%
\left\Vert \hat{\Psi}\left( \hat{X}\right) \right\Vert ^{2}}=\frac{%
\int_{0}^{\infty }\hat{K}D_{p\left( \hat{X},\hat{\lambda}\right) }^{2}\left(
\left( \frac{\left\vert f\left( \hat{X}\right) \right\vert }{\sigma _{\hat{K}%
}^{2}}\right) ^{\frac{1}{2}}\left( \hat{K}+\frac{\sigma _{\hat{K}%
}^{2}F\left( \hat{X},K_{\hat{X}}\right) }{f^{2}\left( \hat{X}\right) }%
\right) \right) d\hat{K}}{\int_{\frac{\sigma _{\hat{K}}^{2}F\left( \hat{X}%
,K_{\hat{X}}\right) }{f^{2}\left( \hat{X}\right) }}^{\infty }\hat{K}%
D_{p\left( \hat{X},\hat{\lambda}\right) }^{2}\left( \left( \frac{\left\vert
f\left( \hat{X}\right) \right\vert }{\sigma _{\hat{K}}^{2}}\right) ^{\frac{1%
}{2}}\hat{K}\right) d\hat{K}}
\end{equation*}%
By a change of variable $\hat{K}+\frac{\sigma _{\hat{K}}^{2}F\left( \hat{X}%
,K_{\hat{X}}\right) }{f^{2}\left( \hat{X}\right) }\rightarrow \hat{K}$ we
can also write: 
\begin{equation*}
K_{\hat{X}}\left\Vert \Psi \left( \hat{X}\right) \right\Vert ^{2}\simeq
C\exp \left( -\frac{\sigma _{X}^{2}u\left( \hat{X},\hat{K}_{\hat{X}}\right) 
}{\sigma _{\hat{K}}^{2}}\right) \int_{\frac{\sigma _{\hat{K}}^{2}F\left( 
\hat{X},K_{\hat{X}}\right) }{f^{2}\left( \hat{X}\right) }}^{\infty }\hat{K}%
D_{p\left( \hat{X},\hat{\lambda}\right) }^{2}\left( \left( \frac{\left\vert
f\left( \hat{X}\right) \right\vert }{\sigma _{\hat{K}}^{2}}\right) ^{\frac{1%
}{2}}\hat{K}\right) d\hat{K}
\end{equation*}%
\begin{equation*}
\left\Vert \hat{\Psi}\left( \hat{X}\right) \right\Vert ^{2}\simeq C\exp
\left( -\frac{\sigma _{X}^{2}u\left( \hat{X},\hat{K}_{\hat{X}}\right) }{%
16\sigma _{\hat{K}}^{2}}\right) \int_{\frac{\sigma _{\hat{K}}^{2}F\left( 
\hat{X},K_{\hat{X}}\right) }{f^{2}\left( \hat{X}\right) }}^{\infty
}D_{p\left( \hat{X},\hat{\lambda}\right) }^{2}\left( \left( \frac{\left\vert
f\left( \hat{X}\right) \right\vert }{\sigma _{\hat{K}}^{2}}\right) ^{\frac{1%
}{2}}\hat{K}\right) d\hat{K}
\end{equation*}%
and by a zeroth order expansion around $0$ of $\hat{K}D_{p\left( \hat{X},%
\hat{\lambda}\right) }^{2}$ and $D_{p\left( \hat{X},\hat{\lambda}\right)
}^{2}$ we have:%
\begin{equation}
K_{\hat{X}}\left\Vert \Psi \left( \hat{X}\right) \right\Vert ^{2}\simeq
C\exp \left( -\frac{\sigma _{X}^{2}u\left( \hat{X},\hat{K}_{\hat{X}}\right) 
}{16\sigma _{\hat{K}}^{2}}\right) \int_{0}^{\infty }\hat{K}D_{p\left( \hat{X}%
,\hat{\lambda}\right) }^{2}\left( \left( \frac{\left\vert f\left( \hat{X}%
\right) \right\vert }{\sigma _{\hat{K}}^{2}}\right) ^{\frac{1}{2}}\hat{K}%
\right) d\hat{K}  \label{mnk}
\end{equation}%
\begin{equation}
\left\Vert \hat{\Psi}\left( \hat{X}\right) \right\Vert ^{2}\simeq C\exp
\left( -\frac{\sigma _{X}^{2}u\left( \hat{X},\hat{K}_{\hat{X}}\right) }{%
16\sigma _{\hat{K}}^{2}}\right) \left( \int_{0}^{\infty }D_{p\left( \hat{X},%
\hat{\lambda}\right) }^{2}\left( \left( \frac{\left\vert f\left( \hat{X}%
\right) \right\vert }{\sigma _{\hat{K}}^{2}}\right) ^{\frac{1}{2}}\hat{K}%
\right) d\hat{K}-\frac{\left( \frac{\left\vert f\left( \hat{X}\right)
\right\vert }{\sigma _{\hat{K}}^{2}}\right) ^{-\frac{1}{2}}2^{\frac{p\left( 
\hat{X},\hat{\lambda}\right) }{2}}\sqrt{\pi }}{\Gamma \left( \frac{1-p\left( 
\hat{X},\hat{\lambda}\right) }{2}\right) }\frac{\sigma _{\hat{K}}^{2}F\left( 
\hat{X},K_{\hat{X}}\right) }{f^{2}\left( \hat{X}\right) }\right)  \label{nrp}
\end{equation}%
To compute $\left\Vert \hat{\Psi}\left( \hat{X}\right) \right\Vert ^{2}$ we
use that the function $D$ satisfies: 
\begin{equation*}
\int D_{p}^{2}=\frac{\sqrt{\pi }}{2^{\frac{3}{2}}}\frac{\func{Psi}\left( 
\frac{1}{2}-\frac{p}{2}\right) -\func{Psi}\left( -\frac{p}{2}\right) }{%
\Gamma \left( -p\right) }
\end{equation*}%
The computation of the norm implies a second change of variable $\hat{K}%
\rightarrow \hat{K}\left( \frac{\left\vert f\left( \hat{X}\right)
\right\vert }{\sigma _{\hat{K}}^{2}}\right) ^{\frac{1}{2}}$ and we obtain
for (\ref{nrp}):%
\begin{eqnarray}
&&\left\Vert \hat{\Psi}\left( \hat{X}\right) \right\Vert ^{2}=\int
\left\Vert \hat{\Psi}_{\lambda ,C}\left( \hat{K},\hat{X}\right) \right\Vert
^{2}d\hat{K}  \label{fmn} \\
&=&C\exp \left( -\frac{\sigma _{X}^{2}u\left( \hat{X},\hat{K}_{\hat{X}%
}\right) }{16\sigma _{\hat{K}}^{2}}\right) \left( \int D_{p\left( \hat{X},%
\hat{\lambda}\right) }^{2}\left( \hat{K}\left( f^{2}\left( \hat{X}\right)
\right) ^{\frac{1}{4}}\right) dK-\frac{\left( \frac{\left\vert f\left( \hat{X%
}\right) \right\vert }{\sigma _{\hat{K}}^{2}}\right) ^{-\frac{1}{2}}2^{\frac{%
p\left( \hat{X},\hat{\lambda}\right) }{2}}\sqrt{\pi }}{\Gamma \left( \frac{%
1-p\left( \hat{X},\hat{\lambda}\right) }{2}\right) }\frac{\sigma _{\hat{K}%
}^{2}F\left( \hat{X},K_{\hat{X}}\right) }{f^{2}\left( \hat{X}\right) }\right)
\notag \\
&=&C\exp \left( -\frac{\sigma _{X}^{2}u\left( \hat{X},\hat{K}_{\hat{X}%
}\right) }{16\sigma _{\hat{K}}^{2}}\right) \left( \frac{\left\vert f\left( 
\hat{X}\right) \right\vert }{\sigma _{\hat{K}}^{2}}\right) ^{-\frac{1}{2}} 
\notag \\
&&\times \left( \frac{\sqrt{\pi }}{2^{\frac{3}{2}}}\frac{\func{Psi}\left( 
\frac{1-p\left( \hat{X},\hat{\lambda}\right) }{2}\right) -\func{Psi}\left( -%
\frac{p\left( \hat{X},\hat{\lambda}\right) }{2}\right) }{\Gamma \left(
-p\left( \hat{X},\hat{\lambda}\right) \right) }-\frac{2^{\frac{p\left( \hat{X%
},\hat{\lambda}\right) }{2}}\sqrt{\pi }}{\Gamma \left( \frac{1-p\left( \hat{X%
},\hat{\lambda}\right) }{2}\right) }\frac{\sigma _{\hat{K}}^{2}F\left( \hat{X%
},K_{\hat{X}}\right) }{f^{2}\left( \hat{X}\right) }\right)  \notag
\end{eqnarray}%
Expression (\ref{mnk}) is computed using that:%
\begin{equation*}
\int_{0}^{\infty }zD_{p}^{2}\left( z\right) dz=\int_{0}^{\infty
}D_{p+1}\left( z\right) D_{p}\left( z\right) dz+p\int_{0}^{\infty
}D_{p-1}\left( z\right) D_{p}\left( z\right) dz
\end{equation*}%
\begin{equation*}
\int_{0}^{\infty }zD_{p}^{2}\left( z\right) dz=\frac{\Gamma \left( -\frac{p+1%
}{2}\right) \Gamma \left( \frac{1-p}{2}\right) -\Gamma \left( -\frac{p}{2}%
\right) \Gamma \left( -\frac{p}{2}\right) }{2^{p+2}\Gamma \left( -p-1\right)
\Gamma \left( -p\right) }+p\frac{\Gamma \left( -\frac{p}{2}\right) \Gamma
\left( \frac{2-p}{2}\right) -\Gamma \left( -\frac{p-1}{2}\right) \Gamma
\left( -\frac{p-1}{2}\right) }{2^{p+1}\Gamma \left( -p\right) \Gamma \left(
-p+1\right) }
\end{equation*}%
and:%
\begin{equation*}
\int \hat{K}D_{p\left( \hat{X},\hat{\lambda}\right) }^{2}\left( \hat{K}%
\left( f^{2}\left( \hat{X}\right) \right) ^{\frac{1}{4}}\right) =\left(
f\left( \hat{X}\right) \right) ^{-1}\int uD_{p\left( \hat{X},\hat{\lambda}%
\right) }^{2}\left( u\right)
\end{equation*}%
We obtain:%
\begin{eqnarray}
K_{\hat{X}}\left\Vert \Psi \left( \hat{X}\right) \right\Vert ^{2} &\simeq
&\exp \left( -\frac{\sigma _{X}^{2}u\left( \hat{X},\hat{K}_{\hat{X}}\right) 
}{16\left( \sigma _{\hat{K}}^{2}\right) ^{2}}\right) \left( \frac{\left\vert
f\left( \hat{X}\right) \right\vert }{\sigma _{\hat{K}}^{2}}\right) ^{-1}C
\label{knm} \\
&&\times \left( \frac{\Gamma \left( -\frac{p+1}{2}\right) \Gamma \left( 
\frac{1-p}{2}\right) -\Gamma \left( -\frac{p}{2}\right) \Gamma \left( \frac{%
-p}{2}\right) }{2^{p+2}\Gamma \left( -p-1\right) \Gamma \left( -p\right) }+p%
\frac{\Gamma \left( -\frac{p}{2}\right) \Gamma \left( \frac{2-p}{2}\right)
-\Gamma \left( -\frac{p-1}{2}\right) \Gamma \left( -\frac{p-1}{2}\right) }{%
2^{p+1}\Gamma \left( -p\right) \Gamma \left( -p+1\right) }\right)  \notag
\end{eqnarray}%
where:%
\begin{equation}
p=p\left( \hat{X},\hat{\lambda}\right)  \label{pft}
\end{equation}%
Ultimately we can compute $\frac{K_{\hat{X}}\left\Vert \Psi \left( \hat{X}%
\right) \right\Vert ^{2}}{\left\Vert \hat{\Psi}\left( \hat{X}\right)
\right\Vert ^{2}}$:%
\begin{eqnarray*}
\frac{K_{\hat{X}}\left\Vert \Psi \left( \hat{X}\right) \right\Vert ^{2}}{%
\left\Vert \hat{\Psi}\left( \hat{X}\right) \right\Vert ^{2}} &\simeq &\left( 
\frac{\left\vert f\left( \hat{X}\right) \right\vert }{\sigma _{\hat{K}}^{2}}%
\right) ^{-\frac{1}{2}}\frac{\frac{\Gamma \left( -\frac{p+1}{2}\right)
\Gamma \left( \frac{1-p}{2}\right) -\Gamma \left( -\frac{p}{2}\right) \Gamma
\left( \frac{-p}{2}\right) }{2^{p+2}\Gamma \left( -p-1\right) \Gamma \left(
-p\right) }+p\frac{\Gamma \left( -\frac{p}{2}\right) \Gamma \left( \frac{2-p%
}{2}\right) -\Gamma \left( -\frac{p-1}{2}\right) \Gamma \left( -\frac{p-1}{2}%
\right) }{2^{p+1}\Gamma \left( -p\right) \Gamma \left( -p+1\right) }}{\frac{%
\sqrt{\pi }}{2^{\frac{3}{2}}}\frac{\func{Psi}\left( \frac{1-p}{2}\right) -%
\func{Psi}\left( -\frac{p}{2}\right) }{\Gamma \left( -p\right) }} \\
&\equiv &\left( \frac{\left\vert f\left( \hat{X}\right) \right\vert }{\sigma
_{\hat{K}}^{2}}\right) ^{-\frac{1}{2}}h\left( p\right) \\
&\simeq &\left( \frac{\left\vert f\left( \hat{X}\right) \right\vert }{\sigma
_{\hat{K}}^{2}}\right) ^{-\frac{1}{2}}\sqrt{p+\frac{1}{2}}
\end{eqnarray*}%
so that:%
\begin{equation}
\exp \left( -\frac{\sigma _{X}^{2}u\left( \hat{X},\hat{K}_{\hat{X}}\right) }{%
\left( \sigma _{\hat{K}}^{2}\right) ^{2}}\right) \simeq \exp \left( -\frac{%
\sigma _{X}^{2}\left( p+\frac{1}{2}\right) ^{2}\left( f^{\prime }\left(
X\right) \right) ^{2}}{96\left\vert f\left( \hat{X}\right) \right\vert ^{3}}%
\right)  \label{fxp}
\end{equation}%
We end this section by finding asymptotic form for $\left\Vert \hat{\Psi}%
\left( \hat{X}\right) \right\Vert ^{2}$ and $K_{\hat{X}}\left\Vert \Psi
\left( \hat{X}\right) \right\Vert ^{2}$

For $\varepsilon <<1$ an asymptotic form yields that:%
\begin{equation}
D_{p\left( \hat{X},\hat{\lambda}\right) }\left( \hat{K}\left( f^{2}\left( 
\hat{X}\right) \right) ^{\frac{1}{4}}\right) \simeq \exp \left( -\frac{\hat{K%
}^{2}\left\vert f\left( \hat{X}\right) \right\vert }{4\sigma _{\hat{K}}^{2}}%
\right) \left( \hat{K}\left( \frac{\left\vert f\left( \hat{X}\right)
\right\vert }{\sigma _{\hat{K}}^{2}}\right) ^{\frac{1}{2}}\right) ^{p\left( 
\hat{X},\hat{\lambda}\right) }  \label{pmn}
\end{equation}%
and we obtain:%
\begin{eqnarray*}
\left\Vert \hat{\Psi}\left( \hat{X}\right) \right\Vert ^{2} &=&C\exp \left( -%
\frac{\sigma _{X}^{2}u\left( \hat{X},\hat{K}_{\hat{X}}\right) }{\sigma _{%
\hat{K}}^{2}}\right) \\
&&\times \int_{0}^{\infty }\exp \left( -\frac{\left( \hat{K}+\frac{\sigma _{%
\hat{K}}^{2}F\left( \hat{X},K_{\hat{X}}\right) }{f^{2}\left( \hat{X}\right) }%
\right) ^{2}\left\vert f\left( \hat{X}\right) \right\vert }{2\sigma _{\hat{K}%
}^{2}}\right) \left( \left( \hat{K}+\frac{\sigma _{\hat{K}}^{2}F\left( \hat{X%
},K_{\hat{X}}\right) }{f^{2}\left( \hat{X}\right) }\right) \left( \frac{%
\left\vert f\left( \hat{X}\right) \right\vert }{\sigma _{\hat{K}}^{2}}%
\right) ^{\frac{1}{2}}\right) ^{2p\left( \hat{X},\hat{\lambda}\right) }d\hat{%
K}
\end{eqnarray*}%
A change of variable $w=\frac{\left( \hat{K}+\frac{\sigma _{\hat{K}%
}^{2}F\left( \hat{X},K_{\hat{X}}\right) }{f^{2}\left( \hat{X}\right) }%
\right) ^{2}\left\vert f\left( \hat{X}\right) \right\vert }{2\sigma _{\hat{K}%
}^{2}}$ leads to:%
\begin{equation}
\left\Vert \hat{\Psi}\left( \hat{X}\right) \right\Vert ^{2}\simeq C\exp
\left( -\frac{\sigma _{X}^{2}u\left( \hat{X},\hat{K}_{\hat{X}}\right) }{%
\sigma _{\hat{K}}^{2}}\right) \left( \frac{\left\vert f\left( \hat{X}\right)
\right\vert }{\sigma _{\hat{K}}^{2}}\right) ^{-\frac{1}{2}}\left( 2^{p\left( 
\hat{X},\hat{\lambda}\right) -\frac{1}{2}}\Gamma \left( p\left( \hat{X},\hat{%
\lambda}\right) +\frac{1}{2}\right) -\frac{2^{\frac{p\left( \hat{X},\hat{%
\lambda}\right) }{2}}\sqrt{\pi }}{\Gamma \left( \frac{1-p\left( \hat{X},\hat{%
\lambda}\right) }{2}\right) }\frac{\sigma _{\hat{K}}^{2}F\left( \hat{X},K_{%
\hat{X}}\right) }{f^{2}\left( \hat{X}\right) }\right)  \label{smn}
\end{equation}

By the same token we can use the asymptotic form (\ref{pmn}) to find $K_{%
\hat{X}}$: 
\begin{eqnarray*}
K_{\hat{X}}\left\Vert \Psi \left( \hat{X}\right) \right\Vert ^{2} &\simeq
&C\exp \left( -\frac{\sigma _{X}^{2}u\left( \hat{X},\hat{K}_{\hat{X}}\right) 
}{\sigma _{\hat{K}}^{2}}\right) \int \hat{K}\exp \left( -\frac{\hat{K}%
^{2}\left\vert f\left( \hat{X}\right) \right\vert }{2\sigma _{\hat{K}}^{2}}%
\right) \left( \hat{K}\left( \frac{\left\vert f\left( \hat{X}\right)
\right\vert }{\sigma _{\hat{K}}^{2}}\right) ^{\frac{1}{2}}\right) ^{2p\left( 
\hat{X},\hat{\lambda}\right) }d\hat{K} \\
&=&\frac{\sigma _{\hat{K}}^{2}C\exp \left( -\frac{\sigma _{X}^{2}u\left( 
\hat{X},\hat{K}_{\hat{X}}\right) }{\sigma _{\hat{K}}^{2}}\right) }{%
\left\vert f\left( \hat{X}\right) \right\vert }\int y\exp \left( -\frac{y^{2}%
}{2}\right) y^{2p\left( \hat{X},\hat{\lambda}\right) }
\end{eqnarray*}%
We set $y=\sqrt{2w}$ and we obtain:%
\begin{eqnarray*}
K_{\hat{X}}\left\Vert \Psi \left( \hat{X}\right) \right\Vert ^{2} &\simeq
&C\exp \left( -\frac{\sigma _{X}^{2}u\left( \hat{X},\hat{K}_{\hat{X}}\right) 
}{\sigma _{\hat{K}}^{2}}\right) 2^{p\left( \hat{X},\hat{\lambda}\right) }%
\frac{\sigma _{\hat{K}}^{2}}{\left\vert f\left( \hat{X}\right) \right\vert }%
\int \exp \left( -w\right) w^{p\left( \hat{X},\hat{\lambda}\right) }dw \\
&=&C\exp \left( -\frac{\sigma _{X}^{2}u\left( \hat{X},\hat{K}_{\hat{X}%
}\right) }{\sigma _{\hat{K}}^{2}}\right) 2^{p\left( \hat{X},\hat{\lambda}%
\right) }\frac{\sigma _{\hat{K}}^{2}}{\left\vert f\left( \hat{X}\right)
\right\vert }\Gamma \left( p\left( \hat{X},\hat{\lambda}\right) +1\right)
\end{eqnarray*}

\paragraph*{A3.1.4.2 \textbf{Computation of} $C$ as a function of $\hat{%
\protect\lambda}$:}

Ultimately, we need to determine \ the value of the Lagrange multiplier $%
\hat{\lambda}$ and of the associated value of $C$. We do so by integrating (%
\ref{spc}) and the result is constrained to be $\hat{N}$, the total number
of agents:%
\begin{equation*}
\hat{N}=\int \left\Vert \hat{\Psi}_{\lambda ,C}\left( \hat{K},\hat{X}\right)
\right\Vert ^{2}d\hat{K}d\hat{X}=\int \left\Vert \hat{\Psi}\left( \hat{X}%
\right) \right\Vert ^{2}d\hat{X}
\end{equation*}%
Using (\ref{fmn}) and (\ref{fxp}), we have:%
\begin{eqnarray}
\hat{N} &=&\int \left\Vert \hat{\Psi}\left( \hat{X}\right) \right\Vert
^{2}\simeq \int C\exp \left( -\frac{\sigma _{X}^{2}u\left( \hat{X},\hat{K}_{%
\hat{X}}\right) }{\sigma _{\hat{K}}^{2}}\right) \left( \frac{\left\vert
f\left( \hat{X}\right) \right\vert }{\sigma _{\hat{K}}^{2}}\right) ^{-\frac{1%
}{2}}  \label{nrb} \\
&&\times \left( \frac{\sqrt{\pi }}{2^{\frac{3}{2}}}\frac{\func{Psi}\left( 
\frac{1-p\left( \hat{X},\hat{\lambda}\right) }{2}\right) -\func{Psi}\left( -%
\frac{p\left( \hat{X},\hat{\lambda}\right) }{2}\right) }{\Gamma \left(
-p\left( \hat{X},\hat{\lambda}\right) \right) }-\frac{2^{\frac{p\left( \hat{X%
},\hat{\lambda}\right) }{2}}\sqrt{\pi }}{\Gamma \left( \frac{1-p\left( \hat{X%
},\hat{\lambda}\right) }{2}\right) }\frac{\sigma _{\hat{K}}^{2}F\left( \hat{X%
},K_{\hat{X}}\right) }{f^{2}\left( \hat{X}\right) }\right) d\hat{X}  \notag
\\
&\simeq &\int C\exp \left( -\frac{\sigma _{X}^{2}\sigma _{\hat{K}}^{2}\left(
p+\frac{1}{2}\right) ^{2}\left( f^{\prime }\left( X\right) \right) ^{2}}{%
96\left\vert f\left( \hat{X}\right) \right\vert ^{3}}\right) \left( \frac{%
\left\vert f\left( \hat{X}\right) \right\vert }{\sigma _{\hat{K}}^{2}}%
\right) ^{-\frac{1}{2}}\frac{\sqrt{\pi }}{2^{\frac{3}{2}}}\frac{\func{Psi}%
\left( \frac{1-p\left( \hat{X},\hat{\lambda}\right) }{2}\right) -\func{Psi}%
\left( -\frac{p\left( \hat{X},\hat{\lambda}\right) }{2}\right) }{\Gamma
\left( -p\left( \hat{X},\hat{\lambda}\right) \right) }d\hat{X}  \notag
\end{eqnarray}%
with $f$ and $g$ given by (\ref{ftf}) and (\ref{gft}). We thus obtain $C$ as
a function of $\hat{\lambda}$. For $f\left( \hat{X}\right) $\ slowly varying
around its average we can replace $\left\vert f\left( \hat{X}\right)
\right\vert $ and $f^{\prime }\left( X\right) $ by $\left\langle \left\vert
f\left( \hat{X}\right) \right\vert \right\rangle $ and $\left\langle
f^{\prime }\left( X\right) \right\rangle $, where the bracket $\left\langle
A\left( \hat{X}\right) \right\rangle $ represents the average of the
quantity $A\left( \hat{X}\right) $ over the sectors space. Given that the
integrated function is of order $\Gamma \left( p\right) $, we can replace
the integral by the maximal values of the integrand. Consequently, we have:%
\begin{equation}
C\left( \bar{p}\left( \hat{\lambda}\right) \right) \simeq \frac{\exp \left( -%
\frac{\sigma _{X}^{2}\sigma _{\hat{K}}^{2}\left( \frac{\left( \bar{p}\left( 
\hat{\lambda}\right) +\frac{1}{2}\right) f^{\prime }\left( X_{0}\right) }{%
f\left( \hat{X}_{0}\right) }\right) ^{2}}{96\left\vert f\left( \hat{X}%
_{0}\right) \right\vert }\right) \hat{N}\Gamma \left( -\bar{p}\left( \hat{%
\lambda}\right) \right) }{\left( \frac{\left\langle \left\vert f\left( \hat{X%
}\right) \right\vert \right\rangle }{\sigma _{\hat{K}}^{2}}\right) ^{-\frac{1%
}{2}}V_{r}\left( \func{Psi}\left( -\frac{\bar{p}\left( \hat{\lambda}\right)
-1}{2}\right) -\func{Psi}\left( -\frac{\bar{p}\left( \hat{\lambda}\right) }{2%
}\right) \right) }  \label{cld}
\end{equation}%
where:%
\begin{equation}
\bar{p}\left( \hat{\lambda}\right) =\left( -\frac{\frac{\left( g\left( \hat{X%
}_{0}\right) \right) ^{2}}{\sigma _{\hat{X}}^{2}}+\left( f\left( \hat{X}%
_{0}\right) +\frac{1}{2}\left\vert f\left( \hat{X}_{0}\right) \right\vert
+\nabla _{\hat{X}}g\left( \hat{X}_{0},K_{\hat{X}_{0}}\right) -\frac{\sigma _{%
\hat{K}}^{2}F^{2}\left( \hat{X}_{0},K_{\hat{X}_{0}}\right) }{2f^{2}\left( 
\hat{X}_{0}\right) }+\hat{\lambda}\right) }{\left\vert f\left( \hat{X}%
_{0}\right) \right\vert }\right)  \label{rpb}
\end{equation}%
and:%
\begin{equation}
\hat{X}_{0}=\arg \min_{\hat{X}}\left( \frac{\sigma _{X}^{2}\sigma _{\hat{K}%
}^{2}\left( \frac{\left( p\left( \hat{\lambda}\right) +\frac{1}{2}\right)
f^{\prime }\left( X\right) }{f\left( \hat{X}\right) }\right) ^{2}}{%
96\left\vert f\left( \hat{X}\right) \right\vert }\right)  \label{dfr}
\end{equation}%
and $V_{r}$ is the volume of the reduced space where the maximum is reached
defined by:%
\begin{equation*}
V_{r}=\sum_{\hat{X}/p\left( \hat{X},\hat{\lambda}\right) =\bar{p}\left( \hat{%
\lambda}\right) }\frac{1}{\left\vert \frac{\left( \left\Vert \hat{\Psi}%
\left( \hat{X}\right) \right\Vert ^{2}\right) ^{\prime \prime }}{C}%
\right\vert }
\end{equation*}%
We thus can replace $C$ by $C\left( \hat{\lambda}\right) $ and we are left
with an infinite number of solutions of (\ref{knq}) parametrized by $\hat{%
\lambda}$ and given by (\ref{spc}). We write $\left\Vert \hat{\Psi}_{\hat{%
\lambda}}\left( \hat{K},\hat{X}\right) \right\Vert ^{2}$ the solution for $%
\hat{\lambda}$.

\paragraph*{A3.1.4.2 Identification equation for $K_{\hat{X}}$}

To each state $\left\Vert \hat{\Psi}_{\hat{\lambda}}\left( \hat{K},\hat{X}%
\right) \right\Vert ^{2}$, we can associate an average level of $K_{\hat{X},%
\hat{\lambda}}$ satisfying (\ref{knm}) rewritten as a function of $\hat{%
\lambda}$. Using (\ref{fxp}) we find:%
\begin{eqnarray}
K_{\hat{X},\hat{\lambda}}\left\Vert \hat{\Psi}_{\hat{\lambda}}\left( \hat{X}%
\right) \right\Vert ^{2} &=&\hat{K}_{\hat{X},\hat{\lambda}}  \label{knl} \\
&=&\exp \left( -\frac{\sigma _{X}^{2}\sigma _{\hat{K}}^{2}\left( p+\frac{1}{2%
}\right) ^{2}\left( f^{\prime }\left( X\right) \right) ^{2}}{96\left\vert
f\left( \hat{X}\right) \right\vert ^{3}}\right) \left( \frac{\left\vert
f\left( \hat{X}\right) \right\vert }{\sigma _{\hat{K}}^{2}}\right) ^{-1} 
\notag \\
&&\times C\left( \bar{p}\left( \hat{\lambda}\right) \right) \left( \frac{%
\Gamma \left( -\frac{p+1}{2}\right) \Gamma \left( \frac{1-p}{2}\right)
-\Gamma \left( -\frac{p}{2}\right) \Gamma \left( \frac{-p}{2}\right) }{%
2^{p+2}\Gamma \left( -p-1\right) \Gamma \left( -p\right) }+p\frac{\Gamma
\left( -\frac{p}{2}\right) \Gamma \left( \frac{2-p}{2}\right) -\Gamma \left(
-\frac{p-1}{2}\right) \Gamma \left( -\frac{p-1}{2}\right) }{2^{p+1}\Gamma
\left( -p\right) \Gamma \left( -p+1\right) }\right)  \notag
\end{eqnarray}

where:%
\begin{equation}
p\left( \hat{X},\hat{\lambda}\right) =-\frac{\left( g\left( \hat{X}\right)
\right) ^{2}+\sigma _{\hat{X}}^{2}\left( f\left( \hat{X}\right) +\nabla _{%
\hat{X}}g\left( \hat{X},K_{\hat{X}}\right) -\frac{\sigma _{\hat{K}%
}^{2}F^{2}\left( \hat{X},K_{\hat{X}}\right) }{2f^{2}\left( \hat{X}\right) }+%
\hat{\lambda}\right) }{\sigma _{\hat{X}}^{2}\sqrt{f^{2}\left( \hat{X}\right) 
}}-\frac{1}{2}
\end{equation}

As explained in the core of the paper, to compute $K_{\hat{X}}$ we must
average (\ref{knl}) over $\hat{\lambda}$ with the weight $\exp \left(
-\left( S_{3}+S_{4}\right) \right) $. Given equation (\ref{fdr}), a solution
(\ref{spc}) for a given $\hat{\lambda}$ and taking into account the
constraint $\left\Vert \hat{\Psi}\left( \hat{K},\hat{X}\right) \right\Vert
^{2}=\hat{N}$, has the associated normalized weight (see (\ref{stT})):

\begin{equation*}
w\left( \left\vert \hat{\lambda}\right\vert \right) =\frac{\exp \left(
-\left( \left\vert \hat{\lambda}\right\vert -M\right) \hat{N}\right) }{%
\int_{\left\vert \hat{\lambda}\right\vert >M}\exp \left( -\left( \left\vert 
\hat{\lambda}\right\vert -M\right) \hat{N}\right) d\left\vert \hat{\lambda}%
\right\vert }
\end{equation*}%
with $M$ is the lower bound for $\left\vert \hat{\lambda}\right\vert $.

This lower bound is found by considering (\ref{nqk}) and adding the term
proportional to $\frac{\sigma _{\hat{X}}^{2}}{2}$:%
\begin{equation}
\frac{\sigma _{\hat{X}}^{2}}{2}\nabla _{\hat{X}}^{2}\hat{\Psi}+\nabla
_{y}^{2}\hat{\Psi}-\left( \sqrt{f^{2}\left( \hat{X}\right) }\frac{y^{2}}{4}+%
\frac{\left( g\left( \hat{X}\right) \right) ^{2}}{\sigma _{\hat{X}}^{2}}%
+\left( f\left( \hat{X}\right) +\nabla _{\hat{X}}g\left( \hat{X},K_{\hat{X}%
}\right) -\frac{\sigma _{\hat{K}}^{2}F^{2}\left( \hat{X},K_{\hat{X}}\right) 
}{2f^{2}\left( \hat{X}\right) }+\hat{\lambda}\right) \right) \Psi
\label{gmk}
\end{equation}%
multiplying (\ref{gmk}) by $\hat{\Psi}^{\dag }$ and integrating. It yields: 
\begin{eqnarray}
0 &=&-\frac{\sigma _{\hat{X}}^{2}}{2}\int \left( \nabla _{\hat{X}}\hat{\Psi}%
^{\dag }\right) \left( \nabla _{\hat{X}}\hat{\Psi}\right)  \label{kmg} \\
&&-\frac{1}{2}\int \sqrt{f^{2}\left( \hat{X}\right) }\left( \left( \nabla
_{y}\hat{\Psi}^{\dag }\right) \left( \nabla _{y}\hat{\Psi}\right) +\hat{\Psi}%
^{\dag }\frac{y^{2}}{4}\hat{\Psi}\right) +\int \hat{\Psi}_{y=0}^{\dag
}\left( \nabla _{y}\hat{\Psi}\right) _{y=0}  \notag \\
&&-\int \hat{\Psi}^{\dag }\left( \sqrt{f^{2}\left( \hat{X}\right) }\frac{%
y^{2}}{4}+\frac{\left( g\left( \hat{X}\right) \right) ^{2}}{\sigma _{\hat{X}%
}^{2}}+\left( f\left( \hat{X}\right) +\nabla _{\hat{X}}g\left( \hat{X},K_{%
\hat{X}}\right) -\frac{\sigma _{\hat{K}}^{2}F^{2}\left( \hat{X},K_{\hat{X}%
}\right) }{2f^{2}\left( \hat{X}\right) }+\hat{\lambda}\right) \right) \Psi 
\notag
\end{eqnarray}%
The first part of the right hand side in (\ref{kmg}):%
\begin{equation}
-\frac{\sigma _{\hat{X}}^{2}}{2}\int \left( \nabla _{\hat{X}}\hat{\Psi}%
^{\dag }\right) \left( \nabla _{\hat{X}}\hat{\Psi}\right) -\int \sqrt{%
f^{2}\left( \hat{X}\right) }\left( \frac{1}{2}\left( \nabla _{y}\hat{\Psi}%
^{\dag }\right) \left( \nabla _{y}\hat{\Psi}\right) +\hat{\Psi}^{\dag }\frac{%
y^{2}}{4}\hat{\Psi}\right)  \label{rhd}
\end{equation}%
includes the hamiltonian of a sum of harmonic oscillators, and thus (\ref%
{rhd}) is lower than $-\frac{\int \hat{\Psi}^{\dag }\sqrt{f^{2}\left( \hat{X}%
\right) }\hat{\Psi}}{2}$. Consequently, we have the inequality for all $\hat{%
X}$:%
\begin{eqnarray*}
&&\hat{\Psi}_{y=0}^{\dag }\left( \nabla _{y}\hat{\Psi}\right) _{y=0}+\int 
\hat{\Psi}^{\dag }\left( \left\vert \hat{\lambda}\right\vert -\frac{\left(
g\left( \hat{X}\right) \right) ^{2}}{\sigma _{\hat{X}}^{2}}-\left( f\left( 
\hat{X}\right) +\nabla _{\hat{X}}g\left( \hat{X},K_{\hat{X}}\right) -\frac{%
\sigma _{\hat{K}}^{2}F^{2}\left( \hat{X},K_{\hat{X}}\right) }{2f^{2}\left( 
\hat{X}\right) }\right) \right) \Psi d\hat{K} \\
&>&\frac{\int \hat{\Psi}^{\dag }\sqrt{f^{2}\left( \hat{X}\right) }\hat{\Psi}d%
\hat{K}}{2}
\end{eqnarray*}%
Since: 
\begin{equation*}
\left\vert \hat{\lambda}\right\vert \int \left\vert \Psi \right\vert ^{2}d%
\hat{K}=\left\vert \hat{\lambda}\right\vert \left\Vert \hat{\Psi}\left( \hat{%
X}\right) \right\Vert ^{2}
\end{equation*}%
and $\hat{\Psi}_{y=0}^{\dag }\left( \nabla _{y}\hat{\Psi}\right) _{y=0}$ is
of\ order $1<<\left\Vert \hat{\Psi}\left( \hat{X}\right) \right\Vert ^{2}$
since it is integrated over $\hat{X}$ only. Consequently, the condition
reduces to:%
\begin{equation*}
\left\vert \hat{\lambda}\right\vert \left\Vert \hat{\Psi}\left( \hat{X}%
\right) \right\Vert ^{2}>\int \hat{\Psi}^{\dag }\left( \frac{\left( g\left( 
\hat{X}\right) \right) ^{2}}{\sigma _{\hat{X}}^{2}}+f\left( \hat{X}\right) +%
\frac{1}{2}\sqrt{f^{2}\left( \hat{X}\right) }+\nabla _{\hat{X}}g\left( \hat{X%
},K_{\hat{X}}\right) -\frac{\sigma _{\hat{K}}^{2}F^{2}\left( \hat{X},K_{\hat{%
X}}\right) }{2f^{2}\left( \hat{X}\right) }\right) \Psi d\hat{K}
\end{equation*}%
that is:%
\begin{equation*}
\left\vert \hat{\lambda}\right\vert >\frac{\left( g\left( \hat{X}\right)
\right) ^{2}}{\sigma _{\hat{X}}^{2}}+f\left( \hat{X}\right) +\frac{1}{2}%
\sqrt{f^{2}\left( \hat{X}\right) }+\nabla _{\hat{X}}g\left( \hat{X},K_{\hat{X%
}}\right) -\frac{\sigma _{\hat{K}}^{2}F^{2}\left( \hat{X},K_{\hat{X}}\right) 
}{2f^{2}\left( \hat{X}\right) }
\end{equation*}%
for each $\hat{X}$, and we have:%
\begin{equation}
M=\max_{\hat{X}}\left( \frac{\left( g\left( \hat{X}\right) \right) ^{2}}{%
\sigma _{\hat{X}}^{2}}+f\left( \hat{X}\right) +\frac{1}{2}\sqrt{f^{2}\left( 
\hat{X}\right) }+\nabla _{\hat{X}}g\left( \hat{X},K_{\hat{X}}\right) -\frac{%
\sigma _{\hat{K}}^{2}F^{2}\left( \hat{X},K_{\hat{X}}\right) }{2f^{2}\left( 
\hat{X}\right) }\right)  \label{mdf}
\end{equation}%
Note that in general, for $\varepsilon <<1$, $f\left( \hat{X}\right) >>1$
and: 
\begin{equation*}
\frac{\sigma _{\hat{K}}^{2}F^{2}\left( \hat{X},K_{\hat{X}}\right) }{%
2f^{2}\left( \hat{X}\right) }<<\frac{\left( g\left( \hat{X}\right) \right)
^{2}}{\sigma _{\hat{X}}^{2}}+f\left( \hat{X}\right) +\frac{1}{2}\sqrt{%
f^{2}\left( \hat{X}\right) }+\nabla _{\hat{X}}g\left( \hat{X},K_{\hat{X}%
}\right)
\end{equation*}%
so that:%
\begin{equation}
M\simeq \max_{\hat{X}}\left( \frac{\left( g\left( \hat{X}\right) \right) ^{2}%
}{\sigma _{\hat{X}}^{2}}+f\left( \hat{X}\right) +\frac{1}{2}\sqrt{%
f^{2}\left( \hat{X}\right) }+\nabla _{\hat{X}}g\left( \hat{X},K_{\hat{X}%
}\right) \right)  \label{mdF}
\end{equation}%
Having found $M$, this yields: 
\begin{equation}
w\left( \left\vert \hat{\lambda}\right\vert \right) =\hat{N}\exp \left(
-\left( \left\vert \hat{\lambda}\right\vert -M\right) \hat{N}\right)
\label{thw}
\end{equation}%
Consequently, averaging equation (\ref{knl}) yields:%
\begin{equation*}
K_{\hat{X}}=\int K_{\hat{X},\hat{\lambda}}\hat{N}\exp \left( -\left(
\left\vert \hat{\lambda}\right\vert -M\right) \hat{N}\right) d\hat{\lambda}
\end{equation*}%
\begin{eqnarray}
K_{\hat{X}}\left\Vert \Psi \left( \hat{X}\right) \right\Vert ^{2} &=&\int
C\left( \hat{\lambda}\right) w\left( \left\vert \hat{\lambda}\right\vert
\right) \exp \left( -\frac{\sigma _{X}^{2}\sigma _{\hat{K}}^{2}\left( p+%
\frac{1}{2}\right) ^{2}\left( f^{\prime }\left( X\right) \right) ^{2}}{%
96\left\vert f\left( \hat{X}\right) \right\vert ^{3}}\right) \left( \frac{%
\left\vert f\left( \hat{X}\right) \right\vert }{\sigma _{\hat{K}}^{2}}%
\right) ^{-1}  \label{kvn} \\
&&\times \left( \frac{\Gamma \left( -\frac{p+1}{2}\right) \Gamma \left( 
\frac{1-p}{2}\right) -\Gamma \left( -\frac{p}{2}\right) \Gamma \left( \frac{%
-p}{2}\right) }{2^{p+2}\Gamma \left( -p-1\right) \Gamma \left( -p\right) }+p%
\frac{\Gamma \left( -\frac{p}{2}\right) \Gamma \left( \frac{2-p}{2}\right)
-\Gamma \left( -\frac{p-1}{2}\right) \Gamma \left( -\frac{p-1}{2}\right) }{%
2^{p+1}\Gamma \left( -p\right) \Gamma \left( -p+1\right) }\right) d\hat{%
\lambda}  \notag
\end{eqnarray}%
with $C\left( \bar{p}\left( \hat{\lambda}\right) \right) $ given by (\ref%
{cld}). \ Given (\ref{thw}), the average value of $\left\vert \hat{\lambda}%
\right\vert $ is $M+\frac{1}{\hat{N}}$ and have:%
\begin{eqnarray}
K_{\hat{X}}\left\Vert \Psi \left( \hat{X}\right) \right\Vert ^{2}\left\vert
f\left( \hat{X}\right) \right\vert &=&C\left( \bar{p}\left( -\left( M-\frac{1%
}{\hat{N}}\right) \right) \right) \sigma _{\hat{K}}^{2}  \label{nkv} \\
&&\times \left( \frac{\Gamma \left( -\frac{p+1}{2}\right) \Gamma \left( 
\frac{1-p}{2}\right) -\Gamma \left( -\frac{p}{2}\right) \Gamma \left( \frac{%
-p}{2}\right) }{2^{p+2}\Gamma \left( -p-1\right) \Gamma \left( -p\right) }+p%
\frac{\Gamma \left( -\frac{p}{2}\right) \Gamma \left( \frac{2-p}{2}\right)
-\Gamma \left( -\frac{p-1}{2}\right) \Gamma \left( -\frac{p-1}{2}\right) }{%
2^{p+1}\Gamma \left( -p\right) \Gamma \left( -p+1\right) }\right)  \notag
\end{eqnarray}

with:%
\begin{equation*}
p=-\frac{\left( g\left( \hat{X}\right) \right) ^{2}+\sigma _{\hat{X}%
}^{2}\left( f\left( \hat{X}\right) +\nabla _{\hat{X}}g\left( \hat{X},K_{\hat{%
X}}\right) -\frac{\sigma _{\hat{K}}^{2}F^{2}\left( \hat{X},K_{\hat{X}%
}\right) }{2f^{2}\left( \hat{X}\right) }-\left( M-\frac{1}{\hat{N}}\right)
\right) }{\sigma _{\hat{X}}^{2}\sqrt{f^{2}\left( \hat{X}\right) }}-\frac{1}{2%
}
\end{equation*}%
We can consider that $\frac{1}{\hat{N}}<<1$ so that $C\left( \bar{p}\left(
-\left( M-\frac{1}{\hat{N}}\right) \right) \right) \simeq C\left( \bar{p}%
\left( -M\right) \right) $. It amounts to consider $\left\vert \hat{\lambda}%
\right\vert =M$. We will also write $\bar{p}\left( -M\right) =\bar{p}$ and
given (\ref{rbp}) we have:%
\begin{equation}
\bar{p}=\left( \frac{M-\frac{\left( g\left( \hat{X}_{0}\right) \right) ^{2}}{%
\sigma _{\hat{X}}^{2}}+\left( f\left( \hat{X}_{0}\right) +\frac{1}{2}%
\left\vert f\left( \hat{X}_{0}\right) \right\vert +\nabla _{\hat{X}}g\left( 
\hat{X}_{0},K_{\hat{X}_{0}}\right) -\frac{\sigma _{\hat{K}}^{2}F^{2}\left( 
\hat{X}_{0},K_{\hat{X}_{0}}\right) }{2f^{2}\left( \hat{X}_{0}\right) }%
\right) }{\left\vert f\left( \hat{X}_{0}\right) \right\vert }\right)
\label{pbr}
\end{equation}%
and:%
\begin{equation}
p=\frac{M-\left( \frac{\left( g\left( \hat{X}\right) \right) ^{2}}{\sigma _{%
\hat{X}}^{2}}+\left( f\left( \hat{X}\right) +\frac{\sqrt{f^{2}\left( \hat{X}%
\right) }}{2}+\nabla _{\hat{X}}g\left( \hat{X},K_{\hat{X}}\right) -\frac{%
\sigma _{\hat{K}}^{2}F^{2}\left( \hat{X},K_{\hat{X}}\right) }{2f^{2}\left( 
\hat{X}\right) }\right) \right) }{\sqrt{f^{2}\left( \hat{X}\right) }}
\label{pdf}
\end{equation}%
Equation (\ref{cld}) rewrites:%
\begin{equation}
C\left( \bar{p}\right) \simeq \frac{\exp \left( -\frac{\sigma _{X}^{2}\sigma
_{\hat{K}}^{2}\left( \frac{\left( \bar{p}\left( \hat{\lambda}\right) +\frac{1%
}{2}\right) f^{\prime }\left( X_{0}\right) }{f\left( \hat{X}_{0}\right) }%
\right) ^{2}}{96\left\vert f\left( \hat{X}_{0}\right) \right\vert }\right) 
\hat{N}\Gamma \left( -\bar{p}\right) }{\left( \frac{\left\langle \left\vert
f\left( \hat{X}\right) \right\vert \right\rangle }{\sigma _{\hat{K}}^{2}}%
\right) ^{-\frac{1}{2}}V_{r}\left( \func{Psi}\left( -\frac{\bar{p}-1}{2}%
\right) -\func{Psi}\left( -\frac{\bar{p}}{2}\right) \right) }  \label{clk}
\end{equation}%
and (\ref{nkv}) reduces to:%
\begin{equation}
K_{\hat{X}}\left\Vert \Psi \left( \hat{X}\right) \right\Vert ^{2}\left\vert
f\left( \hat{X}\right) \right\vert =C\left( \bar{p}\right) \sigma _{\hat{K}%
}^{2}\hat{\Gamma}\left( p+\frac{1}{2}\right)  \label{Nkv}
\end{equation}%
with:%
\begin{eqnarray}
\hat{\Gamma}\left( p+\frac{1}{2}\right) &=&\exp \left( -\frac{\sigma
_{X}^{2}\sigma _{\hat{K}}^{2}\left( p+\frac{1}{2}\right) ^{2}\left(
f^{\prime }\left( X\right) \right) ^{2}}{96\left\vert f\left( \hat{X}\right)
\right\vert ^{3}}\right)  \label{gmh} \\
&&\times \left( \frac{\Gamma \left( -\frac{p+1}{2}\right) \Gamma \left( 
\frac{1-p}{2}\right) -\Gamma \left( -\frac{p}{2}\right) \Gamma \left( \frac{%
-p}{2}\right) }{2^{p+2}\Gamma \left( -p-1\right) \Gamma \left( -p\right) }+p%
\frac{\Gamma \left( -\frac{p}{2}\right) \Gamma \left( \frac{2-p}{2}\right)
-\Gamma \left( -\frac{p-1}{2}\right) \Gamma \left( -\frac{p-1}{2}\right) }{%
2^{p+1}\Gamma \left( -p\right) \Gamma \left( -p+1\right) }\right)  \notag
\end{eqnarray}%
We note that, asymptotically:%
\begin{equation}
\hat{\Gamma}\left( p+\frac{1}{2}\right) \sim _{\infty }\exp \left( -\frac{%
\sigma _{X}^{2}\sigma _{\hat{K}}^{2}\left( p+\frac{1}{2}\right) ^{2}\left(
f^{\prime }\left( X\right) \right) ^{2}}{96\left\vert f\left( \hat{X}\right)
\right\vert ^{3}}\right) \Gamma \left( p+\frac{3}{2}\right)
\end{equation}

\paragraph*{A3.1.4.3 \textbf{Replacing }$\left\Vert \Psi \left( X\right)
\right\Vert ^{2}$\textbf{\ in the} $K_{\hat{X}}$\protect\bigskip e\textbf{%
quation}}

We can isolate $K_{\hat{X}}$ in (\ref{nkv}) by using (\ref{psp}) and (\ref%
{spp}) to rewrite $\left\Vert \Psi \left( \hat{X}\right) \right\Vert ^{2}$:

Using (\ref{nbq}):%
\begin{eqnarray*}
D\left( \left\Vert \Psi \right\Vert ^{2}\right) &=&2\tau \left\Vert \Psi
\left( X\right) \right\Vert ^{2}+\frac{1}{2\sigma _{X}^{2}}\left( \nabla
_{X}R\left( X\right) \right) ^{2}H^{2}\left( \frac{\hat{K}_{X}}{\left\Vert
\Psi \left( X\right) \right\Vert ^{2}}\right) \left( 1-\frac{H^{\prime
}\left( \hat{K}_{X}\right) }{H\left( \hat{K}_{X}\right) }\frac{\hat{K}_{X}}{%
\left\Vert \Psi \left( X\right) \right\Vert ^{2}}\right) \\
&=&2\tau \left\Vert \Psi \left( X\right) \right\Vert ^{2}+\frac{1}{2\sigma
_{X}^{2}}\left( \nabla _{X}R\left( X\right) \right) ^{2}H^{2}\left(
K_{X}\right) \left( 1-\frac{H^{\prime }\left( K_{X}\right) }{H\left(
K_{X}\right) }K_{X}\right)
\end{eqnarray*}%
We rewrite $\left\Vert \Psi \left( X\right) \right\Vert ^{2}$ as a function
of $K_{X}$: 
\begin{equation}
\left\Vert \Psi \left( X\right) \right\Vert ^{2}=\frac{D\left( \left\Vert
\Psi \right\Vert ^{2}\right) -\frac{1}{2\sigma _{X}^{2}}\left( \nabla
_{X}R\left( X\right) \right) ^{2}H^{2}\left( K_{X}\right) \left( 1-\frac{%
H^{\prime }\left( K_{X}\right) }{H\left( K_{X}\right) }K_{X}\right) }{2\tau }%
\equiv D-\bar{H}\left( X,K_{X}\right)  \label{psrd}
\end{equation}%
Ultimately, the equation (\ref{Nkv}) for $K_{\hat{X}}$ can be rewritten: 
\begin{equation}
K_{\hat{X}}\left\vert f\left( \hat{X}\right) \right\vert =\frac{C\left( \bar{%
p}\right) \sigma _{\hat{K}}^{2}}{\left\Vert \Psi \left( X\right) \right\Vert
^{2}}\hat{\Gamma}\left( p+\frac{1}{2}\right) =\frac{C\left( \bar{p}\right)
\sigma _{\hat{K}}^{2}}{D-\bar{H}\left( X,K_{X}\right) }\hat{\Gamma}\left( p+%
\frac{1}{2}\right)  \label{qnk}
\end{equation}%
with $C\left( \bar{p}\right) $\ given by (\ref{clk}), $\hat{\Gamma}\left( p+%
\frac{1}{2}\right) $\ defined in (\ref{gmh}) and $p$ given by (\ref{pdf}).

\section{Appendix 4. Approaches to solutions for $K_{\hat{X}}$}

We present the three approaches to the average capital equation (\ref{qtk})
and their results in the first section. The details of the computations are
given in section two of this appendix.

\section*{A4.1 Solutions for average capital across sectors}

We stated in the text that the final form of the capital equation, (\ref{qtk}%
), cannot be solved analytically, except for some particular cases\footnote{%
These particular cases will be studied in the following sections.} but that
several approaches can be used to study the behaviour of its solutions or
approximate its solutions. Combined, these three approaches confirm and
complete with each other.

\subsection*{A.4.1.1 Stability of average capital and dependency on exoenous
parameters}

One way to better understand equation (\ref{qtk}) is to study its
differential form.

Assume at point $\hat{X}$ of the system, a variation $\delta Y\left( \hat{X}%
\right) $ for any parameter, in which the parameter $Y\left( \hat{X}\right) $
can be either $R\left( X\right) ,$ its gradient, or any parameter arising in
the definition of $f\left( \hat{X}\right) $and $g\left( \hat{X}\right) $.
This variation $\delta Y\left( \hat{X}\right) $ induces in turn a variation $%
\delta K_{\hat{X}}$ in average capital expressed by differentiating (\ref%
{qtk}):

\begin{eqnarray}
\delta K_{\hat{X}} &=&\left( -\left( \frac{\partial \ln f\left( \hat{X},K_{%
\hat{X}}\right) }{\partial K_{\hat{X}}}+\frac{\partial \ln \left\vert \Psi
\left( \hat{X},K_{\hat{X}}\right) \right\vert ^{2}}{\partial K_{\hat{X}}}%
+l\left( \hat{X},K_{\hat{X}}\right) \right) +k\left( p\right) \frac{\partial
p}{\partial K_{\hat{X}}}\right) K_{\hat{X}}\delta K_{\hat{X}}  \label{rvd} \\
&&+\frac{\partial }{\partial Y\left( \hat{X}\right) }\left( \frac{\sigma _{%
\hat{K}}^{2}C\left( \bar{p}\right) 2\hat{\Gamma}\left( p+\frac{1}{2}\right) 
}{\left\vert f\left( \hat{X},K_{\hat{X}}\right) \right\vert \left\Vert \Psi
\left( \hat{X},K_{\hat{X}}\right) \right\Vert ^{2}}\right) \delta Y\left( 
\hat{X}\right)  \notag
\end{eqnarray}%
where the coefficients $l\left( \hat{X},K_{\hat{X}}\right) $ and $k\left(
p\right) $ are computed in appendix 4.2.1. The parameter $l\left( \hat{X},K_{%
\hat{X}}\right) $\ accounts for the variation of the short-term returns
across sectors, while $k\left( p\right) $\ describes the impact of relative
returns variations across sectors.

Equation (\ref{rvd}) will be used to compute the dependency of average
capital per firm in sector $\hat{X}$, i.e. $K_{\hat{X}}$, as a function of
any parameter $Y(\hat{X})$, and more fundamentally to investigate the
stability of the solutions of (\ref{qtk}) with respect to the variations in
parameters.

\subsubsection*{A.4.1.1.1 Local stability}

The differential form given by equation (\ref{rvd}), computes the effect of
a variation $\delta Y\left( \hat{X}\right) $ in the parameters on the
average capital $K_{\hat{X}}$. Moreover, equation (\ref{rvd}) can be
understood as\ the fixed-point equation of a dynamical system of the
following mechanism\textbf{: }each variation $\delta Y\left( \hat{X}\right) $%
\ in the parameters impacts directly the average capital through the second
term in the RHS of (\ref{rvd}). In a second step, the variation $\delta K_{%
\hat{X}}$ impacts the various functions implied in (\ref{qtk}), and
indirectly modifies $K_{\hat{X}}$ through the first term in the rhs of (\ref%
{rvd})\footnote{%
The computations and formula for the dynamics' fixed points are given in
appendix 3.2.1.}.

What matters here is the condition of stability. We show that the fixed
point is stable when: 
\begin{subequations}
\begin{equation}
\left\vert k\left( p\right) \frac{\partial p}{\partial K_{\hat{X}}}-\left( 
\frac{\partial \ln f\left( \hat{X},K_{\hat{X}}\right) }{\partial K_{\hat{X}}}%
+\frac{\partial \ln \left\vert \Psi \left( \hat{X},K_{\hat{X}}\right)
\right\vert ^{2}}{\partial K_{\hat{X}}}+l\left( \hat{X},K_{\hat{X}}\right)
\right) \right\vert <1  \label{sTB}
\end{equation}%
and unstable otherwise.

Thus, two types of solutions emerge for the average capital per firm $K_{%
\hat{X}}$.\ The stable solutions $K_{\hat{X}}$\ can be considered as the
potential equilibrium averages for sector $\hat{X}$.\ However unstable
solutions must rather be considered as thresholds\textbf{:} when $K_{\hat{X}%
} $\ is driven away from this threshold, it may either converge toward a
stable solution of (\ref{qtk}), or diverge towards $0$\ or infinity.

\paragraph*{Remark}

The variation in average capital induced by a change in parameter reveals a
shift $\delta \hat{\Psi}\left( \hat{K},\hat{X}\right) $\ in the background
state $\hat{\Psi}\left( \hat{K},\hat{X}\right) $. The new configuration $%
\hat{\Psi}\left( \hat{K},\hat{X}\right) +\delta \hat{\Psi}\left( \hat{K},%
\hat{X}\right) $\ may not be a minimum of the action functional.\ We must
therefore determine whether the system will settle on a background state,
slightly modified with a different $K_{\hat{X}}$, or be driven towards an
altogether different equilibrium. To this end, we will study the dynamics\
equation for $K_{\hat{X}}$\ in appendix 5.\ 

\subsubsection*{A.4.1.1.2 Dependency of average capital in system's
parameters}

Once the notion of stability understood, we can use equation (\ref{rvd}) to
compute the impact of the variation of any parameter $Y(\hat{X})$ on $\delta
K_{\hat{X}}$. Two applications are of particular interest to us.

\paragraph*{A.4.1.1.2.1 Dependency in relative expected returns}

The main application of equation (\ref{rvd}) is to consider the dependency
of the average capital $K_{\hat{X}}$ in the parameter $p(\hat{X})$ defined
in (\ref{PDF}).\ \ This parameter encompasses sector $\hat{X}$\ relative
expected returns vis-\`{a}-vis its neighbours.

Using (\ref{rvd}), we show (see appendix 4.2.1) that the variation of $K_{%
\hat{X}}$ with respect to $p(\hat{X})$ depends on the notion of equilibrium
stability defined in (\ref{sTB}).

For a stable equilibrium where the expected return $f\left( \hat{X}\right) $%
\ is positive\footnote{%
which is the case of interest for us (see section 11.3.3)}, we find that:%
\textbf{\ } 
\end{subequations}
\begin{equation}
\frac{\delta K_{\hat{X}}}{\delta p\left( \hat{X}\right) }>0  \label{drt}
\end{equation}%
so that $p(\hat{X})$ writes as: 
\begin{equation}
p(\hat{X})=\frac{M-\left( \frac{\left( g\left( \hat{X}\right) \right) ^{2}}{%
\sigma _{\hat{X}}^{2}}+\nabla _{\hat{X}}g\left( \hat{X}\right) -\frac{\sigma
_{\hat{K}}^{2}F^{2}\left( \hat{X}\right) }{2f^{2}\left( \hat{X}\right) }%
\right) }{f\left( \hat{X}\right) }-\frac{3}{2}  \label{fpR}
\end{equation}%
The definitions (\ref{fcg}) and (\ref{VRN}) show that $g\left( \hat{X}%
\right) $\ and $\nabla _{\hat{X}}g\left( \hat{X}\right) $\ are proportional
to $\nabla _{\hat{X}}R\left( \hat{X}\right) $\ and $\nabla _{\hat{X}%
}^{2}R\left( \hat{X}\right) $\ respectively.\ Thus, when the expected
long-term return $R\left( \hat{X}\right) $ is a maximum, $p(\hat{X})$\ is
maximal too\footnote{%
See also section 14.1 for more details.}: under a stable equilibrium,
capital accumulation is maximal in sectors where the expected long-term
return $R\left( \hat{X}\right) $ is maximal.

On the other hand, when the equilibrium is unstable we have:%
\begin{equation}
\frac{\delta K_{\hat{X}}}{\delta p\left( \hat{X}\right) }<0
\end{equation}%
Actually, the capital $K_{\hat{X}}$\ is minimal for $R\left( \hat{X}\right) $%
\ maximal. Actually, as seen above, in the instability range, the average
capital $K_{\hat{X}}$\ acts as a threshold. When, due to variations in the
system's parameters, the average capital per firm is shifted above the
threshold $K_{\hat{X}}$, \ capital will either move to the next stable
equilibrium, possibly zero, or tend to infinity. Our results show that when
the expected long-term return of a sector increases, the threshold $K_{\hat{X%
}}$ decreases, which favours capital accumulation.

\paragraph*{A.4.1.1.2.2 Dependency in short term returns}

A second use of equation (\ref{rvd}) is to consider $Y(\hat{X})$ as any
parameter-function involved in the definition of $f\left( \hat{X},K_{\hat{X}%
}\right) $ that may condition either real short-term returns or the
price-dividend ratio. We show in appendix 4.2.1 that again the result
depends on the stability of the solution.

Around a stable equilibrium, in most cases:%
\begin{equation*}
\frac{\delta K_{\hat{X}}}{\delta f\left( \hat{X}\right) }>0
\end{equation*}

A higher short-term return, decomposed as a sum of dividend and price
variation, induces a higher average capital. This effect is magnified for
larger levels of capital: the third approach will confirm that, in most
cases, the return $f\left( \hat{X}\right) $ is asymptotically a constant $%
c<<1$ when capital is high: $K_{\hat{X}}>>1$.

Turning now to the case of an unstable equilibrium, we find:%
\begin{equation*}
\frac{\delta K_{\hat{X}}}{\delta f\left( \hat{X}\right) }<0
\end{equation*}%
In the instability range, and due to this very instability, an increase in
returns $f\left( \hat{X}\right) $ reduces the threshold of capital
accumulation for low levels of capital. When short-term returns $f\left( 
\hat{X}\right) $\ increase, a lower average capital will trigger capital
accumulation towards an equilibrium. Otherwise, when average capital $K_{%
\hat{X}}$ is below this threshold, it will converge toward $0$.

\subsection*{A.4.1.2 Accumulation points of capital}

The second approach to equation (\ref{qtk}) is to find the average capital
at some particular points $\hat{X}$, and then by first order expansion, the
solutions in the neighbourhood of these particular points. We choose as
particular points $\hat{X}$ those such that $A\left( \hat{X}\right) $
defined in (\ref{DFT}) is maximal. At these points $\left( \hat{X}_{M},K_{%
\hat{X}_{M}}\right) $, we have\footnote{%
See equation (\ref{MDN}).}:%
\begin{equation}
A\left( \hat{X}_{M}\right) =M=\max_{\hat{X}}A\left( \hat{X}\right)
\label{mfl}
\end{equation}%
and $p=0$, given (\ref{fRM}). These sectors are characterized by expected
returns that exhibit a local maximum when compared to their neighboring
sectors. As a result, these sectors serve as points of capital accumulation.

\subsubsection*{A.4.1.2.1 Particular solutions for capital when $p=0$}

For $p=0$ equation (\ref{qtk}) at points $\left( \hat{X}_{M},K_{\hat{X}%
_{M}}\right) $ reduces to:%
\begin{equation}
K_{\hat{X},M}\left\vert f\left( \hat{X},K_{\hat{X},M}\right) \right\vert
\left\Vert \Psi \left( \hat{X},K_{\hat{X},M}\right) \right\Vert ^{2}\simeq
\sigma _{\hat{K}}^{2}C\left( \bar{p}\right) \exp \left( -\frac{\sigma
_{X}^{2}\sigma _{\hat{K}}^{2}\left( f^{\prime }\left( X_{M},K_{\hat{X}%
,M}\right) \right) ^{2}}{384\left\vert f\left( \hat{X}_{M},K_{\hat{X}%
,M}\right) \right\vert ^{3}}\right)  \label{kpt}
\end{equation}%
Given that $\left\Vert \Psi \left( \hat{X}_{M},K_{\hat{X},M}\right)
\right\Vert ^{2}$ is decreasing in $K_{\hat{X}}$ (see (\ref{psrd})), and
assuming $f\left( \hat{X}\right) $ is decreasing too, as is usual for
marginal decreasing returns, equation (\ref{qtk}) has two solutions.

For some particular values of the parameters, an approximate form can be
found for these solutions. Here, we will merely consider a power law for $%
f\left( \hat{X}\right) $:%
\begin{equation}
f\left( \hat{X}\right) \simeq B\left( X\right) K_{\hat{X}}^{\alpha }
\label{fbR}
\end{equation}%
The parameter $B\left( X\right) $ is the productivity in sector $X$, and
equation (\ref{fbR}) shows that the return $f\left( \hat{X}\right) $ is
increasing in $B\left( X\right) $.\smallskip

The stable case corresponds to an intermediate level of capital, $K_{\hat{X}%
}^{\alpha }<<D$.\ In such a case, given the density of producers (\ref{psl}%
), we can assume that this density satisfies $\left\Vert \Psi \left( \hat{X}%
,K_{\hat{X}}\right) \right\Vert ^{2}\simeq D$. The solution to equation (\ref%
{qtk}) is then:%
\begin{equation}
K_{\hat{X}}^{\alpha }=\left( \frac{DB\left( X\right) }{C\left( \bar{p}%
\right) \sigma _{\hat{K}}^{2}}\right) ^{-\frac{\alpha }{\alpha +1}}\exp
\left( W_{0}\left( -\frac{\sigma _{X}^{2}\sigma _{\hat{K}}^{2}\left(
B^{\prime }\left( X\right) \right) ^{2}\alpha }{384\left( B\left( X\right)
\right) ^{3}\left( \alpha +1\right) }\left( \frac{DB\left( X\right) }{%
C\left( \bar{p}\right) \sigma _{\hat{K}}^{2}}\right) ^{\frac{\alpha }{\alpha
+1}}\right) \right)  \label{Kms}
\end{equation}%
where $W_{0}$ is the Lambert $W$ function.

For $B\left( X\right) <<1$, we can check that $K_{\hat{X}}^{\alpha }$ is
increasing with $B\left( X\right) $, i.e. with short-term returns $f\left( 
\hat{X}\right) $\footnote{%
See equation (\ref{fbR}).}, which confirms the results found in the first
approach: in the stable case, capital equilibrium increases with short-term
returns $f\left( \hat{X}\right) $.\smallskip

The unstable case corresponds to a higher level of capital. Given the
density of producers (\ref{psl}), this case amounts to consider, in first
approximation, that capital $K_{X}$\ is concentrated among a small group of
agents, that is $\left\vert \Psi \left( \hat{X},K_{\hat{X}}\right)
\right\vert ^{2}<<1$\textbf{. } Considering a power law for $H^{2}\left(
K_{X}\right) $:%
\begin{equation*}
H^{2}\left( K_{X}\right) =K_{X}^{\alpha }
\end{equation*}%
the solution (\ref{kpt}) can be written as:%
\begin{eqnarray}
K_{\hat{X}}^{\alpha } &\simeq &\frac{2D}{\left( \nabla _{X}R\left( X\right)
\right) ^{2}+\sigma _{X}^{2}\frac{\nabla _{X}^{2}R\left( X\right) }{H\left(
K_{X}\right) }}  \label{Kmn} \\
&&-\left( \frac{\left( \nabla _{X}R\left( X\right) \right) ^{2}+\sigma
_{X}^{2}\frac{\nabla _{X}^{2}R\left( X\right) }{H\left( K_{X}\right) }}{2D}%
\right) ^{\frac{1}{\alpha }}\frac{\sigma _{\hat{K}}^{2}C\left( \bar{p}%
\right) }{DB\left( X\right) }  \notag \\
&&\times \exp \left( -\frac{\sigma _{X}^{2}\sigma _{\hat{K}}^{2}\left(
B^{\prime }\left( X\right) \right) ^{2}}{768D\left( B\left( X\right) \right)
^{3}}\left( \left( \nabla _{X}R\left( X\right) \right) ^{2}+\sigma _{X}^{2}%
\frac{\nabla _{X}^{2}R\left( X\right) }{H\left( K_{X}\right) }\right) \right)
\notag
\end{eqnarray}%
at the first order in $D$, plus corrections of order $\frac{1}{D}$, with:%
\begin{equation*}
f\left( \hat{X}\right) \simeq B\left( X\right) K_{\hat{X}}^{\alpha }
\end{equation*}%
The (in)stability analysis of the previous approach applies. In the range $%
B\left( X\right) <<1$, when $f\left( \hat{X}\right) $\ increases, or which
is equivalent, $B\left( X\right) $ increases, average capital must reduce to
preserve the possibility of unstable equilibria. Likewise, equilibrium
capital is higher when expected returns $R\left( X\right) $ are minimal.
When expected returns increase, the threshold defined by the unstable
equilibrium decreases.

\subsubsection*{A.4.1.2.2 Expansion around particular solutions for $p=0$}

To better understand the behaviour of the solutions of the average capital
equation (\ref{qtk}), we expand this equation around the points $\left( \hat{%
X},K_{\hat{X},M}\right) $ that solve equation (\ref{qtk}). Appendix 4.2.2
computes this expansion at the second order around $\hat{X}_{M}$ and $K_{%
\hat{X},M}$. This yields the form of the solutions of (\ref{qtk}) in the
vincinity of the points $\left( \hat{X},K_{\hat{X},M}\right) $. \ We find:

\begin{eqnarray}
\left( K_{\hat{X}}-K_{\hat{X},M}\right) &=&\frac{1}{D}\left( \sigma
_{X}^{2}\sigma _{\hat{K}}^{2}\frac{3\left( f^{\prime }\left( \hat{X}\right)
\right) ^{3}-2f^{\prime }\left( X\right) f^{\prime \prime }\left( \hat{X}%
\right) \left\vert f\left( \hat{X}\right) \right\vert }{120\left\vert
f\left( \hat{X}\right) \right\vert ^{4}}\right.  \label{pnm} \\
&&\left. -\frac{\frac{\partial f\left( \hat{X},K_{\hat{X}}\right) }{\partial 
\hat{X}}}{f\left( \hat{X},K_{\hat{X}}\right) }-\frac{\frac{\partial
\left\Vert \Psi \left( \hat{X},K_{\hat{X}}\right) \right\Vert ^{2}}{\partial 
\hat{X}}}{\left\Vert \Psi \left( \hat{X},K_{\hat{X}}\right) \right\Vert ^{2}}%
\right) _{K_{\hat{X},M}}\left( \hat{X}-\hat{X}_{M}\right)  \notag \\
&&+\frac{1}{D}\frac{b}{2}\left( \hat{X}-\hat{X}_{M}\right) \nabla _{\hat{X}%
}^{2}\left( \frac{M-A\left( \hat{X}\right) }{f\left( \hat{X}\right) }\right)
_{K_{\hat{X},M}}\left( \hat{X}-\hat{X}_{M}\right)  \notag
\end{eqnarray}%
with $\frac{A\left( \hat{X}\right) }{f\left( \hat{X}\right) }$\ given in
formula (\ref{rlt}), and $b$ and $D$ are coefficients given in the appendix
4.2.2. In fact we recover the analysis of the first approach in terms of
stability. the case $D>0$ corresponds to a stable equilibrium, and $D<0$ to
an unstable one. The expansion (\ref{pnm}) describes the local variations of 
$K_{\hat{X}}$ in the neighbourhood of the points $K_{\hat{X},M}$. This
approximation (\ref{pnm}) suffices to understand the role of the parameters
of the system. The whole analysis is performed in appendix 4.2.2 and
confirms and refines our previous results.

\subsection*{A.4.1.3 Solutions to the capital equation in some particular
cases}

A third approach computes the approximate solutions of (\ref{qtk}) for the
average capital per firm per sector $X$. To do so,\ we choose some general
forms for the three parameter-functions arising in the definition of the
action functional: $f$ that defines short-term returns, that include
dividend and expected long-term price variations, and is given by equation (%
\ref{fcf}); $g$ that describes investors' mobility in the sector space,
given by (\ref{fcg}), and the function $H\left( K_{X}\right) $ involved in
the firms' background field, that describes firms' moves\textbf{\ }in the
sectors space and is given by equation (\ref{psl}).

Once these parameter-functions chosen, the approximate solutions of (\ref%
{qtk}) for average capital per firm per sector can be found. We have already
seen in the second approach that this equation has in general several
solutions. To find them, we must consider several relevant ranges for
average capital, namely a very large level of capital, $K_{X}>>1$, a very
low one, $K_{X}<<1$,\ and an intermediate range $\infty >K_{X}>1$. We will
derive the solutions for $K_{X}$ within these various ranges.\ Details of
the computations are given in appendix 4.2.3.

\subsubsection*{A.4.1.3.1 Choice of parameter functions}

Our choices for the parameter functions $f$,\ $g$ and$\ H^{2}\left(
K_{X}\right) $ are the following.

\subparagraph{Firms' intersectoral moves $H^{2}\left( K_{X}\right) $}

We can choose for $H^{2}\left( K_{X}\right) $ a power function of $K_{X}$,
so that equation (\ref{psl}) rewrites:%
\begin{equation}
\left\Vert \Psi \left( X\right) \right\Vert ^{2}=D-L\left( X\right) \left(
\nabla _{X}R\left( X\right) \right) ^{2}K_{X}^{\eta }  \label{rsp}
\end{equation}%
with $L\left( X\right) $ given in the appendix.

\subparagraph{Short-term returns $f$}

To determine the function $f$, we must first assume a form for $r\left(
K,X\right) $, the physical capital marginal returns, and for $F_{1}$, the
function that measures the impact of expected long-term return on investment
choices.

We assume Cobb-Douglas production functions, i.e. $B\left( X\right)
K^{\alpha }$ with $B\left( X\right) $ a productivity factor.\ We also choose
the expected long-term return $F_{1}$ to be an increasing function of the $%
\arctan $ type, so that investments increase linearly with expected returns
and capital for small-capitalized firms, but is bounded for large values of
capital.

Under these assumption, the short-term return can be written in a compact
form as:%
\begin{equation}
f\left( \hat{X},\Psi ,\hat{\Psi}\right) =B_{1}\left( \hat{X}\right) K_{\hat{X%
}}^{\alpha -1}+B_{2}\left( \hat{X}\right) K_{\hat{X}}^{\alpha }-C\left( \hat{%
X}\right)  \label{stp}
\end{equation}%
The coefficients $B_{1}\left( \hat{X}\right) $, $B_{2}\left( \hat{X}\right) $%
\ and $C\left( \hat{X}\right) $ are given in the appendix 4.2.3.

\subparagraph{investors' mobility in the sector space $g$}

To determine the form of the investors' mobility in the sector space $g$,
given by (\ref{fcg}), we must first choose a form for\textbf{\ }$F_{0}$, the
investors' mobility towards higher long-term returns\footnote{%
See section 4.4.}.

Here again, we choose an $\arctan $ type function of the expected long-term
return, so that the velocity in the sectors' space $g$ increases with
capital, and is bounded and maximal when $K_{\hat{X}}^{\alpha }\rightarrow
\infty $.

Appendix 3.2.2 shows that $g\left( \hat{X},\Psi ,\hat{\Psi}\right) $ can be
written: 
\begin{equation}
g\left( \hat{X},\Psi ,\hat{\Psi}\right) \nabla _{\hat{X}}R\left( \hat{X}%
\right) A\left( \hat{X}\right) K_{\hat{X}}^{\alpha }  \label{gtp}
\end{equation}%
where the function $A\left( \hat{X}\right) $ is given in appendix 4.2.2.

\subsubsection*{A.4.1.3.2 Solutions for the average capital}

Now that the particular functions have been chosen, we can find approximate
solutions to (\ref{qtk}) in several ranges of sector $X$'s average capital:
Very large and stable capitalization, very large and unstable, i.e.
bubble-like, capitalization, large capitalization stable or unstable, the
intermediate case of mid-capitalization and ultimately small
capitalization.\ Besides, we only consider positive short-term returns%
\footnote{%
Solutions for negative returns, $f<0$, are discussed below.}, $f>0$.\ 

We consider the several type of solutions separately.

\begin{case}
Very large and stable capitalization, $K_{\hat{X}}>>>1$
\end{case}

When returns are either slowing or increasing in $\hat{X}$, i.e.\textbf{\ }$%
\left( \nabla _{\hat{X}}R\left( \hat{X}\right) \right) ^{2}\neq 0$, a
solution the capital equation (\ref{qtk}) may exist with $K_{\hat{X}}>>>1$.\
In this case, only a small number of firms are present in the sector.
Indeed, in such a case, the competition-deterent factor $L\left( \hat{X}%
\right) $ in (\ref{rsp}) is very large, and we can assume, in first
approximation, that:%
\begin{equation}
\left\Vert \Psi \left( \hat{X}\right) \right\Vert ^{2}<<1  \label{mlt}
\end{equation}%
A sector in which average capital is very large implies a very high
competition, that act as a barrier to the entry of other firms. In this
case, we can show that $f\left( \hat{X}\right) \simeq c$, for some constant $%
c$. Appendix 4.2.3.2 solves equation (\ref{qtk}) given these assumptions.\
The average capital is given by:%
\begin{eqnarray}
K_{\hat{X}}^{\alpha } &\simeq &\frac{D}{\left( \nabla _{\hat{X}}R\left( \hat{%
X}\right) \right) ^{2}}-\frac{C\left( \bar{p}\right) \sigma _{\hat{K}}^{2}%
\sqrt{\frac{M-c}{c}}}{\left( \nabla _{\hat{X}}R\left( \hat{X}\right) \right)
^{2\left( 1-\frac{1}{\alpha }\right) }D^{\frac{1}{\alpha }}c}  \label{Kln} \\
&&-\frac{d}{R\left( \hat{X}\right) }\frac{\left( \nabla _{\hat{X}}R\left( 
\hat{X}\right) \right) ^{\frac{2}{\alpha }}C\left( \bar{p}\right) \sigma _{%
\hat{K}}^{2}\left( \sqrt{\frac{M-c}{c}}+\frac{\frac{M}{c}+\nabla _{\hat{X}%
}^{2}R\left( \hat{X}\right) \frac{f}{d}}{2\sqrt{\frac{M-c}{c}}}\right) }{%
c^{2}D^{1+\frac{1}{\alpha }}\left( 1-\frac{\left( \nabla _{\hat{X}}R\left( 
\hat{X}\right) \right) ^{\frac{2}{\alpha }}C\left( \bar{p}\right) \sigma _{%
\hat{K}}^{2}}{cD^{1+\frac{1}{\alpha }}}\sqrt{\frac{M-c}{c}}\right) }  \notag
\end{eqnarray}%
which shows that $K_{\hat{X}}^{\alpha }$ is increasing in $f\left( \hat{X}%
\right) $ and $R\left( \hat{X}\right) $ for $K_{\hat{X}}^{\alpha }$ large, $%
f\left( \hat{X}\right) \simeq c<<1$ and $D>>1$.\ Using (\ref{sTB}) shows
that this corresponds to a stable local equilibrium.

\smallskip

\begin{case}
Very large and unstable, i.e. bubble-like, capitalization, $K_{\hat{X}}>>>1$
\end{case}

This case arises when the expected long term returns is a local maximum,
i.e. when $\left( \nabla _{\hat{X}}R\left( \hat{X}\right) \right)
^{2}\rightarrow 0$ and $\nabla _{X}^{2}R\left( K_{X},X\right) <0$\footnote{%
The case $\nabla _{X}^{2}R\left( K_{X},X\right) >0$ i.e. a minimum for the
expected long term return is studied in appendix 3.2.3.2 which shows that
this equilibrium is unlikely and can be discarded.}. This describes a sector
with a large number of firms and very high level of capital. Actually, the
number of firms given in (\ref{rsp}) shows that:%
\begin{equation}
\left\Vert \Psi \left( \hat{X}\right) \right\Vert ^{2}>D>>1  \label{nlt}
\end{equation}%
and appendix 4.2.3.2 shows that the average capital is given by:%
\begin{equation}
K_{\hat{X}}=\left( \frac{C\left( \bar{p}\right) \sigma _{\hat{K}}^{2}}{%
\left\vert \nabla _{X}^{2}R\left( X\right) \right\vert c}\Gamma \left( \frac{%
M-\nabla _{\hat{X}}g\left( \hat{X}\right) }{c}\right) \right) ^{\frac{2}{%
3\alpha }}  \label{KLN}
\end{equation}%
where $f\left( \hat{X}\right) \simeq c<<1$ for some constant $c$ and $D>>1$%
.\ 

The case (\ref{KLN}) is unstable. Actually, in this case $K_{X}$ is
decreasing in $f\left( \hat{X}\right) $. When returns increase, an
equilibrium arises only for a relatively low average capital. Otherwise,
capital tends to accumulate infinitely. When the sector's expected returns
are at a local maximum, the pattern of accumulation becomes unstable. Note
that an equilibrium with $K_{\hat{X}}>>>1$ is merely possible for $c<<1$.\
Otherwise, there is no equilibrium for $R\left( K_{X},X\right) $ maximum.

\begin{case}
Large capitalization, $K_{\hat{X}}>>1$
\end{case}

For a very large and stable capitalization, i.e. when average capital $K_{%
\hat{X}}$ is large but below a given threshold, we can assume in first
approximation that\textbf{\ }the density of firms in sector $X$ (\ref{rsp})\
becomes:%
\begin{equation}
\left\Vert \Psi \left( X\right) \right\Vert ^{2}\simeq D  \label{Bts}
\end{equation}%
Appendix 4.2.3.2 shows that average capital in sector $X$ is :%
\begin{equation}
K_{X}^{\alpha }\simeq \frac{C\left( \bar{p}\right) \sigma _{\hat{K}%
}^{2}\Gamma \left( \frac{M}{c}\right) }{Df\left( X\right) }+\frac{d}{f\left(
X\right) R\left( X\right) }\left( 1+M\func{Psi}\left( \frac{M}{c}\right)
\left( 1+\frac{\nabla _{\hat{X}}^{2}R\left( X\right) }{M}\right) \right)
\label{sshk}
\end{equation}%
where $\func{Psi}\left( x\right) $\ $=\frac{\Gamma ^{\prime }\left( x\right) 
}{\Gamma \left( x\right) }$, and $d$ and $c$ are some constant parameters.
This solution only holds when $f\left( X\right) >0$\textbf{\ }and\textbf{\ }$%
\frac{C\left( \bar{p}\right) \sigma _{\hat{K}}^{2}\Gamma \left( \frac{M}{c}%
\right) }{Df\left( X\right) }>1$.\ 

Formula (\ref{sshk}) shows that this dependency of $K_{\hat{X}}^{\alpha }$
in $R\left( \hat{X}\right) $ depends in turns\ on the sign of the second
term in the rhs of (\ref{sshk}).

When the condition:%
\begin{equation*}
1+M\func{Psi}\left( \frac{M}{c}\right) \left( 1+\frac{\nabla _{\hat{X}%
}^{2}R\left( \hat{X}\right) }{M}\right) >0
\end{equation*}%
holds, average capital in sector $\hat{X}$, $K_{\hat{X}}^{\alpha }$, is a
decreasing function of both returns $R\left( \hat{X}\right) $ and the
short-term returns $f\left( \hat{X}\right) $. The stability analysis (\ref%
{sTB}) thus implies that the solution (\ref{sshk}) is unstable.

On the contrary, when:%
\begin{equation*}
1+M\func{Psi}\left( \frac{M}{c}\right) \left( 1+\frac{\nabla _{\hat{X}%
}^{2}R\left( \hat{X}\right) }{M}\right) <0
\end{equation*}%
a stable equilibrium is possible. In this case, the average capital in
sector $X$, $K_{\hat{X}}^{\alpha }$ , is increasing with both returns $%
R\left( \hat{X}\right) $ and short-term returns $f\left( \hat{X}\right) $.
This case arises when, for already maximum returns, $\nabla _{\hat{X}%
}^{2}R\left( \hat{X}\right) <<0$, a further increase in long-term returns $%
R\left( \hat{X}\right) $ occurs. This increases the number of firms $%
\left\Vert \Psi \left( \hat{X}\right) \right\Vert ^{2}$ in the sector
without impairing average capital per firm. Note that stable equilibrium is
an extreme case of the next case, intermediate level of capital.

\begin{case}
intermediate case, mid-capitalization $\infty >>K_{\hat{X}}>1$
\end{case}

To solve equation (\ref{qtk}) in this general case, we consider that $\sigma
_{X}^{2}<<1$ and the following simplifying assumptions: 
\begin{equation}
f\left( \hat{X}\right) \simeq B_{2}\left( X\right) K_{\hat{X}}^{\alpha }
\label{fRP}
\end{equation}%
and:%
\begin{equation*}
\left\Vert \Psi \left( \hat{X}\right) \right\Vert ^{2}\simeq D
\end{equation*}%
Eventually, appendix 4.2.3.2 shows that:

\begin{eqnarray}
K_{\hat{X}}^{\alpha } &=&\left( \frac{8C\left( \bar{p}\right) }{D}\sqrt{%
\frac{3\sigma _{\hat{K}}^{2}\left\vert B_{2}\left( X\right) \right\vert }{%
\sigma _{X}^{2}\left( B_{2}^{\prime }\left( X\right) \right) ^{2}}\left( \ln
\left( \bar{p}+\frac{1}{2}\right) -1\right) }\right) ^{\frac{2\alpha }{%
1+\alpha }}  \label{kfcw} \\
&&\times \exp \left( -W_{0}\left( -\frac{48\alpha }{1+\alpha }\left( \sqrt{%
\frac{3\sigma _{\hat{K}}^{2}}{\sigma _{X}^{2}}}\frac{8C\left( \bar{p}\right) 
}{D}\right) ^{\frac{2\alpha }{1+\alpha }}\frac{\left\vert B_{2}\left(
X\right) \right\vert ^{3+\frac{\alpha }{1+\alpha }}}{\sigma _{X}^{2}\sigma _{%
\hat{K}}^{2}\left( B_{2}^{\prime }\left( X\right) \right) ^{2+\frac{2\alpha 
}{1+\alpha }}}\left( \ln \left( \bar{p}+\frac{1}{2}\right) -1\right) ^{2+%
\frac{\alpha }{1+\alpha }}\right) \right)  \notag
\end{eqnarray}%
where $W_{0}$ is the Lambert $W$ function and $\bar{p}$ a constant.

In first approximation, equation (\ref{kfcw}) implies that $K_{\hat{X}%
}^{\alpha }$\ is an increasing function of $B_{2}\left( X\right) $. Given
our simplifying assumption (\ref{fRP}), average capital is higher in high
short-term returns sectors.

Moreover, $K_{\hat{X}}^{\alpha }$\ is a decreasing function of $\left(
\nabla _{\hat{X}}R\left( \hat{X}\right) \right) ^{2}$\ and $\nabla _{\hat{X}%
}^{2}R\left( \hat{X}\right) $: capital accumulation is locally maximal when
expected returns $R\left( \hat{X}\right) $ of sector $\hat{X}$\ are at a
local maxima, i.e. $\left( \nabla _{\hat{X}}R\left( \hat{X}\right) \right)
^{2}=0$\ and $\nabla _{\hat{X}}^{2}R\left( \hat{X}\right) <0$.

Thus, in the intermediate case, the average values $K_{\hat{X}}$ are stable.
In addition, both short-term and long term returns matter in the
intermediate range.

\begin{case}
Small capitalization $K_{\hat{X}}<<1$
\end{case}

When average physical capital per firm in sector $\hat{X}$ is very low, we
can use our assumptions about $g\left( \hat{X}\right) $ equation (\ref{gtp}%
), and assume that:%
\begin{equation}
f\left( \hat{X}\right) \simeq B_{1}\left( \hat{X}\right) K_{\hat{X}}^{\alpha
-1}>>1\text{, }g\left( \hat{X}\right) \simeq 0  \label{srt}
\end{equation}%
and:%
\begin{equation*}
\left\Vert \Psi \left( \hat{X}\right) \right\Vert ^{2}\simeq D
\end{equation*}%
For these conditions, the solution of (\ref{qtk}) is locally stable. We show
in appendix 4.2.3.2 that the solution for average capital is at the first
order\footnote{%
Given our hypotheses, $D>>1$\ , which implies that $K_{\hat{X}}<<1$,\ as
needed.}:%
\begin{equation}
K_{\hat{X}}=\left( \frac{C\left( \bar{p}\right) \sigma _{\hat{K}}^{2}\hat{%
\Gamma}\left( -1\right) }{DB_{1}\left( \hat{X}\right) }\right) ^{\frac{1}{%
\alpha }}+\frac{\frac{C\left( \bar{p}\right) \sigma _{\hat{K}}^{2}}{D}\hat{%
\Gamma}^{\prime }\left( -1\right) \left( M-\left( \frac{\left( g\left( \hat{X%
}\right) \right) ^{2}}{\sigma _{\hat{X}}^{2}}+\nabla _{\hat{X}}g\left( \hat{X%
}\right) \right) \right) }{B_{1}^{\frac{1}{\alpha }}\left( \hat{X}\right)
\left( \frac{C\left( \bar{p}\right) \sigma _{\hat{K}}^{2}\hat{\Gamma}\left(
-1\right) }{D}\right) ^{1-\frac{1}{\alpha }}}  \label{RS}
\end{equation}

Equation (\ref{RS}) shows that average capital $K_{\hat{X}}$\ increases with 
$M-\left( \frac{\left( g\left( \hat{X}\right) \right) ^{2}}{\sigma _{\hat{X}%
}^{2}}+\nabla _{\hat{X}}g\left( \hat{X}\right) \right) $: when expected
long-term returns increase, more capital is allocated to the sector.
Equation (\ref{srt}) also shows that average capital $K_{\hat{X}}$\ is
maximal when returns $R\left( \hat{X}\right) $\ are at a local maximum, i.e.
when $\frac{\left( g\left( \hat{X}\right) \right) ^{2}}{\sigma _{\hat{X}}^{2}%
}=0$\ and $\nabla _{\hat{X}}g\left( \hat{X},K_{\hat{X}}\right) <0$.

Inversely, the same equations (\ref{RS}) and (\ref{srt}) show that average
capital $K_{\hat{X}}$ is decreasing in $f\left( \hat{X}\right) $. The
equilibrium is unstable. Recall that in this unstable equilibrium, $K_{\hat{X%
}}$ must be seen as a threshold.\ The rise in $f\left( \hat{X}\right) $
reduces the threshold $K_{\hat{X}}$, which favours\ capital accumulation and
increases the average capital $K_{\hat{X}}$. Actually, when average capital
is very low, i.e. $K_{\hat{X}}<<1$, which is the case studied here, marginal
returns are high.\ Any increase in capital above the threshold $K_{\hat{X}}$%
, or any shift reducing the threshold, widely increases returns, which
drives capital towards the next stable equilibrium, with higher $K_{\hat{X}}$%
.

This case is thus an exception: the dependency of $K_{\hat{X}}$\ in $R\left( 
\hat{X}\right) $\ is stable, but the dependency in $f\left( \hat{X}\right) $%
\ is unstable. This saddle path type of instability may lead the sector,
either towards a higher level of capital (case 4 below) or towards $0$.
where the sector disappears.

\subsubsection*{Remark: The case of negative short-term returns $f<0$}

In the four cases described above, we have only considered the case where a
sector $\hat{X}$\ short-term returns are positive $f\left( \hat{X}\right) >0$%
. We can nonetheless extend our analysis to the case $f\left( \hat{X}\right)
<0$.

In such a case, the equilibria, whether stable or unstable, defined in cases
1, 2 with $K_{\hat{X}}>>1$, and 4 with $K_{\hat{X}}>1$, are still valid, and
capital allocation relies on expectations of high long-term returns. If we
consider that $f\left( \hat{X}\right) <0$\ is an extreme case, where
expectations of large future profits must offset short-term losses. However,
such equilibria become unsustainable when $R\left( \hat{X}\right) $\
decreases to such an extent that it does not compensate for the loss $%
f\left( \hat{X}\right) $\textbf{. }Case 3, $K_{\hat{X}}<1$\ is the only case
that is no longer possible when $f\left( \hat{X}\right) <0$, since the
returns that matter in this case are dividends. If they turn negative, the
equilibrium is no longer sustainable.

\section*{A 4.2 Details of the computations}

\subsubsection*{A 4.2.1 First approach: Differential form of (\protect\ref%
{qtk})}

To understand the behavior of the solutions of (\ref{qtk}), we can write its
differential version. Assume a variation $\delta Y\left( \hat{X}\right) $
for any parameter of the system at point $\hat{X}$. This parameter $Y\left( 
\hat{X}\right) $ can be either $R\left( X\right) $, its gradient, or any
parameter arising in the definition of $f$ and $g$. This induces a variation 
$\delta K_{\hat{X}}$ for the average capital. The equation for $\delta K_{%
\hat{X}}$ is obtained by differentiation of (\ref{qtk}):

\begin{eqnarray}
\delta K_{\hat{X}} &=&\left( -\left( \frac{\frac{\partial f\left( \hat{X},K_{%
\hat{X}}\right) }{\partial K_{\hat{X}}}}{f\left( \hat{X},K_{\hat{X}}\right) }%
+\frac{\frac{\partial \left\Vert \Psi \left( \hat{X},K_{\hat{X}}\right)
\right\Vert ^{2}}{\partial K_{\hat{X}}}}{\left\Vert \Psi \left( \hat{X},K_{%
\hat{X}}\right) \right\Vert ^{2}}+l\left( \hat{X},K_{\hat{X}}\right) \right)
+k\left( p\right) \frac{\partial p}{\partial K_{\hat{X}}}\right) K_{\hat{X}%
}\delta K_{\hat{X}}  \label{dvr} \\
&&+\frac{\partial }{\partial Y\left( \hat{X}\right) }\left( \frac{\sigma _{%
\hat{K}}^{2}C\left( \bar{p}\right) 2\hat{\Gamma}\left( p+\frac{1}{2}\right) 
}{\left\vert f\left( \hat{X},K_{\hat{X}}\right) \right\vert \left\Vert \Psi
\left( \hat{X},K_{\hat{X}}\right) \right\Vert ^{2}}\right) \delta Y\left( 
\hat{X}\right)  \notag
\end{eqnarray}%
where we define:%
\begin{eqnarray*}
l\left( \hat{X},K_{\hat{X}}\right) &=&\frac{\sigma _{X}^{2}\sigma _{\hat{K}%
}^{2}\left( \nabla _{K_{\hat{X}}}\left( f^{\prime }\left( \hat{X}\right)
\right) ^{2}\left\vert f\left( \hat{X}\right) \right\vert -3\left( \nabla
_{K_{\hat{X}}}\left\vert f\left( \hat{X}\right) \right\vert \right) \left(
f^{\prime }\left( \hat{X}\right) \right) ^{2}\right) \left( p+\frac{1}{2}%
\right) ^{2}}{120\left\vert f\left( \hat{X}\right) \right\vert ^{4}} \\
&&+\frac{\partial p}{\partial K_{\hat{X}}}\frac{\sigma _{X}^{2}\sigma _{\hat{%
K}}^{2}\left( p+\frac{1}{2}\right) \left( f^{\prime }\left( X\right) \right)
^{2}}{48\left\vert f\left( \hat{X}\right) \right\vert ^{3}}
\end{eqnarray*}%
\begin{equation}
k\left( p\right) =\frac{\frac{d}{dp}\hat{\Gamma}\left( p+\frac{1}{2}\right) 
}{\hat{\Gamma}\left( p+\frac{1}{2}\right) }\sim _{\infty }\sqrt{\frac{p-%
\frac{1}{2}}{2}}-\frac{\sigma _{X}^{2}\sigma _{\hat{K}}^{2}\left( p+\frac{1}{%
2}\right) \left( f^{\prime }\left( X\right) \right) ^{2}}{48\left\vert
f\left( \hat{X}\right) \right\vert ^{3}}
\end{equation}%
and:%
\begin{equation*}
\frac{\partial p}{\partial K_{\hat{X}}}=\frac{\partial }{\partial K_{\hat{X}}%
}\frac{M-A\left( \hat{X},K_{\hat{X}}\right) }{\left\vert f\left( \hat{X},K_{%
\hat{X}}\right) \right\vert }=-\frac{\partial _{K_{\hat{X}}}\left\vert
f\left( \hat{X},K_{\hat{X}}\right) \right\vert p+\partial _{K_{\hat{X}%
}}A\left( \hat{X},K_{\hat{X}}\right) }{\left\vert f\left( \hat{X},K_{\hat{X}%
}\right) \right\vert }
\end{equation*}%
with:%
\begin{eqnarray*}
A\left( \hat{X},K_{\hat{X}}\right) &=&\frac{\left( g\left( \hat{X},K_{\hat{X}%
}\right) \right) ^{2}}{\sigma _{\hat{X}}^{2}}+\left( f\left( \hat{X},K_{\hat{%
X}}\right) +\frac{\left\vert f\left( \hat{X},K_{\hat{X}}\right) \right\vert 
}{2}+\nabla _{\hat{X}}g\left( \hat{X},K_{\hat{X}}\right) -\frac{\sigma _{%
\hat{K}}^{2}F^{2}\left( \hat{X},K_{\hat{X}}\right) }{2f^{2}\left( \hat{X},K_{%
\hat{X}}\right) }\right) \\
&\simeq &\frac{\left( g\left( \hat{X},K_{\hat{X}}\right) \right) ^{2}}{%
\sigma _{\hat{X}}^{2}}+f\left( \hat{X},K_{\hat{X}}\right) +\frac{\left\vert
f\left( \hat{X},K_{\hat{X}}\right) \right\vert }{2}+\nabla _{\hat{X}}g\left( 
\hat{X},K_{\hat{X}}\right)
\end{eqnarray*}

\paragraph*{A. 4.2.1.1 Expanded form of (\protect\ref{dvr})}

In an expanded form (\ref{dvr}) writes:

\begin{eqnarray*}
\delta K_{\hat{X}} &=&\left( k\left( p\right) \partial _{K_{\hat{X}}}\left(
p\right) \right. \\
&&\left. -\left( \frac{\frac{\partial f\left( \hat{X},K_{\hat{X}}\right) }{%
\partial K_{\hat{X}}}\left( p+\mathcal{H}\left( f\left( \hat{X},K_{\hat{X}%
}\right) \right) +\frac{1}{2}\right) k\left( p\right) }{f\left( \hat{X},K_{%
\hat{X}}\right) }+\frac{\frac{\partial \left\Vert \Psi \left( \hat{X},K_{%
\hat{X}}\right) \right\Vert ^{2}}{\partial K_{\hat{X}}}}{\left\Vert \Psi
\left( \hat{X},K_{\hat{X}}\right) \right\Vert ^{2}}+l\left( \hat{X},K_{\hat{X%
}}\right) \right) \right) K_{\hat{X}}\delta K_{\hat{X}} \\
&&+\frac{\partial }{\partial Y}\left( \frac{\sigma _{\hat{K}}^{2}C\left( 
\bar{p}\right) 2\hat{\Gamma}\left( p+\frac{1}{2}\right) }{\left\vert f\left( 
\hat{X},K_{\hat{X}}\right) \right\vert \left\Vert \Psi \left( \hat{X},K_{%
\hat{X}}\right) \right\Vert ^{2}}\right) \delta Y
\end{eqnarray*}%
with $\mathcal{H}$ the heaviside function. Moreover:%
\begin{eqnarray*}
&&\frac{\partial }{\partial Y}\left( \frac{\sigma _{\hat{K}}^{2}C\left( \bar{%
p}\right) 2\hat{\Gamma}\left( p+\frac{1}{2}\right) }{\left\vert f\left( \hat{%
X},K_{\hat{X}}\right) \right\vert \left\Vert \Psi \left( \hat{X},K_{\hat{X}%
}\right) \right\Vert ^{2}}\right) \delta Y \\
&=&\left( k\left( p\right) \partial _{Y}p\right. \\
&&\left. -\left( \frac{\frac{\partial f\left( \hat{X},K_{\hat{X}}\right) }{%
\partial Y}\left( p+\mathcal{H}\left( f\left( \hat{X},K_{\hat{X}}\right)
\right) +\frac{1}{2}\right) k\left( p\right) }{f\left( \hat{X},K_{\hat{X}%
}\right) }+\frac{\frac{\partial \left\Vert \Psi \left( \hat{X},K_{\hat{X}%
}\right) \right\Vert ^{2}}{\partial Y}}{\left\Vert \Psi \left( \hat{X},K_{%
\hat{X}}\right) \right\Vert ^{2}}+m_{Y}\left( \hat{X},K_{\hat{X}}\right)
\right) \right) K_{\hat{X}}\delta Y
\end{eqnarray*}%
with:%
\begin{eqnarray*}
m_{Y}\left( \hat{X},K_{\hat{X}}\right) &=&\frac{\sigma _{X}^{2}\sigma _{\hat{%
K}}^{2}\left( \nabla _{Y}\left( f^{\prime }\left( \hat{X}\right) \right)
^{2}\left\vert f\left( \hat{X}\right) \right\vert -3\left( \nabla
_{Y}\left\vert f\left( \hat{X}\right) \right\vert \right) \left( f^{\prime
}\left( \hat{X}\right) \right) ^{2}\right) \left( p+\frac{1}{2}\right) ^{2}}{%
120\left\vert f\left( \hat{X}\right) \right\vert ^{4}} \\
&&+\nabla _{Y}p\frac{\sigma _{X}^{2}\sigma _{\hat{K}}^{2}\left( p+\frac{1}{2}%
\right) \left( f^{\prime }\left( X\right) \right) ^{2}}{48\left\vert f\left( 
\hat{X}\right) \right\vert ^{3}}
\end{eqnarray*}%
so that:%
\begin{eqnarray*}
\frac{\delta K_{\hat{X}}}{K_{\hat{X}}} &=&\left( k\left( p\right) \partial
_{Y}\left( p\right) \right. \\
&&\left. -\left( \frac{\frac{\partial f\left( \hat{X},K_{\hat{X}}\right) }{%
\partial Y}\left( p+\mathcal{H}\left( f\left( \hat{X},K_{\hat{X}}\right)
\right) +\frac{1}{2}\right) k\left( p\right) }{f\left( \hat{X},K_{\hat{X}%
}\right) }+\frac{\frac{\partial \left\Vert \Psi \left( \hat{X},K_{\hat{X}%
}\right) \right\Vert ^{2}}{\partial Y}}{\left\Vert \Psi \left( \hat{X},K_{%
\hat{X}}\right) \right\Vert ^{2}}+m_{Y}\left( \hat{X},K_{\hat{X}}\right)
\right) \right) \frac{\delta Y}{D}
\end{eqnarray*}%
with:%
\begin{eqnarray}
D &=&1+\left( \left( \frac{\frac{\partial f\left( \hat{X},K_{\hat{X}}\right) 
}{\partial K_{\hat{X}}}\left( p+\mathcal{H}\left( f\left( \hat{X},K_{\hat{X}%
}\right) \right) +\frac{1}{2}\right) k\left( p\right) }{f\left( \hat{X},K_{%
\hat{X}}\right) }+\frac{\frac{\partial \left\Vert \Psi \left( \hat{X},K_{%
\hat{X}}\right) \right\Vert ^{2}}{\partial K_{\hat{X}}}}{\left\Vert \Psi
\left( \hat{X},K_{\hat{X}}\right) \right\Vert ^{2}}+l\left( \hat{X},K_{\hat{X%
}}\right) \right) \right.  \label{PRD} \\
&&\left. -k\left( p\right) \partial _{K_{\hat{X}}}\left( p\right) \right) K_{%
\hat{X}}  \notag
\end{eqnarray}

\paragraph{A. 4.2.1.2 Local stability}

As explained in the text, equation (\ref{rvd}) can be understood as\ the
fixed-point equation of a dynamical system through the following mechanism%
\textbf{. }

Each variation $\delta Y\left( \hat{X}\right) $\ in the parameters impacts
the average capital, which must then be computed with the new parameters.
The first change induced is written $\delta K_{\hat{X}}^{\left( 1\right) }$:%
\begin{equation}
\delta K_{\hat{X}}^{\left( 1\right) }=\frac{\partial }{\partial Y\left( \hat{%
X}\right) }\left( \frac{\sigma _{\hat{K}}^{2}C\left( \bar{p}\right) 2\hat{%
\Gamma}\left( p+\frac{1}{2}\right) }{\left\vert f\left( \hat{X},K_{\hat{X}%
}\right) \right\vert \left\Vert \Psi \left( \hat{X},K_{\hat{X}}\right)
\right\Vert ^{2}}\right) \delta Y\left( \hat{X}\right)  \label{dk1}
\end{equation}
In a second step, the variation $\delta K_{\hat{X}}$ impacts the various
functions implied in (\ref{qtk}), and indirectly modifies $K_{\hat{X}}$
through the first term in the rhs of (\ref{rvd}):%
\begin{equation}
\left( -\left( \frac{\frac{\partial f\left( \hat{X},K_{\hat{X}}\right) }{%
\partial K_{\hat{X}}}}{f\left( \hat{X},K_{\hat{X}}\right) }+\frac{\frac{%
\partial \left\Vert \Psi \left( \hat{X},K_{\hat{X}}\right) \right\Vert ^{2}}{%
\partial K_{\hat{X}}}}{\left\Vert \Psi \left( \hat{X},K_{\hat{X}}\right)
\right\Vert ^{2}}\right) +k\left( p\right) \frac{\partial p}{\partial K_{%
\hat{X}}}\right) K_{\hat{X}}\delta K_{\hat{X}}^{\left( 1\right) }
\label{dk2}
\end{equation}%
These two effects combined, (\ref{dk1}) and (\ref{dk2}), yield the total
variation $\delta K_{\hat{X}}$.

Importantly, note that if we can interpret $\delta K_{\hat{X}}^{\left(
1\right) }$\ as a variation at time $t$, we can also infer from the indirect
effect (\ref{dk2}) that $\delta K_{\hat{X}}$\ is itself a variation at time $%
t+1$. Equation (\ref{rvd}) can thus be seen as the fixed point equation of a
dynamical system written:%
\begin{eqnarray}
\delta K_{\hat{X}}\left( t+1\right) &=&\left( -\left( \frac{\frac{\partial
f\left( \hat{X},K_{\hat{X}}\right) }{\partial K_{\hat{X}}}}{f\left( \hat{X}%
,K_{\hat{X}}\right) }+\frac{\frac{\partial \left\Vert \Psi \left( \hat{X},K_{%
\hat{X}}\right) \right\Vert ^{2}}{\partial K_{\hat{X}}}}{\left\Vert \Psi
\left( \hat{X},K_{\hat{X}}\right) \right\Vert ^{2}}+l\left( \hat{X},K_{\hat{X%
}}\right) \right) +k\left( p\right) \frac{\partial p}{\partial K_{\hat{X}}}%
\right) K_{\hat{X}}\delta K_{\hat{X}}\left( t\right)  \label{dts} \\
&&+\frac{\partial }{\partial Y\left( \hat{X},t\right) }\left( \frac{\sigma _{%
\hat{K}}^{2}C\left( \bar{p}\right) 2\hat{\Gamma}\left( p+\frac{1}{2}\right) 
}{\left\vert f\left( \hat{X},K_{\hat{X}}\right) \right\vert \left\Vert \Psi
\left( \hat{X},K_{\hat{X}}\right) \right\Vert ^{2}}\right) \delta Y\left( 
\hat{X},t\right)  \notag
\end{eqnarray}%
whose fixed point is the solution of (\ref{rvd}):%
\begin{equation}
\delta K_{\hat{X}}=\frac{\frac{\partial }{\partial Y\left( \hat{X}\right) }%
\left( \frac{\sigma _{\hat{K}}^{2}C\left( \bar{p}\right) 2\hat{\Gamma}\left(
p+\frac{1}{2}\right) }{\left\vert f\left( \hat{X},K_{\hat{X}}\right)
\right\vert \left\Vert \Psi \left( \hat{X},K_{\hat{X}}\right) \right\Vert
^{2}}\right) }{1+\left( \left( \frac{\frac{\partial f\left( \hat{X},K_{\hat{X%
}}\right) }{\partial K_{\hat{X}}}}{f\left( \hat{X},K_{\hat{X}}\right) }+%
\frac{\frac{\partial \left\Vert \Psi \left( \hat{X},K_{\hat{X}}\right)
\right\Vert ^{2}}{\partial K_{\hat{X}}}}{\left\Vert \Psi \left( \hat{X},K_{%
\hat{X}}\right) \right\Vert ^{2}}+l\left( \hat{X},K_{\hat{X}}\right) \right)
-k\left( p\right) \frac{\partial p}{\partial K_{\hat{X}}}\right) K_{\hat{X}}}%
\delta Y\left( \hat{X}\right)
\end{equation}%
This solution (\ref{sol}) is stable when: 
\begin{subequations}
\begin{equation}
\left\vert k\left( p\right) \frac{\partial p}{\partial K_{\hat{X}}}-\left( 
\frac{\frac{\partial f\left( \hat{X},K_{\hat{X}}\right) }{\partial K_{\hat{X}%
}}}{f\left( \hat{X},K_{\hat{X}}\right) }+\frac{\frac{\partial \left\Vert
\Psi \left( \hat{X},K_{\hat{X}}\right) \right\Vert ^{2}}{\partial K_{\hat{X}}%
}}{\left\Vert \Psi \left( \hat{X},K_{\hat{X}}\right) \right\Vert ^{2}}%
+l\left( \hat{X},K_{\hat{X}}\right) \right) \right\vert <1
\end{equation}%
i.e. when $D$, defined in (\ref{D}), is positive, and unstable otherwise. So
that the stability of this average capital depends, in last analysis, on the
sign of $D$.

\paragraph*{A 4.2.1.3 Applications of the differential form: dependency in
expected returns}

The main application of equation (\ref{dtt}) is to consider a parameter
denoted $Y(\hat{X})$, that encompasses the relative expected returns of
sector $X$ vis-\`{a}-vis its neighbouring sectors, and defined as: 
\end{subequations}
\begin{equation}
Y(\hat{X})=p\left( \hat{X}\right)
\end{equation}%
Interpretations are given in the text. To compute the dependency of
averagecapital in this parameter, we use (\ref{dtt}), and we have:\textbf{\ }%
\begin{equation}
\frac{\delta K_{\hat{X}}}{K_{\hat{X}}}=\frac{\frac{k\left( p\right) }{%
f\left( \hat{X},K_{\hat{X}}\right) }}{D}\delta p\left( \hat{X}\right)
\end{equation}%
Given equation (\ref{kdp}), $k\left( p\right) $ is positive at the first
order in $\sigma _{X}^{2}$. \ More precisely, using equation (\ref{kdp}):%
\begin{equation*}
k\left( p\right) \sim _{\infty }\sqrt{\frac{p-\frac{1}{2}}{2}}-\frac{\sigma
_{X}^{2}\sigma _{\hat{K}}^{2}\left( p+\frac{1}{2}\right) \left( f^{\prime
}\left( X\right) \right) ^{2}}{48\left\vert f\left( \hat{X}\right)
\right\vert ^{3}}
\end{equation*}%
along with equation (\ref{plb}), we can infer that $\sqrt{\frac{p-\frac{1}{2}%
}{2}}$\ is of order $\frac{1}{\sigma _{X}}$\ and $\frac{\sigma
_{X}^{2}\sigma _{\hat{K}}^{2}\left( p+\frac{1}{2}\right) \left( f^{\prime
}\left( X\right) \right) ^{2}}{48\left\vert f\left( \hat{X}\right)
\right\vert ^{3}}\sim 1$.

Consequently, in a stable equilibrium, i.e. for $D>0$, equation (\ref{drt})
implies that the dependency of $K_{\hat{X}}$ in the parameter $p\left( \hat{X%
}\right) $ is positive:%
\begin{equation*}
\frac{\delta K_{\hat{X}}}{\delta p\left( \hat{X}\right) }>0
\end{equation*}%
We have seen above that $p\left( \hat{X}\right) $\ is maximal for a maximum
expected long-term return $R\left( \hat{X},K_{\hat{X}}\right) $: when the
equilibrium is stable, capital accumulation is maximal for sectors that are
themselves a local maximum for $R\left( \hat{X},K_{\hat{X}}\right) $.

On the other hand, when the equilibrium is unstable, i.e. for $D<0$, the
capital $K_{\hat{X}}$\ is minimal for $R\left( \hat{X},K_{\hat{X}}\right) $\
maximal.

Actually, as seen above, in the instability range $D<0$\ ,the average
capital $K_{\hat{X}}$\ acts as a threshold. When, due to variations in the
system's parameters, the average capital per firm is shifted above the
threshold $K_{\hat{X}}$, \ capital will either move to the next stable
equilibrium, possibly zero, or tend to infinity. Our results show that when
the expected long-term return of a sector increases, the threshold $K_{\hat{X%
}}$ decreases, which favours capital accumulation.

\paragraph*{A 4.2.1.3 Applications of the differential form: dependency in
short term returns}

A second use of equation (\ref{dtt}) is to consider $Y(\hat{X})$ as any
parameter-function involved in the definition of $f\left( \hat{X},K_{\hat{X}%
}\right) $ that may condition either real short-term returns or the
price-dividend ratio.

We can see that in this case, $Y(\hat{X})$ only impacts $f\left( \hat{X},K_{%
\hat{X}}\right) $, so that equation (\ref{dtt}) simplifies and yields: 
\begin{eqnarray}
\frac{\delta K_{\hat{X}}}{K_{\hat{X}}} &=&-\frac{m_{Y}\left( \hat{X},K_{\hat{%
X}}\right) }{D}\delta Y  \label{dkx} \\
&&-\frac{1}{D}\left( \frac{\frac{\partial f\left( \hat{X},K_{\hat{X}}\right) 
}{\partial Y}\left( 1+\left( p+H\left( f\left( \hat{X},K_{\hat{X}}\right)
\right) +\frac{1}{2}\right) k\left( p\right) \right) }{f\left( \hat{X},K_{%
\hat{X}}\right) }+\frac{\frac{\partial \left\Vert \Psi \left( \hat{X},K_{%
\hat{X}}\right) \right\Vert ^{2}}{\partial Y}}{\left\Vert \Psi \left( \hat{X}%
,K_{\hat{X}}\right) \right\Vert ^{2}}\right) \delta Y  \notag
\end{eqnarray}%
Incidentally, note that $p$ being proportional to $f^{-1}\left( \hat{X}%
\right) $, $m_{Y}\left( \hat{X},K_{\hat{X}}\right) $ rewrites: 
\begin{eqnarray}
-m_{Y}\left( \hat{X},K_{\hat{X}}\right) &=&\frac{\sigma _{X}^{2}\sigma _{%
\hat{K}}^{2}\left( 3\left( \nabla _{Y}\left\vert f\left( \hat{X}\right)
\right\vert \right) \left( f^{\prime }\left( \hat{X}\right) \right)
^{2}-\nabla _{Y}\left( f^{\prime }\left( \hat{X}\right) \right)
^{2}\left\vert f\left( \hat{X}\right) \right\vert \right) \left( p+\frac{1}{2%
}\right) ^{2}}{120\left\vert f\left( \hat{X}\right) \right\vert ^{4}}
\label{mx} \\
&&+\nabla _{Y}\left\vert f\left( \hat{X}\right) \right\vert \frac{\sigma
_{X}^{2}\sigma _{\hat{K}}^{2}p\left( p+\frac{1}{2}\right) \left( f^{\prime
}\left( X\right) \right) ^{2}}{48\left\vert f\left( \hat{X}\right)
\right\vert ^{4}}  \notag
\end{eqnarray}%
The first term in the rhs of (\ref{dkx}) is the impact of an increase in
investors' short-term returns.\ The second is the variation in capital
needed to maintain investors' overall returns.

The sign of $\frac{\delta K_{\hat{X}}}{K_{\hat{X}}}$ given by equation (\ref%
{dkx}) can be studied under two cases: the stable and the unstable
equilibrium.

Let us first consider the case of a stable equilibrium, i.e. $D>0$.

The first term in the rhs of (\ref{dkx}), the variation induced by an
increase in short-term returns, is in general positive for $f^{\prime
}\left( \hat{X}\right) $ proportional to $f\left( \hat{X}\right) $, that is
for instance when the function $f\left( \hat{X}\right) $, that describes
short-term returns and prices, depends on the variable $K_{\hat{X}}$ raised
to some arbitrary power.

Indeed in that case:%
\begin{equation*}
3\left( \nabla _{Y}\left\vert f\left( \hat{X}\right) \right\vert \right)
\left( f^{\prime }\left( \hat{X}\right) \right) ^{2}-\nabla _{Y}\left(
f^{\prime }\left( \hat{X}\right) \right) ^{2}\left\vert f\left( \hat{X}%
\right) \right\vert =\left( \nabla _{Y}\left\vert f\left( \hat{X}\right)
\right\vert \right) \left( f^{\prime }\left( \hat{X}\right) \right) ^{2}
\end{equation*}%
The second term in the rhs of (\ref{dkx}) is in general negative. When $%
\frac{\partial f\left( \hat{X},K_{\hat{X}}\right) }{\partial Y}>0$, i.e.
when returns are increasing in $Y$, a rise in $Y$\ increases returns and
decreases the capital needed to maintain these returns.\textbf{\ }Similarly,
when $\frac{\partial \left\Vert \Psi \left( \hat{X},K_{\hat{X}}\right)
\right\Vert ^{2}}{\partial Y}>0$, i.e. when the number of agents in sector $%
\hat{X}\ $is increasing in $Y$, a rise in $Y$\ increases the number of
agents that move towards point $\hat{X}$, and the average capital per firm
diminishes.

The net variation (\ref{dkx}) of $K_{\hat{X}}$ is the sum of these two
contributions. Considering an expansion of (\ref{dkx}) in powers of $\sigma
_{X}^{2}$, the first contribution $-m_{Y}\left( \hat{X},K_{\hat{X}}\right) $
is of magnitude $\left( \sigma _{X}^{2}\right) ^{-1}$, whereas the second is
proportional to $k\left( p\right) \sim $ $\left( \sigma _{X}\right) ^{-1}$.
The variation $\frac{\delta K_{\hat{X}}}{K_{\hat{X}}}$ is thus positive: $%
\frac{\delta K_{\hat{X}}}{K_{\hat{X}}}>0$. In most cases, a higher
short-term return, decomposed as a sum of dividend and price variation,
induces a higher average capital. This effect is magnified for larger levels
of capital: the third approach will confirm that, in most cases, the return $%
f\left( \hat{X}\right) $ is asymptotically a constant $c<<1$ when capital is
high: $K_{\hat{X}}>>1$.

Turning now to the case of an unstable equilibrium, i.e. $D<0$, the
variation $\frac{\delta K_{\hat{X}}}{K_{\hat{X}}}$ is negative: $\frac{%
\delta K_{\hat{X}}}{K_{\hat{X}}}<0$. In the instability range, and due to
this very instability, an increase in returns $f\left( \hat{X}\right) $
reduces the threshold of capital accumulation for low levels of capital.
When short-term returns $f\left( \hat{X}\right) $\ increase, a lower average
capital will trigger capital accumulation towards an equilibrium. Otherwise,
when average capital $K_{\hat{X}}$ is below this threshold, it will converge
toward $0$.

\subsubsection*{A 4.2.2 Second approach: Expansion around particular
solutions}

As explained in the text, we choose to expand (\ref{Nkv}), or equivalently (%
\ref{qnk}), around solutions with $p=0$.

\paragraph*{A 4.2.2.1 Equation (\protect\ref{qtk}) for $p=0$}

To find the solution with $p=0$, we maximize the function:%
\begin{equation*}
A\left( \hat{X}\right) =\frac{\left( g\left( \hat{X}\right) \right) ^{2}}{%
\sigma _{\hat{X}}^{2}}+f\left( \hat{X}\right) +\frac{1}{2}\sqrt{f^{2}\left( 
\hat{X}\right) }+\nabla _{\hat{X}}g\left( \hat{X},K_{\hat{X}}\right) -\frac{%
\sigma _{\hat{K}}^{2}F^{2}\left( \hat{X},K_{\hat{X}}\right) }{2f^{2}\left( 
\hat{X}\right) }
\end{equation*}%
We write:%
\begin{equation}
M=\max_{\hat{X}}A\left( \hat{X}\right)  \label{Mqn}
\end{equation}%
and denote by $\left( \hat{X}_{M},K_{\hat{X}_{M}}\right) $ the solutions $%
\hat{X}_{M}$ of (\ref{Mqn}) with $K_{\hat{X}_{M}}$ their associated value of
average capital per firm.

Given that 
\begin{equation}
\hat{\Gamma}\left( \frac{1}{2}\right) =\exp \left( -\frac{\sigma
_{X}^{2}\sigma _{\hat{K}}^{2}\left( f^{\prime }\left( X,K_{\hat{X},M}\right)
\right) ^{2}}{384\left\vert f\left( \hat{X},K_{\hat{X},M}\right) \right\vert
^{3}}\right)  \label{Ghf}
\end{equation}
(\ref{Nkv}) becomes at points $\left( \hat{X}_{M},K_{\hat{X}_{M}}\right) $
and $p=0$:%
\begin{equation}
K_{\hat{X},M}\left\vert f\left( \hat{X}_{M},K_{\hat{X}_{M}}\right)
\right\vert \left\Vert \Psi \left( \hat{X}_{M},K_{\hat{X}_{M}}\right)
\right\Vert ^{2}\simeq \sigma _{\hat{K}}^{2}C\left( \bar{p}\right) \exp
\left( -\frac{\sigma _{X}^{2}\sigma _{\hat{K}}^{2}\left( f^{\prime }\left( 
\hat{X}_{M},K_{\hat{X}_{M}}\right) \right) ^{2}}{384\left\vert f\left( \hat{X%
}_{M},K_{\hat{X}_{M}}\right) \right\vert ^{3}}\right)  \label{qMK}
\end{equation}%
This equation has in general several solutions, depending on the assumptions
on $f\left( \hat{X}_{M},K_{\hat{X}_{M}}\right) $.

Note that once a solution $K_{\hat{X}}$ of (\ref{qnk}) is found, the value
of $C\left( \bar{p}\right) $ can be obtained by solving (\ref{pbr}) and
using (\ref{clk}). These solutions are discussed in the text.

The next paragraph computes the expansion of (\ref{Nkv}) around these
solutions with $p=0$. Remark that coming back to (\ref{Nkv}) and (\ref{qnk})
for general values of $p$ defined in (\ref{pdf}), the value of $C\left( \bar{%
p}\right) \sigma _{\hat{K}}^{2}$\ can be replaced by $K_{\hat{X}%
_{M}}\left\vert f\left( \hat{X}_{M},K_{\hat{X}_{M}}\right) \right\vert
\left\Vert \Psi \left( \hat{X}_{M},K_{\hat{X}_{M}}\right) \right\Vert ^{2}$
for any solution $\left( \hat{X}_{M},K_{\hat{X}_{M}}\right) $.

\paragraph*{A 4.2.2.2 Expansion around particular solutions}

To better understand the behavior of the solutions of equation (\ref{qtk}),
we expand this equation around the points $\left( \hat{X},K_{\hat{X}%
,M}\right) $ that solve equation (\ref{qtk}). We can find approximate
solutions to (\ref{Nkv}): 
\begin{equation}
K_{\hat{X}}\left\Vert \Psi \left( \hat{X}\right) \right\Vert ^{2}\left\vert
f\left( \hat{X}\right) \right\vert =C\left( \bar{p}\right) \sigma _{\hat{K}%
}^{2}\hat{\Gamma}\left( p+\frac{1}{2}\right)  \label{Nkvv}
\end{equation}%
with:%
\begin{eqnarray}
\hat{\Gamma}\left( p+\frac{1}{2}\right) &=&\exp \left( -\frac{\sigma
_{X}^{2}\sigma _{\hat{K}}^{2}\left( p+\frac{1}{2}\right) ^{2}\left(
f^{\prime }\left( X\right) \right) ^{2}}{96\left\vert f\left( \hat{X}\right)
\right\vert ^{3}}\right)  \label{gmhh} \\
&&\times \left( \frac{\Gamma \left( -\frac{p+1}{2}\right) \Gamma \left( 
\frac{1-p}{2}\right) -\Gamma \left( -\frac{p}{2}\right) \Gamma \left( \frac{%
-p}{2}\right) }{2^{p+2}\Gamma \left( -p-1\right) \Gamma \left( -p\right) }+p%
\frac{\Gamma \left( -\frac{p}{2}\right) \Gamma \left( \frac{2-p}{2}\right)
-\Gamma \left( -\frac{p-1}{2}\right) \Gamma \left( -\frac{p-1}{2}\right) }{%
2^{p+1}\Gamma \left( -p\right) \Gamma \left( -p+1\right) }\right)  \notag
\end{eqnarray}%
for general form of the functions $f\left( \hat{X}\right) $ and $g\left( 
\hat{X}\right) $ by expanding (\ref{Nkvv}), for each $\hat{X}$,around the
closest point $\hat{X}_{M}$ satisfying (\ref{Nkvv}) with $p=0$. We use that:%
\begin{eqnarray}
&&\left( \frac{\Gamma \left( -\frac{p+1}{2}\right) \Gamma \left( \frac{1-p}{2%
}\right) -\Gamma \left( -\frac{p}{2}\right) \Gamma \left( \frac{-p}{2}%
\right) }{2^{p+2}\Gamma \left( -p-1\right) \Gamma \left( -p\right) }+p\frac{%
\Gamma \left( -\frac{p}{2}\right) \Gamma \left( \frac{2-p}{2}\right) -\Gamma
\left( -\frac{p-1}{2}\right) \Gamma \left( -\frac{p-1}{2}\right) }{%
2^{p+1}\Gamma \left( -p\right) \Gamma \left( -p+1\right) }\right)
\label{dvp} \\
&=&1-p\left( \gamma _{0}+\ln 2-2\right) +o\left( p\right)  \notag
\end{eqnarray}%
with $\gamma _{0}$ the Euler-Mascheroni constant, as well as the following
relations:%
\begin{equation*}
\nabla _{K_{\hat{X}}}\left( -\frac{\sigma _{X}^{2}\sigma _{\hat{K}%
}^{2}h\left( p\right) \left( f^{\prime }\left( \hat{X}\right) \right) ^{2}}{%
96\left\vert f\left( \hat{X}\right) \right\vert ^{3}}\right) _{p=0}\simeq -%
\frac{\nabla _{K_{\hat{X}}}\left( f^{\prime }\left( \hat{X}\right) \right)
^{2}\left\vert f\left( \hat{X}\right) \right\vert -3\left( \nabla _{K_{\hat{X%
}}}\left\vert f\left( \hat{X}\right) \right\vert \right) \left( f^{\prime
}\left( \hat{X}\right) \right) ^{2}}{120\left\vert f\left( \hat{X}\right)
\right\vert ^{4}}
\end{equation*}%
and:%
\begin{eqnarray*}
&&\nabla _{\hat{X}}\left( -\frac{h\left( p\right) \left( f^{\prime }\left( 
\hat{X}\right) \right) ^{2}}{96\left\vert f\left( \hat{X}\right) \right\vert
^{3}}\right) _{p=0} \\
&\simeq &-\frac{2f^{\prime }\left( X\right) f^{\prime \prime }\left( \hat{X}%
\right) \left\vert f\left( \hat{X}\right) \right\vert -3\left( f^{\prime
}\left( \hat{X}\right) \right) ^{3}}{120\left\vert f\left( \hat{X}\right)
\right\vert ^{4}}=\frac{f^{\prime }\left( X\right) \left( 3\left( f^{\prime
}\left( \hat{X}\right) \right) ^{2}-2f^{\prime \prime }\left( X\right)
\left\vert f\left( \hat{X}\right) \right\vert \right) }{120\left\vert
f\left( \hat{X}\right) \right\vert ^{4}}
\end{eqnarray*}%
the expansion of (\ref{Nkvv}) at the lowest order, is:%
\begin{eqnarray*}
&&\left( 1+\frac{\frac{\partial f\left( \hat{X},K_{\hat{X}}\right) }{%
\partial K_{\hat{X}}}}{f\left( \hat{X},K_{\hat{X}}\right) }+\frac{\frac{%
\partial \left\Vert \Psi \left( \hat{X},K_{\hat{X}}\right) \right\Vert ^{2}}{%
\partial K_{\hat{X}}}}{\left\Vert \Psi \left( \hat{X},K_{\hat{X}}\right)
\right\Vert ^{2}}\right) _{K_{\hat{X},M}}\left( K_{\hat{X}}-K_{\hat{X}%
,M}\right) +\left( \frac{\frac{\partial f\left( \hat{X},K_{\hat{X}}\right) }{%
\partial \hat{X}}}{f\left( \hat{X},K_{\hat{X}}\right) }+\frac{\frac{\partial
\left\Vert \Psi \left( \hat{X},K_{\hat{X}}\right) \right\Vert ^{2}}{\partial 
\hat{X}}}{\left\Vert \Psi \left( \hat{X},K_{\hat{X}}\right) \right\Vert ^{2}}%
\right) _{K_{\hat{X},M}}\left( \hat{X}-\hat{X}_{M}\right) \\
&\simeq &-\left( \sigma _{X}^{2}\sigma _{\hat{K}}^{2}\frac{\nabla _{K_{\hat{X%
}}}\left( f^{\prime }\left( \hat{X}\right) \right) ^{2}\left\vert f\left( 
\hat{X}\right) \right\vert -3\left( \nabla _{K_{\hat{X}}}\left\vert f\left( 
\hat{X}\right) \right\vert \right) \left( f^{\prime }\left( \hat{X}\right)
\right) ^{2}}{120\left\vert f\left( \hat{X}\right) \right\vert ^{4}}\right)
_{K_{\hat{X},M}}\left( K_{\hat{X}}-K_{\hat{X},M}\right) \\
&&-\left( \sigma _{X}^{2}\sigma _{\hat{K}}^{2}\frac{2f^{\prime }\left(
X\right) f^{\prime \prime }\left( \hat{X}\right) \left\vert f\left( \hat{X}%
\right) \right\vert -3\left( f^{\prime }\left( \hat{X}\right) \right) ^{3}}{%
120\left\vert f\left( \hat{X}\right) \right\vert ^{4}}\right) _{K_{\hat{X}%
,M}}\left( \hat{X}-\hat{X}_{M}\right) \\
&&-b\frac{\partial _{K_{\hat{X}}}\left( \frac{\left( g\left( \hat{X},K_{\hat{%
X}}\right) \right) ^{2}}{\sigma _{\hat{X}}^{2}}+\left( f\left( \hat{X},K_{%
\hat{X}}\right) +\frac{\left\vert f\left( \hat{X},K_{\hat{X}}\right)
\right\vert }{2}+\nabla _{\hat{X}}g\left( \hat{X},K_{\hat{X}}\right) \right)
\right) \left( K_{\hat{X}}-K_{\hat{X},M}\right) }{\left\vert f\left( \hat{X}%
,K_{\hat{X}}\right) \right\vert } \\
&&-b\frac{\partial _{\hat{X}}\left( \frac{\left( g\left( \hat{X},K_{\hat{X}%
}\right) \right) ^{2}}{\sigma _{\hat{X}}^{2}}+\left( f\left( \hat{X},K_{\hat{%
X}}\right) +\frac{\left\vert f\left( \hat{X},K_{\hat{X}}\right) \right\vert 
}{2}+\nabla _{\hat{X}}g\left( \hat{X},K_{\hat{X}}\right) \right) \right)
\left( \hat{X}-\hat{X}_{M}\right) }{\left\vert f\left( \hat{X},K_{\hat{X}%
}\right) \right\vert }
\end{eqnarray*}%
Given the maximization (\ref{Mqn}), the two last terms in the right hand
side is equal to $0$.%
\begin{eqnarray}
\left( K_{\hat{X}}-K_{\hat{X},M}\right) &=&\frac{1}{D}\left( \sigma
_{X}^{2}\sigma _{\hat{K}}^{2}\frac{3\left( f^{\prime }\left( \hat{X}\right)
\right) ^{3}-2f^{\prime }\left( X\right) f^{\prime \prime }\left( \hat{X}%
\right) \left\vert f\left( \hat{X}\right) \right\vert }{120\left\vert
f\left( \hat{X}\right) \right\vert ^{4}}\right.  \label{prx} \\
&&\left. -\frac{\frac{\partial f\left( \hat{X},K_{\hat{X}}\right) }{\partial 
\hat{X}}}{f\left( \hat{X},K_{\hat{X}}\right) }-\frac{\frac{\partial
\left\Vert \Psi \left( \hat{X},K_{\hat{X}}\right) \right\Vert ^{2}}{\partial 
\hat{X}}}{\left\Vert \Psi \left( \hat{X},K_{\hat{X}}\right) \right\Vert ^{2}}%
\right) _{K_{\hat{X},M}}\left( \hat{X}-\hat{X}_{M}\right)  \notag \\
&&-\frac{1}{D}\frac{b}{2}\left( \hat{X}-\hat{X}_{M}\right) \nabla _{\hat{X}%
}^{2}\left( \frac{\left( g\left( \hat{X},K_{\hat{X}}\right) \right) ^{2}}{%
\sigma _{\hat{X}}^{2}}+\frac{3}{2}f\left( \hat{X},K_{\hat{X}}\right) +\nabla
_{\hat{X}}g\left( \hat{X},K_{\hat{X}}\right) \right) _{K_{\hat{X},M}}\left( 
\hat{X}-\hat{X}_{M}\right)  \notag
\end{eqnarray}%
with:%
\begin{equation*}
D=\left( 1+\frac{\frac{\partial f\left( \hat{X},K_{\hat{X}}\right) }{%
\partial K_{\hat{X}}}}{f\left( \hat{X},K_{\hat{X}}\right) }+\frac{\frac{%
\partial \left\Vert \Psi \left( \hat{X},K_{\hat{X}}\right) \right\Vert ^{2}}{%
\partial K_{\hat{X}}}}{\left\Vert \Psi \left( \hat{X},K_{\hat{X}}\right)
\right\Vert ^{2}}+\frac{\sigma _{X}^{2}\sigma _{\hat{K}}^{2}\left( \nabla
_{K_{\hat{X}}}\left( f^{\prime }\left( \hat{X}\right) \right) ^{2}\left\vert
f\left( \hat{X}\right) \right\vert -3\left( \nabla _{K_{\hat{X}}}\left\vert
f\left( \hat{X}\right) \right\vert \right) \left( f^{\prime }\left( \hat{X}%
\right) \right) ^{2}\right) }{120\left\vert f\left( \hat{X}\right)
\right\vert ^{4}}\right) _{K_{\hat{X}_{M}}}
\end{equation*}%
and $K_{\hat{X}_{M}}$\ solution of:%
\begin{eqnarray*}
&&K_{\hat{X},M}\left\vert f\left( \hat{X},K_{\hat{X},M}\right) \right\vert
\left\Vert \Psi \left( \hat{X},K_{\hat{X},M}\right) \right\Vert ^{2} \\
&\simeq &\sigma _{\hat{K}}^{2}\exp \left( -\frac{\sigma _{X}^{2}\sigma _{%
\hat{K}}^{2}\left( p+\frac{1}{2}\right) ^{2}\left( f^{\prime }\left(
X\right) \right) ^{2}}{96\left\vert f\left( \hat{X}\right) \right\vert ^{3}}%
\right) C\left( \bar{p}\right) \simeq \sigma _{\hat{K}}^{2}C\left( \bar{p}%
\right)
\end{eqnarray*}%
The maximization condition (\ref{Mqn}) cancels the contribution due to: 
\begin{equation*}
\frac{\left( g\left( \hat{X},K_{\hat{X}}\right) \right) ^{2}}{\sigma _{\hat{X%
}}^{2}}+\left( f\left( \hat{X},K_{\hat{X}}\right) +\frac{\left\vert f\left( 
\hat{X},K_{\hat{X}}\right) \right\vert }{2}+\nabla _{\hat{X}}g\left( \hat{X}%
,K_{\hat{X}}\right) \right)
\end{equation*}%
To find a contribution due to this term, we must expand (\ref{Nkvv}) to the
second order. The second order contributions proportional to $\left( K_{\hat{%
X}}-K_{\hat{X},M}\right) ^{2}$ modifies slightly (\ref{prx}) and the term $%
\left( K_{\hat{X}}-K_{\hat{X},M}\right) \left( \hat{X}-\hat{X}_{M}\right) $
shifts $D$ at the first order. Both modifications do not alter the
interpretation for (\ref{prx}). We can thus consider the sole term:%
\begin{equation*}
\frac{C\left( \bar{p}\right) \sigma _{\hat{K}}^{2}\hat{\Gamma}\left( p+\frac{%
1}{2}\right) }{\left\Vert \Psi \left( \hat{X}\right) \right\Vert
^{2}\left\vert f\left( \hat{X}\right) \right\vert }
\end{equation*}%
Due to (\ref{qnk}), for $H\left( K_{\hat{X}}\right) $ slowly varying, the
contribution due to the derivatives of $\left\Vert \Psi \left( \hat{X}%
\right) \right\Vert ^{2}$ can be neglected. Moreover the contribution due to
the derivative of $\left\vert f\left( \hat{X}\right) \right\vert $ are
negligible with respect to the first order terms. We can thus consider only
the second order contributions due to $\hat{\Gamma}\left( p+\frac{1}{2}%
\right) $. In the rhs of (\ref{gmhh}), the second term is dominant.
Moreover, we can check that in the second order expansion of (\ref{dvp}),
the term in $p^{2}$ can be neglected compared to $-p\left( \gamma _{0}+\ln
2-2\right) $. Consequently, the relevant second order correction to (\ref%
{prx}) is :%
\begin{equation*}
b\left( \hat{X}-\hat{X}_{M}\right) \nabla _{\hat{X}}^{2}p\left( \hat{X}-\hat{%
X}_{M}\right) =b\left( \hat{X}-\hat{X}_{M}\right) \nabla _{\hat{X}%
}^{2}\left( \frac{M-\frac{\left( g\left( \hat{X},K_{\hat{X}}\right) \right)
^{2}}{\sigma _{\hat{X}}^{2}}+\frac{3}{2}f\left( \hat{X},K_{\hat{X}}\right)
+\nabla _{\hat{X}}g\left( \hat{X},K_{\hat{X}}\right) }{\left\vert f\left( 
\hat{X}\right) \right\vert }\right) \left( \hat{X}-\hat{X}_{M}\right)
\end{equation*}%
and the relevant contributions to (\ref{prx}) are:

\begin{eqnarray}
\left( K_{\hat{X}}-K_{\hat{X},M}\right) &=&\frac{1}{D}\left( \sigma
_{X}^{2}\sigma _{\hat{K}}^{2}\frac{3\left( f^{\prime }\left( \hat{X}\right)
\right) ^{3}-2f^{\prime }\left( X\right) f^{\prime \prime }\left( \hat{X}%
\right) \left\vert f\left( \hat{X}\right) \right\vert }{120\left\vert
f\left( \hat{X}\right) \right\vert ^{4}}\right.  \label{prZ} \\
&&\left. -\frac{\frac{\partial f\left( \hat{X},K_{\hat{X}}\right) }{\partial 
\hat{X}}}{f\left( \hat{X},K_{\hat{X}}\right) }-\frac{\frac{\partial
\left\Vert \Psi \left( \hat{X},K_{\hat{X}}\right) \right\Vert ^{2}}{\partial 
\hat{X}}}{\left\Vert \Psi \left( \hat{X},K_{\hat{X}}\right) \right\Vert ^{2}}%
\right) _{K_{\hat{X},M}}\left( \hat{X}-\hat{X}_{M}\right)  \notag \\
&&+\frac{1}{D}\frac{b}{2}\left( \hat{X}-\hat{X}_{M}\right) \nabla _{\hat{X}%
}^{2}\left( \frac{M-\frac{\left( g\left( \hat{X},K_{\hat{X}}\right) \right)
^{2}}{\sigma _{\hat{X}}^{2}}+\frac{3}{2}f\left( \hat{X},K_{\hat{X}}\right)
+\nabla _{\hat{X}}g\left( \hat{X},K_{\hat{X}}\right) }{\left\vert f\left( 
\hat{X}\right) \right\vert }\right) _{K_{\hat{X},M}}\left( \hat{X}-\hat{X}%
_{M}\right)  \notag
\end{eqnarray}

\paragraph*{A 4.2.2.3 Interpretation of (\protect\ref{prZ})}

As in the first approach, $D>0$ corresponds to a stable equilibrium, and $%
D<0 $ to an unstable one. The expansion (\ref{pnm}) describes the local
variations of $K_{\hat{X}}$ in the neighbourhood of the points $K_{\hat{X}%
,M} $. This approximation (\ref{pnm}) suffices to understand the role of the
parameters of the system.

We consider the case of stable equilibria, i.e. $D>0$. Note that under
unstable equilibria, $D<0$, the interpretations are inverted, since $K_{\hat{%
X}}$ is interpreted as a threshold\footnote{%
See the first approach.}.

The equation (\ref{pnm}), that expands average capital at sector $\hat{X}%
_{M} $,\ is composed of a first order and a second order contributions.

The\textbf{\ }first order part in the expansion (\ref{pnm}) writes: 
\begin{equation}
\frac{1}{D}\left( \sigma _{X}^{2}\sigma _{\hat{K}}^{2}\frac{3\left(
f^{\prime }\left( \hat{X}\right) \right) ^{3}-2f^{\prime }\left( X\right)
f^{\prime \prime }\left( \hat{X}\right) \left\vert f\left( \hat{X}\right)
\right\vert }{120\left\vert f\left( \hat{X}\right) \right\vert ^{4}}-\frac{%
\frac{\partial f\left( \hat{X},K_{\hat{X}}\right) }{\partial \hat{X}}}{%
f\left( \hat{X},K_{\hat{X}}\right) }-\frac{\frac{\partial \left\Vert \Psi
\left( \hat{X},K_{\hat{X}}\right) \right\Vert ^{2}}{\partial \hat{X}}}{%
\left\Vert \Psi \left( \hat{X},K_{\hat{X}}\right) \right\Vert ^{2}}\right)
_{K_{\hat{X},M}}\left( \hat{X}-\hat{X}_{M}\right)  \label{pnM}
\end{equation}%
It represents the variation of equilibrium capital as a function of its
position. It is decomposed in three contributions:

For $f^{\prime }\left( \hat{X}\right) >0$, the second contribution in (\ref%
{pnM}):

\begin{equation*}
-\frac{\frac{\partial f\left( \hat{X},K_{\hat{X}}\right) }{\partial \hat{X}}%
\left( \hat{X}-\hat{X}_{M}\right) }{f\left( \hat{X},K_{\hat{X}}\right) }
\end{equation*}%
is positive. It represents the decrease in capital needed to reach
equilibrium. Actually, the return is higher at point $\hat{X}$ than at $\hat{%
X}_{M}$: a lower capital will yield the same overall return at point $\hat{X}
$. On the contrary, the first contribution in (\ref{pnM}): 
\begin{equation*}
\frac{\sigma _{X}^{2}\sigma _{\hat{K}}^{2}\frac{3\left( f^{\prime }\left( 
\hat{X}\right) \right) ^{3}-2f^{\prime }\left( X\right) f^{\prime \prime
}\left( \hat{X}\right) \left\vert f\left( \hat{X}\right) \right\vert }{%
120\left\vert f\left( \hat{X}\right) \right\vert ^{4}}\left( \hat{X}-\hat{X}%
_{M}\right) }{D}
\end{equation*}%
describes the "net" variation of capital due to a variation in $f\left(
X\right) $. When returns are decreasing, i.e. when $f^{\prime }\left( \hat{X}%
\right) >0$ and $f^{\prime \prime }\left( \hat{X}\right) <0$, this first
contribution has the sign of $f^{\prime }\left( \hat{X}\right) $. An
increase in returns attracts capital.

The third term in (\ref{pnM}):%
\begin{equation*}
-\frac{1}{D}\left( \frac{\frac{\partial \left\Vert \Psi \left( \hat{X},K_{%
\hat{X}}\right) \right\Vert ^{2}}{\partial \hat{X}}}{\left\Vert \Psi \left( 
\hat{X},K_{\hat{X}}\right) \right\Vert ^{2}}\right) _{K_{\hat{X},M}}\left( 
\hat{X}-\hat{X}_{M}\right)
\end{equation*}%
represents the number effect.\ Actually, when: 
\begin{equation*}
\frac{\frac{\partial \left\Vert \Psi \left( \hat{X},K_{\hat{X}}\right)
\right\Vert ^{2}}{\partial \hat{X}}}{\left\Vert \Psi \left( \hat{X},K_{\hat{X%
}}\right) \right\Vert ^{2}}>0
\end{equation*}%
the number of agents is higher at $\hat{X}$ than at $\hat{X}_{M}$: the
average capital per agent is reduced.

The second order contribution in (\ref{pnm}) represents the effect of the
neighbouring sector space on each sector. Given the first order condition (%
\ref{mfl}): 
\begin{equation*}
\nabla _{\hat{X}}^{2}\left( \frac{M-A\left( \hat{X}\right) }{f\left( \hat{X}%
\right) }\right) _{K_{\hat{X},M}}=\left( \frac{\nabla _{\hat{X}}^{2}\left(
M-A\left( \hat{X}\right) \right) }{f\left( \hat{X}\right) }\right) _{K_{\hat{%
X},M}}
\end{equation*}%
and since $A\left( \hat{X}_{M}\right) $ is a maximum, we have:%
\begin{equation*}
\left( \hat{X}-\hat{X}_{M}\right) \nabla _{\hat{X}}^{2}\left( \frac{%
M-A\left( \hat{X}\right) }{f\left( \hat{X}\right) }\right) _{K_{\hat{X}%
,M}}\left( \hat{X}-\hat{X}_{M}\right) >0
\end{equation*}

When $f\left( \hat{X}\right) $ is constant, $A\left( \hat{X}_{M}\right) $\
is a local maximum, and $K_{\hat{X}_{M}}$\ is a minimum. To put it
differently, $K_{\hat{X}}$ is a decreasing function of $A\left( \hat{X}%
\right) $.\ This is in line with the definition of $A\left( \hat{X}\right) $%
\footnote{%
See discussions after equations (\ref{mqn}) and (\ref{dsc}).}, which
measures the relative attractiveness of sector $\hat{X}$'s\ neighbours: the
higher $A\left( \hat{X}\right) $, the lower the incentive for capital to
stay in sector $\hat{X}$.

\subsubsection*{A 4.2.3 Third approach: Resolution for particular form for
the functions}

As stated in the text, we can find approximate solutions to (\ref{qnk}) by
choosing some forms for the parameters functions. The solutions are then
studied in some ranges for average capital per firm $K_{X}$: $K_{X}>>1$, $%
K_{X}>>>1$, $K_{X}<<1$\textbf{\ }and the intermediate range $\infty >K_{X}>1$
In the case $K_{X}>>>1$, the distinction between stable and unstable cases
has to be made.

\paragraph*{A 4.2.3.1 Function $H^{2}\left( K_{X}\right) $}

We can choose for $H^{2}\left( K_{X}\right) $ a power function of $K_{X}$: 
\begin{equation}
H\left( K_{X}\right) =K_{X}^{\eta }  \label{prb}
\end{equation}%
so that equation (\ref{psl}) rewrites:%
\begin{equation}
\left\Vert \Psi \left( X\right) \right\Vert ^{2}\simeq \frac{D\left(
\left\Vert \Psi \right\Vert ^{2}\right) -\frac{F}{2\sigma _{X}^{2}}\left(
\left( \nabla _{X}R\left( X\right) \right) ^{2}+\frac{2\sigma _{X}^{2}\nabla
_{X}^{2}R\left( K_{X},X\right) }{H\left( K_{X}\right) }\right) K_{X}^{\eta }%
}{2\tau }\equiv D-L\left( X\right) \left( \nabla _{X}R\left( X\right)
\right) ^{2}K_{X}^{\eta }  \label{psr}
\end{equation}

\paragraph*{A 4.2.3.2 Function $f$}

To determine the function $f$, we must first assume a form for $r\left(
K,X\right) $, the physical capital marginal returns, and for $F_{1}$, the
function that measures the impact of expected long-term return on investment
choices.

Assuming the production functions are of Cobb-Douglas type, i.e. $B\left(
X\right) K^{\alpha }$ with $B\left( X\right) $ a productivity factor, we
have for $r\left( K,X\right) $:%
\begin{equation}
r\left( K,X\right) =\frac{\partial r\left( K,X\right) }{\partial K}=\alpha
B\left( X\right) K^{\alpha -1}  \label{rkX}
\end{equation}

For function $F_{1}$, the simplest choice would be a linear form:%
\begin{equation*}
F_{1}\left( \frac{R\left( K_{\hat{X}},\hat{X}\right) }{\int R\left(
K_{X^{\prime }}^{\prime },X^{\prime }\right) \left\Vert \Psi \left(
X^{\prime }\right) \right\Vert ^{2}dX^{\prime }}\right) \simeq F_{1}\left( 
\frac{R\left( K_{\hat{X}},\hat{X}\right) }{\left\langle K_{\hat{X}}^{\alpha
}\right\rangle \left\langle R\left( \hat{X}\right) \right\rangle }\right)
=b\left( \frac{K_{\hat{X}}^{\alpha }R\left( \hat{X}\right) }{\left\langle
K_{X}^{\alpha }\right\rangle \left\langle R\left( X\right) \right\rangle }%
-1\right)
\end{equation*}%
where, for any function $u\left( \hat{X}\right) $, $\left\langle u\left( 
\hat{X}\right) \right\rangle $\ denotes its average over the sector space,\
and $b$\ an arbitrary parameter.

However, when capital$\ K_{\hat{X}}^{\alpha }\rightarrow \infty $ and is
concentrated at $\hat{X}$, we have $\left\langle K_{X}^{\alpha
}\right\rangle \simeq \frac{K_{\hat{X}}^{\alpha }}{N^{\alpha }\left(
X\right) }$, so that $\frac{K_{\hat{X}}^{\alpha }R\left( \hat{X}\right) }{%
\left\langle K_{X}^{\alpha }\right\rangle \left\langle R\left( X\right)
\right\rangle }\rightarrow \frac{N^{\alpha }\left( X\right) R\left( \hat{X}%
\right) }{\left\langle R\left( X\right) \right\rangle }>>1$. To impose some
bound on moves in the sector space we rather choose:%
\begin{equation}
F_{1}\left( \frac{R\left( K_{\hat{X}},\hat{X}\right) }{\left\langle K_{\hat{X%
}}^{\alpha }\right\rangle \left\langle R\left( \hat{X}\right) \right\rangle }%
\right) \simeq b\arctan \left( \frac{K_{\hat{X}}^{\alpha }R\left( \hat{X}%
\right) }{\left\langle K_{X}^{\alpha }\right\rangle \left\langle R\left(
X\right) \right\rangle }-1\right)  \label{prf}
\end{equation}%
so that $F_{1}\left( \frac{R\left( K_{\hat{X}},\hat{X}\right) }{\left\langle
K_{\hat{X}}^{\alpha }\right\rangle \left\langle R\left( \hat{X}\right)
\right\rangle }\right) >0$ when$\ \frac{K_{\hat{X}}^{\alpha }R\left( \hat{X}%
\right) }{\left\langle K_{X}^{\alpha }\right\rangle \left\langle R\left(
X\right) \right\rangle }>1$.

Given the above assumptions, the general formula for $f$\ given in equation (%
\ref{fcf}) rewrites:%
\begin{equation}
f\left( \hat{X},\Psi ,\hat{\Psi}\right) =\frac{1}{\varepsilon }\left(
r\left( \hat{X}\right) K_{\hat{X}}^{\alpha -1}-\gamma \left\Vert \Psi \left( 
\hat{X}\right) \right\Vert ^{2}+b\arctan \left( \frac{K_{\hat{X}}^{\alpha
}R\left( \hat{X}\right) }{\left\langle K_{X}^{\alpha }\right\rangle
\left\langle R\left( X\right) \right\rangle }-1\right) \right)  \label{frt}
\end{equation}%
This general formula can be approximated for $\frac{K_{\hat{X}}^{\alpha
}R\left( \hat{X}\right) }{\left\langle K_{X}^{\alpha }\right\rangle
\left\langle R\left( X\right) \right\rangle }\simeq 1$, when average capital
in sector $\hat{X}$ is close to the average capital of the whole space,
which is usually the case.

Using our choices (\ref{rsp}), (\ref{rkX}) and (\ref{prf}) for $\left\Vert
\Psi \left( X\right) \right\Vert ^{2}$\ $r\left( \hat{X}\right) $\ and $%
F_{1} $\ respectively, the equation (\ref{fcf}) for $f\left( \hat{X},\Psi ,%
\hat{\Psi}\right) $ becomes:%
\begin{equation*}
f\left( \hat{X},\Psi ,\hat{\Psi}\right) =\frac{1}{\varepsilon }\left( \left(
r\left( \hat{X}\right) +\frac{bR\left( \hat{X}\right) K_{\hat{X}}^{\alpha }}{%
\left\langle K_{\hat{X}}^{\alpha }\right\rangle \left\langle R\left( \hat{X}%
\right) \right\rangle }\right) +\gamma L\left( \hat{X}\right) K_{X}^{\eta
}-\gamma D-b\right)
\end{equation*}%
We may assume without impairing the results that $\eta =\alpha $. We thus
have:%
\begin{eqnarray}
f\left( \hat{X},\Psi ,\hat{\Psi}\right) &=&\frac{1}{\varepsilon }\left(
\left( \frac{r\left( \hat{X}\right) }{K_{\hat{X}}^{\alpha }}+\frac{bR\left( 
\hat{X}\right) }{\left\langle K_{\hat{X}}^{\alpha }\right\rangle
\left\langle R\left( \hat{X}\right) \right\rangle }+\gamma L\left( \hat{X}%
\right) \right) K_{\hat{X}}^{\alpha }-\gamma D-b\right)  \label{STP} \\
&\equiv &B_{1}\left( \hat{X}\right) K_{\hat{X}}^{\alpha -1}+B_{2}\left( \hat{%
X}\right) K_{\hat{X}}^{\alpha }-C\left( \hat{X}\right)  \notag
\end{eqnarray}

where:%
\begin{eqnarray*}
B_{1}\left( \hat{X}\right) &=&\frac{\alpha B\left( \hat{X}\right) }{%
\varepsilon } \\
B_{2}\left( \hat{X}\right) &=&\frac{bR\left( \hat{X}\right) }{\varepsilon
\left\langle K_{\hat{X}}^{\alpha }\right\rangle \left\langle R\left( \hat{X}%
\right) \right\rangle }+\frac{\gamma }{\varepsilon } \\
C\left( \hat{X}\right) &=&\gamma D+b
\end{eqnarray*}

\paragraph*{A 4.2.3.3 Function $g$}

To determine the form of function $g$, equation (\ref{fcg}), we must first
choose a form for the function $F_{0}$.\ 

We assume that: 
\begin{equation}
F_{0}\left( R\left( \hat{X},K_{\hat{X}}\right) \right) =a\arctan \left( K_{%
\hat{X}}^{\alpha }R\left( \hat{X}\right) \right)  \label{zrg}
\end{equation}%
where is $a$\ an arbitrary constant.

Combined to our assumption for $F_{1}$, (\ref{prf}), the formula (\ref{fcg})
for $g$ can be written:%
\begin{equation}
g\left( \hat{X},\Psi ,\hat{\Psi}\right) =a\nabla _{\hat{X}}\arctan \left( K_{%
\hat{X}}^{\alpha }R\left( \hat{X}\right) \right) +b\nabla _{\hat{X}}\arctan
\left( \frac{K_{\hat{X}}^{\alpha }R\left( \hat{X}\right) }{\left\langle
K_{X}^{\alpha }\right\rangle \left\langle R\left( X\right) \right\rangle }%
-1\right)  \label{grt}
\end{equation}%
where the $\arctan $ function ensures that the velocity in the sector space $%
g$ increases with capital and is maximal when average capital per firm in
sector $\hat{X}$ tends to infinity, i.e. $K_{\hat{X}}^{\alpha }\rightarrow
\infty $.

This general formula, equation (\ref{tsp}), can be approximated for $\frac{%
K_{\hat{X}}^{\alpha }R\left( \hat{X}\right) }{\left\langle K_{X}^{\alpha
}\right\rangle \left\langle R\left( X\right) \right\rangle }\simeq 1$, when
average capital in sector $\hat{X}$ is close to the average capital of the
whole space. It then reduces to:%
\begin{equation}
g\left( \hat{X},\Psi ,\hat{\Psi}\right) \simeq \frac{K_{\hat{X}}^{\alpha }}{%
\left\langle K_{\hat{X}}^{\alpha }\right\rangle }\nabla _{\hat{X}}R\left( 
\hat{X}\right) \left( 1+\frac{b}{\left\langle R\left( \hat{X}\right)
\right\rangle }\right) \equiv \nabla _{\hat{X}}R\left( \hat{X}\right)
A\left( \hat{X}\right) K_{\hat{X}}^{\alpha }
\end{equation}%
which in turn allows to approximate the gradient of $g$, $\nabla _{\hat{X}%
}g\left( \hat{X},\Psi ,\hat{\Psi}\right) $, by: 
\begin{equation}
\nabla _{\hat{X}}g\left( \hat{X},\Psi ,\hat{\Psi}\right) \simeq \frac{\nabla
_{\hat{X}}^{2}R\left( \hat{X}\right) }{\left\langle K_{\hat{X}}^{\alpha
}\right\rangle }\left( 1+\frac{b}{\left\langle K_{\hat{X}}^{\alpha
}\right\rangle \left\langle R\left( \hat{X}\right) \right\rangle }\right) K_{%
\hat{X}}^{\alpha }\equiv \nabla _{\hat{X}}^{2}R\left( \hat{X}\right) A\left( 
\hat{X}\right) K_{\hat{X}}^{\alpha }
\end{equation}

\paragraph*{A 4.2.3.4 Solving (\protect\ref{qnk})}

Equation (\ref{qnk}) can be studied by considering five cases presented in
the text:

\subparagraph*{Case 1. Very high capital, $K_{\hat{X}}>>>1$, stable case}

In that case, $K_{\hat{X}}>>>1$, and we assume in first approximation that
(discarding the factor $L\left( \hat{X}\right) $):%
\begin{equation}
\left\Vert \Psi \left( \hat{X}\right) \right\Vert ^{2}\simeq D-\left( \nabla
_{X}R\left( \hat{X}\right) \right) ^{2}K_{\hat{X}}^{\alpha }<<1  \label{nLT}
\end{equation}%
This corresponds to a very high level of capital. Consequently, equation (%
\ref{frt})\ implies that the function $f\left( \hat{X}\right) $ can be
rewritten:%
\begin{eqnarray*}
f\left( \hat{X}\right) &=&\frac{1}{\varepsilon }\left( r\left( \hat{X}%
\right) K_{\hat{X}}^{\alpha -1}-\gamma \left\Vert \Psi \left( \hat{X}\right)
\right\Vert ^{2}+b\arctan \left( \frac{K_{\hat{X}}^{\alpha }R\left( \hat{X}%
\right) }{\left\langle K_{X}^{\alpha }\right\rangle \left\langle R\left(
X\right) \right\rangle }-1\right) \right) \\
&\simeq &b\left( \frac{\pi }{2}-\frac{\left\langle K_{X}^{\alpha
}\right\rangle \left\langle R\left( X\right) \right\rangle }{K_{\hat{X}%
}^{\alpha }R\left( \hat{X}\right) }\right) \\
&\equiv &c-\frac{d}{K_{\hat{X}}^{\alpha }R\left( \hat{X}\right) }\simeq c>0
\end{eqnarray*}%
Consequently, the expressions for $f^{\prime }\left( \hat{X}\right) $ $%
g\left( \hat{X}\right) $ and $\nabla _{\hat{X}}g\left( \hat{X}\right) $ (\ref%
{fpr}) and (\ref{prg}) are still valid.

Two different cases arise in the resolution of (\ref{Nkv}).

First, we assume that $\left( \nabla _{\hat{X}}R\left( \hat{X}\right)
\right) ^{2}\neq 0$.

In this case, we will solve (\ref{Nkv}) by using (\ref{nLT}) to replace $K_{%
\hat{X}}\simeq \left( \frac{D}{\left( \nabla _{X}R\left( \hat{X}\right)
\right) ^{2}}\right) ^{\frac{1}{\alpha }}$. We also change the variable $%
\frac{D}{\left( \nabla _{X}R\left( \hat{X}\right) \right) ^{2}}\rightarrow D$
temporarily for the sake of simplicity.

Inequality (\ref{nLT}) along with $K_{\hat{X}}>>>1$ and (\ref{frt}) implies
that only the case $f>0$ has to be considered.

Note that using our results about stability, it is easy to check that in
that case, this solution is locally unstable. A very high level of capital
has the tendency to attract more investments.

Given our assumptions, equation (\ref{qnk}) becomes: 
\begin{equation}
\left( \nabla _{X}R\left( \hat{X}\right) \right) ^{2}D^{\frac{1}{\alpha }%
}\left( D-K_{\hat{X}}^{\alpha }\right) =C\left( \bar{p}\right) \sigma _{\hat{%
K}}^{2}\exp \left( -\frac{\sigma _{X}^{2}\sigma _{\hat{K}}^{2}\left( p+\frac{%
1}{2}\right) ^{2}\left( f^{\prime }\left( X\right) \right) ^{2}}{%
96\left\vert f\left( \hat{X}\right) \right\vert ^{3}}\right) \frac{\Gamma
\left( p+\frac{3}{2}\right) }{\left\vert f\left( \hat{X}\right) \right\vert }
\label{gdr}
\end{equation}%
or equivalently:%
\begin{equation}
K_{\hat{X}}^{\alpha }=D-\frac{C\left( \bar{p}\right) \sigma _{\hat{K}%
}^{2}\exp \left( -\frac{\sigma _{X}^{2}\sigma _{\hat{K}}^{2}\left( p+\frac{1%
}{2}\right) ^{2}\left( f^{\prime }\left( X\right) \right) ^{2}}{96\left\vert
f\left( \hat{X}\right) \right\vert ^{3}}\right) \Gamma \left( p+\frac{3}{2}%
\right) }{\left( \nabla _{X}R\left( \hat{X}\right) \right) ^{2}D^{\frac{1}{%
\alpha }}\left\vert f\left( \hat{X}\right) \right\vert }  \label{lgk}
\end{equation}

Then, defining $V=\frac{1}{K_{\hat{X}}^{\alpha }}$ as in the first case, we
can write (\ref{lgk}) as an equation for $V<<1$ by replacing all quantities
in term of $V$ and then perform a first order expansion.

First, we write (\ref{lgk}) as:%
\begin{equation}
V-\frac{1}{D-\frac{C\left( \bar{p}\right) \sigma _{\hat{K}}^{2}\exp \left( -%
\frac{\sigma _{X}^{2}\sigma _{\hat{K}}^{2}\left( p+\frac{1}{2}\right)
^{2}\left( f^{\prime }\left( X\right) \right) ^{2}}{96\left\vert f\left( 
\hat{X}\right) \right\vert ^{3}}\right) \Gamma \left( p+\frac{3}{2}\right) }{%
\left( \nabla _{X}R\left( \hat{X}\right) \right) ^{2}D^{\frac{1}{\alpha }%
}\left\vert f\left( \hat{X}\right) \right\vert }}=0  \label{Kgl}
\end{equation}%
As in the previous case, the first order expansion in $V$ of $\Gamma \left(
p+\frac{3}{2}\right) $ arising in (\ref{Kgl}) is given by:%
\begin{equation}
\Gamma \left( p+\frac{3}{2}\right) \simeq \Gamma \left( \frac{M}{c}\right) +%
\frac{MV}{c}\left( \frac{\nabla _{\hat{X}}^{2}R\left( \hat{X}\right) f}{%
MR\left( \hat{X}\right) }+\frac{d}{cR\left( \hat{X}\right) }\right) \Gamma
^{\prime }\left( \frac{M}{c}\right)  \label{pqr}
\end{equation}%
Moreover, at the first order:%
\begin{equation*}
\exp \left( -\frac{\sigma _{X}^{2}\sigma _{\hat{K}}^{2}\left( p+\frac{1}{2}%
\right) ^{2}\left( f^{\prime }\left( X\right) \right) ^{2}}{96\left\vert
f\left( \hat{X}\right) \right\vert ^{3}}\right) \simeq 1
\end{equation*}%
and (\ref{Kgl}) becomes:%
\begin{equation*}
V-\frac{\left( \nabla _{\hat{X}}R\left( \hat{X}\right) \right) ^{2}D^{\frac{1%
}{\alpha }}\left\vert f\left( \hat{X}\right) \right\vert }{\left( \nabla _{%
\hat{X}}R\left( \hat{X}\right) \right) ^{2}D^{1+\frac{1}{\alpha }}\left\vert
f\left( \hat{X}\right) \right\vert -C\left( \bar{p}\right) \sigma _{\hat{K}%
}^{2}\Gamma \left( p+\frac{3}{2}\right) }=0
\end{equation*}%
that is:%
\begin{equation}
V-\frac{\left( \nabla _{\hat{X}}R\left( \hat{X}\right) \right) ^{2}D^{\frac{1%
}{\alpha }}\left( c-\frac{dV}{R\left( \hat{X}\right) }\right) }{\left(
\nabla _{\hat{X}}R\left( \hat{X}\right) \right) ^{2}D^{1+\frac{1}{\alpha }%
}\left( c-\frac{dV}{R\left( \hat{X}\right) }\right) -C\left( \bar{p}\right)
\sigma _{\hat{K}}^{2}\Gamma \left( p+\frac{3}{2}\right) }=0  \label{frd}
\end{equation}%
Using (\ref{pqr}) the first order expansion of the dominator in (\ref{frd})
is:%
\begin{eqnarray*}
&&\left( \nabla _{\hat{X}}R\left( \hat{X}\right) \right) ^{2}D^{1+\frac{1}{%
\alpha }}\left( c-\frac{dV}{R\left( \hat{X}\right) }\right) -C\left( \bar{p}%
\right) \sigma _{\hat{K}}^{2}\Gamma \left( p+\frac{3}{2}\right) \\
&=&\left( \nabla _{\hat{X}}R\left( \hat{X}\right) \right) ^{2}D^{1+\frac{1}{%
\alpha }}c-C\left( \bar{p}\right) \sigma _{\hat{K}}^{2}\Gamma \left( \frac{M%
}{c}\right) \\
&&-\left( \left( \nabla _{\hat{X}}R\left( \hat{X}\right) \right) ^{2}D^{1+%
\frac{1}{\alpha }}\frac{d}{R\left( \hat{X}\right) }+\frac{C\left( \bar{p}%
\right) \sigma _{\hat{K}}^{2}M}{c}\left( \frac{\nabla _{\hat{X}}^{2}R\left( 
\hat{X}\right) f}{MR\left( \hat{X}\right) }+\frac{d}{cR\left( \hat{X}\right) 
}\right) \Gamma ^{\prime }\left( \frac{M}{c}\right) \right) V
\end{eqnarray*}%
so that (\ref{frd}) writes:%
\begin{eqnarray}
&&\frac{\left( \nabla _{\hat{X}}R\left( \hat{X}\right) \right) ^{2}D^{\frac{1%
}{\alpha }}c}{\left( \nabla _{\hat{X}}R\left( \hat{X}\right) \right)
^{2}D^{1+\frac{1}{\alpha }}c-C\left( \bar{p}\right) \sigma _{\hat{K}%
}^{2}\Gamma \left( \frac{M}{c}\right) }  \label{frD} \\
&=&\left( 1-\frac{\left( \nabla _{\hat{X}}R\left( \hat{X}\right) \right)
^{2}D^{\frac{1}{\alpha }}c\left( \left( \nabla _{\hat{X}}R\left( \hat{X}%
\right) \right) ^{2}D^{1+\frac{1}{\alpha }}\frac{d}{R\left( \hat{X}\right) }+%
\frac{C\left( \bar{p}\right) \sigma _{\hat{K}}^{2}M}{c}\left( \frac{\nabla _{%
\hat{X}}^{2}R\left( \hat{X}\right) f}{MR\left( \hat{X}\right) }+\frac{d}{%
cR\left( \hat{X}\right) }\right) \Gamma ^{\prime }\left( \frac{M}{c}\right)
\right) }{\left( \left( \nabla _{\hat{X}}R\left( \hat{X}\right) \right)
^{2}D^{1+\frac{1}{\alpha }}c-C\left( \bar{p}\right) \sigma _{\hat{K}%
}^{2}\Gamma \left( \frac{M}{c}\right) \right) ^{2}}\right) V  \notag \\
&&+\frac{\left( \nabla _{\hat{X}}R\left( \hat{X}\right) \right) ^{2}D^{\frac{%
1}{\alpha }}\frac{d}{R\left( \hat{X}\right) }}{\left( \nabla _{\hat{X}%
}R\left( \hat{X}\right) \right) ^{2}D^{1+\frac{1}{\alpha }}c-C\left( \bar{p}%
\right) \sigma _{\hat{K}}^{2}\Gamma \left( \frac{M}{c}\right) }V  \notag
\end{eqnarray}%
Equation (\ref{frD}) can be solved for $V$\ with solution:

\begin{equation*}
\frac{1}{V}=D-\frac{C\left( \bar{p}\right) \sigma _{\hat{K}}^{2}\Gamma
\left( \frac{M}{c}\right) }{\left( \nabla _{\hat{X}}R\left( \hat{X}\right)
\right) ^{2}D^{\frac{1}{\alpha }}c}+\frac{d}{cR\left( \hat{X}\right) }\left(
1-\frac{\left( 1+\frac{C\left( \bar{p}\right) \sigma _{\hat{K}}^{2}M\Gamma
\left( \frac{M}{c}\right) }{c\left( \nabla _{\hat{X}}R\left( \hat{X}\right)
\right) ^{2}D^{1+\frac{1}{\alpha }}}\left( \frac{\nabla _{\hat{X}%
}^{2}R\left( \hat{X}\right) f}{Md}+\frac{1}{c}\right) \func{Psi}\left( \frac{%
M}{c}\right) \right) }{\left( 1-\frac{C\left( \bar{p}\right) \sigma _{\hat{K}%
}^{2}\Gamma \left( \frac{M}{c}\right) }{\left( \nabla _{\hat{X}}R\left( \hat{%
X}\right) \right) ^{2}D^{1+\frac{1}{\alpha }}c}\right) }\right)
\end{equation*}%
Ultimatly, restoring the variable:%
\begin{equation*}
D\rightarrow \frac{D}{\left( \nabla _{\hat{X}}R\left( \hat{X}\right) \right)
^{2}}
\end{equation*}%
we obtain the solution $K_{\hat{X}}^{\alpha }=\frac{1}{V}$:%
\begin{eqnarray}
K_{\hat{X}}^{\alpha } &=&\frac{D}{\left( \nabla _{\hat{X}}R\left( \hat{X}%
\right) \right) ^{2}}-\frac{C\left( \bar{p}\right) \sigma _{\hat{K}%
}^{2}\Gamma \left( \frac{M}{c}\right) }{\left( \nabla _{\hat{X}}R\left( \hat{%
X}\right) \right) ^{2\left( 1-\frac{1}{\alpha }\right) }D^{\frac{1}{\alpha }%
}c}  \label{mGK} \\
&&+\frac{d}{cR\left( \hat{X}\right) }\left( 1-\frac{\left( 1+\frac{C\left( 
\bar{p}\right) \left( \nabla _{\hat{X}}R\left( \hat{X}\right) \right) ^{%
\frac{2}{\alpha }}\sigma _{\hat{K}}^{2}}{cD^{1+\frac{1}{\alpha }}}\left( 
\frac{M}{c}+\frac{\nabla _{\hat{X}}^{2}R\left( \hat{X}\right) f}{d}\right)
\Gamma ^{\prime }\left( \frac{M}{c}\right) \right) }{\left( 1-\frac{\left(
\nabla _{\hat{X}}R\left( \hat{X}\right) \right) ^{\frac{2}{\alpha }}C\left( 
\bar{p}\right) \sigma _{\hat{K}}^{2}}{cD^{1+\frac{1}{\alpha }}}\Gamma \left( 
\frac{M}{c}\right) \right) }\right)  \notag
\end{eqnarray}%
As stated in the text, this is increasing in $c$, i.e. in $f\left( \hat{X}%
\right) $ and in $R\left( \hat{X}\right) $. This corresponds to a stable
level of capital.

\subparagraph*{Case 2. Very high capital, $K_{\hat{X}}>>>1$, unstable case}

In this second case, we consider that $\left( \nabla _{\hat{X}}R\left( \hat{X%
}\right) \right) ^{2}\rightarrow 0$ and formula (\ref{nLT})and (\ref{mGK})
are not valid anymore. Coming back to (\ref{dfg}) leads rather to replace $%
\left( \nabla _{X}R\left( X\right) \right) ^{2}$:%
\begin{equation*}
\left( \nabla _{X}R\left( X\right) \right) ^{2}\rightarrow \left( \nabla
_{X}R\left( X\right) \right) ^{2}+\sigma _{X}^{2}\frac{\nabla
_{X}^{2}R\left( K_{X},X\right) }{H\left( K_{X}\right) }=\sigma _{X}^{2}\frac{%
\nabla _{X}^{2}R\left( K_{X},X\right) }{H\left( K_{X}\right) }
\end{equation*}

Thus, if $\nabla _{X}^{2}R\left( K_{X},X\right) <0$, (\ref{gdr}) is replaced
by:%
\begin{equation*}
K_{\hat{X}}^{\alpha }\left( D+\sigma _{X}^{2}\left\vert \nabla
_{X}^{2}R\left( K_{X},X\right) \right\vert K_{\hat{X}}^{\frac{\alpha }{2}%
}\right) =C\left( \bar{p}\right) \sigma _{\hat{K}}^{2}\frac{\Gamma \left( p+%
\frac{3}{2}\right) }{\left\vert f\left( \hat{X}\right) \right\vert }
\end{equation*}%
with:%
\begin{equation*}
p+\frac{3}{2}\simeq \frac{M-\nabla _{\hat{X}}g\left( \hat{X},K_{\hat{X}%
}\right) }{f\left( \hat{X}\right) }
\end{equation*}%
and the equation for $K_{X}$ writes:%
\begin{equation*}
\sigma _{X}^{2}\left\vert \nabla _{X}^{2}R\left( K_{X},X\right) \right\vert
K_{\hat{X}}^{\frac{3}{2}\alpha }=\frac{C\left( \bar{p}\right) \sigma _{\hat{K%
}}^{2}\Gamma \left( \frac{M-\nabla _{\hat{X}}g\left( \hat{X},K_{\hat{X}%
}\right) }{f\left( \hat{X}\right) }\right) }{\left\vert f\left( \hat{X}%
\right) \right\vert }
\end{equation*}%
Since, given our assumptions $f\left( \hat{X}\right) \rightarrow c$ we find:%
\begin{equation}
K_{\hat{X}}=\left( \frac{C\left( \bar{p}\right) \sigma _{\hat{K}}^{2}}{%
\left\vert \nabla _{X}^{2}R\left( K_{X},X\right) \right\vert c}\Gamma \left( 
\frac{M-\nabla _{\hat{X}}g\left( \hat{X},K_{\hat{X}}\right) }{c}\right)
\right) ^{\frac{2}{3\alpha }}  \label{drsn}
\end{equation}%
Note that given (\ref{drsn}), an equilibrium in the range $K_{\hat{X}}>>>1$
is only possible for $c<<1$ Otherwise, there is no equilibrium for a maximum
of $R\left( K_{X},X\right) $. This equilibrium value of $K_{\hat{X}}$
decreases with $c$, which corresponds to an unstable equilibrium, as
detailed in the text.

On the other hand, if $\nabla _{X}R\left( X\right) =0$ and $\nabla
_{X}^{2}R\left( K_{X},X\right) >0$, expression (\ref{nLT}) becomes:%
\begin{equation*}
\left\Vert \Psi \left( \hat{X}\right) \right\Vert ^{2}\simeq D-\sigma
_{X}^{2}\frac{\nabla _{X}^{2}R\left( K_{X},X\right) }{H\left( K_{X}\right) }%
K_{\hat{X}}^{\alpha }=D-\sigma _{X}^{2}\nabla _{X}^{2}R\left( K_{X},X\right)
K_{\hat{X}}^{\frac{\alpha }{2}}
\end{equation*}%
and thus:%
\begin{equation}
K_{\hat{X}}^{\alpha }\simeq \left( \frac{D}{\sigma _{X}^{2}\nabla
_{X}^{2}R\left( K_{X},X\right) }\right) ^{2}  \label{drsp}
\end{equation}%
However, this solution with $K_{X}>>1$ corresponds to points such that $%
\nabla _{X}^{2}R\left( K_{X},X\right) >0$ and $\nabla _{X}R\left( X\right)
=0 $. Then, these points are minima of $R\left( X\right) $. This equilibrium
may exist only if the level of capital (\ref{drsp}) is high enough to
compensate the weakness of the purely position dependent part of expected
return and match the condition:%
\begin{equation*}
\frac{K_{\hat{X}}^{\alpha }R\left( \hat{X}\right) }{\left\langle
K_{X}^{\alpha }\right\rangle \left\langle R\left( X\right) \right\rangle }%
-1>0
\end{equation*}%
This equilibrium is thus unlikely and may be discarded in general.

\subparagraph*{Case 3. High capital, $K_{\hat{X}}>>1$}

In that case, we assume $K_{\hat{X}}$ relatively large, but bounded, to
ensure that the approximation:%
\begin{equation}
\left\Vert \Psi \left( \hat{X}\right) \right\Vert ^{2}\simeq D  \label{bts}
\end{equation}%
is still valid.

Equations (\ref{frt}) and (\ref{grt}) imply that\ the function $f\left( \hat{%
X}\right) $ is independent of $K_{\hat{X}}$ and that $g\left( \hat{X}\right) 
$ is proportional to $\nabla _{\hat{X}}R\left( \hat{X}\right) $. Given (\ref%
{frt}),\ the function $f\left( \hat{X}\right) $ can be rewritten:%
\begin{eqnarray*}
f\left( \hat{X}\right) &=&\frac{1}{\varepsilon }\left( r\left( \hat{X}%
\right) K_{\hat{X}}^{\alpha -1}-\gamma \left\Vert \Psi \left( \hat{X}\right)
\right\Vert ^{2}+b\arctan \left( \frac{K_{\hat{X}}^{\alpha }R\left( \hat{X}%
\right) }{\left\langle K_{X}^{\alpha }\right\rangle \left\langle R\left(
X\right) \right\rangle }-1\right) \right) \\
&\simeq &b\left( \frac{\pi }{2}-\frac{\left\langle K_{X}^{\alpha
}\right\rangle \left\langle R\left( X\right) \right\rangle }{K_{\hat{X}%
}^{\alpha }R\left( \hat{X}\right) }\right) -\gamma D \\
&\equiv &c-\frac{d}{K_{\hat{X}}^{\alpha }R\left( \hat{X}\right) }-\gamma D
\end{eqnarray*}%
Consequently, the expression for $f^{\prime }\left( \hat{X}\right) $ is:%
\begin{equation}
f^{\prime }\left( \hat{X}\right) \simeq \frac{d\nabla _{\hat{X}}R\left( \hat{%
X}\right) }{K_{\hat{X}}^{\alpha }R^{2}\left( \hat{X}\right) }  \label{fpr}
\end{equation}%
Similarly,we can approximate (\ref{grt}) as:%
\begin{eqnarray}
g\left( \hat{X}\right) &\simeq &-\frac{\nabla _{\hat{X}}R\left( \hat{X}%
\right) f}{K_{\hat{X}}^{\alpha }R\left( \hat{X}\right) }  \label{prg} \\
\nabla _{\hat{X}}g\left( \hat{X}\right) &\simeq &-\frac{\nabla _{\hat{X}%
}^{2}R\left( \hat{X}\right) f}{K_{\hat{X}}^{\alpha }R\left( \hat{X}\right) }
\notag
\end{eqnarray}%
Given (\ref{bts}), and including the constant $\alpha $\ in the definition
of $C\left( \bar{p}\right) $, equation (\ref{qnk}) is:%
\begin{equation}
K_{\hat{X}}D\left\vert f\left( \hat{X}\right) \right\vert =C\left( \bar{p}%
\right) \sigma _{\hat{K}}^{2}\exp \left( -\frac{\sigma _{X}^{2}\sigma _{\hat{%
K}}^{2}\left( p+\frac{1}{2}\right) ^{2}\left( f^{\prime }\left( X\right)
\right) ^{2}}{96\left\vert f\left( \hat{X}\right) \right\vert ^{3}}\right)
\Gamma \left( p+\frac{3}{2}\right)  \label{Lqg}
\end{equation}%
with: 
\begin{equation*}
p+\frac{1}{2}=\frac{M-\left( \frac{\left( g\left( \hat{X}\right) \right) ^{2}%
}{\sigma _{\hat{X}}^{2}}+\left( f\left( \hat{X}\right) +\nabla _{\hat{X}%
}g\left( \hat{X},K_{\hat{X}}\right) -\frac{\sigma _{\hat{K}}^{2}F^{2}\left( 
\hat{X},K_{\hat{X}}\right) }{2f^{2}\left( \hat{X}\right) }\right) \right) }{%
\sqrt{f^{2}\left( \hat{X}\right) }}
\end{equation*}%
Defining $V=\frac{1}{K_{\hat{X}}^{\alpha }}$, we can write (\ref{Lqg}) as an
equation for $V<<1$ by replacing all quantities in term of $V$ and then
perform a first order expansion. To do so, we first, we write (\ref{Lqg}) as:%
\begin{equation}
V-\frac{D\left\vert f\left( \hat{X}\right) \right\vert }{C\left( \bar{p}%
\right) \sigma _{\hat{K}}^{2}\exp \left( -\frac{\sigma _{X}^{2}\sigma _{\hat{%
K}}^{2}\left( p+\frac{1}{2}\right) ^{2}\left( f^{\prime }\left( X\right)
\right) ^{2}}{96\left\vert f\left( \hat{X}\right) \right\vert ^{3}}\right)
\Gamma \left( p+\frac{3}{2}\right) }=0  \label{lqG}
\end{equation}%
and then find an expansion in $V$ for $\Gamma \left( p+\frac{3}{2}\right) $.

The first order expansion in $V$ of $p+\frac{3}{2}$ is: 
\begin{eqnarray*}
p+\frac{3}{2} &=&\frac{M-\left( \frac{\left( g\left( \hat{X}\right) \right)
^{2}}{\sigma _{\hat{X}}^{2}}+\nabla _{\hat{X}}g\left( \hat{X},K_{\hat{X}%
}\right) -\frac{\sigma _{\hat{K}}^{2}F^{2}\left( \hat{X},K_{\hat{X}}\right) 
}{2f^{2}\left( \hat{X}\right) }\right) }{f\left( \hat{X}\right) } \\
&\simeq &\frac{M-\left( \frac{\left( g\left( \hat{X}\right) \right) ^{2}}{%
\sigma _{\hat{X}}^{2}}+\nabla _{\hat{X}}g\left( \hat{X},K_{\hat{X}}\right)
\right) }{c-\frac{d}{K_{\hat{X}}^{\alpha }R\left( \hat{X}\right) }} \\
&=&\frac{M-\left( \frac{\left( \nabla _{\hat{X}}R\left( \hat{X}\right)
\left( -\frac{fV}{R\left( \hat{X}\right) }\right) \right) ^{2}}{\sigma _{%
\hat{X}}^{2}}+\nabla _{\hat{X}}^{2}R\left( \hat{X}\right) \left( -\frac{fV}{%
R\left( \hat{X}\right) }\right) \right) }{c-\frac{dV}{R\left( \hat{X}\right) 
}} \\
&=&\frac{M}{c}+\frac{\nabla _{\hat{X}}^{2}R\left( \hat{X}\right) \frac{fV}{%
R\left( \hat{X}\right) }}{c}+\frac{M\frac{dV}{cR\left( \hat{X}\right) }}{c}
\end{eqnarray*}%
Consequently, $\Gamma \left( p+\frac{3}{2}\right) $ arising in (\ref{Kgl})
is given by:%
\begin{eqnarray*}
\Gamma \left( \frac{M}{c}+\frac{\nabla _{\hat{X}}^{2}R\left( \hat{X}\right) 
\frac{fV}{R\left( \hat{X}\right) }}{c}+\frac{M\frac{dV}{cR\left( \hat{X}%
\right) }}{c}\right) &=&\Gamma \left( \frac{M}{c}\left( 1+\nabla _{\hat{X}%
}^{2}R\left( \hat{X}\right) \frac{fV}{MR\left( \hat{X}\right) }+\frac{dV}{%
cR\left( \hat{X}\right) }\right) \right) \\
&\simeq &\Gamma \left( \frac{M}{c}\right) +\frac{MV}{c}\left( \frac{\nabla _{%
\hat{X}}^{2}R\left( \hat{X}\right) f}{MR\left( \hat{X}\right) }+\frac{d}{%
cR\left( \hat{X}\right) }\right) \Gamma ^{\prime }\left( \frac{M}{c}\right)
\\
&=&\Gamma \left( \frac{M}{c}\right) \left( 1+\frac{MV}{c}\left( \frac{\nabla
_{\hat{X}}^{2}R\left( \hat{X}\right) f}{MR\left( \hat{X}\right) }+\frac{d}{%
cR\left( \hat{X}\right) }\right) \func{Psi}\left( \frac{M}{c}\right) \right)
\end{eqnarray*}%
Ultimately, using that at the first order:%
\begin{equation*}
\exp \left( -\frac{\sigma _{X}^{2}\sigma _{\hat{K}}^{2}\left( p+\frac{1}{2}%
\right) ^{2}\left( f^{\prime }\left( X\right) \right) ^{2}}{96\left\vert
f\left( \hat{X}\right) \right\vert ^{3}}\right) \simeq 1
\end{equation*}%
equation (\ref{lqG}) for $V$ becomes:%
\begin{equation*}
V-\frac{D\left\vert f\left( \hat{X}\right) \right\vert }{C\left( \bar{p}%
\right) \sigma _{\hat{K}}^{2}\Gamma \left( \frac{M}{c}\right) \left( 1+\frac{%
MV}{c}\left( \frac{\nabla _{\hat{X}}^{2}R\left( \hat{X}\right) f}{MR\left( 
\hat{X}\right) }+\frac{d}{cR\left( \hat{X}\right) }\right) \func{Psi}\left( 
\frac{M}{c}\right) \right) }=0
\end{equation*}%
that is:%
\begin{equation*}
V-\frac{D\left( c-\frac{dV}{R\left( \hat{X}\right) }-\gamma D\right) }{%
C\left( \bar{p}\right) \sigma _{\hat{K}}^{2}\Gamma \left( \frac{M}{c}\right)
\left( 1+\frac{MV}{c}\left( \frac{\nabla _{\hat{X}}^{2}R\left( \hat{X}%
\right) f}{MR\left( \hat{X}\right) }+\frac{d}{cR\left( \hat{X}\right) }%
\right) \func{Psi}\left( \frac{M}{c}\right) \right) }=0
\end{equation*}%
And a first order expansion yields:%
\begin{equation*}
V-\frac{D\left( c-\gamma D\right) }{C\left( \bar{p}\right) \sigma _{\hat{K}%
}^{2}\Gamma \left( \frac{M}{c}\right) }\left( 1-\frac{dV}{\left( c-\gamma
D\right) R\left( \hat{X}\right) }-MV\left( \frac{\nabla _{\hat{X}%
}^{2}R\left( \hat{X}\right) f}{MR\left( \hat{X}\right) }+\frac{d}{cR\left( 
\hat{X}\right) }\right) \func{Psi}\left( \frac{M}{c}\right) \right) =0
\end{equation*}%
with solution:%
\begin{equation*}
V=\left( K_{\hat{X}}^{\alpha }\right) ^{-1}=\frac{\frac{D\left( c-\gamma
D\right) }{C\left( \bar{p}\right) \sigma _{\hat{K}}^{2}\Gamma \left( \frac{M%
}{c}\right) }}{1+\frac{D\left( c-\gamma D\right) }{C\left( \bar{p}\right)
\sigma _{\hat{K}}^{2}\Gamma \left( \frac{M}{c}\right) }\left( \frac{d}{%
\left( c-\gamma D\right) R\left( \hat{X}\right) }+M\left( \frac{\nabla _{%
\hat{X}}^{2}R\left( \hat{X}\right) f}{MR\left( \hat{X}\right) }+\frac{d}{%
cR\left( \hat{X}\right) }\right) \func{Psi}\left( \frac{M}{c}\right) \right) 
}
\end{equation*}%
Coming back to $K_{\hat{X}}^{\alpha }$ we have:%
\begin{eqnarray}
K_{\hat{X}}^{\alpha } &=&\frac{C\left( \bar{p}\right) \sigma _{\hat{K}%
}^{2}\Gamma \left( \frac{M}{c}\right) }{D\left( c-\gamma D\right) }+\frac{d}{%
\left( c-\gamma D\right) R\left( \hat{X}\right) }+M\left( \frac{\nabla _{%
\hat{X}}^{2}R\left( \hat{X}\right) f}{MR\left( \hat{X}\right) }+\frac{d}{%
cR\left( \hat{X}\right) }\right) \func{Psi}\left( \frac{M}{c}\right)
\label{shk} \\
&=&\frac{C\left( \bar{p}\right) \sigma _{\hat{K}}^{2}\Gamma \left( \frac{M}{c%
}\right) }{D\left( c-\gamma D\right) }+\frac{d}{\left( c-\gamma D\right)
R\left( \hat{X}\right) }\left( 1+M\func{Psi}\left( \frac{M}{c}\right) \left(
1+\frac{\nabla _{\hat{X}}^{2}R\left( \hat{X}\right) f}{M\left( c-\gamma
D\right) }\right) \right)  \notag
\end{eqnarray}%
This solution satisfies the condition $K_{\hat{X}}>>1$ only if $\frac{%
C\left( \bar{p}\right) \sigma _{\hat{K}}^{2}\sqrt{\frac{M-c}{c}}}{Dc}>>1$:
formula (\ref{shk}) thus shows that the dependency of $K_{\hat{X}}^{\alpha }$
in the return $R\left( \hat{X}\right) $\ depends on the sign of $1+M\func{Psi%
}\left( \frac{M}{c}\right) \left( 1+\frac{\nabla _{\hat{X}}^{2}R\left( \hat{X%
}\right) f}{M\left( c-\gamma D\right) }\right) $. If:%
\begin{equation*}
1+M\func{Psi}\left( \frac{M}{c}\right) \left( 1+\frac{\nabla _{\hat{X}%
}^{2}R\left( \hat{X}\right) f}{M\left( c-\gamma D\right) }\right) >0
\end{equation*}%
then $K_{\hat{X}}^{\alpha }$ decreases with $R\left( \hat{X}\right) $. As
stated in the text, this corresponds to an unstable equilibrium.

If:%
\begin{equation*}
1+M\func{Psi}\left( \frac{M}{c}\right) \left( 1+\frac{\nabla _{\hat{X}%
}^{2}R\left( \hat{X}\right) f}{M\left( c-\gamma D\right) }\right) <0
\end{equation*}%
a stable equilibrium is possible and $K_{\hat{X}}^{\alpha }$ is an
increasing function of $R\left( \hat{X}\right) $ and $f\left( \hat{X}\right) 
$. This corresponds to $\nabla _{\hat{X}}^{2}R\left( \hat{X}\right) <<0$,
which arises for instance for a maximum of $R\left( \hat{X}\right) $. In
such case, an increase in $R\left( \hat{X}\right) $ allows for an increased
number $\left\Vert \Psi \left( \hat{X}\right) \right\Vert ^{2}$ of firms,
without reducing the average capital per firm.

\subparagraph{Case 4. Intermediate capital, $\infty >K_{\hat{X}}>1$:}

We start with asymptotic form of (\ref{Nkv}):%
\begin{equation}
K_{\hat{X}}\left\Vert \Psi \left( \hat{X}\right) \right\Vert ^{2}\left\vert
f\left( \hat{X}\right) \right\vert =C\left( \bar{p}\right) \sigma _{\hat{K}%
}^{2}\exp \left( -\frac{\sigma _{X}^{2}\sigma _{\hat{K}}^{2}\left( p+\frac{1%
}{2}\right) ^{2}\left( f^{\prime }\left( X\right) \right) ^{2}}{96\left\vert
f\left( \hat{X}\right) \right\vert ^{3}}\right) \Gamma \left( p+\frac{3}{2}%
\right)  \label{smP}
\end{equation}%
$\allowbreak \allowbreak $Up to a constant that can be absorbed in the
definition of $C\left( \bar{p}\right) $, we have:%
\begin{equation*}
\Gamma \left( p+\frac{3}{2}\right) \sim _{\infty }\sqrt{p+\frac{1}{2}}\exp
\left( \left( p+\frac{1}{2}\right) \left( \ln \left( p+\frac{1}{2}\right)
-1\right) \right)
\end{equation*}%
and (\ref{smP}) can be rewritten as:%
\begin{equation}
K_{\hat{X}}\left\Vert \Psi \left( \hat{X}\right) \right\Vert ^{2}\left\vert
f\left( \hat{X}\right) \right\vert =C\left( \bar{p}\right) \sigma _{\hat{K}%
}^{2}\sqrt{p+\frac{1}{2}}\exp \left( -\frac{\sigma _{X}^{2}\sigma _{\hat{K}%
}^{2}\left( p+\frac{1}{2}\right) ^{2}\left( f^{\prime }\left( X\right)
\right) ^{2}}{96\left\vert f\left( \hat{X}\right) \right\vert ^{3}}+\left( p+%
\frac{1}{2}\right) \left( \ln \left( p+\frac{1}{2}\right) -1\right) \right)
\label{sMP}
\end{equation}%
Since we are in an intermediate range for the parameters, we can replace, in
first approximation, $\ln \left( p+\frac{1}{2}\right) $ by its average over
this range: $\ln \left( \bar{p}+\frac{1}{2}\right) $. The exponential in (%
\ref{sMP}) thus becomes: 
\begin{equation*}
\exp \left( -\frac{\sigma _{X}^{2}\sigma _{\hat{K}}^{2}\left( p+\frac{1}{2}-%
\frac{48\left\vert f\left( \hat{X}\right) \right\vert ^{3}}{\sigma
_{X}^{2}\sigma _{\hat{K}}^{2}\left( f^{\prime }\left( X\right) \right) ^{2}}%
\left( \ln \left( \bar{p}+\frac{1}{2}\right) -1\right) \right) ^{2}\left(
f^{\prime }\left( X\right) \right) ^{2}}{96\left\vert f\left( \hat{X}\right)
\right\vert ^{3}}+\frac{24\left\vert f\left( \hat{X}\right) \right\vert ^{3}%
}{\sigma _{X}^{2}\sigma _{\hat{K}}^{2}\left( f^{\prime }\left( X\right)
\right) ^{2}}\left( \ln \left( \bar{p}+\frac{1}{2}\right) -1\right)
^{2}\right)
\end{equation*}%
and equation (\ref{sMP}) rewrites:

\begin{eqnarray}
&&K_{\hat{X}}\left\Vert \Psi \left( \hat{X}\right) \right\Vert
^{2}\left\vert f\left( \hat{X}\right) \right\vert \left( \frac{\sigma
_{X}^{2}\sigma _{\hat{K}}^{2}\left( f^{\prime }\left( X\right) \right)
^{2}\exp \left( -\frac{96\left\vert f\left( \hat{X}\right) \right\vert ^{3}}{%
\sigma _{X}^{2}\sigma _{\hat{K}}^{2}\left( f^{\prime }\left( X\right)
\right) ^{2}}\left( \ln \left( \bar{p}+\frac{1}{2}\right) -1\right)
^{2}\right) }{96\left\vert f\left( \hat{X}\right) \right\vert ^{3}}\right) ^{%
\frac{1}{4}}  \label{dvM} \\
&=&C\left( \bar{p}\right) \sigma _{\hat{K}}^{2}\exp \left( -\frac{\sigma
_{X}^{2}\sigma _{\hat{K}}^{2}\left( p+\frac{1}{2}-\frac{48\left\vert f\left( 
\hat{X}\right) \right\vert ^{3}}{\sigma _{X}^{2}\sigma _{\hat{K}}^{2}\left(
f^{\prime }\left( X\right) \right) ^{2}}\left( \ln \left( \bar{p}+\frac{1}{2}%
\right) -1\right) \right) ^{2}\left( f^{\prime }\left( X\right) \right) ^{2}%
}{96\left\vert f\left( \hat{X}\right) \right\vert ^{3}}\right) \sqrt{\left(
p+\frac{1}{2}\right) \sqrt{\frac{\sigma _{X}^{2}\sigma _{\hat{K}}^{2}\left(
f^{\prime }\left( X\right) \right) ^{2}}{96\left\vert f\left( \hat{X}\right)
\right\vert ^{3}}}}  \notag
\end{eqnarray}%
To solve (\ref{dvM}) for $K_{\hat{X}}$, we proceed in two steps.

We first introduce an intermediate variable $W$ and rewrite (\ref{dvM}) as
an equation for $K_{\hat{X}}$ and $W$. We set:%
\begin{equation}
\sqrt{\frac{\sigma _{X}^{2}\sigma _{\hat{K}}^{2}\left( f^{\prime }\left(
X\right) \right) ^{2}}{96\left\vert f\left( \hat{X}\right) \right\vert ^{3}}}%
\left( p+\frac{1}{2}-\frac{48\left\vert f\left( \hat{X}\right) \right\vert
^{3}}{\sigma _{X}^{2}\sigma _{\hat{K}}^{2}\left( f^{\prime }\left( X\right)
\right) ^{2}}\left( \ln \left( \bar{p}+\frac{1}{2}\right) -1\right) \right)
=W  \label{lpw}
\end{equation}%
and rewrite equation (\ref{dvM}) partly in terms of $W$:%
\begin{eqnarray}
&&K_{\hat{X}}\left\Vert \Psi \left( \hat{X}\right) \right\Vert ^{2}\left( 
\frac{\sigma _{X}^{2}\left( f^{\prime }\left( X\right) \right)
^{2}\left\vert f\left( \hat{X}\right) \right\vert \exp \left( -\frac{%
96\left\vert f\left( \hat{X}\right) \right\vert ^{3}}{\sigma _{X}^{2}\sigma
_{\hat{K}}^{2}\left( f^{\prime }\left( X\right) \right) ^{2}}\left( \ln
\left( \bar{p}+\frac{1}{2}\right) -1\right) ^{2}\right) }{96\left( \sigma _{%
\hat{K}}^{2}\right) ^{3}}\right) ^{\frac{1}{4}}  \label{mdg} \\
&=&C\left( \bar{p}\right) \exp \left( -W^{2}\right) \sqrt{W+2\sqrt{\frac{%
96\left\vert f\left( \hat{X}\right) \right\vert ^{3}}{\sigma _{X}^{2}\sigma
_{\hat{K}}^{2}\left( f^{\prime }\left( X\right) \right) ^{2}}}\left( \ln
\left( \bar{p}+\frac{1}{2}\right) -1\right) }  \notag
\end{eqnarray}%
\bigskip

Note that, as seen from (\ref{lpw}), $W$ is a function of $p$ and as such
can be seen as a parameter depending on the shape of the sectors space.

Equation (\ref{mdg}) both depends on $K_{\hat{X}}$ and $W$, and in a second
step, we use (\ref{lpw}) to write $K_{\hat{X}}$ as a function of $W$. To do
so, we use that in the intermediate case $\infty >K_{\hat{X}}>1$, we can
assume that: 
\begin{equation}
f\left( \hat{X}\right) =B_{1}\left( X\right) K_{\hat{X}}^{\alpha
-1}+B_{2}\left( X\right) K_{\hat{X}}^{\alpha }-C\left( \hat{X}\right) \simeq
B_{2}\left( X\right) K_{\hat{X}}^{\alpha }  \label{smT}
\end{equation}%
and that:%
\begin{equation}
\frac{M-\left( \frac{\left( \nabla _{\hat{X}}R\left( \hat{X}\right) A\left( 
\hat{X}\right) K_{\hat{X}}^{\alpha }\right) ^{2}}{\sigma _{\hat{X}}^{2}}%
+\nabla _{\hat{X}}^{2}R\left( \hat{X}\right) A\left( \hat{X}\right) K_{\hat{X%
}}^{\alpha }\right) }{B_{1}\left( X\right) K_{\hat{X}}^{\alpha
-1}+B_{2}\left( X\right) K_{\hat{X}}^{\alpha }-C\left( \hat{X}\right) }-%
\frac{3}{2}\simeq \frac{M-\left( \frac{\left( \nabla _{\hat{X}}R\left( \hat{X%
}\right) A\left( \hat{X}\right) K_{\hat{X}}^{\alpha }\right) ^{2}}{\sigma _{%
\hat{X}}^{2}}+\nabla _{\hat{X}}^{2}R\left( \hat{X}\right) A\left( \hat{X}%
\right) K_{\hat{X}}^{\alpha }\right) }{B_{2}\left( X\right) K_{\hat{X}%
}^{\alpha }}-\frac{3}{2}  \label{ssp}
\end{equation}%
Moreover, we can approximate $\left\Vert \Psi \left( \hat{X}\right)
\right\Vert ^{2}$:%
\begin{equation}
\left\Vert \Psi \left( \hat{X}\right) \right\Vert ^{2}\simeq D  \label{ssP}
\end{equation}

Our assumptions (\ref{smT}), (\ref{ssp}) and (\ref{ssP}) allow to rewrite
the relation (\ref{lpw}) between $K_{\hat{X}}^{\alpha }$ and $W$ as:%
\begin{equation*}
\sqrt{\frac{\sigma _{X}^{2}\sigma _{\hat{K}}^{2}\left( B_{2}^{\prime }\left(
X\right) \right) ^{2}}{96B_{2}^{3}\left( X\right) K_{\hat{X}}^{\alpha }}}%
\left( p+\frac{1}{2}-\frac{48\left\vert f\left( \hat{X}\right) \right\vert
^{3}}{\sigma _{X}^{2}\sigma _{\hat{K}}^{2}\left( f^{\prime }\left( X\right)
\right) ^{2}}\left( \ln \left( \bar{p}+\frac{1}{2}\right) -1\right) \right)
=W
\end{equation*}%
that is:%
\begin{eqnarray}
W &=&\sqrt{\frac{\sigma _{X}^{2}\sigma _{\hat{K}}^{2}\left( B_{2}^{\prime
}\left( X\right) \right) ^{2}}{96B_{2}^{5}\left( X\right) K_{\hat{X}%
}^{3\alpha }}}  \label{knw} \\
&&\times \left( M-\left( \frac{\left( \nabla _{\hat{X}}R\left( \hat{X}%
\right) \right) ^{2}A\left( \hat{X}\right) }{\sigma _{\hat{X}}^{2}}+\frac{%
48B_{2}^{4}\left( X\right) \left( \ln \left( \bar{p}+\frac{1}{2}\right)
-1\right) }{\sigma _{X}^{2}\sigma _{\hat{K}}^{2}\left( B_{2}^{\prime }\left(
X\right) \right) ^{2}}\right) K_{\hat{X}}^{2\alpha }-\left( \nabla _{\hat{X}%
}^{2}R\left( \hat{X}\right) +B_{2}\left( X\right) \right) K_{\hat{X}%
}^{\alpha }\right)  \notag
\end{eqnarray}

To solve this equation for $K_{\hat{X}}^{\alpha }$, we consider $M$ as the
dominant parameter and find an approximate solution of (\ref{knw}). At the
lowest order, we write:%
\begin{equation*}
\sqrt{\frac{\sigma _{X}^{2}\sigma _{\hat{K}}^{2}\left( B_{2}^{\prime }\left(
X\right) \right) ^{2}}{96B_{2}^{5}\left( X\right) K_{\hat{X}}^{3\alpha }}}M=W
\end{equation*}%
with solution:%
\begin{equation*}
K_{\hat{X}}^{\alpha }=\left( \frac{\sigma _{X}^{2}\sigma _{\hat{K}%
}^{2}\left( B_{2}^{\prime }\left( X\right) \right) ^{2}M^{2}}{%
96B_{2}^{5}\left( X\right) W^{2}}\right) ^{\frac{1}{3}}
\end{equation*}%
Considering corrections to this result, the solution to (\ref{knw}) is
decomposed as:%
\begin{equation}
K_{\hat{X}}^{\alpha }=\left( \frac{\sigma _{X}^{2}\sigma _{\hat{K}%
}^{2}\left( B_{2}^{\prime }\left( X\right) \right) ^{2}M^{2}}{%
96B_{2}^{5}\left( X\right) W^{2}}\right) ^{\frac{1}{3}}+\chi  \label{slK}
\end{equation}%
and using the following intermediate results:%
\begin{equation*}
K_{\hat{X}}^{2\alpha }=\left( \frac{\sigma _{X}^{2}\sigma _{\hat{K}%
}^{2}\left( B_{2}^{\prime }\left( X\right) \right) ^{2}M^{2}}{%
96B_{2}^{5}\left( X\right) W^{2}}\right) ^{\frac{2}{3}}\left( 1+2\chi \left( 
\frac{\sigma _{X}^{2}\sigma _{\hat{K}}^{2}\left( B_{2}^{\prime }\left(
X\right) \right) ^{2}M^{2}}{96B_{2}^{5}\left( X\right) W^{2}}\right) ^{-%
\frac{1}{3}}\right)
\end{equation*}%
\begin{equation*}
K_{\hat{X}}^{3\alpha }=\left( \frac{\sigma _{X}^{2}\sigma _{\hat{K}%
}^{2}\left( B_{2}^{\prime }\left( X\right) \right) ^{2}M^{2}}{%
96B_{2}^{5}\left( X\right) W^{2}}\right) \left( 1+3\chi \left( \frac{\sigma
_{X}^{2}\sigma _{\hat{K}}^{2}\left( B_{2}^{\prime }\left( X\right) \right)
^{2}M^{2}}{96B_{2}^{5}\left( X\right) W^{2}}\right) ^{-\frac{1}{3}}\right)
\end{equation*}%
we are led to rewrite (\ref{knw}) as an equation for $\chi $ at first order:%
\begin{eqnarray*}
&&\chi \left( \frac{3}{2}\left( \frac{\sigma _{X}^{2}\sigma _{\hat{K}%
}^{2}\left( B_{2}^{\prime }\left( X\right) \right) ^{2}M^{2}}{%
96B_{2}^{5}\left( X\right) W^{2}}\right) ^{-\frac{1}{3}}W\right. \\
&&\left. +2\frac{W}{M}\left( \frac{\left( \nabla _{\hat{X}}R\left( \hat{X}%
\right) \right) ^{2}A\left( \hat{X}\right) }{\sigma _{\hat{X}}^{2}}+\frac{%
48B_{2}^{4}\left( X\right) \left( \ln \left( \bar{p}+\frac{1}{2}\right)
-1\right) }{\sigma _{X}^{2}\sigma _{\hat{K}}^{2}\left( B_{2}^{\prime }\left(
X\right) \right) ^{2}}\right) \left( \frac{\sigma _{X}^{2}\sigma _{\hat{K}%
}^{2}\left( B_{2}^{\prime }\left( X\right) \right) ^{2}M^{2}}{%
96B_{2}^{5}\left( X\right) W^{2}}\right) ^{\frac{1}{3}}+\frac{W}{M}\left(
\nabla _{\hat{X}}^{2}R\left( \hat{X}\right) +B_{2}\left( X\right) \right)
\right) \\
&=&-\frac{W}{M}\left( \frac{\left( \nabla _{\hat{X}}R\left( \hat{X}\right)
\right) ^{2}A\left( \hat{X}\right) }{\sigma _{\hat{X}}^{2}}+\frac{%
48B_{2}^{4}\left( X\right) \left( \ln \left( \bar{p}+\frac{1}{2}\right)
-1\right) }{\sigma _{X}^{2}\sigma _{\hat{K}}^{2}\left( B_{2}^{\prime }\left(
X\right) \right) ^{2}}\right) \left( \frac{\sigma _{X}^{2}\sigma _{\hat{K}%
}^{2}\left( B_{2}^{\prime }\left( X\right) \right) ^{2}M^{2}}{%
96B_{2}^{5}\left( X\right) W^{2}}\right) ^{\frac{2}{3}} \\
&&-\frac{W}{M}\left( \nabla _{\hat{X}}^{2}R\left( \hat{X}\right)
+B_{2}\left( X\right) \right) \left( \frac{\sigma _{X}^{2}\sigma _{\hat{K}%
}^{2}\left( B_{2}^{\prime }\left( X\right) \right) ^{2}M^{2}}{%
96B_{2}^{5}\left( X\right) W^{2}}\right) ^{\frac{1}{3}}
\end{eqnarray*}%
whose solution is:%
\begin{equation*}
\chi =-\frac{\left( \frac{\left( \nabla _{\hat{X}}R\left( \hat{X}\right)
\right) ^{2}A\left( \hat{X}\right) }{\sigma _{\hat{X}}^{2}}+\frac{%
48B_{2}^{4}\left( X\right) \left( \ln \left( \bar{p}+\frac{1}{2}\right)
-1\right) }{\sigma _{X}^{2}\sigma _{\hat{K}}^{2}\left( B_{2}^{\prime }\left(
X\right) \right) ^{2}}\right) \left( \frac{\sigma _{X}^{2}\sigma _{\hat{K}%
}^{2}\left( B_{2}^{\prime }\left( X\right) \right) ^{2}M^{2}}{%
96B_{2}^{5}\left( X\right) W^{2}}\right) +\left( \nabla _{\hat{X}%
}^{2}R\left( \hat{X}\right) +B_{2}\left( X\right) \right) \left( \frac{%
\sigma _{X}^{2}\sigma _{\hat{K}}^{2}\left( B_{2}^{\prime }\left( X\right)
\right) ^{2}M^{2}}{96B_{2}^{5}\left( X\right) W^{2}}\right) ^{\frac{2}{3}}}{%
\frac{3}{2}M+2\left( \frac{\left( \nabla _{\hat{X}}R\left( \hat{X}\right)
\right) ^{2}A\left( \hat{X}\right) }{\sigma _{\hat{X}}^{2}}+\frac{%
48B_{2}^{4}\left( X\right) \left( \ln \left( \bar{p}+\frac{1}{2}\right)
-1\right) }{\sigma _{X}^{2}\sigma _{\hat{K}}^{2}\left( B_{2}^{\prime }\left(
X\right) \right) ^{2}}\right) \left( \frac{\sigma _{X}^{2}\sigma _{\hat{K}%
}^{2}\left( B_{2}^{\prime }\left( X\right) \right) ^{2}M^{2}}{%
96B_{2}^{5}\left( X\right) W^{2}}\right) ^{\frac{2}{3}}+\left( \nabla _{\hat{%
X}}^{2}R\left( \hat{X}\right) +B_{2}\left( X\right) \right) \left( \frac{%
\sigma _{X}^{2}\sigma _{\hat{K}}^{2}\left( B_{2}^{\prime }\left( X\right)
\right) ^{2}M^{2}}{96B_{2}^{5}\left( X\right) W^{2}}\right) ^{\frac{1}{3}}}
\end{equation*}%
so that (\ref{slK}) yields $K_{\hat{X}}^{\alpha }$:%
\begin{eqnarray}
&&K_{\hat{X}}^{\alpha }-\left( \frac{\sigma _{X}^{2}\sigma _{\hat{K}%
}^{2}\left( B_{2}^{\prime }\left( X\right) \right) ^{2}M^{2}}{%
96B_{2}^{5}\left( X\right) W^{2}}\right) ^{\frac{1}{3}}  \label{kxf} \\
&=&-\frac{\left( \frac{\left( \nabla _{\hat{X}}R\left( \hat{X}\right)
\right) ^{2}A\left( \hat{X}\right) }{\sigma _{\hat{X}}^{2}}+\frac{%
48B_{2}^{4}\left( X\right) \left( \ln \left( \bar{p}+\frac{1}{2}\right)
-1\right) }{\sigma _{X}^{2}\sigma _{\hat{K}}^{2}\left( B_{2}^{\prime }\left(
X\right) \right) ^{2}}\right) \left( \frac{\sigma _{X}^{2}\sigma _{\hat{K}%
}^{2}\left( B_{2}^{\prime }\left( X\right) \right) ^{2}M^{2}}{%
96B_{2}^{5}\left( X\right) W^{2}}\right) +\left( \nabla _{\hat{X}%
}^{2}R\left( \hat{X}\right) +B_{2}\left( X\right) \right) \left( \frac{%
\sigma _{X}^{2}\sigma _{\hat{K}}^{2}\left( B_{2}^{\prime }\left( X\right)
\right) ^{2}M^{2}}{96B_{2}^{5}\left( X\right) W^{2}}\right) ^{\frac{2}{3}}}{%
\frac{3}{2}M+2\left( \frac{\left( \nabla _{\hat{X}}R\left( \hat{X}\right)
\right) ^{2}A\left( \hat{X}\right) }{\sigma _{\hat{X}}^{2}}+\frac{%
48B_{2}^{4}\left( X\right) \left( \ln \left( \bar{p}+\frac{1}{2}\right)
-1\right) }{\sigma _{X}^{2}\sigma _{\hat{K}}^{2}\left( B_{2}^{\prime }\left(
X\right) \right) ^{2}}\right) \left( \frac{\sigma _{X}^{2}\sigma _{\hat{K}%
}^{2}\left( B_{2}^{\prime }\left( X\right) \right) ^{2}M^{2}}{%
96B_{2}^{5}\left( X\right) W^{2}}\right) ^{\frac{2}{3}}+\left( \nabla _{\hat{%
X}}^{2}R\left( \hat{X}\right) +B_{2}\left( X\right) \right) \left( \frac{%
\sigma _{X}^{2}\sigma _{\hat{K}}^{2}\left( B_{2}^{\prime }\left( X\right)
\right) ^{2}M^{2}}{96B_{2}^{5}\left( X\right) W^{2}}\right) ^{\frac{1}{3}}} 
\notag \\
&&\frac{\left( \frac{\left( \nabla _{\hat{X}}R\left( \hat{X}\right) \right)
^{2}A\left( \hat{X}\right) }{\sigma _{\hat{X}}^{2}}+\frac{48B_{2}^{4}\left(
X\right) \left( \ln \left( \bar{p}+\frac{1}{2}\right) -1\right) }{\sigma
_{X}^{2}\sigma _{\hat{K}}^{2}\left( B_{2}^{\prime }\left( X\right) \right)
^{2}}\right) \left( \frac{\sigma _{X}^{2}\sigma _{\hat{K}}^{2}\left(
B_{2}^{\prime }\left( X\right) \right) ^{2}M^{2}}{96B_{2}^{5}\left( X\right)
W^{2}}\right) +\left( \nabla _{\hat{X}}^{2}R\left( \hat{X}\right)
+B_{2}\left( X\right) \right) \left( \frac{\sigma _{X}^{2}\sigma _{\hat{K}%
}^{2}\left( B_{2}^{\prime }\left( X\right) \right) ^{2}M^{2}}{%
96B_{2}^{5}\left( X\right) W^{2}}\right) ^{\frac{2}{3}}}{\frac{3}{2}%
M+2\left( \frac{\left( \nabla _{\hat{X}}R\left( \hat{X}\right) \right)
^{2}A\left( \hat{X}\right) }{\sigma _{\hat{X}}^{2}}+\frac{48B_{2}^{4}\left(
X\right) \left( \ln \left( \bar{p}+\frac{1}{2}\right) -1\right) }{\sigma
_{X}^{2}\sigma _{\hat{K}}^{2}\left( B_{2}^{\prime }\left( X\right) \right)
^{2}}\right) \left( \frac{\sigma _{X}^{2}\sigma _{\hat{K}}^{2}\left(
B_{2}^{\prime }\left( X\right) \right) ^{2}M^{2}}{96B_{2}^{5}\left( X\right)
W^{2}}\right) ^{\frac{2}{3}}+\left( \nabla _{\hat{X}}^{2}R\left( \hat{X}%
\right) +B_{2}\left( X\right) \right) \left( \frac{\sigma _{X}^{2}\sigma _{%
\hat{K}}^{2}\left( B_{2}^{\prime }\left( X\right) \right) ^{2}M^{2}}{%
96B_{2}^{5}\left( X\right) W^{2}}\right) ^{\frac{1}{3}}}  \notag
\end{eqnarray}%
In a third step, we can use equation (\ref{kxf}) to rewrite (\ref{mdg}) in
an approximate form. Actually, expression (\ref{kxf}) implies that in the
intermediate case, where $K_{\hat{X}}^{\alpha }$ is of finite magnitude, we
have $W^{2}\sim \sigma _{X}^{2}\sigma _{\hat{K}}^{2}M^{2}$ and:%
\begin{eqnarray*}
&&\exp \left( -W^{2}+\frac{24\left\vert B_{2}\left( X\right) \right\vert
^{3}K_{\hat{X}}^{\alpha }}{\sigma _{X}^{2}\sigma _{\hat{K}}^{2}\left(
B_{2}^{\prime }\left( X\right) \right) ^{2}}\left( \ln \left( \bar{p}+\frac{1%
}{2}\right) -1\right) ^{2}\right) \\
&\simeq &\exp \left( \frac{24\left\vert B_{2}\left( X\right) \right\vert
^{3}K_{\hat{X}}^{\alpha }}{\sigma _{X}^{2}\sigma _{\hat{K}}^{2}\left(
B_{2}^{\prime }\left( X\right) \right) ^{2}}\left( \ln \left( \bar{p}+\frac{1%
}{2}\right) -1\right) ^{2}\right)
\end{eqnarray*}%
Moreover using that:%
\begin{equation*}
W+2\sqrt{\frac{96\left\vert f\left( \hat{X}\right) \right\vert ^{3}}{\sigma
_{X}^{2}\sigma _{\hat{K}}^{2}\left( f^{\prime }\left( X\right) \right) ^{2}}}%
\left( \ln \left( \bar{p}+\frac{1}{2}\right) -1\right) \simeq 2\sqrt{\frac{%
96\left\vert f\left( \hat{X}\right) \right\vert ^{3}}{\sigma _{X}^{2}\sigma
_{\hat{K}}^{2}\left( f^{\prime }\left( X\right) \right) ^{2}}}\left( \ln
\left( \bar{p}+\frac{1}{2}\right) -1\right)
\end{equation*}%
and that ultimately the left hand side of equation (\ref{mdg}) writes at the
first order:

\begin{eqnarray*}
&&K_{\hat{X}}\left\Vert \Psi \left( \hat{X}\right) \right\Vert ^{2}\left( 
\frac{\sigma _{X}^{2}\left( f^{\prime }\left( X\right) \right)
^{2}\left\vert f\left( \hat{X}\right) \right\vert \exp \left( -\frac{%
96\left\vert f\left( \hat{X}\right) \right\vert ^{3}}{\sigma _{X}^{2}\sigma
_{\hat{K}}^{2}\left( f^{\prime }\left( X\right) \right) ^{2}}\left( \ln
\left( \bar{p}+\frac{1}{2}\right) -1\right) ^{2}\right) }{96\left( \sigma _{%
\hat{K}}^{2}\right) ^{3}}\right) ^{\frac{1}{4}} \\
&=&\left( \frac{\sigma _{X}^{2}\left( B_{2}^{\prime }\left( X\right) \right)
^{2}\left\vert B_{2}\left( X\right) \right\vert }{96\left( \sigma _{\hat{K}%
}^{2}\right) ^{3}}\right) ^{\frac{1}{4}}K_{\hat{X}}^{1+\frac{3\alpha }{4}%
}\left\Vert \Psi \left( \hat{X}\right) \right\Vert ^{2}\exp \left( -\frac{%
24\left\vert f\left( \hat{X}\right) \right\vert ^{3}}{\sigma _{X}^{2}\sigma
_{\hat{K}}^{2}\left( f^{\prime }\left( X\right) \right) ^{2}}\left( \ln
\left( \bar{p}+\frac{1}{2}\right) -1\right) ^{2}\right) \\
&\simeq &D\left( \frac{\sigma _{X}^{2}\left( B_{2}^{\prime }\left( X\right)
\right) ^{2}\left\vert B_{2}\left( X\right) \right\vert }{96\left( \sigma _{%
\hat{K}}^{2}\right) ^{3}}\right) ^{\frac{1}{4}}K_{\hat{X}}^{1+\frac{3\alpha 
}{4}}\exp \left( -\frac{24\left\vert B_{2}\left( X\right) \right\vert ^{3}K_{%
\hat{X}}^{\alpha }}{\sigma _{X}^{2}\sigma _{\hat{K}}^{2}\left( B_{2}^{\prime
}\left( X\right) \right) ^{2}}\left( \ln \left( \bar{p}+\frac{1}{2}\right)
-1\right) ^{2}\right)
\end{eqnarray*}%
equation (\ref{mdg}) writes:%
\begin{eqnarray*}
&&D\left( \frac{\sigma _{X}^{2}\left( B_{2}^{\prime }\left( X\right) \right)
^{2}\left\vert B_{2}\left( X\right) \right\vert }{96\left( \sigma _{\hat{K}%
}^{2}\right) ^{3}}\right) ^{\frac{1}{4}}K_{\hat{X}}^{1+\frac{3\alpha }{4}%
}\exp \left( -\frac{24\left\vert B_{2}\left( X\right) \right\vert ^{3}K_{%
\hat{X}}^{\alpha }}{\sigma _{X}^{2}\sigma _{\hat{K}}^{2}\left( B_{2}^{\prime
}\left( X\right) \right) ^{2}}\left( \ln \left( \bar{p}+\frac{1}{2}\right)
-1\right) ^{2}\right) \\
&=&C\left( \bar{p}\right) \sqrt{2\sqrt{\frac{96\left\vert B_{2}\left(
X\right) \right\vert ^{3}K_{\hat{X}}^{\alpha }}{\sigma _{X}^{2}\sigma _{\hat{%
K}}^{2}\left( B_{2}^{\prime }\left( X\right) \right) ^{2}}}\left( \ln \left( 
\bar{p}+\frac{1}{2}\right) -1\right) }
\end{eqnarray*}%
that is:

\begin{equation}
K_{\hat{X}}^{1+\frac{\alpha }{2}}\exp \left( -\frac{24\left\vert B_{2}\left(
X\right) \right\vert ^{3}K_{\hat{X}}^{\alpha }}{\sigma _{X}^{2}\sigma _{\hat{%
K}}^{2}\left( B_{2}^{\prime }\left( X\right) \right) ^{2}}\left( \ln \left( 
\bar{p}+\frac{1}{2}\right) -1\right) ^{2}\right) =\frac{8C\left( \bar{p}%
\right) }{D}\sqrt{\frac{3\sigma _{\hat{K}}^{2}\left\vert B_{2}\left(
X\right) \right\vert }{\sigma _{X}^{2}\left( B_{2}^{\prime }\left( X\right)
\right) ^{2}}\left( \ln \left( \bar{p}+\frac{1}{2}\right) -1\right) }
\label{qTN}
\end{equation}%
Equation (\ref{qTN}) has the form: 
\begin{equation*}
x^{d}\exp \left( -ax\right) =c
\end{equation*}%
$\allowbreak $with solution:%
\begin{equation*}
x=c^{\frac{1}{d}}\exp \left( -W_{0}\left( -\frac{a}{d}c^{\frac{1}{d}}\right)
\right)
\end{equation*}%
where $W_{0}$ is the Lambert $W$ function with parameter $0$. Applying this
result to our case with:%
\begin{eqnarray*}
d &=&\frac{1+\alpha }{2\alpha } \\
x &=&K_{\hat{X}}^{\alpha } \\
a &=&\frac{24\left\vert B_{2}\left( X\right) \right\vert ^{3}}{\sigma
_{X}^{2}\sigma _{\hat{K}}^{2}\left( B_{2}^{\prime }\left( X\right) \right)
^{2}}\left( \ln \left( \bar{p}+\frac{1}{2}\right) -1\right) ^{2} \\
c &=&\frac{8C\left( \bar{p}\right) }{D}\sqrt{\frac{3\sigma _{\hat{K}%
}^{2}\left\vert B_{2}\left( X\right) \right\vert }{\sigma _{X}^{2}\left(
B_{2}^{\prime }\left( X\right) \right) ^{2}}\left( \ln \left( \bar{p}+\frac{1%
}{2}\right) -1\right) }
\end{eqnarray*}%
we obtain:%
\begin{eqnarray*}
K_{\hat{X}}^{\alpha } &=&\left( \frac{8C\left( \bar{p}\right) }{D}\sqrt{%
\frac{3\sigma _{\hat{K}}^{2}\left\vert B_{2}\left( X\right) \right\vert }{%
\sigma _{X}^{2}\left( B_{2}^{\prime }\left( X\right) \right) ^{2}}\left( \ln
\left( \bar{p}+\frac{1}{2}\right) -1\right) }\right) ^{\frac{2\alpha }{%
1+\alpha }} \\
&&\times \exp \left( -W_{0}\left( -\frac{48\alpha }{1+\alpha }\left( \sqrt{%
\frac{3\sigma _{\hat{K}}^{2}}{\sigma _{X}^{2}}}\frac{8C\left( \bar{p}\right) 
}{D}\right) ^{\frac{2\alpha }{1+\alpha }}\frac{\left\vert B_{2}\left(
X\right) \right\vert ^{3+\frac{\alpha }{1+\alpha }}}{\sigma _{X}^{2}\sigma _{%
\hat{K}}^{2}\left( B_{2}^{\prime }\left( X\right) \right) ^{2+\frac{2\alpha 
}{1+\alpha }}}\left( \ln \left( \bar{p}+\frac{1}{2}\right) -1\right) ^{2+%
\frac{\alpha }{1+\alpha }}\right) \right)
\end{eqnarray*}%
As stated in the text, this is an increasing function of $B_{2}\left(
X\right) $. Moreover, the corrections to this formula, given in (\ref{kxf})
show that $K_{\hat{X}}^{\alpha }$ is a decreasing function of $\left( \nabla
_{\hat{X}}R\left( \hat{X}\right) \right) ^{2}$ and $\nabla _{\hat{X}%
}^{2}R\left( \hat{X}\right) $.

\subparagraph{Case 5. Low capital, $K_{\hat{X}}<<1$:}

When average physical capital per firm in sector $\hat{X}$ is very low, we
can use our assumptions about $g\left( \hat{X},\Psi ,\hat{\Psi}\right) $ and 
$\nabla _{\hat{X}}g\left( \hat{X},\Psi ,\hat{\Psi}\right) $, equations (\ref%
{gtp}) and (\ref{ptg}), and assume that:%
\begin{equation}
f\left( \hat{X}\right) \simeq B_{1}\left( \hat{X}\right) K_{\hat{X}}^{\alpha
-1}>>1
\end{equation}%
and: 
\begin{equation*}
g\left( \hat{X}\right) \simeq 0
\end{equation*}%
and moreover that:%
\begin{equation*}
\left\Vert \Psi \left( \hat{X}\right) \right\Vert ^{2}=D-L\left( X\right)
\left( \nabla _{X}R\left( X\right) \right) ^{2}K_{\hat{X}}^{\alpha }\simeq D
\end{equation*}%
For these conditions, the solution of (\ref{qtc}) is locally stable.

Moreover, the conditions $K_{\hat{X}}<<1$ and the defining equation (\ref%
{STP}) for $f$ imply that $f>0$, and that for $\alpha <1$:%
\begin{equation*}
\frac{\sigma _{X}^{2}\sigma _{\hat{K}}^{2}\left( p+\frac{1}{2}\right)
^{2}\left( f^{\prime }\left( X\right) \right) ^{2}}{96\left\vert f\left( 
\hat{X}\right) \right\vert ^{3}}<<1
\end{equation*}%
Under these assumptions, equation (\ref{qtc}) reduces to:%
\begin{equation}
K_{\hat{X}}D\left\vert f\left( \hat{X}\right) \right\vert \simeq C\left( 
\bar{p}\right) \sigma _{\hat{K}}^{2}\hat{\Gamma}\left( p+\frac{1}{2}\right)
\label{qtC}
\end{equation}%
This equation (\ref{qtC}) can be approximated.\ Actually, using formula (\ref%
{xpr}) for $p$\ yields:%
\begin{equation*}
p+\frac{1}{2}=\frac{M-\left( \frac{\left( g\left( \hat{X}\right) \right) ^{2}%
}{\sigma _{\hat{X}}^{2}}+\nabla _{\hat{X}}g\left( \hat{X},K_{\hat{X}}\right)
\right) }{\sqrt{f^{2}\left( \hat{X}\right) }}-1\simeq -1
\end{equation*}%
and an expansion of $\hat{\Gamma}\left( p+\frac{1}{2}\right) $\ around the
value $p+\frac{1}{2}=-1$\ writes:%
\begin{equation*}
\hat{\Gamma}\left( p+\frac{1}{2}\right) \simeq \hat{\Gamma}\left( -1\right) +%
\hat{\Gamma}^{\prime }\left( -1\right) \frac{M-\left( \frac{\left( g\left( 
\hat{X}\right) \right) ^{2}}{\sigma _{\hat{X}}^{2}}+\nabla _{\hat{X}}g\left( 
\hat{X},K_{\hat{X}}\right) \right) }{\sqrt{f^{2}\left( \hat{X}\right) }}
\end{equation*}%
Consequently, when returns are large, i.e. $f\left( \hat{X}\right) >>1$,
equation (\ref{qtc}) writes:%
\begin{equation*}
K_{\hat{X}}\left( B_{1}\left( \hat{X}\right) K_{\hat{X}}^{\alpha -1}\right)
\simeq \frac{C\left( \bar{p}\right) \sigma _{\hat{K}}^{2}}{D}\left( \hat{%
\Gamma}\left( -1\right) +\hat{\Gamma}^{\prime }\left( -1\right) \frac{%
M-\left( \frac{\left( g\left( \hat{X}\right) \right) ^{2}}{\sigma _{\hat{X}%
}^{2}}+\nabla _{\hat{X}}g\left( \hat{X},K_{\hat{X}}\right) \right) }{%
B_{1}\left( \hat{X}\right) K_{\hat{X}}^{\alpha -1}}\right)
\end{equation*}%
with first order solution\footnote{%
Given our hypotheses, $D>>1$\ , which implies that $K_{\hat{X}}<<1$,\ as
needed.}:%
\begin{equation}
K_{\hat{X}}=\left( \frac{C\left( \bar{p}\right) \sigma _{\hat{K}}^{2}\hat{%
\Gamma}\left( -1\right) }{DB_{1}\left( \hat{X}\right) }\right) ^{\frac{1}{%
\alpha }}+\frac{\frac{C\left( \bar{p}\right) \sigma _{\hat{K}}^{2}}{D}\hat{%
\Gamma}^{\prime }\left( -1\right) \left( M-\left( \frac{\left( g\left( \hat{X%
}\right) \right) ^{2}}{\sigma _{\hat{X}}^{2}}+\nabla _{\hat{X}}g\left( \hat{X%
},K_{\hat{X}}\right) \right) \right) }{B_{1}^{\frac{1}{\alpha }}\left( \hat{X%
}\right) \left( \frac{C\left( \bar{p}\right) \sigma _{\hat{K}}^{2}\hat{\Gamma%
}\left( -1\right) }{D}\right) ^{1-\frac{1}{\alpha }}}  \label{rst}
\end{equation}

Equation (\ref{rst}) shows that average capital $K_{\hat{X}}$\ increases
with $M-\left( \frac{\left( g\left( \hat{X}\right) \right) ^{2}}{\sigma _{%
\hat{X}}^{2}}+\nabla _{\hat{X}}g\left( \hat{X},K_{\hat{X}}\right) \right) $:
when expected long-term returns increase, more capital is allocated to the
sector. Equation (\ref{srt}) also shows that average capital $K_{\hat{X}}$\
is maximal when returns $R\left( \hat{X}\right) $\ are at a local maximum,
i.e. when $\frac{\left( g\left( \hat{X}\right) \right) ^{2}}{\sigma _{\hat{X}%
}^{2}}=0$\ and $\nabla _{\hat{X}}g\left( \hat{X},K_{\hat{X}}\right) <0$.

Inversely, the same equations (\ref{rst}) and (\ref{srt}) show that average
capital $K_{\hat{X}}$ is decreasing in $f\left( \hat{X}\right) $. The
equilibrium is unstable. When average capital is very low, i.e. $K_{\hat{X}%
}<<1$, which is the case studied here, marginal returns are high.\ Any
increase in capital above the threshold widely increases returns, which
drives capital towards the next stable equilibrium, with higher $K_{\hat{X}}$%
. Recall that in this unstable equilibrium, $K_{\hat{X}}$ must be seen as a
threshold.\ The rise in $f\left( \hat{X}\right) $ reduces the threshold $K_{%
\hat{X}}$, which favours\ capital accumulation and increases the average
capital $K_{\hat{X}}$.

This case is thus an exception: the dependency of $K_{\hat{X}}$\ in $R\left( 
\hat{X}\right) $\ is stable, but the dependency in $f\left( \hat{X}\right) $%
\ is unstable. This saddle path type of instability may lead the sector,
either towards a higher level of capital (case 4 below) or towards $0$.
where the sector disappears.

\subsection*{A 4.3 Instability and modification of sectors' space}

\subsubsection*{A 4.3.1 Disappearance of Low average capital sectors}

Average capital is unstable\textbf{\ }when\textbf{\ }$B\left( \hat{X}\right)
<-1$. A shock on average capital can either drive the equilibrium to some
stable value, or worsen the sector's capital landscape.

In the latter case, investors tend to desert the sector, so that both the
average capital and the density of investors tend to $0$: $K_{\hat{X}%
}\rightarrow 0$ and $\left\vert \hat{\Psi}\left( \hat{X},\hat{K}\right)
\right\vert ^{2}\rightarrow 0$.\ Producers remain in the sector but with a
very low capital on average.\ The very lack of capital prevents these firms
to shift towards more attractive sectors in the long run. Assuming physical
capital returns are Cobb-Douglas, marginal productivity is mathematically
high for a very low capital. Thus, short-term returns are very large:\ $%
f\left( \hat{X}\right) \rightarrow \infty $.\textbf{\ }

Note that this type of instability only applies to very low level of average
capital, so that the total capital involved is negligible, and this
instability does not impact the system globally.

\subsubsection*{A 4.3.2 Very high level of average capital and modification
of space}

Average capital is also unstable when $B\left( \hat{X}\right) >1$.\ However,
in this case investors are lured in the sector, so that average capital in
the sector increases quickly $K_{\hat{X}}\rightarrow \infty $, and
short-term returns tend to be small: $f\left( \hat{X}\right) \rightarrow c$
for some constant $c<<1$.\ Consequently, for $K_{\hat{X}}\rightarrow \infty $%
, $\frac{\partial f\left( \hat{X},K_{\hat{X}}\right) }{\partial K_{\hat{X}}}%
\rightarrow 0$, which translates decreasing marginal returns.\ Similarly,
the expected long-term returns will be caped, and $\frac{\partial p}{%
\partial K_{\hat{X}}}\rightarrow 0$, and $l\left( \hat{X},K_{\hat{X}}\right)
\rightarrow 0$.

The instability condition (\ref{nNT}) turns out to be a lower bound for the
sensitivity of firms density relative to average capital:%
\begin{equation}
\frac{\partial \ln \left\vert \Psi \left( \hat{X},K_{\hat{X}}\right)
\right\vert ^{2}}{\partial K_{\hat{X}}}>1  \label{DRP}
\end{equation}%
This lower bound creates a herd effect: the number of firms in sector $\hat{X%
}$ could grow indefinitely with capital: $\left\vert \Psi \left( \hat{X},K_{%
\hat{X}}\right) \right\vert ^{2}\rightarrow \infty $.

However, the fixed number of firms implies that this shift towards sector $%
\hat{X}$\ will necessarily reach a maximum $\left\vert \hat{\Psi}\left( \hat{%
X},\hat{K}\right) \right\vert _{\max }^{2}>>1$. For this maximum density,
the corresponding level of average capital at sector $\hat{X}$\ will be
approximatively:%
\begin{equation*}
K_{\hat{X}}\simeq K_{\max }=\frac{\ln \left( \left\vert \hat{\Psi}\left( 
\hat{X},\hat{K}\right) \right\vert _{\max }^{2}\right) }{r}
\end{equation*}%
This concentration of capital in some sectors directly impacts the amount of
disposable capital along with the instability condition (\ref{nNT}) for the
rest of the system. This occurs in several steps.

First, the disposable average capital for the rest of the system reduces to $%
\left\langle K\right\rangle -\frac{K_{\max }}{V}$, with $V$,\ the volume of
the sector space and $\left\langle K\right\rangle $,\ the average physical
capital in the whole space.

Second, this reduction of average capital negatively impacts the growth
prospects $R\left( \hat{X}\right) $,\ the stock prices $F_{1}\left( X\right) 
$, and consequently the short term returns $f\left( \hat{X}\right) $.

In turn, this modifies the stability condition $\left\vert B\left( \hat{X}%
\right) \right\vert $\ over the whole space. Consequently, some sectors will
move over the instability threshold $B\left( \hat{X}\right) >1$, while
others will move below $B\left( \hat{X}\right) <-1$. Some sectors will
experience a capital increase, others will disappear.

If a stable situation finally emerges, the resulting sectors' space will be
reduced: some sectors will have disappeared, and only sectors with positive
capital will have remained.

\subsection*{A.4.4 Global instability}

This appendix completes the analysis of the solutions of (\ref{qtk}) for
average capital. We have studied the local instability of solutions
previously. However, a second source of instability of the system arises
outside of the equations for average capital per firm per sector, (\ref{qtk}%
), and its differential version, (\ref{rvd}).\ It stems from the sectors'
space expected long-term returns. It is induced by the minimization
equations (\ref{hqn}) and (\ref{nqh}), and is a source of global instability
for the background field.

\subsubsection*{A 4.4.1 Mechanism of global instability:}

In these equations, the Lagrange multiplier $\hat{\lambda}$ is the
eigenvalue of a second-order differential equation. Because there exist an
infinite number of eigenvalues $\hat{\lambda}$, there are an infinite number
of local minimum background fields $\Psi \left( \hat{X},K_{\hat{X}}\right) $%
.\ But the most likely minimum, given in (\ref{mqn}), is obtained for $\hat{%
\lambda}=M$ (see appendix 2).

Yet\textbf{\ }$\hat{\lambda}$ is also the Lagrange multiplier that
implements the constraint of a fixed number $N$\ of agents.\ 

Since the number of investors is computed by:%
\begin{equation*}
\int \left\vert \Psi \left( \hat{X},K_{\hat{X}}\right) \right\vert
^{2}d\left( \hat{X},K_{\hat{X}}\right)
\end{equation*}%
the constraint implemented by $\hat{\lambda}$ is: 
\begin{equation}
\hat{N}=\int \left\vert \Psi \left( \hat{X},K_{\hat{X}}\right) \right\vert
^{2}d\left( \hat{X},K_{\hat{X}}\right)  \label{grandn}
\end{equation}%
since this constraint runs over the whole space, it is a global property of
the system.

Yet equations (\ref{hqn}) and (\ref{nqh}), the minimization equations
defining $\Psi \left( \hat{X},K_{\hat{X}}\right) $, may also be viewed as a
set of local minimization equations at each point $\hat{X}$ of the sector
space. Considered individually, each provide a lower minimum that could be
reached separately for each $\hat{X}$. In other words, provided each
sector's number of agents is fixed independently from the rest of the
system, a stable background field could be reached at every point.

However, our global constraint rules out this set of local minimizations.\
The solutions of (\ref{hqn}) and (\ref{nqh}) are thus a local minimum for
the sole points $\hat{X}$\ such that the lowest value of $\hat{\lambda}$ is
reached at $\hat{X}$, i.e. points such that\footnote{%
See definition (\ref{mfl}) and section 8.2. for a study of such points.
\par
{}}:%
\begin{equation}
A\left( \hat{X}\right) \equiv \frac{\left( g\left( \hat{X}\right) \right)
^{2}}{\sigma _{\hat{X}}^{2}}+f\left( \hat{X}\right) +\frac{1}{2}\sqrt{%
f^{2}\left( \hat{X}\right) }+\nabla _{\hat{X}}g\left( \hat{X}\right) -\frac{%
\sigma _{\hat{K}}^{2}F^{2}\left( \hat{X}\right) }{2f^{2}\left( \hat{X}%
\right) }=M  \label{stl}
\end{equation}%
For points $\hat{X}$ that do not satisfy (\ref{stl}), the solutions $\Psi
\left( \hat{X},K_{\hat{X}}\right) $ and $\Psi ^{\dag }\left( \hat{X},K_{\hat{%
X}}\right) $\ of (\ref{hqn}) and (\ref{nqh}), with $\hat{\lambda}=-M$ are
not global minima, but merely a local one. Any perturbation $\delta \Psi
\left( \hat{X},K_{\hat{X}}\right) $\ due to a change of parameters
destabilizes the whole system: the equilibrium is unstable.

The stability of both the background field and the potential equilibria are
thus determined by $A\left( \hat{X}\right) $, the sector space's overall
shape of returns and expectations. An homogeneous shape, a space such that $%
A\left( \hat{X}\right) $, presents small deviations around $M$ and is more
background-stable than an heterogeneous space.

More importantly, the background fields and associated average capital must
be understood as potential, not actual long-run equilibria: the whole system
is better described as a dynamical system, which is defined in section 5 of
the text, between potential backgrounds where time enters as a
macro-variable. We consider the results of the background field's dynamical
behavior in section 7.

\paragraph*{Removing global instability}

As mentioned above, an homogeneous shape is a space such that the parameter $%
A\left( \hat{X}\right) $ presents small deviations around $M$. In an
heterogeneous shape, the space presents large differences in $A\left( \hat{X}%
\right) $. We find that homogeneous shapes are more background-stable than
heterogeneous ones. This partly results from the global constraint (\ref%
{grandn}) imposed on the number of agents in the model, which ensures that
the number of financial agents in the system is fixed over the whole sector
space.

Relaxing this constraint fully would render the number of agents in sectors
independent.\ The associated background field of each sector could, at each
point, adjust to be minimum and stabilize the system.

To do so, we replace equation (\ref{hqn}), the minimization equation, by a
set of independent equations with independent Lagrange multipliers $\hat{%
\lambda}_{\hat{X}}$ for each sector $\hat{X}$, so that for each $\hat{X}$,
the minimum configuration is reached by setting: 
\begin{equation*}
\hat{\lambda}_{\hat{X}}=\frac{\left( g\left( \hat{X}\right) \right) ^{2}}{%
\sigma _{\hat{X}}^{2}}+f\left( \hat{X}\right) +\frac{1}{2}\sqrt{f^{2}\left( 
\hat{X}\right) }+\nabla _{\hat{X}}g\left( \hat{X}\right) -\frac{\sigma _{%
\hat{K}}^{2}F^{2}\left( \hat{X}\right) }{2f^{2}\left( \hat{X}\right) }
\end{equation*}%
This is similar to the Lagrange multiplier of the minimization equation for
the background field, stripped of the maximum condition $\hat{\lambda}=-M$%
\footnote{%
see discussion following equation (\ref{DFT}).}. This $\hat{X}$\ dependency
of the Lagrange multiplier implies that the average capital equation (\ref%
{qtk}) is replaced by\footnote{%
Expression (\ref{Ghf}) is used to compute $\hat{\Gamma}\left( \frac{1}{2}%
\right) $.}:%
\begin{equation}
K_{\hat{X}}\left\Vert \Psi \left( \hat{X}\right) \right\Vert ^{2}\left\vert
f\left( \hat{X}\right) \right\vert =C\left( \bar{p}\right) \sigma _{\hat{K}%
}^{2}\hat{\Gamma}\left( \frac{1}{2}\right) =C\left( \bar{p}\right) \sigma _{%
\hat{K}}^{2}\exp \left( -\frac{\sigma _{X}^{2}\sigma _{\hat{K}}^{2}\left(
f^{\prime }\left( X\right) \right) ^{2}}{384\left\vert f\left( \hat{X}%
\right) \right\vert ^{3}}\right)  \label{Gbl}
\end{equation}%
This equation is identical to (\ref{kpt}) and has thus at least one locally
stable solution. The solutions are computed in (\ref{Kms}) and (\ref{Kmn}).

Solutions to (\ref{Gbl}) do no longer directly depend on the relative
characteristics of a particular sector, but rather on the returns at point $%
f\left( \hat{X}\right) $ and on the number of firms in the sector, $%
\left\Vert \Psi \left( \hat{X}\right) \right\Vert ^{2}$. Yet this dependency
is only indirect, through the firms' density at sector $\hat{X}$, $%
\left\Vert \Psi \left( \hat{X}\right) \right\Vert ^{2}$, and this quantity
does not vary much in the sector space.

An intermediate situation between (\ref{qtk}) and (\ref{Gbl}) could also be
considered: it would be to assume a constant number of agents in some
regions of the sector space.

Alternatively, limiting the number of investors per sector can be achieved
through some public regulation to maintain a constant flow of investment in
the sector.

\section*{Appendix 5. \textbf{Dynamics for }$K_{\hat{X}}$}

\subsection*{A 5.1 Variation of the defining equation for $K_{\hat{X}}$}

\subsubsection*{A 5.1.1 Compact formulation}

As claimed in the text, we consider the dynamics for $K_{\hat{X}}$ generated
by modification of the parameters. To do so, we compute the variation of
equation (\ref{qnk}). We need the variations of the functions involved in (%
\ref{qnk}) with respect to two dynamical variables $K_{\hat{X}}$ and $%
R\left( X\right) $. Starting with (\ref{qnk}):

\begin{equation}
K_{\hat{X}}\left( D-L\left( \hat{X}\right) K_{\hat{X}}^{\eta }\right) =\frac{%
C\left( \bar{p}\right) \sigma _{\hat{K}}^{2}}{\left\vert f\left( \hat{X}%
\right) \right\vert }\hat{\Gamma}\left( p+\frac{1}{2}\right)  \label{dkn}
\end{equation}%
where:%
\begin{eqnarray}
\hat{\Gamma}\left( p+\frac{1}{2}\right) &=&\exp \left( -\frac{\sigma
_{X}^{2}\sigma _{\hat{K}}^{2}\left( p+\frac{1}{2}\right) ^{2}\left(
f^{\prime }\left( X\right) \right) ^{2}}{96\left\vert f\left( \hat{X}\right)
\right\vert ^{3}}\right) \\
&&\times \left( \frac{\Gamma \left( -\frac{p+1}{2}\right) \Gamma \left( 
\frac{1-p}{2}\right) -\Gamma \left( -\frac{p}{2}\right) \Gamma \left( \frac{%
-p}{2}\right) }{2^{p+2}\Gamma \left( -p-1\right) \Gamma \left( -p\right) }+p%
\frac{\Gamma \left( -\frac{p}{2}\right) \Gamma \left( \frac{2-p}{2}\right)
-\Gamma \left( -\frac{p-1}{2}\right) \Gamma \left( -\frac{p-1}{2}\right) }{%
2^{p+1}\Gamma \left( -p\right) \Gamma \left( -p+1\right) }\right)  \notag
\end{eqnarray}%
We first compute the variations of the right hand side and use that, in
first approximation:

\begin{equation}
\frac{d}{dp}\left( \ln \left( \frac{\Gamma \left( -\frac{p+1}{2}\right)
\Gamma \left( \frac{1-p}{2}\right) -\Gamma \left( -\frac{p}{2}\right) \Gamma
\left( \frac{-p}{2}\right) }{2^{p+2}\Gamma \left( -p-1\right) \Gamma \left(
-p\right) }+p\frac{\Gamma \left( -\frac{p}{2}\right) \Gamma \left( \frac{2-p%
}{2}\right) -\Gamma \left( -\frac{p-1}{2}\right) \Gamma \left( -\frac{p-1}{2}%
\right) }{2^{p+1}\Gamma \left( -p\right) \Gamma \left( -p+1\right) }\right)
\right) \simeq \ln \left( p+\frac{1}{2}\right)
\end{equation}%
and:%
\begin{equation*}
\frac{d}{dp}\left( -\frac{\sigma _{X}^{2}\left( p+\frac{1}{2}\right)
^{2}\left( f^{\prime }\left( X\right) \right) ^{2}}{96\left\vert f\left( 
\hat{X}\right) \right\vert ^{3}}\right) =-\frac{\sigma _{X}^{2}\left( p+%
\frac{1}{2}\right) \left( f^{\prime }\left( X\right) \right) ^{2}}{%
48\left\vert f\left( \hat{X}\right) \right\vert ^{3}}
\end{equation*}%
so that:%
\begin{equation}
\frac{\frac{d}{dp}\hat{\Gamma}\left( p+\frac{1}{2}\right) }{\hat{\Gamma}%
\left( p+\frac{1}{2}\right) }\simeq \ln \left( p+\frac{1}{2}\right) -\frac{%
\sigma _{X}^{2}\sigma _{\hat{K}}^{2}\left( p+\frac{1}{2}\right) \left(
f^{\prime }\left( X\right) \right) ^{2}}{48\left\vert f\left( \hat{X}\right)
\right\vert ^{3}}  \label{pdr}
\end{equation}%
Assuming that $C\left( \bar{p}\right) $ is constant, (\ref{pdr}) allows to
rewrite the variation of of equation (\ref{dkn}):%
\begin{eqnarray*}
\nabla _{\theta }\left( K_{\hat{X}}\left( D-L\left( \hat{X}\right) K_{\hat{X}%
}^{\eta }\right) \right) &=&K_{\hat{X}}\left( D-L\left( \hat{X}\right) K_{%
\hat{X}}^{\eta }\right) \\
&&\times \left( -\frac{\nabla _{\theta }\left\vert f\left( \hat{X}\right)
\right\vert }{\left\vert f\left( \hat{X}\right) \right\vert }+\left( \ln
\left( p+\frac{1}{2}\right) -\frac{\sigma _{X}^{2}\sigma _{\hat{K}%
}^{2}\left( p+\frac{1}{2}\right) \left( f^{\prime }\left( X\right) \right)
^{2}}{48\left\vert f\left( \hat{X}\right) \right\vert ^{3}}\right) \nabla
_{\theta }p\right) \\
&&+\frac{\sigma _{X}^{2}\sigma _{\hat{K}}^{2}\left( p+\frac{1}{2}\right)
^{2}\left( f^{\prime }\left( X\right) \right) ^{2}}{96\left\vert f\left( 
\hat{X}\right) \right\vert ^{3}}\left( \frac{\nabla _{\theta }\left\vert
f\left( \hat{X}\right) \right\vert }{\left\vert f\left( \hat{X}\right)
\right\vert }-\frac{\nabla _{\theta }\left( \frac{f^{\prime }\left( X\right) 
}{f\left( \hat{X}\right) }\right) ^{2}}{\left( \frac{f^{\prime }\left(
X\right) }{f\left( \hat{X}\right) }\right) ^{2}}\right)
\end{eqnarray*}%
and we deduce from this equation, that the dynamic version of equation (\ref%
{dkn}) is:%
\begin{eqnarray}
\frac{\nabla _{\theta }K_{\hat{X}}}{K_{\hat{X}}}-\frac{\nabla _{\theta
}\left( L\left( \hat{X}\right) K_{\hat{X}}^{\eta }\right) }{D-L\left( \hat{X}%
\right) K_{\hat{X}}^{\eta }} &=&\left( \frac{\sigma _{X}^{2}\sigma _{\hat{K}%
}^{2}\left( p+\frac{1}{2}\right) \left( f^{\prime }\left( X\right) \right)
^{2}}{48\left\vert f\left( \hat{X}\right) \right\vert ^{3}}-1\right) \frac{%
\nabla _{\theta }\left\vert f\left( \hat{X}\right) \right\vert }{\left\vert
f\left( \hat{X}\right) \right\vert }  \label{dqt} \\
&&+\left( \ln \left( p+\frac{1}{2}\right) -\frac{\sigma _{X}^{2}\sigma _{%
\hat{K}}^{2}\left( p+\frac{1}{2}\right) \left( f^{\prime }\left( X\right)
\right) ^{2}}{48\left\vert f\left( \hat{X}\right) \right\vert ^{3}}\right)
\nabla _{\theta }p  \notag \\
&&-\frac{\sigma _{X}^{2}\sigma _{\hat{K}}^{2}\left( p+\frac{1}{2}\right) ^{2}%
}{96\left\vert f\left( \hat{X}\right) \right\vert }\left( \nabla _{\theta
}\left( \frac{f^{\prime }\left( X\right) }{f\left( \hat{X}\right) }\right)
^{2}\right)  \notag
\end{eqnarray}

\subsubsection*{A5.1.2 Expanded form of (\protect\ref{dqt})}

To find the dynamic equation for $K_{\hat{X}}$ we expand each side of (\ref%
{dqt}).

The left hand side of (\ref{dqt}) can be developed as: 
\begin{eqnarray*}
&&\left( 1-\eta \frac{L\left( \hat{X}\right) K_{\hat{X}}^{\eta }}{D-L\left( 
\hat{X}\right) K_{\hat{X}}^{\eta }}\right) \frac{\nabla _{\theta }K_{\hat{X}}%
}{K_{\hat{X}}}-\frac{L\left( \hat{X}\right) K_{\hat{X}}^{\eta }}{D-L\left( 
\hat{X}\right) K_{\hat{X}}^{\eta }}\frac{\nabla _{\theta }L\left( \hat{X}%
\right) }{L\left( \hat{X}\right) } \\
&=&\left( 1-\eta \frac{L\left( \hat{X}\right) K_{\hat{X}}^{\eta }}{%
\left\Vert \Psi \left( \hat{X}\right) \right\Vert ^{2}}\right) \frac{\nabla
_{\theta }K_{\hat{X}}}{K_{\hat{X}}}-\frac{L\left( \hat{X}\right) K_{\hat{X}%
}^{\eta }}{\left\Vert \Psi \left( \hat{X}\right) \right\Vert ^{2}}\frac{%
\nabla _{\theta }L\left( \hat{X}\right) }{L\left( \hat{X}\right) } \\
&=&\left( 1-\eta \frac{L\left( \hat{X}\right) K_{\hat{X}}^{\eta }}{%
\left\Vert \Psi \left( \hat{X}\right) \right\Vert ^{2}}\right) \frac{\nabla
_{\theta }K_{\hat{X}}}{K_{\hat{X}}}-\frac{L\left( \hat{X}\right) K_{\hat{X}%
}^{\eta }}{\left\Vert \Psi \left( \hat{X}\right) \right\Vert ^{2}}\frac{%
\nabla _{\theta }\left( \left( \nabla _{X}R\left( X\right) \right)
^{2}+\sigma _{X}^{2}\frac{\nabla _{X}^{2}R\left( K_{X},X\right) }{H\left(
K_{X}\right) }\right) }{\left( \nabla _{X}R\left( X\right) \right)
^{2}+\sigma _{X}^{2}\frac{\nabla _{X}^{2}R\left( K_{X},X\right) }{H\left(
K_{X}\right) }} \\
&\simeq &\left( 1-\eta \frac{L\left( \hat{X}\right) K_{\hat{X}}^{\eta }}{%
\left\Vert \Psi \left( \hat{X}\right) \right\Vert ^{2}}\right) \frac{\nabla
_{\theta }K_{\hat{X}}}{K_{\hat{X}}}-2\frac{D-\left\Vert \Psi \left( \hat{X}%
\right) \right\Vert ^{2}}{\left\Vert \Psi \left( \hat{X}\right) \right\Vert
^{2}}\frac{\nabla _{\theta }\left( \nabla _{X}R\left( X\right) \right) }{%
\nabla _{X}R\left( X\right) }
\end{eqnarray*}%
and (\ref{dqt}) becomes:%
\begin{eqnarray}
&&\left( 1-\eta \frac{D-\left\Vert \Psi \left( \hat{X}\right) \right\Vert
^{2}}{\left\Vert \Psi \left( \hat{X}\right) \right\Vert ^{2}}\right) \frac{%
\nabla _{\theta }K_{\hat{X}}}{K_{\hat{X}}}-2\frac{D-\left\Vert \Psi \left( 
\hat{X}\right) \right\Vert ^{2}}{\left\Vert \Psi \left( \hat{X}\right)
\right\Vert ^{2}}\frac{\nabla _{X}\left( \nabla _{\theta }R\left( X\right)
\right) }{\nabla _{X}R\left( X\right) }  \label{tqd} \\
&=&\left( \frac{\sigma _{X}^{2}\sigma _{\hat{K}}^{2}\left( p+\frac{1}{2}%
\right) ^{2}\left( f^{\prime }\left( X\right) \right) ^{2}}{96\left\vert
f\left( \hat{X}\right) \right\vert ^{3}}-1\right) \frac{\nabla _{\theta
}\left\vert f\left( \hat{X}\right) \right\vert }{\left\vert f\left( \hat{X}%
\right) \right\vert }+\left( \ln \left( p+\frac{1}{2}\right) -\frac{\sigma
_{X}^{2}\sigma _{\hat{K}}^{2}\left( p+\frac{1}{2}\right) \left( f^{\prime
}\left( X\right) \right) ^{2}}{48\left\vert f\left( \hat{X}\right)
\right\vert ^{3}}\right) \nabla _{\theta }p  \notag \\
&&-\frac{\sigma _{X}^{2}\sigma _{\hat{K}}^{2}\left( p+\frac{1}{2}\right) ^{2}%
}{96\left\vert f\left( \hat{X}\right) \right\vert }\nabla _{\theta }\left( 
\frac{f^{\prime }\left( X\right) }{f\left( \hat{X}\right) }\right) ^{2} 
\notag
\end{eqnarray}%
To compute the right hand side of (\ref{tqd}). We use that:

\begin{equation*}
p=-\frac{M-\left( g\left( \hat{X}\right) \right) ^{2}+\sigma _{\hat{X}%
}^{2}\left( \nabla _{\hat{X}}g\left( \hat{X},K_{\hat{X}}\right) \right) }{%
\sigma _{\hat{X}}^{2}f\left( \hat{X}\right) }-\frac{3}{2}
\end{equation*}%
so that, the variation $\nabla _{\theta }p$ is given by: 
\begin{equation*}
\nabla _{\theta }p=-\frac{\nabla _{\theta }\left\vert f\left( \hat{X}\right)
\right\vert }{\left\vert f\left( \hat{X}\right) \right\vert }\left( p+\frac{3%
}{2}\right) -\left( 2\frac{g\left( \hat{X}\right) \nabla _{\theta }g\left( 
\hat{X}\right) }{\sigma _{\hat{X}}^{2}}+\nabla _{\theta }\nabla _{\hat{X}%
}g\left( \hat{X}\right) \right)
\end{equation*}%
To compute $\nabla _{\theta }p$ we must use the form of the functions
defined in Appendix 2. We thus obtain: 
\begin{eqnarray*}
\frac{\nabla _{\theta }g\left( \hat{X}\right) }{g\left( \hat{X}\right) } &=&%
\frac{\nabla _{\theta }\nabla _{\hat{X}}R\left( \hat{X}\right) }{\nabla _{%
\hat{X}}R\left( \hat{X}\right) }+\alpha \frac{\nabla _{\theta }K_{\hat{X}}}{%
K_{\hat{X}}} \\
\frac{\nabla _{\theta }\nabla _{\hat{X}}g\left( \hat{X}\right) }{\nabla _{%
\hat{X}}g\left( \hat{X}\right) } &=&\frac{\nabla _{\theta }\nabla _{\hat{X}%
}^{2}R\left( \hat{X}\right) }{\nabla _{\hat{X}}^{2}R\left( \hat{X}\right) }%
+\alpha \frac{\nabla _{\theta }K_{\hat{X}}}{K_{\hat{X}}}
\end{eqnarray*}%
and as a consequence:%
\begin{equation*}
\nabla _{\theta }p=-\frac{\nabla _{\theta }\left\vert f\left( \hat{X}\right)
\right\vert }{\left\vert f\left( \hat{X}\right) \right\vert }\left( p+\frac{3%
}{2}\right) -\left( 2\frac{g^{2}\left( \hat{X}\right) \left( \frac{\nabla
_{\theta }\nabla _{\hat{X}}R\left( \hat{X}\right) }{\nabla _{\hat{X}}R\left( 
\hat{X}\right) }+\alpha \frac{\nabla _{\theta }K_{\hat{X}}}{K_{\hat{X}}}%
\right) }{\sigma _{\hat{X}}^{2}\left\vert f\left( \hat{X}\right) \right\vert 
}+\frac{\nabla _{\hat{X}}g\left( \hat{X}\right) }{\left\vert f\left( \hat{X}%
\right) \right\vert }\left( \frac{\nabla _{\theta }\nabla _{\hat{X}%
}^{2}R\left( \hat{X}\right) }{\nabla _{\hat{X}}^{2}R\left( \hat{X}\right) }%
+\alpha \frac{\nabla _{\theta }K_{\hat{X}}}{K_{\hat{X}}}\right) \right)
\end{equation*}%
Ultimately, the right hand side of (\ref{tqd}) is given by:%
\begin{eqnarray*}
&&\left( \frac{\sigma _{X}^{2}\sigma _{\hat{K}}^{2}\left( p+\frac{1}{2}%
\right) \left( f^{\prime }\left( X\right) \right) ^{2}}{48\left\vert f\left( 
\hat{X}\right) \right\vert ^{3}}-1\right) \frac{\nabla _{\theta }\left\vert
f\left( \hat{X}\right) \right\vert }{\left\vert f\left( \hat{X}\right)
\right\vert }+\left( \ln \left( p+\frac{1}{2}\right) -\frac{\sigma
_{X}^{2}\sigma _{\hat{K}}^{2}\left( p+\frac{1}{2}\right) \left( f^{\prime
}\left( X\right) \right) ^{2}}{48\left\vert f\left( \hat{X}\right)
\right\vert ^{3}}\right) \nabla _{\theta }p \\
&&-\frac{\sigma _{X}^{2}\sigma _{\hat{K}}^{2}\left( p+\frac{1}{2}\right) ^{2}%
}{96\left\vert f\left( \hat{X}\right) \right\vert }\nabla _{\theta }\left( 
\frac{f^{\prime }\left( X\right) }{f\left( \hat{X}\right) }\right) ^{2} \\
&=&-\frac{\nabla _{\theta }\left\vert f\left( \hat{X}\right) \right\vert }{%
\left\vert f\left( \hat{X}\right) \right\vert }\left( \left( 1-\frac{\sigma
_{X}^{2}\sigma _{\hat{K}}^{2}\left( p+\frac{1}{2}\right) ^{2}\left(
f^{\prime }\left( X\right) \right) ^{2}}{96\left\vert f\left( \hat{X}\right)
\right\vert ^{3}}\right) +\left( p+\frac{3}{2}\right) \left( \ln \left( p+%
\frac{1}{2}\right) -\frac{\sigma _{X}^{2}\sigma _{\hat{K}}^{2}\left( p+\frac{%
1}{2}\right) \left( f^{\prime }\left( X\right) \right) ^{2}}{48\left\vert
f\left( \hat{X}\right) \right\vert ^{3}}\right) \right) \\
&&-\left( 2\frac{g^{2}\left( \hat{X}\right) }{\sigma _{\hat{X}%
}^{2}\left\vert f\left( \hat{X}\right) \right\vert }\frac{\nabla _{\theta
}\nabla _{\hat{X}}R\left( \hat{X}\right) }{\nabla _{\hat{X}}R\left( \hat{X}%
\right) }+\frac{\nabla _{\hat{X}}g\left( \hat{X}\right) }{\left\vert f\left( 
\hat{X}\right) \right\vert }\frac{\nabla _{\theta }\nabla _{\hat{X}%
}^{2}R\left( \hat{X}\right) }{\nabla _{\hat{X}}^{2}R\left( \hat{X}\right) }+%
\frac{\alpha \left( 2\frac{g^{2}\left( \hat{X}\right) }{\sigma _{\hat{X}}^{2}%
}+\nabla _{\hat{X}}g\left( \hat{X}\right) \right) }{\left\vert f\left( \hat{X%
}\right) \right\vert }\frac{\nabla _{\theta }K_{\hat{X}}}{K_{\hat{X}}}\right)
\\
&&\times \left( \ln \left( p+\frac{1}{2}\right) -\frac{\sigma _{X}^{2}\sigma
_{\hat{K}}^{2}\left( p+\frac{1}{2}\right) \left( f^{\prime }\left( X\right)
\right) ^{2}}{48\left\vert f\left( \hat{X}\right) \right\vert ^{3}}\right) -%
\frac{\sigma _{X}^{2}\sigma _{\hat{K}}^{2}\left( p+\frac{1}{2}\right) ^{2}}{%
96\left\vert f\left( \hat{X}\right) \right\vert }\nabla _{\theta }\left( 
\frac{f^{\prime }\left( X\right) }{f\left( \hat{X}\right) }\right) ^{2}
\end{eqnarray*}%
so that the variational equation for $K_{\hat{X}}$ (\ref{tqd}) writes:%
\begin{eqnarray}
&&\left( 1-\eta \frac{D-\left\Vert \Psi \left( \hat{X}\right) \right\Vert
^{2}}{\left\Vert \Psi \left( \hat{X}\right) \right\Vert ^{2}}\right) \frac{%
\nabla _{\theta }K_{\hat{X}}}{K_{\hat{X}}}-2\frac{D-\left\Vert \Psi \left( 
\hat{X}\right) \right\Vert ^{2}}{\left\Vert \Psi \left( \hat{X}\right)
\right\Vert ^{2}}\frac{\nabla _{X}\left( \nabla _{\theta }R\left( X\right)
\right) }{\nabla _{X}R\left( X\right) }  \label{vrq} \\
&=&-C_{3}\left( p,\hat{X}\right) \frac{\nabla _{\theta }\left\vert f\left( 
\hat{X}\right) \right\vert }{\left\vert f\left( \hat{X}\right) \right\vert }%
-C_{1}\left( p,\hat{X}\right) \frac{\nabla _{\theta }\left( \frac{f^{\prime
}\left( X\right) }{f\left( \hat{X}\right) }\right) ^{2}}{\left( \frac{%
f^{\prime }\left( X\right) }{f\left( \hat{X}\right) }\right) ^{2}}  \notag \\
&&-C_{2}\left( p,\hat{X}\right) \left( 2\frac{g^{2}\left( \hat{X}\right) }{%
\sigma _{\hat{X}}^{2}\left\vert f\left( \hat{X}\right) \right\vert }\frac{%
\nabla _{\theta }\nabla _{\hat{X}}R\left( \hat{X}\right) }{\nabla _{\hat{X}%
}R\left( \hat{X}\right) }+\frac{\nabla _{\hat{X}}g\left( \hat{X}\right) }{%
\left\vert f\left( \hat{X}\right) \right\vert }\frac{\nabla _{\theta }\nabla
_{\hat{X}}^{2}R\left( \hat{X}\right) }{\nabla _{\hat{X}}^{2}R\left( \hat{X}%
\right) }+\frac{\alpha \left( 2\frac{g^{2}\left( \hat{X}\right) }{\sigma _{%
\hat{X}}^{2}}+\nabla _{\hat{X}}g\left( \hat{X}\right) \right) }{\left\vert
f\left( \hat{X}\right) \right\vert }\frac{\nabla _{\theta }K_{\hat{X}}}{K_{%
\hat{X}}}\right)  \notag
\end{eqnarray}%
with:%
\begin{eqnarray}
C_{1}\left( p,\hat{X}\right) &=&\frac{\sigma _{X}^{2}\sigma _{\hat{K}%
}^{2}\left( p+\frac{1}{2}\right) ^{2}\left( f^{\prime }\left( X\right)
\right) ^{2}}{96\left\vert f\left( \hat{X}\right) \right\vert ^{3}}
\label{Cqt} \\
C_{2}\left( p,\hat{X}\right) &=&\ln \left( p+\frac{1}{2}\right) -\frac{%
2C_{1}\left( p,\hat{X}\right) }{p+\frac{1}{2}}  \notag \\
C_{3}\left( p,\hat{X}\right) &=&1-C_{1}\left( p,\hat{X}\right) +\left( p+%
\frac{3}{2}\right) C_{2}\left( p,\hat{X}\right)  \notag
\end{eqnarray}%
These term can be reordered and the general dynamic equation for $K_{\hat{X}%
} $ is ultimately written as:%
\begin{eqnarray}
&&\left( 1-\eta \frac{D-\left\Vert \Psi \left( \hat{X}\right) \right\Vert
^{2}}{\left\Vert \Psi \left( \hat{X}\right) \right\Vert ^{2}}+\frac{\alpha
\left( 2\frac{g^{2}\left( \hat{X}\right) }{\sigma _{\hat{X}}^{2}}+\nabla _{%
\hat{X}}g\left( \hat{X}\right) \right) }{\left\vert f\left( \hat{X}\right)
\right\vert }C_{2}\left( p,\hat{X}\right) \right) \frac{\nabla _{\theta }K_{%
\hat{X}}}{K_{\hat{X}}}  \label{gnf} \\
&&+2\left( \frac{g^{2}\left( \hat{X}\right) C_{2}\left( p,\hat{X}\right) }{%
\sigma _{\hat{X}}^{2}\left\vert f\left( \hat{X}\right) \right\vert }-\frac{%
D-\left\Vert \Psi \left( \hat{X}\right) \right\Vert ^{2}}{\left\Vert \Psi
\left( \hat{X}\right) \right\Vert ^{2}}\right) \frac{\nabla _{\theta }\nabla
_{\hat{X}}R\left( \hat{X}\right) }{\nabla _{\hat{X}}R\left( \hat{X}\right) }+%
\frac{\nabla _{\hat{X}}g\left( \hat{X}\right) C_{2}\left( p,\hat{X}\right) }{%
\left\vert f\left( \hat{X}\right) \right\vert }\frac{\nabla _{\theta }\nabla
_{\hat{X}}^{2}R\left( \hat{X}\right) }{\nabla _{\hat{X}}^{2}R\left( \hat{X}%
\right) }  \notag \\
&=&-C_{3}\left( p,\hat{X}\right) \frac{\nabla _{\theta }\left\vert f\left( 
\hat{X}\right) \right\vert }{\left\vert f\left( \hat{X}\right) \right\vert }%
-C_{1}\left( p,\hat{X}\right) \frac{\nabla _{\theta }\left( \frac{f^{\prime
}\left( X\right) }{f\left( \hat{X}\right) }\right) ^{2}}{\left( \frac{%
f^{\prime }\left( X\right) }{f\left( \hat{X}\right) }\right) ^{2}}  \notag
\end{eqnarray}

\subsubsection*{A5.1.3 Dynamic equation for particular forms of $f\left( 
\hat{X},K_{\hat{X}}\right) $ and $\left\Vert \Psi \left( \hat{X}\right)
\right\Vert ^{2}$\protect\bigskip}

We can put equation (\ref{gnf}) in a specific form, by using the explicit
formula for $f\left( \hat{X},K_{\hat{X}}\right) $ and $\left\Vert \Psi
\left( \hat{X}\right) \right\Vert ^{2}$\ given in appendix 2. We have:

\begin{eqnarray*}
\frac{\nabla _{\theta }f\left( \hat{X},K_{\hat{X}}\right) }{f\left( \hat{X}%
,K_{\hat{X}}\right) } &\simeq &\frac{r\left( K_{\hat{X}},\hat{X}\right)
\left( \frac{\nabla _{\theta }r\left( K_{\hat{X}},\hat{X}\right) }{r\left(
K_{\hat{X}},\hat{X}\right) }+\left( \alpha -1\right) \frac{\nabla _{\theta
}K_{\hat{X}}}{K_{\hat{X}}}\right) }{f\left( \hat{X}\right) } \\
&&+\frac{\gamma \left( \eta L\left( \hat{X}\right) K_{\hat{X}}^{\eta }\frac{%
\nabla _{\theta }K_{\hat{X}}}{K_{\hat{X}}}+2L\left( \hat{X}\right) K_{\hat{X}%
}^{\eta }\frac{\nabla _{\theta }\left( \nabla _{X}R\left( X\right) \right) }{%
\nabla _{X}R\left( X\right) }\right) +F_{1}^{\prime }\left( R\left( K_{\hat{X%
}},\hat{X}\right) \right) \frac{\nabla _{\theta }R\left( K_{\hat{X}},\hat{X}%
\right) }{R\left( K_{\hat{X}},\hat{X}\right) }}{f\left( \hat{X}\right) } \\
&\simeq &\frac{r\left( K_{\hat{X}},\hat{X}\right) \left( \frac{\nabla
_{\theta }r\left( \hat{X}\right) }{r\left( K_{\hat{X}},\hat{X}\right) }%
+\left( \alpha -1\right) \frac{\nabla _{\theta }K_{\hat{X}}}{K_{\hat{X}}}%
\right) }{f\left( \hat{X}\right) } \\
&&+\frac{\gamma \left( \eta L\left( \hat{X}\right) K_{\hat{X}}^{\eta }\frac{%
\nabla _{\theta }K_{\hat{X}}}{K_{\hat{X}}}+2L\left( \hat{X}\right) K_{\hat{X}%
}^{\eta }\frac{\nabla _{\theta }\left( \nabla _{X}R\left( X\right) \right) }{%
\nabla _{X}R\left( X\right) }\right) +\varsigma F_{1}\left( R\left( K_{\hat{X%
}},\hat{X}\right) \right) \frac{\nabla _{\theta }R\left( K_{\hat{X}},\hat{X}%
\right) }{R\left( K_{\hat{X}},\hat{X}\right) }}{f\left( \hat{X}\right) } \\
&=&\frac{r\left( K_{\hat{X}},\hat{X}\right) \frac{\nabla _{\theta }r\left( 
\hat{X}\right) }{r\left( K_{\hat{X}},\hat{X}\right) }}{f\left( \hat{X}%
\right) } \\
&&+\frac{\left( \gamma \eta L\left( \hat{X}\right) K_{\hat{X}}^{\eta
}+\left( \alpha -1\right) \right) \frac{\nabla _{\theta }K_{\hat{X}}}{K_{%
\hat{X}}}+\varsigma F_{1}\left( R\left( K_{\hat{X}},\hat{X}\right) \right) 
\frac{\nabla _{\theta }R\left( K_{\hat{X}},\hat{X}\right) }{R\left( K_{\hat{X%
}},\hat{X}\right) }+2\gamma L\left( \hat{X}\right) K_{\hat{X}}^{\eta }\frac{%
\nabla _{\theta }\left( \nabla _{X}R\left( X\right) \right) }{\nabla
_{X}R\left( X\right) }}{f\left( \hat{X}\right) }
\end{eqnarray*}%
To compute $\nabla _{\theta }\ln \left( \frac{f^{\prime }\left( X\right) }{%
f\left( \hat{X}\right) }\right) ^{2}$ arising in (\ref{gnf}), we use that in
first approximation, for relatively large $K_{\hat{X}}$:%
\begin{equation*}
\left( \frac{f^{\prime }\left( X\right) }{f\left( \hat{X}\right) }\right)
^{2}\simeq \left( \frac{F_{1}^{\prime }\left( R\left( K_{\hat{X}},\hat{X}%
\right) \right) \nabla _{\hat{X}}R\left( K_{\hat{X}},\hat{X}\right) }{%
F_{1}\left( R\left( K_{\hat{X}},\hat{X}\right) \right) }\right) ^{2}\simeq
\left( \varsigma \nabla _{\hat{X}}R\left( K_{\hat{X}},\hat{X}\right) \right)
^{2}
\end{equation*}%
that can be considered int the sequel negligible at the first order.

Consequently, for the chosen forms of the parameter functions, the dynamics
equation (\ref{gnf}) becomes ultimately:%
\begin{equation}
k\frac{\nabla _{\theta }K_{\hat{X}}}{K_{\hat{X}}}+l\frac{\nabla _{\theta
}R\left( \hat{X}\right) }{R\left( \hat{X}\right) }-2m\frac{\nabla _{\hat{X}%
}\nabla _{\theta }R\left( \hat{X}\right) }{\nabla _{\hat{X}}R\left( \hat{X}%
\right) }+n\frac{\nabla _{\hat{X}}^{2}\nabla _{\theta }R\left( \hat{X}%
\right) }{\nabla _{\hat{X}}^{2}R\left( \hat{X}\right) }=-C_{3}\left( p,\hat{X%
}\right) \frac{\nabla _{\theta }r\left( \hat{X}\right) }{f\left( \hat{X}%
\right) }  \label{cnd}
\end{equation}%
with:%
\begin{eqnarray}
k &=&1-\eta \left( 1-\frac{\gamma C_{3}\left( p,\hat{X}\right) }{\left\vert
f\left( \hat{X}\right) \right\vert }\right) \frac{D-\left\Vert \Psi \left( 
\hat{X}\right) \right\Vert ^{2}}{\left\Vert \Psi \left( \hat{X}\right)
\right\Vert ^{2}}  \label{Cfn} \\
&&+\frac{\alpha \left( 2\frac{g^{2}\left( \hat{X}\right) }{\sigma _{\hat{X}%
}^{2}}+\nabla _{\hat{X}}g\left( \hat{X}\right) \right) C_{2}\left( p,\hat{X}%
\right) -\left( 1-\alpha \right) C_{3}\left( p,\hat{X}\right) }{\left\vert
f\left( \hat{X}\right) \right\vert }  \notag \\
l &=&\frac{\varsigma F_{1}\left( R\left( K_{\hat{X}},\hat{X}\right) \right)
C_{3}\left( p,\hat{X}\right) }{f\left( \hat{X}\right) }  \notag \\
m &=&\left( 1-\frac{\gamma C_{3}\left( p,\hat{X}\right) }{f\left( \hat{X}%
\right) }\right) \frac{D-\left\Vert \Psi \left( \hat{X}\right) \right\Vert
^{2}}{\left\Vert \Psi \left( \hat{X}\right) \right\Vert ^{2}}-\frac{%
g^{2}\left( \hat{X}\right) C_{2}\left( p,\hat{X}\right) }{\sigma _{\hat{X}%
}^{2}}  \notag \\
n &=&\frac{\nabla _{\hat{X}}g\left( \hat{X}\right) C_{2}\left( p,\hat{X}%
\right) }{\left\vert f\left( \hat{X}\right) \right\vert }  \notag
\end{eqnarray}

\subsection*{A5.2 Full dynamical system}

To make the system self-consistent, we introduce also a dynamics for $R$.

We assume that $R$ depends on $K_{\hat{X}},\hat{X}$ and $\nabla _{\theta }K_{%
\hat{X}}$, that leads to write: $R\left( K_{\hat{X}},\hat{X},\nabla _{\theta
}K_{\hat{X}}\right) $. The variation is assumed to follow a diffusion
process:%
\begin{eqnarray*}
\nabla _{\theta }R\left( \theta ,\hat{X}\right) &=&\int_{\theta ^{\prime
}<\theta }G_{1}\left( \left( \theta ,\hat{X}\right) ,\left( \theta ^{\prime
},\hat{X}^{\prime }\right) \right) \nabla _{\theta ^{\prime }}R\left( \theta
^{\prime },\hat{X}^{\prime }\right) d\left( \theta ^{\prime },\hat{X}%
^{\prime }\right) \\
&&+\int_{\theta ^{\prime }<\theta }G_{2}\left( \left( \theta ,\hat{X}\right)
,\left( \theta ^{\prime },\hat{X}^{\prime }\right) \right) \nabla _{\theta
^{\prime }}K_{\hat{X}^{\prime }}d\left( \theta ^{\prime },\hat{X}^{\prime
}\right)
\end{eqnarray*}%
The first orders expansion of the right hand side leads to the following
form for $\nabla _{\theta }R\left( \theta ,\hat{X}\right) $:

\begin{eqnarray}
\nabla _{\theta }R\left( \theta ,\hat{X}\right) &=&\int \left( \hat{X}-\hat{X%
}^{\prime }\right) \left( G_{1}\left( \left( \theta ,\hat{X}\right) ,\left(
\theta ^{\prime },\hat{X}^{\prime }\right) \right) \nabla _{\hat{X}}\nabla
_{\theta }R\left( \theta ,\hat{X}\right) +G_{2}\left( \left( \theta ,\hat{X}%
\right) ,\left( \theta ^{\prime },\hat{X}^{\prime }\right) \right) \nabla _{%
\hat{X}}\nabla _{\theta }K_{\hat{X}}\right)  \label{dtr} \\
&&+\frac{1}{2}\int \left( \hat{X}-\hat{X}^{\prime }\right) ^{2}\left(
G_{1}\left( \left( \theta ,\hat{X}\right) ,\left( \theta ^{\prime },\hat{X}%
^{\prime }\right) \right) \nabla _{\hat{X}}^{2}\nabla _{\theta }R\left(
\theta ,\hat{X}\right) +G_{2}\left( \left( \theta ,\hat{X}\right) ,\left(
\theta ^{\prime },\hat{X}^{\prime }\right) \right) \nabla _{\hat{X}%
}^{2}\nabla _{\theta }K_{\hat{X}}\right)  \notag \\
&&+\int \left( \theta -\theta ^{\prime }\right) \left( G_{1}\left( \left(
\theta ,\hat{X}\right) ,\left( \theta ^{\prime },\hat{X}^{\prime }\right)
\right) \nabla _{\theta }\nabla _{\theta }R\left( \theta ,\hat{X}\right)
+G_{2}\left( \left( \theta ,\hat{X}\right) ,\left( \theta ^{\prime },\hat{X}%
^{\prime }\right) \right) \nabla _{\theta }\nabla _{\theta }K_{\hat{X}%
}\right)  \notag \\
&&+\frac{1}{2}\int \left( \theta -\theta ^{\prime }\right) ^{2}\left(
G_{1}\left( \left( \theta ,\hat{X}\right) ,\left( \theta ^{\prime },\hat{X}%
^{\prime }\right) \right) \nabla _{\theta }^{2}\nabla _{\theta }R\left(
\theta ,\hat{X}\right) +G_{2}\left( \left( \theta ,\hat{X}\right) ,\left(
\theta ^{\prime },\hat{X}^{\prime }\right) \right) \nabla _{\theta
}^{2}\nabla _{\theta }K_{\hat{X}}\right)  \notag \\
&&+...  \notag
\end{eqnarray}%
where the crossed derivatives have been discarded for the sake of
simplicity. We assume $G_{1}\left( \left( \theta ,\hat{X}\right) ,\left(
\theta ,\hat{X}\right) \right) =0$ to avoid auto-interaction.

Performing the integrals yields:

\begin{eqnarray}
\nabla _{\theta }R\left( \theta ,\hat{X}\right) &=&a_{0}\left( \hat{X}%
\right) \nabla _{\theta }K_{\hat{X}}+a\left( \hat{X}\right) \nabla _{\hat{X}%
}\nabla _{\theta }K_{\hat{X}}+b\left( \hat{X}\right) \nabla _{\hat{X}%
}^{2}\nabla _{\theta }K_{\hat{X}}  \label{dtR} \\
&&+c\left( \hat{X}\right) \nabla _{\theta }\left( \nabla _{\theta }K_{\hat{X}%
}\right) +d\left( \hat{X}\right) \nabla _{\theta }^{2}\left( \nabla _{\theta
}K_{\hat{X}}\right)  \notag \\
&&+e\left( \hat{X}\right) \nabla _{\hat{X}}\left( \nabla _{\theta }R\left(
\theta ,\hat{X}\right) \right) +f\left( \hat{X}\right) \nabla _{\hat{X}%
}^{2}\left( \nabla _{\theta }R\left( \theta ,\hat{X}\right) \right)  \notag
\\
&&+g\left( \hat{X}\right) \nabla _{\theta }\left( \nabla _{\theta }R\left(
\theta ,\hat{X}\right) \right) +h\left( \hat{X}\right) \nabla _{\theta
}^{2}\left( \nabla _{\theta }R\left( \theta ,\hat{X}\right) \right)  \notag
\\
&&+u\left( \hat{X}\right) \nabla _{\hat{X}}\nabla _{\theta }\left( \nabla
_{\theta }K_{\hat{X}}\right) +v\left( \hat{X}\right) \nabla _{\hat{X}}\nabla
_{\theta }\left( \nabla _{\theta }R\left( \theta ,\hat{X}\right) \right) 
\notag
\end{eqnarray}%
We ssume that the coefficients are slowly varying, since their are obtained
by averages.

Gathering the dynamics (\ref{cnd}) and (\ref{dtR})\ for $\nabla _{\theta }K_{%
\hat{X}}$ and $\nabla _{\theta }R\left( \theta ,\hat{X}\right) $ leads to a
matricial system: 
\begin{eqnarray}
&&0=\left( 
\begin{array}{cc}
\frac{k}{K_{\hat{X}}} & \frac{l}{R\left( \hat{X}\right) } \\ 
-a_{0}\left( \hat{X}\right) & 1%
\end{array}%
\right) \left( 
\begin{array}{c}
\nabla _{\theta }K_{\hat{X}} \\ 
\nabla _{\theta }R%
\end{array}%
\right)  \label{mtr} \\
&&-\left( 
\begin{array}{cc}
0 & \frac{2m}{\nabla _{\hat{X}}R\left( \hat{X}\right) }\nabla _{\hat{X}} \\ 
a\left( \hat{X}\right) \nabla _{\hat{X}}+c\left( \hat{X}\right) \nabla
_{\theta } & e\left( \hat{X}\right) \nabla _{\hat{X}}+g\left( \hat{X}\right)
\nabla _{\theta }%
\end{array}%
\right) \left( 
\begin{array}{c}
\nabla _{\theta }K_{\hat{X}} \\ 
\nabla _{\theta }R%
\end{array}%
\right)  \notag \\
&&-\left( 
\begin{array}{cc}
0 & -\frac{n}{\nabla _{\hat{X}}^{2}R\left( \hat{X}\right) }\nabla _{\hat{X}%
}^{2} \\ 
d\left( \hat{X}\right) \nabla _{\theta }^{2}+b\left( \hat{X}\right) \nabla _{%
\hat{X}}^{2}+u\left( \hat{X}\right) \nabla _{\hat{X}}\nabla _{\theta } & 
e\left( \hat{X}\right) \nabla _{\theta }^{2}+f\left( \hat{X}\right) \nabla _{%
\hat{X}}^{2}+v\nabla _{\hat{X}}\nabla _{\theta }%
\end{array}%
\right) \left( 
\begin{array}{c}
\nabla _{\theta }K_{\hat{X}} \\ 
\nabla _{\theta }R%
\end{array}%
\right)  \notag
\end{eqnarray}

\subsection*{A5.3 Oscillatory solutions\protect\bigskip}

We look for a solution of (\ref{drk}) of the form:%
\begin{equation*}
\left( 
\begin{array}{c}
\nabla _{\theta }K_{\hat{X}} \\ 
\nabla _{\theta }R\left( \hat{X}\right)%
\end{array}%
\right) =\exp \left( i\Omega \left( \hat{X}\right) \theta +iG\left( \hat{X}%
\right) \hat{X}\right) \left( 
\begin{array}{c}
\nabla _{\theta }K_{0} \\ 
\nabla _{\theta }R_{0}%
\end{array}%
\right)
\end{equation*}%
with $G\left( \hat{X}\right) $ and $\Omega \left( \hat{X}\right) $ slowly
varying. Consequently, the system (\ref{mtr}) writes:%
\begin{equation}
\left( 
\begin{array}{cc}
\frac{k}{K_{\hat{X}}} & \frac{l}{R\left( \hat{X}\right) }-i\frac{2m}{\nabla
_{\hat{X}}R\left( \hat{X}\right) }G-\frac{n}{\nabla _{\hat{X}}^{2}R\left( 
\hat{X}\right) }G^{2} \\ 
\begin{array}{c}
-a_{0}\left( \hat{X}\right) -ia\left( \hat{X}\right) G-ic\left( \hat{X}%
\right) \Omega \\ 
+d\Omega ^{2}+bG^{2}+u\Omega G%
\end{array}
& 
\begin{array}{c}
1-ie\left( \hat{X}\right) G-ig\left( \hat{X}\right) \Omega +e\Omega ^{2} \\ 
+fG^{2}+u\Omega G%
\end{array}%
\end{array}%
\right) \left( 
\begin{array}{c}
\nabla _{\theta }K_{\hat{X}} \\ 
\nabla _{\theta }R%
\end{array}%
\right) =0  \label{drk}
\end{equation}%
By canceling the determinant of the system, we are led to the following
relation between $\Omega \left( \hat{X}\right) $ and $G\left( \hat{X}\right) 
$: 
\begin{eqnarray*}
0 &=&\frac{k}{K_{\hat{X}}}\left( 1-ieG-ig\Omega \right) +\left( \frac{l}{%
R\left( \hat{X}\right) }-i\frac{2m}{\nabla _{\hat{X}}R\left( \hat{X}\right) }%
G\right) \left( a_{0}+iaG+ic\Omega \right) \\
&&-\frac{l}{R\left( \hat{X}\right) }\left( d\Omega ^{2}+bG^{2}+u\Omega
G\right) +\frac{k}{K_{\hat{X}}}\left( e\Omega ^{2}+fG^{2}+v\Omega G\right)
\end{eqnarray*}%
In the sequel, we restrict to the first order terms, which yields the
expression for $\Omega $:%
\begin{eqnarray*}
\Omega &=&\frac{i}{\left( \frac{lc}{R\left( \hat{X}\right) }-i\frac{2mc}{%
\nabla _{\hat{X}}R\left( \hat{X}\right) }G\right) -\frac{kg}{K_{\hat{X}}}}%
\left( \frac{k}{K_{\hat{X}}}\left( 1-ieG\right) +\left( \frac{l}{R\left( 
\hat{X}\right) }-i\frac{2m}{\nabla _{\hat{X}}R\left( \hat{X}\right) }%
G\right) \left( a_{0}+iaG\right) \right) \\
&=&\frac{\left( \frac{lc}{R\left( \hat{X}\right) }-\frac{kg}{K_{\hat{X}}}%
\right) +i\frac{2mc}{\nabla _{\hat{X}}R\left( \hat{X}\right) }G}{\left( 
\frac{lc}{R\left( \hat{X}\right) }-\frac{kg}{K_{\hat{X}}}\right) ^{2}+\left( 
\frac{2mc}{\nabla _{\hat{X}}R\left( \hat{X}\right) }G\right) ^{2}} \\
&&\times \left( \left( \frac{ke}{K_{\hat{X}}}+\left( \frac{2ma_{0}}{\nabla _{%
\hat{X}}R\left( \hat{X}\right) }-\frac{la}{R\left( \hat{X}\right) }\right)
\right) G+i\left( \frac{k}{K_{\hat{X}}}+\frac{a_{0}l}{R\left( \hat{X}\right) 
}+\frac{2ma}{\nabla _{\hat{X}}R\left( \hat{X}\right) }G^{2}\right) \right)
\end{eqnarray*}%
Or equivalently:%
\begin{eqnarray*}
&&\Omega =\frac{\left( \frac{lc}{R\left( \hat{X}\right) }-\frac{kg}{K_{\hat{X%
}}}\right) \left( \frac{ke}{K_{\hat{X}}}+\left( \frac{2ma_{0}}{\nabla _{\hat{%
X}}R\left( \hat{X}\right) }-\frac{la}{R\left( \hat{X}\right) }\right)
\right) G-\frac{2mc}{\nabla _{\hat{X}}R\left( \hat{X}\right) }G\left( \frac{k%
}{K_{\hat{X}}}+\frac{a_{0}l}{R\left( \hat{X}\right) }+\frac{2ma}{\nabla _{%
\hat{X}}R\left( \hat{X}\right) }G^{2}\right) }{\left( \frac{lc}{R\left( \hat{%
X}\right) }-\frac{kg}{K_{\hat{X}}}\right) ^{2}+\left( \frac{2mc}{\nabla _{%
\hat{X}}R\left( \hat{X}\right) }G\right) ^{2}} \\
&&+i\frac{\left( \frac{lc}{R\left( \hat{X}\right) }-\frac{kg}{K_{\hat{X}}}%
\right) \left( \frac{k}{K_{\hat{X}}}+\frac{a_{0}l}{R\left( \hat{X}\right) }+%
\frac{2ma}{\nabla _{\hat{X}}R\left( \hat{X}\right) }G^{2}\right) +\frac{2mc}{%
\nabla _{\hat{X}}R\left( \hat{X}\right) }\left( \frac{ke}{K_{\hat{X}}}%
+\left( \frac{2ma_{0}}{\nabla _{\hat{X}}R\left( \hat{X}\right) }-\frac{la}{%
R\left( \hat{X}\right) }\right) \right) G^{2}}{\left( \frac{lc}{R\left( \hat{%
X}\right) }-\frac{kg}{K_{\hat{X}}}\right) ^{2}+\left( \frac{2mc}{\nabla _{%
\hat{X}}R\left( \hat{X}\right) }G\right) ^{2}}
\end{eqnarray*}%
We focus on the influence of time variations of $\nabla _{\theta }K_{\hat{X}%
} $ \ on $\nabla _{\theta }R$, and we can assume $g\simeq 0$ so that there
is no self influence of $\nabla _{\theta }R$ on itself: $\nabla _{\theta }R$
depends on the variations of $\nabla _{\theta }K_{\hat{X}}$ as well as the
neighboorhood sectors variations of $\nabla _{\theta }R$. Moreover, the
coefficients $e$ and $a$, being obtained by integration or first order
expansion, can be considered as nul.

Consequently, the equation for $\Omega $ reduces to:%
\begin{equation*}
\Omega =\frac{\frac{lc}{R\left( \hat{X}\right) }\left( \frac{2ma_{0}}{\nabla
_{\hat{X}}R\left( \hat{X}\right) }\right) G-\frac{2mc}{\nabla _{\hat{X}%
}R\left( \hat{X}\right) }G\left( \frac{k}{K_{\hat{X}}}+\frac{a_{0}l}{R\left( 
\hat{X}\right) }\right) }{\left( \frac{lc}{R\left( \hat{X}\right) }\right)
^{2}+\left( \frac{2mc}{\nabla _{\hat{X}}R\left( \hat{X}\right) }G\right) ^{2}%
}+i\frac{\frac{lc}{R\left( \hat{X}\right) }\left( \frac{k}{K_{\hat{X}}}+%
\frac{a_{0}l}{R\left( \hat{X}\right) }\right) +\frac{2mc}{\nabla _{\hat{X}%
}R\left( \hat{X}\right) }\left( \frac{2ma_{0}}{\nabla _{\hat{X}}R\left( \hat{%
X}\right) }\right) G^{2}}{\left( \frac{lc}{R\left( \hat{X}\right) }\right)
^{2}+\left( \frac{2mc}{\nabla _{\hat{X}}R\left( \hat{X}\right) }G\right) ^{2}%
}
\end{equation*}

\subsection*{A5.4 Stability}

The system is stable and the dynamics is dampening if:%
\begin{equation}
\frac{lc}{R\left( \hat{X}\right) }\left( \frac{k}{K_{\hat{X}}}+\frac{a_{0}l}{%
R\left( \hat{X}\right) }\right) +\frac{4m^{2}ca_{0}}{\left( \nabla _{\hat{X}%
}R\left( \hat{X}\right) \right) ^{2}}G^{2}>0  \label{stB}
\end{equation}%
To study the sign of (\ref{stB}) we need to estimate the coefficient $k$.

\subsubsection*{A5.4.1 Estimation of the coefficients $k$, $l$ and $m$}

We can estimate $k$ and $l$ by computing the factors $C_{i}\left( p,\hat{X}%
\right) $, for $i=1,2,3$.

This is done by estimating $p+\frac{1}{2}$. We start with the asymptotic
form of $\hat{\Gamma}\left( p+\frac{1}{2}\right) $:

\begin{equation*}
\hat{\Gamma}\left( p+\frac{1}{2}\right) \simeq \sqrt{p+\frac{1}{2}}\exp
\left( -\frac{\sigma _{X}^{2}\left( p+\frac{1}{2}\right) ^{2}\left(
f^{\prime }\left( X\right) \right) ^{2}}{96\left\vert f\left( \hat{X}\right)
\right\vert ^{3}}\right)
\end{equation*}%
and rewriting the equation for $K_{\hat{X}}$ as: 
\begin{equation}
K_{\hat{X}}\left\Vert \Psi \left( \hat{X}\right) \right\Vert ^{2}\left\vert
f\left( \hat{X}\right) \right\vert \left( \frac{\sigma _{X}^{2}\left(
f^{\prime }\left( X\right) \right) ^{2}}{96\left\vert f\left( \hat{X}\right)
\right\vert ^{3}}\right) ^{\frac{1}{4}}=C\left( \bar{p}\right) \sigma _{\hat{%
K}}^{2}\exp \left( -\frac{\sigma _{X}^{2}\left( p+\frac{1}{2}\right)
^{2}\left( f^{\prime }\left( X\right) \right) ^{2}}{96\left\vert f\left( 
\hat{X}\right) \right\vert ^{3}}\right) \sqrt{\left( p+\frac{1}{2}\right) 
\sqrt{\frac{\sigma _{X}^{2}\left( f^{\prime }\left( X\right) \right) ^{2}}{%
96\left\vert f\left( \hat{X}\right) \right\vert ^{3}}}}  \label{vrK}
\end{equation}%
Then, using (\ref{Cqt}), we set:%
\begin{equation}
\left( p+\frac{1}{2}\right) \sqrt{\frac{\sigma _{X}^{2}\left( f^{\prime
}\left( X\right) \right) ^{2}}{96\left\vert f\left( \hat{X}\right)
\right\vert ^{3}}}=\sqrt{C_{1}\left( p,\hat{X}\right) }
\end{equation}%
Equation (\ref{vrK}) writes:%
\begin{equation}
\frac{K_{\hat{X}}\left\Vert \Psi \left( \hat{X}\right) \right\Vert
^{2}\left\vert f\left( \hat{X}\right) \right\vert \left( \frac{\sigma
_{X}^{2}\left( f^{\prime }\left( X\right) \right) ^{2}}{96\left\vert f\left( 
\hat{X}\right) \right\vert ^{3}}\right) ^{\frac{1}{4}}}{C\left( \bar{p}%
\right) \sigma _{\hat{K}}^{2}}=\exp \left( -C_{1}\left( p,\hat{X}\right)
\right) \left( C_{1}\left( p,\hat{X}\right) \right) ^{\frac{1}{4}}
\label{Cpq}
\end{equation}%
$\allowbreak $and the solution to (\ref{Cpq}) is:%
\begin{eqnarray}
C_{1}\left( p,\hat{X}\right) &=&\frac{\sigma _{X}^{2}\left( p+\frac{1}{2}%
\right) ^{2}\left( f^{\prime }\left( X\right) \right) ^{2}}{96\left\vert
f\left( \hat{X}\right) \right\vert ^{3}}  \label{Cpo} \\
&=&C_{0}\left( \hat{X},K_{\hat{X}}\right) \exp \left( -W\left(
k,-4C_{0}\left( \hat{X},K_{\hat{X}}\right) \right) \right)  \notag
\end{eqnarray}%
with:%
\begin{equation*}
C_{0}\left( \hat{X},K_{\hat{X}}\right) =\left( \frac{K_{\hat{X}}\left\Vert
\Psi \left( \hat{X}\right) \right\Vert ^{2}\left\vert f\left( \hat{X}\right)
\right\vert }{C\left( \bar{p}\right) \sigma _{\hat{K}}^{2}}\right) ^{4}\frac{%
\sigma _{X}^{2}\left( f^{\prime }\left( X\right) \right) ^{2}}{96\left\vert
f\left( \hat{X}\right) \right\vert ^{3}}
\end{equation*}%
and where $W\left( k,x\right) $ is the Lambert $W$ function. The parameter $%
k=0$ for the stable case with low $K_{\hat{X}}$ and $k=-1$ for the unstable
case with $K_{\hat{X}}$ large.

We can deduce $p+\frac{1}{2}$ from (\ref{Cpo}): 
\begin{equation}
p+\frac{1}{2}=\frac{\sqrt{C_{1}\left( p,\hat{X}\right) }}{\sqrt{\frac{\sigma
_{X}^{2}\left( f^{\prime }\left( X\right) \right) ^{2}}{96\left\vert f\left( 
\hat{X}\right) \right\vert ^{3}}}}  \label{Pon}
\end{equation}%
$\allowbreak $

and $2\frac{C_{1}\left( p,\hat{X}\right) }{p+\frac{1}{2}}$:%
\begin{equation*}
2\frac{C_{1}\left( p,\hat{X}\right) }{p+\frac{1}{2}}=\sqrt{C_{1}\left( p,%
\hat{X}\right) }\frac{\sigma _{X}^{2}\left( f^{\prime }\left( X\right)
\right) ^{2}}{48\left\vert f\left( \hat{X}\right) \right\vert ^{3}}
\end{equation*}%
From (\ref{Pon}) and (\ref{Cpd}) we deduce:%
\begin{eqnarray}
C_{2}\left( p,\hat{X}\right) &=&\ln \left( p+\frac{1}{2}\right) -\frac{%
2C_{1}\left( p,\hat{X}\right) }{p+\frac{1}{2}}  \label{CpD} \\
&=&\frac{1}{2}\ln \frac{C_{1}\left( p,\hat{X}\right) }{\frac{\sigma
_{X}^{2}\left( f^{\prime }\left( X\right) \right) ^{2}}{96\left\vert f\left( 
\hat{X}\right) \right\vert ^{3}}}-\sqrt{C_{1}\left( p,\hat{X}\right) }\frac{%
\sigma _{X}^{2}\left( f^{\prime }\left( X\right) \right) ^{2}}{48\left\vert
f\left( \hat{X}\right) \right\vert ^{3}}\simeq \frac{1}{2}\ln \frac{%
96\left\vert f\left( \hat{X}\right) \right\vert ^{3}}{\sigma _{X}^{2}\left(
f^{\prime }\left( X\right) \right) ^{2}}  \notag
\end{eqnarray}%
We can also compute:%
\begin{eqnarray*}
\left( p+\frac{3}{2}\right) \ln \left( p+\frac{1}{2}\right) &\simeq &\frac{%
48\left\vert f\left( \hat{X}\right) \right\vert ^{3}C_{1}\left( p,\hat{X}%
\right) }{\sigma _{X}^{2}\left( f^{\prime }\left( X\right) \right) ^{2}}\ln 
\frac{96\left\vert f\left( \hat{X}\right) \right\vert ^{3}}{\sigma
_{X}^{2}\left( f^{\prime }\left( X\right) \right) ^{2}} \\
&=&\frac{1}{2}\left( \frac{K_{\hat{X}}\left\Vert \Psi \left( \hat{X}\right)
\right\Vert ^{2}\left\vert f\left( \hat{X}\right) \right\vert }{C\left( \bar{%
p}\right) \sigma _{\hat{K}}^{2}}\right) ^{4}\ln \frac{96\left\vert f\left( 
\hat{X}\right) \right\vert ^{3}}{\sigma _{X}^{2}\left( f^{\prime }\left(
X\right) \right) ^{2}} \\
&&\times \exp \left( -W\left( k,-4C_{0}\left( \hat{X},K_{\hat{X}}\right)
\right) \right)
\end{eqnarray*}%
so that:%
\begin{eqnarray}
C_{3}\left( p,\hat{X}\right) &=&1-C_{1}\left( p,\hat{X}\right) +\left( p+%
\frac{3}{2}\right) C_{2}\left( p,\hat{X}\right)  \label{Cpt} \\
&=&1-C_{1}\left( p,\hat{X}\right) +\left( p+\frac{3}{2}\right) \left( \ln
\left( p+\frac{1}{2}\right) -\frac{2C_{1}\left( p,\hat{X}\right) }{p+\frac{1%
}{2}}\right)  \notag \\
&\simeq &1+\left( \frac{48\left\vert f\left( \hat{X}\right) \right\vert ^{3}%
}{\sigma _{X}^{2}\left( f^{\prime }\left( X\right) \right) ^{2}}\ln \frac{%
96\left\vert f\left( \hat{X}\right) \right\vert ^{3}}{\sigma _{X}^{2}\left(
f^{\prime }\left( X\right) \right) ^{2}}-1\right) C_{1}\left( p,\hat{X}%
\right)  \notag \\
&\simeq &1+\frac{1}{2}\left( \frac{K_{\hat{X}}\left\Vert \Psi \left( \hat{X}%
\right) \right\Vert ^{2}\left\vert f\left( \hat{X}\right) \right\vert }{%
C\left( \bar{p}\right) \sigma _{\hat{K}}^{2}}\right) ^{4}\exp \left(
-W\left( k,-4C_{0}\left( \hat{X},K_{\hat{X}}\right) \right) \right) \ln 
\frac{96\left\vert f\left( \hat{X}\right) \right\vert ^{3}}{\sigma
_{X}^{2}\left( f^{\prime }\left( X\right) \right) ^{2}}  \notag
\end{eqnarray}%
Given that our assumptions $\sigma _{X}^{2}<1$ and in most cases $\frac{%
96\left\vert f\left( \hat{X}\right) \right\vert ^{3}}{\sigma _{X}^{2}\left(
f^{\prime }\left( X\right) \right) ^{2}}>>1$, then $\frac{96\left\vert
f\left( \hat{X}\right) \right\vert ^{3}}{\sigma _{X}^{2}\left( f^{\prime
}\left( X\right) \right) ^{2}}>>1$ and $C_{3}\left( p,\hat{X}\right) >>1$.

These computations allow to estimate $k$ and $l$. We start with $k$. Given
that (see (\ref{Cfn})): 
\begin{eqnarray*}
k &=&1-\eta \left( 1-\frac{\gamma C_{3}\left( p,\hat{X}\right) }{\left\vert
f\left( \hat{X}\right) \right\vert }\right) \frac{D-\left\Vert \Psi \left( 
\hat{X}\right) \right\Vert ^{2}}{\left\Vert \Psi \left( \hat{X}\right)
\right\Vert ^{2}} \\
&&+\frac{\alpha \left( 2\frac{g^{2}\left( \hat{X}\right) }{\sigma _{\hat{X}%
}^{2}}+\nabla _{\hat{X}}g\left( \hat{X}\right) \right) C_{2}\left( p,\hat{X}%
\right) -\left( 1-\alpha \right) C_{3}\left( p,\hat{X}\right) }{\left\vert
f\left( \hat{X}\right) \right\vert } \\
l &=&\frac{\varsigma F_{1}\left( R\left( K_{\hat{X}},\hat{X}\right) \right)
C_{3}\left( p,\hat{X}\right) }{f\left( \hat{X}\right) } \\
m &=&\left( 1-\frac{\gamma C_{3}\left( p,\hat{X}\right) }{f\left( \hat{X}%
\right) }\right) \frac{D-\left\Vert \Psi \left( \hat{X}\right) \right\Vert
^{2}}{\left\Vert \Psi \left( \hat{X}\right) \right\Vert ^{2}}
\end{eqnarray*}%
the sign of $k$ and $l$\ depend on the magnitude of $K_{\hat{X}}$.

\paragraph*{A5.4.1.1 $K_{\hat{X}}>>1$}

For $K_{\hat{X}}>>1$, using (\ref{mgk}) and:%
\begin{equation*}
\left\Vert \Psi \left( \hat{X}\right) \right\Vert ^{2}=D-\left( \nabla
_{X}R\left( \hat{X}\right) \right) ^{2}K_{\hat{X}}^{\alpha }
\end{equation*}%
we have:%
\begin{equation*}
K_{\hat{X}}^{\alpha }\simeq \frac{D}{\left( \nabla _{\hat{X}}R\left( \hat{X}%
\right) \right) ^{2}}-\frac{C\left( \bar{p}\right) \sigma _{\hat{K}}^{2}%
\sqrt{\frac{M-c}{c}}}{\left( \nabla _{\hat{X}}R\left( \hat{X}\right) \right)
^{2\left( 1-\frac{1}{\alpha }\right) }D^{\frac{1}{\alpha }}c}
\end{equation*}%
and:%
\begin{equation*}
\frac{D-\left\Vert \Psi \left( \hat{X}\right) \right\Vert ^{2}}{\left\Vert
\Psi \left( \hat{X}\right) \right\Vert ^{2}}\simeq \frac{D^{1+\frac{1}{%
\alpha }}c}{C\left( \bar{p}\right) \sigma _{\hat{K}}^{2}\sqrt{\frac{M-c}{c}}%
\left( \nabla _{\hat{X}}R\left( \hat{X}\right) \right) ^{\frac{2}{\alpha }}}
\end{equation*}%
The constant $c$ has been defined in appendix 3, and satisfies $c<<1$. As a
consequence:%
\begin{eqnarray*}
k &\simeq &\eta \frac{\gamma C_{3}\left( p,\hat{X}\right) }{\left\vert
f\left( \hat{X}\right) \right\vert }\frac{D-\left\Vert \Psi \left( \hat{X}%
\right) \right\Vert ^{2}}{\left\Vert \Psi \left( \hat{X}\right) \right\Vert
^{2}}-\left( 1-\alpha \right) \frac{C_{3}\left( p,\hat{X}\right) }{%
\left\vert f\left( \hat{X}\right) \right\vert } \\
&\simeq &\left( \frac{\eta \gamma D^{1+\frac{1}{\alpha }}c}{C\left( \bar{p}%
\right) \sigma _{\hat{K}}^{2}\sqrt{\frac{M-c}{c}}\left( \nabla _{\hat{X}%
}R\left( \hat{X}\right) \right) ^{\frac{2}{\alpha }}}-\left( 1-\alpha
\right) \right) \frac{C_{3}\left( p,\hat{X}\right) }{\left\vert f\left( \hat{%
X}\right) \right\vert }
\end{eqnarray*}%
This may be negative or positive depending on the relative magnitude of $%
\frac{\eta \gamma D^{1+\frac{1}{\alpha }}c}{C\left( \bar{p}\right) \sigma _{%
\hat{K}}^{2}\sqrt{\frac{M-c}{c}}\left( \nabla _{\hat{X}}R\left( \hat{X}%
\right) \right) ^{\frac{2}{\alpha }}}$ and $\left( 1-\alpha \right) $. The
first case correspond to the stable equilibrium with large $K_{\hat{X}}$ and
the second case to the stable case with large $K_{\hat{X}}$ studied in
appendix 2.

\subparagraph{Unstable case}

This case corresponds to:%
\begin{equation*}
\frac{D^{1+\frac{1}{\alpha }}c}{C\left( \bar{p}\right) \sigma _{\hat{K}}^{2}%
\sqrt{\frac{M-c}{c}}\left( \nabla _{\hat{X}}R\left( \hat{X}\right) \right) ^{%
\frac{2}{\alpha }}}>>1
\end{equation*}%
Moreover, using (\ref{Cpt}) and the following estimation, we have:%
\begin{equation}
k\simeq \eta \frac{\gamma C_{3}\left( p,\hat{X}\right) }{\left\vert f\left( 
\hat{X}\right) \right\vert }\frac{\eta \gamma D^{1+\frac{1}{\alpha }}c}{%
C\left( \bar{p}\right) \sigma _{\hat{K}}^{2}\sqrt{\frac{M-c}{c}}\left(
\nabla _{\hat{X}}R\left( \hat{X}\right) \right) ^{\frac{2}{\alpha }}}>>1
\label{kfK}
\end{equation}%
We can also estimate $\left\vert \frac{k}{K_{\hat{X}}}\right\vert $. In this
case:%
\begin{equation}
\frac{k}{K_{\hat{X}}}\simeq \eta \frac{\gamma C_{3}\left( p,\hat{X}\right) }{%
\left\vert f\left( \hat{X}\right) \right\vert }\frac{\eta \gamma D^{\frac{1}{%
\alpha }}c}{C\left( \bar{p}\right) \sigma _{\hat{K}}^{2}\sqrt{\frac{M-c}{c}}%
\left( \nabla _{\hat{X}}R\left( \hat{X}\right) \right) ^{\frac{2}{\alpha }}}%
>>1  \label{stK}
\end{equation}%
We can estimate $l$ by the same token: 
\begin{equation*}
l=\frac{\varsigma F_{1}\left( R\left( K_{\hat{X}},\hat{X}\right) \right)
C_{3}\left( p,\hat{X}\right) }{f\left( \hat{X}\right) }>>1
\end{equation*}%
and using (\ref{stK}) we have:%
\begin{equation*}
\left\vert \frac{k}{K_{\hat{X}}}\right\vert >>l
\end{equation*}%
The coefficient $m$ is obtained by using that in this case:%
\begin{equation*}
m\simeq \left( 1-\frac{\gamma C_{3}\left( p,\hat{X}\right) }{f\left( \hat{X}%
\right) }\right) \frac{D-\left\Vert \Psi \left( \hat{X}\right) \right\Vert
^{2}}{\left\Vert \Psi \left( \hat{X}\right) \right\Vert ^{2}}\simeq -\frac{1%
}{\eta }k
\end{equation*}

\subparagraph{Stable case}

For the stable case we have:%
\begin{equation*}
\frac{\eta \gamma D^{1+\frac{1}{\alpha }}c}{C\left( \bar{p}\right) \sigma _{%
\hat{K}}^{2}\sqrt{\frac{M-c}{c}}\left( \nabla _{\hat{X}}R\left( \hat{X}%
\right) \right) ^{\frac{2}{\alpha }}}-\left( 1-\alpha \right) <0
\end{equation*}%
and we write:%
\begin{equation*}
k\simeq -\left( 1-\alpha \right) \frac{C_{3}\left( p,\hat{X}\right) }{%
\left\vert f\left( \hat{X}\right) \right\vert }<0
\end{equation*}%
We have:%
\begin{equation*}
\left\vert k\right\vert >>1
\end{equation*}%
and moreover:%
\begin{equation}
\left\vert \frac{k}{K_{\hat{X}}}\right\vert \simeq \left( 1-\alpha \right) 
\frac{C_{3}\left( p,\hat{X}\right) }{K_{\hat{X}}\left\vert f\left( \hat{X}%
\right) \right\vert }=\frac{1-\alpha }{\varsigma F_{1}\left( R\left( K_{\hat{%
X}},\hat{X}\right) \right) K_{\hat{X}}}l<<l  \label{nst}
\end{equation}%
The coefficient $m$ is obtained by using that in the stable case:%
\begin{equation*}
m\simeq -\frac{\gamma }{\varsigma F_{1}\left( R\left( K_{\hat{X}},\hat{X}%
\right) \right) }l
\end{equation*}

\paragraph*{A5.4.1.2 $K_{\hat{X}}<<1$}

On the other hand, for $K_{\hat{X}}\leqslant 1$, we have:%
\begin{equation}
\frac{D-\left\Vert \Psi \left( \hat{X}\right) \right\Vert ^{2}}{\left\Vert
\Psi \left( \hat{X}\right) \right\Vert ^{2}}<<1  \label{ngb}
\end{equation}%
so that:%
\begin{equation*}
k\simeq 1+\frac{\alpha \left( 2\frac{g^{2}\left( \hat{X}\right) }{\sigma _{%
\hat{X}}^{2}}+\nabla _{\hat{X}}g\left( \hat{X}\right) \right) C_{2}\left( p,%
\hat{X}\right) -\left( 1-\alpha \right) C_{3}\left( p,\hat{X}\right) }{%
\left\vert f\left( \hat{X}\right) \right\vert }
\end{equation*}%
Given (\ref{CpD}) and (\ref{Cpt}), this yields:%
\begin{equation}
k\simeq -\frac{\left( 1-\alpha \right) C_{3}\left( p,\hat{X}\right) }{%
\left\vert f\left( \hat{X}\right) \right\vert }<0  \label{sgK}
\end{equation}%
and, as in the previous case:%
\begin{eqnarray*}
\left\vert k\right\vert &>&>1 \\
l &>&>1
\end{eqnarray*}%
Moreover, given that $K_{\hat{X}}<<1$:%
\begin{equation}
\left\vert \frac{k}{K_{\hat{X}}}\right\vert >>1  \label{vlk}
\end{equation}%
and:%
\begin{equation}
\left\vert \frac{k}{K_{\hat{X}}}\right\vert >>l
\end{equation}%
Moreover, given (\ref{ngb}):%
\begin{equation*}
\left\vert m\right\vert =\left\vert 1-\frac{\gamma C_{3}\left( p,\hat{X}%
\right) }{f\left( \hat{X}\right) }\right\vert \frac{D-\left\Vert \Psi \left( 
\hat{X}\right) \right\Vert ^{2}}{\left\Vert \Psi \left( \hat{X}\right)
\right\Vert ^{2}}<<\left\vert \frac{\gamma C_{3}\left( p,\hat{X}\right) }{%
f\left( \hat{X}\right) }\right\vert
\end{equation*}%
and:%
\begin{equation*}
\left\vert m\right\vert <<l
\end{equation*}

\paragraph*{A5.4.1.3 Intermediate case\protect\bigskip}

In this case, we can consider that $\frac{D-\left\Vert \Psi \left( \hat{X}%
\right) \right\Vert ^{2}}{\left\Vert \Psi \left( \hat{X}\right) \right\Vert
^{2}}$ is of order $1$:%
\begin{equation}
\frac{D-\left\Vert \Psi \left( \hat{X}\right) \right\Vert ^{2}}{\left\Vert
\Psi \left( \hat{X}\right) \right\Vert ^{2}}=O\left( 1\right)  \label{ngl}
\end{equation}
Assuming that $\gamma <<1$ we have:%
\begin{eqnarray*}
k &\simeq &1+\frac{\alpha \left( 2\frac{g^{2}\left( \hat{X}\right) }{\sigma
_{\hat{X}}^{2}}+\nabla _{\hat{X}}g\left( \hat{X}\right) \right) C_{2}\left(
p,\hat{X}\right) -\left( 1-\alpha \right) C_{3}\left( p,\hat{X}\right) }{%
\left\vert f\left( \hat{X}\right) \right\vert } \\
&\simeq &1+\frac{\frac{\alpha }{2}\left( 2\frac{g^{2}\left( \hat{X}\right) }{%
\sigma _{\hat{X}}^{2}}+\nabla _{\hat{X}}g\left( \hat{X}\right) \right) -%
\frac{1-\alpha }{2}\left( \frac{K_{\hat{X}}\left\Vert \Psi \left( \hat{X}%
\right) \right\Vert ^{2}\left\vert f\left( \hat{X}\right) \right\vert }{%
C\left( \bar{p}\right) \sigma _{\hat{K}}^{2}}\right) ^{4}\exp \left(
-W\left( k,-4C_{0}\left( \hat{X},K_{\hat{X}}\right) \right) \right) }{%
\left\vert f\left( \hat{X}\right) \right\vert }\ln \frac{96\left\vert
f\left( \hat{X}\right) \right\vert ^{3}}{\sigma _{X}^{2}\left( f^{\prime
}\left( X\right) \right) ^{2}}
\end{eqnarray*}%
Given that the intermediate case is stable (see appendix 2), the relation
between $K_{\hat{X}}$ and $R\left( \hat{X}\right) $ is positive, we can
assume that $k<0$ and:%
\begin{eqnarray*}
k &\simeq &1+\frac{\alpha \left( 2\frac{g^{2}\left( \hat{X}\right) }{\sigma
_{\hat{X}}^{2}}+\nabla _{\hat{X}}g\left( \hat{X}\right) \right) C_{2}\left(
p,\hat{X}\right) -\left( 1-\alpha \right) C_{3}\left( p,\hat{X}\right) }{%
\left\vert f\left( \hat{X}\right) \right\vert } \\
&\simeq &-\frac{\frac{1-\alpha }{2}\left( \frac{K_{\hat{X}}\left\Vert \Psi
\left( \hat{X}\right) \right\Vert ^{2}\left\vert f\left( \hat{X}\right)
\right\vert }{C\left( \bar{p}\right) \sigma _{\hat{K}}^{2}}\right) ^{4}\exp
\left( -W\left( 0,-4C_{0}\left( \hat{X},K_{\hat{X}}\right) \right) \right) }{%
\left\vert f\left( \hat{X}\right) \right\vert }\ln \frac{96\left\vert
f\left( \hat{X}\right) \right\vert ^{3}}{\sigma _{X}^{2}\left( f^{\prime
}\left( X\right) \right) ^{2}}
\end{eqnarray*}%
and:%
\begin{eqnarray*}
l &=&\frac{\varsigma F_{1}\left( R\left( K_{\hat{X}},\hat{X}\right) \right)
C_{3}\left( p,\hat{X}\right) }{f\left( \hat{X}\right) }\simeq l \\
&=&\frac{\varsigma F_{1}\left( R\left( K_{\hat{X}},\hat{X}\right) \right)
\left( \frac{K_{\hat{X}}\left\Vert \Psi \left( \hat{X}\right) \right\Vert
^{2}\left\vert f\left( \hat{X}\right) \right\vert }{C\left( \bar{p}\right)
\sigma _{\hat{K}}^{2}}\right) ^{4}\exp \left( -W\left( 0,-4C_{0}\left( \hat{X%
},K_{\hat{X}}\right) \right) \right) }{f\left( \hat{X}\right) }\ln \frac{%
96\left\vert f\left( \hat{X}\right) \right\vert ^{3}}{\sigma _{X}^{2}\left(
f^{\prime }\left( X\right) \right) ^{2}}
\end{eqnarray*}%
Note that in this case:%
\begin{equation*}
k\simeq -\frac{1-\alpha }{\varsigma F_{1}\left( R\left( K_{\hat{X}},\hat{X}%
\right) \right) }l
\end{equation*}%
and, given (\ref{ngl}):%
\begin{equation*}
m\simeq -\gamma \frac{D-\left\Vert \Psi \left( \hat{X}\right) \right\Vert
^{2}}{\left\Vert \Psi \left( \hat{X}\right) \right\Vert ^{2}\varsigma
F_{1}\left( R\left( K_{\hat{X}},\hat{X}\right) \right) }l
\end{equation*}

\subsubsection*{A5.4.2 Stability conditions}

This appendix presents the computations leading to the stability conditions
for the three ranges of capital considered. Apart from the intermediate
case, interpretations are detailed in the text.

\paragraph*{A5.4.2.1 Case $K_{\hat{X}}>>1$}

\subparagraph{Stable case}

As shown above, $k<0$, $\left\vert k\right\vert >>1$, $l>>1$ and $\left\vert 
\frac{k}{K_{\hat{X}}}\right\vert <<l$. Coefficients $l$ and $m$ are of the
same order. Thus (\ref{stB}) becomes:%
\begin{equation*}
\frac{l^{2}a_{0}c}{\left( R\left( \hat{X}\right) \right) ^{2}}+\frac{%
4m^{2}ca_{0}}{\left( \nabla _{\hat{X}}R\left( \hat{X}\right) \right) ^{2}}%
G^{2}>0
\end{equation*}%
That is, for $c>0$ the oscillations are stable, whereas for $c<0$ they are
unstable.

\subparagraph{Unstable case}

In this case, $k>0$, $\left\vert k\right\vert >>1$, $l>>1$ and $\left\vert 
\frac{k}{K_{\hat{X}}}\right\vert >>l$. We have also $m\simeq -\frac{1}{\eta }%
k$ and (\ref{stB}) writes: 
\begin{equation}
\frac{cl}{R\left( \hat{X}\right) }\frac{k}{K_{\hat{X}}}+\frac{4k^{2}ca_{0}}{%
\eta ^{2}\left( \nabla _{\hat{X}}R\left( \hat{X}\right) \right) ^{2}}G^{2}>0
\end{equation}%
That is, for $c>0$ the oscillations are stable, whereas for $c<0$ they are
unstable.

\paragraph*{A5.4.2.1.2 Case $K_{\hat{X}}<<1$}

Equations (\ref{sgK}) and (\ref{vlk}) show that $k<0$, $\left\vert
k\right\vert >>1$, $l>>1$, $\left\vert m\right\vert <<l$ and $\left\vert 
\frac{k}{K_{\hat{X}}}\right\vert >>l$. Equation (\ref{stB}) thus writes: 
\begin{equation}
\frac{cl}{R\left( \hat{X}\right) }\frac{k}{K_{\hat{X}}}>0
\end{equation}%
That is, for $c>0$ the oscillations are unstable, whereas for $c<0$ they are
stable.

\paragraph*{A5.4.2.3 Intermediate case}

In this case, we have seen above that $k<0$:%
\begin{equation*}
k\simeq -\frac{1-\alpha }{\varsigma F_{1}\left( R\left( K_{\hat{X}},\hat{X}%
\right) \right) }l
\end{equation*}%
and:%
\begin{equation*}
m\simeq -\gamma \frac{D-\left\Vert \Psi \left( \hat{X}\right) \right\Vert
^{2}}{\left\Vert \Psi \left( \hat{X}\right) \right\Vert ^{2}\varsigma
F_{1}\left( R\left( K_{\hat{X}},\hat{X}\right) \right) }
\end{equation*}%
Consequently, equation (\ref{stB}) particularizes as:

\begin{equation*}
\frac{l^{2}c}{R\left( \hat{X}\right) }\left( \frac{a_{0}}{R\left( \hat{X}%
\right) }-\frac{1-\alpha }{\varsigma K_{\hat{X}}F_{1}\left( R\left( K_{\hat{X%
}},\hat{X}\right) \right) }\right) +4ca_{0}\left( \frac{\gamma \left(
D-\left\Vert \Psi \left( \hat{X}\right) \right\Vert ^{2}\right) }{\varsigma
\nabla _{\hat{X}}R\left( \hat{X}\right) F_{1}\left( R\left( K_{\hat{X}},\hat{%
X}\right) \right) \left\Vert \Psi \left( \hat{X}\right) \right\Vert ^{2}}%
\right) ^{2}G^{2}>0
\end{equation*}%
Given the definition of $a_{0}$ and the stability of the intermediate case,
we assume $a_{0}>0$. Thus, 2 possibilities arise.

\subparagraph{Coefficient $c>0$}

In this case, the oscillations are stable if: 
\begin{equation*}
\frac{a_{0}}{R\left( \hat{X}\right) }-\frac{1-\alpha }{\varsigma K_{\hat{X}%
}F_{1}\left( R\left( K_{\hat{X}},\hat{X}\right) \right) }>0
\end{equation*}%
or if:%
\begin{equation*}
\frac{a_{0}}{R\left( \hat{X}\right) }-\frac{1-\alpha }{\varsigma K_{\hat{X}%
}F_{1}\left( R\left( K_{\hat{X}},\hat{X}\right) \right) }<0
\end{equation*}%
and:%
\begin{equation*}
G^{2}>\frac{l^{2}\left( \nabla _{\hat{X}}R\left( \hat{X}\right) \right) ^{2}%
}{4a_{0}R\left( \hat{X}\right) }\left( \frac{\varsigma \left( \nabla _{\hat{X%
}}R\left( \hat{X}\right) F_{1}\left( R\left( K_{\hat{X}},\hat{X}\right)
\right) \left\Vert \Psi \left( \hat{X}\right) \right\Vert ^{2}\right) }{%
\gamma \left( D-\left\Vert \Psi \left( \hat{X}\right) \right\Vert
^{2}\right) }\right) ^{2}\left\vert \frac{a_{0}}{R\left( \hat{X}\right) }-%
\frac{1-\alpha }{\varsigma F_{1}\left( R\left( K_{\hat{X}},\hat{X}\right)
\right) }\right\vert
\end{equation*}%
Otherwise, the oscillations are unstable.

The constant $\varsigma $\ is irrelevant here, although it arises in
appendix 3 to estimate short-term returns. The function $F_{1}$,\ defined in
(\ref{pr}), determines the stock's prices evolution. The coefficient $\alpha 
$ is the Cobb-Douglas power arising in the dividend part of short-term
returns. The constant $D$, defined in (\ref{psl}), determines the relation
between number of firms and average capital at sector $\hat{X}$.

We recover the large average capital case. A relatively high reactivity of
expectations to fluctuations in capital allows to maintain the capital at
its equilibrium value. This stability is favoured for sectors with large
average capital when $G$\ is relatively large, i.e. when this sectors
present large discrepancies in capital with their neighbours.

\subparagraph{Coefficient $c<0$}

The oscillations are stable if:%
\begin{equation}
\frac{a_{0}}{R\left( \hat{X}\right) }-\frac{1-\alpha }{\varsigma K_{\hat{X}%
}F_{1}\left( R\left( K_{\hat{X}},\hat{X}\right) \right) }<0  \label{stc}
\end{equation}%
and:%
\begin{equation}
G^{2}<\frac{l^{2}\left( \nabla _{\hat{X}}R\left( \hat{X}\right) \right) ^{2}%
}{4a_{0}R\left( \hat{X}\right) }\left( \frac{\varsigma \left( \nabla _{\hat{X%
}}R\left( \hat{X}\right) F_{1}\left( R\left( K_{\hat{X}},\hat{X}\right)
\right) \left\Vert \Psi \left( \hat{X}\right) \right\Vert ^{2}\right) }{%
\gamma \left( D-\left\Vert \Psi \left( \hat{X}\right) \right\Vert
^{2}\right) }\right) ^{2}\left\vert \frac{a_{0}}{R\left( \hat{X}\right) }-%
\frac{1-\alpha }{\varsigma F_{1}\left( R\left( K_{\hat{X}},\hat{X}\right)
\right) }\right\vert  \label{sTC}
\end{equation}

Conditions (\ref{stc}) and (\ref{sTC}) correspond to the case of relatively
low capital for which a stability in the oscillations may be reached when
expectations are moderately reactive to variation in capital. The condition (%
\ref{sTC}) shows that the stability in oscillations is reached for moderate
values of $G$, i.e. relatively small discrepancy between neighbouring
sectors.

We recover the large average capital case. A relatively high reactivity of
expectations to fluctuations in capital allows to maintain the capital at
its equilibrium value. This stability is favoured for sectors with large
average capital when $G$\ is relatively large, i.e. when this sectors
present large discrepancies in capital with their neighbours.

\section*{Appendix 6 Computation of effective action at the second order}

We compute the second-order derivatives for the real and the financial
economy respectively.

\subsection*{A6.1 Real economy}

In first approximation:

\begin{eqnarray}
&&\frac{\delta ^{2}\left( S_{1}+S_{2}\right) }{\delta \Psi ^{\dag }\left(
Z,\theta \right) \delta \Psi \left( Z,\theta \right) }  \label{SCD} \\
&\simeq &-\int \delta \Psi ^{\dag }\left( K,X\right) \left( \nabla
_{X}\left( \frac{\sigma _{X}^{2}}{2}\nabla _{X}-\nabla _{X}R\left(
K,X\right) H\left( K\right) \right) -4\tau \left( \left\vert \Psi _{0}\left(
X\right) \right\vert ^{2}\right) \right.  \notag \\
&&+\left. \nabla _{K}\left( \frac{\sigma _{K}^{2}}{2}\nabla _{K}+u\left(
K,X,\Psi _{0},\hat{\Psi}_{0}\right) \right) \right) \delta \Psi \left(
K,X\right) dKdX  \notag
\end{eqnarray}%
where:%
\begin{equation*}
\left\vert \Psi _{0}\left( X\right) \right\vert ^{2}=\int \left\vert \Psi
_{0}\left( K^{\prime },X\right) \right\vert ^{2}dK^{\prime }
\end{equation*}%
and:%
\begin{equation}
u\left( K,X,\Psi _{0},\hat{\Psi}_{0}\right) \rightarrow \frac{1}{\varepsilon 
}\left( K-\int \hat{F}_{2}\left( s,R\left( K,X\right) \right) \hat{K}%
\left\Vert \hat{\Psi}_{0}\left( \hat{K},X\right) \right\Vert ^{2}d\hat{K}%
\right) =\frac{1}{\varepsilon }\left( K-\hat{F}_{2}\left( s,R\left(
K,X\right) \right) K_{X}d\hat{K}\right)  \label{FR}
\end{equation}%
In equation (\ref{FR}), we used the notation:%
\begin{equation*}
\int \hat{F}_{2}\left( s,R\left( K,X\right) \right) \hat{K}\left\Vert \hat{%
\Psi}_{0}\left( \hat{K},X\right) \right\Vert ^{2}d\hat{K}=\hat{F}_{2}\left(
s,R\left( K,X\right) \right) \hat{K}_{X}
\end{equation*}%
We perform a change of variables in (\ref{SCD}):%
\begin{eqnarray}
\Delta \Psi \left( K,X\right) &=&\exp \left( \int^{X}\frac{\nabla
_{X}R\left( X\right) }{\sigma _{X}^{2}}H\left( \frac{\int \hat{K}\left\Vert 
\hat{\Psi}\left( \hat{K},X\right) \right\Vert ^{2}d\hat{K}}{\left\Vert \Psi
\left( X\right) \right\Vert ^{2}}\right) \right)  \label{chn} \\
&&\times \exp \left( \int \left( K-\frac{F_{2}\left( R\left( K,X\right)
\right) K_{X}}{F_{2}\left( R\left( K_{X},X\right) \right) }\right) dK\right)
\delta \Psi \left( K,X\right)  \notag \\
\Delta \Psi ^{\dag }\left( K,X\right) &=&\exp \left( -\int^{X}\frac{\nabla
_{X}R\left( X\right) }{\sigma _{X}^{2}}H\left( \frac{\int \hat{K}\left\Vert 
\hat{\Psi}\left( \hat{K},X\right) \right\Vert ^{2}d\hat{K}}{\left\Vert \Psi
\left( X\right) \right\Vert ^{2}}\right) \right)  \notag \\
&&\times \exp \left( -\int \left( K-\frac{F_{2}\left( R\left( K,X\right)
\right) K_{X}}{F_{2}\left( R\left( K_{X},X\right) \right) }\right) dK\right)
\delta \Psi ^{\dag }\left( K,X\right)  \notag
\end{eqnarray}%
where $K_{X}$, the average invested capital per firm in sector $X$:%
\begin{equation}
K_{X}=\frac{\int \hat{K}\left\Vert \hat{\Psi}\left( \hat{K},X\right)
\right\Vert ^{2}d\hat{K}}{\left\Vert \Psi \left( X\right) \right\Vert ^{2}}
\end{equation}%
so that the effective action (\ref{SCD}) for the real economy\textbf{\ }%
becomes:

\begin{eqnarray}
&&\Delta \Psi ^{\dag }\left( Z,\theta \right) \left( \frac{\delta ^{2}\left(
S_{1}+S_{2}\right) }{\delta \Psi ^{\dag }\left( Z,\theta \right) \delta \Psi
\left( Z,\theta \right) }\right) _{\Psi \left( Z,\theta \right) =\Psi
_{0}\left( Z,\theta \right) }\Delta \Psi \left( Z,\theta \right)  \label{SCN}
\\
&=&\int \Delta \Psi ^{\dag }\left( Z,\theta \right) \left( -\frac{\sigma
_{X}^{2}}{2}\nabla _{X}^{2}+\frac{\left( \nabla _{X}R\left( K,X\right)
H\left( K_{X}\right) \right) ^{2}}{2\sigma _{X}^{2}}+\frac{\nabla
_{X}^{2}R\left( K,X\right) }{2}H\left( K\right) +4\tau \left\vert \Psi
\left( X\right) \right\vert ^{2}\right) \Delta \Psi \left( Z,\theta \right) 
\notag \\
&&+\int \Delta \Psi ^{\dag }\left( Z,\theta \right) \left( -\frac{\sigma
_{K}^{2}}{2}\nabla _{K}^{2}+\frac{1}{2\sigma _{K}^{2}}\left( K-\hat{F}%
_{2}\left( s,R\left( K,X\right) \right) K_{X}\right) ^{2}+\frac{1-\nabla _{K}%
\hat{F}_{2}\left( s,R\left( K,X\right) \right) K_{X}}{2}\right) \Delta \Psi
\left( Z,\theta \right)  \notag
\end{eqnarray}%
As explained in section 10.1.2, the effects of competition can be refined by%
\textbf{\ }considering repulsive forces that are capital dependent\textbf{. }%
It amounts to replace in (\ref{SCN}), the term:%
\begin{equation*}
\int \Delta \Psi ^{\dag }\left( Z,\theta \right) \left( 2\tau \left\vert
\Psi \left( X\right) \right\vert ^{2}\right) \Delta \Psi \left( Z,\theta
\right)
\end{equation*}%
by the term:%
\begin{eqnarray}
&&\int \Delta \Psi ^{\dag }\left( K,X\right) \left( 2\tau \frac{\int
K^{\prime }\left\vert \Psi \left( K^{\prime },X\right) \right\vert
^{2}dK^{\prime }}{K}\right) \Delta \Psi \left( K,\theta \right)  \label{NT}
\\
&=&\int \Delta \Psi ^{\dag }\left( K,X\right) \left( 2\tau \frac{\left\vert
\Psi \left( X\right) \right\vert ^{2}K_{X}}{K}\right) \Delta \Psi \left(
K,\theta \right)  \notag
\end{eqnarray}%
with:%
\begin{eqnarray*}
\left\vert \Psi \left( X\right) \right\vert ^{2} &=&\int \left\vert \Psi
\left( K^{\prime },X\right) \right\vert ^{2}dK^{\prime } \\
K_{X} &=&\frac{\int K^{\prime }\left\vert \Psi \left( K^{\prime },X\right)
\right\vert ^{2}dK^{\prime }}{\left\vert \Psi \left( X\right) \right\vert
^{2}}
\end{eqnarray*}%
This models repulsive forces that are inversely proportional to capital and
mainly affect low-capital firms. Note that this change in the interaction
does not modify the collective state, since by setting $K=K_{X}$, we recover
the previous repulsive term. Ultimately, using:%
\begin{equation}
\left\Vert \Psi \left( X\right) \right\Vert ^{2}=\left( 2\tau \right)
^{-1}\left( D\left( \left\Vert \Psi \right\Vert ^{2}\right) -\frac{1}{%
2\sigma _{X}^{2}}\left( \left( \nabla _{X}R\left( X\right) \right) ^{2}+%
\frac{\sigma _{X}^{2}\nabla _{X}^{2}R\left( K_{X},X\right) }{H\left(
K_{X}\right) }\right) H^{2}\left( K_{X}\right) \left( 1-\frac{H^{\prime
}\left( \hat{K}_{X}\right) K_{X}}{H\left( \hat{K}_{X}\right) }\right) \right)
\label{PFR}
\end{equation}%
the interaction term (\ref{NT}) becomes:%
\begin{eqnarray*}
&&\int \Delta \Psi ^{\dag }\left( K,X\right) \frac{1}{2}\left( \left( \nabla
_{X}R\left( X\right) \right) ^{2}+\frac{\sigma _{X}^{2}\nabla
_{X}^{2}R\left( K_{X},X\right) }{H\left( K_{X}\right) }\right) \\
&&\times H^{2}\left( K_{X}\right) \left( 1-\frac{H^{\prime }\left( \hat{K}%
_{X}\right) K_{X}}{H\left( \hat{K}_{X}\right) }\right) +2\tau \frac{%
\left\vert \Psi \left( X\right) \right\vert ^{2}K_{X}}{K}\Delta \Psi \left(
K,\theta \right) \\
&=&\int \Delta \Psi ^{\dag }\left( K,X\right) \left( D\left( \left\Vert \Psi
\right\Vert ^{2}\right) +2\tau \frac{\left\vert \Psi \left( X\right)
\right\vert ^{2}\left( K_{X}-K\right) }{K}\right) \Delta \Psi \left(
K,\theta \right)
\end{eqnarray*}%
When the above expression is used to rewrite (\ref{SCN}), it yields the
formula:%
\begin{eqnarray}
\frac{\delta ^{2}\left( S_{1}+S_{2}\right) }{\delta \Psi ^{\dag }\left(
Z,\theta \right) \delta \Psi \left( Z,\theta \right) } &=&-\frac{\sigma
_{X}^{2}}{2}\nabla _{X}^{2}-\frac{\sigma _{K}^{2}}{2}\nabla _{K}^{2}+\left(
D\left( \left\Vert \Psi \right\Vert ^{2}\right) +2\tau \frac{\left\vert \Psi
\left( X\right) \right\vert ^{2}\left( K_{X}-K\right) }{K}\right)
\label{SCR} \\
&&+\frac{1}{2\sigma _{K}^{2}}\left( K-\hat{F}_{2}\left( s,R\left( K,X\right)
\right) K_{X}\right) ^{2}+\frac{1-\nabla _{K}\hat{F}_{2}\left( s,R\left(
K,X\right) \right) K_{X}}{2}  \notag
\end{eqnarray}%
as stated in the text.

\subsection*{A6.2 Financial economy}

For the financial sector, we consider the field-action for $\hat{\Psi}^{\dag
}\left( \hat{K},\hat{X}\right) $: 
\begin{equation}
S_{3}+S_{4}=-\int \hat{\Psi}^{\dag }\left( \hat{K},\hat{X}\right) \left(
\nabla _{\hat{K}}\left( \frac{\sigma _{\hat{K}}^{2}}{2}\nabla _{\hat{K}}-%
\hat{K}f\left( \hat{X},K_{\hat{X}}\right) \right) +\nabla _{\hat{X}}\left( 
\frac{\sigma _{\hat{X}}^{2}}{2}\nabla _{\hat{X}}-g\left( \hat{X},K_{\hat{X}%
}\right) \right) \right) \hat{\Psi}\left( \hat{K},\hat{X}\right)  \label{STF}
\end{equation}%
with:

\begin{eqnarray}
f\left( \hat{X},K_{\hat{X}}\right) &=&\frac{1}{\varepsilon }\left( r\left(
K_{\hat{X}},\hat{X}\right) -\gamma \left\Vert \Psi \left( \hat{X}\right)
\right\Vert ^{2}+F_{1}\left( \frac{R\left( K_{\hat{X}},\hat{X}\right) }{\int
R\left( K_{X^{\prime }}^{\prime },X^{\prime }\right) \left\Vert \Psi \left(
X^{\prime }\right) \right\Vert ^{2}dX^{\prime }}\right) \right) \\
g\left( \hat{X},K_{\hat{X}}\right) &=&\left( \frac{\nabla _{\hat{X}%
}F_{0}\left( R\left( K_{\hat{X}},\hat{X}\right) \right) }{\left\Vert \nabla
_{\hat{X}}R\left( K_{\hat{X}},\hat{X}\right) \right\Vert }+\nu \nabla _{\hat{%
X}}F_{1}\left( \frac{R\left( K_{\hat{X}},\hat{X}\right) }{\int R\left(
K_{X^{\prime }}^{\prime },X^{\prime }\right) \left\Vert \Psi \left(
X^{\prime }\right) \right\Vert ^{2}dX^{\prime }}\right) \right)
\end{eqnarray}%
Using a change of variable (see appendix 3.1):%
\begin{eqnarray}
\hat{\Psi} &\rightarrow &\exp \left( \frac{1}{\sigma _{\hat{X}}^{2}}\int
g\left( \hat{X}\right) d\hat{X}+\frac{\hat{K}^{2}}{\sigma _{\hat{K}}^{2}}%
f\left( \hat{X}\right) \right) \hat{\Psi}  \label{chg} \\
\hat{\Psi}^{\dag } &\rightarrow &\exp \left( \frac{1}{\sigma _{\hat{X}}^{2}}%
\int g\left( \hat{X}\right) d\hat{X}+\frac{\hat{K}^{2}}{\sigma _{\hat{K}}^{2}%
}f\left( \hat{X}\right) \right) \hat{\Psi}^{\dag }  \notag
\end{eqnarray}%
the action (\ref{STF}) becomes:%
\begin{eqnarray}
&&S_{3}+S_{4}=-\int \hat{\Psi}^{\dag }\left( \frac{\sigma _{\hat{X}}^{2}}{2}%
\nabla _{\hat{X}}^{2}-\frac{1}{2\sigma _{\hat{X}}^{2}}\left( g\left( \hat{X}%
,K_{\hat{X}}\right) \right) ^{2}-\frac{1}{2}\nabla _{\hat{X}}g\left( \hat{X}%
,K_{\hat{X}}\right) \right) \hat{\Psi}  \label{stm} \\
&&-\int \hat{\Psi}^{\dag }\left( \nabla _{\hat{K}}\left( \frac{\sigma _{\hat{%
K}}^{2}}{2}\nabla _{\hat{K}}-\hat{K}f\left( \hat{X},K_{\hat{X}}\right)
\right) \right) \hat{\Psi}  \notag
\end{eqnarray}

To obtain the second-order expansion of the field's action, we start by the
first derivative of (\ref{stm}) arising in the minimization equation in
(Gosselin Lotz Wambst 2022):%
\begin{eqnarray}
\frac{\delta \left( S_{3}\left( \Psi \right) +S_{4}\left( \Psi \right)
\right) }{\delta \hat{\Psi}^{\dag }\left( Z,\theta \right) } &=&-\frac{%
\sigma _{\hat{X}}^{2}}{2}\nabla _{\hat{X}}^{2}\hat{\Psi}-\frac{\sigma _{\hat{%
K}}^{2}}{2}\nabla _{\hat{K}}^{2}\hat{\Psi}+\frac{1}{2\sigma _{\hat{X}}^{2}}%
\left( g\left( \hat{X},K_{\hat{X}}\right) \right) ^{2}+\frac{1}{2}\nabla _{%
\hat{X}}g\left( \hat{X},K_{\hat{X}}\right) \hat{\Psi}  \label{SCP} \\
&&+\frac{\hat{K}^{2}}{2\sigma _{\hat{K}}^{2}}f^{2}\left( \hat{X}\right) +%
\frac{1}{2}f\left( \hat{X},K_{\hat{X}}\right) \hat{\Psi}+F\left( \hat{X},K_{%
\hat{X}}\right) \hat{K}\hat{\Psi}  \notag
\end{eqnarray}%
with:%
\begin{eqnarray}
F\left( \hat{X},K_{\hat{X}}\right) &=&\nabla _{K_{\hat{X}}}\left( \frac{%
\left( g\left( \hat{X},K_{\hat{X}}\right) \right) ^{2}}{2\sigma _{\hat{X}%
}^{2}}+\frac{1}{2}\nabla _{\hat{X}}g\left( \hat{X},K_{\hat{X}}\right)
+f\left( \hat{X},K_{\hat{X}}\right) \right) \frac{\left\Vert \hat{\Psi}%
\left( \hat{X}\right) \right\Vert ^{2}}{\left\Vert \Psi \left( \hat{X}%
\right) \right\Vert ^{2}} \\
&&+\frac{\nabla _{K_{\hat{X}}}f^{2}\left( \hat{X},K_{\hat{X}}\right) }{%
\sigma _{\hat{K}}^{2}\left\Vert \Psi \left( \hat{X}\right) \right\Vert ^{2}}%
\left\langle \hat{K}^{2}\right\rangle _{\hat{X}}  \notag
\end{eqnarray}%
so that :%
\begin{eqnarray*}
\frac{\delta ^{2}\left( S_{3}\left( \Psi \right) +S_{4}\left( \Psi \right)
\right) }{\delta \hat{\Psi}^{\dag }\left( Z,\theta \right) \delta \hat{\Psi}%
\left( Z,\theta \right) } &=&-\frac{\sigma _{\hat{X}}^{2}}{2}\nabla _{\hat{X}%
}^{2}-\frac{\sigma _{\hat{K}}^{2}}{2}\nabla _{\hat{K}}^{2}+\frac{1}{2\sigma
_{\hat{X}}^{2}}\left( g\left( \hat{X},K_{\hat{X}}\right) \right) ^{2}+\frac{1%
}{2}\nabla _{\hat{X}}g\left( \hat{X},K_{\hat{X}}\right) \\
&&+\frac{\hat{K}^{2}}{2\sigma _{\hat{K}}^{2}}f^{2}\left( \hat{X}\right) +%
\frac{1}{2}f\left( \hat{X},K_{\hat{X}}\right) +F\left( \hat{X},K_{\hat{X}%
}\right) \hat{K}-\hat{\Psi}^{\dag }\frac{\delta F\left( \hat{X},K_{\hat{X}%
}\right) }{\delta \left\Vert \hat{\Psi}\left( \hat{K},\hat{X}\right)
\right\Vert ^{2}}\hat{K}\hat{\Psi}
\end{eqnarray*}%
where the last term is given by:%
\begin{eqnarray*}
\frac{\delta F\left( \hat{X},K_{\hat{X}}\right) }{\delta \hat{\Psi}\left(
Z,\theta \right) } &\simeq &\nabla _{K_{\hat{X}}}\left( \frac{\left( g\left( 
\hat{X},K_{\hat{X}}\right) \right) ^{2}}{2\sigma _{\hat{X}}^{2}}+\frac{1}{2}%
\nabla _{\hat{X}}g\left( \hat{X},K_{\hat{X}}\right) +f\left( \hat{X},K_{\hat{%
X}}\right) \right) \frac{\hat{\Psi}^{\dagger }\left( \hat{K},\hat{X}\right) 
}{\left\Vert \Psi \left( \hat{X}\right) \right\Vert ^{2}} \\
&&+\frac{\nabla _{K_{\hat{X}}}f^{2}\left( \hat{X},K_{\hat{X}}\right) }{%
\sigma _{\hat{K}}^{2}\left\Vert \Psi \left( \hat{X}\right) \right\Vert ^{2}}%
\hat{K}^{2}\hat{\Psi}^{\dagger }\left( \hat{K},\hat{X}\right)
\end{eqnarray*}%
Following (Gosselin Lotz Wambst 2022) we neglect in first approximation the
derivatives with respect to $K_{\hat{X}}$ , and define the new variable:%
\begin{equation}
y=\frac{\hat{K}+\frac{\sigma _{\hat{K}}^{2}F\left( \hat{X},K_{\hat{X}%
}\right) }{f^{2}\left( \hat{X}\right) }}{\sqrt{\sigma _{\hat{K}}^{2}}}\left(
f^{2}\left( \hat{X}\right) \right) ^{\frac{1}{4}}  \label{cvr}
\end{equation}%
\begin{equation}
\frac{\delta ^{2}\left( S_{3}\left( \Psi \right) +S_{4}\left( \Psi \right)
\right) }{\delta \hat{\Psi}^{\dag }\left( Z,\theta \right) \delta \hat{\Psi}%
\left( Z,\theta \right) }=-\frac{\sigma _{\hat{X}}^{2}}{2}\nabla _{\hat{X}%
}^{2}-\nabla _{y}^{2}+\left( \frac{y^{2}}{4}+\frac{\left( g\left( \hat{X}%
\right) \right) ^{2}+\sigma _{\hat{X}}^{2}\left( f\left( \hat{X}\right)
+\nabla _{\hat{X}}g\left( \hat{X},K_{\hat{X}}\right) -\frac{\sigma _{\hat{K}%
}^{2}F^{2}\left( \hat{X},K_{\hat{X}}\right) }{2f^{2}\left( \hat{X}\right) }%
\right) }{\sigma _{\hat{X}}^{2}\sqrt{f^{2}\left( \hat{X}\right) }}\right)
\label{SCQ}
\end{equation}%
This leads to:%
\begin{eqnarray*}
&&\Delta \hat{\Psi}^{\dag }\left( Z,\theta \right) \frac{\delta ^{2}\left(
S_{3}\left( \Psi \right) +S_{4}\left( \Psi \right) \right) }{\delta \hat{\Psi%
}^{\dag }\left( Z,\theta \right) \delta \hat{\Psi}\left( Z,\theta \right) }%
\Delta \hat{\Psi}\left( Z,\theta \right) \\
&=&\Delta \hat{\Psi}^{\dag }\left( Z,\theta \right) \left( -\frac{\sigma _{%
\hat{X}}^{2}}{2}\nabla _{\hat{X}}^{2}+\frac{\left( g\left( \hat{X}\right)
\right) ^{2}+\sigma _{\hat{X}}^{2}\left( f\left( \hat{X}\right) +\nabla _{%
\hat{X}}g\left( \hat{X},K_{\hat{X}}\right) -\frac{\sigma _{\hat{K}%
}^{2}F^{2}\left( \hat{X},K_{\hat{X}}\right) }{2f^{2}\left( \hat{X}\right) }%
\right) }{\sigma _{\hat{X}}^{2}\sqrt{f^{2}\left( \hat{X}\right) }}\right. \\
&&\left. -\frac{\sigma _{\hat{K}}^{2}}{2\sqrt{f^{2}\left( \hat{X}\right) }}%
\nabla _{\hat{K}}^{2}+\left( \frac{\sqrt{f^{2}\left( \hat{X}\right) }\left( 
\hat{K}+\frac{\sigma _{\hat{K}}^{2}F\left( \hat{X},K_{\hat{X}}\right) }{%
f^{2}\left( \hat{X}\right) }\right) ^{2}}{4\sigma _{\hat{K}}^{2}}\right)
\right) \Delta \hat{\Psi}\left( Z,\theta \right)
\end{eqnarray*}

\section*{Appendix 7 Higher order corrections to the effective action}

The higher-order corrections are obtained by expanding at higher-orders in $%
\Delta \Psi \left( Z,\theta \right) $ and $\Delta \hat{\Psi}\left( Z,\theta
\right) $. These variations around the background fields can be considered
to be orthogonals to $\Psi _{0}\left( Z,\theta \right) $ and $\hat{\Psi}%
_{0}\left( Z,\theta \right) $.

\subsection*{A7.1 Third order terms}

The orthogonality condition implies that the third-order terms in the
expansion can be neglected. Actually, in first approximation the third-order
terms arising in the expansion of $S$ have the form:%
\begin{eqnarray}
&&2\tau \int \Delta \Psi \left( K^{\prime },X\right) \Psi _{0}^{\dag }\left(
K^{\prime },X^{\prime }\right) dK^{\prime }\left\vert \Delta \Psi \left(
K,X\right) \right\vert ^{2}dKdX  \label{CLC} \\
&&-\int \Delta \Psi ^{\dag }\left( K,X\right) \Psi _{0}^{\dag }\left(
K^{\prime },X^{\prime }\right) \nabla _{K}\frac{\delta u\left( K,X,\Psi ,%
\hat{\Psi}\right) }{\delta \left\vert \Psi \left( K^{\prime },X\right)
\right\vert ^{2}}\Delta \Psi \left( K^{\prime },X^{\prime }\right) \Delta
\Psi \left( K,X\right)  \notag \\
&&-\int \Delta \Psi ^{\dagger }\left( K,\theta \right) \hat{\Psi}%
_{0}^{\dagger }\left( \hat{K},\theta \right) \nabla _{K}\frac{\delta u\left(
K,X,\Psi ,\hat{\Psi}\right) }{\delta \left\vert \hat{\Psi}\left( \hat{K},%
\hat{X}\right) \right\vert ^{2}}\Delta \hat{\Psi}\left( \hat{K},\theta
\right) \Delta \Psi \left( K,\theta \right)  \notag \\
&&-\int \Delta \hat{\Psi}^{\dag }\left( \hat{K},\hat{X}\right) \Psi
_{0}^{\dag }\left( K^{\prime },\theta \right) \left\{ \nabla _{\hat{K}}\frac{%
\hat{K}\delta ^{2}f\left( \hat{X},\Psi ,\hat{\Psi}\right) }{\delta
\left\vert \Psi \left( K^{\prime },X\right) \right\vert ^{2}}+\nabla _{\hat{X%
}}\frac{\delta g\left( \hat{X},\Psi ,\hat{\Psi}\right) }{\delta \left\vert
\Psi \left( K^{\prime },X\right) \right\vert ^{2}}\right\} \Delta \Psi
\left( K^{\prime },X^{\prime }\right) \Delta \hat{\Psi}\left( \hat{K},\hat{X}%
\right)  \notag \\
&&+H.C.  \notag
\end{eqnarray}%
where the notation $H.C.$ stands for the hermitian conjugate of the
expression. Replacing the terms:%
\begin{equation*}
\nabla _{K}\frac{\delta u\left( K,X,\Psi ,\hat{\Psi}\right) }{\delta
\left\vert \Psi \left( K^{\prime },X\right) \right\vert ^{2}}\text{ , }%
\nabla _{K}\frac{\delta u\left( K,X,\Psi ,\hat{\Psi}\right) }{\delta
\left\vert \hat{\Psi}\left( \hat{K},\hat{X}\right) \right\vert ^{2}}
\end{equation*}%
and:%
\begin{equation*}
\nabla _{\hat{K}}\frac{\hat{K}\delta ^{2}f\left( \hat{X},\Psi ,\hat{\Psi}%
\right) }{\delta \left\vert \Psi \left( K^{\prime },X\right) \right\vert ^{2}%
}+\nabla _{\hat{X}}\frac{\delta g\left( \hat{X},\Psi ,\hat{\Psi}\right) }{%
\delta \left\vert \Psi \left( K^{\prime },X\right) \right\vert ^{2}}
\end{equation*}%
by their averages in (\ref{CLC}), and using the orthogonality conditions:%
\begin{equation*}
\int \hat{\Psi}_{0}^{\dagger }\left( \hat{K},\theta \right) \Delta \hat{\Psi}%
\left( \hat{K},\theta \right) =\int \Psi _{0}^{\dag }\left( K^{\prime
},X^{\prime }\right) \Delta \Psi \left( K^{\prime },X^{\prime }\right) =0
\end{equation*}%
leads to neglect the third-order terms in first approximation.

\subsection*{A7.2 Fourth order terms}

\subsubsection*{A7.2.1 General formula}

Considering the fourth-order in the action expansion yields quartic
corrections. Using that in average:

\begin{equation*}
\frac{\delta ^{2}f\left( \hat{X},\Psi ,\hat{\Psi}\right) }{\delta \hat{\Psi}%
\left( \hat{K},\hat{X}\right) \delta \hat{\Psi}^{\dagger }\left( \hat{K},%
\hat{X}\right) }\simeq 0
\end{equation*}%
\begin{equation*}
\frac{\delta ^{2}g\left( \hat{X},\Psi ,\hat{\Psi}\right) }{\delta \hat{\Psi}%
\left( \hat{K},\hat{X}\right) \delta \hat{\Psi}^{\dagger }\left( \hat{K},%
\hat{X}\right) }\simeq 0
\end{equation*}%
the fourth-order terms in the fields' action become:%
\begin{eqnarray}
&&2\tau \int \left\vert \Delta \Psi \left( K^{\prime },X\right) \right\vert
^{2}dK^{\prime }\left\vert \Delta \Psi \left( K,X\right) \right\vert ^{2}dKdX
\label{FRT} \\
&&-\Delta \Psi ^{\dag }\left( K,X\right) \Delta \Psi ^{\dag }\left(
K^{\prime },X^{\prime }\right) \nabla _{K}\frac{\delta ^{2}u\left( K,X,\Psi ,%
\hat{\Psi}\right) }{\delta \Psi \left( K^{\prime },X\right) \delta \Psi
^{\dagger }\left( K^{\prime },X\right) }\Delta \Psi \left( K^{\prime
},X^{\prime }\right) \Delta \Psi \left( K,X\right)  \notag \\
&&-\Delta \Psi ^{\dagger }\left( K,\theta \right) \Delta \hat{\Psi}^{\dagger
}\left( \hat{K},\theta \right) \nabla _{K}\frac{\delta ^{2}u\left( K,X,\Psi ,%
\hat{\Psi}\right) }{\delta \hat{\Psi}\left( \hat{K},\hat{X}\right) \delta 
\hat{\Psi}^{\dagger }\left( \hat{K},\hat{X}\right) }\Delta \hat{\Psi}\left( 
\hat{K},\theta \right) \Delta \Psi \left( K,\theta \right)  \notag \\
&&-\Delta \hat{\Psi}^{\dag }\left( \hat{K},\hat{X}\right) \Delta \Psi ^{\dag
}\left( K^{\prime },\theta \right) \left\{ \nabla _{\hat{K}}\frac{\hat{K}%
\delta ^{2}f\left( \hat{X},\Psi ,\hat{\Psi}\right) }{\delta \Psi \left(
K^{\prime },X\right) \delta \Psi ^{\dagger }\left( K^{\prime },X\right) }%
+\nabla _{\hat{X}}\frac{\delta ^{2}g\left( \hat{X},\Psi ,\hat{\Psi}\right) }{%
\delta \Psi \left( K^{\prime },X\right) \delta \Psi ^{\dagger }\left(
K^{\prime },X\right) }\right\} \Delta \Psi \left( K^{\prime },X^{\prime
}\right) \Delta \hat{\Psi}\left( \hat{K},\hat{X}\right)  \notag
\end{eqnarray}

\subsubsection*{A72.2 Estimation of the various terms}

The three last terms in the rhs of (\ref{FRT}) can be evaluated. The second
term is given by:%
\begin{eqnarray*}
&&\Delta \Psi ^{\dag }\left( K,X\right) \Delta \Psi ^{\dag }\left( K^{\prime
},X^{\prime }\right) \frac{\delta ^{2}u\left( K,X,\Psi ,\hat{\Psi}\right) }{%
\delta \Psi \left( K^{\prime },X\right) \delta \Psi ^{\dagger }\left(
K^{\prime },X\right) }\Delta \Psi \left( K,X\right) \Delta \Psi \left(
K^{\prime },\theta \right) \\
&=&\Delta \Psi ^{\dag }\left( K,\theta \right) \Delta \Psi ^{\dag }\left(
K^{\prime },\theta \right) \left\{ \int \hat{F}_{2}\left( s,R\left(
K,X\right) \right) \hat{F}_{2}\left( s^{\prime },R\left( K^{\prime
},X^{\prime }\right) \right) \hat{K}\left\Vert \hat{\Psi}\left( \hat{K}%
,X\right) \right\Vert ^{2}d\hat{K}\right\} \Delta \Psi \left( K,\theta
\right) \Delta \Psi \left( K^{\prime },\theta \right) \\
&&-2\Delta \Psi ^{\dag }\left( K,\theta \right) \Delta \Psi ^{\dag }\left(
K^{\prime },\theta \right) \\
&&\times \left\{ \int \Psi _{0}^{\dag }\left( K^{\prime },X\right) \frac{%
\hat{F}_{2}\left( s,R\left( K,X\right) \right) \hat{F}_{2}\left( s^{\prime
},R\left( K^{\prime },X^{\prime }\right) \right) }{\int F_{2}\left(
s^{\prime },R\left( K^{\prime },X\right) \right) \left\Vert \Psi \left(
K^{\prime },X\right) \right\Vert ^{2}dK^{\prime }}\Psi _{0}\left( K,\hat{X}%
\right) \hat{K}\left\Vert \hat{\Psi}\left( \hat{K},X\right) \right\Vert ^{2}d%
\hat{K}\right\} \Delta \Psi \left( K,\theta \right) \Delta \Psi \left(
K^{\prime },\theta \right) \\
&\simeq &\Delta \Psi ^{\dag }\left( K,\theta \right) \Delta \Psi ^{\dag
}\left( K^{\prime },\theta \right) \left\{ \int \hat{F}_{2}\left( s,R\left(
K,X\right) \right) \hat{F}_{2}\left( s^{\prime },R\left( K^{\prime
},X^{\prime }\right) \right) \hat{K}\left\Vert \hat{\Psi}\left( \hat{K}%
,X\right) \right\Vert ^{2}d\hat{K}\right\} \Delta \Psi \left( K,\theta
\right) \Delta \Psi \left( K^{\prime },\theta \right)
\end{eqnarray*}

The second term in the rhs of (\ref{FRT}) is equal to:%
\begin{eqnarray*}
&&\Delta \Psi ^{\dagger }\left( K,\theta \right) \Delta \hat{\Psi}^{\dagger
}\left( \hat{K},\theta \right) \frac{\delta ^{2}u\left( K,X,\Psi ,\hat{\Psi}%
\right) }{\delta \hat{\Psi}\left( \hat{K},\hat{X}\right) \delta \hat{\Psi}%
^{\dagger }\left( \hat{K},\hat{X}\right) }\Delta \Psi \left( K,\theta
\right) \Delta \hat{\Psi}\left( \hat{K},\theta \right) \\
&=&-\Delta \Psi ^{\dagger }\left( K,\theta \right) \Delta \hat{\Psi}%
^{\dagger }\left( \hat{K},\theta \right) \frac{1}{\varepsilon }\hat{F}%
_{2}\left( s,R\left( K,X\right) \right) \hat{K}\Delta \Psi \left( K,\theta
\right) \Delta \hat{\Psi}\left( \hat{K},\theta \right)
\end{eqnarray*}%
Ultimately, the last term in the rhs of (\ref{FRT}):%
\begin{equation*}
\Delta \hat{\Psi}^{\dag }\left( \hat{K},\hat{X}\right) \Delta \Psi ^{\dag
}\left( K^{\prime },\theta \right) \left\{ \nabla _{\hat{K}}\frac{\hat{K}%
\delta ^{2}f\left( \hat{X},\Psi ,\hat{\Psi}\right) }{\delta \Psi \left(
K^{\prime },X\right) \delta \Psi ^{\dagger }\left( K^{\prime },X\right) }%
+\nabla _{\hat{X}}\frac{\delta ^{2}g\left( \hat{X},\Psi ,\hat{\Psi}\right) }{%
\delta \Psi \left( K^{\prime },X\right) \delta \Psi ^{\dagger }\left(
K^{\prime },X\right) }\right\} \Delta \Psi \left( K^{\prime },X^{\prime
}\right) \Delta \hat{\Psi}\left( \hat{K},\hat{X}\right)
\end{equation*}%
is obtained by using the expressions of $f\left( \hat{X},\Psi ,\hat{\Psi}%
\right) $ and $g\left( \hat{X},\Psi ,\hat{\Psi}\right) $ that compute
short-term and long-term returns, respectively:

\begin{eqnarray*}
f\left( \hat{X},\Psi ,\hat{\Psi}\right) &=&\frac{1}{\varepsilon }\int \left(
r\left( K,X\right) -\gamma \frac{\int K^{\prime }\left\Vert \Psi \left(
K^{\prime },X\right) \right\Vert ^{2}}{K}+F_{1}\left( \frac{R\left(
K,X\right) }{\int R\left( K^{\prime },X^{\prime }\right) \left\Vert \Psi
\left( K^{\prime },X^{\prime }\right) \right\Vert ^{2}d\left( K^{\prime
},X^{\prime }\right) },\Gamma \left( K,X\right) \right) \right) \\
&&\times \hat{F}_{2}\left( s,R\left( K,X\right) \right) \left\Vert \Psi
\left( K,\hat{X}\right) \right\Vert ^{2}dK \\
g\left( K,\hat{X},\Psi ,\hat{\Psi}\right) &=&\int \left( \nabla _{\hat{X}%
}F_{0}\left( R\left( K,\hat{X}\right) \right) +\nu \nabla _{\hat{X}%
}F_{1}\left( \frac{R\left( K,\hat{X}\right) }{\int R\left( K^{\prime
},X^{\prime }\right) \left\Vert \Psi \left( K^{\prime },X^{\prime }\right)
\right\Vert ^{2}d\left( K^{\prime },X^{\prime }\right) }\right) \right) \\
&&\times \frac{\left\Vert \Psi \left( K,\hat{X}\right) \right\Vert ^{2}dK}{%
\int \left\Vert \Psi \left( K^{\prime },\hat{X}\right) \right\Vert
^{2}dK^{\prime }}
\end{eqnarray*}%
We find:%
\begin{eqnarray}
&&\frac{\delta ^{2}f\left( \hat{X},\Psi ,\hat{\Psi}\right) }{\delta \Psi
\left( K^{\prime },X\right) \delta \Psi ^{\dagger }\left( K^{\prime
},X\right) }  \label{DLT} \\
&=&\frac{1}{\varepsilon }\Delta \left( r\left( K^{\prime },X\right) -\gamma 
\frac{\int K^{\prime }\left\Vert \Psi \left( K^{\prime },X\right)
\right\Vert ^{2}}{K^{\prime }}+F_{1}\left( \frac{R\left( K^{\prime
},X\right) }{\int R\left( K^{\prime },X^{\prime }\right) \left\Vert \Psi
\left( K^{\prime },X^{\prime }\right) \right\Vert ^{2}d\left( K^{\prime
},X^{\prime }\right) },\Gamma \left( K,X\right) \right) \right)  \notag \\
&&\times \frac{F_{2}\left( s^{\prime },R\left( K^{\prime },\hat{X}\right)
\right) }{\int F_{2}\left( s^{\prime }R\left( K^{\prime },\hat{X}\right)
\right) \left\Vert \Psi \left( K^{\prime },\hat{X}\right) \right\Vert
^{2}dK^{\prime }}  \notag \\
&&-\frac{1}{\varepsilon }\int \left( \gamma \frac{K^{\prime }}{K}+\frac{%
R\left( K^{\prime },X\right) R\left( K_{\hat{X}},X\right) }{\left( \int
R\left( K^{\prime },X^{\prime }\right) \left\Vert \Psi \left( K^{\prime
},X^{\prime }\right) \right\Vert ^{2}d\left( K^{\prime },X^{\prime }\right)
\right) ^{2}}F_{1}^{\prime }\left( \frac{R\left( K,X\right) }{\int R\left(
K^{\prime },X^{\prime }\right) \left\Vert \Psi \left( K^{\prime },X^{\prime
}\right) \right\Vert ^{2}d\left( K^{\prime },X^{\prime }\right) },\Gamma
\left( K,X\right) \right) \right)  \notag \\
&&\times \hat{F}_{2}\left( s,R\left( K,X\right) \right) \left\Vert \Psi
\left( K,\hat{X}\right) \right\Vert ^{2}  \notag
\end{eqnarray}%
where we define the deviation $\Delta Y$ of a quantity by the difference:%
\begin{equation}
\Delta Y=Y-\left\langle Y\right\rangle  \label{RV}
\end{equation}%
with $\left\langle Y\right\rangle $, the average of $Y$:%
\begin{equation*}
\left\langle Y\right\rangle =\int Y\left( K,X\right) dKdX
\end{equation*}%
\textbf{\ }Thus we write:%
\begin{eqnarray*}
&&\Delta \left( r\left( K^{\prime },X\right) -\gamma \frac{\int K^{\prime
}\left\Vert \Psi \left( K^{\prime },X\right) \right\Vert ^{2}}{K^{\prime }}%
+F_{1}\left( \frac{R\left( K^{\prime },X\right) }{\int R\left( K^{\prime
},X^{\prime }\right) \left\Vert \Psi \left( K^{\prime },X^{\prime }\right)
\right\Vert ^{2}d\left( K^{\prime },X^{\prime }\right) },\Gamma \left(
K,X\right) \right) \right) \\
&=&\left( r\left( K^{\prime },X\right) -\gamma \frac{\int K^{\prime
}\left\Vert \Psi \left( K^{\prime },X\right) \right\Vert ^{2}}{K^{\prime }}%
+F_{1}\left( \frac{R\left( K^{\prime },X\right) }{\int R\left( K^{\prime
},X^{\prime }\right) \left\Vert \Psi \left( K^{\prime },X^{\prime }\right)
\right\Vert ^{2}d\left( K^{\prime },X^{\prime }\right) },\Gamma \left(
K,X\right) \right) \right) \\
&&-\left\langle \left( r\left( K^{\prime },X\right) -\gamma \frac{\int
K^{\prime }\left\Vert \Psi \left( K^{\prime },X\right) \right\Vert ^{2}}{%
K^{\prime }}+F_{1}\left( \frac{R\left( K^{\prime },X\right) }{\int R\left(
K^{\prime },X^{\prime }\right) \left\Vert \Psi \left( K^{\prime },X^{\prime
}\right) \right\Vert ^{2}d\left( K^{\prime },X^{\prime }\right) },\Gamma
\left( K,X\right) \right) \right) \right\rangle
\end{eqnarray*}%
and in first approximation, (\ref{DLT}) reduces to:%
\begin{eqnarray*}
&&\frac{\delta ^{2}f\left( \hat{X},\Psi ,\hat{\Psi}\right) }{\delta \Psi
\left( K^{\prime },X\right) \delta \Psi ^{\dagger }\left( K^{\prime
},X\right) } \\
&\simeq &\frac{1}{\varepsilon }\left( \Delta \left( r\left( K^{\prime
},X\right) -\gamma \frac{K_{X}}{K^{\prime }}+F_{1}\left( \frac{R\left(
K^{\prime },X\right) }{\int R\left( K^{\prime },X^{\prime }\right)
\left\Vert \Psi \left( K^{\prime },X^{\prime }\right) \right\Vert
^{2}d\left( K^{\prime },X^{\prime }\right) },\Gamma \left( K,X\right)
\right) \right) -\gamma \frac{K^{\prime }}{K_{X}}\right) \\
&\simeq &\frac{1}{\varepsilon }\left( \Delta f\left( K^{\prime },\hat{X}%
,\Psi ,\hat{\Psi}\right) -\gamma \frac{K^{\prime }}{K_{X}}\right)
\end{eqnarray*}%
where:%
\begin{equation*}
\Delta f\left( K^{\prime },\hat{X},\Psi ,\hat{\Psi}\right) =f\left(
K^{\prime },\hat{X},\Psi ,\hat{\Psi}\right) -f\left( K_{\hat{X}},\hat{X}%
,\Psi ,\hat{\Psi}\right)
\end{equation*}%
is the relative short-term return for firm with capital $K^{\prime }$ at
sector $\hat{X}$.

Similarly, the second derivative for $g\left( \hat{X},\Psi ,\hat{\Psi}%
\right) $ is:%
\begin{eqnarray*}
&&\frac{\delta ^{2}g\left( \hat{X},\Psi ,\hat{\Psi}\right) }{\delta \Psi
\left( K^{\prime },X\right) \delta \Psi ^{\dagger }\left( K^{\prime
},X\right) } \\
&=&\frac{1}{\int \left\Vert \Psi \left( K^{\prime },\hat{X}\right)
\right\Vert ^{2}dK^{\prime }}\Delta \left( \nabla _{\hat{X}}F_{0}\left(
R\left( K^{\prime },\hat{X}\right) \right) +\nu \nabla _{\hat{X}}F_{1}\left( 
\frac{R\left( K^{\prime },\hat{X}\right) }{\int R\left( K^{\prime
},X^{\prime }\right) \left\Vert \Psi \left( K^{\prime },X^{\prime }\right)
\right\Vert ^{2}d\left( K^{\prime },X^{\prime }\right) }\right) \right) \\
&=&\frac{1}{\int \left\Vert \Psi \left( K^{\prime },\hat{X}\right)
\right\Vert ^{2}dK^{\prime }}\Delta \left( g\left( K^{\prime },\hat{X},\Psi ,%
\hat{\Psi}\right) \right)
\end{eqnarray*}%
with:%
\begin{eqnarray*}
&&\Delta \left( \nabla _{\hat{X}}F_{0}\left( R\left( K^{\prime },\hat{X}%
\right) \right) +\nu \nabla _{\hat{X}}F_{1}\left( \frac{R\left( K^{\prime },%
\hat{X}\right) }{\int R\left( K^{\prime },X^{\prime }\right) \left\Vert \Psi
\left( K^{\prime },X^{\prime }\right) \right\Vert ^{2}d\left( K^{\prime
},X^{\prime }\right) }\right) \right) \\
&=&\left( \nabla _{\hat{X}}F_{0}\left( R\left( K^{\prime },\hat{X}\right)
\right) +\nu \nabla _{\hat{X}}F_{1}\left( \frac{R\left( K^{\prime },\hat{X}%
\right) }{\int R\left( K^{\prime },X^{\prime }\right) \left\Vert \Psi \left(
K^{\prime },X^{\prime }\right) \right\Vert ^{2}d\left( K^{\prime },X^{\prime
}\right) }\right) \right) \\
&&-\left\langle \nabla _{\hat{X}}F_{0}\left( R\left( K^{\prime },\hat{X}%
\right) \right) +\nu \nabla _{\hat{X}}F_{1}\left( \frac{R\left( K^{\prime },%
\hat{X}\right) }{\int R\left( K^{\prime },X^{\prime }\right) \left\Vert \Psi
\left( K^{\prime },X^{\prime }\right) \right\Vert ^{2}d\left( K^{\prime
},X^{\prime }\right) }\right) \right\rangle
\end{eqnarray*}%
in other words:%
\begin{equation*}
\Delta g\left( K^{\prime },\hat{X},\Psi ,\hat{\Psi}\right) =g\left(
K^{\prime },\hat{X},\Psi ,\hat{\Psi}\right) -g\left( \hat{X},\Psi ,\hat{\Psi}%
\right)
\end{equation*}%
is the relative long-term return for firm with capital $K^{\prime }$ at
sector $\hat{X}$.

\section*{Appendix 8: "free" transition functions}

Given the second-order operator arising in the expansion for the fields'
action:%
\begin{equation}
O\left( \Psi _{0}\left( Z,\theta \right) \right) \simeq \left( 
\begin{array}{cc}
\frac{\delta ^{2}\left( S_{1}+S_{2}\right) }{\delta \Psi ^{\dag }\left(
Z,\theta \right) \delta \Psi \left( Z,\theta \right) } & 0 \\ 
0 & \frac{\delta ^{2}\left( S_{3}\left( \Psi \right) +S_{4}\left( \Psi
\right) \right) }{\delta \hat{\Psi}^{\dag }\left( Z,\theta \right) \delta 
\hat{\Psi}\left( Z,\theta \right) }%
\end{array}%
\right) _{\substack{ \Psi \left( Z,\theta \right) =\Psi _{0}\left( Z,\theta
\right)  \\ \hat{\Psi}\left( Z,\theta \right) =\hat{\Psi}_{0}\left( Z,\theta
\right) }}
\end{equation}%
The transition functions for the individual firms:

\begin{equation*}
G_{1}\left( \left( K_{f},X_{f}\right) ,\left( X_{i},K_{i}\right) ,\alpha
\right)
\end{equation*}%
and investors:%
\begin{equation*}
G_{2}\left( \left( \hat{K}_{f},\hat{X}_{f}\right) ,\left( \hat{X}_{i},\hat{K}%
_{i}\right) ,\alpha \right)
\end{equation*}%
satisfy:%
\begin{eqnarray*}
\left( \frac{\delta ^{2}\left( S_{1}+S_{2}\right) }{\delta \Psi ^{\dag
}\left( Z,\theta \right) \delta \Psi \left( Z,\theta \right) }+\alpha
\right) G_{1}\left( \left( K_{f},X_{f}\right) ,\left( X_{i},K_{i}\right)
,\alpha \right) &=&\delta \left( \left( K_{f},X_{f}\right) -\left(
X_{i},K_{i}\right) \right) \\
\left( \frac{\delta ^{2}\left( S_{3}\left( \Psi \right) +S_{4}\left( \Psi
\right) \right) }{\delta \hat{\Psi}^{\dag }\left( Z,\theta \right) \delta 
\hat{\Psi}\left( Z,\theta \right) }+\alpha \right) G_{2}\left( \left( \hat{K}%
_{f},\hat{X}_{f}\right) ,\left( \hat{X}_{i},\hat{K}_{i}\right) ,\alpha
\right) &=&\delta \left( \left( \hat{K}_{f},\hat{X}_{f}\right) -\left( \hat{X%
}_{i},\hat{K}_{i}\right) \right)
\end{eqnarray*}%
The functions $G_{1}\left( \left( K_{f},X_{f}\right) ,\left(
X_{i},K_{i}\right) ,\alpha \right) $ and $G_{2}\left( \left( \hat{K}_{f},%
\hat{X}_{f}\right) ,\left( \hat{X}_{i},\hat{K}_{i}\right) ,\alpha \right) $\
are the Laplace transforms of the following transition functions:%
\begin{equation*}
T_{1}\left( \left( K_{f},X_{f}\right) ,\left( X_{i},K_{i}\right) ,t\right)
\end{equation*}%
\begin{equation*}
T_{2}\left( \left( \hat{K}_{f},\hat{X}_{f}\right) ,\left( \hat{X}_{i},\hat{K}%
_{i}\right) ,t\right)
\end{equation*}%
satisfying:%
\begin{equation}
-\frac{\partial }{\partial t}T_{1}\left( \left( K_{f},X_{f}\right) ,\left(
X_{i},K_{i}\right) ,t\right) =\left( \frac{\delta ^{2}\left(
S_{1}+S_{2}\right) }{\delta \Psi ^{\dag }\left( Z,\theta \right) \delta \Psi
\left( Z,\theta \right) }\right) T_{1}\left( \left( K_{f},X_{f}\right)
,\left( X_{i},K_{i}\right) ,t\right)  \label{TNQ}
\end{equation}%
\begin{equation}
-\frac{\partial }{\partial t}T_{2}\left( \left( \hat{K}_{f},\hat{X}%
_{f}\right) ,\left( \hat{X}_{i},\hat{K}_{i}\right) ,t\right) =\left( \frac{%
\delta ^{2}\left( S_{3}\left( \Psi \right) +S_{4}\left( \Psi \right) \right) 
}{\delta \hat{\Psi}^{\dag }\left( Z,\theta \right) \delta \hat{\Psi}\left(
Z,\theta \right) }\right) T_{2}\left( \left( \hat{K}_{f},\hat{X}_{f}\right)
,\left( \hat{X}_{i},\hat{K}_{i}\right) ,t\right)  \label{TSQ}
\end{equation}

\subsection*{A8.1 Approximations to and (\protect\ref{TNQ}) \ and (\protect
\ref{TSQ})}

We consider some approximations to find the solutions of equations (\ref{SCR}%
) and (\ref{SCQ}). We first assume that:%
\begin{equation*}
\frac{\nabla _{K}\frac{F_{2}\left( R\left( K,X\right) \right) }{\left\langle
F_{2}\left( R\left( K,X\right) \right) \right\rangle _{K}}K_{X}}{2}<<1
\end{equation*}%
so that:%
\begin{eqnarray*}
K-\hat{F}_{2}\left( s,R\left( K,X\right) \right) K_{X} &\simeq &K-\hat{F}%
_{2}\left( R\left( K_{X},X\right) \right) K_{X}-\nabla _{K_{X}}\hat{F}%
_{2}\left( R\left( K_{X},X\right) \right) \left( K-K_{X}\right) \\
&\simeq &K-\hat{F}_{2}\left( R\left( K_{X},X\right) \right) K_{X}
\end{eqnarray*}%
Equation (\ref{SCR}) then simplifies as:%
\begin{eqnarray}
\frac{\delta ^{2}\left( S_{1}+S_{2}\right) }{\delta \Psi ^{\dag }\left(
Z,\theta \right) \delta \Psi \left( Z,\theta \right) } &=&-\frac{\sigma
_{X}^{2}}{2}\nabla _{X}^{2}-\frac{\sigma _{K}^{2}}{2}\nabla _{K}^{2}+\left(
D\left( \left\Vert \Psi \right\Vert ^{2}\right) +2\tau \frac{\left\vert \Psi
\left( X\right) \right\vert ^{2}\left( K_{X}-K\right) }{K}\right) \\
&&+\frac{1}{2\sigma _{K}^{2}}\left( K-\hat{F}_{2}\left( s,R\left(
K_{X},X\right) \right) K_{X}\right) ^{2}  \notag
\end{eqnarray}%
and equation (\ref{TNQ}) becomes:%
\begin{eqnarray}
&&-\frac{\partial }{\partial t}T_{1}\left( \left( K_{f},X_{f}\right) ,\left(
X_{i},K_{i}\right) ,t\right)  \label{TNV} \\
&=&\left( -\frac{\sigma _{X}^{2}}{2}\nabla _{X}^{2}+D\left( \left\Vert \Psi
\right\Vert ^{2}\right) +2\tau \frac{K_{X}-K}{K}\left\Vert \Psi \left(
X\right) \right\Vert ^{2}\right) T_{1}\left( \left( K_{f},X_{f}\right)
,\left( X_{i},K_{i}\right) ,t\right)  \notag \\
&&+\left( -\frac{\sigma _{K}^{2}}{2}\nabla _{K}^{2}+\frac{1}{2\sigma _{K}^{2}%
}\left( K-\hat{F}_{2}\left( R\left( K_{X},X\right) \right) K_{X}\right)
^{2}\right) T_{1}\left( \left( K_{f},X_{f}\right) ,\left( X_{i},K_{i}\right)
,t\right)  \notag
\end{eqnarray}%
Second, we assumed from the beginning that the motion of firms in the
sectors space is at slower pace than capital fluctuations. Moreover, we may
assume that in average $\left\vert \frac{K_{X}-K}{K}\right\vert <<1$. As a
consequence, along the path from the initial point $\left(
X_{i},K_{i}\right) $ to the final point $\left( K_{f},X_{f}\right) $, we can
consider that:%
\begin{equation*}
\frac{K_{X}-K}{K}\left\Vert \Psi \left( X\right) \right\Vert ^{2}
\end{equation*}%
is slowly varying and can be replaced by its average.

The equation for $T_{1}$ thus rewrites:

\begin{eqnarray}
&&-\frac{\partial }{\partial t}T_{1}\left( \left( K_{f},X_{f}\right) ,\left(
X_{i},K_{i}\right) \right)  \label{trnv} \\
&=&\left( -\frac{\sigma _{X}^{2}}{2}\nabla _{X}^{2}+D\left( \left\Vert \Psi
\right\Vert ^{2}\right) +\tau \left( \frac{\left\vert \Psi \left(
X_{f}\right) \right\vert ^{2}\left( K_{X_{f}}-K_{f}\right) }{K_{f}}+\frac{%
\left\vert \Psi \left( X_{i}\right) \right\vert ^{2}\left(
K_{X_{i}}-K_{i}\right) }{K_{i}}\right) \right) T_{1}\left( \left(
K_{f},X_{f}\right) ,\left( X_{i},K_{i}\right) \right)  \notag \\
&&+\left( -\frac{\sigma _{K}^{2}}{2}\nabla _{K}^{2}+\frac{1}{2\sigma _{K}^{2}%
}\left( K-\hat{F}_{2}\left( R\left( K_{X},X\right) \right) K_{X}\right)
^{2}\right) T_{1}\left( \left( K_{f},X_{f}\right) ,\left( X_{i},K_{i}\right)
\right)  \notag
\end{eqnarray}%
On the other hand, the derivation of the equation for $T_{2}$ yields
directly: 
\begin{eqnarray}
&&-\frac{\partial }{\partial t}T_{2}\left( \left( \hat{K}_{f},\hat{X}%
_{f}\right) ,\left( \hat{X}_{i},\hat{K}_{i}\right) \right)  \label{TRV} \\
&=&\left( -\frac{\sigma _{\hat{X}}^{2}}{2}\nabla _{\hat{X}}^{2}-\nabla
_{y}^{2}\right) T_{2}\left( \left( \hat{K}_{f},\hat{X}_{f}\right) ,\left( 
\hat{X}_{i},\hat{K}_{i}\right) \right)  \notag \\
&&+\left( \frac{y^{2}}{4}+\frac{\left( g\left( \hat{X}\right) \right)
^{2}+\sigma _{\hat{X}}^{2}\left( f\left( \hat{X}\right) +\nabla _{\hat{X}%
}g\left( \hat{X},K_{\hat{X}}\right) -\frac{\sigma _{\hat{K}}^{2}F^{2}\left( 
\hat{X},K_{\hat{X}}\right) }{2f^{2}\left( \hat{X}\right) }\right) }{\sigma _{%
\hat{X}}^{2}\sqrt{f^{2}\left( \hat{X}\right) }}\right) T_{2}\left( \left( 
\hat{K}_{f},\hat{X}_{f}\right) ,\left( \hat{X}_{i},\hat{K}_{i}\right) \right)
\notag
\end{eqnarray}

\subsection*{A8.2 Computation of $T_{1}$}

\subsubsection{A8.2.1 Solution of (\protect\ref{trnv})}

We first rewrite the competition term in (\ref{trnv}) as:%
\begin{eqnarray*}
&&\frac{\left\vert \Psi \left( X_{f}\right) \right\vert ^{2}\left(
K_{X_{f}}-K_{f}\right) }{K_{f}}+\frac{\left\vert \Psi \left( X_{i}\right)
\right\vert ^{2}\left( K_{X_{i}}-K_{i}\right) }{K_{i}} \\
&=&\left( \frac{\left\vert \Psi \left( X_{f}\right) \right\vert ^{2}\left(
K_{X_{f}}-K_{f}\right) }{K_{f}}-\frac{\left\vert \Psi \left( X_{i}\right)
\right\vert ^{2}\left( K_{X_{i}}-K_{i}\right) }{K_{i}}\right) +\frac{%
\left\vert \Psi \left( X_{i}\right) \right\vert ^{2}\left(
K_{X_{i}}-K_{i}\right) }{K_{i}}
\end{eqnarray*}%
Then, we normalize the transition functions by factoring the solution of (%
\ref{trnv}):%
\begin{eqnarray}
&&T_{1}\left( \left( K_{f},X_{f}\right) ,\left( X_{i},K_{i}\right) \right)
\label{PNL} \\
&=&\exp \left( -t\left( D\left( \left\Vert \Psi \right\Vert ^{2}\right)
+\tau \left( \frac{\left\vert \Psi \left( X_{f}\right) \right\vert
^{2}\left( K_{X_{f}}-K_{f}\right) }{K_{f}}+\frac{\left\vert \Psi \left(
X_{i}\right) \right\vert ^{2}\left( K_{X_{i}}-K_{i}\right) }{K_{i}}\right)
\right) \right) \hat{T}_{1}\left( \left( K_{f},X_{f}\right) ,\left(
X_{i},K_{i}\right) \right)  \notag
\end{eqnarray}%
so that the transition equation writes:%
\begin{eqnarray}
&&-\frac{\partial }{\partial t}\hat{T}_{1}\left( \left( K_{f},X_{f}\right)
,\left( X_{i},K_{i}\right) \right)  \label{TRN} \\
&=&\left( -\frac{\sigma _{X}^{2}}{2}\nabla _{X}^{2}-\frac{\sigma _{K}^{2}}{2}%
\nabla _{K}^{2}+\frac{1}{2\sigma _{K}^{2}}\left( K-\hat{F}_{2}\left( R\left(
K,X\right) \right) K_{X}\right) ^{2}\right) \hat{T}_{1}\left( \left(
K_{f},X_{f}\right) ,\left( X_{i},K_{i}\right) \right)  \notag
\end{eqnarray}%
Note that, given the exponential factor, if $K_{i}<<K_{X_{i}}$, $\frac{%
\left\vert \Psi \left( X_{i}\right) \right\vert ^{2}\left(
K_{X_{i}}-K_{i}\right) }{K_{i}}<0$ and the probability to move away from $%
X_{i}$ is very low. The same applies for $\frac{\left\vert \Psi \left(
X_{f}\right) \right\vert ^{2}\left( K_{X_{f}}-K_{f}\right) }{K_{f}}>0$.

The transition function $\hat{T}_{1}\left( \left( K_{f},X_{f}\right) ,\left(
X_{i},K_{i}\right) \right) $ can be found by using our assumption that
shifts in sectors space are slower than the fluctuations in capital. In (\ref%
{TRN}) we can thus consider in first approximation that the term:%
\begin{equation*}
K-\hat{F}_{2}\left( R\left( K,X\right) \right) K_{X}
\end{equation*}%
shifts the inital and final values of capital:%
\begin{eqnarray*}
K_{i} &\rightarrow &K_{i}-\hat{F}_{2}\left( s,R\left( K_{X_{i}},X_{i}\right)
\right) K_{X_{i}}=K_{i}^{\prime } \\
K_{f} &\rightarrow &K_{f}-\hat{F}_{2}\left( s,R\left( K_{X_{f}},X_{f}\right)
\right) K_{X_{f}}=K_{f}^{\prime }
\end{eqnarray*}%
So that we have:%
\begin{equation}
\hat{T}_{1}\left( \left( K_{f},X_{f}\right) ,\left( X_{i},K_{i}\right)
\right) \simeq \tilde{T}_{1}\left( \left( K_{f}-\hat{F}_{2}\left( s,R\left(
K_{X_{f}},X_{f}\right) \right) K_{X_{f}},X_{f}\right) ,\left( K_{i}-\hat{F}%
_{2}\left( R\left( s,K_{X_{i}},X_{i}\right) \right) K_{X_{i}},K_{i}\right)
\right)  \label{NRM}
\end{equation}%
where $\tilde{T}_{1}$ satisfies:%
\begin{equation}
-\frac{\partial }{\partial t}\tilde{T}_{1}=\left( -\frac{\sigma _{X}^{2}}{2}%
\nabla _{X}^{2}-\frac{\sigma _{K}^{2}}{2}\nabla _{K}^{2}+\frac{1}{2\sigma
_{K}^{2}}K^{2}\right) \tilde{T}_{1}  \label{NRP}
\end{equation}%
Up to a normalization factor, the solution of (\ref{NRP}) is:%
\begin{equation*}
\tilde{T}_{1}\left( \left( K_{f}^{\prime },X_{f}\right) ,\left(
X_{i},K_{i}^{\prime }\right) \right) =\exp \left( -\left( \frac{\left(
X_{f}-X_{i}\right) ^{2}}{2t\sigma _{X}^{2}}+\frac{\left( K_{f}^{\prime
}-K_{i}^{\prime }\right) ^{2}}{2t\sigma _{K}^{2}}-\frac{t\sigma _{K}^{2}}{2}%
K_{f}^{\prime }K_{f}^{\prime }\right) \right)
\end{equation*}%
Using (\ref{NRM}) and (\ref{PNL}), we find the solution of (\ref{trm}):%
\begin{eqnarray}
&&\exp \left( -t\left( D\left( \left\Vert \Psi \right\Vert ^{2}\right) +\tau
\left( \frac{\left\vert \Psi \left( X_{f}\right) \right\vert ^{2}\left(
K_{X_{f}}-K_{f}\right) }{K_{f}}+\frac{\left\vert \Psi \left( X_{i}\right)
\right\vert ^{2}\left( K_{X_{i}}-K_{i}\right) }{K_{i}}\right) \right) \right)
\label{SPR} \\
&&\times \exp \left( -\left( \frac{\left( X_{f}-X_{i}\right) ^{2}}{2t\sigma
_{X}^{2}}+\frac{\left( K_{f}^{\prime }-K_{i}^{\prime }\right) ^{2}}{2t\sigma
_{K}^{2}}-\frac{t\sigma _{K}^{2}}{2}K_{f}^{\prime }K_{f}^{\prime }\right)
\right)  \notag
\end{eqnarray}

\subsubsection{A8.2.2 Full transition function}

\bigskip

To obtain the full transition function, recall that (\ref{SPR}) has been
obtained by a change of variable (\ref{chn}). To come back to the initial
variables we have to introduce an other exponential factor to account for
the trend of the transition, and we find:%
\begin{eqnarray*}
&&T_{1}\left( \left( K_{f},X_{f}\right) ,\left( X_{i},K_{i}\right) \right) \\
&\simeq &\exp \left( \int_{X_{i}}^{X_{f}}\frac{\nabla _{X}R\left(
K_{X},X\right) }{\sigma _{X}^{2}}H\left( K_{X}\right) -t\left( D\left(
\left\Vert \Psi \right\Vert ^{2}\right) +\tau \left( \frac{\left\vert \Psi
\left( X_{f}\right) \right\vert ^{2}\left( K_{X_{f}}-K_{f}\right) }{K_{f}}+%
\frac{\left\vert \Psi \left( X_{i}\right) \right\vert ^{2}\left(
K_{X_{i}}-K_{i}\right) }{K_{i}}\right) \right) \right) \\
&&\times \exp \left( -\int^{K_{f}}\left( K-\hat{F}_{2}\left( R\left(
s,K,X_{f}\right) \right) K_{X_{f}}\right) dK+\int^{K_{i}}\left( K-\hat{F}%
_{2}\left( s,R\left( K,X_{i}\right) \right) K_{X_{i}}\right) dK\right) \\
&&\times \exp \left( -\left( \frac{\left( X_{f}-X_{i}\right) ^{2}}{2t\sigma
_{X}^{2}}+\frac{\left( \tilde{K}_{f}-\tilde{K}_{i}\right) ^{2}}{2t\sigma
_{K}^{2}}\right) \right) \\
&&\times \exp \left( -\frac{t\sigma _{K}^{2}}{2}\left( K_{f}-\hat{F}%
_{2}\left( s,R\left( K_{f},\bar{X}\right) \right) K_{\bar{X}}\right) \left(
K_{i}-\hat{F}_{2}\left( s,R\left( K_{i},\bar{X}\right) \right) K_{\bar{X}%
}\right) \right)
\end{eqnarray*}%
\begin{eqnarray}
&&T_{1}\left( \left( K_{f},X_{f}\right) ,\left( X_{i},K_{i}\right) \right) \\
&\simeq &\exp \left( \int_{X_{i}}^{X_{f}}\frac{\nabla _{X}R\left(
K_{X},X\right) }{\sigma _{X}^{2}}H\left( K_{X}\right) -t\left( D\left(
\left\Vert \Psi \right\Vert ^{2}\right) +\tau \left( \frac{\left\vert \Psi
\left( X_{f}\right) \right\vert ^{2}\left( K_{X_{f}}-K_{f}\right) }{K_{f}}+%
\frac{\left\vert \Psi \left( X_{i}\right) \right\vert ^{2}\left(
K_{X_{i}}-K_{i}\right) }{K_{i}}\right) \right) \right)  \notag \\
&&\times \exp \left( -\int^{K_{f}}\left( K-\hat{F}_{2}\left( s,R\left(
K,X_{f}\right) \right) K_{X_{f}}\right) dK+\int^{K_{i}}\left( K-\hat{F}%
_{2}\left( s,R\left( K,X_{i}\right) \right) K_{X_{i}}\right) dK\right) 
\notag \\
&&\times \exp \left( -\left( \frac{\left( X_{f}-X_{i}\right) ^{2}}{2t\sigma
_{X}^{2}}+\frac{\left( K_{f}^{\prime }-K_{i}^{\prime }\right) ^{2}}{2t\sigma
_{K}^{2}}-\frac{t\sigma _{K}^{2}}{2}K_{f}^{\prime }K_{f}^{\prime }\right)
\right)  \notag
\end{eqnarray}%
with:%
\begin{eqnarray*}
K_{i}^{\prime } &=&K_{i}-\hat{F}_{2}\left( R\left( K_{X_{i}},X_{i}\right)
\right) K_{X_{i}} \\
K_{f}^{\prime } &=&K_{f}-\hat{F}_{2}\left( R\left( K_{X_{f}},X_{f}\right)
\right) K_{X_{f}}
\end{eqnarray*}%
The Laplace transform of this function is the transition function given in
the text.

\subsection*{A8.3 Computation of $T_{2}$}

\subsubsection{A8.3.1 Solution ot (\protect\ref{TRV})}

Solving (\ref{TRV}) is straightforward, and similar to the derivation $T_{1} 
$.

We first introduce a change of variable:%
\begin{eqnarray*}
&&T_{2}\left( \left( \hat{K}_{f},\hat{X}_{f}\right) ,\left( \hat{X}_{i},\hat{%
K}_{i}\right) \right) \\
&=&\exp \left( -t\int_{\hat{X}_{i}}^{\hat{X}_{f}}\frac{\left( g\left( \hat{X}%
\right) \right) ^{2}+\sigma _{\hat{X}}^{2}\left( f\left( \hat{X}\right)
+\nabla _{\hat{X}}g\left( \hat{X},K_{\hat{X}}\right) -\frac{\sigma _{\hat{K}%
}^{2}F^{2}\left( \hat{X},K_{\hat{X}}\right) }{2f^{2}\left( \hat{X}\right) }%
\right) }{\left\Vert \hat{X}_{f}-\hat{X}_{i}\right\Vert \sigma _{\hat{X}}^{2}%
\sqrt{f^{2}\left( \hat{X}\right) }}\right) \hat{T}_{2}\left( \left( \hat{K}%
_{f},\hat{X}_{f}\right) ,\left( \hat{X}_{i},\hat{K}_{i}\right) \right)
\end{eqnarray*}%
The term in the exponential is the average of the relative return:%
\begin{equation*}
\frac{\left( g\left( \hat{X}\right) \right) ^{2}+\sigma _{\hat{X}}^{2}\left(
f\left( \hat{X}\right) +\nabla _{\hat{X}}g\left( \hat{X},K_{\hat{X}}\right) -%
\frac{\sigma _{\hat{K}}^{2}F^{2}\left( \hat{X},K_{\hat{X}}\right) }{%
2f^{2}\left( \hat{X}\right) }\right) }{\sigma _{\hat{X}}^{2}\sqrt{%
f^{2}\left( \hat{X}\right) }}
\end{equation*}%
along the average path, considered as a straight line, from $\hat{X}_{i}$ to 
$\hat{X}_{f}$. We have assumed slow shifts in the sectors space, so that $%
\hat{T}_{2}$ satisfies the following equation in first approximation: 
\begin{eqnarray}
&&-\frac{\partial }{\partial t}\hat{T}_{2}\left( \left( \hat{K}_{f},\hat{X}%
_{f}\right) ,\left( \hat{X}_{i},\hat{K}_{i}\right) \right)  \label{TRZ} \\
&\simeq &\left( -\frac{\sigma _{\hat{X}}^{2}}{2}\nabla _{\hat{X}}^{2}-\nabla
_{y}^{2}\right) \hat{T}_{2}\left( \left( \hat{K}_{f},\hat{X}_{f}\right)
,\left( \hat{X}_{i},\hat{K}_{i}\right) \right) +\frac{y^{2}}{4}\hat{T}%
_{2}\left( \left( \hat{K}_{f},\hat{X}_{f}\right) ,\left( \hat{X}_{i},\hat{K}%
_{i}\right) \right)  \notag
\end{eqnarray}%
Given (\ref{cvr}), we can assume that $y$ is independent from $\hat{X}$ in
first approximation. Thus, solving (\ref{TRZ}) yields:

\begin{eqnarray}
&&\hat{T}_{2}\left( \left( \hat{K}_{f},\hat{X}_{f}\right) ,\left( \hat{X}%
_{i},\hat{K}_{i}\right) \right)  \label{PRF} \\
&\simeq &\exp \left( -\left( \frac{\sigma _{\hat{X}}^{2}}{2}t\left( \hat{K}%
_{f}+\frac{\sigma _{\hat{K}}^{2}F\left( \hat{X}_{f},K_{\hat{X}_{f}}\right) }{%
f^{2}\left( \hat{X}_{f}\right) }\right) \left( \hat{K}_{i}+\frac{\sigma _{%
\hat{K}}^{2}F\left( \hat{X}_{i},K_{\hat{X}_{i}}\right) }{f^{2}\left( \hat{X}%
_{i}\right) }\right) \right) \right)  \notag \\
&&\times \exp \left( -\frac{\sqrt{f^{2}\left( \frac{\hat{X}_{f}+\hat{X}_{i}}{%
2}\right) }}{2t\sigma _{\hat{X}}^{2}}\left( \left( \hat{K}_{f}+\frac{\sigma
_{\hat{K}}^{2}F\left( \hat{X}_{f},K_{\hat{X}_{f}}\right) }{f^{2}\left( \hat{X%
}_{f}\right) }\right) -\left( \hat{K}_{i}+\frac{\sigma _{\hat{K}}^{2}F\left( 
\hat{X}_{i},K_{\hat{X}_{i}}\right) }{f^{2}\left( \hat{X}_{i}\right) }\right)
\right) ^{2}\right)  \notag
\end{eqnarray}

\subsubsection{A8.3.2 Full transition function}

Reintroducing the change of variables (\ref{chg}) amounts to introduce a
factor:%
\begin{equation*}
\exp \left( \frac{1}{\sigma _{\hat{X}}^{2}}\int_{\hat{X}_{i}}^{\hat{X}%
_{f}}g\left( \hat{X}\right) d\hat{X}+\frac{\hat{K}_{f}^{2}}{\sigma _{\hat{K}%
}^{2}}f\left( \hat{X}_{f}\right) -\frac{\hat{K}_{i}^{2}}{\sigma _{\hat{K}%
}^{2}}f\left( \hat{X}_{i}\right) \right)
\end{equation*}%
in the formula for $T_{2}$\ and this leads to the full formula for the
transition function:%
\begin{eqnarray}
&&T_{2}\left( \left( \hat{K}_{f},\hat{X}_{f}\right) ,\left( \hat{X}_{i},\hat{%
K}_{i}\right) \right)  \label{TRC} \\
&\simeq &\exp \left( -t\int_{\hat{X}_{i}}^{\hat{X}_{f}}\frac{\left( g\left( 
\hat{X}\right) \right) ^{2}+\sigma _{\hat{X}}^{2}\left( f\left( \hat{X}%
\right) +\nabla _{\hat{X}}g\left( \hat{X},K_{\hat{X}}\right) -\frac{\sigma _{%
\hat{K}}^{2}F^{2}\left( \hat{X},K_{\hat{X}}\right) }{2f^{2}\left( \hat{X}%
\right) }\right) }{\left\Vert \hat{X}_{f}-\hat{X}_{i}\right\Vert \sigma _{%
\hat{X}}^{2}\sqrt{f^{2}\left( \hat{X}\right) }}\right)  \notag \\
&&\times \exp \left( \frac{1}{\sigma _{\hat{X}}^{2}}\int_{\hat{X}_{i}}^{\hat{%
X}_{f}}g\left( \hat{X}\right) d\hat{X}+\frac{\hat{K}_{f}^{2}}{\sigma _{\hat{K%
}}^{2}}f\left( \hat{X}_{f}\right) -\frac{\hat{K}_{i}^{2}}{\sigma _{\hat{K}%
}^{2}}f\left( \hat{X}_{i}\right) \right) \hat{T}_{2}\left( \left( \hat{K}%
_{f},\hat{X}_{f}\right) ,\left( \hat{X}_{i},\hat{K}_{i}\right) \right) 
\notag
\end{eqnarray}%
with:%
\begin{equation*}
\hat{T}_{2}\left( \left( \hat{K}_{f},\hat{X}_{f}\right) ,\left( \hat{X}_{i},%
\hat{K}_{i}\right) \right)
\end{equation*}%
given by (\ref{PRF}).

The Laplace transform of (\ref{TRC}) is the formula presented in the text.

\section*{Appendix 9}

We write the series expansion in $\Delta S_{\text{fourth order}}$ of $\exp
\left( -S\left( \Psi \right) \right) $: 
\begin{eqnarray*}
\exp \left( -S\left( \Psi \right) \right) &=&\exp \left( -\left( S\left(
\Psi _{0},\hat{\Psi}_{0}\right) +\int \left( \Delta \Psi ^{\dag }\left(
Z,\theta \right) ,\Delta \hat{\Psi}^{\dag }\left( Z,\theta \right) \right)
\left( Z,\theta \right) O\left( \Psi _{0}\left( Z,\theta \right) \right)
\left( 
\begin{array}{c}
\Delta \Psi \left( Z,\theta \right) \\ 
\Delta \hat{\Psi}\left( Z,\theta \right)%
\end{array}%
\right) \right) \right) \\
&&\left( 1+\sum_{n\geqslant 1}\frac{\left( -\Delta S_{\text{fourth order}%
}\left( \Psi ,\hat{\Psi}\right) \right) ^{n}}{n!}\right)
\end{eqnarray*}%
where $O\left( \Psi _{0}\left( Z,\theta \right) \right) $ is defined in (\ref%
{Dfr}).

Then, we decompose $\Delta S_{\text{fourth order}}\left( \Psi ,\hat{\Psi}%
\right) $ as a sum of two combinations:%
\begin{eqnarray*}
\Delta S_{\text{fourth order}}\left( \Psi ,\hat{\Psi}\right) &=&\int \Delta
\Psi ^{\dag }\left( K,X\right) \Delta \Psi ^{\dag }\left( K^{\prime
},X^{\prime }\right) \Delta S_{11}\Delta \Psi \left( K^{\prime },X^{\prime
}\right) \Delta \Psi \left( K,X\right) \\
&&+\Delta \Psi ^{\dagger }\left( K^{\prime },X^{\prime }\right) \Delta \hat{%
\Psi}^{\dagger }\left( \hat{K},\hat{X}\right) \Delta S_{12}\Delta \Psi
\left( K^{\prime },X^{\prime }\right) \Delta \hat{\Psi}\left( \hat{K},\hat{X}%
\right)
\end{eqnarray*}%
with:%
\begin{equation}
\Delta S_{11}=\left( 2\tau -\nabla _{K}\frac{\delta ^{2}u\left( K,X,\Psi ,%
\hat{\Psi}\right) }{\delta \Psi \left( K^{\prime },X\right) \delta \Psi
^{\dagger }\left( K^{\prime },X\right) }\right) \delta \left( X-X^{\prime
}\right)  \notag
\end{equation}%
and:%
\begin{equation}
\Delta S_{12}=-\left( \nabla _{K}\frac{\delta ^{2}u\left( K,X,\Psi ,\hat{\Psi%
}\right) }{\delta \hat{\Psi}\left( \hat{K},\hat{X}\right) \delta \hat{\Psi}%
^{\dagger }\left( \hat{K},\hat{X}\right) }+\left\{ \nabla _{\hat{K}}\frac{%
\hat{K}\delta ^{2}f\left( \hat{X},\Psi ,\hat{\Psi}\right) }{\delta \Psi
\left( K^{\prime },X\right) \delta \Psi ^{\dagger }\left( K^{\prime
},X\right) }+\nabla _{\hat{X}}\frac{\delta ^{2}g\left( \hat{X},\Psi ,\hat{%
\Psi}\right) }{\delta \Psi \left( K^{\prime },X\right) \delta \Psi ^{\dagger
}\left( K^{\prime },X\right) }\right\} \right) \delta \left( X-X^{\prime
}\right)  \notag
\end{equation}%
Application of (\ref{trg}) leads to the following form of the transition
functions:%
\begin{eqnarray}
&&G_{ij}\left( \left[ \left( K_{f},X_{f}\right) ,\left( K_{f},X_{f}\right)
^{\prime }\right] ,\left[ \left( X_{i},K_{i}\right) ,\left(
X_{i},K_{i}\right) ^{\prime }\right] \right)  \label{frg} \\
&=&G_{i}\left( \left( K_{f},X_{f}\right) ,\left( X_{i},K_{i}\right) \right)
G_{j}\left( \left( K_{f},X_{f}\right) ^{\prime },\left( X_{i},K_{i}\right)
^{\prime }\right)  \notag \\
&&+\sum_{p\geqslant 1}\frac{\left( -1\right) ^{p}}{p!}\int G_{i}\left(
\left( K_{f},X_{f}\right) ,\left( X_{p},K_{p}\right) \right) G_{j}\left(
\left( K_{f},X_{f}\right) ^{\prime },\left( X_{p},K_{p}\right) ^{\prime
}\right) \Delta S_{ij}\left( \left( X_{p},K_{p}\right) ,\left(
X_{p},K_{p}\right) ^{\prime }\right)  \notag \\
&&\times G_{i}\left( \left( X_{p},K_{p}\right) ,\left(
X_{p-1},K_{p-1}\right) \right) G_{j}\left( \left( X_{p},K_{p}\right)
^{\prime },\left( X_{p-1},K_{p-1}\right) ^{\prime }\right) \Delta
S_{ij}\left( \left( X_{p-1},K_{p-1}\right) ,\left( X_{p-1},K_{p-1}\right)
^{\prime }\right)  \notag \\
&&...\times \Delta S_{ij}\left( \left( K_{1},X_{1}\right) ,\left(
K_{1},X_{1}\right) ^{\prime }\right) G_{1}\left( \left( K_{1},X_{1}\right)
,\left( X_{i},K_{i}\right) \right) G_{1}\left( \left( K_{1},X_{1}\right)
^{\prime },\left( X_{i},K_{i}\right) ^{\prime }\right)
\prod\limits_{k\leqslant p}d\left( \left( X_{k},K_{k}\right) ,\left(
X_{k},K_{k}\right) ^{\prime }\right)  \notag
\end{eqnarray}

These corrections modify the $n$ agents Green functions and can be computed
using graphs expansion. In the sequel we will focus only on the first order
corrections to the four agents Green functions. This is sufficient to stress
the impact of interactions of agents in the background state.

The term $\Delta S_{11}$ measures the interaction between firms, and $\Delta
S_{12}$ the firms-investors interactions. There is no term $\Delta S_{22}$
of investors-investors interaction. In our model all interactions depend on
firms.

To estimate the impact of interactions, we can assume the paths from $\left(
\left( X_{i},K_{i}\right) ,\left( K_{f},X_{f}\right) \right) $ to $\left(
\left( X_{i},K_{i}\right) ,\left( K_{f},X_{f}\right) ^{\prime }\right) $
cross each other one time at some $X$ and approximate the terms $\Delta
S_{ij}$ by their average value estimated on the average paths from $%
K_{i},K_{i}^{\prime }$ to $K_{f},K_{f}^{\prime }$,

In this approximation, we find:%
\begin{eqnarray*}
&&G_{ij}\left( \left[ \left( K_{f},X_{f}\right) ,\left( K_{f},X_{f}\right)
^{\prime }\right] ,\left[ \left( X_{i},K_{i}\right) ,\left(
X_{i},K_{i}\right) ^{\prime }\right] \right) \\
&\simeq &G_{i}\left( \left( K_{f},X_{f}\right) ,\left( X_{i},K_{i}\right)
\right) G_{j}\left( \left( K_{f},X_{f}\right) ^{\prime },\left(
X_{i},K_{i}\right) ^{\prime }\right) \\
&&-G_{i}\left( \left( K_{f},X_{f}\right) ,\left( X,K\right) \right)
G_{j}\left( \left( K_{f},X_{f}\right) ^{\prime },\left( X,K\right) ^{\prime
}\right) \\
&&\times \Delta S_{ij}\left( \left( X,\bar{K}\right) ,\left( X,\bar{K}%
\right) ^{\prime }\right) G_{1}\left( \left( X,K\right) ,\left(
X_{i},K_{i}\right) \right) G_{1}\left( \left( X,K\right) ^{\prime },\left(
X_{i},K_{i}\right) ^{\prime }\right) \\
&\simeq &G_{i}\left( \left( K_{f},X_{f}\right) ,\left( X_{i},K_{i}\right)
\right) G_{j}\left( \left( K_{f},X_{f}\right) ^{\prime },\left(
X_{i},K_{i}\right) ^{\prime }\right) \\
&&-\Delta S_{ij}\left( \left( \bar{X},\bar{K}\right) ,\left( \bar{X},\bar{K}%
\right) ^{\prime }\right) \hat{G}_{i}\left( \left( K_{f},X_{f}\right)
,\left( X,K\right) \right) \hat{G}_{j}\left( \left( K_{f},X_{f}\right)
^{\prime },\left( X,K\right) ^{\prime }\right)
\end{eqnarray*}%
with:%
\begin{eqnarray*}
\left( \bar{X},\bar{K}\right) &=&\frac{\left( K_{f},X_{f}\right) +\left(
X_{i},K_{i}\right) }{2} \\
\left( \bar{X},\bar{K}\right) ^{\prime } &=&\frac{\left( K_{f},X_{f}\right)
^{\prime }+\left( X_{i},K_{i}\right) ^{\prime }}{2}
\end{eqnarray*}%
and:%
\begin{equation*}
\hat{G}_{i}\left( \left( K_{f},X_{f}\right) ,\left( X,K\right) \right) \hat{G%
}_{j}\left( \left( K_{f},X_{f}\right) ^{\prime },\left( X,K\right) ^{\prime
}\right)
\end{equation*}%
is the transition function computed on path that cross once. Applied to the
three transition functions for two agents yields the results of the text.

\section*{Appendix 10}

\section*{One agent transition functions}

\subsection*{A10.1 Firms transition function}

We interpret the various term involved in (\ref{Gn}) and their influence on
firms individual dynamics.

\subsubsection*{A10.1.1 Drift term}

\paragraph*{The three contributions}

The first term in (\ref{Gn}):%
\begin{equation*}
D\left( \left( K_{f},X_{f}\right) ,\left( X_{i},K_{i}\right) \right)
\end{equation*}%
is a drift term between $\left( X_{i},K_{i}\right) $ and $\left(
K_{f},X_{f}\right) $. It is composed of three contributions (see (\ref{frd}%
)):

The first term of (\ref{frd}):

\begin{equation*}
\int_{X_{i}}^{X_{f}}\frac{\nabla _{X}R\left( K_{X},X\right) }{\sigma _{X}^{2}%
}H\left( K_{X}\right)
\end{equation*}%
models the shift of producers towards sectors that have the highest
long-term returns.

To interpret the second contribution to $D\left( \left( K_{f},X_{f}\right)
,\left( X_{i},K_{i}\right) \right) $:

\begin{equation}
\int_{K_{i}}^{K_{f}}\left( K-\hat{F}_{2}\left( s,R\left( K,\bar{X}\right)
\right) K_{\bar{X}}\right) dK  \label{scn}
\end{equation}%
, recall that $\frac{F_{2}\left( R\left( K,\bar{X}\right) \right) }{%
F_{2}\left( R\left( K_{\bar{X}},\bar{X}\right) \right) }$ models the
relative expectations of returns of the firm along its path from $\left(
X_{i},K_{i}\right) $ to $\left( K_{f},X_{f}\right) $ based on their returns'
expectation $R\left( K,\bar{X}\right) $ and that $\frac{F_{2}\left( R\left(
K,\bar{X}\right) \right) K_{\bar{X}}}{F_{2}\left( R\left( K_{\bar{X}},\bar{X}%
\right) \right) }$ represents the capital investors are ready to invest in
the firm along this path. Along the path from $\left( X_{i},K_{i}\right) $
to $\left( K_{f},X_{f}\right) $, the capital invested in the firm will
increase as long as the investors expect growth and as long as additional
investment is likely to increase the firm's returns. Once the level of
capital reaches their expectations, that is 
\begin{equation}
K_{T}\left( \bar{X}\right) -\hat{F}_{2}\left( R\left( K_{T},\bar{X}\right)
\right) K_{\bar{X}}=0  \label{hl}
\end{equation}%
i.e., when the firm has reached the capital threshold, investment stops.

However, this condition is not always fulfilled. The shape of $F_{2}$ is
critical. If $F_{2}$ is above the line $Y=K$, then for $K<K_{T}$, the
threshold $K_{T}$ will be reached gradually. In this case, $K_{T}$ is an
equilibrium point. If, on the contrary, $F_{2}$ is below the line $Y=K$,
then for $K<K_{T}$, the threshold $K_{T}$ will never be reached, and $%
K\rightarrow 0$. If $K>K_{T}$, $K$ can increase indefinitely. This
corresponds to firms whose profitability is perceived as boundless as long
as more capital is invested in.

The third term in (\ref{frd}):

\begin{equation*}
\int_{K_{i}}^{K_{f}}\left( \left( \frac{X_{f}-X_{i}}{2}\right) \nabla _{X}%
\hat{F}_{2}\left( s,R\left( K,\bar{X}\right) \right) K_{\bar{X}}\right) dK
\end{equation*}%
induces firms to move towards more appropriate sectors, according to
investors and given the capital of the firm. The firm does not solely move
according to the new investors it could attract but must also take into
account its current investors. If it moves, it risks losing the investors it
has already attracted.

\paragraph*{Trade-off between terms}

There is a trade-off between the first and the third terms: firms want to
move towards sectors with higher returns, but differences in average capital
between sectors could make a firm unattractive in a new sector. The loss of
investors \textbf{incured }during a shift of sector must be compared with
the number of investors possibly attracted in the new sector: the level of
attractiveness may decrease for a given amount of capital.

The second contribution to $D$ is an indicator of the firm's growth
potential in a given sector.\ It depends on the firm's level of capital
compared to the threshold capital requirement and its dynamics in this
sector.

A move along sectors due to the terms 1 and 3 modifies the firm's relative
capital, which is sector-dependant: $F_{2}$, measuring the firm
attractiveness in the sector and indirectly the threshold of capital $K_{T}$
defined in (\ref{hl}) are modified by the shift from one sector to another.
Therefore, a firm could be below the value of $K_{T}$ in one sector, then
above in the next sector, which will reverse its capital dynamics. The
firm's capital dynamics remains the same as long as its relative
attractiveness in a sector does not change significantly.

\subsubsection*{A10.1.2 Effective time of transition}

The following term depends on the competition in a sector:%
\begin{eqnarray}
\alpha _{eff}\left( \Psi ,\left( K_{f},X_{f}\right) ,\left(
X_{i},K_{i}\right) \right) &=&\alpha +D\left( \left\Vert \Psi \right\Vert
^{2}\right) +\frac{\tau }{2}\left( \frac{\left\vert \Psi \left( X_{f}\right)
\right\vert ^{2}\left( K_{X_{f}}-K_{f}\right) }{K_{f}}-\frac{\left\vert \Psi
\left( X_{i}\right) \right\vert ^{2}\left( K_{X_{i}}-K_{i}\right) }{K_{i}}%
\right)  \label{fct} \\
&&+\frac{\sigma _{K}^{2}}{2}\left( K_{f}-\hat{F}_{2}\left( s,R\left( K_{f},%
\bar{X}\right) \right) K_{\bar{X}}\right) \left( K_{i}-\hat{F}_{2}\left(
s,R\left( K_{i},\bar{X}\right) \right) K_{\bar{X}}\right)  \notag
\end{eqnarray}

Recall that the constant $\alpha $ is the inverse of the average lifetime of
the agents. The larger it is, the lower the probability of transition. In
the transition functions, $\alpha $ is shifted by two path-dependent terms
and replaced by $\alpha _{eff}$. Thus, $\alpha _{eff}$ is the inverse
mobility of the firm during its transition from one point to another.

Therefore, the likelihood of shifts in capital and sectors depends not only
on the average lifespan of the firm, but also on terms that are directly
related to the collective state.

The first correction to $\alpha $ is $D\left( \left\Vert \Psi \right\Vert
^{2}\right) $ that is related to competition. We have shown in (Gosselin
Lotz Wambst 2022):

\begin{equation*}
D\left( \left\Vert \Psi \right\Vert ^{2}\right) \simeq 2\tau \frac{N}{V-V_{0}%
}+\frac{1}{2\sigma _{X}^{2}}\left\langle \left( \nabla _{X}R\left( X\right)
\right) ^{2}\right\rangle _{V/V_{0}}H^{2}\left( \frac{\left\langle \hat{K}%
\right\rangle }{N}\right) \left( 1-\frac{H^{\prime }\left( \frac{%
\left\langle \hat{K}\right\rangle }{N}\right) }{H\left( \frac{\left\langle 
\hat{K}\right\rangle }{N}\right) }\frac{\left\langle \hat{K}\right\rangle }{N%
}\right)
\end{equation*}%
where $V$ is the volume of the sectors space and $V_{0}$ is the locus where $%
\left\Vert \Psi \left( X\right) \right\Vert ^{2}=0$. As a consequence, the
stronger the competition, i.e., the larger $\tau $, the greater $D\left(
\left\Vert \Psi \right\Vert ^{2}\right) $, and the less possibilities of
shifting from a sector to another.

The third term in (\ref{fct}):%
\begin{equation}
\frac{\tau }{2}\left( \frac{\left\vert \Psi \left( X_{f}\right) \right\vert
^{2}\left( K_{X_{f}}-K_{f}\right) }{K_{f}}-\frac{\left\vert \Psi \left(
X_{i}\right) \right\vert ^{2}\left( K_{X_{i}}-K_{i}\right) }{K_{i}}\right)
\label{stv}
\end{equation}%
is also linked to the competition, but depends on the level of capital of
the agent, and the number of d agents in the sectors crossed. This term
measures the strength of agent's shift from one sector to another.

It is negative when $K_{f}>K_{X_{f}}$, and when $K_{i}<K_{X_{i}}$. In other
words, when a firm has less capital than the average in its initial sector,
and when, it ends up in a sector in which it has more capital than the
average, the probability of transition from $K_{i}$ to $K_{f}$ is high. In
other words, under-average capital favors the exit from a sector.
Above-average capital promotes entry into the sector. Shifts from high
average capital sectors to lower-average-capital sectors are favoured.

This phenomenon is amplified by the number of agents. The greater the
competition in a sector, i.e., the more firms in the sector, the greater the
probability for a lower-than average capitalized firm to be ousted from the
sector by higher-than average capitalized firms that enter the sector.
Thus,\ the density of producers $\left\vert \Psi \left( X\right) \right\vert
^{2}$ along the movement enhances competition and favours high capitalized
firm to move towards higher capitalized sectors, and drives low capitalized
firms toward low capitalized sectors.

The last term in (\ref{fct}):%
\begin{equation*}
\frac{\sigma _{K}^{2}}{2}\left( K_{f}-\hat{F}_{2}\left( s,R\left( K_{f},\bar{%
X}\right) \right) K_{\bar{X}}\right) \left( K_{i}-\hat{F}_{2}\left(
s,R\left( K_{i},\bar{X}\right) \right) K_{\bar{X}}\right)
\end{equation*}%
shows that in average, shifts from the initial point to the final point is
done respecting $K=\hat{F}_{2}\left( s,R\left( K,\bar{X}\right) \right) K_{%
\bar{X}}$, the capital investors allocate to the firm. Actually, if $K_{i}-%
\hat{F}_{2}\left( s,R\left( K_{i},\bar{X}\right) \right) K_{\bar{X}}<0$,
there is a higher probability to reach $K_{f}-\hat{F}_{2}\left( s,R\left(
K_{f},\bar{X}\right) \right) K_{\bar{X}}>0$. If $K_{i}-\hat{F}_{2}\left(
s,R\left( K_{i},\bar{X}\right) \right) K_{\bar{X}}>0$, there is a higher
probability to reach $K_{f}-\hat{F}_{2}\left( s,R\left( K_{f},\bar{X}\right)
\right) K_{\bar{X}}<0$.

If a firm starts with a capital lower than $\hat{F}_{2}\left( s,R\left( K,%
\bar{X}\right) \right) K_{\bar{X}}$, it is more likely to end up with
capital above the new $\hat{F}_{2}\left( s,R\left( K,\bar{X}\right) \right)
K_{\bar{X}}$, in another sector. The movement induces a change in the $F_{2}$%
, but firms tend to fluctuate around the $F_{2}$ of transition.

\subsubsection*{A10.1.3 Fluctuation terms}

The term:%
\begin{equation*}
\sqrt{\frac{\left( X_{f}-X_{i}\right) ^{2}}{2\sigma _{X}^{2}}+\frac{\left( 
\tilde{K}_{f}-\tilde{K}_{i}\right) ^{2}}{2\sigma _{K}^{2}}}
\end{equation*}%
describes oscillations around:%
\begin{equation*}
K_{f}-\hat{F}_{2}\left( s,R\left( K_{f},\bar{X}\right) \right) K_{\bar{X}%
}+K_{i}-\hat{F}_{2}\left( s,R\left( K_{i},\bar{X}\right) \right) K_{\bar{X}%
}=0
\end{equation*}%
On average, during a transition from the initial point to the final point,
the firm's capital is governed by the equation $K=K_{T}=\hat{F}_{2}\left(
R\left( K_{i},\bar{X}\right) \right) K_{\bar{X}}$. If a firm starts with
less capital than the threshold set by $F_{2}$, it is more likely to end up
with a capital above the new $F_{2}$, in another sector. The transition from
one sector to another involves oscillations around the sector-dependent
threshold $F_{2}$. However, these oscillations may affect the final
destination of the transition. Starting with a capital level above $K_{T}$,
i.e. $K>K_{T}$, the firm may shift towards sectors with higher perspectives,
i.e. with a higher threshold $K_{T}$. This favors in turn accumulation and
faster transition to other sectors. Finally, if $\nabla _{K}\hat{F}%
_{2}\left( s,R\left( \frac{K_{f}+K_{i}}{2},\bar{X}\right) \right) >0$, $%
F_{2} $ is highly responsive to changes in capital, and larger moves are
favored.

\subsection*{A10.2 Investors transition functions}

We interpret the different contributions in (\ref{Gt}).

\subsubsection*{A10.2.1 Drift term}

The term $D^{\prime }\left( \left( \hat{K}_{f},\hat{X}_{f}\right) ,\left( 
\hat{X}_{i},\hat{K}_{i}\right) \right) $ is composed of two contributions.
The first one:%
\begin{equation*}
\frac{1}{\sigma _{\hat{X}}^{2}}\int_{\hat{X}_{i}}^{\hat{X}_{f}}g\left( \hat{X%
}\right) d\hat{X}+\frac{\hat{K}_{f}^{2}}{\sigma _{\hat{K}}^{2}}f\left( \hat{X%
}_{f}\right) -\frac{\hat{K}_{i}^{2}}{\sigma _{\hat{K}}^{2}}f\left( \hat{X}%
_{i}\right)
\end{equation*}%
is composed of two elements.

The first element is $\frac{1}{\sigma _{\hat{X}}^{2}}\int_{\hat{X}_{i}}^{%
\hat{X}_{f}}g\left( \hat{X}\right) d\hat{X}$. Since the function $g\left( 
\hat{X}\right) $ is the anticipation of higher returns and rising stock
prices, investors move towards sectors were they anticipate the highest
returns and stock prices increase.

The second element $\frac{\hat{K}_{f}^{2}}{\sigma _{\hat{K}}^{2}}f\left( 
\hat{X}_{f}\right) -\frac{\hat{K}_{i}^{2}}{\sigma _{\hat{K}}^{2}}f\left( 
\hat{X}_{i}\right) $: this element can be rewritten as two bits:

\begin{equation*}
\frac{\hat{K}_{f}^{2}-\hat{K}_{i}^{2}}{\sigma _{\hat{K}}^{2}}f\left( \frac{%
\hat{X}_{f}+\hat{X}_{i}}{2}\right) +\frac{\left( \hat{K}_{f}^{2}+\hat{K}%
_{i}^{2}\right) }{2\sigma _{\hat{K}}^{2}}\left( \hat{X}_{f}-\hat{X}%
_{i}\right) \nabla _{X}f\left( \frac{\hat{X}_{f}+\hat{X}_{i}}{2}\right)
\end{equation*}%
The first bit, $\frac{\hat{K}_{f}^{2}-\hat{K}_{i}^{2}}{\sigma _{\hat{K}}^{2}}%
f\left( \frac{\hat{X}_{f}+\hat{X}_{i}}{2}\right) $ shows that the highest
the short-term return, the more probable is the increase in capital. The
second bit, which is equal to $\frac{\left( \hat{K}_{f}^{2}+\hat{K}%
_{i}^{2}\right) }{2\sigma _{\hat{K}}^{2}}\left( \hat{X}_{f}-\hat{X}%
_{i}\right) \nabla _{X}f\left( \frac{\hat{X}_{f}+\hat{X}_{i}}{2}\right) $\
indicates that investors move towards sectors with highest returns.\ The
more capital they have, the fastest the shift.

The second term arising in $D^{\prime }\left( \left( \hat{K}_{f},\hat{X}%
_{f}\right) ,\left( \hat{X}_{i},\hat{K}_{i}\right) \right) $:

\begin{equation*}
-\frac{1}{\sigma _{\hat{X}}^{2}}\int_{\hat{X}_{i}}^{\hat{X}_{f}}\frac{\left(
g\left( \hat{X}\right) \right) ^{2}+\sigma _{\hat{X}}^{2}\left( f\left( \hat{%
X}\right) +\nabla _{\hat{X}}g\left( \hat{X},K_{\hat{X}}\right) -\frac{\sigma
_{\hat{K}}^{2}F^{2}\left( \hat{X},K_{\hat{X}}\right) }{2f^{2}\left( \hat{X}%
\right) }\right) }{\sigma _{\hat{X}}^{2}\sqrt{f^{2}\left( \hat{X}\right) }}d%
\hat{X}
\end{equation*}%
is similar to the determinant of capital accumulation in a collective state
and has the same interpretation. There is a tradeoff between long-term and
short-term returns. It further shows the importance of relative long-term
return. Investors move towards relative long-term returns. Mathematically,
we can measure the dependence of agents' capital accumulation in neighboring
sectors using the integrand: 
\begin{equation}
p=\frac{-\left( \frac{\left( g\left( \hat{X},K_{\hat{X}_{M}}\right) \right)
^{2}}{\sigma _{\hat{X}}^{2}}+\nabla _{\hat{X}}g\left( \hat{X},K_{\hat{X}%
_{M}}\right) -\frac{\sigma _{\hat{K}}^{2}F^{2}\left( \hat{X},K_{\hat{X}%
}\right) }{2f^{2}\left( \hat{X}\right) }\right) }{f\left( \hat{X}\right) }
\label{nPR}
\end{equation}%
The function $p$ represents the relative attractivity of a sector vis-a-vis
its neighbours and depends on the gradients of long-term returns $R\left( 
\hat{X}\right) $ through the function $g\left( \hat{X}\right) $, the capital
mobility at sector $\hat{X}$.\ \ This function $g\left( \hat{X}\right) $,
which depicts investors' propensity to seek higher returns across
sectors,and is indeed proportional to $\nabla _{\hat{X}}R\left( \hat{X}%
\right) $. The gradient of $g$, $\nabla _{\hat{X}}g$, is proportional to $%
\nabla _{\hat{X}}^{2}R\left( \hat{X}\right) $: it measures the position of
the sector relative to its neighbours.\ At a local maximum, the second
derivative of $R\left( \hat{X}\right) $ is negative: $\nabla _{\hat{X}%
}^{2}R\left( \hat{X}\right) <0$. At a minimum, it is positive.

The last term, $\frac{\sigma _{\hat{K}}^{2}F^{2}\left( \hat{X},K_{\hat{X}%
}\right) }{2f^{2}\left( \hat{X}\right) }$, involved in the definition of $Y(%
\hat{X})$ and $p$ is a smoothing factor between neighbouring sectors. It
reduces differences between sectors: it increases when the relative
attractivity with respect to $K_{\hat{X}}$ decreases. The number of
investors and capital will increase in sectors\ that are in the
neighbourhood of significantly more attractive sectors, i.e. with higher
average capital and number of investors. It slows down the transitions.

\subsubsection*{A10.2.2 Fluctuation terms}

The last term involved in (\ref{Gt}):

\begin{equation*}
\alpha _{eff}^{\prime }\left( \left( \hat{K}_{f},\hat{X}_{f}\right) ,\left( 
\hat{X}_{i},\hat{K}_{i}\right) \right) \sqrt{\frac{\left\vert f\left( \frac{%
\hat{X}_{f}+\hat{X}_{i}}{2}\right) \right\vert }{2\sigma _{\hat{X}}^{2}}}%
\left\vert \left( \hat{K}_{f}+\frac{\sigma _{\hat{K}}^{2}F\left( \hat{X}%
_{f},K_{\hat{X}_{f}}\right) }{f^{2}\left( \hat{X}_{f}\right) }\right)
-\left( \hat{K}_{i}+\frac{\sigma _{\hat{K}}^{2}F\left( \hat{X}_{i},K_{\hat{X}%
_{i}}\right) }{f^{2}\left( \hat{X}_{i}\right) }\right) \right\vert
\end{equation*}%
depicts the possible oscillations around averages (that) occur (within)in a
definite timespan, so(such) that the transition probability decrases with $%
\hat{K}_{f}-\hat{K}_{i}$ and $X_{f}-X_{i}$. However, this probability
decreases with short-term returns: the higher the returns, the lower the
incentive to switch from one sector to another.

\end{document}